\documentclass[a4paper,11pt,serif]{article}

\usepackage{latexsym,amssymb,amsmath,amsthm}
\usepackage{graphicx}
\usepackage{fancyhdr}
\usepackage{enumerate}
\usepackage{amsbsy, natbib}
\usepackage{amsfonts, bm}

\author{Dhananjoy Dey$^{(1)}$ Prasanna Raghaw Mishra$^{(1)}$ Indranath Sengupta$^{(2)}$\\\\$^{(1)}$SAG, Metcalfe House, Delhi-110 054, INDIA.\\$^{(2)}$School of Mathematical Sciences, RKM Vivekananda University, \\Belur Math, Howrah, WB 711 202, INDIA.\\(On Lien) Department of Mathematics, Jadavpur University, \\Kolkata, WB 700 032, INDIA.\\email: $\{$ddey06, prasanna.r.mishra, indranathsg$\}$ @gmail.com}

\title{HF-hash : Hash Functions Using Restricted HFE Challenge-1} 
\date{}

\def\Z{\mathbb Z}

\begin{document}

\maketitle
\begin{abstract}
Vulnerability of dedicated hash functions to various attacks has made the task of designing hash function much more challenging. This provides us a strong motivation to design a new cryptographic hash function viz. \textit{HF-hash}. This is a hash function, whose compression function is designed by using first $32$ polynomials of HFE Challenge-$1$ ~\cite{bib:cou} with $64$ variables by forcing remaining $16$ variables as zero. \textit{HF-hash} gives $256$ bits message digest and is as efficient as SHA-$256$. It is secure against the differential attack proposed by Chabaud and Joux in~\cite{bib:cj} as well as by Wang et. al. in~\cite{bib:wyy}  applied to SHA-$0$ and SHA-$1$. 

\end{abstract}

\maketitle

\section{Introduction}
The majority of dedicated hash functions published are more or less designed using ideas inspired by hash functions MD$4$~\cite{bib:riv} and MD$5$~\cite{bib:riv1}. Not only the hash functions HAVAL~\cite{bib:zps}, RIPEMD~\cite{bib:bp}, RIPEMD-$160$~\cite{bib:pbd} but also SHA-$0$~\cite{bib:nist},  SHA-$1$~\cite{bib:nist1} and SHA-$2$ family~\cite{bib:nist2} are designed using the similar ideas. The hash functions HAS-$160$~\cite{bib:tak} and HAS-V~\cite{bib:phl} both exhibit strong resemblance with SHA-$1$. 

While comparing compression functions of the aforementioned hash functions it is easy to observe that all of them have the three fundamental parts viz. \textit{the message expansion algorithm} which is required for creating more disturbance pattern for the input to the compression function, \textit{the iteration of the step transformation} which is required for taking arbitrary length of input and \textit{the state feed-forward operation} which is required for updating the chaining variables or the internal hash value. 

The most commonly used dedicated hash functions are MD$5$ and SHA-$1$. The first member of the MD family, viz. MD$4$ was published in $1990$. After one year, an attack on the last two out of three rounds has been presented in~\cite{bib:bb}. After that Rivest designed the improved version of MD$4$, called MD$5$. Later, Vaudenay showed that the first two rounds of MD$4$ are not collision-resistant and it is possible to get near-collisions for the full MD$4$~\cite{bib:vau}.

In $1993$, Boer and Bosselaers showed that it is possible to find pseudo-collisions for the compression function of MD$5$, i.e. they showed a way of finding two different values of the initial value $\mathcal{IV}$ for the same message $M$ such that MD$5$-compress($\mathcal{IV}, M$) = MD$5$-compress($\mathcal{IV}', M$) ~\cite{bib:bb1}. This was the first attack on MD$5$. This did not threaten the usual applications of MD$5$, since in normal situations one cannot control inputs of chaining variables. 

A major step forward in the analysis of MD-based designs was made by H. Dobbertin who developed a general method of attacking designs similar to MD$4$ in $1996$. His method aims at finding collisions and is based on describing the function as a system of complicated, non-linear equations that represent the function. With this method he successfully attacked MD$4$ showing that one can find collisions using computational effort of around $2^{20}$ hash evaluations~\cite{bib:dou}. He also showed collisions for the compression function of MD$5$ with a chosen $\mathcal{IV}$~\cite{bib:dou1}.

The other family of dedicated hash function is SHA family. The first version of the Secure Hash Algorithm (SHA) i.e. SHA-$0$ was presented by NIST in $1993$. Two years later, this function was slightly modified and an updated version of the standard was issued in $1995$. Indeed, in $1998$ Chabaud and Joux presented a differential attack on the initially proposed function, SHA-$0$, that can be used to find collisions with complexity of $2^{61}$ hash evaluations. Since SHA-$0$ and SHA-$1$ are different by a small change in the message expansion algorithms, it is quite natural question to ask whether it is possible to extend the original attack of Chabaud and Joux to the improved design of SHA-$1$. Due to the same round structure, the same technique used to attack SHA-$0$ could be applied to launch an attack on SHA-$1$ provided there exists a good enough differential pattern. Novel ideas of Wang et al. contributed a lot in opening new avenues of analysis of SHA-$1$. It seems the ability to influence the value of the new word of the state in each step combined with rather weak message expansion algorithms is the fundamental weakness of designs of that family that can be exploited that way or another.

In August $2002$, NIST announced a new standard FIPS $180$-$2$ that introduced three new cryptographic hash functions viz. SHA-$256$, SHA-$384$ and SHA-$512$. In $2004$ the specification was updated with one more hash, SHA-$224$. All these algorithms are very closely related. In fact SHA-$224$ is just SHA-$256$ with truncated hash and SHA-$384$ is a truncated version of SHA-$512$. These are called the SHA-$2$ family of hashes. The design of SHA-$512$ is very similar to SHA-$256$, but it uses $64$-bit words and some parameters are different to accommodate for this change. Clearly, the fundamental design of this family is SHA-$256$ and all the other algorithms are variations of that one, so the question of the security of SHA-$256$ is an extremely interesting one. 

We have designed a new hash function \textit{HF-hash} using the restricted version of HFE Challenge-$1$ as the compression function which gives $256$ bits message digest. We have used the first $32$ equations of HFE Challenge-$1$ with first $64$ variables by setting remaining $16$ variables to zero. Although the first proposal of designing hash function using quadratic or higher degree multivariate polynomials over a finite field as the compression function was given by Billet et. al.~\cite{bib:brp} as well as by Ding and Yang~\cite{bib:dy} in $2007$, they did not present how to design a secure hash function. In these papers they have used multivariate polynomials for both cases viz. message expansion as well as message compression. 

In this paper we present a complete description of \textit{HF-hash}, and its analysis in the subsequent sections. 

\section{\textit{HF-hash}}
\textit{HF-hash} function can take arbitrary length ($<2^{64}$) of input and gives $256$ bits output. We have designed an iterative hash function which uses restricted HFE Challenge-$1$~\cite{bib:cou} as compression function. The hash value of a message $M$ of length $l$ bits can be computed in the following manner:

\noindent
\textbf{Padding:} First we append $1$ to the end of the message $M$. Let $k$ be the number of zeros added for padding. The $64$-bit representation of $l$ is appended to the end of $k$ zeros. The padded message $M$ is shown in the following figure. Now $k$ will be the smallest positive integer satisfying the following condition:
\noindent
\begin{eqnarray*}
l+1+k+64&\equiv & 0 \;mod \;448\\
i.e.,\;k+l&\equiv & 383 \;mod \;448
\end{eqnarray*} 

\begin{center}
	\includegraphics[width=0.80\textwidth]{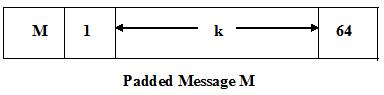}
\end{center}

\noindent
\textbf{Parsing:} Let $l'$ be the length of the padded message. Divide the padded message into $n (\;=\;\frac{l'}{448} )$ $448$-bit block i.e. $14$ $32$-bit words. Let $M^{(i)}$ denote the $i^{th}$ block of the padded message, where $1\le i\le n$ and each word of $i^{th}$ block is denoted by $M^{(i)}_j$ for $1\le j\le 14$.

\noindent
\textbf{Initial Value:} Take the first $256$ bits initial value i.e. $8$ $32$-bit words from the expansion of the fractional part of $\pi$ and hexadecimal value of these $8$ words are given below:

\begin{center}
	\begin{tabular}{lcl}
	$h^{(0)}_0$ &= &$243F6A88$\\
	$h^{(0)}_1$ &= &$85A308D3$\\
	$h^{(0)}_2$ &= &$13198A2E$\\
	$h^{(0)}_3$ &= &$03707344$\\
	$h^{(0)}_4$ &= &$A4093822$\\
	$h^{(0)}_5$ &= &$299F31D0$\\
	$h^{(0)}_6$ &= &$082EFA98$\\
	$h^{(0)}_7$ &= &$EC4E6C89$\\
	\end{tabular}
\end{center}

\noindent
\textbf{Hash Computation:} For each $448$-bit block $M^{(1)}$, $M^{(2)} \ldots$, $M^{(n)}$, the following four steps are executed for all the values of $i$ from $1$ to $n$.

\begin{enumerate}

	\item \textbf{Initialization}
	
	$H_0=h^{(i-1)}_0$, $H_1=h^{(i-1)}_1$, $H_2=h^{(i-1)}_2$, $H_3=h^{(i-1)}_3$, $H_4=h^{(i-1)}_4$, $H_5=h^{(i-1)}_5$, $H_6=h^{(i-1)}_6$ \& $H_7=h^{(i-1)}_7$.

	\item \textbf{Expansion}

\begin{enumerate}[i.]
	\item $W_0=H_0$
	\item $W_j=M^{(i)}_j$ for $1\le j\le 14$
	\item $W_{15}=H_7$
	\item $W_j=rotl_3(W_{j-16}\oplus W_{j-14}\oplus W_{j-8}\oplus W_{j-1})$  for $16\le j\le 63$,
	where $rotl_k$ denotes the left rotation by $k$
\end{enumerate}
 
This is the expansion of the message blocks without padding. In the last block we apply padding rule. If $(l+1)>384$ bits, then we have two extra blocks in the padded message. Otherwise we have one extra block in the padded message. In both the cases, we apply the following expansion rule for the last block so that the length of the message appears in the end of the padded message.

\begin{enumerate}[i.]
	
	\item $W_0=H_0$
	\item $W_{1}=H_7$
	\item $W_j=M^{(i)}_j$ for $2\le j\le 15$	
	\item $W_j=rotl_3(W_{j-16}\oplus W_{j-14}\oplus W_{j-8}\oplus W_{j-1})$ for $16\le j\le 63$
\end{enumerate}

	\item \textbf{Iteration}
	For $j=0$ to $63$
	
\begin{enumerate}[i.]
	\item $T_1=H_1+H_2+p(H_3||H_0)+K_j$\footnote{the operation $||$ denotes the concatenation and $+$ denotes the addition mod $2^{32}$}
	
	\item $T_2=H_4+H_5+p(H_7||H_6)+W_j$
	
	\item $H_7=H_6$
	
	\item $H_6=H_5$
	
	\item $H_5=H_4$
	
	\item $H_4=rotl_5(H_3+T_1)$
	
	\item $H_3=H_2$
	
	\item $H_2=H_1$
	
	\item $H_1=H_0$
	
	\item $H_0=T_1+T_2$,\\
	
\end{enumerate}

\noindent
where $T_1$ and $T_2$ are two temporary variables and $p \;: \; \Z_{2^{64}} \rightarrow \Z_{2^{32}}$ be a function defined by 

$$p(x)=2^{31}.p_1(x_1,\dots, x_{64})+2^{30}.p_2(x_1,\dots, x_{64})+\dots +1.p_{32}(x_1,\dots, x_{64}),$$ 

Since any element $x\in\Z_{2^{64}}$ can be represented by $x_1x_2\dots x_{64}$, where $x_1x_2\dots x_{64}$ denotes the bits of $x$ in decreasing order of their significance. $p_i(x_1,\dots, x_{64})$ denotes the $i^{th}$ polynomial of HFE challenge-$1$ with $64$ variables by setting the remaining $16$ variables to zero for $1\le i\le 32$. The $64$ constants $K_j$ taken from the fractional part of $e$ are given in Table $1$.
\end{enumerate}	

\begin{table*}[htbp]	
\begin{center}
\footnotesize\begin{tabular}{llll}
	 $K_{0}= AC211BEC$  &$K_{1}= 5FEFE110$  &$K_{2}= 112276F8$  &$K_{3}= 8AE122A4$ \\ 
	 $K_{4}= 18B3488B$  &$K_{5}= 00921A36$  &$K_{6}= 40C045F8$  &$K_{7}= C8C0A3DA$ \\ 
	 $K_{8}= C4ABF676$  &$K_{9}= 6A68C750$  &$K_{10}= A37AFE0F$ &$K_{11}= 732806F3$ \\ 
	 $K_{12}= 25722CB7$ &$K_{13}= 3FF43825$ &$K_{14}= ACDF96D7$ &$K_{15}= 9B53BCD3$ \\ 
	 $K_{16}= E34950DE$ &$K_{17}= D9780CCB$ &$K_{18}= 8B5F9BB7$ &$K_{19}= 3D1182ED$ \\ 
	 $K_{20}= 1921B44A$ &$K_{21}= 7003F30D$ &$K_{22}= 42657E31$ &$K_{23}= 231E7B55$ \\ 
	 $K_{24}= 91E3A28E$ &$K_{25}= 95CD4AB0$ &$K_{26}= 0A0AC2E3$ &$K_{27}= FCDEBE5E$ \\ 
	 $K_{28}= FCF1E321$ &$K_{29}= 1D136560$ &$K_{30}= 2974BF63$ &$K_{31}= 70963992$ \\ 
	 $K_{32}= 4F5B5107$ &$K_{33}= 0072C0C1$ &$K_{34}= C99F3C1D$ &$K_{35}= C56598D9$ \\ 
	 $K_{36}= 77A1D027$ &$K_{37}= 36675FB6$ &$K_{38}= A40C34E8$ &$K_{39}= 46764EAD$ \\ 
	 $K_{40}= F8823861$ &$K_{41}= 19F66E64$ &$K_{42}= 87E10299$ &$K_{43}= 4311C8C2$ \\ 
	 $K_{44}= 07C102B9$ &$K_{45}= 9F4EC8CE$ &$K_{46}= 29D81EBA$ &$K_{47}= 992744F9$ \\ 
	 $K_{48}= 4CDA6790$ &$K_{49}= 13DA5357$ &$K_{50}= BA6D7772$ &$K_{51}= 80673F08$ \\ 
	 $K_{52}= B049EE4C$ &$K_{53}= 839F8647$ &$K_{54}= 736F658B$ &$K_{55}= EBE90F9B$ \\ 
	 $K_{56}= FA6DC4D1$ &$K_{57}= E951630E$ &$K_{58}= AFC453E4$ &$K_{59}= 159B7483$ \\ 
	 $K_{60}= 45EABF9D$ &$K_{61}= 4292A60E$ &$K_{62}= 17AA0ABD$ &$K_{63}= 94E81C30$ \\
\end{tabular}
	\caption{\textbf{64 Constants}}
\end{center}
\end{table*}

\normalsize
\begin{enumerate}[4.]
	\item \textbf{Intermediate Hash Value}
	
	The $i^{th}$ intermediate hash value 
\begin{center}
$h^{(i)}=h^{(i)}_0||h^{(i)}_1||h^{(i)}_2||h^{(i)}_3||h^{(i)}_4||h^{(i)}_5||h^{(i)}_6||h^{(i)}_7$
\end{center}
 where $h^{(i)}_j=H_j$ for $0\le j\le 7$. This $h^{(i)}$ will be the initial value for the message block $M^{(i+1)}$.
\end{enumerate}

The final hash value of the message $M$ will be 
\begin{center}
$h^{(n)}_0||h^{(n)}_1||h^{(n)}_2||h^{(n)}_3||h^{(n)}_4||h^{(n)}_5||h^{(n)}_6||h^{(n)}_7,$
\end{center}

where $h^{(n)}_i=H_i$ for $0\le i\le 7$.

\begin{flushleft}
\textbf{Process of Implementation}
\end{flushleft}

Suppose we have to compute \textit{HF-hash}$(M)$. First we apply the padding rule and then padded message is divided  into $448$-bit blocks. Now each $448$-bit block is divided  into $14$ $32$-bit words and each $32$-bit word is read in little endian format. For example, suppose we have to read \textit{`abcd'} from a file, it will be read as $0$x$64636261$.     

\begin{flushleft}
\textbf{Test Value of \textit{HF-hash}}
\end{flushleft}

Test values of the three inputs are given below:
\begin{displaymath}
\begin{array}{lcllll}

\textit{HF-hash}(a) &= &04EAF5F6 &B215D974 &B827FCC2 &5ECA45C3\\
 &&031524E8 &472617D1 &C14D9C85 &6ACD1DC3\\

\textit{HF-hash}(ab) &= &F2DD83C8 &34E96291 &E39040B9 &BCD3E624\\
 &&BA01846E &0D5E5083 &492DC4BF &C0720235\\

\textit{HF-hash}(abc) &= &E9582019 &216033AA &346E8D46 &11D131A7\\
 &&D0635A5E &92D5B13D &2DC481B8 &836774B6\\

\end{array}
\end{displaymath}
  
\section{Analysis of \textit{HF-hash}}
In this section we will present the complete analysis of \textit{HF-hash} which includes properties, efficiency as well as the security analysis of this function.
\subsection{Properties of \textit{HF-hash}}
This subsection describes the properties of \textit{HF-hash} required for cryptographic applications. 

\begin{enumerate}

	\item \textbf{Easy to compute:} For any given value $x$ it is easy to compute \textit{HF-hash(x)} and the efficiency of this hash function is given in section $3.2$. 
	\item \textbf{One-wayness:} Suppose one knows the \textit{HF-hash(x)} for an input $x$. Now to find the value of $x$, (s)he has to solve the system of polynomial equations consisting of $32$ polynomials with $64$ variables for each round operation. Since this system of equations are underdefined therefore $XL$~\cite{bib:ckps} method or any variant of $XL$~\cite{bib:yc} cannot be applied to solve this system. 
	
Now if one wants to solve this system of equations using the Algorithm A\footnote{which is the best algorithm for solving our system of equations among Algorithms A, B $\&$ C} given by Courtois et. al. in~\cite{bib:cgmt}, then at least $2^{25}$ operations are required to solve for one round of \textit{HF-hash}. Since \textit{HF-hash} has 64 rounds one has to compute $2^{25\times 64}$ operations to get back the value of $x$ for given \textit{HF-hash(x)}. This is far beyond the today's computation power. Thus, for any given \textit{HF-hash(x)} it is difficult to find the input $x$.
  
\item \textbf{Randomness:} We have taken an input file $M$ consisting of $448$ bits and computed \textit{HF-hash}$(M)$. $448$ flies $M_i$ are generated by changing the $i^{th}$ bit of $M$ for $1\le i\le 448$. Then computed \textit{HF-hash}$(M_i)$ of all the $448$ files and calculated the Hamming distance $d_i$ between \textit{HF-hash}$(M)$ and  \textit{HF-hash}$(M_i)$  for $1\le i\le 448$ as well as the distances between corresponding $8$ $32$-bit words of the hash values. Table $2$ shows maximum, minimum, mode and mean of the above distances. 

\begin{table*}[htbp]	
\begin{center}
	\begin{tabular}{|c|c|c|c|c|c|c|c|c|c|}
	\hline
	Changes &$W_1$&$W_2$&$W_3$&$W_4$&$W_5$&$W_6$&$W_7$&$W_8$&\textit{HF-hash}\\
	\hline
	Max	&25	&24	&24	&26	&25	&23	&23	&24		&149\\
	\hline
	Min	&6	&7	&7	&8	&7	&8	&9	&8		&103\\
	\hline
	Mode	&14	&17	&17	&16	&16	&17	&16	&15		&132\\
	\hline
	Mean	&16	&16	&16	&16	&16	&16	&16	&16		&128\\
	\hline
	\end{tabular}
	\caption{\textbf{Hamming Distances}}
\end{center}
\end{table*}

For ideal case $d_i$ should be $128$ for $1\le i\le 448$. But we have found that $d_i'$s were lying between $103$ and $149$ for the above files. The following bar chart and the table show the distribution of above $448$ files with respect to their distances.  

\begin{figure*}[htbp]
\begin{center}
	\includegraphics[width=0.75\textwidth]{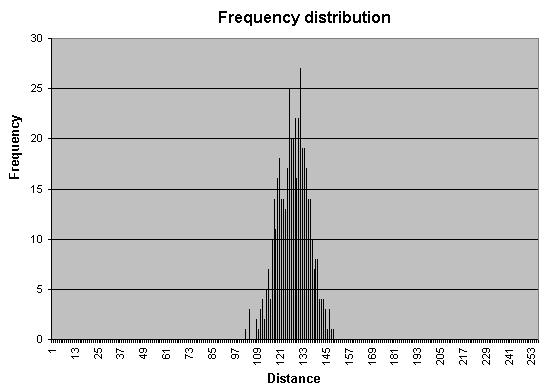}
\end{center}
\end{figure*}

\begin{center}
	\begin{tabular}{|c|c|c|}
		\hline
		Range &No. of &Percentage\\
		of Distance&Files&\\
		\hline
		$128\pm5$ &$215$ &$47.99$\\
		\hline
		$128\pm10$ &$362$ &$80.80$\\
		\hline
		$128\pm15$ &$421$ &$93.97$\\
		\hline
		$128\pm20$ &$443$ &$98.88$\\
		\hline		
	\end{tabular}
\end{center}

\end{enumerate}

The above analyses show that \textit{HF-hash} exhibits a reasonably good avalanche effect. Thus it can be used for cryptographic applications.   
\subsection{Efficiency of \textit{HF-hash}}
The following table gives a comparative study in the efficiency of \textit{HF-hash} with SHA-$256$ in HP Pentium - D with $3$ GHz processor and $512$ MB RAM. 
\begin{center}
	\begin{tabular}{|c|c|c|}
	\hline
	File Size &\textit{HF-hash} &SHA-$256$\\
	(in MB) &(in Sec.) &(in Sec.)\\
	\hline
	$1.4$ &$20.02$ &$18.64$\\
	\hline
	$4.84$ &$67.72$ &$60.08$\\
	\hline
	$7.48$ &$109.73$ &$103.59$\\
	\hline
	$12.94$ &$181.01$ &$169.19$\\
	\hline
	$24.3$ &$345.53$ &$313.53$\\
	\hline
	\end{tabular}
\end{center}

Although, SHA-$256$ is little bit faster than \textit{HF-hash} but \textit{HF-hash} is more secure than SHA-$256$ in case of either collision search or differential attack. Since the design principle of SHA-$256$ is almost similar to that of SHA-$1$, therefore all the attacks applied to SHA-$1$ can also be extended to SHA-$256$. 

\subsection{Security Analysis}
In this paper we have applied a new method for expanding a $512$-bit message block into $2048$-bit block. For this purpose we have to change the padding rule and the procedure of parsing a padded message. In case of MD-$5$, SHA-$1$ \& SHA-$256$,  the padded message is divided into $512$-bit blocks whereas in  case of HF-hash, the padded message is divided into $448$-bit blocks.Then two $32$-bit words are added to construct a $512$-bit block as the input for each iteration, where these two words depend on the previous internal hash updates or chaining variables. So, in each iteration, the $512$-bit blocks are not independent from the previous message blocks as in the case of MD-$5$, SHA-$1$ or SHA-$256$. Thus, differential attack by Chabaud \& Joux is not applicable to our hash function because one does not have any control over two $32$-bit words coming from the previous internal hash updates. Moreover, a $1$-bit difference in any one of $14$ initial $32$-bit word propagates itself to at least $165$ bits of the expanded message since we have taken the $64$ round operations. Less than $75$ bit difference in expanded message and input message is obtained by changing $1$-bit input when $32$ or $48$ round operation are performed. That is why we have taken $64$ round operations for \textit{HF-hash} function.  This makes it impossible to find corrective patterns used by Chabaud and Joux in ~\cite{bib:cj}, due to the reason that differences propagate to other positions.

The idea of Wang et. al. for finding collision in SHA-$0$ ~\cite{bib:wyy1} and SHA-$1$ ~\cite{bib:wyy} is to find out the disturbance vectors with low Hamming weight first and then to construct a differential path. To construct a valid differential path, it is important to control the difference propagation in each chaining variable. After identifying the wanted and unwanted differences one can apply the Boolean functions (mainly IF) and the carry effect to cancel out these differences. The key of these attacks was the Boolean functions used in compression function which in combination with carry effect facilitate the differential attack. As we have replaced the Boolean functions with restricted hidden field polynomials, it is evident that these attacks are not applicable to our hash function.   

Thus the compression function of \textit{HF-hash} is collision-resistant against Wang et. al. attack. Since $\mathcal{IV}$ of \textit{HF-hash} is fixed and the padding procedure of \textit{HF-hash} includes the length of the message, therefore by Merkle-Damg{\aa}rd theorem~\cite{bib:dam}~\cite{bib:mer} we can say that \textit{HF-hash} is collision-resistant against existing attacks.

\section{Conclusions}
In this paper a dedicated hash function \textit{HF-hash} has been presented. The differential attack applied by Chabaud and Joux in SHA-$0$ as well as collision search for SHA-$1$ by Wang et. al. are not applicable to this hash function. The main differences of \textit{HF-hash} with MD family and SHA family lie in the the procedure of message expansion  and the compression function. A system of multivariate polynomials taken from HFE challenge-$1$ (restricted form) is used for designing the compression function of this hash function. Analysis of this hash functions viz. randomness as well as security proof are also described here.

The system of equations in HFE challenge-$1$ are neither regular system nor the minimal set of polynomials. Presently we are looking at the behaviour of \textit{HF-hash} when the minimal system or the Gr$\ddot{o}$bner basis of the ideal generated by the above system or randomly selected $32$ polynomials with $64$ variables is taken.

\begin{center}
\textbf{Appendix:List of Polynomials}
\end{center}

\tiny 
$y_{1} = x_{1}x_{2} + x_{1}x_{6} + x_{1}x_{7} + x_{1}x_{8} + x_{1}x_{9} + x_{1}x_{10} + x_{1}x_{12} + x_{1}x_{16} + x_{1}x_{18} + x_{1}x_{20} + x_{1}x_{21} + x_{1}x_{22} + x_{1}x_{24} + x_{1}x_{25} + x_{1}x_{27} + x_{1}x_{28} + x_{1}x_{29} + x_{1}x_{34} + x_{1}x_{35} + x_{1}x_{39} + x_{1}x_{40} + x_{1}x_{42} + x_{1}x_{43} + x_{1}x_{44} + x_{1}x_{45} + x_{1}x_{50} + x_{1}x_{51} + x_{1}x_{52} + x_{1}x_{54} + x_{1}x_{56} + x_{1}x_{57} + x_{1}x_{61} + x_{1}x_{62} + x_{1}x_{63} + x_{2}x_{3} + x_{2}x_{4} + x_{2}x_{5} + x_{2}x_{7} + x_{2}x_{10} + x_{2}x_{11} + x_{2}x_{13} + x_{2}x_{16} + x_{2}x_{19} + x_{2}x_{20} + x_{2}x_{23} + x_{2}x_{26} + x_{2}x_{28} + x_{2}x_{30} + x_{2}x_{31} + x_{2}x_{38} + x_{2}x_{42} + x_{2}x_{44} + x_{2}x_{49} + x_{2}x_{50} + x_{2}x_{51} + x_{2}x_{53} + x_{2}x_{54} + x_{2}x_{55} + x_{2}x_{56} + x_{2}x_{64} + x_{3}x_{6} + x_{3}x_{15} + x_{3}x_{18} + x_{3}x_{21} + x_{3}x_{25} + x_{3}x_{26} + x_{3}x_{27} + x_{3}x_{28} + x_{3}x_{30} + x_{3}x_{31} + x_{3}x_{32} + x_{3}x_{35} + x_{3}x_{37} + x_{3}x_{40} + x_{3}x_{41} + x_{3}x_{42} + x_{3}x_{44} + x_{3}x_{47} + x_{3}x_{49} + x_{3}x_{50} + x_{3}x_{51} + x_{3}x_{53} + x_{3}x_{54} + x_{3}x_{55} + x_{3}x_{57} + x_{3}x_{61} + x_{4}x_{5} + x_{4}x_{6} + x_{4}x_{8} + x_{4}x_{10} + x_{4}x_{14} + x_{4}x_{17} + x_{4}x_{20} + x_{4}x_{24} + x_{4}x_{25} + x_{4}x_{28} + x_{4}x_{29} + x_{4}x_{31} + x_{4}x_{33} + x_{4}x_{34} + x_{4}x_{36} + x_{4}x_{37} + x_{4}x_{38} + x_{4}x_{42} + x_{4}x_{47} + x_{4}x_{49} + x_{4}x_{50} + x_{4}x_{52} + x_{4}x_{55} + x_{4}x_{56} + x_{4}x_{57} + x_{4}x_{58} + x_{4}x_{61} + x_{4}x_{63} + x_{5}x_{6} + x_{5}x_{8} + x_{5}x_{9} + x_{5}x_{13} + x_{5}x_{19} + x_{5}x_{22} + x_{5}x_{23} + x_{5}x_{26} + x_{5}x_{29} + x_{5}x_{32} + x_{5}x_{33} + x_{5}x_{36} + x_{5}x_{37} + x_{5}x_{39} + x_{5}x_{41} + x_{5}x_{42} + x_{5}x_{44} + x_{5}x_{47} + x_{5}x_{48} + x_{5}x_{49} + x_{5}x_{50} + x_{5}x_{51} + x_{5}x_{53} + x_{5}x_{55} + x_{5}x_{57} + x_{5}x_{59} + x_{6}x_{7} + x_{6}x_{8} + x_{6}x_{11} + x_{6}x_{12} + x_{6}x_{13} + x_{6}x_{16} + x_{6}x_{20} + x_{6}x_{21} + x_{6}x_{22} + x_{6}x_{24} + x_{6}x_{25} + x_{6}x_{27} + x_{6}x_{29} + x_{6}x_{30} + x_{6}x_{31} + x_{6}x_{33} + x_{6}x_{34} + x_{6}x_{37} + x_{6}x_{40} + x_{6}x_{42} + x_{6}x_{43} + x_{6}x_{45} + x_{6}x_{47} + x_{6}x_{48} + x_{6}x_{53} + x_{6}x_{55} + x_{6}x_{57} + x_{6}x_{59} + x_{6}x_{60} + x_{6}x_{61} + x_{6}x_{62} + x_{6}x_{64} + x_{7}x_{8} + x_{7}x_{9} + x_{7}x_{10} + x_{7}x_{13} + x_{7}x_{18} + x_{7}x_{19} + x_{7}x_{21} + x_{7}x_{25} + x_{7}x_{26} + x_{7}x_{28} + x_{7}x_{32} + x_{7}x_{33} + x_{7}x_{35} + x_{7}x_{36} + x_{7}x_{39} + x_{7}x_{42} + x_{7}x_{44} + x_{7}x_{47} + x_{7}x_{51} + x_{7}x_{52} + x_{7}x_{54} + x_{7}x_{57} + x_{7}x_{60} + x_{7}x_{61} + x_{7}x_{62} + x_{7}x_{64} + x_{8}x_{10} + x_{8}x_{11} + x_{8}x_{13} + x_{8}x_{14} + x_{8}x_{16} + x_{8}x_{17} + x_{8}x_{19} + x_{8}x_{23} + x_{8}x_{26} + x_{8}x_{27} + x_{8}x_{31} + x_{8}x_{33} + x_{8}x_{37} + x_{8}x_{38} + x_{8}x_{43} + x_{8}x_{44} + x_{8}x_{46} + x_{8}x_{48} + x_{8}x_{51} + x_{8}x_{53} + x_{8}x_{55} + x_{8}x_{56} + x_{8}x_{58} + x_{8}x_{64} + x_{9}x_{11} + x_{9}x_{12} + x_{9}x_{14} + x_{9}x_{17} + x_{9}x_{18} + x_{9}x_{19} + x_{9}x_{20} + x_{9}x_{21} + x_{9}x_{23} + x_{9}x_{25} + x_{9}x_{26} + x_{9}x_{27} + x_{9}x_{28} + x_{9}x_{30} + x_{9}x_{31} + x_{9}x_{34} + x_{9}x_{35} + x_{9}x_{36} + x_{9}x_{37} + x_{9}x_{38} + x_{9}x_{39} + x_{9}x_{41} + x_{9}x_{43} + x_{9}x_{45} + x_{9}x_{46} + x_{9}x_{50} + x_{9}x_{54} + x_{9}x_{55} + x_{9}x_{64} + x_{10}x_{11} + x_{10}x_{12} + x_{10}x_{14} + x_{10}x_{16} + x_{10}x_{17} + x_{10}x_{18} + x_{10}x_{20} + x_{10}x_{21} + x_{10}x_{24} + x_{10}x_{25} + x_{10}x_{26} + x_{10}x_{27} + x_{10}x_{28} + x_{10}x_{29} + x_{10}x_{30} + x_{10}x_{32} + x_{10}x_{37} + x_{10}x_{40} + x_{10}x_{44} + x_{10}x_{46} + x_{10}x_{49} + x_{10}x_{50} + x_{10}x_{51} + x_{10}x_{59} + x_{10}x_{60} + x_{10}x_{61} + x_{10}x_{62} + x_{11}x_{13} + x_{11}x_{14} + x_{11}x_{17} + x_{11}x_{19} + x_{11}x_{23} + x_{11}x_{25} + x_{11}x_{29} + x_{11}x_{30} + x_{11}x_{31} + x_{11}x_{32} + x_{11}x_{33} + x_{11}x_{35} + x_{11}x_{40} + x_{11}x_{42} + x_{11}x_{44} + x_{11}x_{46} + x_{11}x_{47} + x_{11}x_{50} + x_{11}x_{55} + x_{11}x_{56} + x_{11}x_{57} + x_{11}x_{59} + x_{11}x_{61} + x_{11}x_{63} + x_{12}x_{13} + x_{12}x_{18} + x_{12}x_{21} + x_{12}x_{23} + x_{12}x_{26} + x_{12}x_{34} + x_{12}x_{38} + x_{12}x_{39} + x_{12}x_{42} + x_{12}x_{43} + x_{12}x_{45} + x_{12}x_{47} + x_{12}x_{49} + x_{12}x_{51} + x_{12}x_{53} + x_{12}x_{55} + x_{12}x_{56} + x_{12}x_{57} + x_{12}x_{58} + x_{12}x_{60} + x_{12}x_{64} + x_{13}x_{14} + x_{13}x_{15} + x_{13}x_{16} + x_{13}x_{20} + x_{13}x_{21} + x_{13}x_{23} + x_{13}x_{26} + x_{13}x_{27} + x_{13}x_{29} + x_{13}x_{30} + x_{13}x_{32} + x_{13}x_{33} + x_{13}x_{35} + x_{13}x_{36} + x_{13}x_{40} + x_{13}x_{41} + x_{13}x_{42} + x_{13}x_{43} + x_{13}x_{44} + x_{13}x_{45} + x_{13}x_{46} + x_{13}x_{49} + x_{13}x_{50} + x_{13}x_{52} + x_{13}x_{53} + x_{13}x_{54} + x_{13}x_{55} + x_{13}x_{63} + x_{14}x_{15} + x_{14}x_{17} + x_{14}x_{23} + x_{14}x_{28} + x_{14}x_{31} + x_{14}x_{34} + x_{14}x_{36} + x_{14}x_{38} + x_{14}x_{39} + x_{14}x_{42} + x_{14}x_{43} + x_{14}x_{44} + x_{14}x_{45} + x_{14}x_{47} + x_{14}x_{48} + x_{14}x_{50} + x_{14}x_{51} + x_{14}x_{52} + x_{14}x_{53} + x_{14}x_{54} + x_{14}x_{55} + x_{14}x_{56} + x_{14}x_{58} + x_{14}x_{60} + x_{14}x_{63} + x_{15}x_{16} + x_{15}x_{17} + x_{15}x_{18} + x_{15}x_{20} + x_{15}x_{22} + x_{15}x_{24} + x_{15}x_{25} + x_{15}x_{27} + x_{15}x_{30} + x_{15}x_{36} + x_{15}x_{38} + x_{15}x_{39} + x_{15}x_{44} + x_{15}x_{45} + x_{15}x_{48} + x_{15}x_{49} + x_{15}x_{50} + x_{15}x_{52} + x_{15}x_{56} + x_{15}x_{57} + x_{15}x_{58} + x_{15}x_{59} + x_{15}x_{60} + x_{16}x_{18} + x_{16}x_{19} + x_{16}x_{22} + x_{16}x_{25} + x_{16}x_{26} + x_{16}x_{28} + x_{16}x_{31} + x_{16}x_{32} + x_{16}x_{33} + x_{16}x_{35} + x_{16}x_{40} + x_{16}x_{41} + x_{16}x_{42} + x_{16}x_{43} + x_{16}x_{46} + x_{16}x_{47} + x_{16}x_{49} + x_{16}x_{53} + x_{16}x_{54} + x_{16}x_{57} + x_{16}x_{59} + x_{16}x_{62} + x_{17}x_{18} + x_{17}x_{19} + x_{17}x_{20} + x_{17}x_{21} + x_{17}x_{23} + x_{17}x_{24} + x_{17}x_{25} + x_{17}x_{27} + x_{17}x_{30} + x_{17}x_{31} + x_{17}x_{33} + x_{17}x_{34} + x_{17}x_{37} + x_{17}x_{38} + x_{17}x_{40} + x_{17}x_{42} + x_{17}x_{44} + x_{17}x_{45} + x_{17}x_{48} + x_{17}x_{53} + x_{17}x_{54} + x_{17}x_{56} + x_{17}x_{59} + x_{17}x_{61} + x_{17}x_{63} + x_{17}x_{64} + x_{18}x_{19} + x_{18}x_{21} + x_{18}x_{22} + x_{18}x_{24} + x_{18}x_{28} + x_{18}x_{29} + x_{18}x_{31} + x_{18}x_{32} + x_{18}x_{33} + x_{18}x_{34} + x_{18}x_{39} + x_{18}x_{40} + x_{18}x_{41} + x_{18}x_{45} + x_{18}x_{48} + x_{18}x_{50} + x_{18}x_{51} + x_{18}x_{52} + x_{18}x_{54} + x_{18}x_{55} + x_{18}x_{56} + x_{18}x_{57} + x_{18}x_{58} + x_{18}x_{59} + x_{18}x_{60} + x_{18}x_{62} + x_{19}x_{20} + x_{19}x_{21} + x_{19}x_{27} + x_{19}x_{28} + x_{19}x_{29} + x_{19}x_{30} + x_{19}x_{33} + x_{19}x_{37} + x_{19}x_{38} + x_{19}x_{39} + x_{19}x_{41} + x_{19}x_{43} + x_{19}x_{44} + x_{19}x_{45} + x_{19}x_{46} + x_{19}x_{48} + x_{19}x_{49} + x_{19}x_{51} + x_{19}x_{52} + x_{19}x_{55} + x_{19}x_{56} + x_{19}x_{57} + x_{19}x_{59} + x_{19}x_{60} + x_{19}x_{63} + x_{19}x_{64} + x_{20}x_{21} + x_{20}x_{24} + x_{20}x_{26} + x_{20}x_{27} + x_{20}x_{28} + x_{20}x_{29} + x_{20}x_{30} + x_{20}x_{32} + x_{20}x_{35} + x_{20}x_{36} + x_{20}x_{38} + x_{20}x_{40} + x_{20}x_{41} + x_{20}x_{42} + x_{20}x_{44} + x_{20}x_{45} + x_{20}x_{46} + x_{20}x_{47} + x_{20}x_{48} + x_{20}x_{51} + x_{20}x_{54} + x_{20}x_{55} + x_{20}x_{57} + x_{20}x_{60} + x_{20}x_{61} + x_{20}x_{64} + x_{21}x_{24} + x_{21}x_{25} + x_{21}x_{26} + x_{21}x_{27} + x_{21}x_{32} + x_{21}x_{33} + x_{21}x_{37} + x_{21}x_{40} + x_{21}x_{41} + x_{21}x_{42} + x_{21}x_{43} + x_{21}x_{45} + x_{21}x_{46} + x_{21}x_{49} + x_{21}x_{51} + x_{21}x_{52} + x_{21}x_{54} + x_{21}x_{56} + x_{21}x_{57} + x_{21}x_{60} + x_{21}x_{62} + x_{21}x_{63} + x_{21}x_{64} + x_{22}x_{23} + x_{22}x_{25} + x_{22}x_{29} + x_{22}x_{34} + x_{22}x_{35} + x_{22}x_{38} + x_{22}x_{39} + x_{22}x_{40} + x_{22}x_{44} + x_{22}x_{49} + x_{22}x_{50} + x_{22}x_{51} + x_{22}x_{52} + x_{22}x_{53} + x_{22}x_{54} + x_{22}x_{55} + x_{22}x_{57} + x_{22}x_{58} + x_{22}x_{60} + x_{22}x_{62} + x_{23}x_{26} + x_{23}x_{32} + x_{23}x_{34} + x_{23}x_{35} + x_{23}x_{36} + x_{23}x_{38} + x_{23}x_{40} + x_{23}x_{41} + x_{23}x_{42} + x_{23}x_{44} + x_{23}x_{45} + x_{23}x_{46} + x_{23}x_{48} + x_{23}x_{49} + x_{23}x_{51} + x_{23}x_{54} + x_{23}x_{55} + x_{23}x_{56} + x_{23}x_{57} + x_{23}x_{58} + x_{23}x_{60} + x_{23}x_{62} + x_{24}x_{25} + x_{24}x_{26} + x_{24}x_{28} + x_{24}x_{33} + x_{24}x_{35} + x_{24}x_{37} + x_{24}x_{39} + x_{24}x_{40} + x_{24}x_{41} + x_{24}x_{45} + x_{24}x_{48} + x_{24}x_{49} + x_{24}x_{50} + x_{24}x_{51} + x_{24}x_{53} + x_{24}x_{54} + x_{24}x_{55} + x_{24}x_{57} + x_{24}x_{58} + x_{24}x_{59} + x_{24}x_{60} + x_{24}x_{61} + x_{24}x_{63} + x_{24}x_{64} + x_{25}x_{27} + x_{25}x_{29} + x_{25}x_{31} + x_{25}x_{32} + x_{25}x_{33} + x_{25}x_{37} + x_{25}x_{43} + x_{25}x_{44} + x_{25}x_{46} + x_{25}x_{47} + x_{25}x_{51} + x_{25}x_{53} + x_{25}x_{54} + x_{25}x_{56} + x_{25}x_{57} + x_{25}x_{58} + x_{25}x_{60} + x_{25}x_{61} + x_{25}x_{62} + x_{25}x_{64} + x_{26}x_{27} + x_{26}x_{28} + x_{26}x_{29} + x_{26}x_{30} + x_{26}x_{35} + x_{26}x_{36} + x_{26}x_{39} + x_{26}x_{40} + x_{26}x_{42} + x_{26}x_{45} + x_{26}x_{46} + x_{26}x_{47} + x_{26}x_{48} + x_{26}x_{52} + x_{26}x_{54} + x_{26}x_{56} + x_{26}x_{57} + x_{26}x_{64} + x_{27}x_{28} + x_{27}x_{29} + x_{27}x_{30} + x_{27}x_{32} + x_{27}x_{33} + x_{27}x_{37} + x_{27}x_{38} + x_{27}x_{39} + x_{27}x_{44} + x_{27}x_{47} + x_{27}x_{49} + x_{27}x_{51} + x_{27}x_{52} + x_{27}x_{53} + x_{27}x_{54} + x_{27}x_{61} + x_{28}x_{29} + x_{28}x_{30} + x_{28}x_{32} + x_{28}x_{34} + x_{28}x_{36} + x_{28}x_{38} + x_{28}x_{39} + x_{28}x_{40} + x_{28}x_{46} + x_{28}x_{48} + x_{28}x_{51} + x_{28}x_{52} + x_{28}x_{55} + x_{28}x_{56} + x_{28}x_{57} + x_{28}x_{58} + x_{28}x_{60} + x_{28}x_{61} + x_{28}x_{62} + x_{28}x_{64} + x_{29}x_{30} + x_{29}x_{34} + x_{29}x_{36} + x_{29}x_{37} + x_{29}x_{38} + x_{29}x_{39} + x_{29}x_{40} + x_{29}x_{41} + x_{29}x_{46} + x_{29}x_{48} + x_{29}x_{50} + x_{29}x_{52} + x_{29}x_{53} + x_{29}x_{54} + x_{29}x_{56} + x_{29}x_{57} + x_{29}x_{58} + x_{29}x_{59} + x_{29}x_{60} + x_{29}x_{62} + x_{30}x_{33} + x_{30}x_{42} + x_{30}x_{44} + x_{30}x_{45} + x_{30}x_{46} + x_{30}x_{47} + x_{30}x_{49} + x_{30}x_{50} + x_{30}x_{52} + x_{30}x_{54} + x_{30}x_{55} + x_{30}x_{56} + x_{30}x_{58} + x_{30}x_{62} + x_{30}x_{63} + x_{30}x_{64} + x_{31}x_{32} + x_{31}x_{33} + x_{31}x_{35} + x_{31}x_{36} + x_{31}x_{37} + x_{31}x_{38} + x_{31}x_{42} + x_{31}x_{43} + x_{31}x_{44} + x_{31}x_{46} + x_{31}x_{47} + x_{31}x_{50} + x_{31}x_{52} + x_{31}x_{54} + x_{31}x_{55} + x_{31}x_{59} + x_{32}x_{33} + x_{32}x_{35} + x_{32}x_{37} + x_{32}x_{38} + x_{32}x_{39} + x_{32}x_{40} + x_{32}x_{41} + x_{32}x_{42} + x_{32}x_{44} + x_{32}x_{46} + x_{32}x_{48} + x_{32}x_{50} + x_{32}x_{51} + x_{32}x_{53} + x_{32}x_{56} + x_{32}x_{57} + x_{32}x_{60} + x_{32}x_{61} + x_{32}x_{62} + x_{32}x_{63} + x_{33}x_{34} + x_{33}x_{37} + x_{33}x_{41} + x_{33}x_{43} + x_{33}x_{44} + x_{33}x_{48} + x_{33}x_{50} + x_{33}x_{53} + x_{33}x_{56} + x_{33}x_{57} + x_{33}x_{61} + x_{33}x_{62} + x_{33}x_{63} + x_{34}x_{39} + x_{34}x_{41} + x_{34}x_{45} + x_{34}x_{49} + x_{34}x_{53} + x_{34}x_{54} + x_{34}x_{57} + x_{34}x_{58} + x_{34}x_{59} + x_{34}x_{61} + x_{34}x_{62} + x_{34}x_{63} + x_{34}x_{64} + x_{35}x_{38} + x_{35}x_{39} + x_{35}x_{40} + x_{35}x_{42} + x_{35}x_{45} + x_{35}x_{48} + x_{35}x_{49} + x_{35}x_{53} + x_{35}x_{54} + x_{35}x_{55} + x_{35}x_{56} + x_{35}x_{58} + x_{35}x_{63} + x_{36}x_{37} + x_{36}x_{38} + x_{36}x_{39} + x_{36}x_{41} + x_{36}x_{44} + x_{36}x_{45} + x_{36}x_{46} + x_{36}x_{50} + x_{36}x_{53} + x_{36}x_{54} + x_{36}x_{59} + x_{36}x_{60} + x_{36}x_{61} + x_{36}x_{62} + x_{36}x_{63} + x_{37}x_{38} + x_{37}x_{40} + x_{37}x_{42} + x_{37}x_{43} + x_{37}x_{44} + x_{37}x_{45} + x_{37}x_{46} + x_{37}x_{47} + x_{37}x_{48} + x_{37}x_{49} + x_{37}x_{52} + x_{37}x_{56} + x_{37}x_{57} + x_{37}x_{60} + x_{37}x_{62} + x_{38}x_{41} + x_{38}x_{42} + x_{38}x_{44} + x_{38}x_{46} + x_{38}x_{50} + x_{38}x_{54} + x_{38}x_{56} + x_{38}x_{57} + x_{38}x_{58} + x_{38}x_{59} + x_{38}x_{60} + x_{38}x_{62} + x_{38}x_{63} + x_{38}x_{64} + x_{39}x_{41} + x_{39}x_{44} + x_{39}x_{54} + x_{39}x_{56} + x_{39}x_{58} + x_{39}x_{59} + x_{39}x_{61} + x_{39}x_{62} + x_{39}x_{64} + x_{40}x_{43} + x_{40}x_{47} + x_{40}x_{52} + x_{40}x_{53} + x_{40}x_{57} + x_{40}x_{58} + x_{40}x_{59} + x_{40}x_{61} + x_{40}x_{62} + x_{41}x_{43} + x_{41}x_{45} + x_{41}x_{48} + x_{41}x_{50} + x_{41}x_{51} + x_{41}x_{54} + x_{41}x_{56} + x_{41}x_{57} + x_{41}x_{59} + x_{41}x_{60} + x_{41}x_{62} + x_{42}x_{45} + x_{42}x_{47} + x_{42}x_{48} + x_{42}x_{49} + x_{42}x_{52} + x_{42}x_{57} + x_{42}x_{58} + x_{42}x_{63} + x_{42}x_{64} + x_{43}x_{45} + x_{43}x_{46} + x_{43}x_{50} + x_{43}x_{51} + x_{43}x_{52} + x_{43}x_{53} + x_{43}x_{54} + x_{43}x_{55} + x_{43}x_{56} + x_{43}x_{57} + x_{43}x_{58} + x_{43}x_{59} + x_{43}x_{61} + x_{43}x_{63} + x_{43}x_{64} + x_{44}x_{45} + x_{44}x_{47} + x_{44}x_{48} + x_{44}x_{49} + x_{44}x_{50} + x_{44}x_{51} + x_{44}x_{53} + x_{44}x_{55} + x_{44}x_{56} + x_{44}x_{57} + x_{44}x_{58} + x_{44}x_{59} + x_{44}x_{60} + x_{44}x_{61} + x_{44}x_{62} + x_{44}x_{64} + x_{45}x_{46} + x_{45}x_{49} + x_{45}x_{50} + x_{45}x_{55} + x_{45}x_{56} + x_{45}x_{59} + x_{45}x_{60} + x_{45}x_{63} + x_{45}x_{64} + x_{46}x_{47} + x_{46}x_{48} + x_{46}x_{49} + x_{46}x_{51} + x_{46}x_{52} + x_{46}x_{54} + x_{46}x_{55} + x_{46}x_{57} + x_{46}x_{58} + x_{46}x_{60} + x_{46}x_{61} + x_{46}x_{62} + x_{46}x_{64} + x_{47}x_{49} + x_{47}x_{51} + x_{47}x_{52} + x_{47}x_{53} + x_{47}x_{54} + x_{47}x_{55} + x_{47}x_{57} + x_{47}x_{58} + x_{47}x_{59} + x_{47}x_{63} + x_{48}x_{53} + x_{48}x_{55} + x_{48}x_{61} + x_{48}x_{63} + x_{49}x_{50} + x_{49}x_{53} + x_{49}x_{54} + x_{49}x_{55} + x_{49}x_{59} + x_{49}x_{61} + x_{49}x_{63} + x_{49}x_{64} + x_{50}x_{51} + x_{50}x_{52} + x_{50}x_{53} + x_{50}x_{54} + x_{50}x_{57} + x_{50}x_{61} + x_{50}x_{62} + x_{50}x_{63} + x_{51}x_{52} + x_{51}x_{55} + x_{51}x_{56} + x_{51}x_{58} + x_{51}x_{62} + x_{51}x_{63} + x_{51}x_{64} + x_{52}x_{53} + x_{52}x_{55} + x_{52}x_{58} + x_{52}x_{59} + x_{52}x_{60} + x_{52}x_{64} + x_{53}x_{54} + x_{53}x_{55} + x_{53}x_{56} + x_{53}x_{64} + x_{54}x_{55} + x_{54}x_{57} + x_{54}x_{61} + x_{55}x_{57} + x_{55}x_{58} + x_{55}x_{62} + x_{55}x_{64} + x_{56}x_{57} + x_{56}x_{59} + x_{56}x_{60} + x_{56}x_{61} + x_{56}x_{62} + x_{57}x_{58} + x_{57}x_{61} + x_{58}x_{62} + x_{58}x_{63} + x_{58}x_{64} + x_{59}x_{61} + x_{59}x_{62} + x_{59}x_{63} + x_{59}x_{64} + x_{60}x_{62} + x_{61}x_{63} + x_{61}x_{64} + x_{62}x_{63} + x_{62}x_{64} + x_{1} + x_{3} + x_{4} + x_{5} + x_{6} + x_{8} + x_{10} + x_{14} + x_{17} + x_{18} + x_{20} + x_{21} + x_{27} + x_{28} + x_{30} + x_{31} + x_{33} + x_{35} + x_{39} + x_{40} + x_{42} + x_{44} + x_{45} + x_{46} + x_{48} + x_{49} + x_{51} + x_{52} + x_{53} + x_{54} + x_{56} + x_{58} + x_{64} + 1$

$y_{2} = x_{1}x_{3} + x_{1}x_{4} + x_{1}x_{5} + x_{1}x_{6} + x_{1}x_{9} + x_{1}x_{10} + x_{1}x_{14} + x_{1}x_{15} + x_{1}x_{16} + x_{1}x_{17} + x_{1}x_{18} + x_{1}x_{19} + x_{1}x_{24} + x_{1}x_{25} + x_{1}x_{26} + x_{1}x_{27} + x_{1}x_{29} + x_{1}x_{30} + x_{1}x_{31} + x_{1}x_{32} + x_{1}x_{33} + x_{1}x_{35} + x_{1}x_{37} + x_{1}x_{38} + x_{1}x_{39} + x_{1}x_{41} + x_{1}x_{43} + x_{1}x_{46} + x_{1}x_{47} + x_{1}x_{51} + x_{1}x_{52} + x_{1}x_{53} + x_{1}x_{55} + x_{1}x_{57} + x_{1}x_{60} + x_{1}x_{62} + x_{1}x_{64} + x_{2}x_{4} + x_{2}x_{5} + x_{2}x_{7} + x_{2}x_{8} + x_{2}x_{12} + x_{2}x_{13} + x_{2}x_{14} + x_{2}x_{16} + x_{2}x_{18} + x_{2}x_{20} + x_{2}x_{22} + x_{2}x_{23} + x_{2}x_{24} + x_{2}x_{25} + x_{2}x_{27} + x_{2}x_{31} + x_{2}x_{32} + x_{2}x_{33} + x_{2}x_{35} + x_{2}x_{37} + x_{2}x_{38} + x_{2}x_{41} + x_{2}x_{45} + x_{2}x_{46} + x_{2}x_{49} + x_{2}x_{50} + x_{2}x_{60} + x_{2}x_{61} + x_{2}x_{62} + x_{2}x_{63} + x_{2}x_{64} + x_{3}x_{7} + x_{3}x_{9} + x_{3}x_{12} + x_{3}x_{14} + x_{3}x_{15} + x_{3}x_{20} + x_{3}x_{22} + x_{3}x_{26} + x_{3}x_{29} + x_{3}x_{30} + x_{3}x_{32} + x_{3}x_{35} + x_{3}x_{36} + x_{3}x_{38} + x_{3}x_{40} + x_{3}x_{41} + x_{3}x_{42} + x_{3}x_{44} + x_{3}x_{47} + x_{3}x_{50} + x_{3}x_{56} + x_{3}x_{57} + x_{3}x_{59} + x_{3}x_{61} + x_{3}x_{62} + x_{3}x_{64} + x_{4}x_{5} + x_{4}x_{6} + x_{4}x_{11} + x_{4}x_{12} + x_{4}x_{13} + x_{4}x_{17} + x_{4}x_{19} + x_{4}x_{21} + x_{4}x_{23} + x_{4}x_{24} + x_{4}x_{25} + x_{4}x_{30} + x_{4}x_{31} + x_{4}x_{32} + x_{4}x_{33} + x_{4}x_{34} + x_{4}x_{39} + x_{4}x_{40} + x_{4}x_{42} + x_{4}x_{44} + x_{4}x_{48} + x_{4}x_{50} + x_{4}x_{51} + x_{4}x_{52} + x_{4}x_{56} + x_{4}x_{61} + x_{4}x_{62} + x_{4}x_{63} + x_{5}x_{6} + x_{5}x_{9} + x_{5}x_{10} + x_{5}x_{11} + x_{5}x_{12} + x_{5}x_{14} + x_{5}x_{15} + x_{5}x_{16} + x_{5}x_{17} + x_{5}x_{19} + x_{5}x_{22} + x_{5}x_{23} + x_{5}x_{25} + x_{5}x_{26} + x_{5}x_{33} + x_{5}x_{35} + x_{5}x_{36} + x_{5}x_{39} + x_{5}x_{40} + x_{5}x_{41} + x_{5}x_{43} + x_{5}x_{44} + x_{5}x_{46} + x_{5}x_{47} + x_{5}x_{48} + x_{5}x_{49} + x_{5}x_{51} + x_{5}x_{54} + x_{5}x_{55} + x_{5}x_{56} + x_{5}x_{58} + x_{5}x_{60} + x_{6}x_{8} + x_{6}x_{10} + x_{6}x_{11} + x_{6}x_{12} + x_{6}x_{14} + x_{6}x_{17} + x_{6}x_{19} + x_{6}x_{23} + x_{6}x_{24} + x_{6}x_{25} + x_{6}x_{31} + x_{6}x_{32} + x_{6}x_{37} + x_{6}x_{38} + x_{6}x_{39} + x_{6}x_{43} + x_{6}x_{44} + x_{6}x_{46} + x_{6}x_{49} + x_{6}x_{50} + x_{6}x_{55} + x_{6}x_{57} + x_{6}x_{59} + x_{6}x_{60} + x_{6}x_{61} + x_{6}x_{64} + x_{7}x_{8} + x_{7}x_{10} + x_{7}x_{11} + x_{7}x_{12} + x_{7}x_{14} + x_{7}x_{15} + x_{7}x_{17} + x_{7}x_{18} + x_{7}x_{19} + x_{7}x_{20} + x_{7}x_{24} + x_{7}x_{27} + x_{7}x_{28} + x_{7}x_{29} + x_{7}x_{30} + x_{7}x_{32} + x_{7}x_{33} + x_{7}x_{34} + x_{7}x_{36} + x_{7}x_{37} + x_{7}x_{38} + x_{7}x_{40} + x_{7}x_{43} + x_{7}x_{46} + x_{7}x_{47} + x_{7}x_{51} + x_{7}x_{53} + x_{7}x_{54} + x_{7}x_{56} + x_{7}x_{57} + x_{7}x_{59} + x_{7}x_{62} + x_{7}x_{63} + x_{8}x_{9} + x_{8}x_{10} + x_{8}x_{12} + x_{8}x_{13} + x_{8}x_{17} + x_{8}x_{19} + x_{8}x_{21} + x_{8}x_{23} + x_{8}x_{27} + x_{8}x_{28} + x_{8}x_{30} + x_{8}x_{31} + x_{8}x_{32} + x_{8}x_{33} + x_{8}x_{34} + x_{8}x_{36} + x_{8}x_{38} + x_{8}x_{39} + x_{8}x_{40} + x_{8}x_{41} + x_{8}x_{43} + x_{8}x_{45} + x_{8}x_{46} + x_{8}x_{47} + x_{8}x_{50} + x_{8}x_{54} + x_{8}x_{55} + x_{8}x_{56} + x_{8}x_{62} + x_{8}x_{63} + x_{8}x_{64} + x_{9}x_{11} + x_{9}x_{14} + x_{9}x_{16} + x_{9}x_{18} + x_{9}x_{23} + x_{9}x_{25} + x_{9}x_{26} + x_{9}x_{30} + x_{9}x_{32} + x_{9}x_{35} + x_{9}x_{36} + x_{9}x_{37} + x_{9}x_{40} + x_{9}x_{42} + x_{9}x_{43} + x_{9}x_{44} + x_{9}x_{45} + x_{9}x_{46} + x_{9}x_{47} + x_{9}x_{48} + x_{9}x_{49} + x_{9}x_{50} + x_{9}x_{51} + x_{9}x_{54} + x_{9}x_{56} + x_{9}x_{60} + x_{9}x_{62} + x_{10}x_{11} + x_{10}x_{12} + x_{10}x_{13} + x_{10}x_{14} + x_{10}x_{18} + x_{10}x_{19} + x_{10}x_{22} + x_{10}x_{23} + x_{10}x_{24} + x_{10}x_{25} + x_{10}x_{28} + x_{10}x_{29} + x_{10}x_{34} + x_{10}x_{35} + x_{10}x_{36} + x_{10}x_{37} + x_{10}x_{40} + x_{10}x_{41} + x_{10}x_{43} + x_{10}x_{44} + x_{10}x_{47} + x_{10}x_{48} + x_{10}x_{50} + x_{10}x_{52} + x_{10}x_{54} + x_{10}x_{56} + x_{10}x_{57} + x_{10}x_{62} + x_{10}x_{63} + x_{11}x_{12} + x_{11}x_{13} + x_{11}x_{18} + x_{11}x_{21} + x_{11}x_{22} + x_{11}x_{23} + x_{11}x_{25} + x_{11}x_{26} + x_{11}x_{29} + x_{11}x_{30} + x_{11}x_{31} + x_{11}x_{33} + x_{11}x_{34} + x_{11}x_{35} + x_{11}x_{36} + x_{11}x_{42} + x_{11}x_{51} + x_{11}x_{52} + x_{11}x_{53} + x_{11}x_{56} + x_{11}x_{59} + x_{11}x_{60} + x_{11}x_{64} + x_{12}x_{15} + x_{12}x_{16} + x_{12}x_{19} + x_{12}x_{20} + x_{12}x_{23} + x_{12}x_{25} + x_{12}x_{26} + x_{12}x_{27} + x_{12}x_{29} + x_{12}x_{30} + x_{12}x_{31} + x_{12}x_{32} + x_{12}x_{33} + x_{12}x_{34} + x_{12}x_{38} + x_{12}x_{39} + x_{12}x_{40} + x_{12}x_{42} + x_{12}x_{43} + x_{12}x_{44} + x_{12}x_{45} + x_{12}x_{48} + x_{12}x_{51} + x_{12}x_{53} + x_{12}x_{54} + x_{12}x_{55} + x_{12}x_{56} + x_{12}x_{59} + x_{12}x_{60} + x_{12}x_{61} + x_{12}x_{62} + x_{12}x_{64} + x_{13}x_{14} + x_{13}x_{15} + x_{13}x_{16} + x_{13}x_{20} + x_{13}x_{21} + x_{13}x_{23} + x_{13}x_{25} + x_{13}x_{27} + x_{13}x_{28} + x_{13}x_{30} + x_{13}x_{31} + x_{13}x_{33} + x_{13}x_{36} + x_{13}x_{37} + x_{13}x_{39} + x_{13}x_{45} + x_{13}x_{46} + x_{13}x_{47} + x_{13}x_{48} + x_{13}x_{51} + x_{13}x_{55} + x_{13}x_{57} + x_{13}x_{61} + x_{13}x_{62} + x_{13}x_{63} + x_{14}x_{15} + x_{14}x_{18} + x_{14}x_{21} + x_{14}x_{22} + x_{14}x_{24} + x_{14}x_{25} + x_{14}x_{26} + x_{14}x_{27} + x_{14}x_{31} + x_{14}x_{33} + x_{14}x_{34} + x_{14}x_{36} + x_{14}x_{38} + x_{14}x_{39} + x_{14}x_{40} + x_{14}x_{41} + x_{14}x_{43} + x_{14}x_{44} + x_{14}x_{46} + x_{14}x_{47} + x_{14}x_{48} + x_{14}x_{50} + x_{14}x_{51} + x_{14}x_{55} + x_{14}x_{56} + x_{14}x_{57} + x_{14}x_{59} + x_{14}x_{60} + x_{14}x_{61} + x_{14}x_{63} + x_{14}x_{64} + x_{15}x_{20} + x_{15}x_{22} + x_{15}x_{23} + x_{15}x_{24} + x_{15}x_{25} + x_{15}x_{27} + x_{15}x_{28} + x_{15}x_{31} + x_{15}x_{32} + x_{15}x_{33} + x_{15}x_{34} + x_{15}x_{39} + x_{15}x_{41} + x_{15}x_{43} + x_{15}x_{44} + x_{15}x_{45} + x_{15}x_{46} + x_{15}x_{48} + x_{15}x_{49} + x_{15}x_{50} + x_{15}x_{51} + x_{15}x_{53} + x_{15}x_{55} + x_{15}x_{57} + x_{15}x_{64} + x_{16}x_{18} + x_{16}x_{19} + x_{16}x_{20} + x_{16}x_{21} + x_{16}x_{22} + x_{16}x_{23} + x_{16}x_{24} + x_{16}x_{25} + x_{16}x_{33} + x_{16}x_{34} + x_{16}x_{36} + x_{16}x_{43} + x_{16}x_{44} + x_{16}x_{46} + x_{16}x_{47} + x_{16}x_{50} + x_{16}x_{53} + x_{16}x_{54} + x_{16}x_{55} + x_{16}x_{59} + x_{16}x_{61} + x_{16}x_{64} + x_{17}x_{21} + x_{17}x_{22} + x_{17}x_{24} + x_{17}x_{26} + x_{17}x_{27} + x_{17}x_{28} + x_{17}x_{30} + x_{17}x_{31} + x_{17}x_{33} + x_{17}x_{34} + x_{17}x_{37} + x_{17}x_{39} + x_{17}x_{41} + x_{17}x_{43} + x_{17}x_{46} + x_{17}x_{47} + x_{17}x_{51} + x_{17}x_{53} + x_{17}x_{55} + x_{17}x_{57} + x_{17}x_{61} + x_{17}x_{63} + x_{18}x_{19} + x_{18}x_{20} + x_{18}x_{21} + x_{18}x_{22} + x_{18}x_{24} + x_{18}x_{27} + x_{18}x_{28} + x_{18}x_{32} + x_{18}x_{35} + x_{18}x_{38} + x_{18}x_{39} + x_{18}x_{40} + x_{18}x_{41} + x_{18}x_{42} + x_{18}x_{46} + x_{18}x_{47} + x_{18}x_{50} + x_{18}x_{51} + x_{18}x_{55} + x_{18}x_{56} + x_{18}x_{57} + x_{18}x_{60} + x_{18}x_{61} + x_{18}x_{63} + x_{18}x_{64} + x_{19}x_{22} + x_{19}x_{27} + x_{19}x_{28} + x_{19}x_{29} + x_{19}x_{32} + x_{19}x_{36} + x_{19}x_{37} + x_{19}x_{41} + x_{19}x_{44} + x_{19}x_{45} + x_{19}x_{48} + x_{19}x_{49} + x_{19}x_{50} + x_{19}x_{52} + x_{19}x_{53} + x_{19}x_{56} + x_{19}x_{58} + x_{19}x_{59} + x_{19}x_{60} + x_{19}x_{64} + x_{20}x_{21} + x_{20}x_{24} + x_{20}x_{26} + x_{20}x_{27} + x_{20}x_{31} + x_{20}x_{32} + x_{20}x_{33} + x_{20}x_{34} + x_{20}x_{35} + x_{20}x_{38} + x_{20}x_{39} + x_{20}x_{40} + x_{20}x_{49} + x_{20}x_{51} + x_{20}x_{52} + x_{20}x_{53} + x_{20}x_{59} + x_{20}x_{61} + x_{21}x_{22} + x_{21}x_{23} + x_{21}x_{24} + x_{21}x_{27} + x_{21}x_{28} + x_{21}x_{33} + x_{21}x_{36} + x_{21}x_{38} + x_{21}x_{39} + x_{21}x_{42} + x_{21}x_{43} + x_{21}x_{44} + x_{21}x_{55} + x_{21}x_{56} + x_{21}x_{59} + x_{21}x_{60} + x_{21}x_{61} + x_{21}x_{62} + x_{21}x_{64} + x_{22}x_{23} + x_{22}x_{26} + x_{22}x_{28} + x_{22}x_{29} + x_{22}x_{30} + x_{22}x_{31} + x_{22}x_{32} + x_{22}x_{36} + x_{22}x_{38} + x_{22}x_{39} + x_{22}x_{40} + x_{22}x_{42} + x_{22}x_{44} + x_{22}x_{48} + x_{22}x_{49} + x_{22}x_{50} + x_{22}x_{51} + x_{22}x_{52} + x_{22}x_{53} + x_{22}x_{54} + x_{22}x_{57} + x_{22}x_{60} + x_{22}x_{61} + x_{22}x_{62} + x_{23}x_{24} + x_{23}x_{25} + x_{23}x_{27} + x_{23}x_{31} + x_{23}x_{33} + x_{23}x_{35} + x_{23}x_{36} + x_{23}x_{40} + x_{23}x_{41} + x_{23}x_{46} + x_{23}x_{50} + x_{23}x_{51} + x_{23}x_{53} + x_{23}x_{54} + x_{23}x_{57} + x_{23}x_{58} + x_{23}x_{59} + x_{23}x_{62} + x_{23}x_{64} + x_{24}x_{29} + x_{24}x_{30} + x_{24}x_{32} + x_{24}x_{33} + x_{24}x_{34} + x_{24}x_{37} + x_{24}x_{38} + x_{24}x_{45} + x_{24}x_{46} + x_{24}x_{49} + x_{24}x_{51} + x_{24}x_{52} + x_{24}x_{55} + x_{24}x_{56} + x_{24}x_{59} + x_{24}x_{62} + x_{24}x_{64} + x_{25}x_{27} + x_{25}x_{37} + x_{25}x_{40} + x_{25}x_{41} + x_{25}x_{43} + x_{25}x_{44} + x_{25}x_{47} + x_{25}x_{49} + x_{25}x_{50} + x_{25}x_{51} + x_{25}x_{53} + x_{25}x_{54} + x_{25}x_{56} + x_{25}x_{58} + x_{25}x_{60} + x_{25}x_{61} + x_{25}x_{62} + x_{25}x_{63} + x_{26}x_{27} + x_{26}x_{29} + x_{26}x_{32} + x_{26}x_{33} + x_{26}x_{34} + x_{26}x_{35} + x_{26}x_{36} + x_{26}x_{37} + x_{26}x_{38} + x_{26}x_{39} + x_{26}x_{40} + x_{26}x_{42} + x_{26}x_{45} + x_{26}x_{49} + x_{26}x_{50} + x_{26}x_{51} + x_{26}x_{52} + x_{26}x_{54} + x_{26}x_{57} + x_{26}x_{58} + x_{26}x_{60} + x_{26}x_{61} + x_{26}x_{63} + x_{26}x_{64} + x_{27}x_{28} + x_{27}x_{31} + x_{27}x_{32} + x_{27}x_{33} + x_{27}x_{36} + x_{27}x_{41} + x_{27}x_{42} + x_{27}x_{45} + x_{27}x_{47} + x_{27}x_{48} + x_{27}x_{49} + x_{27}x_{50} + x_{27}x_{51} + x_{27}x_{53} + x_{27}x_{54} + x_{27}x_{55} + x_{27}x_{56} + x_{27}x_{57} + x_{27}x_{59} + x_{27}x_{60} + x_{27}x_{62} + x_{28}x_{29} + x_{28}x_{30} + x_{28}x_{31} + x_{28}x_{35} + x_{28}x_{36} + x_{28}x_{37} + x_{28}x_{40} + x_{28}x_{41} + x_{28}x_{43} + x_{28}x_{48} + x_{28}x_{49} + x_{28}x_{50} + x_{28}x_{53} + x_{28}x_{54} + x_{28}x_{56} + x_{28}x_{57} + x_{28}x_{58} + x_{28}x_{59} + x_{28}x_{61} + x_{28}x_{62} + x_{28}x_{63} + x_{29}x_{33} + x_{29}x_{36} + x_{29}x_{40} + x_{29}x_{41} + x_{29}x_{42} + x_{29}x_{47} + x_{29}x_{48} + x_{29}x_{49} + x_{29}x_{51} + x_{29}x_{53} + x_{29}x_{55} + x_{29}x_{58} + x_{29}x_{60} + x_{29}x_{62} + x_{30}x_{31} + x_{30}x_{32} + x_{30}x_{37} + x_{30}x_{38} + x_{30}x_{39} + x_{30}x_{43} + x_{30}x_{44} + x_{30}x_{45} + x_{30}x_{47} + x_{30}x_{49} + x_{30}x_{53} + x_{30}x_{55} + x_{30}x_{57} + x_{30}x_{61} + x_{30}x_{63} + x_{30}x_{64} + x_{31}x_{32} + x_{31}x_{33} + x_{31}x_{34} + x_{31}x_{35} + x_{31}x_{37} + x_{31}x_{38} + x_{31}x_{39} + x_{31}x_{45} + x_{31}x_{46} + x_{31}x_{49} + x_{31}x_{55} + x_{31}x_{56} + x_{31}x_{57} + x_{31}x_{59} + x_{31}x_{61} + x_{31}x_{62} + x_{31}x_{63} + x_{32}x_{33} + x_{32}x_{34} + x_{32}x_{36} + x_{32}x_{38} + x_{32}x_{39} + x_{32}x_{42} + x_{32}x_{43} + x_{32}x_{45} + x_{32}x_{48} + x_{32}x_{49} + x_{32}x_{50} + x_{32}x_{51} + x_{32}x_{52} + x_{32}x_{53} + x_{32}x_{55} + x_{32}x_{57} + x_{32}x_{58} + x_{32}x_{59} + x_{32}x_{61} + x_{32}x_{62} + x_{32}x_{63} + x_{32}x_{64} + x_{33}x_{34} + x_{33}x_{35} + x_{33}x_{36} + x_{33}x_{37} + x_{33}x_{39} + x_{33}x_{41} + x_{33}x_{43} + x_{33}x_{45} + x_{33}x_{48} + x_{33}x_{49} + x_{33}x_{51} + x_{33}x_{52} + x_{33}x_{58} + x_{33}x_{61} + x_{33}x_{63} + x_{34}x_{35} + x_{34}x_{37} + x_{34}x_{38} + x_{34}x_{39} + x_{34}x_{40} + x_{34}x_{43} + x_{34}x_{48} + x_{34}x_{51} + x_{34}x_{53} + x_{34}x_{55} + x_{34}x_{56} + x_{34}x_{57} + x_{34}x_{59} + x_{34}x_{60} + x_{34}x_{62} + x_{34}x_{64} + x_{35}x_{36} + x_{35}x_{37} + x_{35}x_{38} + x_{35}x_{40} + x_{35}x_{41} + x_{35}x_{42} + x_{35}x_{43} + x_{35}x_{45} + x_{35}x_{46} + x_{35}x_{49} + x_{35}x_{50} + x_{35}x_{52} + x_{35}x_{53} + x_{35}x_{58} + x_{35}x_{60} + x_{35}x_{63} + x_{35}x_{64} + x_{36}x_{37} + x_{36}x_{38} + x_{36}x_{39} + x_{36}x_{40} + x_{36}x_{41} + x_{36}x_{47} + x_{36}x_{48} + x_{36}x_{50} + x_{36}x_{51} + x_{36}x_{52} + x_{36}x_{53} + x_{36}x_{54} + x_{36}x_{55} + x_{36}x_{56} + x_{36}x_{59} + x_{36}x_{60} + x_{36}x_{61} + x_{36}x_{62} + x_{36}x_{64} + x_{37}x_{40} + x_{37}x_{42} + x_{37}x_{43} + x_{37}x_{44} + x_{37}x_{49} + x_{37}x_{51} + x_{37}x_{52} + x_{37}x_{53} + x_{37}x_{56} + x_{37}x_{61} + x_{37}x_{62} + x_{37}x_{63} + x_{37}x_{64} + x_{38}x_{39} + x_{38}x_{41} + x_{38}x_{43} + x_{38}x_{44} + x_{38}x_{45} + x_{38}x_{47} + x_{38}x_{51} + x_{38}x_{52} + x_{38}x_{53} + x_{38}x_{55} + x_{38}x_{56} + x_{38}x_{57} + x_{38}x_{59} + x_{38}x_{60} + x_{38}x_{62} + x_{39}x_{41} + x_{39}x_{43} + x_{39}x_{44} + x_{39}x_{49} + x_{39}x_{50} + x_{39}x_{51} + x_{39}x_{53} + x_{39}x_{54} + x_{39}x_{55} + x_{39}x_{59} + x_{39}x_{60} + x_{39}x_{62} + x_{39}x_{64} + x_{40}x_{41} + x_{40}x_{48} + x_{40}x_{51} + x_{40}x_{55} + x_{40}x_{58} + x_{40}x_{59} + x_{40}x_{60} + x_{40}x_{62} + x_{40}x_{64} + x_{41}x_{45} + x_{41}x_{47} + x_{41}x_{48} + x_{41}x_{49} + x_{41}x_{50} + x_{41}x_{51} + x_{41}x_{53} + x_{41}x_{56} + x_{41}x_{59} + x_{41}x_{61} + x_{41}x_{63} + x_{41}x_{64} + x_{42}x_{43} + x_{42}x_{44} + x_{42}x_{46} + x_{42}x_{48} + x_{42}x_{50} + x_{42}x_{51} + x_{42}x_{56} + x_{42}x_{57} + x_{42}x_{61} + x_{42}x_{63} + x_{43}x_{45} + x_{43}x_{46} + x_{43}x_{47} + x_{43}x_{49} + x_{43}x_{50} + x_{43}x_{51} + x_{43}x_{52} + x_{43}x_{53} + x_{43}x_{54} + x_{43}x_{56} + x_{43}x_{59} + x_{43}x_{62} + x_{43}x_{63} + x_{44}x_{45} + x_{44}x_{46} + x_{44}x_{47} + x_{44}x_{49} + x_{44}x_{50} + x_{44}x_{51} + x_{44}x_{52} + x_{44}x_{53} + x_{44}x_{54} + x_{44}x_{56} + x_{44}x_{57} + x_{44}x_{61} + x_{44}x_{63} + x_{44}x_{64} + x_{45}x_{46} + x_{45}x_{47} + x_{45}x_{50} + x_{45}x_{53} + x_{45}x_{54} + x_{45}x_{56} + x_{45}x_{58} + x_{45}x_{59} + x_{45}x_{61} + x_{46}x_{48} + x_{46}x_{50} + x_{46}x_{55} + x_{46}x_{56} + x_{46}x_{57} + x_{46}x_{58} + x_{46}x_{60} + x_{46}x_{61} + x_{46}x_{64} + x_{47}x_{48} + x_{47}x_{50} + x_{47}x_{51} + x_{47}x_{52} + x_{47}x_{55} + x_{47}x_{56} + x_{47}x_{57} + x_{47}x_{59} + x_{47}x_{61} + x_{47}x_{62} + x_{47}x_{64} + x_{48}x_{49} + x_{48}x_{51} + x_{48}x_{54} + x_{48}x_{56} + x_{48}x_{59} + x_{48}x_{60} + x_{48}x_{61} + x_{48}x_{63} + x_{48}x_{64} + x_{49}x_{51} + x_{49}x_{52} + x_{49}x_{54} + x_{49}x_{55} + x_{49}x_{56} + x_{49}x_{59} + x_{49}x_{62} + x_{49}x_{63} + x_{50}x_{51} + x_{50}x_{53} + x_{50}x_{57} + x_{50}x_{58} + x_{50}x_{59} + x_{50}x_{60} + x_{50}x_{61} + x_{50}x_{62} + x_{50}x_{63} + x_{50}x_{64} + x_{51}x_{53} + x_{51}x_{54} + x_{51}x_{56} + x_{51}x_{58} + x_{51}x_{60} + x_{51}x_{64} + x_{52}x_{53} + x_{52}x_{54} + x_{52}x_{55} + x_{52}x_{56} + x_{52}x_{58} + x_{52}x_{59} + x_{52}x_{62} + x_{52}x_{63} + x_{53}x_{54} + x_{53}x_{55} + x_{53}x_{58} + x_{53}x_{60} + x_{53}x_{61} + x_{53}x_{62} + x_{53}x_{63} + x_{54}x_{55} + x_{54}x_{58} + x_{54}x_{60} + x_{54}x_{62} + x_{54}x_{63} + x_{55}x_{56} + x_{55}x_{57} + x_{55}x_{58} + x_{55}x_{61} + x_{55}x_{63} + x_{56}x_{57} + x_{56}x_{60} + x_{56}x_{62} + x_{56}x_{63} + x_{57}x_{58} + x_{57}x_{59} + x_{57}x_{60} + x_{57}x_{63} + x_{58}x_{59} + x_{58}x_{61} + x_{58}x_{62} + x_{58}x_{63} + x_{59}x_{62} + x_{60}x_{63} + x_{61}x_{62} + x_{62}x_{63} + x_{1} + x_{2} + x_{4} + x_{5} + x_{9} + x_{10} + x_{11} + x_{12} + x_{14} + x_{17} + x_{18} + x_{19} + x_{20} + x_{21} + x_{22} + x_{23} + x_{24} + x_{25} + x_{27} + x_{31} + x_{32} + x_{38} + x_{40} + x_{41} + x_{42} + x_{43} + x_{45} + x_{47} + x_{50} + x_{52} + x_{54} + x_{56} + x_{60} + x_{63} + 1$

$y_{3} = x_{1}x_{2} + x_{1}x_{3} + x_{1}x_{5} + x_{1}x_{7} + x_{1}x_{13} + x_{1}x_{18} + x_{1}x_{22} + x_{1}x_{24} + x_{1}x_{25} + x_{1}x_{26} + x_{1}x_{27} + x_{1}x_{30} + x_{1}x_{31} + x_{1}x_{33} + x_{1}x_{34} + x_{1}x_{35} + x_{1}x_{37} + x_{1}x_{38} + x_{1}x_{39} + x_{1}x_{42} + x_{1}x_{44} + x_{1}x_{45} + x_{1}x_{47} + x_{1}x_{48} + x_{1}x_{49} + x_{1}x_{50} + x_{1}x_{52} + x_{1}x_{54} + x_{1}x_{55} + x_{1}x_{56} + x_{1}x_{57} + x_{1}x_{58} + x_{1}x_{61} + x_{1}x_{64} + x_{2}x_{3} + x_{2}x_{7} + x_{2}x_{8} + x_{2}x_{9} + x_{2}x_{10} + x_{2}x_{13} + x_{2}x_{14} + x_{2}x_{16} + x_{2}x_{18} + x_{2}x_{21} + x_{2}x_{22} + x_{2}x_{23} + x_{2}x_{24} + x_{2}x_{26} + x_{2}x_{27} + x_{2}x_{29} + x_{2}x_{36} + x_{2}x_{37} + x_{2}x_{41} + x_{2}x_{42} + x_{2}x_{43} + x_{2}x_{45} + x_{2}x_{47} + x_{2}x_{48} + x_{2}x_{49} + x_{2}x_{50} + x_{2}x_{56} + x_{2}x_{58} + x_{2}x_{59} + x_{2}x_{62} + x_{2}x_{63} + x_{3}x_{5} + x_{3}x_{7} + x_{3}x_{8} + x_{3}x_{9} + x_{3}x_{10} + x_{3}x_{13} + x_{3}x_{15} + x_{3}x_{18} + x_{3}x_{20} + x_{3}x_{21} + x_{3}x_{22} + x_{3}x_{23} + x_{3}x_{24} + x_{3}x_{31} + x_{3}x_{32} + x_{3}x_{33} + x_{3}x_{34} + x_{3}x_{35} + x_{3}x_{38} + x_{3}x_{41} + x_{3}x_{42} + x_{3}x_{43} + x_{3}x_{44} + x_{3}x_{46} + x_{3}x_{47} + x_{3}x_{48} + x_{3}x_{50} + x_{3}x_{52} + x_{3}x_{53} + x_{3}x_{56} + x_{3}x_{57} + x_{3}x_{58} + x_{3}x_{59} + x_{3}x_{61} + x_{3}x_{62} + x_{3}x_{64} + x_{4}x_{5} + x_{4}x_{6} + x_{4}x_{12} + x_{4}x_{14} + x_{4}x_{15} + x_{4}x_{21} + x_{4}x_{23} + x_{4}x_{26} + x_{4}x_{27} + x_{4}x_{30} + x_{4}x_{33} + x_{4}x_{35} + x_{4}x_{36} + x_{4}x_{37} + x_{4}x_{38} + x_{4}x_{39} + x_{4}x_{40} + x_{4}x_{42} + x_{4}x_{43} + x_{4}x_{44} + x_{4}x_{46} + x_{4}x_{47} + x_{4}x_{50} + x_{4}x_{51} + x_{4}x_{52} + x_{4}x_{54} + x_{4}x_{56} + x_{4}x_{58} + x_{4}x_{59} + x_{4}x_{60} + x_{4}x_{62} + x_{5}x_{6} + x_{5}x_{7} + x_{5}x_{8} + x_{5}x_{12} + x_{5}x_{15} + x_{5}x_{16} + x_{5}x_{17} + x_{5}x_{18} + x_{5}x_{21} + x_{5}x_{26} + x_{5}x_{27} + x_{5}x_{29} + x_{5}x_{32} + x_{5}x_{34} + x_{5}x_{36} + x_{5}x_{38} + x_{5}x_{39} + x_{5}x_{40} + x_{5}x_{41} + x_{5}x_{43} + x_{5}x_{44} + x_{5}x_{46} + x_{5}x_{47} + x_{5}x_{48} + x_{5}x_{49} + x_{5}x_{51} + x_{5}x_{54} + x_{5}x_{55} + x_{5}x_{58} + x_{5}x_{62} + x_{5}x_{63} + x_{6}x_{7} + x_{6}x_{8} + x_{6}x_{10} + x_{6}x_{12} + x_{6}x_{16} + x_{6}x_{20} + x_{6}x_{24} + x_{6}x_{25} + x_{6}x_{26} + x_{6}x_{30} + x_{6}x_{33} + x_{6}x_{34} + x_{6}x_{35} + x_{6}x_{38} + x_{6}x_{39} + x_{6}x_{41} + x_{6}x_{46} + x_{6}x_{47} + x_{6}x_{48} + x_{6}x_{52} + x_{6}x_{53} + x_{6}x_{54} + x_{6}x_{56} + x_{6}x_{57} + x_{6}x_{59} + x_{6}x_{60} + x_{6}x_{61} + x_{6}x_{63} + x_{6}x_{64} + x_{7}x_{8} + x_{7}x_{9} + x_{7}x_{10} + x_{7}x_{12} + x_{7}x_{14} + x_{7}x_{16} + x_{7}x_{17} + x_{7}x_{18} + x_{7}x_{20} + x_{7}x_{21} + x_{7}x_{23} + x_{7}x_{28} + x_{7}x_{29} + x_{7}x_{30} + x_{7}x_{31} + x_{7}x_{32} + x_{7}x_{33} + x_{7}x_{36} + x_{7}x_{37} + x_{7}x_{41} + x_{7}x_{42} + x_{7}x_{44} + x_{7}x_{48} + x_{7}x_{49} + x_{7}x_{53} + x_{7}x_{54} + x_{7}x_{55} + x_{7}x_{56} + x_{7}x_{58} + x_{7}x_{59} + x_{7}x_{61} + x_{7}x_{62} + x_{7}x_{63} + x_{8}x_{11} + x_{8}x_{12} + x_{8}x_{15} + x_{8}x_{20} + x_{8}x_{21} + x_{8}x_{22} + x_{8}x_{23} + x_{8}x_{24} + x_{8}x_{26} + x_{8}x_{29} + x_{8}x_{31} + x_{8}x_{32} + x_{8}x_{33} + x_{8}x_{34} + x_{8}x_{35} + x_{8}x_{37} + x_{8}x_{40} + x_{8}x_{43} + x_{8}x_{45} + x_{8}x_{47} + x_{8}x_{52} + x_{8}x_{53} + x_{8}x_{54} + x_{8}x_{59} + x_{8}x_{63} + x_{8}x_{64} + x_{9}x_{10} + x_{9}x_{12} + x_{9}x_{14} + x_{9}x_{15} + x_{9}x_{25} + x_{9}x_{26} + x_{9}x_{30} + x_{9}x_{35} + x_{9}x_{38} + x_{9}x_{39} + x_{9}x_{41} + x_{9}x_{46} + x_{9}x_{48} + x_{9}x_{53} + x_{9}x_{54} + x_{9}x_{55} + x_{9}x_{56} + x_{9}x_{58} + x_{9}x_{59} + x_{9}x_{61} + x_{9}x_{63} + x_{9}x_{64} + x_{10}x_{12} + x_{10}x_{13} + x_{10}x_{14} + x_{10}x_{15} + x_{10}x_{16} + x_{10}x_{17} + x_{10}x_{18} + x_{10}x_{23} + x_{10}x_{25} + x_{10}x_{28} + x_{10}x_{29} + x_{10}x_{33} + x_{10}x_{34} + x_{10}x_{36} + x_{10}x_{37} + x_{10}x_{39} + x_{10}x_{40} + x_{10}x_{42} + x_{10}x_{45} + x_{10}x_{46} + x_{10}x_{47} + x_{10}x_{48} + x_{10}x_{49} + x_{10}x_{50} + x_{10}x_{51} + x_{10}x_{53} + x_{10}x_{54} + x_{10}x_{56} + x_{10}x_{57} + x_{10}x_{58} + x_{10}x_{63} + x_{11}x_{12} + x_{11}x_{15} + x_{11}x_{16} + x_{11}x_{17} + x_{11}x_{18} + x_{11}x_{19} + x_{11}x_{20} + x_{11}x_{21} + x_{11}x_{24} + x_{11}x_{27} + x_{11}x_{29} + x_{11}x_{32} + x_{11}x_{33} + x_{11}x_{35} + x_{11}x_{36} + x_{11}x_{37} + x_{11}x_{41} + x_{11}x_{43} + x_{11}x_{46} + x_{11}x_{47} + x_{11}x_{50} + x_{11}x_{54} + x_{11}x_{55} + x_{11}x_{57} + x_{11}x_{58} + x_{11}x_{63} + x_{11}x_{64} + x_{12}x_{14} + x_{12}x_{15} + x_{12}x_{16} + x_{12}x_{18} + x_{12}x_{19} + x_{12}x_{20} + x_{12}x_{22} + x_{12}x_{25} + x_{12}x_{26} + x_{12}x_{27} + x_{12}x_{28} + x_{12}x_{29} + x_{12}x_{31} + x_{12}x_{35} + x_{12}x_{37} + x_{12}x_{38} + x_{12}x_{39} + x_{12}x_{43} + x_{12}x_{44} + x_{12}x_{47} + x_{12}x_{48} + x_{12}x_{50} + x_{12}x_{51} + x_{12}x_{52} + x_{12}x_{55} + x_{12}x_{58} + x_{12}x_{59} + x_{12}x_{60} + x_{12}x_{61} + x_{13}x_{14} + x_{13}x_{15} + x_{13}x_{17} + x_{13}x_{18} + x_{13}x_{19} + x_{13}x_{21} + x_{13}x_{22} + x_{13}x_{25} + x_{13}x_{27} + x_{13}x_{28} + x_{13}x_{29} + x_{13}x_{30} + x_{13}x_{33} + x_{13}x_{34} + x_{13}x_{35} + x_{13}x_{36} + x_{13}x_{37} + x_{13}x_{38} + x_{13}x_{40} + x_{13}x_{45} + x_{13}x_{47} + x_{13}x_{49} + x_{13}x_{50} + x_{13}x_{52} + x_{13}x_{53} + x_{13}x_{54} + x_{13}x_{56} + x_{13}x_{57} + x_{13}x_{58} + x_{13}x_{62} + x_{13}x_{63} + x_{13}x_{64} + x_{14}x_{20} + x_{14}x_{21} + x_{14}x_{22} + x_{14}x_{24} + x_{14}x_{25} + x_{14}x_{28} + x_{14}x_{29} + x_{14}x_{36} + x_{14}x_{38} + x_{14}x_{41} + x_{14}x_{42} + x_{14}x_{43} + x_{14}x_{48} + x_{14}x_{50} + x_{14}x_{51} + x_{14}x_{52} + x_{14}x_{54} + x_{14}x_{56} + x_{14}x_{58} + x_{14}x_{59} + x_{14}x_{61} + x_{15}x_{18} + x_{15}x_{22} + x_{15}x_{25} + x_{15}x_{29} + x_{15}x_{31} + x_{15}x_{33} + x_{15}x_{34} + x_{15}x_{39} + x_{15}x_{41} + x_{15}x_{45} + x_{15}x_{47} + x_{15}x_{48} + x_{15}x_{49} + x_{15}x_{50} + x_{15}x_{51} + x_{15}x_{54} + x_{15}x_{56} + x_{15}x_{57} + x_{15}x_{58} + x_{15}x_{60} + x_{15}x_{61} + x_{15}x_{64} + x_{16}x_{17} + x_{16}x_{23} + x_{16}x_{27} + x_{16}x_{28} + x_{16}x_{29} + x_{16}x_{33} + x_{16}x_{35} + x_{16}x_{36} + x_{16}x_{39} + x_{16}x_{41} + x_{16}x_{48} + x_{16}x_{49} + x_{16}x_{51} + x_{16}x_{57} + x_{16}x_{59} + x_{16}x_{60} + x_{16}x_{63} + x_{17}x_{20} + x_{17}x_{21} + x_{17}x_{22} + x_{17}x_{23} + x_{17}x_{24} + x_{17}x_{26} + x_{17}x_{28} + x_{17}x_{30} + x_{17}x_{32} + x_{17}x_{33} + x_{17}x_{34} + x_{17}x_{35} + x_{17}x_{38} + x_{17}x_{40} + x_{17}x_{41} + x_{17}x_{43} + x_{17}x_{46} + x_{17}x_{49} + x_{17}x_{51} + x_{17}x_{52} + x_{17}x_{53} + x_{17}x_{54} + x_{17}x_{55} + x_{17}x_{57} + x_{17}x_{58} + x_{17}x_{62} + x_{17}x_{64} + x_{18}x_{20} + x_{18}x_{23} + x_{18}x_{25} + x_{18}x_{32} + x_{18}x_{33} + x_{18}x_{34} + x_{18}x_{37} + x_{18}x_{38} + x_{18}x_{43} + x_{18}x_{44} + x_{18}x_{45} + x_{18}x_{47} + x_{18}x_{49} + x_{18}x_{54} + x_{18}x_{55} + x_{18}x_{56} + x_{18}x_{57} + x_{18}x_{58} + x_{18}x_{59} + x_{18}x_{60} + x_{18}x_{63} + x_{18}x_{64} + x_{19}x_{21} + x_{19}x_{23} + x_{19}x_{24} + x_{19}x_{25} + x_{19}x_{26} + x_{19}x_{28} + x_{19}x_{30} + x_{19}x_{35} + x_{19}x_{36} + x_{19}x_{40} + x_{19}x_{43} + x_{19}x_{48} + x_{19}x_{49} + x_{19}x_{51} + x_{19}x_{53} + x_{19}x_{54} + x_{19}x_{55} + x_{19}x_{57} + x_{19}x_{58} + x_{19}x_{60} + x_{19}x_{61} + x_{19}x_{64} + x_{20}x_{22} + x_{20}x_{23} + x_{20}x_{25} + x_{20}x_{27} + x_{20}x_{30} + x_{20}x_{31} + x_{20}x_{32} + x_{20}x_{33} + x_{20}x_{36} + x_{20}x_{37} + x_{20}x_{40} + x_{20}x_{42} + x_{20}x_{43} + x_{20}x_{45} + x_{20}x_{47} + x_{20}x_{53} + x_{20}x_{54} + x_{20}x_{57} + x_{20}x_{58} + x_{20}x_{61} + x_{20}x_{62} + x_{20}x_{63} + x_{21}x_{23} + x_{21}x_{28} + x_{21}x_{30} + x_{21}x_{33} + x_{21}x_{34} + x_{21}x_{37} + x_{21}x_{38} + x_{21}x_{40} + x_{21}x_{41} + x_{21}x_{43} + x_{21}x_{44} + x_{21}x_{45} + x_{21}x_{46} + x_{21}x_{51} + x_{21}x_{53} + x_{21}x_{58} + x_{21}x_{59} + x_{21}x_{60} + x_{21}x_{62} + x_{21}x_{63} + x_{21}x_{64} + x_{22}x_{23} + x_{22}x_{26} + x_{22}x_{27} + x_{22}x_{28} + x_{22}x_{32} + x_{22}x_{33} + x_{22}x_{34} + x_{22}x_{35} + x_{22}x_{37} + x_{22}x_{41} + x_{22}x_{42} + x_{22}x_{46} + x_{22}x_{48} + x_{22}x_{49} + x_{22}x_{50} + x_{22}x_{52} + x_{22}x_{55} + x_{22}x_{58} + x_{22}x_{61} + x_{22}x_{63} + x_{23}x_{24} + x_{23}x_{25} + x_{23}x_{26} + x_{23}x_{27} + x_{23}x_{28} + x_{23}x_{30} + x_{23}x_{33} + x_{23}x_{35} + x_{23}x_{38} + x_{23}x_{42} + x_{23}x_{43} + x_{23}x_{45} + x_{23}x_{48} + x_{23}x_{49} + x_{23}x_{52} + x_{23}x_{55} + x_{23}x_{56} + x_{23}x_{59} + x_{23}x_{60} + x_{23}x_{62} + x_{23}x_{63} + x_{23}x_{64} + x_{24}x_{25} + x_{24}x_{27} + x_{24}x_{31} + x_{24}x_{34} + x_{24}x_{36} + x_{24}x_{38} + x_{24}x_{40} + x_{24}x_{41} + x_{24}x_{43} + x_{24}x_{44} + x_{24}x_{45} + x_{24}x_{46} + x_{24}x_{47} + x_{24}x_{48} + x_{24}x_{50} + x_{24}x_{52} + x_{24}x_{55} + x_{24}x_{59} + x_{24}x_{60} + x_{24}x_{61} + x_{25}x_{26} + x_{25}x_{30} + x_{25}x_{31} + x_{25}x_{32} + x_{25}x_{34} + x_{25}x_{39} + x_{25}x_{40} + x_{25}x_{43} + x_{25}x_{46} + x_{25}x_{47} + x_{25}x_{49} + x_{25}x_{51} + x_{25}x_{52} + x_{25}x_{54} + x_{25}x_{57} + x_{25}x_{58} + x_{25}x_{59} + x_{25}x_{60} + x_{26}x_{32} + x_{26}x_{36} + x_{26}x_{37} + x_{26}x_{38} + x_{26}x_{39} + x_{26}x_{40} + x_{26}x_{43} + x_{26}x_{44} + x_{26}x_{53} + x_{26}x_{54} + x_{26}x_{55} + x_{26}x_{56} + x_{26}x_{57} + x_{26}x_{59} + x_{26}x_{60} + x_{26}x_{63} + x_{27}x_{29} + x_{27}x_{30} + x_{27}x_{33} + x_{27}x_{34} + x_{27}x_{35} + x_{27}x_{36} + x_{27}x_{39} + x_{27}x_{40} + x_{27}x_{43} + x_{27}x_{44} + x_{27}x_{47} + x_{27}x_{48} + x_{27}x_{49} + x_{27}x_{50} + x_{27}x_{51} + x_{27}x_{54} + x_{27}x_{55} + x_{27}x_{58} + x_{27}x_{60} + x_{27}x_{61} + x_{27}x_{62} + x_{28}x_{29} + x_{28}x_{30} + x_{28}x_{32} + x_{28}x_{35} + x_{28}x_{37} + x_{28}x_{38} + x_{28}x_{39} + x_{28}x_{41} + x_{28}x_{42} + x_{28}x_{47} + x_{28}x_{48} + x_{28}x_{52} + x_{28}x_{55} + x_{28}x_{59} + x_{28}x_{61} + x_{28}x_{62} + x_{28}x_{64} + x_{29}x_{31} + x_{29}x_{32} + x_{29}x_{33} + x_{29}x_{36} + x_{29}x_{37} + x_{29}x_{39} + x_{29}x_{41} + x_{29}x_{43} + x_{29}x_{44} + x_{29}x_{46} + x_{29}x_{47} + x_{29}x_{50} + x_{29}x_{52} + x_{29}x_{53} + x_{29}x_{58} + x_{29}x_{60} + x_{29}x_{61} + x_{30}x_{31} + x_{30}x_{32} + x_{30}x_{36} + x_{30}x_{37} + x_{30}x_{39} + x_{30}x_{41} + x_{30}x_{47} + x_{30}x_{49} + x_{30}x_{50} + x_{30}x_{51} + x_{30}x_{53} + x_{30}x_{56} + x_{30}x_{57} + x_{30}x_{58} + x_{30}x_{60} + x_{30}x_{61} + x_{30}x_{62} + x_{31}x_{33} + x_{31}x_{34} + x_{31}x_{36} + x_{31}x_{39} + x_{31}x_{41} + x_{31}x_{43} + x_{31}x_{46} + x_{31}x_{47} + x_{31}x_{48} + x_{31}x_{53} + x_{31}x_{55} + x_{31}x_{57} + x_{31}x_{60} + x_{31}x_{62} + x_{32}x_{33} + x_{32}x_{36} + x_{32}x_{37} + x_{32}x_{38} + x_{32}x_{41} + x_{32}x_{43} + x_{32}x_{44} + x_{32}x_{45} + x_{32}x_{46} + x_{32}x_{47} + x_{32}x_{49} + x_{32}x_{51} + x_{32}x_{52} + x_{32}x_{53} + x_{32}x_{55} + x_{32}x_{57} + x_{32}x_{58} + x_{32}x_{60} + x_{32}x_{61} + x_{33}x_{36} + x_{33}x_{37} + x_{33}x_{39} + x_{33}x_{42} + x_{33}x_{44} + x_{33}x_{45} + x_{33}x_{47} + x_{33}x_{48} + x_{33}x_{50} + x_{33}x_{53} + x_{33}x_{56} + x_{33}x_{57} + x_{33}x_{58} + x_{33}x_{59} + x_{33}x_{60} + x_{33}x_{64} + x_{34}x_{38} + x_{34}x_{39} + x_{34}x_{40} + x_{34}x_{41} + x_{34}x_{43} + x_{34}x_{44} + x_{34}x_{46} + x_{34}x_{50} + x_{34}x_{53} + x_{34}x_{58} + x_{34}x_{59} + x_{34}x_{61} + x_{34}x_{62} + x_{34}x_{63} + x_{35}x_{37} + x_{35}x_{40} + x_{35}x_{42} + x_{35}x_{44} + x_{35}x_{47} + x_{35}x_{54} + x_{35}x_{55} + x_{35}x_{56} + x_{35}x_{57} + x_{35}x_{58} + x_{35}x_{59} + x_{35}x_{60} + x_{35}x_{61} + x_{35}x_{62} + x_{36}x_{40} + x_{36}x_{42} + x_{36}x_{49} + x_{36}x_{52} + x_{36}x_{55} + x_{36}x_{59} + x_{36}x_{60} + x_{36}x_{63} + x_{37}x_{38} + x_{37}x_{39} + x_{37}x_{40} + x_{37}x_{44} + x_{37}x_{45} + x_{37}x_{53} + x_{37}x_{54} + x_{37}x_{55} + x_{37}x_{60} + x_{37}x_{61} + x_{37}x_{62} + x_{37}x_{63} + x_{38}x_{45} + x_{38}x_{46} + x_{38}x_{48} + x_{38}x_{49} + x_{38}x_{50} + x_{38}x_{51} + x_{38}x_{54} + x_{38}x_{59} + x_{38}x_{63} + x_{38}x_{64} + x_{39}x_{41} + x_{39}x_{42} + x_{39}x_{47} + x_{39}x_{49} + x_{39}x_{52} + x_{39}x_{54} + x_{39}x_{57} + x_{39}x_{58} + x_{39}x_{60} + x_{39}x_{63} + x_{40}x_{41} + x_{40}x_{42} + x_{40}x_{44} + x_{40}x_{45} + x_{40}x_{46} + x_{40}x_{47} + x_{40}x_{48} + x_{40}x_{49} + x_{40}x_{50} + x_{40}x_{54} + x_{40}x_{58} + x_{40}x_{59} + x_{40}x_{63} + x_{41}x_{42} + x_{41}x_{45} + x_{41}x_{46} + x_{41}x_{47} + x_{41}x_{49} + x_{41}x_{50} + x_{41}x_{51} + x_{41}x_{52} + x_{41}x_{54} + x_{41}x_{58} + x_{41}x_{59} + x_{41}x_{60} + x_{41}x_{61} + x_{41}x_{63} + x_{41}x_{64} + x_{42}x_{46} + x_{42}x_{49} + x_{42}x_{52} + x_{42}x_{54} + x_{42}x_{55} + x_{42}x_{56} + x_{42}x_{57} + x_{42}x_{58} + x_{42}x_{62} + x_{43}x_{46} + x_{43}x_{48} + x_{43}x_{49} + x_{43}x_{50} + x_{43}x_{51} + x_{43}x_{52} + x_{43}x_{53} + x_{43}x_{55} + x_{43}x_{56} + x_{43}x_{58} + x_{43}x_{59} + x_{43}x_{60} + x_{43}x_{62} + x_{44}x_{45} + x_{44}x_{49} + x_{44}x_{51} + x_{44}x_{52} + x_{44}x_{53} + x_{44}x_{54} + x_{44}x_{56} + x_{44}x_{58} + x_{44}x_{59} + x_{44}x_{60} + x_{44}x_{61} + x_{44}x_{62} + x_{45}x_{46} + x_{45}x_{47} + x_{45}x_{48} + x_{45}x_{50} + x_{45}x_{51} + x_{45}x_{55} + x_{45}x_{57} + x_{45}x_{60} + x_{45}x_{63} + x_{45}x_{64} + x_{46}x_{48} + x_{46}x_{51} + x_{46}x_{52} + x_{46}x_{53} + x_{46}x_{54} + x_{46}x_{57} + x_{46}x_{62} + x_{46}x_{63} + x_{46}x_{64} + x_{47}x_{48} + x_{47}x_{52} + x_{47}x_{53} + x_{47}x_{54} + x_{47}x_{55} + x_{47}x_{56} + x_{47}x_{57} + x_{47}x_{63} + x_{48}x_{50} + x_{48}x_{51} + x_{48}x_{52} + x_{48}x_{53} + x_{48}x_{56} + x_{48}x_{58} + x_{48}x_{63} + x_{49}x_{50} + x_{49}x_{53} + x_{49}x_{54} + x_{49}x_{56} + x_{49}x_{57} + x_{49}x_{58} + x_{49}x_{59} + x_{49}x_{64} + x_{50}x_{54} + x_{50}x_{57} + x_{50}x_{58} + x_{50}x_{61} + x_{51}x_{52} + x_{51}x_{53} + x_{51}x_{57} + x_{51}x_{59} + x_{51}x_{61} + x_{52}x_{56} + x_{52}x_{58} + x_{52}x_{59} + x_{52}x_{60} + x_{52}x_{61} + x_{52}x_{63} + x_{53}x_{56} + x_{53}x_{58} + x_{53}x_{61} + x_{53}x_{64} + x_{54}x_{56} + x_{54}x_{58} + x_{54}x_{59} + x_{54}x_{61} + x_{54}x_{64} + x_{55}x_{56} + x_{55}x_{59} + x_{55}x_{60} + x_{55}x_{61} + x_{55}x_{62} + x_{55}x_{63} + x_{56}x_{57} + x_{56}x_{60} + x_{56}x_{61} + x_{56}x_{64} + x_{57}x_{61} + x_{58}x_{59} + x_{58}x_{60} + x_{58}x_{61} + x_{58}x_{62} + x_{58}x_{63} + x_{59}x_{60} + x_{59}x_{62} + x_{59}x_{64} + x_{60}x_{61} + x_{60}x_{63} + x_{60}x_{64} + x_{61}x_{63} + x_{61}x_{64} + x_{62}x_{64} + x_{63}x_{64} + x_{1} + x_{2} + x_{7} + x_{8} + x_{13} + x_{14} + x_{23} + x_{25} + x_{26} + x_{28} + x_{32} + x_{35} + x_{37} + x_{39} + x_{40} + x_{41} + x_{42} + x_{44} + x_{45} + x_{46} + x_{47} + x_{48} + x_{49} + x_{50} + x_{51} + x_{53} + x_{54} + x_{55} + x_{57} + x_{58} + x_{59} + x_{60} + x_{61}$

$y_{4} = x_{1}x_{3} + x_{1}x_{4} + x_{1}x_{5} + x_{1}x_{7} + x_{1}x_{8} + x_{1}x_{9} + x_{1}x_{13} + x_{1}x_{14} + x_{1}x_{15} + x_{1}x_{17} + x_{1}x_{20} + x_{1}x_{22} + x_{1}x_{25} + x_{1}x_{27} + x_{1}x_{33} + x_{1}x_{34} + x_{1}x_{35} + x_{1}x_{36} + x_{1}x_{37} + x_{1}x_{40} + x_{1}x_{41} + x_{1}x_{42} + x_{1}x_{45} + x_{1}x_{46} + x_{1}x_{47} + x_{1}x_{49} + x_{1}x_{50} + x_{1}x_{51} + x_{1}x_{54} + x_{1}x_{58} + x_{1}x_{61} + x_{2}x_{4} + x_{2}x_{5} + x_{2}x_{6} + x_{2}x_{7} + x_{2}x_{8} + x_{2}x_{9} + x_{2}x_{10} + x_{2}x_{12} + x_{2}x_{13} + x_{2}x_{14} + x_{2}x_{16} + x_{2}x_{17} + x_{2}x_{18} + x_{2}x_{19} + x_{2}x_{20} + x_{2}x_{22} + x_{2}x_{24} + x_{2}x_{25} + x_{2}x_{26} + x_{2}x_{28} + x_{2}x_{30} + x_{2}x_{35} + x_{2}x_{37} + x_{2}x_{39} + x_{2}x_{40} + x_{2}x_{41} + x_{2}x_{43} + x_{2}x_{44} + x_{2}x_{47} + x_{2}x_{50} + x_{2}x_{51} + x_{2}x_{52} + x_{2}x_{53} + x_{2}x_{55} + x_{2}x_{58} + x_{2}x_{64} + x_{3}x_{4} + x_{3}x_{7} + x_{3}x_{8} + x_{3}x_{10} + x_{3}x_{13} + x_{3}x_{15} + x_{3}x_{17} + x_{3}x_{18} + x_{3}x_{20} + x_{3}x_{22} + x_{3}x_{23} + x_{3}x_{27} + x_{3}x_{28} + x_{3}x_{29} + x_{3}x_{31} + x_{3}x_{36} + x_{3}x_{38} + x_{3}x_{39} + x_{3}x_{44} + x_{3}x_{45} + x_{3}x_{49} + x_{3}x_{53} + x_{3}x_{54} + x_{3}x_{56} + x_{3}x_{57} + x_{3}x_{59} + x_{3}x_{60} + x_{3}x_{62} + x_{3}x_{63} + x_{3}x_{64} + x_{4}x_{6} + x_{4}x_{10} + x_{4}x_{12} + x_{4}x_{17} + x_{4}x_{18} + x_{4}x_{19} + x_{4}x_{20} + x_{4}x_{26} + x_{4}x_{27} + x_{4}x_{28} + x_{4}x_{30} + x_{4}x_{31} + x_{4}x_{32} + x_{4}x_{33} + x_{4}x_{35} + x_{4}x_{36} + x_{4}x_{42} + x_{4}x_{43} + x_{4}x_{44} + x_{4}x_{45} + x_{4}x_{51} + x_{4}x_{52} + x_{4}x_{57} + x_{4}x_{59} + x_{4}x_{60} + x_{4}x_{61} + x_{4}x_{63} + x_{5}x_{7} + x_{5}x_{8} + x_{5}x_{9} + x_{5}x_{11} + x_{5}x_{12} + x_{5}x_{15} + x_{5}x_{16} + x_{5}x_{17} + x_{5}x_{19} + x_{5}x_{21} + x_{5}x_{22} + x_{5}x_{23} + x_{5}x_{24} + x_{5}x_{25} + x_{5}x_{26} + x_{5}x_{27} + x_{5}x_{29} + x_{5}x_{30} + x_{5}x_{31} + x_{5}x_{32} + x_{5}x_{33} + x_{5}x_{34} + x_{5}x_{35} + x_{5}x_{36} + x_{5}x_{37} + x_{5}x_{38} + x_{5}x_{42} + x_{5}x_{45} + x_{5}x_{47} + x_{5}x_{49} + x_{5}x_{50} + x_{5}x_{53} + x_{5}x_{54} + x_{5}x_{57} + x_{5}x_{58} + x_{5}x_{59} + x_{5}x_{61} + x_{5}x_{63} + x_{5}x_{64} + x_{6}x_{7} + x_{6}x_{8} + x_{6}x_{9} + x_{6}x_{10} + x_{6}x_{11} + x_{6}x_{18} + x_{6}x_{19} + x_{6}x_{20} + x_{6}x_{23} + x_{6}x_{24} + x_{6}x_{26} + x_{6}x_{27} + x_{6}x_{28} + x_{6}x_{29} + x_{6}x_{30} + x_{6}x_{33} + x_{6}x_{34} + x_{6}x_{35} + x_{6}x_{37} + x_{6}x_{40} + x_{6}x_{44} + x_{6}x_{45} + x_{6}x_{49} + x_{6}x_{52} + x_{6}x_{57} + x_{6}x_{59} + x_{6}x_{60} + x_{6}x_{61} + x_{6}x_{62} + x_{7}x_{9} + x_{7}x_{10} + x_{7}x_{12} + x_{7}x_{14} + x_{7}x_{16} + x_{7}x_{17} + x_{7}x_{24} + x_{7}x_{26} + x_{7}x_{30} + x_{7}x_{32} + x_{7}x_{34} + x_{7}x_{35} + x_{7}x_{36} + x_{7}x_{39} + x_{7}x_{40} + x_{7}x_{41} + x_{7}x_{42} + x_{7}x_{46} + x_{7}x_{48} + x_{7}x_{49} + x_{7}x_{51} + x_{7}x_{56} + x_{7}x_{58} + x_{8}x_{14} + x_{8}x_{15} + x_{8}x_{17} + x_{8}x_{18} + x_{8}x_{21} + x_{8}x_{24} + x_{8}x_{25} + x_{8}x_{29} + x_{8}x_{30} + x_{8}x_{32} + x_{8}x_{36} + x_{8}x_{40} + x_{8}x_{42} + x_{8}x_{43} + x_{8}x_{49} + x_{8}x_{50} + x_{8}x_{51} + x_{8}x_{59} + x_{8}x_{61} + x_{9}x_{11} + x_{9}x_{14} + x_{9}x_{16} + x_{9}x_{18} + x_{9}x_{19} + x_{9}x_{21} + x_{9}x_{22} + x_{9}x_{23} + x_{9}x_{30} + x_{9}x_{33} + x_{9}x_{34} + x_{9}x_{36} + x_{9}x_{42} + x_{9}x_{44} + x_{9}x_{45} + x_{9}x_{47} + x_{9}x_{48} + x_{9}x_{49} + x_{9}x_{52} + x_{9}x_{54} + x_{9}x_{55} + x_{9}x_{57} + x_{9}x_{59} + x_{9}x_{60} + x_{9}x_{62} + x_{9}x_{63} + x_{9}x_{64} + x_{10}x_{15} + x_{10}x_{16} + x_{10}x_{18} + x_{10}x_{20} + x_{10}x_{21} + x_{10}x_{23} + x_{10}x_{24} + x_{10}x_{26} + x_{10}x_{29} + x_{10}x_{31} + x_{10}x_{33} + x_{10}x_{36} + x_{10}x_{39} + x_{10}x_{43} + x_{10}x_{44} + x_{10}x_{45} + x_{10}x_{46} + x_{10}x_{48} + x_{10}x_{54} + x_{10}x_{55} + x_{10}x_{56} + x_{10}x_{58} + x_{10}x_{59} + x_{10}x_{60} + x_{11}x_{14} + x_{11}x_{15} + x_{11}x_{16} + x_{11}x_{18} + x_{11}x_{19} + x_{11}x_{20} + x_{11}x_{21} + x_{11}x_{22} + x_{11}x_{23} + x_{11}x_{24} + x_{11}x_{27} + x_{11}x_{31} + x_{11}x_{32} + x_{11}x_{34} + x_{11}x_{35} + x_{11}x_{37} + x_{11}x_{39} + x_{11}x_{40} + x_{11}x_{44} + x_{11}x_{47} + x_{11}x_{49} + x_{11}x_{51} + x_{11}x_{52} + x_{11}x_{53} + x_{11}x_{54} + x_{11}x_{56} + x_{11}x_{57} + x_{11}x_{58} + x_{11}x_{61} + x_{11}x_{64} + x_{12}x_{13} + x_{12}x_{15} + x_{12}x_{19} + x_{12}x_{20} + x_{12}x_{22} + x_{12}x_{24} + x_{12}x_{25} + x_{12}x_{27} + x_{12}x_{28} + x_{12}x_{30} + x_{12}x_{31} + x_{12}x_{33} + x_{12}x_{38} + x_{12}x_{39} + x_{12}x_{41} + x_{12}x_{42} + x_{12}x_{43} + x_{12}x_{45} + x_{12}x_{48} + x_{12}x_{49} + x_{12}x_{51} + x_{12}x_{58} + x_{12}x_{60} + x_{13}x_{15} + x_{13}x_{16} + x_{13}x_{21} + x_{13}x_{23} + x_{13}x_{24} + x_{13}x_{28} + x_{13}x_{32} + x_{13}x_{33} + x_{13}x_{34} + x_{13}x_{36} + x_{13}x_{38} + x_{13}x_{41} + x_{13}x_{45} + x_{13}x_{49} + x_{13}x_{52} + x_{13}x_{54} + x_{13}x_{60} + x_{13}x_{61} + x_{13}x_{62} + x_{13}x_{64} + x_{14}x_{15} + x_{14}x_{16} + x_{14}x_{19} + x_{14}x_{27} + x_{14}x_{29} + x_{14}x_{30} + x_{14}x_{31} + x_{14}x_{32} + x_{14}x_{33} + x_{14}x_{35} + x_{14}x_{38} + x_{14}x_{39} + x_{14}x_{40} + x_{14}x_{41} + x_{14}x_{42} + x_{14}x_{43} + x_{14}x_{44} + x_{14}x_{46} + x_{14}x_{47} + x_{14}x_{48} + x_{14}x_{50} + x_{14}x_{51} + x_{14}x_{52} + x_{14}x_{53} + x_{14}x_{55} + x_{14}x_{60} + x_{14}x_{63} + x_{14}x_{64} + x_{15}x_{16} + x_{15}x_{18} + x_{15}x_{20} + x_{15}x_{23} + x_{15}x_{26} + x_{15}x_{27} + x_{15}x_{29} + x_{15}x_{30} + x_{15}x_{32} + x_{15}x_{33} + x_{15}x_{40} + x_{15}x_{43} + x_{15}x_{44} + x_{15}x_{45} + x_{15}x_{46} + x_{15}x_{49} + x_{15}x_{51} + x_{15}x_{55} + x_{15}x_{56} + x_{15}x_{57} + x_{15}x_{64} + x_{16}x_{17} + x_{16}x_{19} + x_{16}x_{26} + x_{16}x_{27} + x_{16}x_{28} + x_{16}x_{29} + x_{16}x_{30} + x_{16}x_{31} + x_{16}x_{32} + x_{16}x_{33} + x_{16}x_{37} + x_{16}x_{39} + x_{16}x_{42} + x_{16}x_{43} + x_{16}x_{44} + x_{16}x_{46} + x_{16}x_{49} + x_{16}x_{50} + x_{16}x_{54} + x_{16}x_{55} + x_{16}x_{57} + x_{16}x_{62} + x_{17}x_{18} + x_{17}x_{20} + x_{17}x_{23} + x_{17}x_{25} + x_{17}x_{28} + x_{17}x_{29} + x_{17}x_{33} + x_{17}x_{34} + x_{17}x_{35} + x_{17}x_{36} + x_{17}x_{37} + x_{17}x_{39} + x_{17}x_{42} + x_{17}x_{44} + x_{17}x_{45} + x_{17}x_{46} + x_{17}x_{47} + x_{17}x_{48} + x_{17}x_{50} + x_{17}x_{51} + x_{17}x_{52} + x_{17}x_{53} + x_{17}x_{55} + x_{17}x_{57} + x_{17}x_{60} + x_{17}x_{64} + x_{18}x_{19} + x_{18}x_{20} + x_{18}x_{23} + x_{18}x_{25} + x_{18}x_{26} + x_{18}x_{27} + x_{18}x_{28} + x_{18}x_{34} + x_{18}x_{36} + x_{18}x_{44} + x_{18}x_{46} + x_{18}x_{49} + x_{18}x_{51} + x_{18}x_{52} + x_{18}x_{53} + x_{18}x_{55} + x_{18}x_{56} + x_{18}x_{58} + x_{18}x_{59} + x_{18}x_{60} + x_{18}x_{61} + x_{18}x_{63} + x_{19}x_{20} + x_{19}x_{22} + x_{19}x_{25} + x_{19}x_{26} + x_{19}x_{27} + x_{19}x_{28} + x_{19}x_{33} + x_{19}x_{36} + x_{19}x_{40} + x_{19}x_{41} + x_{19}x_{43} + x_{19}x_{48} + x_{19}x_{49} + x_{19}x_{52} + x_{19}x_{55} + x_{19}x_{56} + x_{19}x_{57} + x_{19}x_{59} + x_{19}x_{60} + x_{19}x_{61} + x_{19}x_{62} + x_{19}x_{63} + x_{19}x_{64} + x_{20}x_{21} + x_{20}x_{24} + x_{20}x_{25} + x_{20}x_{27} + x_{20}x_{28} + x_{20}x_{29} + x_{20}x_{32} + x_{20}x_{33} + x_{20}x_{35} + x_{20}x_{43} + x_{20}x_{46} + x_{20}x_{49} + x_{20}x_{50} + x_{20}x_{52} + x_{20}x_{53} + x_{20}x_{54} + x_{20}x_{55} + x_{20}x_{57} + x_{20}x_{59} + x_{20}x_{60} + x_{20}x_{64} + x_{21}x_{22} + x_{21}x_{26} + x_{21}x_{28} + x_{21}x_{29} + x_{21}x_{30} + x_{21}x_{32} + x_{21}x_{33} + x_{21}x_{34} + x_{21}x_{36} + x_{21}x_{38} + x_{21}x_{39} + x_{21}x_{43} + x_{21}x_{44} + x_{21}x_{46} + x_{21}x_{50} + x_{21}x_{51} + x_{21}x_{53} + x_{21}x_{54} + x_{21}x_{58} + x_{21}x_{59} + x_{21}x_{60} + x_{21}x_{62} + x_{21}x_{64} + x_{22}x_{26} + x_{22}x_{27} + x_{22}x_{29} + x_{22}x_{30} + x_{22}x_{31} + x_{22}x_{33} + x_{22}x_{34} + x_{22}x_{35} + x_{22}x_{36} + x_{22}x_{39} + x_{22}x_{40} + x_{22}x_{42} + x_{22}x_{43} + x_{22}x_{44} + x_{22}x_{45} + x_{22}x_{46} + x_{22}x_{48} + x_{22}x_{50} + x_{22}x_{51} + x_{22}x_{56} + x_{22}x_{57} + x_{22}x_{58} + x_{22}x_{59} + x_{22}x_{60} + x_{22}x_{61} + x_{22}x_{62} + x_{22}x_{63} + x_{23}x_{25} + x_{23}x_{26} + x_{23}x_{27} + x_{23}x_{29} + x_{23}x_{33} + x_{23}x_{37} + x_{23}x_{38} + x_{23}x_{39} + x_{23}x_{40} + x_{23}x_{42} + x_{23}x_{45} + x_{23}x_{47} + x_{23}x_{48} + x_{23}x_{49} + x_{23}x_{51} + x_{23}x_{52} + x_{23}x_{53} + x_{23}x_{55} + x_{23}x_{56} + x_{23}x_{59} + x_{23}x_{60} + x_{23}x_{61} + x_{23}x_{63} + x_{23}x_{64} + x_{24}x_{25} + x_{24}x_{27} + x_{24}x_{28} + x_{24}x_{33} + x_{24}x_{35} + x_{24}x_{36} + x_{24}x_{38} + x_{24}x_{39} + x_{24}x_{40} + x_{24}x_{43} + x_{24}x_{44} + x_{24}x_{45} + x_{24}x_{46} + x_{24}x_{47} + x_{24}x_{48} + x_{24}x_{54} + x_{24}x_{55} + x_{24}x_{56} + x_{24}x_{57} + x_{24}x_{58} + x_{24}x_{59} + x_{24}x_{61} + x_{24}x_{62} + x_{25}x_{26} + x_{25}x_{27} + x_{25}x_{28} + x_{25}x_{29} + x_{25}x_{32} + x_{25}x_{34} + x_{25}x_{35} + x_{25}x_{37} + x_{25}x_{38} + x_{25}x_{39} + x_{25}x_{41} + x_{25}x_{45} + x_{25}x_{49} + x_{25}x_{50} + x_{25}x_{51} + x_{25}x_{53} + x_{25}x_{54} + x_{25}x_{57} + x_{25}x_{62} + x_{25}x_{63} + x_{25}x_{64} + x_{26}x_{29} + x_{26}x_{31} + x_{26}x_{36} + x_{26}x_{40} + x_{26}x_{42} + x_{26}x_{48} + x_{26}x_{51} + x_{26}x_{52} + x_{26}x_{54} + x_{26}x_{57} + x_{26}x_{58} + x_{26}x_{59} + x_{26}x_{60} + x_{26}x_{61} + x_{26}x_{62} + x_{26}x_{63} + x_{26}x_{64} + x_{27}x_{28} + x_{27}x_{29} + x_{27}x_{30} + x_{27}x_{31} + x_{27}x_{33} + x_{27}x_{34} + x_{27}x_{37} + x_{27}x_{42} + x_{27}x_{43} + x_{27}x_{44} + x_{27}x_{46} + x_{27}x_{48} + x_{27}x_{49} + x_{27}x_{51} + x_{27}x_{53} + x_{27}x_{54} + x_{27}x_{56} + x_{27}x_{58} + x_{27}x_{59} + x_{27}x_{63} + x_{27}x_{64} + x_{28}x_{33} + x_{28}x_{34} + x_{28}x_{35} + x_{28}x_{36} + x_{28}x_{37} + x_{28}x_{39} + x_{28}x_{40} + x_{28}x_{42} + x_{28}x_{43} + x_{28}x_{44} + x_{28}x_{45} + x_{28}x_{51} + x_{28}x_{52} + x_{28}x_{53} + x_{28}x_{54} + x_{28}x_{55} + x_{28}x_{58} + x_{28}x_{60} + x_{28}x_{61} + x_{28}x_{63} + x_{28}x_{64} + x_{29}x_{30} + x_{29}x_{34} + x_{29}x_{35} + x_{29}x_{39} + x_{29}x_{40} + x_{29}x_{42} + x_{29}x_{43} + x_{29}x_{44} + x_{29}x_{45} + x_{29}x_{46} + x_{29}x_{47} + x_{29}x_{49} + x_{29}x_{53} + x_{29}x_{62} + x_{29}x_{63} + x_{30}x_{33} + x_{30}x_{34} + x_{30}x_{35} + x_{30}x_{36} + x_{30}x_{37} + x_{30}x_{38} + x_{30}x_{41} + x_{30}x_{43} + x_{30}x_{45} + x_{30}x_{48} + x_{30}x_{49} + x_{30}x_{51} + x_{30}x_{53} + x_{30}x_{56} + x_{30}x_{58} + x_{30}x_{61} + x_{30}x_{62} + x_{30}x_{64} + x_{31}x_{32} + x_{31}x_{34} + x_{31}x_{36} + x_{31}x_{40} + x_{31}x_{42} + x_{31}x_{45} + x_{31}x_{47} + x_{31}x_{52} + x_{31}x_{53} + x_{31}x_{54} + x_{31}x_{56} + x_{32}x_{33} + x_{32}x_{36} + x_{32}x_{38} + x_{32}x_{39} + x_{32}x_{41} + x_{32}x_{42} + x_{32}x_{43} + x_{32}x_{44} + x_{32}x_{45} + x_{32}x_{47} + x_{32}x_{50} + x_{32}x_{51} + x_{32}x_{52} + x_{32}x_{55} + x_{32}x_{56} + x_{32}x_{58} + x_{32}x_{59} + x_{33}x_{36} + x_{33}x_{41} + x_{33}x_{42} + x_{33}x_{45} + x_{33}x_{46} + x_{33}x_{48} + x_{33}x_{50} + x_{33}x_{54} + x_{33}x_{55} + x_{33}x_{61} + x_{33}x_{62} + x_{33}x_{63} + x_{33}x_{64} + x_{34}x_{35} + x_{34}x_{36} + x_{34}x_{37} + x_{34}x_{39} + x_{34}x_{52} + x_{34}x_{53} + x_{34}x_{54} + x_{34}x_{55} + x_{34}x_{57} + x_{34}x_{60} + x_{34}x_{62} + x_{34}x_{63} + x_{34}x_{64} + x_{35}x_{36} + x_{35}x_{38} + x_{35}x_{44} + x_{35}x_{45} + x_{35}x_{46} + x_{35}x_{50} + x_{35}x_{51} + x_{35}x_{53} + x_{35}x_{56} + x_{35}x_{57} + x_{35}x_{59} + x_{35}x_{60} + x_{35}x_{61} + x_{35}x_{62} + x_{35}x_{64} + x_{36}x_{37} + x_{36}x_{38} + x_{36}x_{39} + x_{36}x_{40} + x_{36}x_{47} + x_{36}x_{50} + x_{36}x_{53} + x_{36}x_{54} + x_{36}x_{56} + x_{36}x_{63} + x_{37}x_{38} + x_{37}x_{39} + x_{37}x_{41} + x_{37}x_{43} + x_{37}x_{44} + x_{37}x_{47} + x_{37}x_{49} + x_{37}x_{53} + x_{37}x_{58} + x_{37}x_{60} + x_{38}x_{39} + x_{38}x_{41} + x_{38}x_{43} + x_{38}x_{45} + x_{38}x_{46} + x_{38}x_{51} + x_{38}x_{52} + x_{38}x_{55} + x_{38}x_{58} + x_{38}x_{59} + x_{38}x_{60} + x_{38}x_{63} + x_{39}x_{40} + x_{39}x_{42} + x_{39}x_{45} + x_{39}x_{46} + x_{39}x_{47} + x_{39}x_{48} + x_{39}x_{50} + x_{39}x_{52} + x_{39}x_{55} + x_{39}x_{60} + x_{40}x_{41} + x_{40}x_{42} + x_{40}x_{44} + x_{40}x_{45} + x_{40}x_{55} + x_{40}x_{58} + x_{40}x_{60} + x_{40}x_{63} + x_{41}x_{42} + x_{41}x_{43} + x_{41}x_{49} + x_{41}x_{55} + x_{41}x_{56} + x_{41}x_{58} + x_{41}x_{62} + x_{41}x_{63} + x_{41}x_{64} + x_{42}x_{43} + x_{42}x_{45} + x_{42}x_{46} + x_{42}x_{47} + x_{42}x_{48} + x_{42}x_{50} + x_{42}x_{55} + x_{42}x_{56} + x_{42}x_{57} + x_{42}x_{62} + x_{43}x_{45} + x_{43}x_{46} + x_{43}x_{47} + x_{43}x_{49} + x_{43}x_{50} + x_{43}x_{51} + x_{43}x_{60} + x_{43}x_{62} + x_{43}x_{63} + x_{43}x_{64} + x_{44}x_{45} + x_{44}x_{47} + x_{44}x_{49} + x_{44}x_{52} + x_{44}x_{53} + x_{44}x_{59} + x_{44}x_{60} + x_{44}x_{62} + x_{45}x_{47} + x_{45}x_{48} + x_{45}x_{49} + x_{45}x_{51} + x_{45}x_{54} + x_{45}x_{55} + x_{45}x_{57} + x_{45}x_{58} + x_{45}x_{63} + x_{46}x_{50} + x_{46}x_{51} + x_{46}x_{52} + x_{46}x_{54} + x_{46}x_{55} + x_{46}x_{58} + x_{46}x_{62} + x_{46}x_{63} + x_{46}x_{64} + x_{47}x_{48} + x_{47}x_{52} + x_{47}x_{55} + x_{47}x_{58} + x_{47}x_{59} + x_{47}x_{60} + x_{48}x_{49} + x_{48}x_{50} + x_{48}x_{54} + x_{48}x_{55} + x_{48}x_{59} + x_{49}x_{53} + x_{49}x_{56} + x_{49}x_{57} + x_{49}x_{59} + x_{49}x_{61} + x_{49}x_{62} + x_{49}x_{63} + x_{50}x_{53} + x_{50}x_{55} + x_{50}x_{59} + x_{50}x_{62} + x_{50}x_{63} + x_{50}x_{64} + x_{51}x_{52} + x_{51}x_{53} + x_{51}x_{54} + x_{51}x_{55} + x_{51}x_{58} + x_{51}x_{60} + x_{51}x_{61} + x_{51}x_{63} + x_{51}x_{64} + x_{52}x_{53} + x_{52}x_{54} + x_{52}x_{55} + x_{52}x_{56} + x_{52}x_{57} + x_{52}x_{60} + x_{52}x_{61} + x_{52}x_{62} + x_{52}x_{63} + x_{53}x_{54} + x_{53}x_{56} + x_{53}x_{57} + x_{53}x_{60} + x_{54}x_{57} + x_{54}x_{58} + x_{55}x_{56} + x_{55}x_{57} + x_{55}x_{58} + x_{55}x_{60} + x_{55}x_{63} + x_{56}x_{57} + x_{56}x_{58} + x_{57}x_{59} + x_{57}x_{60} + x_{57}x_{61} + x_{57}x_{62} + x_{58}x_{60} + x_{58}x_{61} + x_{58}x_{63} + x_{58}x_{64} + x_{59}x_{60} + x_{59}x_{63} + x_{60}x_{61} + x_{60}x_{64} + x_{61}x_{63} + x_{61}x_{64} + x_{62}x_{63} + x_{63}x_{64} + x_{1} + x_{8} + x_{11} + x_{21} + x_{27} + x_{29} + x_{31} + x_{32} + x_{33} + x_{35} + x_{37} + x_{40} + x_{45} + x_{46} + x_{47} + x_{48} + x_{49} + x_{50} + x_{52} + x_{53} + x_{54} + x_{56} + x_{57} + x_{59} + x_{60} + x_{61} + x_{62} + x_{63} + 1$

$y_{5} = x_{1}x_{3} + x_{1}x_{6} + x_{1}x_{8} + x_{1}x_{10} + x_{1}x_{11} + x_{1}x_{12} + x_{1}x_{14} + x_{1}x_{18} + x_{1}x_{20} + x_{1}x_{22} + x_{1}x_{23} + x_{1}x_{25} + x_{1}x_{30} + x_{1}x_{31} + x_{1}x_{33} + x_{1}x_{34} + x_{1}x_{35} + x_{1}x_{36} + x_{1}x_{37} + x_{1}x_{38} + x_{1}x_{42} + x_{1}x_{46} + x_{1}x_{47} + x_{1}x_{49} + x_{1}x_{50} + x_{1}x_{53} + x_{1}x_{56} + x_{1}x_{57} + x_{1}x_{60} + x_{1}x_{61} + x_{1}x_{63} + x_{2}x_{4} + x_{2}x_{5} + x_{2}x_{6} + x_{2}x_{9} + x_{2}x_{10} + x_{2}x_{11} + x_{2}x_{12} + x_{2}x_{13} + x_{2}x_{14} + x_{2}x_{16} + x_{2}x_{19} + x_{2}x_{20} + x_{2}x_{21} + x_{2}x_{25} + x_{2}x_{31} + x_{2}x_{32} + x_{2}x_{35} + x_{2}x_{38} + x_{2}x_{39} + x_{2}x_{40} + x_{2}x_{45} + x_{2}x_{46} + x_{2}x_{50} + x_{2}x_{51} + x_{2}x_{53} + x_{2}x_{54} + x_{2}x_{55} + x_{2}x_{60} + x_{2}x_{62} + x_{2}x_{63} + x_{3}x_{9} + x_{3}x_{11} + x_{3}x_{13} + x_{3}x_{15} + x_{3}x_{18} + x_{3}x_{19} + x_{3}x_{21} + x_{3}x_{23} + x_{3}x_{24} + x_{3}x_{25} + x_{3}x_{26} + x_{3}x_{27} + x_{3}x_{28} + x_{3}x_{31} + x_{3}x_{32} + x_{3}x_{34} + x_{3}x_{37} + x_{3}x_{40} + x_{3}x_{42} + x_{3}x_{43} + x_{3}x_{45} + x_{3}x_{46} + x_{3}x_{48} + x_{3}x_{50} + x_{3}x_{51} + x_{3}x_{55} + x_{3}x_{57} + x_{3}x_{58} + x_{3}x_{60} + x_{3}x_{61} + x_{3}x_{62} + x_{3}x_{63} + x_{3}x_{64} + x_{4}x_{8} + x_{4}x_{9} + x_{4}x_{14} + x_{4}x_{16} + x_{4}x_{18} + x_{4}x_{21} + x_{4}x_{22} + x_{4}x_{23} + x_{4}x_{32} + x_{4}x_{33} + x_{4}x_{34} + x_{4}x_{36} + x_{4}x_{38} + x_{4}x_{39} + x_{4}x_{41} + x_{4}x_{43} + x_{4}x_{44} + x_{4}x_{45} + x_{4}x_{47} + x_{4}x_{50} + x_{4}x_{53} + x_{4}x_{54} + x_{4}x_{55} + x_{4}x_{56} + x_{4}x_{57} + x_{5}x_{6} + x_{5}x_{7} + x_{5}x_{8} + x_{5}x_{9} + x_{5}x_{12} + x_{5}x_{13} + x_{5}x_{15} + x_{5}x_{16} + x_{5}x_{17} + x_{5}x_{18} + x_{5}x_{19} + x_{5}x_{21} + x_{5}x_{25} + x_{5}x_{26} + x_{5}x_{27} + x_{5}x_{28} + x_{5}x_{29} + x_{5}x_{30} + x_{5}x_{32} + x_{5}x_{33} + x_{5}x_{34} + x_{5}x_{36} + x_{5}x_{37} + x_{5}x_{38} + x_{5}x_{43} + x_{5}x_{47} + x_{5}x_{51} + x_{5}x_{53} + x_{5}x_{54} + x_{5}x_{56} + x_{5}x_{61} + x_{5}x_{62} + x_{5}x_{63} + x_{6}x_{10} + x_{6}x_{11} + x_{6}x_{14} + x_{6}x_{23} + x_{6}x_{25} + x_{6}x_{26} + x_{6}x_{27} + x_{6}x_{30} + x_{6}x_{32} + x_{6}x_{35} + x_{6}x_{37} + x_{6}x_{38} + x_{6}x_{39} + x_{6}x_{42} + x_{6}x_{44} + x_{6}x_{45} + x_{6}x_{47} + x_{6}x_{50} + x_{6}x_{57} + x_{6}x_{58} + x_{6}x_{60} + x_{6}x_{61} + x_{6}x_{63} + x_{6}x_{64} + x_{7}x_{8} + x_{7}x_{13} + x_{7}x_{14} + x_{7}x_{17} + x_{7}x_{18} + x_{7}x_{20} + x_{7}x_{21} + x_{7}x_{22} + x_{7}x_{23} + x_{7}x_{24} + x_{7}x_{27} + x_{7}x_{29} + x_{7}x_{30} + x_{7}x_{34} + x_{7}x_{36} + x_{7}x_{40} + x_{7}x_{42} + x_{7}x_{43} + x_{7}x_{44} + x_{7}x_{45} + x_{7}x_{46} + x_{7}x_{48} + x_{7}x_{49} + x_{7}x_{51} + x_{7}x_{56} + x_{7}x_{57} + x_{7}x_{61} + x_{7}x_{64} + x_{8}x_{9} + x_{8}x_{15} + x_{8}x_{17} + x_{8}x_{18} + x_{8}x_{19} + x_{8}x_{20} + x_{8}x_{21} + x_{8}x_{24} + x_{8}x_{25} + x_{8}x_{27} + x_{8}x_{28} + x_{8}x_{29} + x_{8}x_{30} + x_{8}x_{31} + x_{8}x_{32} + x_{8}x_{33} + x_{8}x_{34} + x_{8}x_{35} + x_{8}x_{38} + x_{8}x_{41} + x_{8}x_{42} + x_{8}x_{45} + x_{8}x_{46} + x_{8}x_{47} + x_{8}x_{48} + x_{8}x_{49} + x_{8}x_{51} + x_{8}x_{52} + x_{8}x_{54} + x_{8}x_{57} + x_{8}x_{58} + x_{8}x_{62} + x_{8}x_{64} + x_{9}x_{11} + x_{9}x_{13} + x_{9}x_{14} + x_{9}x_{15} + x_{9}x_{19} + x_{9}x_{21} + x_{9}x_{22} + x_{9}x_{25} + x_{9}x_{26} + x_{9}x_{27} + x_{9}x_{30} + x_{9}x_{31} + x_{9}x_{32} + x_{9}x_{37} + x_{9}x_{39} + x_{9}x_{41} + x_{9}x_{42} + x_{9}x_{43} + x_{9}x_{44} + x_{9}x_{45} + x_{9}x_{46} + x_{9}x_{47} + x_{9}x_{51} + x_{9}x_{52} + x_{9}x_{56} + x_{9}x_{57} + x_{9}x_{58} + x_{9}x_{59} + x_{9}x_{61} + x_{9}x_{64} + x_{10}x_{11} + x_{10}x_{12} + x_{10}x_{14} + x_{10}x_{19} + x_{10}x_{22} + x_{10}x_{24} + x_{10}x_{26} + x_{10}x_{27} + x_{10}x_{30} + x_{10}x_{32} + x_{10}x_{34} + x_{10}x_{35} + x_{10}x_{36} + x_{10}x_{37} + x_{10}x_{41} + x_{10}x_{42} + x_{10}x_{43} + x_{10}x_{44} + x_{10}x_{46} + x_{10}x_{47} + x_{10}x_{48} + x_{10}x_{49} + x_{10}x_{50} + x_{10}x_{51} + x_{10}x_{52} + x_{10}x_{55} + x_{11}x_{13} + x_{11}x_{15} + x_{11}x_{17} + x_{11}x_{19} + x_{11}x_{22} + x_{11}x_{23} + x_{11}x_{24} + x_{11}x_{26} + x_{11}x_{27} + x_{11}x_{28} + x_{11}x_{29} + x_{11}x_{30} + x_{11}x_{32} + x_{11}x_{33} + x_{11}x_{42} + x_{11}x_{43} + x_{11}x_{44} + x_{11}x_{47} + x_{11}x_{48} + x_{11}x_{52} + x_{11}x_{53} + x_{11}x_{55} + x_{11}x_{56} + x_{11}x_{59} + x_{11}x_{60} + x_{11}x_{62} + x_{11}x_{64} + x_{12}x_{14} + x_{12}x_{21} + x_{12}x_{23} + x_{12}x_{24} + x_{12}x_{26} + x_{12}x_{27} + x_{12}x_{28} + x_{12}x_{32} + x_{12}x_{35} + x_{12}x_{37} + x_{12}x_{39} + x_{12}x_{41} + x_{12}x_{43} + x_{12}x_{48} + x_{12}x_{49} + x_{12}x_{50} + x_{12}x_{52} + x_{12}x_{53} + x_{12}x_{54} + x_{12}x_{55} + x_{12}x_{56} + x_{12}x_{63} + x_{13}x_{15} + x_{13}x_{16} + x_{13}x_{18} + x_{13}x_{20} + x_{13}x_{22} + x_{13}x_{23} + x_{13}x_{24} + x_{13}x_{25} + x_{13}x_{27} + x_{13}x_{30} + x_{13}x_{35} + x_{13}x_{37} + x_{13}x_{38} + x_{13}x_{39} + x_{13}x_{42} + x_{13}x_{43} + x_{13}x_{46} + x_{13}x_{48} + x_{13}x_{54} + x_{13}x_{56} + x_{13}x_{59} + x_{13}x_{60} + x_{13}x_{62} + x_{14}x_{17} + x_{14}x_{20} + x_{14}x_{23} + x_{14}x_{24} + x_{14}x_{25} + x_{14}x_{26} + x_{14}x_{27} + x_{14}x_{30} + x_{14}x_{31} + x_{14}x_{33} + x_{14}x_{35} + x_{14}x_{37} + x_{14}x_{38} + x_{14}x_{39} + x_{14}x_{40} + x_{14}x_{42} + x_{14}x_{44} + x_{14}x_{46} + x_{14}x_{47} + x_{14}x_{48} + x_{14}x_{50} + x_{14}x_{57} + x_{14}x_{59} + x_{14}x_{60} + x_{14}x_{64} + x_{15}x_{18} + x_{15}x_{19} + x_{15}x_{21} + x_{15}x_{22} + x_{15}x_{24} + x_{15}x_{25} + x_{15}x_{26} + x_{15}x_{27} + x_{15}x_{28} + x_{15}x_{29} + x_{15}x_{31} + x_{15}x_{32} + x_{15}x_{34} + x_{15}x_{35} + x_{15}x_{37} + x_{15}x_{38} + x_{15}x_{39} + x_{15}x_{42} + x_{15}x_{43} + x_{15}x_{44} + x_{15}x_{45} + x_{15}x_{46} + x_{15}x_{47} + x_{15}x_{49} + x_{15}x_{51} + x_{15}x_{52} + x_{15}x_{57} + x_{15}x_{59} + x_{15}x_{60} + x_{15}x_{62} + x_{15}x_{63} + x_{16}x_{19} + x_{16}x_{20} + x_{16}x_{21} + x_{16}x_{23} + x_{16}x_{25} + x_{16}x_{26} + x_{16}x_{29} + x_{16}x_{30} + x_{16}x_{31} + x_{16}x_{32} + x_{16}x_{34} + x_{16}x_{37} + x_{16}x_{39} + x_{16}x_{43} + x_{16}x_{44} + x_{16}x_{46} + x_{16}x_{47} + x_{16}x_{49} + x_{16}x_{50} + x_{16}x_{51} + x_{16}x_{53} + x_{16}x_{54} + x_{16}x_{60} + x_{16}x_{63} + x_{16}x_{64} + x_{17}x_{19} + x_{17}x_{20} + x_{17}x_{25} + x_{17}x_{27} + x_{17}x_{28} + x_{17}x_{31} + x_{17}x_{34} + x_{17}x_{37} + x_{17}x_{38} + x_{17}x_{39} + x_{17}x_{41} + x_{17}x_{42} + x_{17}x_{45} + x_{17}x_{48} + x_{17}x_{49} + x_{17}x_{50} + x_{17}x_{51} + x_{17}x_{57} + x_{17}x_{58} + x_{17}x_{60} + x_{17}x_{64} + x_{18}x_{20} + x_{18}x_{22} + x_{18}x_{23} + x_{18}x_{25} + x_{18}x_{29} + x_{18}x_{32} + x_{18}x_{34} + x_{18}x_{35} + x_{18}x_{36} + x_{18}x_{40} + x_{18}x_{41} + x_{18}x_{42} + x_{18}x_{43} + x_{18}x_{44} + x_{18}x_{46} + x_{18}x_{47} + x_{18}x_{48} + x_{18}x_{50} + x_{18}x_{51} + x_{18}x_{54} + x_{18}x_{56} + x_{18}x_{64} + x_{19}x_{20} + x_{19}x_{25} + x_{19}x_{27} + x_{19}x_{28} + x_{19}x_{31} + x_{19}x_{32} + x_{19}x_{33} + x_{19}x_{35} + x_{19}x_{37} + x_{19}x_{38} + x_{19}x_{40} + x_{19}x_{42} + x_{19}x_{45} + x_{19}x_{48} + x_{19}x_{50} + x_{19}x_{52} + x_{19}x_{53} + x_{19}x_{58} + x_{19}x_{63} + x_{20}x_{21} + x_{20}x_{24} + x_{20}x_{25} + x_{20}x_{27} + x_{20}x_{28} + x_{20}x_{30} + x_{20}x_{33} + x_{20}x_{43} + x_{20}x_{44} + x_{20}x_{45} + x_{20}x_{48} + x_{20}x_{50} + x_{20}x_{54} + x_{20}x_{57} + x_{20}x_{59} + x_{20}x_{61} + x_{20}x_{63} + x_{20}x_{64} + x_{21}x_{23} + x_{21}x_{25} + x_{21}x_{27} + x_{21}x_{29} + x_{21}x_{30} + x_{21}x_{31} + x_{21}x_{32} + x_{21}x_{33} + x_{21}x_{34} + x_{21}x_{36} + x_{21}x_{42} + x_{21}x_{43} + x_{21}x_{45} + x_{21}x_{46} + x_{21}x_{49} + x_{21}x_{51} + x_{21}x_{54} + x_{21}x_{56} + x_{21}x_{58} + x_{21}x_{59} + x_{21}x_{60} + x_{21}x_{61} + x_{21}x_{62} + x_{21}x_{64} + x_{22}x_{23} + x_{22}x_{27} + x_{22}x_{28} + x_{22}x_{31} + x_{22}x_{32} + x_{22}x_{33} + x_{22}x_{35} + x_{22}x_{36} + x_{22}x_{38} + x_{22}x_{39} + x_{22}x_{40} + x_{22}x_{41} + x_{22}x_{43} + x_{22}x_{45} + x_{22}x_{46} + x_{22}x_{47} + x_{22}x_{48} + x_{22}x_{53} + x_{22}x_{54} + x_{22}x_{55} + x_{22}x_{56} + x_{22}x_{59} + x_{22}x_{60} + x_{22}x_{61} + x_{22}x_{62} + x_{22}x_{63} + x_{23}x_{25} + x_{23}x_{26} + x_{23}x_{27} + x_{23}x_{33} + x_{23}x_{36} + x_{23}x_{38} + x_{23}x_{40} + x_{23}x_{42} + x_{23}x_{45} + x_{23}x_{47} + x_{23}x_{48} + x_{23}x_{49} + x_{23}x_{52} + x_{23}x_{53} + x_{23}x_{54} + x_{23}x_{55} + x_{23}x_{56} + x_{23}x_{61} + x_{23}x_{62} + x_{23}x_{63} + x_{24}x_{27} + x_{24}x_{29} + x_{24}x_{31} + x_{24}x_{33} + x_{24}x_{38} + x_{24}x_{39} + x_{24}x_{41} + x_{24}x_{42} + x_{24}x_{44} + x_{24}x_{45} + x_{24}x_{46} + x_{24}x_{52} + x_{24}x_{53} + x_{24}x_{55} + x_{24}x_{57} + x_{24}x_{64} + x_{25}x_{27} + x_{25}x_{30} + x_{25}x_{32} + x_{25}x_{33} + x_{25}x_{35} + x_{25}x_{37} + x_{25}x_{38} + x_{25}x_{39} + x_{25}x_{40} + x_{25}x_{41} + x_{25}x_{43} + x_{25}x_{45} + x_{25}x_{48} + x_{25}x_{57} + x_{25}x_{59} + x_{25}x_{64} + x_{26}x_{28} + x_{26}x_{29} + x_{26}x_{33} + x_{26}x_{35} + x_{26}x_{36} + x_{26}x_{38} + x_{26}x_{40} + x_{26}x_{45} + x_{26}x_{47} + x_{26}x_{48} + x_{26}x_{50} + x_{26}x_{52} + x_{26}x_{55} + x_{26}x_{57} + x_{26}x_{58} + x_{26}x_{59} + x_{26}x_{60} + x_{27}x_{28} + x_{27}x_{30} + x_{27}x_{31} + x_{27}x_{32} + x_{27}x_{37} + x_{27}x_{40} + x_{27}x_{41} + x_{27}x_{43} + x_{27}x_{44} + x_{27}x_{46} + x_{27}x_{53} + x_{27}x_{56} + x_{27}x_{57} + x_{27}x_{59} + x_{27}x_{62} + x_{27}x_{63} + x_{28}x_{30} + x_{28}x_{31} + x_{28}x_{32} + x_{28}x_{33} + x_{28}x_{34} + x_{28}x_{35} + x_{28}x_{38} + x_{28}x_{41} + x_{28}x_{43} + x_{28}x_{44} + x_{28}x_{46} + x_{28}x_{47} + x_{28}x_{48} + x_{28}x_{50} + x_{28}x_{51} + x_{28}x_{52} + x_{28}x_{55} + x_{28}x_{57} + x_{28}x_{58} + x_{28}x_{59} + x_{28}x_{60} + x_{28}x_{61} + x_{28}x_{63} + x_{28}x_{64} + x_{29}x_{34} + x_{29}x_{35} + x_{29}x_{37} + x_{29}x_{39} + x_{29}x_{41} + x_{29}x_{44} + x_{29}x_{45} + x_{29}x_{47} + x_{29}x_{48} + x_{29}x_{49} + x_{29}x_{55} + x_{29}x_{56} + x_{29}x_{58} + x_{29}x_{59} + x_{29}x_{63} + x_{29}x_{64} + x_{30}x_{31} + x_{30}x_{32} + x_{30}x_{36} + x_{30}x_{37} + x_{30}x_{43} + x_{30}x_{45} + x_{30}x_{47} + x_{30}x_{48} + x_{30}x_{49} + x_{30}x_{50} + x_{30}x_{52} + x_{30}x_{55} + x_{30}x_{57} + x_{30}x_{58} + x_{30}x_{59} + x_{30}x_{61} + x_{30}x_{62} + x_{30}x_{64} + x_{31}x_{33} + x_{31}x_{35} + x_{31}x_{37} + x_{31}x_{40} + x_{31}x_{43} + x_{31}x_{44} + x_{31}x_{47} + x_{31}x_{48} + x_{31}x_{51} + x_{31}x_{52} + x_{31}x_{54} + x_{31}x_{55} + x_{31}x_{56} + x_{31}x_{58} + x_{31}x_{60} + x_{31}x_{61} + x_{32}x_{35} + x_{32}x_{36} + x_{32}x_{41} + x_{32}x_{42} + x_{32}x_{43} + x_{32}x_{45} + x_{32}x_{46} + x_{32}x_{49} + x_{32}x_{50} + x_{32}x_{52} + x_{32}x_{53} + x_{32}x_{55} + x_{32}x_{61} + x_{32}x_{62} + x_{33}x_{34} + x_{33}x_{37} + x_{33}x_{39} + x_{33}x_{40} + x_{33}x_{44} + x_{33}x_{45} + x_{33}x_{47} + x_{33}x_{48} + x_{33}x_{50} + x_{33}x_{52} + x_{33}x_{55} + x_{33}x_{56} + x_{33}x_{57} + x_{33}x_{58} + x_{33}x_{63} + x_{33}x_{64} + x_{34}x_{37} + x_{34}x_{39} + x_{34}x_{42} + x_{34}x_{45} + x_{34}x_{46} + x_{34}x_{47} + x_{34}x_{52} + x_{34}x_{53} + x_{34}x_{57} + x_{34}x_{59} + x_{34}x_{63} + x_{34}x_{64} + x_{35}x_{36} + x_{35}x_{37} + x_{35}x_{39} + x_{35}x_{40} + x_{35}x_{42} + x_{35}x_{44} + x_{35}x_{45} + x_{35}x_{47} + x_{35}x_{51} + x_{35}x_{53} + x_{35}x_{54} + x_{35}x_{55} + x_{35}x_{56} + x_{35}x_{58} + x_{35}x_{60} + x_{35}x_{61} + x_{35}x_{63} + x_{36}x_{39} + x_{36}x_{41} + x_{36}x_{42} + x_{36}x_{44} + x_{36}x_{46} + x_{36}x_{48} + x_{36}x_{50} + x_{36}x_{51} + x_{36}x_{52} + x_{36}x_{53} + x_{36}x_{54} + x_{36}x_{55} + x_{36}x_{59} + x_{36}x_{60} + x_{36}x_{64} + x_{37}x_{38} + x_{37}x_{39} + x_{37}x_{40} + x_{37}x_{42} + x_{37}x_{43} + x_{37}x_{45} + x_{37}x_{46} + x_{37}x_{47} + x_{37}x_{48} + x_{37}x_{51} + x_{37}x_{53} + x_{37}x_{55} + x_{37}x_{58} + x_{37}x_{62} + x_{38}x_{40} + x_{38}x_{42} + x_{38}x_{43} + x_{38}x_{45} + x_{38}x_{49} + x_{38}x_{50} + x_{38}x_{53} + x_{38}x_{54} + x_{38}x_{55} + x_{38}x_{56} + x_{38}x_{58} + x_{38}x_{59} + x_{38}x_{60} + x_{38}x_{61} + x_{38}x_{62} + x_{38}x_{64} + x_{39}x_{40} + x_{39}x_{43} + x_{39}x_{45} + x_{39}x_{46} + x_{39}x_{49} + x_{39}x_{50} + x_{39}x_{51} + x_{39}x_{52} + x_{39}x_{56} + x_{39}x_{58} + x_{39}x_{60} + x_{39}x_{62} + x_{39}x_{63} + x_{40}x_{44} + x_{40}x_{48} + x_{40}x_{49} + x_{40}x_{51} + x_{40}x_{54} + x_{40}x_{55} + x_{40}x_{58} + x_{40}x_{59} + x_{40}x_{62} + x_{40}x_{63} + x_{41}x_{44} + x_{41}x_{47} + x_{41}x_{48} + x_{41}x_{49} + x_{41}x_{50} + x_{41}x_{52} + x_{41}x_{54} + x_{41}x_{58} + x_{41}x_{59} + x_{41}x_{61} + x_{41}x_{62} + x_{41}x_{64} + x_{42}x_{43} + x_{42}x_{46} + x_{42}x_{48} + x_{42}x_{50} + x_{42}x_{51} + x_{42}x_{56} + x_{42}x_{57} + x_{42}x_{61} + x_{42}x_{64} + x_{43}x_{45} + x_{43}x_{49} + x_{43}x_{50} + x_{43}x_{54} + x_{43}x_{57} + x_{43}x_{58} + x_{43}x_{59} + x_{43}x_{60} + x_{43}x_{61} + x_{43}x_{64} + x_{44}x_{47} + x_{44}x_{49} + x_{44}x_{53} + x_{44}x_{55} + x_{44}x_{60} + x_{44}x_{61} + x_{44}x_{63} + x_{45}x_{46} + x_{45}x_{47} + x_{45}x_{50} + x_{45}x_{51} + x_{45}x_{52} + x_{45}x_{54} + x_{45}x_{55} + x_{45}x_{56} + x_{45}x_{60} + x_{45}x_{61} + x_{46}x_{47} + x_{46}x_{48} + x_{46}x_{49} + x_{46}x_{53} + x_{46}x_{56} + x_{46}x_{59} + x_{46}x_{60} + x_{46}x_{63} + x_{46}x_{64} + x_{47}x_{49} + x_{47}x_{50} + x_{47}x_{52} + x_{47}x_{53} + x_{47}x_{54} + x_{47}x_{55} + x_{47}x_{58} + x_{47}x_{59} + x_{47}x_{60} + x_{47}x_{61} + x_{47}x_{62} + x_{47}x_{63} + x_{47}x_{64} + x_{48}x_{49} + x_{48}x_{52} + x_{48}x_{53} + x_{48}x_{54} + x_{48}x_{55} + x_{48}x_{56} + x_{48}x_{60} + x_{48}x_{61} + x_{48}x_{63} + x_{48}x_{64} + x_{49}x_{52} + x_{49}x_{53} + x_{49}x_{54} + x_{49}x_{58} + x_{49}x_{60} + x_{49}x_{64} + x_{50}x_{51} + x_{50}x_{52} + x_{50}x_{53} + x_{50}x_{55} + x_{50}x_{57} + x_{50}x_{58} + x_{50}x_{59} + x_{50}x_{60} + x_{50}x_{61} + x_{50}x_{64} + x_{51}x_{55} + x_{51}x_{58} + x_{51}x_{59} + x_{51}x_{62} + x_{52}x_{53} + x_{52}x_{55} + x_{52}x_{58} + x_{52}x_{61} + x_{52}x_{62} + x_{52}x_{63} + x_{53}x_{55} + x_{53}x_{58} + x_{53}x_{60} + x_{53}x_{61} + x_{53}x_{62} + x_{53}x_{63} + x_{54}x_{55} + x_{54}x_{56} + x_{54}x_{58} + x_{54}x_{61} + x_{54}x_{62} + x_{54}x_{63} + x_{54}x_{64} + x_{55}x_{56} + x_{55}x_{57} + x_{55}x_{62} + x_{55}x_{63} + x_{56}x_{60} + x_{56}x_{63} + x_{57}x_{61} + x_{57}x_{62} + x_{58}x_{60} + x_{58}x_{62} + x_{58}x_{63} + x_{59}x_{60} + x_{59}x_{61} + x_{59}x_{63} + x_{59}x_{64} + x_{60}x_{61} + x_{60}x_{63} + x_{61}x_{63} + x_{61}x_{64} + x_{62}x_{63} + x_{63}x_{64} + x_{1} + x_{3} + x_{5} + x_{6} + x_{7} + x_{11} + x_{14} + x_{17} + x_{19} + x_{20} + x_{21} + x_{22} + x_{24} + x_{29} + x_{33} + x_{34} + x_{39} + x_{40} + x_{43} + x_{44} + x_{45} + x_{46} + x_{47} + x_{55} + x_{56} + x_{57} + x_{62} + 1$

$y_{6} = x_{1}x_{2} + x_{1}x_{4} + x_{1}x_{7} + x_{1}x_{11} + x_{1}x_{15} + x_{1}x_{16} + x_{1}x_{18} + x_{1}x_{21} + x_{1}x_{22} + x_{1}x_{23} + x_{1}x_{29} + x_{1}x_{30} + x_{1}x_{33} + x_{1}x_{35} + x_{1}x_{38} + x_{1}x_{39} + x_{1}x_{40} + x_{1}x_{41} + x_{1}x_{42} + x_{1}x_{44} + x_{1}x_{47} + x_{1}x_{49} + x_{1}x_{50} + x_{1}x_{51} + x_{1}x_{52} + x_{1}x_{54} + x_{1}x_{55} + x_{1}x_{56} + x_{1}x_{57} + x_{1}x_{58} + x_{1}x_{59} + x_{1}x_{60} + x_{1}x_{61} + x_{1}x_{63} + x_{2}x_{4} + x_{2}x_{10} + x_{2}x_{13} + x_{2}x_{16} + x_{2}x_{18} + x_{2}x_{19} + x_{2}x_{20} + x_{2}x_{22} + x_{2}x_{26} + x_{2}x_{28} + x_{2}x_{30} + x_{2}x_{31} + x_{2}x_{32} + x_{2}x_{33} + x_{2}x_{39} + x_{2}x_{41} + x_{2}x_{43} + x_{2}x_{46} + x_{2}x_{48} + x_{2}x_{50} + x_{2}x_{55} + x_{2}x_{57} + x_{2}x_{61} + x_{2}x_{62} + x_{2}x_{63} + x_{2}x_{64} + x_{3}x_{8} + x_{3}x_{9} + x_{3}x_{10} + x_{3}x_{12} + x_{3}x_{21} + x_{3}x_{24} + x_{3}x_{25} + x_{3}x_{28} + x_{3}x_{29} + x_{3}x_{30} + x_{3}x_{31} + x_{3}x_{35} + x_{3}x_{36} + x_{3}x_{37} + x_{3}x_{38} + x_{3}x_{42} + x_{3}x_{50} + x_{3}x_{51} + x_{3}x_{52} + x_{3}x_{55} + x_{3}x_{59} + x_{3}x_{61} + x_{3}x_{64} + x_{4}x_{5} + x_{4}x_{7} + x_{4}x_{8} + x_{4}x_{9} + x_{4}x_{14} + x_{4}x_{15} + x_{4}x_{18} + x_{4}x_{20} + x_{4}x_{22} + x_{4}x_{23} + x_{4}x_{25} + x_{4}x_{26} + x_{4}x_{28} + x_{4}x_{30} + x_{4}x_{32} + x_{4}x_{34} + x_{4}x_{36} + x_{4}x_{37} + x_{4}x_{40} + x_{4}x_{43} + x_{4}x_{51} + x_{4}x_{52} + x_{4}x_{62} + x_{4}x_{63} + x_{5}x_{6} + x_{5}x_{8} + x_{5}x_{10} + x_{5}x_{12} + x_{5}x_{13} + x_{5}x_{15} + x_{5}x_{17} + x_{5}x_{18} + x_{5}x_{20} + x_{5}x_{21} + x_{5}x_{22} + x_{5}x_{23} + x_{5}x_{25} + x_{5}x_{29} + x_{5}x_{35} + x_{5}x_{37} + x_{5}x_{39} + x_{5}x_{41} + x_{5}x_{45} + x_{5}x_{47} + x_{5}x_{49} + x_{5}x_{50} + x_{5}x_{53} + x_{5}x_{55} + x_{5}x_{57} + x_{5}x_{58} + x_{5}x_{60} + x_{5}x_{63} + x_{6}x_{7} + x_{6}x_{9} + x_{6}x_{12} + x_{6}x_{14} + x_{6}x_{17} + x_{6}x_{18} + x_{6}x_{19} + x_{6}x_{20} + x_{6}x_{22} + x_{6}x_{23} + x_{6}x_{24} + x_{6}x_{28} + x_{6}x_{29} + x_{6}x_{30} + x_{6}x_{31} + x_{6}x_{34} + x_{6}x_{35} + x_{6}x_{40} + x_{6}x_{41} + x_{6}x_{44} + x_{6}x_{49} + x_{6}x_{53} + x_{6}x_{57} + x_{6}x_{59} + x_{6}x_{61} + x_{6}x_{64} + x_{7}x_{8} + x_{7}x_{10} + x_{7}x_{11} + x_{7}x_{12} + x_{7}x_{13} + x_{7}x_{14} + x_{7}x_{15} + x_{7}x_{16} + x_{7}x_{17} + x_{7}x_{19} + x_{7}x_{21} + x_{7}x_{22} + x_{7}x_{28} + x_{7}x_{32} + x_{7}x_{34} + x_{7}x_{35} + x_{7}x_{37} + x_{7}x_{39} + x_{7}x_{40} + x_{7}x_{42} + x_{7}x_{43} + x_{7}x_{47} + x_{7}x_{48} + x_{7}x_{50} + x_{7}x_{51} + x_{7}x_{52} + x_{7}x_{53} + x_{7}x_{54} + x_{7}x_{56} + x_{7}x_{57} + x_{7}x_{63} + x_{8}x_{9} + x_{8}x_{11} + x_{8}x_{19} + x_{8}x_{20} + x_{8}x_{24} + x_{8}x_{26} + x_{8}x_{27} + x_{8}x_{28} + x_{8}x_{29} + x_{8}x_{30} + x_{8}x_{31} + x_{8}x_{33} + x_{8}x_{34} + x_{8}x_{35} + x_{8}x_{37} + x_{8}x_{38} + x_{8}x_{43} + x_{8}x_{44} + x_{8}x_{45} + x_{8}x_{46} + x_{8}x_{49} + x_{8}x_{50} + x_{8}x_{51} + x_{8}x_{52} + x_{8}x_{53} + x_{8}x_{54} + x_{8}x_{56} + x_{8}x_{58} + x_{8}x_{59} + x_{8}x_{62} + x_{8}x_{63} + x_{8}x_{64} + x_{9}x_{11} + x_{9}x_{12} + x_{9}x_{13} + x_{9}x_{14} + x_{9}x_{16} + x_{9}x_{17} + x_{9}x_{20} + x_{9}x_{24} + x_{9}x_{25} + x_{9}x_{30} + x_{9}x_{31} + x_{9}x_{33} + x_{9}x_{36} + x_{9}x_{38} + x_{9}x_{40} + x_{9}x_{41} + x_{9}x_{42} + x_{9}x_{43} + x_{9}x_{48} + x_{9}x_{51} + x_{9}x_{52} + x_{9}x_{54} + x_{9}x_{57} + x_{9}x_{58} + x_{9}x_{59} + x_{9}x_{60} + x_{9}x_{61} + x_{9}x_{62} + x_{9}x_{63} + x_{10}x_{13} + x_{10}x_{14} + x_{10}x_{22} + x_{10}x_{23} + x_{10}x_{24} + x_{10}x_{26} + x_{10}x_{28} + x_{10}x_{29} + x_{10}x_{30} + x_{10}x_{31} + x_{10}x_{33} + x_{10}x_{37} + x_{10}x_{38} + x_{10}x_{40} + x_{10}x_{44} + x_{10}x_{45} + x_{10}x_{46} + x_{10}x_{51} + x_{10}x_{56} + x_{10}x_{57} + x_{10}x_{58} + x_{10}x_{59} + x_{10}x_{61} + x_{10}x_{62} + x_{11}x_{14} + x_{11}x_{19} + x_{11}x_{20} + x_{11}x_{22} + x_{11}x_{23} + x_{11}x_{24} + x_{11}x_{25} + x_{11}x_{27} + x_{11}x_{31} + x_{11}x_{32} + x_{11}x_{36} + x_{11}x_{37} + x_{11}x_{39} + x_{11}x_{40} + x_{11}x_{44} + x_{11}x_{46} + x_{11}x_{49} + x_{11}x_{51} + x_{11}x_{57} + x_{11}x_{59} + x_{11}x_{63} + x_{11}x_{64} + x_{12}x_{13} + x_{12}x_{15} + x_{12}x_{17} + x_{12}x_{18} + x_{12}x_{20} + x_{12}x_{21} + x_{12}x_{22} + x_{12}x_{23} + x_{12}x_{24} + x_{12}x_{25} + x_{12}x_{26} + x_{12}x_{27} + x_{12}x_{29} + x_{12}x_{30} + x_{12}x_{31} + x_{12}x_{32} + x_{12}x_{33} + x_{12}x_{34} + x_{12}x_{35} + x_{12}x_{37} + x_{12}x_{38} + x_{12}x_{42} + x_{12}x_{43} + x_{12}x_{45} + x_{12}x_{46} + x_{12}x_{47} + x_{12}x_{48} + x_{12}x_{51} + x_{12}x_{53} + x_{12}x_{54} + x_{12}x_{55} + x_{12}x_{56} + x_{12}x_{57} + x_{12}x_{60} + x_{12}x_{61} + x_{12}x_{62} + x_{12}x_{63} + x_{13}x_{14} + x_{13}x_{15} + x_{13}x_{18} + x_{13}x_{27} + x_{13}x_{28} + x_{13}x_{39} + x_{13}x_{40} + x_{13}x_{44} + x_{13}x_{47} + x_{13}x_{48} + x_{13}x_{49} + x_{13}x_{50} + x_{13}x_{51} + x_{13}x_{53} + x_{13}x_{56} + x_{13}x_{57} + x_{13}x_{62} + x_{14}x_{17} + x_{14}x_{21} + x_{14}x_{22} + x_{14}x_{24} + x_{14}x_{25} + x_{14}x_{26} + x_{14}x_{28} + x_{14}x_{31} + x_{14}x_{32} + x_{14}x_{33} + x_{14}x_{36} + x_{14}x_{37} + x_{14}x_{39} + x_{14}x_{42} + x_{14}x_{44} + x_{14}x_{47} + x_{14}x_{48} + x_{14}x_{49} + x_{14}x_{50} + x_{14}x_{52} + x_{14}x_{53} + x_{14}x_{54} + x_{14}x_{56} + x_{14}x_{58} + x_{14}x_{60} + x_{14}x_{61} + x_{14}x_{64} + x_{15}x_{16} + x_{15}x_{18} + x_{15}x_{20} + x_{15}x_{21} + x_{15}x_{22} + x_{15}x_{24} + x_{15}x_{27} + x_{15}x_{29} + x_{15}x_{30} + x_{15}x_{32} + x_{15}x_{33} + x_{15}x_{34} + x_{15}x_{38} + x_{15}x_{40} + x_{15}x_{41} + x_{15}x_{43} + x_{15}x_{44} + x_{15}x_{47} + x_{15}x_{49} + x_{15}x_{51} + x_{15}x_{53} + x_{15}x_{55} + x_{15}x_{58} + x_{15}x_{59} + x_{15}x_{60} + x_{15}x_{61} + x_{15}x_{62} + x_{16}x_{17} + x_{16}x_{20} + x_{16}x_{22} + x_{16}x_{24} + x_{16}x_{26} + x_{16}x_{28} + x_{16}x_{30} + x_{16}x_{33} + x_{16}x_{35} + x_{16}x_{37} + x_{16}x_{40} + x_{16}x_{41} + x_{16}x_{42} + x_{16}x_{45} + x_{16}x_{49} + x_{16}x_{52} + x_{16}x_{53} + x_{16}x_{55} + x_{16}x_{58} + x_{16}x_{59} + x_{16}x_{60} + x_{16}x_{61} + x_{16}x_{62} + x_{17}x_{18} + x_{17}x_{19} + x_{17}x_{20} + x_{17}x_{24} + x_{17}x_{27} + x_{17}x_{29} + x_{17}x_{32} + x_{17}x_{33} + x_{17}x_{35} + x_{17}x_{36} + x_{17}x_{39} + x_{17}x_{40} + x_{17}x_{41} + x_{17}x_{42} + x_{17}x_{45} + x_{17}x_{48} + x_{17}x_{52} + x_{17}x_{55} + x_{17}x_{57} + x_{17}x_{60} + x_{17}x_{62} + x_{17}x_{64} + x_{18}x_{20} + x_{18}x_{22} + x_{18}x_{23} + x_{18}x_{24} + x_{18}x_{26} + x_{18}x_{29} + x_{18}x_{30} + x_{18}x_{31} + x_{18}x_{32} + x_{18}x_{33} + x_{18}x_{34} + x_{18}x_{36} + x_{18}x_{37} + x_{18}x_{39} + x_{18}x_{42} + x_{18}x_{45} + x_{18}x_{47} + x_{18}x_{50} + x_{18}x_{53} + x_{18}x_{56} + x_{18}x_{59} + x_{19}x_{20} + x_{19}x_{21} + x_{19}x_{22} + x_{19}x_{24} + x_{19}x_{25} + x_{19}x_{26} + x_{19}x_{27} + x_{19}x_{30} + x_{19}x_{31} + x_{19}x_{33} + x_{19}x_{34} + x_{19}x_{38} + x_{19}x_{39} + x_{19}x_{40} + x_{19}x_{41} + x_{19}x_{42} + x_{19}x_{44} + x_{19}x_{45} + x_{19}x_{51} + x_{19}x_{54} + x_{19}x_{55} + x_{19}x_{56} + x_{19}x_{57} + x_{19}x_{60} + x_{19}x_{61} + x_{19}x_{62} + x_{19}x_{64} + x_{20}x_{23} + x_{20}x_{24} + x_{20}x_{26} + x_{20}x_{27} + x_{20}x_{28} + x_{20}x_{29} + x_{20}x_{30} + x_{20}x_{32} + x_{20}x_{33} + x_{20}x_{34} + x_{20}x_{35} + x_{20}x_{37} + x_{20}x_{38} + x_{20}x_{43} + x_{20}x_{44} + x_{20}x_{45} + x_{20}x_{46} + x_{20}x_{47} + x_{20}x_{48} + x_{20}x_{49} + x_{20}x_{51} + x_{20}x_{52} + x_{20}x_{53} + x_{20}x_{54} + x_{20}x_{56} + x_{20}x_{64} + x_{21}x_{24} + x_{21}x_{25} + x_{21}x_{26} + x_{21}x_{27} + x_{21}x_{28} + x_{21}x_{32} + x_{21}x_{35} + x_{21}x_{38} + x_{21}x_{41} + x_{21}x_{43} + x_{21}x_{44} + x_{21}x_{47} + x_{21}x_{48} + x_{21}x_{52} + x_{21}x_{53} + x_{21}x_{54} + x_{21}x_{55} + x_{21}x_{56} + x_{21}x_{58} + x_{21}x_{59} + x_{21}x_{60} + x_{21}x_{61} + x_{21}x_{63} + x_{21}x_{64} + x_{22}x_{24} + x_{22}x_{26} + x_{22}x_{27} + x_{22}x_{33} + x_{22}x_{34} + x_{22}x_{35} + x_{22}x_{36} + x_{22}x_{37} + x_{22}x_{39} + x_{22}x_{40} + x_{22}x_{41} + x_{22}x_{42} + x_{22}x_{44} + x_{22}x_{46} + x_{22}x_{48} + x_{22}x_{51} + x_{22}x_{53} + x_{22}x_{54} + x_{22}x_{55} + x_{22}x_{56} + x_{22}x_{57} + x_{22}x_{58} + x_{22}x_{61} + x_{22}x_{63} + x_{22}x_{64} + x_{23}x_{25} + x_{23}x_{33} + x_{23}x_{34} + x_{23}x_{36} + x_{23}x_{38} + x_{23}x_{40} + x_{23}x_{41} + x_{23}x_{46} + x_{23}x_{47} + x_{23}x_{48} + x_{23}x_{50} + x_{23}x_{51} + x_{23}x_{52} + x_{23}x_{53} + x_{23}x_{58} + x_{24}x_{26} + x_{24}x_{29} + x_{24}x_{30} + x_{24}x_{31} + x_{24}x_{35} + x_{24}x_{37} + x_{24}x_{44} + x_{24}x_{45} + x_{24}x_{47} + x_{24}x_{52} + x_{24}x_{54} + x_{24}x_{56} + x_{24}x_{58} + x_{24}x_{59} + x_{24}x_{60} + x_{24}x_{62} + x_{24}x_{64} + x_{25}x_{29} + x_{25}x_{31} + x_{25}x_{32} + x_{25}x_{36} + x_{25}x_{38} + x_{25}x_{39} + x_{25}x_{41} + x_{25}x_{42} + x_{25}x_{43} + x_{25}x_{44} + x_{25}x_{50} + x_{25}x_{52} + x_{25}x_{53} + x_{25}x_{56} + x_{25}x_{59} + x_{25}x_{62} + x_{26}x_{28} + x_{26}x_{30} + x_{26}x_{32} + x_{26}x_{33} + x_{26}x_{34} + x_{26}x_{35} + x_{26}x_{37} + x_{26}x_{38} + x_{26}x_{39} + x_{26}x_{40} + x_{26}x_{41} + x_{26}x_{43} + x_{26}x_{44} + x_{26}x_{46} + x_{26}x_{47} + x_{26}x_{49} + x_{26}x_{53} + x_{26}x_{54} + x_{26}x_{57} + x_{26}x_{59} + x_{26}x_{60} + x_{26}x_{63} + x_{26}x_{64} + x_{27}x_{29} + x_{27}x_{35} + x_{27}x_{37} + x_{27}x_{42} + x_{27}x_{46} + x_{27}x_{47} + x_{27}x_{50} + x_{27}x_{51} + x_{27}x_{55} + x_{27}x_{56} + x_{27}x_{57} + x_{27}x_{58} + x_{27}x_{59} + x_{27}x_{60} + x_{27}x_{61} + x_{27}x_{63} + x_{27}x_{64} + x_{28}x_{29} + x_{28}x_{32} + x_{28}x_{33} + x_{28}x_{37} + x_{28}x_{38} + x_{28}x_{40} + x_{28}x_{42} + x_{28}x_{43} + x_{28}x_{46} + x_{28}x_{48} + x_{28}x_{49} + x_{28}x_{51} + x_{28}x_{55} + x_{28}x_{56} + x_{28}x_{61} + x_{29}x_{30} + x_{29}x_{31} + x_{29}x_{32} + x_{29}x_{34} + x_{29}x_{37} + x_{29}x_{38} + x_{29}x_{41} + x_{29}x_{42} + x_{29}x_{43} + x_{29}x_{44} + x_{29}x_{48} + x_{29}x_{49} + x_{29}x_{51} + x_{29}x_{52} + x_{29}x_{55} + x_{29}x_{57} + x_{29}x_{58} + x_{29}x_{62} + x_{29}x_{63} + x_{29}x_{64} + x_{30}x_{32} + x_{30}x_{33} + x_{30}x_{36} + x_{30}x_{38} + x_{30}x_{41} + x_{30}x_{44} + x_{30}x_{45} + x_{30}x_{46} + x_{30}x_{48} + x_{30}x_{49} + x_{30}x_{50} + x_{30}x_{53} + x_{30}x_{54} + x_{30}x_{55} + x_{30}x_{58} + x_{30}x_{61} + x_{31}x_{32} + x_{31}x_{33} + x_{31}x_{37} + x_{31}x_{40} + x_{31}x_{43} + x_{31}x_{47} + x_{31}x_{48} + x_{31}x_{50} + x_{31}x_{51} + x_{31}x_{52} + x_{31}x_{54} + x_{31}x_{55} + x_{31}x_{58} + x_{31}x_{61} + x_{31}x_{63} + x_{32}x_{36} + x_{32}x_{42} + x_{32}x_{43} + x_{32}x_{44} + x_{32}x_{45} + x_{32}x_{49} + x_{32}x_{50} + x_{32}x_{53} + x_{32}x_{56} + x_{32}x_{57} + x_{32}x_{58} + x_{32}x_{61} + x_{32}x_{62} + x_{33}x_{34} + x_{33}x_{35} + x_{33}x_{40} + x_{33}x_{41} + x_{33}x_{42} + x_{33}x_{44} + x_{33}x_{45} + x_{33}x_{47} + x_{33}x_{48} + x_{33}x_{50} + x_{33}x_{51} + x_{33}x_{52} + x_{33}x_{53} + x_{33}x_{54} + x_{33}x_{55} + x_{33}x_{56} + x_{33}x_{58} + x_{33}x_{62} + x_{33}x_{64} + x_{34}x_{35} + x_{34}x_{39} + x_{34}x_{40} + x_{34}x_{41} + x_{34}x_{42} + x_{34}x_{43} + x_{34}x_{44} + x_{34}x_{45} + x_{34}x_{47} + x_{34}x_{48} + x_{34}x_{49} + x_{34}x_{51} + x_{34}x_{52} + x_{34}x_{53} + x_{34}x_{54} + x_{34}x_{55} + x_{34}x_{57} + x_{34}x_{59} + x_{34}x_{63} + x_{35}x_{42} + x_{35}x_{43} + x_{35}x_{44} + x_{35}x_{48} + x_{35}x_{49} + x_{35}x_{50} + x_{35}x_{52} + x_{35}x_{53} + x_{35}x_{55} + x_{35}x_{57} + x_{35}x_{58} + x_{35}x_{59} + x_{35}x_{60} + x_{35}x_{61} + x_{35}x_{62} + x_{35}x_{63} + x_{36}x_{37} + x_{36}x_{38} + x_{36}x_{41} + x_{36}x_{42} + x_{36}x_{48} + x_{36}x_{49} + x_{36}x_{50} + x_{36}x_{52} + x_{36}x_{53} + x_{36}x_{55} + x_{36}x_{56} + x_{36}x_{57} + x_{36}x_{59} + x_{36}x_{62} + x_{36}x_{63} + x_{37}x_{39} + x_{37}x_{41} + x_{37}x_{42} + x_{37}x_{44} + x_{37}x_{45} + x_{37}x_{46} + x_{37}x_{49} + x_{37}x_{51} + x_{37}x_{53} + x_{37}x_{55} + x_{37}x_{60} + x_{37}x_{61} + x_{38}x_{40} + x_{38}x_{41} + x_{38}x_{42} + x_{38}x_{44} + x_{38}x_{45} + x_{38}x_{46} + x_{38}x_{49} + x_{38}x_{50} + x_{38}x_{52} + x_{38}x_{53} + x_{38}x_{54} + x_{38}x_{55} + x_{38}x_{58} + x_{38}x_{59} + x_{38}x_{60} + x_{38}x_{62} + x_{38}x_{63} + x_{39}x_{40} + x_{39}x_{44} + x_{39}x_{45} + x_{39}x_{46} + x_{39}x_{48} + x_{39}x_{49} + x_{39}x_{50} + x_{39}x_{52} + x_{39}x_{53} + x_{39}x_{54} + x_{39}x_{56} + x_{39}x_{58} + x_{39}x_{61} + x_{39}x_{62} + x_{39}x_{64} + x_{40}x_{41} + x_{40}x_{45} + x_{40}x_{47} + x_{40}x_{49} + x_{40}x_{50} + x_{40}x_{51} + x_{40}x_{52} + x_{40}x_{54} + x_{40}x_{56} + x_{40}x_{58} + x_{40}x_{61} + x_{40}x_{62} + x_{40}x_{63} + x_{41}x_{42} + x_{41}x_{44} + x_{41}x_{46} + x_{41}x_{47} + x_{41}x_{48} + x_{41}x_{49} + x_{41}x_{51} + x_{41}x_{54} + x_{41}x_{55} + x_{41}x_{56} + x_{41}x_{57} + x_{41}x_{58} + x_{41}x_{62} + x_{41}x_{63} + x_{41}x_{64} + x_{42}x_{43} + x_{42}x_{44} + x_{42}x_{45} + x_{42}x_{46} + x_{42}x_{48} + x_{42}x_{52} + x_{42}x_{57} + x_{42}x_{63} + x_{43}x_{44} + x_{43}x_{45} + x_{43}x_{46} + x_{43}x_{48} + x_{43}x_{49} + x_{43}x_{51} + x_{43}x_{53} + x_{43}x_{56} + x_{43}x_{58} + x_{43}x_{60} + x_{43}x_{62} + x_{44}x_{47} + x_{44}x_{48} + x_{44}x_{49} + x_{44}x_{51} + x_{44}x_{55} + x_{44}x_{56} + x_{44}x_{57} + x_{44}x_{58} + x_{44}x_{59} + x_{44}x_{61} + x_{44}x_{62} + x_{44}x_{64} + x_{45}x_{48} + x_{45}x_{52} + x_{45}x_{54} + x_{45}x_{56} + x_{45}x_{57} + x_{45}x_{58} + x_{45}x_{59} + x_{45}x_{63} + x_{45}x_{64} + x_{46}x_{48} + x_{46}x_{49} + x_{46}x_{51} + x_{46}x_{52} + x_{46}x_{53} + x_{46}x_{55} + x_{46}x_{56} + x_{46}x_{58} + x_{46}x_{59} + x_{46}x_{62} + x_{46}x_{63} + x_{46}x_{64} + x_{47}x_{50} + x_{47}x_{52} + x_{47}x_{54} + x_{47}x_{56} + x_{47}x_{57} + x_{47}x_{59} + x_{47}x_{61} + x_{48}x_{51} + x_{48}x_{52} + x_{48}x_{53} + x_{48}x_{54} + x_{48}x_{58} + x_{48}x_{62} + x_{48}x_{63} + x_{49}x_{51} + x_{49}x_{54} + x_{49}x_{56} + x_{49}x_{58} + x_{49}x_{59} + x_{49}x_{60} + x_{49}x_{62} + x_{50}x_{53} + x_{50}x_{55} + x_{50}x_{56} + x_{50}x_{58} + x_{50}x_{60} + x_{50}x_{62} + x_{50}x_{64} + x_{51}x_{52} + x_{51}x_{53} + x_{51}x_{55} + x_{51}x_{57} + x_{51}x_{58} + x_{51}x_{59} + x_{51}x_{60} + x_{52}x_{58} + x_{52}x_{61} + x_{52}x_{62} + x_{52}x_{63} + x_{52}x_{64} + x_{53}x_{54} + x_{53}x_{56} + x_{53}x_{58} + x_{53}x_{60} + x_{53}x_{61} + x_{53}x_{62} + x_{53}x_{63} + x_{54}x_{56} + x_{54}x_{57} + x_{54}x_{62} + x_{54}x_{63} + x_{54}x_{64} + x_{55}x_{56} + x_{55}x_{57} + x_{55}x_{58} + x_{55}x_{59} + x_{55}x_{61} + x_{55}x_{63} + x_{55}x_{64} + x_{56}x_{57} + x_{56}x_{62} + x_{56}x_{64} + x_{57}x_{58} + x_{57}x_{60} + x_{57}x_{63} + x_{58}x_{59} + x_{58}x_{60} + x_{58}x_{61} + x_{58}x_{64} + x_{59}x_{60} + x_{59}x_{62} + x_{60}x_{63} + x_{60}x_{64} + x_{61}x_{62} + x_{61}x_{64} + x_{3} + x_{6} + x_{7} + x_{11} + x_{13} + x_{19} + x_{20} + x_{22} + x_{24} + x_{26} + x_{29} + x_{31} + x_{32} + x_{33} + x_{34} + x_{37} + x_{40} + x_{42} + x_{43} + x_{44} + x_{45} + x_{46} + x_{47} + x_{48} + x_{50} + x_{51} + x_{53} + x_{57} + x_{59} + x_{60} + x_{61} + x_{62} + 1$

$y_{7} = x_{1}x_{2} + x_{1}x_{4} + x_{1}x_{7} + x_{1}x_{8} + x_{1}x_{10} + x_{1}x_{12} + x_{1}x_{14} + x_{1}x_{15} + x_{1}x_{17} + x_{1}x_{18} + x_{1}x_{19} + x_{1}x_{21} + x_{1}x_{22} + x_{1}x_{24} + x_{1}x_{27} + x_{1}x_{34} + x_{1}x_{35} + x_{1}x_{36} + x_{1}x_{40} + x_{1}x_{42} + x_{1}x_{43} + x_{1}x_{45} + x_{1}x_{47} + x_{1}x_{48} + x_{1}x_{54} + x_{1}x_{56} + x_{1}x_{57} + x_{1}x_{59} + x_{1}x_{62} + x_{1}x_{63} + x_{2}x_{7} + x_{2}x_{11} + x_{2}x_{13} + x_{2}x_{14} + x_{2}x_{16} + x_{2}x_{17} + x_{2}x_{18} + x_{2}x_{24} + x_{2}x_{25} + x_{2}x_{26} + x_{2}x_{29} + x_{2}x_{41} + x_{2}x_{44} + x_{2}x_{45} + x_{2}x_{46} + x_{2}x_{47} + x_{2}x_{49} + x_{2}x_{52} + x_{2}x_{53} + x_{2}x_{58} + x_{2}x_{59} + x_{2}x_{62} + x_{2}x_{63} + x_{2}x_{64} + x_{3}x_{4} + x_{3}x_{6} + x_{3}x_{7} + x_{3}x_{8} + x_{3}x_{9} + x_{3}x_{10} + x_{3}x_{14} + x_{3}x_{15} + x_{3}x_{16} + x_{3}x_{17} + x_{3}x_{18} + x_{3}x_{19} + x_{3}x_{20} + x_{3}x_{22} + x_{3}x_{23} + x_{3}x_{24} + x_{3}x_{28} + x_{3}x_{29} + x_{3}x_{30} + x_{3}x_{31} + x_{3}x_{33} + x_{3}x_{34} + x_{3}x_{35} + x_{3}x_{37} + x_{3}x_{38} + x_{3}x_{40} + x_{3}x_{42} + x_{3}x_{45} + x_{3}x_{50} + x_{3}x_{51} + x_{3}x_{53} + x_{3}x_{55} + x_{3}x_{56} + x_{3}x_{58} + x_{3}x_{61} + x_{3}x_{62} + x_{3}x_{64} + x_{4}x_{11} + x_{4}x_{12} + x_{4}x_{13} + x_{4}x_{15} + x_{4}x_{16} + x_{4}x_{20} + x_{4}x_{21} + x_{4}x_{23} + x_{4}x_{24} + x_{4}x_{27} + x_{4}x_{28} + x_{4}x_{30} + x_{4}x_{33} + x_{4}x_{35} + x_{4}x_{36} + x_{4}x_{37} + x_{4}x_{39} + x_{4}x_{41} + x_{4}x_{43} + x_{4}x_{44} + x_{4}x_{48} + x_{4}x_{49} + x_{4}x_{51} + x_{4}x_{52} + x_{4}x_{53} + x_{4}x_{54} + x_{4}x_{56} + x_{4}x_{58} + x_{4}x_{60} + x_{4}x_{61} + x_{4}x_{62} + x_{5}x_{6} + x_{5}x_{7} + x_{5}x_{11} + x_{5}x_{15} + x_{5}x_{17} + x_{5}x_{18} + x_{5}x_{19} + x_{5}x_{20} + x_{5}x_{22} + x_{5}x_{23} + x_{5}x_{24} + x_{5}x_{25} + x_{5}x_{26} + x_{5}x_{27} + x_{5}x_{29} + x_{5}x_{30} + x_{5}x_{34} + x_{5}x_{35} + x_{5}x_{36} + x_{5}x_{40} + x_{5}x_{42} + x_{5}x_{43} + x_{5}x_{45} + x_{5}x_{49} + x_{5}x_{50} + x_{5}x_{51} + x_{5}x_{54} + x_{5}x_{57} + x_{5}x_{59} + x_{5}x_{60} + x_{5}x_{61} + x_{5}x_{63} + x_{6}x_{7} + x_{6}x_{8} + x_{6}x_{9} + x_{6}x_{10} + x_{6}x_{11} + x_{6}x_{12} + x_{6}x_{13} + x_{6}x_{14} + x_{6}x_{15} + x_{6}x_{16} + x_{6}x_{17} + x_{6}x_{18} + x_{6}x_{19} + x_{6}x_{20} + x_{6}x_{22} + x_{6}x_{23} + x_{6}x_{24} + x_{6}x_{26} + x_{6}x_{28} + x_{6}x_{30} + x_{6}x_{31} + x_{6}x_{32} + x_{6}x_{34} + x_{6}x_{35} + x_{6}x_{36} + x_{6}x_{41} + x_{6}x_{42} + x_{6}x_{46} + x_{6}x_{47} + x_{6}x_{48} + x_{6}x_{49} + x_{6}x_{50} + x_{6}x_{51} + x_{6}x_{52} + x_{6}x_{54} + x_{6}x_{56} + x_{6}x_{59} + x_{6}x_{63} + x_{7}x_{10} + x_{7}x_{11} + x_{7}x_{13} + x_{7}x_{19} + x_{7}x_{20} + x_{7}x_{23} + x_{7}x_{24} + x_{7}x_{25} + x_{7}x_{26} + x_{7}x_{27} + x_{7}x_{31} + x_{7}x_{36} + x_{7}x_{38} + x_{7}x_{40} + x_{7}x_{41} + x_{7}x_{43} + x_{7}x_{44} + x_{7}x_{46} + x_{7}x_{47} + x_{7}x_{48} + x_{7}x_{51} + x_{7}x_{52} + x_{7}x_{53} + x_{7}x_{56} + x_{7}x_{57} + x_{7}x_{58} + x_{7}x_{60} + x_{7}x_{61} + x_{8}x_{9} + x_{8}x_{10} + x_{8}x_{11} + x_{8}x_{12} + x_{8}x_{13} + x_{8}x_{14} + x_{8}x_{15} + x_{8}x_{17} + x_{8}x_{18} + x_{8}x_{19} + x_{8}x_{20} + x_{8}x_{22} + x_{8}x_{24} + x_{8}x_{27} + x_{8}x_{28} + x_{8}x_{31} + x_{8}x_{36} + x_{8}x_{37} + x_{8}x_{38} + x_{8}x_{43} + x_{8}x_{45} + x_{8}x_{46} + x_{8}x_{49} + x_{8}x_{51} + x_{8}x_{53} + x_{8}x_{54} + x_{8}x_{56} + x_{8}x_{57} + x_{8}x_{58} + x_{8}x_{59} + x_{8}x_{61} + x_{8}x_{63} + x_{8}x_{64} + x_{9}x_{10} + x_{9}x_{13} + x_{9}x_{14} + x_{9}x_{15} + x_{9}x_{17} + x_{9}x_{19} + x_{9}x_{20} + x_{9}x_{22} + x_{9}x_{23} + x_{9}x_{26} + x_{9}x_{27} + x_{9}x_{30} + x_{9}x_{31} + x_{9}x_{33} + x_{9}x_{36} + x_{9}x_{38} + x_{9}x_{40} + x_{9}x_{46} + x_{9}x_{47} + x_{9}x_{48} + x_{9}x_{49} + x_{9}x_{50} + x_{9}x_{52} + x_{9}x_{53} + x_{9}x_{58} + x_{9}x_{60} + x_{9}x_{63} + x_{10}x_{11} + x_{10}x_{12} + x_{10}x_{15} + x_{10}x_{17} + x_{10}x_{19} + x_{10}x_{22} + x_{10}x_{23} + x_{10}x_{24} + x_{10}x_{25} + x_{10}x_{26} + x_{10}x_{27} + x_{10}x_{28} + x_{10}x_{30} + x_{10}x_{32} + x_{10}x_{33} + x_{10}x_{34} + x_{10}x_{36} + x_{10}x_{37} + x_{10}x_{38} + x_{10}x_{39} + x_{10}x_{43} + x_{10}x_{46} + x_{10}x_{47} + x_{10}x_{49} + x_{10}x_{50} + x_{10}x_{51} + x_{10}x_{52} + x_{10}x_{53} + x_{10}x_{54} + x_{10}x_{55} + x_{10}x_{56} + x_{10}x_{57} + x_{10}x_{58} + x_{10}x_{59} + x_{10}x_{60} + x_{10}x_{61} + x_{10}x_{63} + x_{10}x_{64} + x_{11}x_{12} + x_{11}x_{13} + x_{11}x_{14} + x_{11}x_{16} + x_{11}x_{17} + x_{11}x_{19} + x_{11}x_{20} + x_{11}x_{25} + x_{11}x_{26} + x_{11}x_{29} + x_{11}x_{31} + x_{11}x_{35} + x_{11}x_{36} + x_{11}x_{40} + x_{11}x_{41} + x_{11}x_{48} + x_{11}x_{50} + x_{11}x_{54} + x_{11}x_{56} + x_{11}x_{57} + x_{11}x_{60} + x_{11}x_{63} + x_{11}x_{64} + x_{12}x_{13} + x_{12}x_{16} + x_{12}x_{17} + x_{12}x_{19} + x_{12}x_{21} + x_{12}x_{22} + x_{12}x_{24} + x_{12}x_{28} + x_{12}x_{30} + x_{12}x_{34} + x_{12}x_{35} + x_{12}x_{36} + x_{12}x_{38} + x_{12}x_{45} + x_{12}x_{47} + x_{12}x_{49} + x_{12}x_{50} + x_{12}x_{53} + x_{12}x_{55} + x_{12}x_{56} + x_{12}x_{57} + x_{12}x_{58} + x_{12}x_{60} + x_{12}x_{61} + x_{12}x_{62} + x_{12}x_{63} + x_{13}x_{14} + x_{13}x_{16} + x_{13}x_{18} + x_{13}x_{20} + x_{13}x_{21} + x_{13}x_{22} + x_{13}x_{24} + x_{13}x_{25} + x_{13}x_{26} + x_{13}x_{30} + x_{13}x_{32} + x_{13}x_{33} + x_{13}x_{35} + x_{13}x_{36} + x_{13}x_{37} + x_{13}x_{39} + x_{13}x_{41} + x_{13}x_{43} + x_{13}x_{44} + x_{13}x_{46} + x_{13}x_{48} + x_{13}x_{49} + x_{13}x_{50} + x_{13}x_{53} + x_{13}x_{57} + x_{13}x_{61} + x_{13}x_{62} + x_{13}x_{64} + x_{14}x_{15} + x_{14}x_{20} + x_{14}x_{21} + x_{14}x_{22} + x_{14}x_{23} + x_{14}x_{24} + x_{14}x_{25} + x_{14}x_{28} + x_{14}x_{31} + x_{14}x_{36} + x_{14}x_{38} + x_{14}x_{40} + x_{14}x_{41} + x_{14}x_{42} + x_{14}x_{43} + x_{14}x_{45} + x_{14}x_{46} + x_{14}x_{47} + x_{14}x_{48} + x_{14}x_{49} + x_{14}x_{54} + x_{14}x_{55} + x_{14}x_{56} + x_{14}x_{57} + x_{14}x_{61} + x_{15}x_{17} + x_{15}x_{18} + x_{15}x_{20} + x_{15}x_{21} + x_{15}x_{22} + x_{15}x_{26} + x_{15}x_{28} + x_{15}x_{29} + x_{15}x_{31} + x_{15}x_{33} + x_{15}x_{34} + x_{15}x_{35} + x_{15}x_{40} + x_{15}x_{41} + x_{15}x_{43} + x_{15}x_{49} + x_{15}x_{51} + x_{15}x_{52} + x_{15}x_{55} + x_{15}x_{58} + x_{15}x_{63} + x_{15}x_{64} + x_{16}x_{19} + x_{16}x_{23} + x_{16}x_{24} + x_{16}x_{27} + x_{16}x_{29} + x_{16}x_{31} + x_{16}x_{33} + x_{16}x_{34} + x_{16}x_{38} + x_{16}x_{39} + x_{16}x_{42} + x_{16}x_{45} + x_{16}x_{46} + x_{16}x_{47} + x_{16}x_{48} + x_{16}x_{49} + x_{16}x_{51} + x_{16}x_{55} + x_{16}x_{57} + x_{16}x_{59} + x_{16}x_{60} + x_{16}x_{61} + x_{16}x_{62} + x_{16}x_{64} + x_{17}x_{21} + x_{17}x_{22} + x_{17}x_{23} + x_{17}x_{24} + x_{17}x_{26} + x_{17}x_{27} + x_{17}x_{32} + x_{17}x_{33} + x_{17}x_{35} + x_{17}x_{37} + x_{17}x_{40} + x_{17}x_{41} + x_{17}x_{42} + x_{17}x_{44} + x_{17}x_{45} + x_{17}x_{46} + x_{17}x_{47} + x_{17}x_{49} + x_{17}x_{50} + x_{17}x_{52} + x_{17}x_{53} + x_{17}x_{54} + x_{17}x_{57} + x_{17}x_{58} + x_{17}x_{59} + x_{17}x_{61} + x_{17}x_{62} + x_{17}x_{63} + x_{18}x_{20} + x_{18}x_{21} + x_{18}x_{22} + x_{18}x_{24} + x_{18}x_{27} + x_{18}x_{29} + x_{18}x_{31} + x_{18}x_{35} + x_{18}x_{36} + x_{18}x_{39} + x_{18}x_{40} + x_{18}x_{41} + x_{18}x_{43} + x_{18}x_{44} + x_{18}x_{48} + x_{18}x_{49} + x_{18}x_{53} + x_{18}x_{54} + x_{18}x_{55} + x_{18}x_{56} + x_{18}x_{59} + x_{18}x_{63} + x_{18}x_{64} + x_{19}x_{22} + x_{19}x_{23} + x_{19}x_{26} + x_{19}x_{31} + x_{19}x_{35} + x_{19}x_{36} + x_{19}x_{37} + x_{19}x_{40} + x_{19}x_{41} + x_{19}x_{42} + x_{19}x_{43} + x_{19}x_{48} + x_{19}x_{49} + x_{19}x_{50} + x_{19}x_{52} + x_{19}x_{55} + x_{19}x_{57} + x_{19}x_{61} + x_{19}x_{63} + x_{20}x_{22} + x_{20}x_{23} + x_{20}x_{24} + x_{20}x_{26} + x_{20}x_{29} + x_{20}x_{30} + x_{20}x_{37} + x_{20}x_{38} + x_{20}x_{43} + x_{20}x_{44} + x_{20}x_{46} + x_{20}x_{48} + x_{20}x_{51} + x_{20}x_{53} + x_{20}x_{54} + x_{20}x_{55} + x_{20}x_{56} + x_{20}x_{57} + x_{20}x_{58} + x_{20}x_{59} + x_{20}x_{60} + x_{20}x_{61} + x_{20}x_{62} + x_{21}x_{26} + x_{21}x_{27} + x_{21}x_{31} + x_{21}x_{33} + x_{21}x_{35} + x_{21}x_{39} + x_{21}x_{43} + x_{21}x_{45} + x_{21}x_{46} + x_{21}x_{47} + x_{21}x_{48} + x_{21}x_{49} + x_{21}x_{51} + x_{21}x_{52} + x_{21}x_{54} + x_{21}x_{55} + x_{21}x_{56} + x_{21}x_{59} + x_{21}x_{62} + x_{21}x_{63} + x_{21}x_{64} + x_{22}x_{24} + x_{22}x_{27} + x_{22}x_{28} + x_{22}x_{31} + x_{22}x_{34} + x_{22}x_{36} + x_{22}x_{37} + x_{22}x_{38} + x_{22}x_{41} + x_{22}x_{43} + x_{22}x_{44} + x_{22}x_{45} + x_{22}x_{46} + x_{22}x_{49} + x_{22}x_{51} + x_{22}x_{54} + x_{22}x_{55} + x_{22}x_{56} + x_{22}x_{58} + x_{22}x_{59} + x_{22}x_{62} + x_{23}x_{24} + x_{23}x_{25} + x_{23}x_{29} + x_{23}x_{30} + x_{23}x_{32} + x_{23}x_{34} + x_{23}x_{36} + x_{23}x_{37} + x_{23}x_{38} + x_{23}x_{42} + x_{23}x_{45} + x_{23}x_{48} + x_{23}x_{49} + x_{23}x_{52} + x_{23}x_{53} + x_{23}x_{57} + x_{23}x_{58} + x_{23}x_{59} + x_{23}x_{60} + x_{23}x_{61} + x_{23}x_{62} + x_{23}x_{63} + x_{23}x_{64} + x_{24}x_{30} + x_{24}x_{32} + x_{24}x_{35} + x_{24}x_{38} + x_{24}x_{41} + x_{24}x_{43} + x_{24}x_{45} + x_{24}x_{48} + x_{24}x_{52} + x_{24}x_{54} + x_{24}x_{56} + x_{24}x_{57} + x_{24}x_{58} + x_{24}x_{60} + x_{24}x_{62} + x_{24}x_{63} + x_{25}x_{27} + x_{25}x_{33} + x_{25}x_{37} + x_{25}x_{38} + x_{25}x_{39} + x_{25}x_{42} + x_{25}x_{52} + x_{25}x_{59} + x_{25}x_{62} + x_{25}x_{64} + x_{26}x_{28} + x_{26}x_{29} + x_{26}x_{41} + x_{26}x_{42} + x_{26}x_{45} + x_{26}x_{48} + x_{26}x_{49} + x_{26}x_{50} + x_{26}x_{51} + x_{26}x_{54} + x_{26}x_{55} + x_{26}x_{59} + x_{26}x_{60} + x_{26}x_{63} + x_{26}x_{64} + x_{27}x_{29} + x_{27}x_{30} + x_{27}x_{32} + x_{27}x_{34} + x_{27}x_{36} + x_{27}x_{37} + x_{27}x_{42} + x_{27}x_{44} + x_{27}x_{45} + x_{27}x_{49} + x_{27}x_{51} + x_{27}x_{52} + x_{27}x_{53} + x_{27}x_{54} + x_{27}x_{55} + x_{27}x_{57} + x_{27}x_{59} + x_{27}x_{62} + x_{28}x_{31} + x_{28}x_{33} + x_{28}x_{34} + x_{28}x_{35} + x_{28}x_{36} + x_{28}x_{39} + x_{28}x_{40} + x_{28}x_{41} + x_{28}x_{42} + x_{28}x_{48} + x_{28}x_{49} + x_{28}x_{52} + x_{28}x_{56} + x_{28}x_{57} + x_{28}x_{63} + x_{29}x_{30} + x_{29}x_{31} + x_{29}x_{33} + x_{29}x_{34} + x_{29}x_{37} + x_{29}x_{38} + x_{29}x_{41} + x_{29}x_{42} + x_{29}x_{46} + x_{29}x_{48} + x_{29}x_{56} + x_{29}x_{59} + x_{29}x_{61} + x_{29}x_{62} + x_{30}x_{31} + x_{30}x_{32} + x_{30}x_{33} + x_{30}x_{34} + x_{30}x_{36} + x_{30}x_{37} + x_{30}x_{39} + x_{30}x_{41} + x_{30}x_{42} + x_{30}x_{43} + x_{30}x_{46} + x_{30}x_{48} + x_{30}x_{50} + x_{30}x_{52} + x_{30}x_{54} + x_{30}x_{55} + x_{30}x_{57} + x_{30}x_{58} + x_{30}x_{60} + x_{30}x_{62} + x_{31}x_{38} + x_{31}x_{40} + x_{31}x_{41} + x_{31}x_{42} + x_{31}x_{43} + x_{31}x_{44} + x_{31}x_{45} + x_{31}x_{46} + x_{31}x_{47} + x_{31}x_{51} + x_{31}x_{53} + x_{31}x_{54} + x_{31}x_{57} + x_{31}x_{58} + x_{31}x_{59} + x_{31}x_{60} + x_{31}x_{61} + x_{31}x_{63} + x_{32}x_{33} + x_{32}x_{34} + x_{32}x_{35} + x_{32}x_{37} + x_{32}x_{40} + x_{32}x_{41} + x_{32}x_{43} + x_{32}x_{46} + x_{32}x_{47} + x_{32}x_{51} + x_{32}x_{53} + x_{32}x_{58} + x_{32}x_{60} + x_{32}x_{61} + x_{32}x_{64} + x_{33}x_{34} + x_{33}x_{35} + x_{33}x_{36} + x_{33}x_{37} + x_{33}x_{38} + x_{33}x_{39} + x_{33}x_{40} + x_{33}x_{41} + x_{33}x_{42} + x_{33}x_{43} + x_{33}x_{45} + x_{33}x_{47} + x_{33}x_{49} + x_{33}x_{50} + x_{33}x_{51} + x_{33}x_{53} + x_{33}x_{54} + x_{33}x_{55} + x_{33}x_{56} + x_{33}x_{58} + x_{33}x_{60} + x_{33}x_{61} + x_{33}x_{63} + x_{34}x_{39} + x_{34}x_{40} + x_{34}x_{41} + x_{34}x_{44} + x_{34}x_{46} + x_{34}x_{47} + x_{34}x_{49} + x_{34}x_{51} + x_{34}x_{52} + x_{34}x_{53} + x_{34}x_{54} + x_{34}x_{56} + x_{34}x_{57} + x_{34}x_{58} + x_{34}x_{60} + x_{35}x_{38} + x_{35}x_{41} + x_{35}x_{44} + x_{35}x_{50} + x_{35}x_{51} + x_{35}x_{52} + x_{35}x_{53} + x_{35}x_{55} + x_{35}x_{56} + x_{35}x_{58} + x_{35}x_{62} + x_{35}x_{63} + x_{35}x_{64} + x_{36}x_{37} + x_{36}x_{38} + x_{36}x_{39} + x_{36}x_{40} + x_{36}x_{41} + x_{36}x_{42} + x_{36}x_{43} + x_{36}x_{46} + x_{36}x_{49} + x_{36}x_{50} + x_{36}x_{52} + x_{36}x_{53} + x_{36}x_{57} + x_{36}x_{60} + x_{36}x_{61} + x_{36}x_{62} + x_{36}x_{64} + x_{37}x_{38} + x_{37}x_{39} + x_{37}x_{40} + x_{37}x_{41} + x_{37}x_{42} + x_{37}x_{43} + x_{37}x_{44} + x_{37}x_{45} + x_{37}x_{47} + x_{37}x_{48} + x_{37}x_{49} + x_{37}x_{50} + x_{37}x_{53} + x_{37}x_{54} + x_{37}x_{56} + x_{37}x_{59} + x_{37}x_{60} + x_{37}x_{63} + x_{37}x_{64} + x_{38}x_{40} + x_{38}x_{43} + x_{38}x_{44} + x_{38}x_{45} + x_{38}x_{50} + x_{38}x_{53} + x_{38}x_{56} + x_{38}x_{58} + x_{38}x_{61} + x_{38}x_{63} + x_{39}x_{41} + x_{39}x_{42} + x_{39}x_{44} + x_{39}x_{48} + x_{39}x_{51} + x_{39}x_{54} + x_{39}x_{58} + x_{39}x_{62} + x_{40}x_{41} + x_{40}x_{42} + x_{40}x_{43} + x_{40}x_{44} + x_{40}x_{48} + x_{40}x_{49} + x_{40}x_{50} + x_{40}x_{51} + x_{40}x_{52} + x_{40}x_{53} + x_{40}x_{55} + x_{40}x_{57} + x_{40}x_{60} + x_{40}x_{62} + x_{41}x_{43} + x_{41}x_{44} + x_{41}x_{47} + x_{41}x_{49} + x_{41}x_{51} + x_{41}x_{56} + x_{41}x_{57} + x_{41}x_{58} + x_{41}x_{64} + x_{42}x_{43} + x_{42}x_{45} + x_{42}x_{48} + x_{42}x_{49} + x_{42}x_{52} + x_{42}x_{54} + x_{42}x_{56} + x_{42}x_{57} + x_{42}x_{59} + x_{42}x_{62} + x_{43}x_{44} + x_{43}x_{46} + x_{43}x_{49} + x_{43}x_{51} + x_{43}x_{52} + x_{43}x_{53} + x_{43}x_{57} + x_{43}x_{58} + x_{43}x_{60} + x_{43}x_{62} + x_{43}x_{64} + x_{44}x_{45} + x_{44}x_{46} + x_{44}x_{48} + x_{44}x_{49} + x_{44}x_{51} + x_{44}x_{54} + x_{44}x_{55} + x_{44}x_{58} + x_{44}x_{59} + x_{44}x_{60} + x_{44}x_{62} + x_{45}x_{47} + x_{45}x_{49} + x_{45}x_{51} + x_{45}x_{52} + x_{45}x_{56} + x_{45}x_{59} + x_{45}x_{60} + x_{45}x_{61} + x_{45}x_{64} + x_{46}x_{49} + x_{46}x_{50} + x_{46}x_{51} + x_{46}x_{52} + x_{46}x_{55} + x_{46}x_{57} + x_{46}x_{58} + x_{46}x_{59} + x_{46}x_{60} + x_{46}x_{61} + x_{46}x_{63} + x_{46}x_{64} + x_{47}x_{48} + x_{47}x_{49} + x_{47}x_{51} + x_{47}x_{52} + x_{47}x_{53} + x_{47}x_{55} + x_{47}x_{56} + x_{47}x_{58} + x_{47}x_{61} + x_{48}x_{50} + x_{48}x_{53} + x_{48}x_{54} + x_{48}x_{55} + x_{48}x_{56} + x_{48}x_{59} + x_{48}x_{60} + x_{48}x_{64} + x_{49}x_{52} + x_{49}x_{54} + x_{49}x_{56} + x_{49}x_{57} + x_{49}x_{63} + x_{50}x_{56} + x_{50}x_{58} + x_{50}x_{60} + x_{50}x_{62} + x_{50}x_{64} + x_{51}x_{54} + x_{51}x_{56} + x_{51}x_{61} + x_{51}x_{63} + x_{52}x_{55} + x_{52}x_{57} + x_{52}x_{58} + x_{52}x_{59} + x_{52}x_{61} + x_{52}x_{62} + x_{52}x_{63} + x_{52}x_{64} + x_{53}x_{55} + x_{53}x_{57} + x_{53}x_{58} + x_{53}x_{60} + x_{53}x_{61} + x_{53}x_{62} + x_{53}x_{63} + x_{53}x_{64} + x_{54}x_{55} + x_{54}x_{56} + x_{54}x_{57} + x_{54}x_{58} + x_{54}x_{61} + x_{54}x_{63} + x_{54}x_{64} + x_{55}x_{56} + x_{55}x_{57} + x_{55}x_{58} + x_{55}x_{61} + x_{55}x_{62} + x_{55}x_{63} + x_{56}x_{57} + x_{56}x_{58} + x_{56}x_{61} + x_{56}x_{62} + x_{56}x_{64} + x_{57}x_{60} + x_{57}x_{61} + x_{57}x_{62} + x_{58}x_{61} + x_{58}x_{64} + x_{59}x_{60} + x_{60}x_{61} + x_{61}x_{63} + x_{62}x_{63} + x_{62}x_{64} + x_{2} + x_{3} + x_{4} + x_{5} + x_{10} + x_{11} + x_{12} + x_{15} + x_{16} + x_{17} + x_{18} + x_{24} + x_{25} + x_{27} + x_{31} + x_{32} + x_{35} + x_{42} + x_{44} + x_{46} + x_{47} + x_{53} + x_{55} + x_{59} + 1$

$y_{8} = x_{1}x_{2} + x_{1}x_{3} + x_{1}x_{9} + x_{1}x_{10} + x_{1}x_{11} + x_{1}x_{14} + x_{1}x_{15} + x_{1}x_{16} + x_{1}x_{18} + x_{1}x_{19} + x_{1}x_{22} + x_{1}x_{23} + x_{1}x_{27} + x_{1}x_{29} + x_{1}x_{31} + x_{1}x_{32} + x_{1}x_{36} + x_{1}x_{37} + x_{1}x_{38} + x_{1}x_{40} + x_{1}x_{41} + x_{1}x_{45} + x_{1}x_{46} + x_{1}x_{47} + x_{1}x_{50} + x_{1}x_{52} + x_{1}x_{53} + x_{1}x_{54} + x_{1}x_{55} + x_{1}x_{56} + x_{1}x_{57} + x_{1}x_{63} + x_{1}x_{64} + x_{2}x_{3} + x_{2}x_{4} + x_{2}x_{6} + x_{2}x_{7} + x_{2}x_{8} + x_{2}x_{10} + x_{2}x_{11} + x_{2}x_{13} + x_{2}x_{15} + x_{2}x_{17} + x_{2}x_{19} + x_{2}x_{20} + x_{2}x_{21} + x_{2}x_{23} + x_{2}x_{24} + x_{2}x_{25} + x_{2}x_{26} + x_{2}x_{27} + x_{2}x_{28} + x_{2}x_{29} + x_{2}x_{30} + x_{2}x_{33} + x_{2}x_{37} + x_{2}x_{39} + x_{2}x_{45} + x_{2}x_{47} + x_{2}x_{49} + x_{2}x_{50} + x_{2}x_{52} + x_{2}x_{53} + x_{2}x_{54} + x_{2}x_{56} + x_{2}x_{58} + x_{2}x_{61} + x_{2}x_{62} + x_{3}x_{5} + x_{3}x_{7} + x_{3}x_{8} + x_{3}x_{9} + x_{3}x_{10} + x_{3}x_{11} + x_{3}x_{12} + x_{3}x_{13} + x_{3}x_{14} + x_{3}x_{15} + x_{3}x_{16} + x_{3}x_{20} + x_{3}x_{21} + x_{3}x_{25} + x_{3}x_{28} + x_{3}x_{29} + x_{3}x_{30} + x_{3}x_{33} + x_{3}x_{35} + x_{3}x_{39} + x_{3}x_{40} + x_{3}x_{41} + x_{3}x_{43} + x_{3}x_{48} + x_{3}x_{49} + x_{3}x_{50} + x_{3}x_{51} + x_{3}x_{52} + x_{3}x_{53} + x_{3}x_{55} + x_{3}x_{56} + x_{3}x_{57} + x_{3}x_{59} + x_{3}x_{62} + x_{3}x_{63} + x_{3}x_{64} + x_{4}x_{6} + x_{4}x_{7} + x_{4}x_{8} + x_{4}x_{10} + x_{4}x_{11} + x_{4}x_{13} + x_{4}x_{16} + x_{4}x_{19} + x_{4}x_{20} + x_{4}x_{24} + x_{4}x_{26} + x_{4}x_{27} + x_{4}x_{28} + x_{4}x_{30} + x_{4}x_{33} + x_{4}x_{34} + x_{4}x_{35} + x_{4}x_{36} + x_{4}x_{37} + x_{4}x_{39} + x_{4}x_{40} + x_{4}x_{44} + x_{4}x_{46} + x_{4}x_{48} + x_{4}x_{50} + x_{4}x_{52} + x_{4}x_{53} + x_{4}x_{54} + x_{4}x_{55} + x_{4}x_{56} + x_{4}x_{58} + x_{4}x_{59} + x_{4}x_{60} + x_{4}x_{62} + x_{4}x_{64} + x_{5}x_{9} + x_{5}x_{10} + x_{5}x_{12} + x_{5}x_{14} + x_{5}x_{18} + x_{5}x_{20} + x_{5}x_{24} + x_{5}x_{26} + x_{5}x_{27} + x_{5}x_{30} + x_{5}x_{31} + x_{5}x_{32} + x_{5}x_{36} + x_{5}x_{39} + x_{5}x_{41} + x_{5}x_{43} + x_{5}x_{45} + x_{5}x_{50} + x_{5}x_{52} + x_{5}x_{53} + x_{5}x_{55} + x_{5}x_{56} + x_{5}x_{57} + x_{5}x_{60} + x_{5}x_{62} + x_{5}x_{64} + x_{6}x_{7} + x_{6}x_{9} + x_{6}x_{10} + x_{6}x_{13} + x_{6}x_{14} + x_{6}x_{15} + x_{6}x_{16} + x_{6}x_{17} + x_{6}x_{21} + x_{6}x_{23} + x_{6}x_{24} + x_{6}x_{27} + x_{6}x_{29} + x_{6}x_{30} + x_{6}x_{31} + x_{6}x_{32} + x_{6}x_{35} + x_{6}x_{37} + x_{6}x_{38} + x_{6}x_{39} + x_{6}x_{40} + x_{6}x_{43} + x_{6}x_{51} + x_{6}x_{52} + x_{6}x_{53} + x_{6}x_{56} + x_{6}x_{60} + x_{6}x_{61} + x_{6}x_{64} + x_{7}x_{8} + x_{7}x_{12} + x_{7}x_{15} + x_{7}x_{16} + x_{7}x_{18} + x_{7}x_{20} + x_{7}x_{24} + x_{7}x_{27} + x_{7}x_{28} + x_{7}x_{31} + x_{7}x_{35} + x_{7}x_{36} + x_{7}x_{37} + x_{7}x_{38} + x_{7}x_{39} + x_{7}x_{41} + x_{7}x_{43} + x_{7}x_{44} + x_{7}x_{45} + x_{7}x_{48} + x_{7}x_{50} + x_{7}x_{52} + x_{7}x_{54} + x_{7}x_{56} + x_{7}x_{57} + x_{7}x_{60} + x_{7}x_{63} + x_{8}x_{9} + x_{8}x_{11} + x_{8}x_{14} + x_{8}x_{15} + x_{8}x_{17} + x_{8}x_{22} + x_{8}x_{24} + x_{8}x_{29} + x_{8}x_{31} + x_{8}x_{32} + x_{8}x_{33} + x_{8}x_{34} + x_{8}x_{35} + x_{8}x_{36} + x_{8}x_{40} + x_{8}x_{41} + x_{8}x_{42} + x_{8}x_{43} + x_{8}x_{46} + x_{8}x_{47} + x_{8}x_{48} + x_{8}x_{51} + x_{8}x_{53} + x_{8}x_{54} + x_{8}x_{55} + x_{8}x_{59} + x_{8}x_{61} + x_{9}x_{11} + x_{9}x_{12} + x_{9}x_{13} + x_{9}x_{14} + x_{9}x_{15} + x_{9}x_{22} + x_{9}x_{25} + x_{9}x_{28} + x_{9}x_{29} + x_{9}x_{30} + x_{9}x_{32} + x_{9}x_{33} + x_{9}x_{34} + x_{9}x_{35} + x_{9}x_{36} + x_{9}x_{37} + x_{9}x_{38} + x_{9}x_{39} + x_{9}x_{42} + x_{9}x_{43} + x_{9}x_{46} + x_{9}x_{47} + x_{9}x_{51} + x_{9}x_{56} + x_{9}x_{57} + x_{9}x_{59} + x_{9}x_{60} + x_{9}x_{61} + x_{9}x_{63} + x_{10}x_{14} + x_{10}x_{16} + x_{10}x_{18} + x_{10}x_{19} + x_{10}x_{21} + x_{10}x_{22} + x_{10}x_{27} + x_{10}x_{29} + x_{10}x_{31} + x_{10}x_{33} + x_{10}x_{44} + x_{10}x_{45} + x_{10}x_{50} + x_{10}x_{52} + x_{10}x_{53} + x_{10}x_{54} + x_{10}x_{55} + x_{10}x_{57} + x_{10}x_{58} + x_{10}x_{59} + x_{10}x_{60} + x_{10}x_{61} + x_{10}x_{62} + x_{10}x_{63} + x_{10}x_{64} + x_{11}x_{12} + x_{11}x_{13} + x_{11}x_{14} + x_{11}x_{18} + x_{11}x_{20} + x_{11}x_{24} + x_{11}x_{25} + x_{11}x_{26} + x_{11}x_{27} + x_{11}x_{29} + x_{11}x_{31} + x_{11}x_{34} + x_{11}x_{35} + x_{11}x_{37} + x_{11}x_{39} + x_{11}x_{40} + x_{11}x_{41} + x_{11}x_{42} + x_{11}x_{45} + x_{11}x_{48} + x_{11}x_{49} + x_{11}x_{50} + x_{11}x_{51} + x_{11}x_{52} + x_{11}x_{53} + x_{11}x_{54} + x_{11}x_{56} + x_{11}x_{57} + x_{11}x_{59} + x_{11}x_{61} + x_{11}x_{62} + x_{11}x_{63} + x_{12}x_{13} + x_{12}x_{15} + x_{12}x_{16} + x_{12}x_{18} + x_{12}x_{21} + x_{12}x_{23} + x_{12}x_{24} + x_{12}x_{26} + x_{12}x_{28} + x_{12}x_{30} + x_{12}x_{34} + x_{12}x_{36} + x_{12}x_{38} + x_{12}x_{40} + x_{12}x_{41} + x_{12}x_{42} + x_{12}x_{44} + x_{12}x_{45} + x_{12}x_{46} + x_{12}x_{49} + x_{12}x_{51} + x_{12}x_{53} + x_{12}x_{54} + x_{12}x_{57} + x_{12}x_{59} + x_{12}x_{62} + x_{12}x_{63} + x_{12}x_{64} + x_{13}x_{20} + x_{13}x_{21} + x_{13}x_{22} + x_{13}x_{23} + x_{13}x_{24} + x_{13}x_{25} + x_{13}x_{28} + x_{13}x_{30} + x_{13}x_{33} + x_{13}x_{39} + x_{13}x_{41} + x_{13}x_{43} + x_{13}x_{44} + x_{13}x_{49} + x_{13}x_{50} + x_{13}x_{52} + x_{13}x_{53} + x_{13}x_{54} + x_{13}x_{55} + x_{13}x_{56} + x_{13}x_{57} + x_{13}x_{61} + x_{14}x_{17} + x_{14}x_{18} + x_{14}x_{19} + x_{14}x_{20} + x_{14}x_{21} + x_{14}x_{24} + x_{14}x_{25} + x_{14}x_{26} + x_{14}x_{27} + x_{14}x_{28} + x_{14}x_{30} + x_{14}x_{36} + x_{14}x_{37} + x_{14}x_{39} + x_{14}x_{40} + x_{14}x_{41} + x_{14}x_{42} + x_{14}x_{43} + x_{14}x_{45} + x_{14}x_{47} + x_{14}x_{51} + x_{14}x_{52} + x_{14}x_{53} + x_{14}x_{54} + x_{14}x_{55} + x_{14}x_{56} + x_{14}x_{60} + x_{14}x_{64} + x_{15}x_{16} + x_{15}x_{17} + x_{15}x_{18} + x_{15}x_{23} + x_{15}x_{25} + x_{15}x_{26} + x_{15}x_{27} + x_{15}x_{29} + x_{15}x_{31} + x_{15}x_{35} + x_{15}x_{36} + x_{15}x_{38} + x_{15}x_{42} + x_{15}x_{45} + x_{15}x_{47} + x_{15}x_{51} + x_{15}x_{52} + x_{15}x_{53} + x_{15}x_{55} + x_{15}x_{56} + x_{15}x_{59} + x_{16}x_{17} + x_{16}x_{18} + x_{16}x_{19} + x_{16}x_{21} + x_{16}x_{22} + x_{16}x_{23} + x_{16}x_{30} + x_{16}x_{31} + x_{16}x_{33} + x_{16}x_{35} + x_{16}x_{38} + x_{16}x_{39} + x_{16}x_{40} + x_{16}x_{41} + x_{16}x_{42} + x_{16}x_{44} + x_{16}x_{46} + x_{16}x_{47} + x_{16}x_{48} + x_{16}x_{49} + x_{16}x_{53} + x_{16}x_{57} + x_{16}x_{59} + x_{16}x_{61} + x_{16}x_{62} + x_{17}x_{19} + x_{17}x_{20} + x_{17}x_{21} + x_{17}x_{23} + x_{17}x_{24} + x_{17}x_{25} + x_{17}x_{29} + x_{17}x_{30} + x_{17}x_{34} + x_{17}x_{35} + x_{17}x_{37} + x_{17}x_{40} + x_{17}x_{42} + x_{17}x_{43} + x_{17}x_{45} + x_{17}x_{46} + x_{17}x_{47} + x_{17}x_{48} + x_{17}x_{50} + x_{17}x_{53} + x_{17}x_{54} + x_{17}x_{55} + x_{17}x_{56} + x_{17}x_{57} + x_{17}x_{59} + x_{17}x_{60} + x_{17}x_{61} + x_{17}x_{62} + x_{18}x_{20} + x_{18}x_{21} + x_{18}x_{22} + x_{18}x_{24} + x_{18}x_{25} + x_{18}x_{26} + x_{18}x_{28} + x_{18}x_{29} + x_{18}x_{33} + x_{18}x_{34} + x_{18}x_{35} + x_{18}x_{38} + x_{18}x_{42} + x_{18}x_{44} + x_{18}x_{45} + x_{18}x_{46} + x_{18}x_{47} + x_{18}x_{51} + x_{18}x_{54} + x_{18}x_{56} + x_{18}x_{57} + x_{18}x_{58} + x_{18}x_{60} + x_{18}x_{61} + x_{18}x_{62} + x_{18}x_{64} + x_{19}x_{20} + x_{19}x_{21} + x_{19}x_{23} + x_{19}x_{27} + x_{19}x_{29} + x_{19}x_{33} + x_{19}x_{38} + x_{19}x_{40} + x_{19}x_{41} + x_{19}x_{43} + x_{19}x_{44} + x_{19}x_{45} + x_{19}x_{46} + x_{19}x_{47} + x_{19}x_{48} + x_{19}x_{51} + x_{19}x_{52} + x_{19}x_{55} + x_{19}x_{56} + x_{19}x_{59} + x_{19}x_{61} + x_{19}x_{62} + x_{19}x_{63} + x_{20}x_{24} + x_{20}x_{25} + x_{20}x_{28} + x_{20}x_{29} + x_{20}x_{34} + x_{20}x_{36} + x_{20}x_{38} + x_{20}x_{39} + x_{20}x_{43} + x_{20}x_{44} + x_{20}x_{45} + x_{20}x_{46} + x_{20}x_{51} + x_{20}x_{53} + x_{20}x_{54} + x_{20}x_{55} + x_{20}x_{58} + x_{20}x_{62} + x_{21}x_{22} + x_{21}x_{24} + x_{21}x_{25} + x_{21}x_{26} + x_{21}x_{27} + x_{21}x_{28} + x_{21}x_{29} + x_{21}x_{30} + x_{21}x_{31} + x_{21}x_{34} + x_{21}x_{37} + x_{21}x_{38} + x_{21}x_{39} + x_{21}x_{40} + x_{21}x_{41} + x_{21}x_{42} + x_{21}x_{43} + x_{21}x_{47} + x_{21}x_{52} + x_{21}x_{53} + x_{21}x_{54} + x_{21}x_{59} + x_{21}x_{62} + x_{21}x_{63} + x_{21}x_{64} + x_{22}x_{24} + x_{22}x_{25} + x_{22}x_{27} + x_{22}x_{29} + x_{22}x_{32} + x_{22}x_{33} + x_{22}x_{34} + x_{22}x_{36} + x_{22}x_{39} + x_{22}x_{42} + x_{22}x_{43} + x_{22}x_{44} + x_{22}x_{46} + x_{22}x_{47} + x_{22}x_{48} + x_{22}x_{49} + x_{22}x_{51} + x_{22}x_{53} + x_{22}x_{54} + x_{22}x_{55} + x_{22}x_{58} + x_{22}x_{59} + x_{22}x_{62} + x_{23}x_{25} + x_{23}x_{26} + x_{23}x_{27} + x_{23}x_{30} + x_{23}x_{31} + x_{23}x_{32} + x_{23}x_{36} + x_{23}x_{37} + x_{23}x_{40} + x_{23}x_{41} + x_{23}x_{42} + x_{23}x_{43} + x_{23}x_{46} + x_{23}x_{48} + x_{23}x_{50} + x_{23}x_{51} + x_{23}x_{54} + x_{23}x_{58} + x_{23}x_{62} + x_{23}x_{63} + x_{23}x_{64} + x_{24}x_{25} + x_{24}x_{27} + x_{24}x_{28} + x_{24}x_{29} + x_{24}x_{31} + x_{24}x_{35} + x_{24}x_{36} + x_{24}x_{39} + x_{24}x_{42} + x_{24}x_{43} + x_{24}x_{47} + x_{24}x_{48} + x_{24}x_{50} + x_{24}x_{52} + x_{24}x_{53} + x_{24}x_{54} + x_{24}x_{56} + x_{24}x_{58} + x_{24}x_{59} + x_{24}x_{60} + x_{24}x_{61} + x_{25}x_{28} + x_{25}x_{29} + x_{25}x_{30} + x_{25}x_{32} + x_{25}x_{33} + x_{25}x_{39} + x_{25}x_{40} + x_{25}x_{42} + x_{25}x_{43} + x_{25}x_{45} + x_{25}x_{47} + x_{25}x_{48} + x_{25}x_{52} + x_{25}x_{54} + x_{25}x_{56} + x_{25}x_{57} + x_{25}x_{59} + x_{25}x_{60} + x_{25}x_{61} + x_{25}x_{62} + x_{25}x_{63} + x_{26}x_{28} + x_{26}x_{30} + x_{26}x_{31} + x_{26}x_{33} + x_{26}x_{35} + x_{26}x_{36} + x_{26}x_{38} + x_{26}x_{40} + x_{26}x_{44} + x_{26}x_{45} + x_{26}x_{48} + x_{26}x_{50} + x_{26}x_{51} + x_{26}x_{54} + x_{26}x_{55} + x_{26}x_{59} + x_{26}x_{60} + x_{26}x_{62} + x_{26}x_{63} + x_{27}x_{29} + x_{27}x_{30} + x_{27}x_{31} + x_{27}x_{33} + x_{27}x_{35} + x_{27}x_{39} + x_{27}x_{40} + x_{27}x_{42} + x_{27}x_{43} + x_{27}x_{45} + x_{27}x_{46} + x_{27}x_{47} + x_{27}x_{48} + x_{27}x_{49} + x_{27}x_{50} + x_{27}x_{53} + x_{27}x_{55} + x_{27}x_{58} + x_{27}x_{60} + x_{27}x_{61} + x_{27}x_{62} + x_{27}x_{63} + x_{27}x_{64} + x_{28}x_{29} + x_{28}x_{31} + x_{28}x_{35} + x_{28}x_{36} + x_{28}x_{37} + x_{28}x_{39} + x_{28}x_{40} + x_{28}x_{41} + x_{28}x_{43} + x_{28}x_{45} + x_{28}x_{46} + x_{28}x_{50} + x_{28}x_{52} + x_{28}x_{53} + x_{28}x_{55} + x_{28}x_{57} + x_{28}x_{58} + x_{28}x_{59} + x_{28}x_{60} + x_{28}x_{62} + x_{28}x_{63} + x_{28}x_{64} + x_{29}x_{30} + x_{29}x_{32} + x_{29}x_{34} + x_{29}x_{35} + x_{29}x_{36} + x_{29}x_{40} + x_{29}x_{42} + x_{29}x_{43} + x_{29}x_{45} + x_{29}x_{48} + x_{29}x_{49} + x_{29}x_{52} + x_{29}x_{53} + x_{29}x_{56} + x_{29}x_{60} + x_{29}x_{61} + x_{29}x_{62} + x_{29}x_{63} + x_{30}x_{31} + x_{30}x_{35} + x_{30}x_{36} + x_{30}x_{38} + x_{30}x_{39} + x_{30}x_{40} + x_{30}x_{43} + x_{30}x_{44} + x_{30}x_{45} + x_{30}x_{47} + x_{30}x_{48} + x_{30}x_{52} + x_{30}x_{53} + x_{30}x_{54} + x_{30}x_{56} + x_{30}x_{58} + x_{30}x_{61} + x_{30}x_{62} + x_{30}x_{63} + x_{30}x_{64} + x_{31}x_{33} + x_{31}x_{34} + x_{31}x_{36} + x_{31}x_{38} + x_{31}x_{40} + x_{31}x_{41} + x_{31}x_{44} + x_{31}x_{45} + x_{31}x_{49} + x_{31}x_{50} + x_{31}x_{52} + x_{31}x_{53} + x_{31}x_{55} + x_{31}x_{56} + x_{31}x_{58} + x_{31}x_{62} + x_{31}x_{64} + x_{32}x_{33} + x_{32}x_{34} + x_{32}x_{35} + x_{32}x_{37} + x_{32}x_{39} + x_{32}x_{42} + x_{32}x_{44} + x_{32}x_{46} + x_{32}x_{50} + x_{32}x_{51} + x_{32}x_{52} + x_{32}x_{55} + x_{32}x_{60} + x_{32}x_{61} + x_{32}x_{62} + x_{32}x_{64} + x_{33}x_{36} + x_{33}x_{39} + x_{33}x_{42} + x_{33}x_{44} + x_{33}x_{49} + x_{33}x_{52} + x_{33}x_{53} + x_{33}x_{55} + x_{33}x_{58} + x_{33}x_{59} + x_{33}x_{61} + x_{33}x_{63} + x_{33}x_{64} + x_{34}x_{36} + x_{34}x_{39} + x_{34}x_{40} + x_{34}x_{44} + x_{34}x_{45} + x_{34}x_{48} + x_{34}x_{49} + x_{34}x_{50} + x_{34}x_{51} + x_{34}x_{52} + x_{34}x_{55} + x_{34}x_{57} + x_{34}x_{59} + x_{34}x_{61} + x_{34}x_{64} + x_{35}x_{39} + x_{35}x_{41} + x_{35}x_{42} + x_{35}x_{43} + x_{35}x_{45} + x_{35}x_{47} + x_{35}x_{48} + x_{35}x_{49} + x_{35}x_{50} + x_{35}x_{52} + x_{35}x_{55} + x_{35}x_{56} + x_{35}x_{57} + x_{35}x_{58} + x_{35}x_{63} + x_{36}x_{38} + x_{36}x_{40} + x_{36}x_{46} + x_{36}x_{49} + x_{36}x_{50} + x_{36}x_{53} + x_{36}x_{54} + x_{36}x_{56} + x_{36}x_{58} + x_{36}x_{61} + x_{36}x_{62} + x_{36}x_{63} + x_{36}x_{64} + x_{37}x_{38} + x_{37}x_{40} + x_{37}x_{43} + x_{37}x_{45} + x_{37}x_{47} + x_{37}x_{51} + x_{37}x_{52} + x_{37}x_{55} + x_{37}x_{57} + x_{37}x_{58} + x_{37}x_{59} + x_{37}x_{60} + x_{37}x_{64} + x_{38}x_{42} + x_{38}x_{43} + x_{38}x_{48} + x_{38}x_{49} + x_{38}x_{51} + x_{38}x_{52} + x_{38}x_{53} + x_{38}x_{55} + x_{38}x_{59} + x_{38}x_{60} + x_{38}x_{61} + x_{38}x_{62} + x_{38}x_{63} + x_{39}x_{42} + x_{39}x_{45} + x_{39}x_{46} + x_{39}x_{50} + x_{39}x_{52} + x_{39}x_{54} + x_{39}x_{56} + x_{39}x_{57} + x_{39}x_{60} + x_{39}x_{63} + x_{40}x_{43} + x_{40}x_{45} + x_{40}x_{46} + x_{40}x_{47} + x_{40}x_{49} + x_{40}x_{50} + x_{40}x_{51} + x_{40}x_{54} + x_{40}x_{55} + x_{40}x_{57} + x_{40}x_{59} + x_{40}x_{60} + x_{40}x_{61} + x_{41}x_{43} + x_{41}x_{44} + x_{41}x_{45} + x_{41}x_{46} + x_{41}x_{48} + x_{41}x_{51} + x_{41}x_{52} + x_{41}x_{56} + x_{41}x_{58} + x_{41}x_{59} + x_{41}x_{61} + x_{42}x_{45} + x_{42}x_{46} + x_{42}x_{49} + x_{42}x_{50} + x_{42}x_{51} + x_{42}x_{54} + x_{42}x_{55} + x_{42}x_{57} + x_{42}x_{58} + x_{43}x_{44} + x_{43}x_{49} + x_{43}x_{51} + x_{43}x_{52} + x_{43}x_{53} + x_{43}x_{56} + x_{43}x_{58} + x_{43}x_{59} + x_{43}x_{64} + x_{44}x_{45} + x_{44}x_{47} + x_{44}x_{48} + x_{44}x_{53} + x_{44}x_{55} + x_{44}x_{57} + x_{44}x_{60} + x_{44}x_{62} + x_{45}x_{46} + x_{45}x_{47} + x_{45}x_{49} + x_{45}x_{51} + x_{45}x_{52} + x_{45}x_{59} + x_{45}x_{60} + x_{45}x_{61} + x_{45}x_{63} + x_{46}x_{47} + x_{46}x_{50} + x_{46}x_{52} + x_{46}x_{56} + x_{46}x_{59} + x_{46}x_{60} + x_{46}x_{63} + x_{46}x_{64} + x_{47}x_{50} + x_{47}x_{53} + x_{47}x_{58} + x_{47}x_{63} + x_{48}x_{51} + x_{48}x_{52} + x_{48}x_{53} + x_{48}x_{55} + x_{48}x_{57} + x_{48}x_{59} + x_{48}x_{61} + x_{49}x_{51} + x_{49}x_{54} + x_{49}x_{57} + x_{49}x_{59} + x_{49}x_{60} + x_{49}x_{61} + x_{49}x_{62} + x_{50}x_{51} + x_{50}x_{52} + x_{50}x_{53} + x_{50}x_{54} + x_{50}x_{57} + x_{50}x_{58} + x_{50}x_{60} + x_{50}x_{61} + x_{50}x_{63} + x_{51}x_{52} + x_{51}x_{57} + x_{51}x_{58} + x_{51}x_{59} + x_{51}x_{61} + x_{51}x_{63} + x_{51}x_{64} + x_{52}x_{54} + x_{52}x_{58} + x_{52}x_{61} + x_{52}x_{62} + x_{52}x_{63} + x_{53}x_{55} + x_{53}x_{56} + x_{53}x_{57} + x_{53}x_{58} + x_{53}x_{61} + x_{53}x_{63} + x_{53}x_{64} + x_{54}x_{56} + x_{54}x_{57} + x_{54}x_{58} + x_{54}x_{59} + x_{54}x_{60} + x_{54}x_{61} + x_{54}x_{62} + x_{54}x_{64} + x_{55}x_{56} + x_{55}x_{57} + x_{55}x_{58} + x_{55}x_{59} + x_{55}x_{60} + x_{55}x_{63} + x_{55}x_{64} + x_{56}x_{57} + x_{56}x_{61} + x_{56}x_{63} + x_{57}x_{58} + x_{57}x_{60} + x_{57}x_{62} + x_{58}x_{62} + x_{58}x_{63} + x_{58}x_{64} + x_{59}x_{60} + x_{59}x_{61} + x_{59}x_{64} + x_{60}x_{61} + x_{60}x_{63} + x_{60}x_{64} + x_{61}x_{64} + x_{62}x_{63} + x_{63}x_{64} + x_{1} + x_{4} + x_{7} + x_{8} + x_{9} + x_{11} + x_{12} + x_{14} + x_{15} + x_{16} + x_{18} + x_{19} + x_{22} + x_{24} + x_{25} + x_{26} + x_{30} + x_{35} + x_{36} + x_{38} + x_{40} + x_{41} + x_{48} + x_{49} + x_{50} + x_{51} + x_{54} + x_{55} + x_{56} + x_{57} + x_{59} + x_{61} + x_{62} + 1$

$y_{9} = x_{1}x_{2} + x_{1}x_{3} + x_{1}x_{5} + x_{1}x_{6} + x_{1}x_{9} + x_{1}x_{10} + x_{1}x_{11} + x_{1}x_{12} + x_{1}x_{14} + x_{1}x_{19} + x_{1}x_{22} + x_{1}x_{24} + x_{1}x_{28} + x_{1}x_{29} + x_{1}x_{35} + x_{1}x_{37} + x_{1}x_{38} + x_{1}x_{39} + x_{1}x_{44} + x_{1}x_{46} + x_{1}x_{48} + x_{1}x_{52} + x_{1}x_{53} + x_{1}x_{55} + x_{1}x_{57} + x_{1}x_{59} + x_{1}x_{60} + x_{1}x_{61} + x_{1}x_{63} + x_{2}x_{3} + x_{2}x_{6} + x_{2}x_{8} + x_{2}x_{12} + x_{2}x_{16} + x_{2}x_{17} + x_{2}x_{23} + x_{2}x_{24} + x_{2}x_{25} + x_{2}x_{34} + x_{2}x_{37} + x_{2}x_{38} + x_{2}x_{39} + x_{2}x_{42} + x_{2}x_{44} + x_{2}x_{46} + x_{2}x_{50} + x_{2}x_{51} + x_{2}x_{52} + x_{2}x_{54} + x_{2}x_{56} + x_{2}x_{57} + x_{2}x_{59} + x_{2}x_{62} + x_{2}x_{63} + x_{2}x_{64} + x_{3}x_{5} + x_{3}x_{6} + x_{3}x_{8} + x_{3}x_{10} + x_{3}x_{12} + x_{3}x_{15} + x_{3}x_{20} + x_{3}x_{21} + x_{3}x_{22} + x_{3}x_{24} + x_{3}x_{26} + x_{3}x_{30} + x_{3}x_{32} + x_{3}x_{33} + x_{3}x_{37} + x_{3}x_{38} + x_{3}x_{39} + x_{3}x_{42} + x_{3}x_{43} + x_{3}x_{44} + x_{3}x_{48} + x_{3}x_{50} + x_{3}x_{51} + x_{3}x_{52} + x_{3}x_{61} + x_{3}x_{63} + x_{4}x_{5} + x_{4}x_{6} + x_{4}x_{7} + x_{4}x_{10} + x_{4}x_{12} + x_{4}x_{15} + x_{4}x_{16} + x_{4}x_{19} + x_{4}x_{20} + x_{4}x_{22} + x_{4}x_{23} + x_{4}x_{25} + x_{4}x_{26} + x_{4}x_{27} + x_{4}x_{29} + x_{4}x_{30} + x_{4}x_{31} + x_{4}x_{32} + x_{4}x_{34} + x_{4}x_{37} + x_{4}x_{38} + x_{4}x_{41} + x_{4}x_{47} + x_{4}x_{50} + x_{4}x_{51} + x_{4}x_{54} + x_{4}x_{55} + x_{4}x_{57} + x_{4}x_{58} + x_{4}x_{59} + x_{4}x_{60} + x_{4}x_{64} + x_{5}x_{6} + x_{5}x_{9} + x_{5}x_{11} + x_{5}x_{13} + x_{5}x_{14} + x_{5}x_{18} + x_{5}x_{19} + x_{5}x_{20} + x_{5}x_{22} + x_{5}x_{26} + x_{5}x_{27} + x_{5}x_{31} + x_{5}x_{35} + x_{5}x_{36} + x_{5}x_{40} + x_{5}x_{41} + x_{5}x_{44} + x_{5}x_{45} + x_{5}x_{46} + x_{5}x_{48} + x_{5}x_{49} + x_{5}x_{50} + x_{5}x_{52} + x_{5}x_{56} + x_{5}x_{57} + x_{5}x_{59} + x_{5}x_{60} + x_{5}x_{61} + x_{5}x_{63} + x_{5}x_{64} + x_{6}x_{7} + x_{6}x_{9} + x_{6}x_{10} + x_{6}x_{14} + x_{6}x_{17} + x_{6}x_{19} + x_{6}x_{22} + x_{6}x_{24} + x_{6}x_{25} + x_{6}x_{37} + x_{6}x_{39} + x_{6}x_{42} + x_{6}x_{44} + x_{6}x_{45} + x_{6}x_{46} + x_{6}x_{47} + x_{6}x_{48} + x_{6}x_{51} + x_{6}x_{52} + x_{6}x_{54} + x_{6}x_{57} + x_{6}x_{58} + x_{6}x_{59} + x_{6}x_{60} + x_{6}x_{62} + x_{6}x_{63} + x_{6}x_{64} + x_{7}x_{8} + x_{7}x_{9} + x_{7}x_{10} + x_{7}x_{11} + x_{7}x_{12} + x_{7}x_{16} + x_{7}x_{17} + x_{7}x_{19} + x_{7}x_{20} + x_{7}x_{22} + x_{7}x_{23} + x_{7}x_{24} + x_{7}x_{26} + x_{7}x_{27} + x_{7}x_{29} + x_{7}x_{30} + x_{7}x_{31} + x_{7}x_{32} + x_{7}x_{36} + x_{7}x_{37} + x_{7}x_{38} + x_{7}x_{39} + x_{7}x_{41} + x_{7}x_{42} + x_{7}x_{43} + x_{7}x_{44} + x_{7}x_{48} + x_{7}x_{49} + x_{7}x_{53} + x_{7}x_{54} + x_{7}x_{55} + x_{7}x_{57} + x_{7}x_{58} + x_{7}x_{59} + x_{7}x_{63} + x_{7}x_{64} + x_{8}x_{12} + x_{8}x_{13} + x_{8}x_{14} + x_{8}x_{15} + x_{8}x_{16} + x_{8}x_{19} + x_{8}x_{22} + x_{8}x_{25} + x_{8}x_{26} + x_{8}x_{27} + x_{8}x_{29} + x_{8}x_{30} + x_{8}x_{31} + x_{8}x_{33} + x_{8}x_{34} + x_{8}x_{35} + x_{8}x_{37} + x_{8}x_{38} + x_{8}x_{39} + x_{8}x_{40} + x_{8}x_{42} + x_{8}x_{44} + x_{8}x_{47} + x_{8}x_{48} + x_{8}x_{50} + x_{8}x_{51} + x_{8}x_{54} + x_{8}x_{55} + x_{8}x_{57} + x_{8}x_{59} + x_{8}x_{61} + x_{8}x_{63} + x_{9}x_{11} + x_{9}x_{15} + x_{9}x_{16} + x_{9}x_{17} + x_{9}x_{19} + x_{9}x_{20} + x_{9}x_{24} + x_{9}x_{25} + x_{9}x_{26} + x_{9}x_{28} + x_{9}x_{29} + x_{9}x_{30} + x_{9}x_{31} + x_{9}x_{33} + x_{9}x_{34} + x_{9}x_{35} + x_{9}x_{37} + x_{9}x_{39} + x_{9}x_{40} + x_{9}x_{41} + x_{9}x_{49} + x_{9}x_{55} + x_{9}x_{58} + x_{9}x_{60} + x_{9}x_{61} + x_{10}x_{11} + x_{10}x_{13} + x_{10}x_{18} + x_{10}x_{19} + x_{10}x_{21} + x_{10}x_{23} + x_{10}x_{27} + x_{10}x_{32} + x_{10}x_{35} + x_{10}x_{36} + x_{10}x_{37} + x_{10}x_{38} + x_{10}x_{39} + x_{10}x_{40} + x_{10}x_{41} + x_{10}x_{42} + x_{10}x_{43} + x_{10}x_{46} + x_{10}x_{49} + x_{10}x_{50} + x_{10}x_{51} + x_{10}x_{55} + x_{10}x_{56} + x_{10}x_{57} + x_{10}x_{58} + x_{10}x_{59} + x_{10}x_{61} + x_{10}x_{62} + x_{10}x_{63} + x_{10}x_{64} + x_{11}x_{14} + x_{11}x_{15} + x_{11}x_{19} + x_{11}x_{24} + x_{11}x_{27} + x_{11}x_{28} + x_{11}x_{29} + x_{11}x_{31} + x_{11}x_{36} + x_{11}x_{37} + x_{11}x_{38} + x_{11}x_{42} + x_{11}x_{44} + x_{11}x_{45} + x_{11}x_{51} + x_{11}x_{54} + x_{11}x_{56} + x_{11}x_{57} + x_{11}x_{59} + x_{11}x_{60} + x_{11}x_{63} + x_{12}x_{13} + x_{12}x_{15} + x_{12}x_{17} + x_{12}x_{22} + x_{12}x_{23} + x_{12}x_{24} + x_{12}x_{26} + x_{12}x_{27} + x_{12}x_{28} + x_{12}x_{29} + x_{12}x_{30} + x_{12}x_{31} + x_{12}x_{32} + x_{12}x_{34} + x_{12}x_{36} + x_{12}x_{39} + x_{12}x_{40} + x_{12}x_{43} + x_{12}x_{44} + x_{12}x_{45} + x_{12}x_{47} + x_{12}x_{49} + x_{12}x_{52} + x_{12}x_{53} + x_{12}x_{54} + x_{12}x_{55} + x_{12}x_{57} + x_{12}x_{59} + x_{12}x_{64} + x_{13}x_{15} + x_{13}x_{16} + x_{13}x_{17} + x_{13}x_{22} + x_{13}x_{26} + x_{13}x_{29} + x_{13}x_{30} + x_{13}x_{31} + x_{13}x_{35} + x_{13}x_{38} + x_{13}x_{39} + x_{13}x_{40} + x_{13}x_{43} + x_{13}x_{44} + x_{13}x_{46} + x_{13}x_{48} + x_{13}x_{49} + x_{13}x_{50} + x_{13}x_{52} + x_{13}x_{54} + x_{13}x_{55} + x_{13}x_{59} + x_{13}x_{63} + x_{13}x_{64} + x_{14}x_{15} + x_{14}x_{16} + x_{14}x_{18} + x_{14}x_{22} + x_{14}x_{23} + x_{14}x_{26} + x_{14}x_{27} + x_{14}x_{29} + x_{14}x_{32} + x_{14}x_{34} + x_{14}x_{38} + x_{14}x_{39} + x_{14}x_{41} + x_{14}x_{43} + x_{14}x_{44} + x_{14}x_{45} + x_{14}x_{46} + x_{14}x_{50} + x_{14}x_{51} + x_{14}x_{52} + x_{14}x_{54} + x_{14}x_{57} + x_{14}x_{58} + x_{14}x_{61} + x_{14}x_{64} + x_{15}x_{17} + x_{15}x_{18} + x_{15}x_{19} + x_{15}x_{20} + x_{15}x_{23} + x_{15}x_{24} + x_{15}x_{28} + x_{15}x_{30} + x_{15}x_{33} + x_{15}x_{34} + x_{15}x_{36} + x_{15}x_{38} + x_{15}x_{41} + x_{15}x_{43} + x_{15}x_{45} + x_{15}x_{48} + x_{15}x_{51} + x_{15}x_{53} + x_{15}x_{55} + x_{15}x_{56} + x_{15}x_{58} + x_{15}x_{59} + x_{15}x_{60} + x_{15}x_{61} + x_{15}x_{62} + x_{15}x_{63} + x_{15}x_{64} + x_{16}x_{19} + x_{16}x_{26} + x_{16}x_{30} + x_{16}x_{31} + x_{16}x_{33} + x_{16}x_{37} + x_{16}x_{39} + x_{16}x_{40} + x_{16}x_{42} + x_{16}x_{43} + x_{16}x_{46} + x_{16}x_{47} + x_{16}x_{48} + x_{16}x_{50} + x_{16}x_{51} + x_{16}x_{53} + x_{16}x_{55} + x_{16}x_{56} + x_{16}x_{58} + x_{16}x_{63} + x_{16}x_{64} + x_{17}x_{18} + x_{17}x_{20} + x_{17}x_{21} + x_{17}x_{23} + x_{17}x_{25} + x_{17}x_{26} + x_{17}x_{28} + x_{17}x_{32} + x_{17}x_{33} + x_{17}x_{35} + x_{17}x_{37} + x_{17}x_{38} + x_{17}x_{39} + x_{17}x_{40} + x_{17}x_{41} + x_{17}x_{42} + x_{17}x_{44} + x_{17}x_{45} + x_{17}x_{46} + x_{17}x_{49} + x_{17}x_{51} + x_{17}x_{52} + x_{17}x_{53} + x_{17}x_{54} + x_{17}x_{56} + x_{17}x_{58} + x_{17}x_{60} + x_{17}x_{61} + x_{17}x_{63} + x_{17}x_{64} + x_{18}x_{19} + x_{18}x_{20} + x_{18}x_{24} + x_{18}x_{25} + x_{18}x_{26} + x_{18}x_{28} + x_{18}x_{30} + x_{18}x_{34} + x_{18}x_{35} + x_{18}x_{38} + x_{18}x_{39} + x_{18}x_{40} + x_{18}x_{45} + x_{18}x_{46} + x_{18}x_{48} + x_{18}x_{49} + x_{18}x_{51} + x_{18}x_{52} + x_{18}x_{53} + x_{18}x_{55} + x_{18}x_{59} + x_{18}x_{60} + x_{19}x_{24} + x_{19}x_{25} + x_{19}x_{28} + x_{19}x_{29} + x_{19}x_{30} + x_{19}x_{32} + x_{19}x_{33} + x_{19}x_{40} + x_{19}x_{41} + x_{19}x_{42} + x_{19}x_{44} + x_{19}x_{47} + x_{19}x_{48} + x_{19}x_{52} + x_{19}x_{53} + x_{19}x_{54} + x_{19}x_{55} + x_{19}x_{57} + x_{19}x_{59} + x_{19}x_{60} + x_{19}x_{62} + x_{19}x_{63} + x_{19}x_{64} + x_{20}x_{22} + x_{20}x_{23} + x_{20}x_{25} + x_{20}x_{26} + x_{20}x_{28} + x_{20}x_{29} + x_{20}x_{33} + x_{20}x_{34} + x_{20}x_{35} + x_{20}x_{36} + x_{20}x_{37} + x_{20}x_{41} + x_{20}x_{42} + x_{20}x_{45} + x_{20}x_{46} + x_{20}x_{51} + x_{20}x_{55} + x_{20}x_{57} + x_{20}x_{59} + x_{20}x_{61} + x_{20}x_{63} + x_{21}x_{25} + x_{21}x_{27} + x_{21}x_{30} + x_{21}x_{31} + x_{21}x_{32} + x_{21}x_{33} + x_{21}x_{34} + x_{21}x_{36} + x_{21}x_{40} + x_{21}x_{44} + x_{21}x_{47} + x_{21}x_{49} + x_{21}x_{51} + x_{21}x_{52} + x_{21}x_{53} + x_{21}x_{54} + x_{21}x_{55} + x_{21}x_{57} + x_{21}x_{58} + x_{21}x_{60} + x_{21}x_{61} + x_{21}x_{63} + x_{21}x_{64} + x_{22}x_{23} + x_{22}x_{24} + x_{22}x_{25} + x_{22}x_{26} + x_{22}x_{27} + x_{22}x_{28} + x_{22}x_{33} + x_{22}x_{35} + x_{22}x_{36} + x_{22}x_{37} + x_{22}x_{39} + x_{22}x_{40} + x_{22}x_{41} + x_{22}x_{42} + x_{22}x_{44} + x_{22}x_{45} + x_{22}x_{48} + x_{22}x_{50} + x_{22}x_{52} + x_{22}x_{53} + x_{22}x_{54} + x_{22}x_{55} + x_{22}x_{59} + x_{22}x_{60} + x_{22}x_{61} + x_{22}x_{62} + x_{23}x_{24} + x_{23}x_{25} + x_{23}x_{26} + x_{23}x_{27} + x_{23}x_{29} + x_{23}x_{30} + x_{23}x_{32} + x_{23}x_{33} + x_{23}x_{34} + x_{23}x_{38} + x_{23}x_{39} + x_{23}x_{41} + x_{23}x_{46} + x_{23}x_{48} + x_{23}x_{49} + x_{23}x_{50} + x_{23}x_{51} + x_{23}x_{52} + x_{23}x_{56} + x_{23}x_{57} + x_{23}x_{58} + x_{23}x_{59} + x_{24}x_{25} + x_{24}x_{26} + x_{24}x_{27} + x_{24}x_{28} + x_{24}x_{30} + x_{24}x_{31} + x_{24}x_{34} + x_{24}x_{35} + x_{24}x_{36} + x_{24}x_{37} + x_{24}x_{39} + x_{24}x_{41} + x_{24}x_{43} + x_{24}x_{44} + x_{24}x_{45} + x_{24}x_{48} + x_{24}x_{51} + x_{24}x_{56} + x_{24}x_{57} + x_{24}x_{59} + x_{24}x_{61} + x_{24}x_{62} + x_{25}x_{33} + x_{25}x_{36} + x_{25}x_{38} + x_{25}x_{41} + x_{25}x_{43} + x_{25}x_{45} + x_{25}x_{51} + x_{25}x_{52} + x_{25}x_{53} + x_{25}x_{61} + x_{25}x_{62} + x_{25}x_{63} + x_{26}x_{29} + x_{26}x_{30} + x_{26}x_{33} + x_{26}x_{35} + x_{26}x_{41} + x_{26}x_{43} + x_{26}x_{45} + x_{26}x_{47} + x_{26}x_{48} + x_{26}x_{49} + x_{26}x_{50} + x_{26}x_{51} + x_{26}x_{55} + x_{26}x_{56} + x_{26}x_{58} + x_{26}x_{59} + x_{26}x_{64} + x_{27}x_{29} + x_{27}x_{31} + x_{27}x_{33} + x_{27}x_{36} + x_{27}x_{40} + x_{27}x_{41} + x_{27}x_{42} + x_{27}x_{43} + x_{27}x_{44} + x_{27}x_{45} + x_{27}x_{46} + x_{27}x_{47} + x_{27}x_{48} + x_{27}x_{51} + x_{27}x_{54} + x_{27}x_{55} + x_{27}x_{60} + x_{27}x_{61} + x_{27}x_{64} + x_{28}x_{30} + x_{28}x_{31} + x_{28}x_{34} + x_{28}x_{35} + x_{28}x_{38} + x_{28}x_{41} + x_{28}x_{45} + x_{28}x_{47} + x_{28}x_{49} + x_{28}x_{55} + x_{28}x_{56} + x_{28}x_{59} + x_{28}x_{60} + x_{28}x_{61} + x_{28}x_{63} + x_{28}x_{64} + x_{29}x_{30} + x_{29}x_{35} + x_{29}x_{36} + x_{29}x_{42} + x_{29}x_{43} + x_{29}x_{45} + x_{29}x_{49} + x_{29}x_{50} + x_{29}x_{51} + x_{29}x_{52} + x_{29}x_{57} + x_{29}x_{58} + x_{29}x_{59} + x_{29}x_{60} + x_{29}x_{62} + x_{30}x_{32} + x_{30}x_{35} + x_{30}x_{36} + x_{30}x_{37} + x_{30}x_{38} + x_{30}x_{40} + x_{30}x_{41} + x_{30}x_{45} + x_{30}x_{49} + x_{30}x_{51} + x_{30}x_{52} + x_{30}x_{55} + x_{30}x_{57} + x_{30}x_{61} + x_{30}x_{62} + x_{31}x_{33} + x_{31}x_{35} + x_{31}x_{37} + x_{31}x_{38} + x_{31}x_{39} + x_{31}x_{40} + x_{31}x_{43} + x_{31}x_{45} + x_{31}x_{47} + x_{31}x_{48} + x_{31}x_{49} + x_{31}x_{50} + x_{31}x_{53} + x_{31}x_{54} + x_{31}x_{56} + x_{31}x_{57} + x_{31}x_{58} + x_{31}x_{60} + x_{32}x_{34} + x_{32}x_{37} + x_{32}x_{38} + x_{32}x_{40} + x_{32}x_{42} + x_{32}x_{43} + x_{32}x_{47} + x_{32}x_{48} + x_{32}x_{51} + x_{32}x_{55} + x_{32}x_{56} + x_{32}x_{58} + x_{32}x_{62} + x_{33}x_{38} + x_{33}x_{40} + x_{33}x_{41} + x_{33}x_{44} + x_{33}x_{45} + x_{33}x_{46} + x_{33}x_{48} + x_{33}x_{55} + x_{33}x_{56} + x_{33}x_{57} + x_{33}x_{58} + x_{33}x_{59} + x_{33}x_{60} + x_{33}x_{62} + x_{33}x_{63} + x_{33}x_{64} + x_{34}x_{38} + x_{34}x_{44} + x_{34}x_{45} + x_{34}x_{46} + x_{34}x_{47} + x_{34}x_{48} + x_{34}x_{49} + x_{34}x_{50} + x_{34}x_{51} + x_{34}x_{52} + x_{34}x_{53} + x_{34}x_{58} + x_{34}x_{59} + x_{34}x_{63} + x_{35}x_{41} + x_{35}x_{44} + x_{35}x_{45} + x_{35}x_{46} + x_{35}x_{49} + x_{35}x_{50} + x_{35}x_{53} + x_{35}x_{54} + x_{35}x_{55} + x_{35}x_{57} + x_{35}x_{61} + x_{35}x_{63} + x_{36}x_{37} + x_{36}x_{39} + x_{36}x_{40} + x_{36}x_{42} + x_{36}x_{46} + x_{36}x_{49} + x_{36}x_{52} + x_{36}x_{54} + x_{36}x_{55} + x_{36}x_{57} + x_{36}x_{60} + x_{36}x_{62} + x_{36}x_{63} + x_{36}x_{64} + x_{37}x_{40} + x_{37}x_{41} + x_{37}x_{43} + x_{37}x_{47} + x_{37}x_{49} + x_{37}x_{50} + x_{37}x_{54} + x_{37}x_{55} + x_{37}x_{56} + x_{37}x_{58} + x_{37}x_{59} + x_{37}x_{60} + x_{37}x_{61} + x_{37}x_{64} + x_{38}x_{39} + x_{38}x_{41} + x_{38}x_{43} + x_{38}x_{45} + x_{38}x_{47} + x_{38}x_{48} + x_{38}x_{51} + x_{38}x_{52} + x_{38}x_{55} + x_{38}x_{57} + x_{38}x_{58} + x_{38}x_{60} + x_{38}x_{62} + x_{38}x_{64} + x_{39}x_{40} + x_{39}x_{48} + x_{39}x_{51} + x_{39}x_{52} + x_{39}x_{53} + x_{39}x_{54} + x_{39}x_{55} + x_{39}x_{56} + x_{39}x_{59} + x_{39}x_{60} + x_{39}x_{64} + x_{40}x_{41} + x_{40}x_{42} + x_{40}x_{43} + x_{40}x_{44} + x_{40}x_{45} + x_{40}x_{49} + x_{40}x_{53} + x_{40}x_{55} + x_{40}x_{59} + x_{40}x_{60} + x_{40}x_{61} + x_{40}x_{62} + x_{40}x_{63} + x_{40}x_{64} + x_{41}x_{46} + x_{41}x_{47} + x_{41}x_{48} + x_{41}x_{50} + x_{41}x_{51} + x_{41}x_{53} + x_{41}x_{54} + x_{41}x_{56} + x_{41}x_{57} + x_{41}x_{58} + x_{41}x_{60} + x_{41}x_{61} + x_{41}x_{62} + x_{41}x_{63} + x_{41}x_{64} + x_{42}x_{46} + x_{42}x_{51} + x_{42}x_{52} + x_{42}x_{54} + x_{42}x_{55} + x_{42}x_{57} + x_{42}x_{59} + x_{42}x_{60} + x_{42}x_{62} + x_{42}x_{63} + x_{42}x_{64} + x_{43}x_{44} + x_{43}x_{45} + x_{43}x_{46} + x_{43}x_{48} + x_{43}x_{49} + x_{43}x_{50} + x_{43}x_{51} + x_{43}x_{52} + x_{43}x_{53} + x_{43}x_{54} + x_{43}x_{56} + x_{43}x_{58} + x_{43}x_{59} + x_{43}x_{63} + x_{43}x_{64} + x_{44}x_{47} + x_{44}x_{48} + x_{44}x_{49} + x_{44}x_{54} + x_{44}x_{58} + x_{44}x_{59} + x_{44}x_{60} + x_{44}x_{62} + x_{44}x_{64} + x_{45}x_{46} + x_{45}x_{47} + x_{45}x_{48} + x_{45}x_{51} + x_{45}x_{55} + x_{45}x_{56} + x_{45}x_{57} + x_{45}x_{58} + x_{45}x_{60} + x_{45}x_{61} + x_{45}x_{64} + x_{46}x_{47} + x_{46}x_{48} + x_{46}x_{51} + x_{46}x_{53} + x_{46}x_{54} + x_{46}x_{57} + x_{46}x_{58} + x_{46}x_{63} + x_{47}x_{48} + x_{47}x_{50} + x_{47}x_{52} + x_{47}x_{54} + x_{47}x_{57} + x_{47}x_{59} + x_{47}x_{63} + x_{48}x_{52} + x_{48}x_{57} + x_{48}x_{58} + x_{48}x_{63} + x_{49}x_{50} + x_{49}x_{53} + x_{49}x_{55} + x_{49}x_{56} + x_{49}x_{57} + x_{49}x_{59} + x_{49}x_{60} + x_{49}x_{61} + x_{49}x_{62} + x_{49}x_{63} + x_{49}x_{64} + x_{50}x_{51} + x_{50}x_{55} + x_{50}x_{58} + x_{50}x_{59} + x_{50}x_{60} + x_{50}x_{61} + x_{50}x_{62} + x_{50}x_{63} + x_{50}x_{64} + x_{51}x_{54} + x_{51}x_{55} + x_{51}x_{58} + x_{51}x_{60} + x_{51}x_{61} + x_{51}x_{62} + x_{51}x_{63} + x_{51}x_{64} + x_{52}x_{54} + x_{52}x_{55} + x_{52}x_{61} + x_{52}x_{64} + x_{53}x_{54} + x_{53}x_{56} + x_{53}x_{59} + x_{53}x_{61} + x_{53}x_{62} + x_{53}x_{63} + x_{54}x_{56} + x_{54}x_{57} + x_{54}x_{61} + x_{54}x_{62} + x_{55}x_{57} + x_{55}x_{58} + x_{55}x_{62} + x_{55}x_{64} + x_{56}x_{57} + x_{56}x_{59} + x_{56}x_{61} + x_{56}x_{64} + x_{57}x_{58} + x_{57}x_{59} + x_{57}x_{60} + x_{57}x_{61} + x_{57}x_{62} + x_{57}x_{63} + x_{57}x_{64} + x_{58}x_{60} + x_{58}x_{61} + x_{58}x_{64} + x_{59}x_{60} + x_{59}x_{61} + x_{59}x_{62} + x_{59}x_{63} + x_{60}x_{61} + x_{60}x_{64} + x_{61}x_{63} + x_{61}x_{64} + x_{62}x_{63} + x_{63}x_{64} + x_{2} + x_{3} + x_{4} + x_{8} + x_{9} + x_{10} + x_{11} + x_{12} + x_{13} + x_{15} + x_{16} + x_{21} + x_{27} + x_{28} + x_{30} + x_{32} + x_{33} + x_{35} + x_{36} + x_{37} + x_{38} + x_{43} + x_{44} + x_{45} + x_{46} + x_{47} + x_{48} + x_{49} + x_{51} + x_{55} + x_{56} + x_{62} + x_{64} + 1$

$y_{10} = x_{1}x_{3} + x_{1}x_{6} + x_{1}x_{9} + x_{1}x_{13} + x_{1}x_{15} + x_{1}x_{17} + x_{1}x_{20} + x_{1}x_{21} + x_{1}x_{22} + x_{1}x_{23} + x_{1}x_{25} + x_{1}x_{26} + x_{1}x_{27} + x_{1}x_{29} + x_{1}x_{30} + x_{1}x_{31} + x_{1}x_{33} + x_{1}x_{38} + x_{1}x_{41} + x_{1}x_{46} + x_{1}x_{47} + x_{1}x_{51} + x_{1}x_{52} + x_{1}x_{53} + x_{1}x_{54} + x_{1}x_{56} + x_{1}x_{57} + x_{1}x_{62} + x_{1}x_{64} + x_{2}x_{4} + x_{2}x_{7} + x_{2}x_{9} + x_{2}x_{12} + x_{2}x_{16} + x_{2}x_{17} + x_{2}x_{20} + x_{2}x_{24} + x_{2}x_{29} + x_{2}x_{32} + x_{2}x_{36} + x_{2}x_{39} + x_{2}x_{40} + x_{2}x_{42} + x_{2}x_{44} + x_{2}x_{45} + x_{2}x_{46} + x_{2}x_{47} + x_{2}x_{48} + x_{2}x_{49} + x_{2}x_{51} + x_{2}x_{52} + x_{2}x_{53} + x_{2}x_{54} + x_{2}x_{59} + x_{2}x_{60} + x_{2}x_{64} + x_{3}x_{4} + x_{3}x_{5} + x_{3}x_{6} + x_{3}x_{7} + x_{3}x_{8} + x_{3}x_{10} + x_{3}x_{11} + x_{3}x_{14} + x_{3}x_{16} + x_{3}x_{19} + x_{3}x_{21} + x_{3}x_{23} + x_{3}x_{25} + x_{3}x_{27} + x_{3}x_{28} + x_{3}x_{29} + x_{3}x_{30} + x_{3}x_{33} + x_{3}x_{35} + x_{3}x_{36} + x_{3}x_{37} + x_{3}x_{39} + x_{3}x_{41} + x_{3}x_{42} + x_{3}x_{43} + x_{3}x_{45} + x_{3}x_{47} + x_{3}x_{48} + x_{3}x_{53} + x_{3}x_{57} + x_{3}x_{60} + x_{3}x_{62} + x_{3}x_{64} + x_{4}x_{6} + x_{4}x_{7} + x_{4}x_{8} + x_{4}x_{10} + x_{4}x_{11} + x_{4}x_{12} + x_{4}x_{16} + x_{4}x_{20} + x_{4}x_{21} + x_{4}x_{22} + x_{4}x_{25} + x_{4}x_{30} + x_{4}x_{33} + x_{4}x_{36} + x_{4}x_{38} + x_{4}x_{40} + x_{4}x_{42} + x_{4}x_{43} + x_{4}x_{45} + x_{4}x_{46} + x_{4}x_{47} + x_{4}x_{48} + x_{4}x_{49} + x_{4}x_{51} + x_{4}x_{52} + x_{4}x_{55} + x_{4}x_{57} + x_{4}x_{58} + x_{4}x_{61} + x_{4}x_{62} + x_{4}x_{64} + x_{5}x_{6} + x_{5}x_{9} + x_{5}x_{10} + x_{5}x_{12} + x_{5}x_{14} + x_{5}x_{16} + x_{5}x_{19} + x_{5}x_{22} + x_{5}x_{23} + x_{5}x_{24} + x_{5}x_{25} + x_{5}x_{26} + x_{5}x_{27} + x_{5}x_{28} + x_{5}x_{31} + x_{5}x_{33} + x_{5}x_{38} + x_{5}x_{39} + x_{5}x_{44} + x_{5}x_{46} + x_{5}x_{49} + x_{5}x_{51} + x_{5}x_{56} + x_{5}x_{57} + x_{5}x_{58} + x_{5}x_{60} + x_{5}x_{61} + x_{5}x_{62} + x_{5}x_{63} + x_{6}x_{9} + x_{6}x_{11} + x_{6}x_{13} + x_{6}x_{17} + x_{6}x_{19} + x_{6}x_{20} + x_{6}x_{24} + x_{6}x_{25} + x_{6}x_{26} + x_{6}x_{27} + x_{6}x_{29} + x_{6}x_{31} + x_{6}x_{33} + x_{6}x_{34} + x_{6}x_{35} + x_{6}x_{38} + x_{6}x_{41} + x_{6}x_{43} + x_{6}x_{44} + x_{6}x_{45} + x_{6}x_{47} + x_{6}x_{51} + x_{6}x_{54} + x_{6}x_{56} + x_{6}x_{57} + x_{6}x_{58} + x_{6}x_{60} + x_{7}x_{9} + x_{7}x_{12} + x_{7}x_{13} + x_{7}x_{16} + x_{7}x_{18} + x_{7}x_{19} + x_{7}x_{21} + x_{7}x_{25} + x_{7}x_{26} + x_{7}x_{29} + x_{7}x_{31} + x_{7}x_{33} + x_{7}x_{34} + x_{7}x_{38} + x_{7}x_{39} + x_{7}x_{40} + x_{7}x_{41} + x_{7}x_{42} + x_{7}x_{44} + x_{7}x_{45} + x_{7}x_{48} + x_{7}x_{49} + x_{7}x_{51} + x_{7}x_{53} + x_{7}x_{55} + x_{7}x_{61} + x_{7}x_{63} + x_{8}x_{9} + x_{8}x_{10} + x_{8}x_{11} + x_{8}x_{12} + x_{8}x_{16} + x_{8}x_{17} + x_{8}x_{18} + x_{8}x_{19} + x_{8}x_{21} + x_{8}x_{23} + x_{8}x_{25} + x_{8}x_{26} + x_{8}x_{30} + x_{8}x_{32} + x_{8}x_{36} + x_{8}x_{37} + x_{8}x_{38} + x_{8}x_{39} + x_{8}x_{42} + x_{8}x_{43} + x_{8}x_{44} + x_{8}x_{45} + x_{8}x_{46} + x_{8}x_{48} + x_{8}x_{51} + x_{8}x_{52} + x_{8}x_{53} + x_{8}x_{56} + x_{8}x_{57} + x_{8}x_{58} + x_{8}x_{61} + x_{8}x_{64} + x_{9}x_{10} + x_{9}x_{11} + x_{9}x_{16} + x_{9}x_{17} + x_{9}x_{18} + x_{9}x_{19} + x_{9}x_{21} + x_{9}x_{26} + x_{9}x_{28} + x_{9}x_{30} + x_{9}x_{31} + x_{9}x_{33} + x_{9}x_{34} + x_{9}x_{35} + x_{9}x_{36} + x_{9}x_{40} + x_{9}x_{43} + x_{9}x_{46} + x_{9}x_{49} + x_{9}x_{54} + x_{9}x_{56} + x_{9}x_{57} + x_{9}x_{58} + x_{9}x_{59} + x_{9}x_{60} + x_{9}x_{61} + x_{9}x_{62} + x_{10}x_{13} + x_{10}x_{14} + x_{10}x_{16} + x_{10}x_{17} + x_{10}x_{18} + x_{10}x_{22} + x_{10}x_{23} + x_{10}x_{24} + x_{10}x_{25} + x_{10}x_{27} + x_{10}x_{28} + x_{10}x_{29} + x_{10}x_{30} + x_{10}x_{32} + x_{10}x_{35} + x_{10}x_{37} + x_{10}x_{41} + x_{10}x_{42} + x_{10}x_{44} + x_{10}x_{46} + x_{10}x_{48} + x_{10}x_{50} + x_{10}x_{51} + x_{10}x_{52} + x_{10}x_{54} + x_{10}x_{55} + x_{10}x_{59} + x_{10}x_{60} + x_{10}x_{61} + x_{10}x_{62} + x_{10}x_{64} + x_{11}x_{12} + x_{11}x_{13} + x_{11}x_{15} + x_{11}x_{20} + x_{11}x_{23} + x_{11}x_{25} + x_{11}x_{26} + x_{11}x_{32} + x_{11}x_{35} + x_{11}x_{36} + x_{11}x_{37} + x_{11}x_{38} + x_{11}x_{40} + x_{11}x_{43} + x_{11}x_{44} + x_{11}x_{46} + x_{11}x_{51} + x_{11}x_{53} + x_{11}x_{54} + x_{11}x_{60} + x_{11}x_{63} + x_{12}x_{17} + x_{12}x_{18} + x_{12}x_{20} + x_{12}x_{24} + x_{12}x_{30} + x_{12}x_{32} + x_{12}x_{33} + x_{12}x_{34} + x_{12}x_{40} + x_{12}x_{41} + x_{12}x_{44} + x_{12}x_{45} + x_{12}x_{46} + x_{12}x_{50} + x_{12}x_{51} + x_{12}x_{55} + x_{12}x_{64} + x_{13}x_{14} + x_{13}x_{15} + x_{13}x_{17} + x_{13}x_{19} + x_{13}x_{20} + x_{13}x_{22} + x_{13}x_{23} + x_{13}x_{24} + x_{13}x_{26} + x_{13}x_{28} + x_{13}x_{29} + x_{13}x_{35} + x_{13}x_{36} + x_{13}x_{41} + x_{13}x_{45} + x_{13}x_{48} + x_{13}x_{49} + x_{13}x_{50} + x_{13}x_{51} + x_{13}x_{54} + x_{13}x_{55} + x_{13}x_{58} + x_{13}x_{60} + x_{13}x_{61} + x_{13}x_{62} + x_{13}x_{64} + x_{14}x_{15} + x_{14}x_{20} + x_{14}x_{21} + x_{14}x_{23} + x_{14}x_{24} + x_{14}x_{26} + x_{14}x_{28} + x_{14}x_{31} + x_{14}x_{32} + x_{14}x_{33} + x_{14}x_{34} + x_{14}x_{36} + x_{14}x_{37} + x_{14}x_{39} + x_{14}x_{40} + x_{14}x_{43} + x_{14}x_{45} + x_{14}x_{46} + x_{14}x_{49} + x_{14}x_{51} + x_{14}x_{53} + x_{14}x_{55} + x_{14}x_{56} + x_{14}x_{57} + x_{14}x_{58} + x_{14}x_{60} + x_{14}x_{61} + x_{14}x_{64} + x_{15}x_{17} + x_{15}x_{19} + x_{15}x_{21} + x_{15}x_{22} + x_{15}x_{24} + x_{15}x_{29} + x_{15}x_{31} + x_{15}x_{33} + x_{15}x_{34} + x_{15}x_{35} + x_{15}x_{36} + x_{15}x_{39} + x_{15}x_{44} + x_{15}x_{45} + x_{15}x_{46} + x_{15}x_{48} + x_{15}x_{51} + x_{15}x_{54} + x_{15}x_{59} + x_{15}x_{62} + x_{15}x_{63} + x_{15}x_{64} + x_{16}x_{17} + x_{16}x_{20} + x_{16}x_{21} + x_{16}x_{25} + x_{16}x_{26} + x_{16}x_{27} + x_{16}x_{28} + x_{16}x_{29} + x_{16}x_{30} + x_{16}x_{32} + x_{16}x_{34} + x_{16}x_{38} + x_{16}x_{39} + x_{16}x_{40} + x_{16}x_{42} + x_{16}x_{46} + x_{16}x_{48} + x_{16}x_{51} + x_{16}x_{52} + x_{16}x_{53} + x_{16}x_{54} + x_{16}x_{57} + x_{16}x_{58} + x_{16}x_{60} + x_{16}x_{62} + x_{16}x_{63} + x_{16}x_{64} + x_{17}x_{18} + x_{17}x_{20} + x_{17}x_{22} + x_{17}x_{23} + x_{17}x_{24} + x_{17}x_{25} + x_{17}x_{27} + x_{17}x_{28} + x_{17}x_{29} + x_{17}x_{30} + x_{17}x_{31} + x_{17}x_{32} + x_{17}x_{34} + x_{17}x_{35} + x_{17}x_{36} + x_{17}x_{39} + x_{17}x_{41} + x_{17}x_{42} + x_{17}x_{44} + x_{17}x_{45} + x_{17}x_{47} + x_{17}x_{50} + x_{17}x_{53} + x_{17}x_{54} + x_{17}x_{55} + x_{17}x_{56} + x_{17}x_{61} + x_{17}x_{62} + x_{17}x_{64} + x_{18}x_{20} + x_{18}x_{21} + x_{18}x_{27} + x_{18}x_{28} + x_{18}x_{29} + x_{18}x_{30} + x_{18}x_{32} + x_{18}x_{33} + x_{18}x_{34} + x_{18}x_{35} + x_{18}x_{38} + x_{18}x_{40} + x_{18}x_{41} + x_{18}x_{42} + x_{18}x_{44} + x_{18}x_{46} + x_{18}x_{47} + x_{18}x_{57} + x_{18}x_{58} + x_{18}x_{61} + x_{18}x_{64} + x_{19}x_{22} + x_{19}x_{23} + x_{19}x_{26} + x_{19}x_{27} + x_{19}x_{28} + x_{19}x_{31} + x_{19}x_{33} + x_{19}x_{34} + x_{19}x_{35} + x_{19}x_{36} + x_{19}x_{39} + x_{19}x_{42} + x_{19}x_{43} + x_{19}x_{46} + x_{19}x_{48} + x_{19}x_{50} + x_{19}x_{51} + x_{19}x_{52} + x_{19}x_{53} + x_{19}x_{56} + x_{19}x_{58} + x_{19}x_{62} + x_{20}x_{22} + x_{20}x_{24} + x_{20}x_{31} + x_{20}x_{32} + x_{20}x_{34} + x_{20}x_{35} + x_{20}x_{36} + x_{20}x_{37} + x_{20}x_{39} + x_{20}x_{40} + x_{20}x_{44} + x_{20}x_{45} + x_{20}x_{46} + x_{20}x_{49} + x_{20}x_{50} + x_{20}x_{53} + x_{20}x_{55} + x_{20}x_{56} + x_{20}x_{58} + x_{20}x_{59} + x_{20}x_{61} + x_{20}x_{63} + x_{21}x_{22} + x_{21}x_{23} + x_{21}x_{25} + x_{21}x_{26} + x_{21}x_{27} + x_{21}x_{29} + x_{21}x_{30} + x_{21}x_{32} + x_{21}x_{33} + x_{21}x_{35} + x_{21}x_{39} + x_{21}x_{41} + x_{21}x_{42} + x_{21}x_{43} + x_{21}x_{44} + x_{21}x_{46} + x_{21}x_{47} + x_{21}x_{49} + x_{21}x_{53} + x_{21}x_{54} + x_{21}x_{56} + x_{21}x_{57} + x_{21}x_{58} + x_{21}x_{61} + x_{21}x_{64} + x_{22}x_{24} + x_{22}x_{25} + x_{22}x_{26} + x_{22}x_{27} + x_{22}x_{29} + x_{22}x_{30} + x_{22}x_{32} + x_{22}x_{35} + x_{22}x_{38} + x_{22}x_{42} + x_{22}x_{45} + x_{22}x_{46} + x_{22}x_{50} + x_{22}x_{52} + x_{22}x_{54} + x_{22}x_{55} + x_{22}x_{56} + x_{22}x_{58} + x_{22}x_{59} + x_{22}x_{61} + x_{22}x_{63} + x_{22}x_{64} + x_{23}x_{26} + x_{23}x_{28} + x_{23}x_{30} + x_{23}x_{31} + x_{23}x_{34} + x_{23}x_{35} + x_{23}x_{36} + x_{23}x_{37} + x_{23}x_{39} + x_{23}x_{41} + x_{23}x_{43} + x_{23}x_{44} + x_{23}x_{47} + x_{23}x_{48} + x_{23}x_{50} + x_{23}x_{51} + x_{23}x_{53} + x_{23}x_{54} + x_{23}x_{56} + x_{23}x_{57} + x_{23}x_{58} + x_{23}x_{59} + x_{23}x_{61} + x_{23}x_{62} + x_{24}x_{26} + x_{24}x_{29} + x_{24}x_{30} + x_{24}x_{32} + x_{24}x_{34} + x_{24}x_{35} + x_{24}x_{39} + x_{24}x_{42} + x_{24}x_{44} + x_{24}x_{46} + x_{24}x_{47} + x_{24}x_{51} + x_{24}x_{52} + x_{24}x_{53} + x_{24}x_{54} + x_{24}x_{56} + x_{24}x_{57} + x_{24}x_{59} + x_{24}x_{63} + x_{25}x_{26} + x_{25}x_{28} + x_{25}x_{30} + x_{25}x_{31} + x_{25}x_{33} + x_{25}x_{34} + x_{25}x_{40} + x_{25}x_{44} + x_{25}x_{45} + x_{25}x_{46} + x_{25}x_{48} + x_{25}x_{49} + x_{25}x_{50} + x_{25}x_{51} + x_{25}x_{53} + x_{25}x_{54} + x_{25}x_{55} + x_{25}x_{56} + x_{25}x_{59} + x_{25}x_{60} + x_{25}x_{61} + x_{25}x_{63} + x_{26}x_{29} + x_{26}x_{33} + x_{26}x_{35} + x_{26}x_{36} + x_{26}x_{38} + x_{26}x_{39} + x_{26}x_{42} + x_{26}x_{43} + x_{26}x_{44} + x_{26}x_{45} + x_{26}x_{46} + x_{26}x_{50} + x_{26}x_{52} + x_{26}x_{54} + x_{26}x_{55} + x_{26}x_{56} + x_{26}x_{59} + x_{26}x_{60} + x_{26}x_{61} + x_{26}x_{62} + x_{26}x_{63} + x_{27}x_{28} + x_{27}x_{29} + x_{27}x_{31} + x_{27}x_{32} + x_{27}x_{34} + x_{27}x_{35} + x_{27}x_{36} + x_{27}x_{39} + x_{27}x_{42} + x_{27}x_{43} + x_{27}x_{45} + x_{27}x_{46} + x_{27}x_{47} + x_{27}x_{54} + x_{27}x_{55} + x_{27}x_{56} + x_{27}x_{58} + x_{27}x_{59} + x_{27}x_{60} + x_{27}x_{61} + x_{27}x_{63} + x_{27}x_{64} + x_{28}x_{29} + x_{28}x_{31} + x_{28}x_{34} + x_{28}x_{38} + x_{28}x_{40} + x_{28}x_{41} + x_{28}x_{45} + x_{28}x_{48} + x_{28}x_{51} + x_{28}x_{53} + x_{28}x_{56} + x_{28}x_{57} + x_{28}x_{62} + x_{28}x_{63} + x_{28}x_{64} + x_{29}x_{31} + x_{29}x_{32} + x_{29}x_{36} + x_{29}x_{38} + x_{29}x_{40} + x_{29}x_{42} + x_{29}x_{44} + x_{29}x_{45} + x_{29}x_{48} + x_{29}x_{49} + x_{29}x_{50} + x_{29}x_{51} + x_{29}x_{52} + x_{29}x_{53} + x_{29}x_{54} + x_{29}x_{56} + x_{29}x_{57} + x_{29}x_{58} + x_{29}x_{59} + x_{29}x_{61} + x_{29}x_{62} + x_{29}x_{64} + x_{30}x_{31} + x_{30}x_{35} + x_{30}x_{38} + x_{30}x_{39} + x_{30}x_{41} + x_{30}x_{43} + x_{30}x_{44} + x_{30}x_{45} + x_{30}x_{47} + x_{30}x_{50} + x_{30}x_{51} + x_{30}x_{53} + x_{30}x_{54} + x_{30}x_{56} + x_{30}x_{57} + x_{30}x_{58} + x_{30}x_{61} + x_{30}x_{63} + x_{31}x_{32} + x_{31}x_{33} + x_{31}x_{34} + x_{31}x_{36} + x_{31}x_{37} + x_{31}x_{38} + x_{31}x_{39} + x_{31}x_{43} + x_{31}x_{44} + x_{31}x_{47} + x_{31}x_{49} + x_{31}x_{53} + x_{31}x_{56} + x_{31}x_{59} + x_{31}x_{61} + x_{31}x_{63} + x_{31}x_{64} + x_{32}x_{33} + x_{32}x_{34} + x_{32}x_{38} + x_{32}x_{39} + x_{32}x_{43} + x_{32}x_{49} + x_{32}x_{53} + x_{32}x_{55} + x_{32}x_{56} + x_{32}x_{58} + x_{32}x_{59} + x_{32}x_{61} + x_{32}x_{63} + x_{33}x_{34} + x_{33}x_{35} + x_{33}x_{38} + x_{33}x_{42} + x_{33}x_{43} + x_{33}x_{44} + x_{33}x_{46} + x_{33}x_{51} + x_{33}x_{52} + x_{33}x_{53} + x_{33}x_{54} + x_{33}x_{55} + x_{33}x_{56} + x_{33}x_{57} + x_{33}x_{58} + x_{33}x_{60} + x_{33}x_{61} + x_{33}x_{62} + x_{34}x_{39} + x_{34}x_{40} + x_{34}x_{42} + x_{34}x_{43} + x_{34}x_{44} + x_{34}x_{45} + x_{34}x_{46} + x_{34}x_{50} + x_{34}x_{51} + x_{34}x_{52} + x_{34}x_{57} + x_{34}x_{59} + x_{34}x_{63} + x_{35}x_{36} + x_{35}x_{41} + x_{35}x_{45} + x_{35}x_{48} + x_{35}x_{50} + x_{35}x_{55} + x_{35}x_{56} + x_{35}x_{58} + x_{35}x_{60} + x_{35}x_{61} + x_{36}x_{40} + x_{36}x_{41} + x_{36}x_{44} + x_{36}x_{47} + x_{36}x_{50} + x_{36}x_{54} + x_{36}x_{55} + x_{36}x_{57} + x_{36}x_{60} + x_{36}x_{61} + x_{36}x_{63} + x_{36}x_{64} + x_{37}x_{38} + x_{37}x_{39} + x_{37}x_{40} + x_{37}x_{45} + x_{37}x_{46} + x_{37}x_{47} + x_{37}x_{50} + x_{37}x_{54} + x_{37}x_{55} + x_{37}x_{56} + x_{37}x_{57} + x_{37}x_{58} + x_{37}x_{60} + x_{37}x_{63} + x_{38}x_{40} + x_{38}x_{42} + x_{38}x_{44} + x_{38}x_{45} + x_{38}x_{46} + x_{38}x_{47} + x_{38}x_{53} + x_{38}x_{55} + x_{38}x_{56} + x_{38}x_{57} + x_{38}x_{58} + x_{38}x_{59} + x_{38}x_{64} + x_{39}x_{40} + x_{39}x_{46} + x_{39}x_{48} + x_{39}x_{49} + x_{39}x_{50} + x_{39}x_{51} + x_{39}x_{53} + x_{39}x_{54} + x_{39}x_{55} + x_{39}x_{56} + x_{39}x_{57} + x_{39}x_{58} + x_{39}x_{63} + x_{40}x_{41} + x_{40}x_{42} + x_{40}x_{44} + x_{40}x_{47} + x_{40}x_{49} + x_{40}x_{54} + x_{40}x_{56} + x_{40}x_{57} + x_{40}x_{58} + x_{40}x_{60} + x_{40}x_{61} + x_{40}x_{64} + x_{41}x_{45} + x_{41}x_{47} + x_{41}x_{48} + x_{41}x_{49} + x_{41}x_{51} + x_{41}x_{53} + x_{41}x_{57} + x_{41}x_{60} + x_{41}x_{61} + x_{41}x_{63} + x_{42}x_{44} + x_{42}x_{46} + x_{42}x_{49} + x_{42}x_{53} + x_{42}x_{55} + x_{42}x_{58} + x_{42}x_{59} + x_{42}x_{61} + x_{43}x_{44} + x_{43}x_{45} + x_{43}x_{47} + x_{43}x_{49} + x_{43}x_{50} + x_{43}x_{51} + x_{43}x_{55} + x_{43}x_{57} + x_{43}x_{58} + x_{43}x_{64} + x_{44}x_{45} + x_{44}x_{49} + x_{44}x_{54} + x_{44}x_{55} + x_{44}x_{56} + x_{44}x_{57} + x_{44}x_{58} + x_{44}x_{61} + x_{44}x_{64} + x_{45}x_{46} + x_{45}x_{52} + x_{45}x_{53} + x_{45}x_{55} + x_{45}x_{56} + x_{45}x_{57} + x_{45}x_{63} + x_{46}x_{47} + x_{46}x_{49} + x_{46}x_{50} + x_{46}x_{51} + x_{46}x_{53} + x_{46}x_{57} + x_{46}x_{60} + x_{46}x_{61} + x_{46}x_{64} + x_{47}x_{48} + x_{47}x_{49} + x_{47}x_{52} + x_{47}x_{56} + x_{47}x_{58} + x_{47}x_{63} + x_{47}x_{64} + x_{48}x_{50} + x_{48}x_{51} + x_{48}x_{54} + x_{48}x_{56} + x_{48}x_{59} + x_{48}x_{60} + x_{48}x_{62} + x_{48}x_{63} + x_{48}x_{64} + x_{49}x_{50} + x_{49}x_{51} + x_{49}x_{52} + x_{49}x_{54} + x_{49}x_{64} + x_{50}x_{54} + x_{50}x_{55} + x_{50}x_{59} + x_{50}x_{62} + x_{50}x_{64} + x_{51}x_{54} + x_{51}x_{56} + x_{51}x_{58} + x_{51}x_{59} + x_{51}x_{61} + x_{51}x_{64} + x_{52}x_{57} + x_{52}x_{58} + x_{52}x_{61} + x_{52}x_{62} + x_{52}x_{63} + x_{53}x_{54} + x_{53}x_{55} + x_{53}x_{57} + x_{53}x_{58} + x_{53}x_{59} + x_{53}x_{60} + x_{54}x_{56} + x_{54}x_{58} + x_{54}x_{59} + x_{54}x_{60} + x_{54}x_{62} + x_{54}x_{64} + x_{55}x_{56} + x_{55}x_{61} + x_{55}x_{62} + x_{56}x_{57} + x_{56}x_{59} + x_{56}x_{60} + x_{56}x_{62} + x_{56}x_{64} + x_{57}x_{61} + x_{57}x_{63} + x_{57}x_{64} + x_{58}x_{59} + x_{58}x_{60} + x_{58}x_{61} + x_{59}x_{62} + x_{59}x_{63} + x_{60}x_{64} + x_{61}x_{64} + x_{62}x_{64} + x_{63}x_{64} + x_{3} + x_{4} + x_{7} + x_{8} + x_{11} + x_{12} + x_{15} + x_{16} + x_{18} + x_{20} + x_{21} + x_{22} + x_{25} + x_{30} + x_{33} + x_{37} + x_{38} + x_{39} + x_{40} + x_{45} + x_{46} + x_{47} + x_{51} + x_{53} + x_{54} + x_{55} + x_{56} + x_{57} + x_{60} + x_{61} + x_{62} + 1$

$y_{11} = x_{1}x_{6} + x_{1}x_{9} + x_{1}x_{13} + x_{1}x_{15} + x_{1}x_{18} + x_{1}x_{20} + x_{1}x_{21} + x_{1}x_{22} + x_{1}x_{23} + x_{1}x_{24} + x_{1}x_{26} + x_{1}x_{30} + x_{1}x_{32} + x_{1}x_{33} + x_{1}x_{38} + x_{1}x_{39} + x_{1}x_{41} + x_{1}x_{43} + x_{1}x_{44} + x_{1}x_{47} + x_{1}x_{48} + x_{1}x_{50} + x_{1}x_{52} + x_{1}x_{55} + x_{1}x_{56} + x_{1}x_{58} + x_{1}x_{61} + x_{1}x_{63} + x_{2}x_{4} + x_{2}x_{7} + x_{2}x_{8} + x_{2}x_{9} + x_{2}x_{10} + x_{2}x_{11} + x_{2}x_{13} + x_{2}x_{16} + x_{2}x_{17} + x_{2}x_{19} + x_{2}x_{20} + x_{2}x_{24} + x_{2}x_{25} + x_{2}x_{28} + x_{2}x_{30} + x_{2}x_{31} + x_{2}x_{33} + x_{2}x_{34} + x_{2}x_{38} + x_{2}x_{44} + x_{2}x_{45} + x_{2}x_{46} + x_{2}x_{47} + x_{2}x_{51} + x_{2}x_{54} + x_{2}x_{55} + x_{2}x_{57} + x_{2}x_{58} + x_{2}x_{60} + x_{2}x_{61} + x_{2}x_{62} + x_{2}x_{64} + x_{3}x_{4} + x_{3}x_{5} + x_{3}x_{7} + x_{3}x_{9} + x_{3}x_{12} + x_{3}x_{13} + x_{3}x_{14} + x_{3}x_{15} + x_{3}x_{17} + x_{3}x_{19} + x_{3}x_{20} + x_{3}x_{22} + x_{3}x_{23} + x_{3}x_{31} + x_{3}x_{32} + x_{3}x_{33} + x_{3}x_{34} + x_{3}x_{35} + x_{3}x_{38} + x_{3}x_{39} + x_{3}x_{40} + x_{3}x_{41} + x_{3}x_{43} + x_{3}x_{46} + x_{3}x_{47} + x_{3}x_{48} + x_{3}x_{49} + x_{3}x_{50} + x_{3}x_{51} + x_{3}x_{52} + x_{3}x_{53} + x_{3}x_{55} + x_{3}x_{59} + x_{3}x_{64} + x_{4}x_{5} + x_{4}x_{6} + x_{4}x_{9} + x_{4}x_{11} + x_{4}x_{12} + x_{4}x_{14} + x_{4}x_{15} + x_{4}x_{17} + x_{4}x_{18} + x_{4}x_{20} + x_{4}x_{21} + x_{4}x_{24} + x_{4}x_{25} + x_{4}x_{28} + x_{4}x_{30} + x_{4}x_{31} + x_{4}x_{36} + x_{4}x_{39} + x_{4}x_{40} + x_{4}x_{41} + x_{4}x_{44} + x_{4}x_{48} + x_{4}x_{51} + x_{4}x_{54} + x_{4}x_{55} + x_{4}x_{57} + x_{4}x_{58} + x_{4}x_{63} + x_{4}x_{64} + x_{5}x_{6} + x_{5}x_{8} + x_{5}x_{13} + x_{5}x_{15} + x_{5}x_{18} + x_{5}x_{19} + x_{5}x_{21} + x_{5}x_{22} + x_{5}x_{25} + x_{5}x_{26} + x_{5}x_{28} + x_{5}x_{29} + x_{5}x_{30} + x_{5}x_{31} + x_{5}x_{33} + x_{5}x_{34} + x_{5}x_{36} + x_{5}x_{37} + x_{5}x_{38} + x_{5}x_{39} + x_{5}x_{40} + x_{5}x_{42} + x_{5}x_{43} + x_{5}x_{47} + x_{5}x_{54} + x_{5}x_{56} + x_{5}x_{61} + x_{5}x_{62} + x_{5}x_{63} + x_{6}x_{7} + x_{6}x_{13} + x_{6}x_{17} + x_{6}x_{20} + x_{6}x_{23} + x_{6}x_{24} + x_{6}x_{25} + x_{6}x_{27} + x_{6}x_{28} + x_{6}x_{30} + x_{6}x_{31} + x_{6}x_{32} + x_{6}x_{36} + x_{6}x_{37} + x_{6}x_{38} + x_{6}x_{40} + x_{6}x_{41} + x_{6}x_{42} + x_{6}x_{49} + x_{6}x_{50} + x_{6}x_{55} + x_{6}x_{58} + x_{6}x_{61} + x_{6}x_{63} + x_{7}x_{9} + x_{7}x_{10} + x_{7}x_{16} + x_{7}x_{17} + x_{7}x_{21} + x_{7}x_{23} + x_{7}x_{25} + x_{7}x_{31} + x_{7}x_{33} + x_{7}x_{34} + x_{7}x_{35} + x_{7}x_{38} + x_{7}x_{42} + x_{7}x_{44} + x_{7}x_{45} + x_{7}x_{48} + x_{7}x_{51} + x_{7}x_{54} + x_{7}x_{55} + x_{7}x_{57} + x_{7}x_{59} + x_{7}x_{61} + x_{7}x_{63} + x_{7}x_{64} + x_{8}x_{10} + x_{8}x_{12} + x_{8}x_{13} + x_{8}x_{14} + x_{8}x_{16} + x_{8}x_{18} + x_{8}x_{24} + x_{8}x_{25} + x_{8}x_{27} + x_{8}x_{31} + x_{8}x_{32} + x_{8}x_{34} + x_{8}x_{35} + x_{8}x_{36} + x_{8}x_{37} + x_{8}x_{41} + x_{8}x_{42} + x_{8}x_{46} + x_{8}x_{47} + x_{8}x_{48} + x_{8}x_{50} + x_{8}x_{51} + x_{8}x_{52} + x_{8}x_{55} + x_{8}x_{56} + x_{8}x_{61} + x_{8}x_{64} + x_{9}x_{10} + x_{9}x_{11} + x_{9}x_{14} + x_{9}x_{15} + x_{9}x_{16} + x_{9}x_{20} + x_{9}x_{21} + x_{9}x_{23} + x_{9}x_{24} + x_{9}x_{25} + x_{9}x_{27} + x_{9}x_{30} + x_{9}x_{31} + x_{9}x_{38} + x_{9}x_{40} + x_{9}x_{42} + x_{9}x_{44} + x_{9}x_{45} + x_{9}x_{46} + x_{9}x_{48} + x_{9}x_{49} + x_{9}x_{50} + x_{9}x_{51} + x_{9}x_{54} + x_{9}x_{55} + x_{9}x_{56} + x_{9}x_{57} + x_{9}x_{58} + x_{9}x_{61} + x_{9}x_{62} + x_{9}x_{63} + x_{9}x_{64} + x_{10}x_{11} + x_{10}x_{13} + x_{10}x_{15} + x_{10}x_{18} + x_{10}x_{19} + x_{10}x_{20} + x_{10}x_{21} + x_{10}x_{22} + x_{10}x_{24} + x_{10}x_{26} + x_{10}x_{30} + x_{10}x_{32} + x_{10}x_{34} + x_{10}x_{36} + x_{10}x_{41} + x_{10}x_{43} + x_{10}x_{44} + x_{10}x_{45} + x_{10}x_{51} + x_{10}x_{54} + x_{10}x_{56} + x_{10}x_{57} + x_{10}x_{59} + x_{10}x_{60} + x_{10}x_{61} + x_{10}x_{62} + x_{10}x_{64} + x_{11}x_{13} + x_{11}x_{15} + x_{11}x_{19} + x_{11}x_{25} + x_{11}x_{29} + x_{11}x_{30} + x_{11}x_{31} + x_{11}x_{32} + x_{11}x_{33} + x_{11}x_{34} + x_{11}x_{36} + x_{11}x_{37} + x_{11}x_{38} + x_{11}x_{40} + x_{11}x_{41} + x_{11}x_{42} + x_{11}x_{45} + x_{11}x_{46} + x_{11}x_{47} + x_{11}x_{50} + x_{11}x_{52} + x_{11}x_{53} + x_{11}x_{54} + x_{11}x_{55} + x_{11}x_{56} + x_{11}x_{58} + x_{11}x_{59} + x_{11}x_{60} + x_{11}x_{61} + x_{11}x_{63} + x_{12}x_{13} + x_{12}x_{14} + x_{12}x_{15} + x_{12}x_{16} + x_{12}x_{17} + x_{12}x_{19} + x_{12}x_{24} + x_{12}x_{26} + x_{12}x_{27} + x_{12}x_{29} + x_{12}x_{31} + x_{12}x_{32} + x_{12}x_{35} + x_{12}x_{37} + x_{12}x_{38} + x_{12}x_{39} + x_{12}x_{41} + x_{12}x_{42} + x_{12}x_{44} + x_{12}x_{48} + x_{12}x_{49} + x_{12}x_{51} + x_{12}x_{52} + x_{12}x_{55} + x_{12}x_{57} + x_{12}x_{58} + x_{12}x_{59} + x_{12}x_{61} + x_{12}x_{62} + x_{13}x_{14} + x_{13}x_{15} + x_{13}x_{18} + x_{13}x_{20} + x_{13}x_{22} + x_{13}x_{25} + x_{13}x_{29} + x_{13}x_{30} + x_{13}x_{31} + x_{13}x_{32} + x_{13}x_{33} + x_{13}x_{34} + x_{13}x_{35} + x_{13}x_{37} + x_{13}x_{38} + x_{13}x_{40} + x_{13}x_{42} + x_{13}x_{43} + x_{13}x_{46} + x_{13}x_{47} + x_{13}x_{48} + x_{13}x_{49} + x_{13}x_{51} + x_{13}x_{52} + x_{13}x_{55} + x_{13}x_{56} + x_{13}x_{57} + x_{13}x_{58} + x_{13}x_{62} + x_{14}x_{17} + x_{14}x_{18} + x_{14}x_{22} + x_{14}x_{26} + x_{14}x_{27} + x_{14}x_{28} + x_{14}x_{29} + x_{14}x_{30} + x_{14}x_{31} + x_{14}x_{32} + x_{14}x_{33} + x_{14}x_{34} + x_{14}x_{40} + x_{14}x_{43} + x_{14}x_{50} + x_{14}x_{51} + x_{14}x_{53} + x_{14}x_{55} + x_{14}x_{56} + x_{14}x_{61} + x_{14}x_{62} + x_{14}x_{64} + x_{15}x_{25} + x_{15}x_{28} + x_{15}x_{31} + x_{15}x_{32} + x_{15}x_{33} + x_{15}x_{35} + x_{15}x_{36} + x_{15}x_{37} + x_{15}x_{38} + x_{15}x_{41} + x_{15}x_{44} + x_{15}x_{45} + x_{15}x_{47} + x_{15}x_{53} + x_{15}x_{54} + x_{15}x_{55} + x_{15}x_{56} + x_{15}x_{57} + x_{15}x_{60} + x_{15}x_{61} + x_{15}x_{62} + x_{15}x_{63} + x_{16}x_{19} + x_{16}x_{20} + x_{16}x_{25} + x_{16}x_{26} + x_{16}x_{29} + x_{16}x_{32} + x_{16}x_{36} + x_{16}x_{37} + x_{16}x_{40} + x_{16}x_{41} + x_{16}x_{44} + x_{16}x_{46} + x_{16}x_{48} + x_{16}x_{49} + x_{16}x_{51} + x_{16}x_{53} + x_{16}x_{54} + x_{16}x_{56} + x_{16}x_{57} + x_{16}x_{60} + x_{16}x_{61} + x_{17}x_{18} + x_{17}x_{19} + x_{17}x_{20} + x_{17}x_{24} + x_{17}x_{26} + x_{17}x_{28} + x_{17}x_{30} + x_{17}x_{32} + x_{17}x_{33} + x_{17}x_{36} + x_{17}x_{46} + x_{17}x_{48} + x_{17}x_{53} + x_{17}x_{57} + x_{17}x_{61} + x_{17}x_{64} + x_{18}x_{19} + x_{18}x_{20} + x_{18}x_{21} + x_{18}x_{22} + x_{18}x_{23} + x_{18}x_{29} + x_{18}x_{30} + x_{18}x_{31} + x_{18}x_{32} + x_{18}x_{35} + x_{18}x_{38} + x_{18}x_{39} + x_{18}x_{44} + x_{18}x_{46} + x_{18}x_{47} + x_{18}x_{49} + x_{18}x_{53} + x_{18}x_{54} + x_{18}x_{56} + x_{18}x_{58} + x_{18}x_{60} + x_{18}x_{62} + x_{18}x_{63} + x_{19}x_{21} + x_{19}x_{22} + x_{19}x_{23} + x_{19}x_{24} + x_{19}x_{25} + x_{19}x_{26} + x_{19}x_{29} + x_{19}x_{30} + x_{19}x_{32} + x_{19}x_{34} + x_{19}x_{37} + x_{19}x_{38} + x_{19}x_{39} + x_{19}x_{40} + x_{19}x_{42} + x_{19}x_{43} + x_{19}x_{45} + x_{19}x_{46} + x_{19}x_{47} + x_{19}x_{49} + x_{19}x_{50} + x_{19}x_{54} + x_{19}x_{55} + x_{19}x_{57} + x_{19}x_{58} + x_{19}x_{59} + x_{19}x_{62} + x_{19}x_{64} + x_{20}x_{22} + x_{20}x_{26} + x_{20}x_{27} + x_{20}x_{28} + x_{20}x_{29} + x_{20}x_{30} + x_{20}x_{31} + x_{20}x_{32} + x_{20}x_{33} + x_{20}x_{34} + x_{20}x_{38} + x_{20}x_{39} + x_{20}x_{45} + x_{20}x_{46} + x_{20}x_{47} + x_{20}x_{48} + x_{20}x_{49} + x_{20}x_{50} + x_{20}x_{57} + x_{20}x_{58} + x_{20}x_{60} + x_{20}x_{61} + x_{20}x_{62} + x_{20}x_{63} + x_{21}x_{22} + x_{21}x_{24} + x_{21}x_{26} + x_{21}x_{30} + x_{21}x_{31} + x_{21}x_{33} + x_{21}x_{34} + x_{21}x_{36} + x_{21}x_{37} + x_{21}x_{40} + x_{21}x_{42} + x_{21}x_{43} + x_{21}x_{45} + x_{21}x_{48} + x_{21}x_{49} + x_{21}x_{50} + x_{21}x_{53} + x_{21}x_{58} + x_{21}x_{59} + x_{21}x_{60} + x_{21}x_{61} + x_{21}x_{64} + x_{22}x_{23} + x_{22}x_{25} + x_{22}x_{27} + x_{22}x_{29} + x_{22}x_{30} + x_{22}x_{31} + x_{22}x_{32} + x_{22}x_{33} + x_{22}x_{34} + x_{22}x_{36} + x_{22}x_{40} + x_{22}x_{43} + x_{22}x_{45} + x_{22}x_{48} + x_{22}x_{51} + x_{22}x_{53} + x_{22}x_{58} + x_{22}x_{59} + x_{22}x_{60} + x_{22}x_{62} + x_{23}x_{26} + x_{23}x_{28} + x_{23}x_{30} + x_{23}x_{31} + x_{23}x_{34} + x_{23}x_{35} + x_{23}x_{36} + x_{23}x_{40} + x_{23}x_{41} + x_{23}x_{46} + x_{23}x_{47} + x_{23}x_{48} + x_{23}x_{53} + x_{23}x_{54} + x_{23}x_{55} + x_{23}x_{61} + x_{23}x_{62} + x_{24}x_{28} + x_{24}x_{30} + x_{24}x_{32} + x_{24}x_{40} + x_{24}x_{43} + x_{24}x_{45} + x_{24}x_{50} + x_{24}x_{53} + x_{24}x_{54} + x_{24}x_{56} + x_{24}x_{57} + x_{24}x_{60} + x_{24}x_{64} + x_{25}x_{29} + x_{25}x_{30} + x_{25}x_{31} + x_{25}x_{32} + x_{25}x_{34} + x_{25}x_{35} + x_{25}x_{36} + x_{25}x_{39} + x_{25}x_{40} + x_{25}x_{41} + x_{25}x_{42} + x_{25}x_{43} + x_{25}x_{44} + x_{25}x_{49} + x_{25}x_{51} + x_{25}x_{53} + x_{25}x_{54} + x_{25}x_{58} + x_{25}x_{64} + x_{26}x_{27} + x_{26}x_{32} + x_{26}x_{33} + x_{26}x_{34} + x_{26}x_{35} + x_{26}x_{36} + x_{26}x_{37} + x_{26}x_{39} + x_{26}x_{42} + x_{26}x_{44} + x_{26}x_{46} + x_{26}x_{49} + x_{26}x_{52} + x_{26}x_{53} + x_{26}x_{54} + x_{26}x_{58} + x_{26}x_{61} + x_{26}x_{63} + x_{27}x_{28} + x_{27}x_{29} + x_{27}x_{30} + x_{27}x_{31} + x_{27}x_{32} + x_{27}x_{33} + x_{27}x_{35} + x_{27}x_{36} + x_{27}x_{41} + x_{27}x_{43} + x_{27}x_{44} + x_{27}x_{45} + x_{27}x_{46} + x_{27}x_{47} + x_{27}x_{51} + x_{27}x_{52} + x_{27}x_{55} + x_{27}x_{58} + x_{27}x_{59} + x_{27}x_{62} + x_{27}x_{63} + x_{27}x_{64} + x_{28}x_{30} + x_{28}x_{33} + x_{28}x_{35} + x_{28}x_{36} + x_{28}x_{37} + x_{28}x_{39} + x_{28}x_{42} + x_{28}x_{43} + x_{28}x_{44} + x_{28}x_{47} + x_{28}x_{50} + x_{28}x_{51} + x_{28}x_{54} + x_{28}x_{57} + x_{28}x_{58} + x_{28}x_{59} + x_{28}x_{61} + x_{28}x_{62} + x_{29}x_{30} + x_{29}x_{31} + x_{29}x_{34} + x_{29}x_{37} + x_{29}x_{40} + x_{29}x_{41} + x_{29}x_{43} + x_{29}x_{44} + x_{29}x_{48} + x_{29}x_{49} + x_{29}x_{52} + x_{29}x_{56} + x_{29}x_{57} + x_{29}x_{58} + x_{29}x_{59} + x_{29}x_{62} + x_{30}x_{35} + x_{30}x_{36} + x_{30}x_{38} + x_{30}x_{39} + x_{30}x_{41} + x_{30}x_{44} + x_{30}x_{45} + x_{30}x_{46} + x_{30}x_{47} + x_{30}x_{50} + x_{30}x_{52} + x_{30}x_{53} + x_{30}x_{54} + x_{30}x_{55} + x_{30}x_{57} + x_{30}x_{59} + x_{30}x_{61} + x_{30}x_{63} + x_{31}x_{33} + x_{31}x_{34} + x_{31}x_{35} + x_{31}x_{36} + x_{31}x_{37} + x_{31}x_{39} + x_{31}x_{44} + x_{31}x_{45} + x_{31}x_{48} + x_{31}x_{49} + x_{31}x_{50} + x_{31}x_{55} + x_{31}x_{56} + x_{31}x_{60} + x_{31}x_{63} + x_{31}x_{64} + x_{32}x_{33} + x_{32}x_{34} + x_{32}x_{35} + x_{32}x_{37} + x_{32}x_{38} + x_{32}x_{39} + x_{32}x_{40} + x_{32}x_{42} + x_{32}x_{43} + x_{32}x_{45} + x_{32}x_{47} + x_{32}x_{48} + x_{32}x_{52} + x_{32}x_{53} + x_{32}x_{55} + x_{32}x_{59} + x_{32}x_{60} + x_{32}x_{63} + x_{33}x_{36} + x_{33}x_{38} + x_{33}x_{39} + x_{33}x_{40} + x_{33}x_{41} + x_{33}x_{42} + x_{33}x_{48} + x_{33}x_{53} + x_{33}x_{54} + x_{33}x_{55} + x_{33}x_{57} + x_{33}x_{58} + x_{33}x_{62} + x_{33}x_{64} + x_{34}x_{36} + x_{34}x_{37} + x_{34}x_{38} + x_{34}x_{43} + x_{34}x_{44} + x_{34}x_{46} + x_{34}x_{50} + x_{34}x_{52} + x_{34}x_{53} + x_{34}x_{54} + x_{34}x_{58} + x_{34}x_{60} + x_{34}x_{62} + x_{35}x_{38} + x_{35}x_{39} + x_{35}x_{42} + x_{35}x_{43} + x_{35}x_{44} + x_{35}x_{46} + x_{35}x_{48} + x_{35}x_{54} + x_{35}x_{56} + x_{35}x_{57} + x_{35}x_{58} + x_{36}x_{37} + x_{36}x_{42} + x_{36}x_{43} + x_{36}x_{48} + x_{36}x_{49} + x_{36}x_{55} + x_{36}x_{57} + x_{36}x_{58} + x_{36}x_{59} + x_{36}x_{60} + x_{36}x_{62} + x_{36}x_{64} + x_{37}x_{39} + x_{37}x_{41} + x_{37}x_{42} + x_{37}x_{44} + x_{37}x_{48} + x_{37}x_{51} + x_{37}x_{53} + x_{37}x_{56} + x_{37}x_{57} + x_{37}x_{58} + x_{37}x_{64} + x_{38}x_{39} + x_{38}x_{43} + x_{38}x_{44} + x_{38}x_{46} + x_{38}x_{50} + x_{38}x_{51} + x_{38}x_{54} + x_{38}x_{56} + x_{38}x_{63} + x_{39}x_{40} + x_{39}x_{41} + x_{39}x_{46} + x_{39}x_{48} + x_{39}x_{49} + x_{39}x_{53} + x_{39}x_{54} + x_{39}x_{56} + x_{39}x_{61} + x_{39}x_{62} + x_{40}x_{42} + x_{40}x_{44} + x_{40}x_{45} + x_{40}x_{47} + x_{40}x_{50} + x_{40}x_{51} + x_{40}x_{52} + x_{40}x_{54} + x_{40}x_{56} + x_{40}x_{59} + x_{40}x_{61} + x_{40}x_{62} + x_{41}x_{42} + x_{41}x_{43} + x_{41}x_{48} + x_{41}x_{51} + x_{41}x_{54} + x_{41}x_{55} + x_{41}x_{56} + x_{41}x_{57} + x_{41}x_{59} + x_{41}x_{60} + x_{41}x_{61} + x_{41}x_{62} + x_{41}x_{63} + x_{41}x_{64} + x_{42}x_{43} + x_{42}x_{44} + x_{42}x_{45} + x_{42}x_{48} + x_{42}x_{49} + x_{42}x_{50} + x_{42}x_{54} + x_{42}x_{58} + x_{42}x_{60} + x_{42}x_{61} + x_{42}x_{64} + x_{43}x_{46} + x_{43}x_{48} + x_{43}x_{49} + x_{43}x_{52} + x_{43}x_{53} + x_{43}x_{54} + x_{43}x_{55} + x_{43}x_{58} + x_{43}x_{59} + x_{43}x_{61} + x_{44}x_{46} + x_{44}x_{47} + x_{44}x_{50} + x_{44}x_{52} + x_{44}x_{56} + x_{44}x_{57} + x_{44}x_{58} + x_{44}x_{59} + x_{44}x_{60} + x_{44}x_{63} + x_{44}x_{64} + x_{45}x_{47} + x_{45}x_{51} + x_{45}x_{53} + x_{45}x_{56} + x_{45}x_{60} + x_{45}x_{61} + x_{45}x_{63} + x_{45}x_{64} + x_{46}x_{47} + x_{46}x_{48} + x_{46}x_{49} + x_{46}x_{51} + x_{46}x_{52} + x_{46}x_{55} + x_{46}x_{56} + x_{46}x_{57} + x_{46}x_{58} + x_{46}x_{62} + x_{46}x_{63} + x_{46}x_{64} + x_{47}x_{48} + x_{47}x_{51} + x_{47}x_{54} + x_{47}x_{55} + x_{47}x_{57} + x_{47}x_{58} + x_{47}x_{60} + x_{47}x_{61} + x_{47}x_{63} + x_{47}x_{64} + x_{48}x_{54} + x_{48}x_{57} + x_{48}x_{58} + x_{48}x_{59} + x_{48}x_{60} + x_{48}x_{62} + x_{48}x_{63} + x_{49}x_{50} + x_{49}x_{51} + x_{49}x_{54} + x_{49}x_{57} + x_{49}x_{61} + x_{50}x_{51} + x_{50}x_{53} + x_{50}x_{55} + x_{50}x_{56} + x_{50}x_{61} + x_{50}x_{62} + x_{51}x_{52} + x_{51}x_{56} + x_{51}x_{57} + x_{51}x_{60} + x_{51}x_{61} + x_{51}x_{62} + x_{51}x_{64} + x_{52}x_{58} + x_{52}x_{59} + x_{52}x_{60} + x_{52}x_{64} + x_{53}x_{55} + x_{53}x_{57} + x_{53}x_{58} + x_{53}x_{59} + x_{53}x_{60} + x_{53}x_{61} + x_{53}x_{62} + x_{53}x_{63} + x_{53}x_{64} + x_{54}x_{56} + x_{54}x_{58} + x_{54}x_{59} + x_{54}x_{61} + x_{54}x_{63} + x_{54}x_{64} + x_{55}x_{58} + x_{55}x_{59} + x_{55}x_{63} + x_{56}x_{58} + x_{56}x_{59} + x_{56}x_{61} + x_{56}x_{63} + x_{56}x_{64} + x_{57}x_{58} + x_{57}x_{59} + x_{57}x_{62} + x_{57}x_{64} + x_{58}x_{59} + x_{58}x_{61} + x_{58}x_{63} + x_{59}x_{60} + x_{61}x_{62} + x_{61}x_{64} + x_{62}x_{63} + x_{5} + x_{8} + x_{9} + x_{11} + x_{12} + x_{13} + x_{21} + x_{22} + x_{24} + x_{25} + x_{26} + x_{27} + x_{30} + x_{31} + x_{32} + x_{33} + x_{34} + x_{38} + x_{39} + x_{41} + x_{42} + x_{46} + x_{48} + x_{50} + x_{54} + x_{55} + x_{56} + x_{57} + x_{59} + x_{60} + x_{61} + x_{62} + x_{63} + 1$

$y_{12} = x_{1}x_{2} + x_{1}x_{3} + x_{1}x_{5} + x_{1}x_{6} + x_{1}x_{11} + x_{1}x_{13} + x_{1}x_{14} + x_{1}x_{21} + x_{1}x_{22} + x_{1}x_{26} + x_{1}x_{27} + x_{1}x_{28} + x_{1}x_{29} + x_{1}x_{30} + x_{1}x_{33} + x_{1}x_{34} + x_{1}x_{35} + x_{1}x_{36} + x_{1}x_{38} + x_{1}x_{40} + x_{1}x_{41} + x_{1}x_{42} + x_{1}x_{43} + x_{1}x_{46} + x_{1}x_{47} + x_{1}x_{48} + x_{1}x_{50} + x_{1}x_{55} + x_{1}x_{56} + x_{1}x_{60} + x_{1}x_{61} + x_{1}x_{62} + x_{1}x_{64} + x_{2}x_{3} + x_{2}x_{4} + x_{2}x_{5} + x_{2}x_{7} + x_{2}x_{11} + x_{2}x_{12} + x_{2}x_{14} + x_{2}x_{15} + x_{2}x_{16} + x_{2}x_{17} + x_{2}x_{19} + x_{2}x_{21} + x_{2}x_{22} + x_{2}x_{23} + x_{2}x_{25} + x_{2}x_{28} + x_{2}x_{29} + x_{2}x_{34} + x_{2}x_{35} + x_{2}x_{36} + x_{2}x_{38} + x_{2}x_{39} + x_{2}x_{40} + x_{2}x_{42} + x_{2}x_{44} + x_{2}x_{50} + x_{2}x_{53} + x_{2}x_{54} + x_{2}x_{55} + x_{2}x_{56} + x_{2}x_{59} + x_{2}x_{63} + x_{3}x_{5} + x_{3}x_{8} + x_{3}x_{9} + x_{3}x_{10} + x_{3}x_{11} + x_{3}x_{17} + x_{3}x_{20} + x_{3}x_{22} + x_{3}x_{23} + x_{3}x_{28} + x_{3}x_{30} + x_{3}x_{31} + x_{3}x_{37} + x_{3}x_{39} + x_{3}x_{40} + x_{3}x_{44} + x_{3}x_{47} + x_{3}x_{48} + x_{3}x_{49} + x_{3}x_{50} + x_{3}x_{52} + x_{3}x_{53} + x_{3}x_{54} + x_{3}x_{55} + x_{4}x_{6} + x_{4}x_{8} + x_{4}x_{9} + x_{4}x_{13} + x_{4}x_{15} + x_{4}x_{16} + x_{4}x_{17} + x_{4}x_{18} + x_{4}x_{19} + x_{4}x_{20} + x_{4}x_{21} + x_{4}x_{24} + x_{4}x_{25} + x_{4}x_{27} + x_{4}x_{33} + x_{4}x_{35} + x_{4}x_{36} + x_{4}x_{38} + x_{4}x_{39} + x_{4}x_{40} + x_{4}x_{42} + x_{4}x_{43} + x_{4}x_{44} + x_{4}x_{48} + x_{4}x_{49} + x_{4}x_{54} + x_{4}x_{55} + x_{4}x_{56} + x_{4}x_{57} + x_{4}x_{58} + x_{4}x_{60} + x_{4}x_{61} + x_{4}x_{62} + x_{4}x_{64} + x_{5}x_{6} + x_{5}x_{7} + x_{5}x_{8} + x_{5}x_{11} + x_{5}x_{14} + x_{5}x_{17} + x_{5}x_{21} + x_{5}x_{23} + x_{5}x_{24} + x_{5}x_{27} + x_{5}x_{28} + x_{5}x_{31} + x_{5}x_{32} + x_{5}x_{35} + x_{5}x_{36} + x_{5}x_{40} + x_{5}x_{41} + x_{5}x_{44} + x_{5}x_{48} + x_{5}x_{49} + x_{5}x_{50} + x_{5}x_{51} + x_{5}x_{53} + x_{5}x_{54} + x_{5}x_{58} + x_{5}x_{59} + x_{5}x_{61} + x_{5}x_{62} + x_{6}x_{7} + x_{6}x_{9} + x_{6}x_{10} + x_{6}x_{11} + x_{6}x_{14} + x_{6}x_{15} + x_{6}x_{17} + x_{6}x_{18} + x_{6}x_{23} + x_{6}x_{24} + x_{6}x_{26} + x_{6}x_{30} + x_{6}x_{31} + x_{6}x_{33} + x_{6}x_{36} + x_{6}x_{37} + x_{6}x_{38} + x_{6}x_{40} + x_{6}x_{41} + x_{6}x_{42} + x_{6}x_{43} + x_{6}x_{49} + x_{6}x_{51} + x_{6}x_{53} + x_{6}x_{55} + x_{6}x_{58} + x_{6}x_{59} + x_{6}x_{60} + x_{6}x_{61} + x_{6}x_{62} + x_{7}x_{9} + x_{7}x_{12} + x_{7}x_{13} + x_{7}x_{14} + x_{7}x_{15} + x_{7}x_{16} + x_{7}x_{17} + x_{7}x_{19} + x_{7}x_{21} + x_{7}x_{23} + x_{7}x_{24} + x_{7}x_{26} + x_{7}x_{34} + x_{7}x_{35} + x_{7}x_{36} + x_{7}x_{37} + x_{7}x_{38} + x_{7}x_{39} + x_{7}x_{40} + x_{7}x_{41} + x_{7}x_{43} + x_{7}x_{46} + x_{7}x_{47} + x_{7}x_{51} + x_{7}x_{56} + x_{7}x_{57} + x_{7}x_{60} + x_{7}x_{61} + x_{7}x_{62} + x_{7}x_{63} + x_{7}x_{64} + x_{8}x_{11} + x_{8}x_{14} + x_{8}x_{15} + x_{8}x_{16} + x_{8}x_{18} + x_{8}x_{21} + x_{8}x_{22} + x_{8}x_{23} + x_{8}x_{26} + x_{8}x_{27} + x_{8}x_{30} + x_{8}x_{31} + x_{8}x_{34} + x_{8}x_{37} + x_{8}x_{40} + x_{8}x_{42} + x_{8}x_{44} + x_{8}x_{46} + x_{8}x_{47} + x_{8}x_{52} + x_{8}x_{55} + x_{8}x_{60} + x_{9}x_{11} + x_{9}x_{12} + x_{9}x_{13} + x_{9}x_{14} + x_{9}x_{15} + x_{9}x_{22} + x_{9}x_{24} + x_{9}x_{25} + x_{9}x_{26} + x_{9}x_{28} + x_{9}x_{30} + x_{9}x_{31} + x_{9}x_{32} + x_{9}x_{34} + x_{9}x_{36} + x_{9}x_{37} + x_{9}x_{38} + x_{9}x_{40} + x_{9}x_{41} + x_{9}x_{47} + x_{9}x_{49} + x_{9}x_{54} + x_{9}x_{55} + x_{9}x_{56} + x_{9}x_{57} + x_{9}x_{58} + x_{9}x_{60} + x_{9}x_{62} + x_{9}x_{63} + x_{9}x_{64} + x_{10}x_{11} + x_{10}x_{12} + x_{10}x_{15} + x_{10}x_{17} + x_{10}x_{18} + x_{10}x_{23} + x_{10}x_{28} + x_{10}x_{29} + x_{10}x_{30} + x_{10}x_{32} + x_{10}x_{36} + x_{10}x_{37} + x_{10}x_{38} + x_{10}x_{39} + x_{10}x_{41} + x_{10}x_{42} + x_{10}x_{44} + x_{10}x_{45} + x_{10}x_{46} + x_{10}x_{48} + x_{10}x_{49} + x_{10}x_{50} + x_{10}x_{52} + x_{10}x_{53} + x_{10}x_{57} + x_{10}x_{59} + x_{10}x_{60} + x_{10}x_{61} + x_{10}x_{62} + x_{10}x_{64} + x_{11}x_{12} + x_{11}x_{15} + x_{11}x_{17} + x_{11}x_{21} + x_{11}x_{23} + x_{11}x_{25} + x_{11}x_{32} + x_{11}x_{33} + x_{11}x_{35} + x_{11}x_{36} + x_{11}x_{38} + x_{11}x_{41} + x_{11}x_{42} + x_{11}x_{46} + x_{11}x_{48} + x_{11}x_{49} + x_{11}x_{52} + x_{11}x_{53} + x_{11}x_{54} + x_{11}x_{56} + x_{11}x_{58} + x_{11}x_{62} + x_{11}x_{63} + x_{12}x_{13} + x_{12}x_{19} + x_{12}x_{21} + x_{12}x_{23} + x_{12}x_{27} + x_{12}x_{29} + x_{12}x_{37} + x_{12}x_{40} + x_{12}x_{42} + x_{12}x_{43} + x_{12}x_{44} + x_{12}x_{45} + x_{12}x_{46} + x_{12}x_{47} + x_{12}x_{48} + x_{12}x_{49} + x_{12}x_{50} + x_{12}x_{52} + x_{12}x_{54} + x_{12}x_{56} + x_{12}x_{60} + x_{12}x_{62} + x_{12}x_{63} + x_{13}x_{14} + x_{13}x_{15} + x_{13}x_{18} + x_{13}x_{19} + x_{13}x_{20} + x_{13}x_{23} + x_{13}x_{27} + x_{13}x_{28} + x_{13}x_{31} + x_{13}x_{34} + x_{13}x_{35} + x_{13}x_{36} + x_{13}x_{39} + x_{13}x_{45} + x_{13}x_{48} + x_{13}x_{49} + x_{13}x_{50} + x_{13}x_{53} + x_{13}x_{54} + x_{13}x_{55} + x_{13}x_{57} + x_{13}x_{59} + x_{13}x_{60} + x_{13}x_{62} + x_{13}x_{63} + x_{14}x_{18} + x_{14}x_{19} + x_{14}x_{20} + x_{14}x_{22} + x_{14}x_{25} + x_{14}x_{26} + x_{14}x_{27} + x_{14}x_{36} + x_{14}x_{37} + x_{14}x_{40} + x_{14}x_{41} + x_{14}x_{43} + x_{14}x_{48} + x_{14}x_{51} + x_{14}x_{54} + x_{14}x_{57} + x_{14}x_{58} + x_{14}x_{59} + x_{14}x_{60} + x_{14}x_{61} + x_{14}x_{62} + x_{14}x_{63} + x_{14}x_{64} + x_{15}x_{22} + x_{15}x_{23} + x_{15}x_{25} + x_{15}x_{27} + x_{15}x_{28} + x_{15}x_{30} + x_{15}x_{32} + x_{15}x_{33} + x_{15}x_{36} + x_{15}x_{37} + x_{15}x_{38} + x_{15}x_{39} + x_{15}x_{40} + x_{15}x_{42} + x_{15}x_{44} + x_{15}x_{46} + x_{15}x_{48} + x_{15}x_{49} + x_{15}x_{50} + x_{15}x_{51} + x_{15}x_{52} + x_{15}x_{58} + x_{15}x_{60} + x_{15}x_{63} + x_{16}x_{18} + x_{16}x_{25} + x_{16}x_{26} + x_{16}x_{27} + x_{16}x_{28} + x_{16}x_{29} + x_{16}x_{31} + x_{16}x_{33} + x_{16}x_{34} + x_{16}x_{35} + x_{16}x_{41} + x_{16}x_{42} + x_{16}x_{44} + x_{16}x_{50} + x_{16}x_{52} + x_{16}x_{53} + x_{16}x_{54} + x_{16}x_{55} + x_{16}x_{57} + x_{16}x_{58} + x_{16}x_{60} + x_{16}x_{61} + x_{16}x_{62} + x_{16}x_{63} + x_{16}x_{64} + x_{17}x_{18} + x_{17}x_{19} + x_{17}x_{20} + x_{17}x_{21} + x_{17}x_{23} + x_{17}x_{24} + x_{17}x_{25} + x_{17}x_{27} + x_{17}x_{29} + x_{17}x_{33} + x_{17}x_{36} + x_{17}x_{38} + x_{17}x_{39} + x_{17}x_{41} + x_{17}x_{44} + x_{17}x_{45} + x_{17}x_{46} + x_{17}x_{48} + x_{17}x_{49} + x_{17}x_{51} + x_{17}x_{55} + x_{17}x_{59} + x_{17}x_{60} + x_{17}x_{61} + x_{18}x_{19} + x_{18}x_{20} + x_{18}x_{21} + x_{18}x_{24} + x_{18}x_{25} + x_{18}x_{26} + x_{18}x_{29} + x_{18}x_{30} + x_{18}x_{32} + x_{18}x_{34} + x_{18}x_{35} + x_{18}x_{36} + x_{18}x_{37} + x_{18}x_{43} + x_{18}x_{45} + x_{18}x_{46} + x_{18}x_{49} + x_{18}x_{50} + x_{18}x_{53} + x_{18}x_{54} + x_{18}x_{55} + x_{18}x_{59} + x_{18}x_{60} + x_{18}x_{61} + x_{18}x_{64} + x_{19}x_{21} + x_{19}x_{22} + x_{19}x_{24} + x_{19}x_{27} + x_{19}x_{28} + x_{19}x_{33} + x_{19}x_{35} + x_{19}x_{36} + x_{19}x_{38} + x_{19}x_{40} + x_{19}x_{41} + x_{19}x_{43} + x_{19}x_{44} + x_{19}x_{46} + x_{19}x_{47} + x_{19}x_{48} + x_{19}x_{49} + x_{19}x_{51} + x_{19}x_{54} + x_{19}x_{58} + x_{19}x_{59} + x_{19}x_{60} + x_{19}x_{63} + x_{19}x_{64} + x_{20}x_{21} + x_{20}x_{23} + x_{20}x_{25} + x_{20}x_{29} + x_{20}x_{31} + x_{20}x_{33} + x_{20}x_{34} + x_{20}x_{38} + x_{20}x_{40} + x_{20}x_{41} + x_{20}x_{42} + x_{20}x_{43} + x_{20}x_{44} + x_{20}x_{45} + x_{20}x_{46} + x_{20}x_{49} + x_{20}x_{52} + x_{20}x_{55} + x_{20}x_{57} + x_{20}x_{59} + x_{20}x_{60} + x_{20}x_{62} + x_{20}x_{63} + x_{21}x_{26} + x_{21}x_{28} + x_{21}x_{29} + x_{21}x_{31} + x_{21}x_{33} + x_{21}x_{37} + x_{21}x_{38} + x_{21}x_{40} + x_{21}x_{43} + x_{21}x_{47} + x_{21}x_{49} + x_{21}x_{53} + x_{21}x_{54} + x_{21}x_{55} + x_{21}x_{58} + x_{21}x_{59} + x_{21}x_{60} + x_{21}x_{61} + x_{21}x_{62} + x_{21}x_{63} + x_{22}x_{24} + x_{22}x_{27} + x_{22}x_{30} + x_{22}x_{31} + x_{22}x_{32} + x_{22}x_{34} + x_{22}x_{36} + x_{22}x_{37} + x_{22}x_{39} + x_{22}x_{40} + x_{22}x_{41} + x_{22}x_{43} + x_{22}x_{44} + x_{22}x_{45} + x_{22}x_{46} + x_{22}x_{50} + x_{22}x_{51} + x_{22}x_{52} + x_{22}x_{60} + x_{22}x_{62} + x_{22}x_{63} + x_{23}x_{26} + x_{23}x_{27} + x_{23}x_{28} + x_{23}x_{32} + x_{23}x_{33} + x_{23}x_{37} + x_{23}x_{38} + x_{23}x_{46} + x_{23}x_{47} + x_{23}x_{48} + x_{23}x_{51} + x_{23}x_{52} + x_{23}x_{53} + x_{23}x_{54} + x_{23}x_{55} + x_{23}x_{58} + x_{23}x_{59} + x_{24}x_{27} + x_{24}x_{29} + x_{24}x_{32} + x_{24}x_{34} + x_{24}x_{36} + x_{24}x_{38} + x_{24}x_{42} + x_{24}x_{44} + x_{24}x_{46} + x_{24}x_{48} + x_{24}x_{51} + x_{24}x_{52} + x_{24}x_{55} + x_{24}x_{58} + x_{24}x_{59} + x_{24}x_{62} + x_{24}x_{63} + x_{24}x_{64} + x_{25}x_{26} + x_{25}x_{28} + x_{25}x_{33} + x_{25}x_{35} + x_{25}x_{43} + x_{25}x_{44} + x_{25}x_{45} + x_{25}x_{46} + x_{25}x_{48} + x_{25}x_{50} + x_{25}x_{51} + x_{25}x_{53} + x_{25}x_{59} + x_{25}x_{61} + x_{25}x_{62} + x_{25}x_{64} + x_{26}x_{30} + x_{26}x_{32} + x_{26}x_{34} + x_{26}x_{37} + x_{26}x_{42} + x_{26}x_{43} + x_{26}x_{44} + x_{26}x_{46} + x_{26}x_{47} + x_{26}x_{49} + x_{26}x_{50} + x_{26}x_{57} + x_{26}x_{58} + x_{26}x_{60} + x_{26}x_{61} + x_{26}x_{62} + x_{27}x_{28} + x_{27}x_{32} + x_{27}x_{37} + x_{27}x_{44} + x_{27}x_{46} + x_{27}x_{48} + x_{27}x_{49} + x_{27}x_{51} + x_{27}x_{52} + x_{27}x_{53} + x_{27}x_{54} + x_{27}x_{55} + x_{27}x_{58} + x_{27}x_{59} + x_{27}x_{63} + x_{27}x_{64} + x_{28}x_{30} + x_{28}x_{31} + x_{28}x_{32} + x_{28}x_{34} + x_{28}x_{37} + x_{28}x_{38} + x_{28}x_{39} + x_{28}x_{40} + x_{28}x_{41} + x_{28}x_{43} + x_{28}x_{44} + x_{28}x_{47} + x_{28}x_{49} + x_{28}x_{50} + x_{28}x_{56} + x_{28}x_{57} + x_{28}x_{60} + x_{28}x_{61} + x_{28}x_{63} + x_{28}x_{64} + x_{29}x_{33} + x_{29}x_{34} + x_{29}x_{35} + x_{29}x_{39} + x_{29}x_{40} + x_{29}x_{44} + x_{29}x_{45} + x_{29}x_{52} + x_{29}x_{54} + x_{29}x_{58} + x_{29}x_{59} + x_{29}x_{60} + x_{29}x_{61} + x_{29}x_{63} + x_{29}x_{64} + x_{30}x_{31} + x_{30}x_{36} + x_{30}x_{37} + x_{30}x_{42} + x_{30}x_{43} + x_{30}x_{44} + x_{30}x_{45} + x_{30}x_{46} + x_{30}x_{47} + x_{30}x_{49} + x_{30}x_{54} + x_{30}x_{55} + x_{30}x_{58} + x_{30}x_{62} + x_{31}x_{33} + x_{31}x_{34} + x_{31}x_{35} + x_{31}x_{36} + x_{31}x_{37} + x_{31}x_{38} + x_{31}x_{39} + x_{31}x_{41} + x_{31}x_{46} + x_{31}x_{47} + x_{31}x_{49} + x_{31}x_{51} + x_{31}x_{53} + x_{31}x_{56} + x_{31}x_{59} + x_{31}x_{62} + x_{31}x_{63} + x_{31}x_{64} + x_{32}x_{37} + x_{32}x_{39} + x_{32}x_{40} + x_{32}x_{43} + x_{32}x_{44} + x_{32}x_{47} + x_{32}x_{48} + x_{32}x_{51} + x_{32}x_{52} + x_{32}x_{54} + x_{32}x_{55} + x_{32}x_{56} + x_{32}x_{58} + x_{32}x_{59} + x_{32}x_{60} + x_{32}x_{62} + x_{32}x_{64} + x_{33}x_{34} + x_{33}x_{35} + x_{33}x_{38} + x_{33}x_{42} + x_{33}x_{43} + x_{33}x_{44} + x_{33}x_{45} + x_{33}x_{46} + x_{33}x_{51} + x_{33}x_{52} + x_{33}x_{54} + x_{33}x_{57} + x_{33}x_{58} + x_{33}x_{59} + x_{33}x_{61} + x_{33}x_{63} + x_{34}x_{35} + x_{34}x_{39} + x_{34}x_{40} + x_{34}x_{41} + x_{34}x_{46} + x_{34}x_{50} + x_{34}x_{52} + x_{34}x_{58} + x_{34}x_{59} + x_{34}x_{60} + x_{34}x_{63} + x_{34}x_{64} + x_{35}x_{37} + x_{35}x_{38} + x_{35}x_{40} + x_{35}x_{41} + x_{35}x_{42} + x_{35}x_{46} + x_{35}x_{47} + x_{35}x_{48} + x_{35}x_{49} + x_{35}x_{50} + x_{35}x_{53} + x_{35}x_{54} + x_{35}x_{56} + x_{35}x_{57} + x_{35}x_{59} + x_{35}x_{60} + x_{36}x_{37} + x_{36}x_{38} + x_{36}x_{40} + x_{36}x_{42} + x_{36}x_{43} + x_{36}x_{46} + x_{36}x_{48} + x_{36}x_{50} + x_{36}x_{52} + x_{36}x_{54} + x_{36}x_{55} + x_{36}x_{56} + x_{36}x_{57} + x_{36}x_{58} + x_{36}x_{60} + x_{37}x_{42} + x_{37}x_{44} + x_{37}x_{46} + x_{37}x_{49} + x_{37}x_{51} + x_{37}x_{52} + x_{37}x_{53} + x_{37}x_{54} + x_{37}x_{55} + x_{37}x_{56} + x_{37}x_{57} + x_{37}x_{59} + x_{37}x_{64} + x_{38}x_{39} + x_{38}x_{40} + x_{38}x_{44} + x_{38}x_{45} + x_{38}x_{47} + x_{38}x_{48} + x_{38}x_{50} + x_{38}x_{52} + x_{38}x_{53} + x_{38}x_{55} + x_{38}x_{58} + x_{38}x_{63} + x_{39}x_{43} + x_{39}x_{44} + x_{39}x_{46} + x_{39}x_{47} + x_{39}x_{48} + x_{39}x_{49} + x_{39}x_{51} + x_{39}x_{55} + x_{39}x_{57} + x_{39}x_{60} + x_{39}x_{61} + x_{39}x_{64} + x_{40}x_{42} + x_{40}x_{45} + x_{40}x_{46} + x_{40}x_{49} + x_{40}x_{50} + x_{40}x_{51} + x_{40}x_{54} + x_{40}x_{55} + x_{40}x_{58} + x_{40}x_{59} + x_{40}x_{63} + x_{40}x_{64} + x_{41}x_{42} + x_{41}x_{44} + x_{41}x_{46} + x_{41}x_{48} + x_{41}x_{49} + x_{41}x_{53} + x_{41}x_{56} + x_{41}x_{57} + x_{41}x_{58} + x_{41}x_{61} + x_{41}x_{62} + x_{41}x_{64} + x_{42}x_{45} + x_{42}x_{47} + x_{42}x_{49} + x_{42}x_{50} + x_{42}x_{51} + x_{42}x_{52} + x_{42}x_{55} + x_{42}x_{56} + x_{42}x_{57} + x_{42}x_{58} + x_{42}x_{64} + x_{43}x_{44} + x_{43}x_{45} + x_{43}x_{46} + x_{43}x_{51} + x_{43}x_{55} + x_{43}x_{56} + x_{43}x_{58} + x_{43}x_{61} + x_{43}x_{62} + x_{43}x_{63} + x_{43}x_{64} + x_{44}x_{45} + x_{44}x_{47} + x_{44}x_{48} + x_{44}x_{49} + x_{44}x_{51} + x_{44}x_{53} + x_{44}x_{55} + x_{44}x_{56} + x_{44}x_{59} + x_{44}x_{61} + x_{45}x_{46} + x_{45}x_{48} + x_{45}x_{51} + x_{45}x_{53} + x_{45}x_{54} + x_{45}x_{56} + x_{45}x_{57} + x_{45}x_{60} + x_{45}x_{64} + x_{46}x_{47} + x_{46}x_{48} + x_{46}x_{52} + x_{46}x_{57} + x_{46}x_{58} + x_{46}x_{59} + x_{46}x_{61} + x_{46}x_{64} + x_{47}x_{48} + x_{47}x_{51} + x_{47}x_{54} + x_{47}x_{55} + x_{47}x_{59} + x_{47}x_{61} + x_{47}x_{63} + x_{47}x_{64} + x_{48}x_{50} + x_{48}x_{51} + x_{48}x_{53} + x_{48}x_{54} + x_{48}x_{57} + x_{48}x_{58} + x_{48}x_{59} + x_{48}x_{62} + x_{48}x_{63} + x_{49}x_{50} + x_{49}x_{51} + x_{49}x_{54} + x_{49}x_{56} + x_{49}x_{60} + x_{49}x_{61} + x_{49}x_{64} + x_{50}x_{56} + x_{50}x_{58} + x_{50}x_{60} + x_{50}x_{62} + x_{50}x_{63} + x_{51}x_{52} + x_{51}x_{53} + x_{51}x_{54} + x_{51}x_{55} + x_{51}x_{56} + x_{51}x_{57} + x_{51}x_{58} + x_{51}x_{59} + x_{51}x_{61} + x_{51}x_{62} + x_{51}x_{64} + x_{52}x_{53} + x_{52}x_{54} + x_{52}x_{59} + x_{52}x_{60} + x_{52}x_{64} + x_{53}x_{54} + x_{53}x_{56} + x_{53}x_{57} + x_{53}x_{58} + x_{53}x_{59} + x_{53}x_{60} + x_{53}x_{62} + x_{53}x_{63} + x_{53}x_{64} + x_{54}x_{57} + x_{54}x_{59} + x_{54}x_{61} + x_{54}x_{63} + x_{55}x_{57} + x_{55}x_{58} + x_{55}x_{59} + x_{55}x_{61} + x_{55}x_{62} + x_{55}x_{64} + x_{56}x_{60} + x_{57}x_{58} + x_{58}x_{59} + x_{58}x_{60} + x_{58}x_{62} + x_{58}x_{64} + x_{59}x_{60} + x_{59}x_{62} + x_{59}x_{63} + x_{59}x_{64} + x_{60}x_{62} + x_{60}x_{63} + x_{60}x_{64} + x_{62}x_{63} + x_{63}x_{64} + x_{1} + x_{2} + x_{3} + x_{4} + x_{5} + x_{6} + x_{7} + x_{9} + x_{10} + x_{11} + x_{12} + x_{13} + x_{15} + x_{17} + x_{19} + x_{21} + x_{24} + x_{25} + x_{27} + x_{29} + x_{30} + x_{31} + x_{33} + x_{36} + x_{39} + x_{40} + x_{41} + x_{42} + x_{43} + x_{44} + x_{45} + x_{46} + x_{50} + x_{52} + x_{53} + x_{55} + x_{58} + x_{59} + x_{61} + x_{64}$

$y_{13} = x_{1}x_{2} + x_{1}x_{3} + x_{1}x_{4} + x_{1}x_{7} + x_{1}x_{8} + x_{1}x_{9} + x_{1}x_{10} + x_{1}x_{12} + x_{1}x_{13} + x_{1}x_{14} + x_{1}x_{16} + x_{1}x_{22} + x_{1}x_{23} + x_{1}x_{24} + x_{1}x_{25} + x_{1}x_{26} + x_{1}x_{29} + x_{1}x_{30} + x_{1}x_{35} + x_{1}x_{36} + x_{1}x_{37} + x_{1}x_{39} + x_{1}x_{42} + x_{1}x_{43} + x_{1}x_{44} + x_{1}x_{45} + x_{1}x_{47} + x_{1}x_{48} + x_{1}x_{49} + x_{1}x_{53} + x_{1}x_{55} + x_{1}x_{56} + x_{1}x_{57} + x_{1}x_{59} + x_{1}x_{63} + x_{2}x_{4} + x_{2}x_{5} + x_{2}x_{7} + x_{2}x_{8} + x_{2}x_{10} + x_{2}x_{11} + x_{2}x_{14} + x_{2}x_{16} + x_{2}x_{18} + x_{2}x_{22} + x_{2}x_{23} + x_{2}x_{24} + x_{2}x_{26} + x_{2}x_{27} + x_{2}x_{29} + x_{2}x_{30} + x_{2}x_{31} + x_{2}x_{32} + x_{2}x_{33} + x_{2}x_{34} + x_{2}x_{36} + x_{2}x_{37} + x_{2}x_{41} + x_{2}x_{45} + x_{2}x_{46} + x_{2}x_{49} + x_{2}x_{50} + x_{2}x_{51} + x_{2}x_{52} + x_{2}x_{53} + x_{2}x_{54} + x_{2}x_{57} + x_{2}x_{59} + x_{2}x_{61} + x_{2}x_{64} + x_{3}x_{5} + x_{3}x_{6} + x_{3}x_{7} + x_{3}x_{8} + x_{3}x_{9} + x_{3}x_{13} + x_{3}x_{16} + x_{3}x_{17} + x_{3}x_{18} + x_{3}x_{20} + x_{3}x_{21} + x_{3}x_{24} + x_{3}x_{26} + x_{3}x_{27} + x_{3}x_{30} + x_{3}x_{31} + x_{3}x_{34} + x_{3}x_{35} + x_{3}x_{36} + x_{3}x_{37} + x_{3}x_{39} + x_{3}x_{40} + x_{3}x_{45} + x_{3}x_{47} + x_{3}x_{48} + x_{3}x_{49} + x_{3}x_{51} + x_{3}x_{55} + x_{3}x_{57} + x_{3}x_{58} + x_{3}x_{59} + x_{3}x_{61} + x_{3}x_{64} + x_{4}x_{5} + x_{4}x_{8} + x_{4}x_{10} + x_{4}x_{11} + x_{4}x_{12} + x_{4}x_{13} + x_{4}x_{14} + x_{4}x_{17} + x_{4}x_{19} + x_{4}x_{21} + x_{4}x_{24} + x_{4}x_{26} + x_{4}x_{30} + x_{4}x_{32} + x_{4}x_{33} + x_{4}x_{35} + x_{4}x_{36} + x_{4}x_{40} + x_{4}x_{42} + x_{4}x_{44} + x_{4}x_{45} + x_{4}x_{49} + x_{4}x_{50} + x_{4}x_{52} + x_{4}x_{53} + x_{4}x_{54} + x_{4}x_{56} + x_{4}x_{60} + x_{4}x_{62} + x_{4}x_{63} + x_{5}x_{6} + x_{5}x_{7} + x_{5}x_{8} + x_{5}x_{9} + x_{5}x_{10} + x_{5}x_{11} + x_{5}x_{12} + x_{5}x_{15} + x_{5}x_{17} + x_{5}x_{18} + x_{5}x_{19} + x_{5}x_{20} + x_{5}x_{21} + x_{5}x_{28} + x_{5}x_{31} + x_{5}x_{32} + x_{5}x_{34} + x_{5}x_{35} + x_{5}x_{38} + x_{5}x_{39} + x_{5}x_{40} + x_{5}x_{41} + x_{5}x_{43} + x_{5}x_{46} + x_{5}x_{47} + x_{5}x_{49} + x_{5}x_{52} + x_{5}x_{54} + x_{5}x_{55} + x_{5}x_{56} + x_{5}x_{58} + x_{5}x_{60} + x_{5}x_{62} + x_{5}x_{63} + x_{5}x_{64} + x_{6}x_{7} + x_{6}x_{8} + x_{6}x_{10} + x_{6}x_{11} + x_{6}x_{16} + x_{6}x_{17} + x_{6}x_{18} + x_{6}x_{20} + x_{6}x_{22} + x_{6}x_{24} + x_{6}x_{26} + x_{6}x_{27} + x_{6}x_{28} + x_{6}x_{32} + x_{6}x_{33} + x_{6}x_{34} + x_{6}x_{36} + x_{6}x_{38} + x_{6}x_{39} + x_{6}x_{46} + x_{6}x_{47} + x_{6}x_{48} + x_{6}x_{50} + x_{6}x_{51} + x_{6}x_{57} + x_{6}x_{58} + x_{6}x_{60} + x_{6}x_{61} + x_{6}x_{62} + x_{6}x_{63} + x_{6}x_{64} + x_{7}x_{9} + x_{7}x_{10} + x_{7}x_{11} + x_{7}x_{13} + x_{7}x_{16} + x_{7}x_{20} + x_{7}x_{21} + x_{7}x_{22} + x_{7}x_{23} + x_{7}x_{24} + x_{7}x_{27} + x_{7}x_{29} + x_{7}x_{30} + x_{7}x_{31} + x_{7}x_{33} + x_{7}x_{38} + x_{7}x_{42} + x_{7}x_{43} + x_{7}x_{46} + x_{7}x_{47} + x_{7}x_{48} + x_{7}x_{50} + x_{7}x_{53} + x_{7}x_{54} + x_{7}x_{56} + x_{7}x_{58} + x_{7}x_{59} + x_{7}x_{60} + x_{7}x_{61} + x_{7}x_{64} + x_{8}x_{9} + x_{8}x_{10} + x_{8}x_{11} + x_{8}x_{12} + x_{8}x_{13} + x_{8}x_{17} + x_{8}x_{19} + x_{8}x_{27} + x_{8}x_{28} + x_{8}x_{30} + x_{8}x_{31} + x_{8}x_{32} + x_{8}x_{35} + x_{8}x_{41} + x_{8}x_{45} + x_{8}x_{47} + x_{8}x_{51} + x_{8}x_{55} + x_{8}x_{56} + x_{8}x_{58} + x_{8}x_{59} + x_{8}x_{63} + x_{9}x_{11} + x_{9}x_{14} + x_{9}x_{16} + x_{9}x_{20} + x_{9}x_{23} + x_{9}x_{25} + x_{9}x_{30} + x_{9}x_{33} + x_{9}x_{35} + x_{9}x_{36} + x_{9}x_{39} + x_{9}x_{42} + x_{9}x_{43} + x_{9}x_{45} + x_{9}x_{46} + x_{9}x_{47} + x_{9}x_{54} + x_{9}x_{55} + x_{9}x_{57} + x_{9}x_{59} + x_{10}x_{11} + x_{10}x_{13} + x_{10}x_{17} + x_{10}x_{18} + x_{10}x_{22} + x_{10}x_{23} + x_{10}x_{28} + x_{10}x_{29} + x_{10}x_{31} + x_{10}x_{34} + x_{10}x_{35} + x_{10}x_{36} + x_{10}x_{37} + x_{10}x_{38} + x_{10}x_{39} + x_{10}x_{40} + x_{10}x_{41} + x_{10}x_{49} + x_{10}x_{52} + x_{10}x_{53} + x_{10}x_{57} + x_{10}x_{58} + x_{10}x_{59} + x_{10}x_{60} + x_{10}x_{63} + x_{11}x_{12} + x_{11}x_{13} + x_{11}x_{14} + x_{11}x_{16} + x_{11}x_{19} + x_{11}x_{20} + x_{11}x_{21} + x_{11}x_{22} + x_{11}x_{23} + x_{11}x_{24} + x_{11}x_{26} + x_{11}x_{29} + x_{11}x_{31} + x_{11}x_{32} + x_{11}x_{33} + x_{11}x_{35} + x_{11}x_{36} + x_{11}x_{37} + x_{11}x_{38} + x_{11}x_{39} + x_{11}x_{42} + x_{11}x_{44} + x_{11}x_{47} + x_{11}x_{50} + x_{11}x_{52} + x_{11}x_{53} + x_{11}x_{55} + x_{11}x_{56} + x_{11}x_{57} + x_{11}x_{58} + x_{11}x_{62} + x_{12}x_{13} + x_{12}x_{14} + x_{12}x_{18} + x_{12}x_{20} + x_{12}x_{23} + x_{12}x_{24} + x_{12}x_{26} + x_{12}x_{28} + x_{12}x_{29} + x_{12}x_{32} + x_{12}x_{33} + x_{12}x_{36} + x_{12}x_{37} + x_{12}x_{40} + x_{12}x_{43} + x_{12}x_{45} + x_{12}x_{47} + x_{12}x_{48} + x_{12}x_{51} + x_{12}x_{53} + x_{12}x_{54} + x_{12}x_{55} + x_{12}x_{56} + x_{12}x_{58} + x_{12}x_{60} + x_{12}x_{64} + x_{13}x_{16} + x_{13}x_{18} + x_{13}x_{20} + x_{13}x_{21} + x_{13}x_{22} + x_{13}x_{23} + x_{13}x_{24} + x_{13}x_{26} + x_{13}x_{27} + x_{13}x_{29} + x_{13}x_{30} + x_{13}x_{35} + x_{13}x_{36} + x_{13}x_{37} + x_{13}x_{43} + x_{13}x_{44} + x_{13}x_{45} + x_{13}x_{46} + x_{13}x_{48} + x_{13}x_{50} + x_{13}x_{51} + x_{13}x_{52} + x_{13}x_{53} + x_{13}x_{54} + x_{13}x_{56} + x_{13}x_{57} + x_{13}x_{59} + x_{13}x_{60} + x_{13}x_{62} + x_{13}x_{63} + x_{13}x_{64} + x_{14}x_{15} + x_{14}x_{16} + x_{14}x_{19} + x_{14}x_{21} + x_{14}x_{22} + x_{14}x_{25} + x_{14}x_{27} + x_{14}x_{28} + x_{14}x_{29} + x_{14}x_{32} + x_{14}x_{33} + x_{14}x_{34} + x_{14}x_{35} + x_{14}x_{41} + x_{14}x_{43} + x_{14}x_{45} + x_{14}x_{47} + x_{14}x_{52} + x_{14}x_{57} + x_{14}x_{60} + x_{14}x_{61} + x_{14}x_{64} + x_{15}x_{16} + x_{15}x_{17} + x_{15}x_{19} + x_{15}x_{20} + x_{15}x_{21} + x_{15}x_{22} + x_{15}x_{23} + x_{15}x_{24} + x_{15}x_{25} + x_{15}x_{27} + x_{15}x_{28} + x_{15}x_{30} + x_{15}x_{32} + x_{15}x_{35} + x_{15}x_{37} + x_{15}x_{39} + x_{15}x_{41} + x_{15}x_{42} + x_{15}x_{46} + x_{15}x_{48} + x_{15}x_{53} + x_{15}x_{55} + x_{15}x_{56} + x_{15}x_{59} + x_{15}x_{60} + x_{15}x_{62} + x_{15}x_{63} + x_{15}x_{64} + x_{16}x_{17} + x_{16}x_{20} + x_{16}x_{21} + x_{16}x_{23} + x_{16}x_{24} + x_{16}x_{25} + x_{16}x_{26} + x_{16}x_{27} + x_{16}x_{31} + x_{16}x_{37} + x_{16}x_{39} + x_{16}x_{41} + x_{16}x_{43} + x_{16}x_{44} + x_{16}x_{45} + x_{16}x_{46} + x_{16}x_{48} + x_{16}x_{50} + x_{16}x_{51} + x_{16}x_{53} + x_{16}x_{54} + x_{16}x_{56} + x_{16}x_{58} + x_{16}x_{60} + x_{16}x_{61} + x_{16}x_{64} + x_{17}x_{18} + x_{17}x_{26} + x_{17}x_{29} + x_{17}x_{31} + x_{17}x_{33} + x_{17}x_{35} + x_{17}x_{36} + x_{17}x_{37} + x_{17}x_{39} + x_{17}x_{45} + x_{17}x_{46} + x_{17}x_{47} + x_{17}x_{48} + x_{17}x_{57} + x_{17}x_{58} + x_{17}x_{60} + x_{17}x_{61} + x_{17}x_{63} + x_{17}x_{64} + x_{18}x_{19} + x_{18}x_{20} + x_{18}x_{22} + x_{18}x_{23} + x_{18}x_{25} + x_{18}x_{27} + x_{18}x_{29} + x_{18}x_{33} + x_{18}x_{35} + x_{18}x_{37} + x_{18}x_{39} + x_{18}x_{40} + x_{18}x_{41} + x_{18}x_{42} + x_{18}x_{44} + x_{18}x_{46} + x_{18}x_{50} + x_{18}x_{51} + x_{18}x_{52} + x_{18}x_{55} + x_{18}x_{56} + x_{18}x_{57} + x_{18}x_{58} + x_{18}x_{59} + x_{19}x_{20} + x_{19}x_{22} + x_{19}x_{24} + x_{19}x_{25} + x_{19}x_{28} + x_{19}x_{29} + x_{19}x_{30} + x_{19}x_{33} + x_{19}x_{34} + x_{19}x_{39} + x_{19}x_{40} + x_{19}x_{41} + x_{19}x_{42} + x_{19}x_{43} + x_{19}x_{44} + x_{19}x_{45} + x_{19}x_{46} + x_{19}x_{47} + x_{19}x_{48} + x_{19}x_{50} + x_{19}x_{52} + x_{19}x_{53} + x_{19}x_{54} + x_{19}x_{55} + x_{19}x_{58} + x_{19}x_{60} + x_{19}x_{61} + x_{19}x_{62} + x_{20}x_{25} + x_{20}x_{27} + x_{20}x_{28} + x_{20}x_{30} + x_{20}x_{32} + x_{20}x_{33} + x_{20}x_{37} + x_{20}x_{40} + x_{20}x_{45} + x_{20}x_{47} + x_{20}x_{48} + x_{20}x_{53} + x_{20}x_{54} + x_{20}x_{57} + x_{20}x_{60} + x_{20}x_{62} + x_{20}x_{63} + x_{21}x_{24} + x_{21}x_{29} + x_{21}x_{30} + x_{21}x_{32} + x_{21}x_{33} + x_{21}x_{34} + x_{21}x_{39} + x_{21}x_{40} + x_{21}x_{41} + x_{21}x_{42} + x_{21}x_{43} + x_{21}x_{45} + x_{21}x_{46} + x_{21}x_{48} + x_{21}x_{51} + x_{21}x_{52} + x_{21}x_{54} + x_{21}x_{57} + x_{21}x_{60} + x_{21}x_{62} + x_{22}x_{25} + x_{22}x_{26} + x_{22}x_{27} + x_{22}x_{32} + x_{22}x_{33} + x_{22}x_{34} + x_{22}x_{35} + x_{22}x_{39} + x_{22}x_{40} + x_{22}x_{41} + x_{22}x_{42} + x_{22}x_{43} + x_{22}x_{44} + x_{22}x_{47} + x_{22}x_{49} + x_{22}x_{51} + x_{22}x_{52} + x_{22}x_{54} + x_{22}x_{55} + x_{22}x_{56} + x_{22}x_{57} + x_{22}x_{58} + x_{22}x_{59} + x_{22}x_{61} + x_{23}x_{25} + x_{23}x_{28} + x_{23}x_{29} + x_{23}x_{31} + x_{23}x_{35} + x_{23}x_{40} + x_{23}x_{42} + x_{23}x_{44} + x_{23}x_{46} + x_{23}x_{47} + x_{23}x_{48} + x_{23}x_{49} + x_{23}x_{51} + x_{23}x_{53} + x_{23}x_{56} + x_{23}x_{60} + x_{23}x_{61} + x_{23}x_{63} + x_{24}x_{25} + x_{24}x_{26} + x_{24}x_{31} + x_{24}x_{33} + x_{24}x_{37} + x_{24}x_{40} + x_{24}x_{44} + x_{24}x_{45} + x_{24}x_{46} + x_{24}x_{49} + x_{24}x_{51} + x_{24}x_{53} + x_{24}x_{54} + x_{24}x_{56} + x_{24}x_{57} + x_{24}x_{58} + x_{24}x_{64} + x_{25}x_{26} + x_{25}x_{28} + x_{25}x_{31} + x_{25}x_{33} + x_{25}x_{36} + x_{25}x_{38} + x_{25}x_{42} + x_{25}x_{45} + x_{25}x_{47} + x_{25}x_{49} + x_{25}x_{51} + x_{25}x_{53} + x_{25}x_{55} + x_{25}x_{56} + x_{25}x_{57} + x_{25}x_{59} + x_{25}x_{60} + x_{25}x_{64} + x_{26}x_{28} + x_{26}x_{30} + x_{26}x_{31} + x_{26}x_{33} + x_{26}x_{35} + x_{26}x_{38} + x_{26}x_{39} + x_{26}x_{42} + x_{26}x_{44} + x_{26}x_{45} + x_{26}x_{46} + x_{26}x_{52} + x_{26}x_{53} + x_{26}x_{56} + x_{26}x_{58} + x_{26}x_{59} + x_{26}x_{64} + x_{27}x_{29} + x_{27}x_{30} + x_{27}x_{31} + x_{27}x_{35} + x_{27}x_{39} + x_{27}x_{40} + x_{27}x_{44} + x_{27}x_{45} + x_{27}x_{47} + x_{27}x_{51} + x_{27}x_{53} + x_{27}x_{56} + x_{27}x_{57} + x_{27}x_{58} + x_{27}x_{59} + x_{27}x_{60} + x_{27}x_{61} + x_{27}x_{62} + x_{27}x_{63} + x_{28}x_{34} + x_{28}x_{36} + x_{28}x_{39} + x_{28}x_{40} + x_{28}x_{43} + x_{28}x_{47} + x_{28}x_{48} + x_{28}x_{51} + x_{28}x_{56} + x_{28}x_{61} + x_{28}x_{64} + x_{29}x_{30} + x_{29}x_{31} + x_{29}x_{32} + x_{29}x_{33} + x_{29}x_{38} + x_{29}x_{39} + x_{29}x_{41} + x_{29}x_{47} + x_{29}x_{48} + x_{29}x_{50} + x_{29}x_{52} + x_{29}x_{53} + x_{29}x_{54} + x_{29}x_{56} + x_{29}x_{58} + x_{29}x_{60} + x_{29}x_{62} + x_{29}x_{64} + x_{30}x_{31} + x_{30}x_{32} + x_{30}x_{33} + x_{30}x_{35} + x_{30}x_{36} + x_{30}x_{38} + x_{30}x_{39} + x_{30}x_{40} + x_{30}x_{41} + x_{30}x_{42} + x_{30}x_{43} + x_{30}x_{45} + x_{30}x_{47} + x_{30}x_{49} + x_{30}x_{53} + x_{30}x_{56} + x_{30}x_{63} + x_{30}x_{64} + x_{31}x_{33} + x_{31}x_{34} + x_{31}x_{36} + x_{31}x_{39} + x_{31}x_{41} + x_{31}x_{43} + x_{31}x_{44} + x_{31}x_{45} + x_{31}x_{46} + x_{31}x_{47} + x_{31}x_{48} + x_{31}x_{49} + x_{31}x_{51} + x_{31}x_{54} + x_{31}x_{60} + x_{31}x_{61} + x_{31}x_{62} + x_{32}x_{35} + x_{32}x_{37} + x_{32}x_{39} + x_{32}x_{40} + x_{32}x_{41} + x_{32}x_{43} + x_{32}x_{46} + x_{32}x_{47} + x_{32}x_{49} + x_{32}x_{51} + x_{32}x_{52} + x_{32}x_{53} + x_{32}x_{56} + x_{32}x_{57} + x_{32}x_{58} + x_{32}x_{59} + x_{32}x_{60} + x_{33}x_{34} + x_{33}x_{35} + x_{33}x_{43} + x_{33}x_{44} + x_{33}x_{45} + x_{33}x_{46} + x_{33}x_{51} + x_{33}x_{55} + x_{33}x_{56} + x_{33}x_{57} + x_{33}x_{60} + x_{33}x_{61} + x_{33}x_{63} + x_{34}x_{35} + x_{34}x_{36} + x_{34}x_{37} + x_{34}x_{38} + x_{34}x_{40} + x_{34}x_{41} + x_{34}x_{43} + x_{34}x_{44} + x_{34}x_{52} + x_{34}x_{56} + x_{34}x_{59} + x_{34}x_{63} + x_{35}x_{37} + x_{35}x_{39} + x_{35}x_{40} + x_{35}x_{51} + x_{35}x_{52} + x_{35}x_{53} + x_{35}x_{58} + x_{35}x_{62} + x_{36}x_{37} + x_{36}x_{38} + x_{36}x_{39} + x_{36}x_{41} + x_{36}x_{42} + x_{36}x_{43} + x_{36}x_{47} + x_{36}x_{48} + x_{36}x_{49} + x_{36}x_{50} + x_{36}x_{52} + x_{36}x_{53} + x_{36}x_{54} + x_{36}x_{55} + x_{36}x_{58} + x_{36}x_{59} + x_{37}x_{41} + x_{37}x_{42} + x_{37}x_{45} + x_{37}x_{48} + x_{37}x_{50} + x_{37}x_{51} + x_{37}x_{52} + x_{37}x_{53} + x_{37}x_{55} + x_{37}x_{62} + x_{37}x_{64} + x_{38}x_{39} + x_{38}x_{40} + x_{38}x_{41} + x_{38}x_{43} + x_{38}x_{46} + x_{38}x_{48} + x_{38}x_{49} + x_{38}x_{52} + x_{38}x_{54} + x_{38}x_{57} + x_{38}x_{58} + x_{38}x_{61} + x_{38}x_{62} + x_{38}x_{63} + x_{38}x_{64} + x_{39}x_{40} + x_{39}x_{42} + x_{39}x_{46} + x_{39}x_{47} + x_{39}x_{50} + x_{39}x_{52} + x_{39}x_{55} + x_{39}x_{57} + x_{39}x_{59} + x_{39}x_{60} + x_{39}x_{62} + x_{39}x_{63} + x_{40}x_{42} + x_{40}x_{43} + x_{40}x_{45} + x_{40}x_{46} + x_{40}x_{52} + x_{40}x_{53} + x_{40}x_{54} + x_{40}x_{55} + x_{40}x_{57} + x_{40}x_{59} + x_{40}x_{60} + x_{40}x_{62} + x_{40}x_{63} + x_{40}x_{64} + x_{41}x_{42} + x_{41}x_{44} + x_{41}x_{47} + x_{41}x_{49} + x_{41}x_{51} + x_{41}x_{52} + x_{41}x_{53} + x_{41}x_{57} + x_{41}x_{58} + x_{41}x_{62} + x_{41}x_{64} + x_{42}x_{45} + x_{42}x_{47} + x_{42}x_{53} + x_{42}x_{55} + x_{42}x_{57} + x_{42}x_{58} + x_{42}x_{60} + x_{42}x_{63} + x_{43}x_{44} + x_{43}x_{46} + x_{43}x_{47} + x_{43}x_{48} + x_{43}x_{49} + x_{43}x_{51} + x_{43}x_{53} + x_{43}x_{54} + x_{43}x_{57} + x_{43}x_{59} + x_{43}x_{60} + x_{43}x_{61} + x_{43}x_{62} + x_{43}x_{63} + x_{43}x_{64} + x_{44}x_{48} + x_{44}x_{49} + x_{44}x_{51} + x_{44}x_{52} + x_{44}x_{54} + x_{44}x_{58} + x_{44}x_{63} + x_{45}x_{48} + x_{45}x_{49} + x_{45}x_{50} + x_{45}x_{52} + x_{45}x_{53} + x_{45}x_{56} + x_{45}x_{57} + x_{45}x_{59} + x_{45}x_{61} + x_{45}x_{64} + x_{46}x_{47} + x_{46}x_{48} + x_{46}x_{57} + x_{46}x_{60} + x_{46}x_{61} + x_{46}x_{63} + x_{46}x_{64} + x_{47}x_{49} + x_{47}x_{51} + x_{47}x_{57} + x_{47}x_{58} + x_{47}x_{59} + x_{47}x_{60} + x_{47}x_{64} + x_{48}x_{49} + x_{48}x_{50} + x_{48}x_{52} + x_{48}x_{53} + x_{48}x_{56} + x_{48}x_{57} + x_{48}x_{58} + x_{48}x_{61} + x_{48}x_{63} + x_{49}x_{52} + x_{49}x_{53} + x_{49}x_{56} + x_{49}x_{57} + x_{49}x_{58} + x_{49}x_{59} + x_{49}x_{62} + x_{49}x_{64} + x_{50}x_{53} + x_{50}x_{54} + x_{50}x_{56} + x_{50}x_{58} + x_{50}x_{60} + x_{50}x_{61} + x_{50}x_{62} + x_{50}x_{63} + x_{50}x_{64} + x_{51}x_{52} + x_{51}x_{53} + x_{51}x_{54} + x_{51}x_{55} + x_{51}x_{56} + x_{51}x_{57} + x_{51}x_{58} + x_{51}x_{60} + x_{52}x_{53} + x_{52}x_{58} + x_{52}x_{60} + x_{52}x_{61} + x_{52}x_{62} + x_{52}x_{63} + x_{53}x_{55} + x_{53}x_{57} + x_{53}x_{58} + x_{53}x_{61} + x_{53}x_{62} + x_{53}x_{63} + x_{54}x_{55} + x_{54}x_{58} + x_{54}x_{60} + x_{55}x_{56} + x_{55}x_{57} + x_{55}x_{59} + x_{55}x_{60} + x_{55}x_{62} + x_{55}x_{64} + x_{56}x_{58} + x_{56}x_{59} + x_{56}x_{61} + x_{56}x_{63} + x_{56}x_{64} + x_{57}x_{58} + x_{57}x_{59} + x_{57}x_{60} + x_{57}x_{63} + x_{58}x_{59} + x_{58}x_{60} + x_{58}x_{61} + x_{58}x_{62} + x_{58}x_{64} + x_{59}x_{60} + x_{59}x_{64} + x_{60}x_{61} + x_{61}x_{63} + x_{2} + x_{3} + x_{4} + x_{5} + x_{6} + x_{7} + x_{10} + x_{11} + x_{12} + x_{14} + x_{18} + x_{19} + x_{20} + x_{22} + x_{25} + x_{26} + x_{31} + x_{34} + x_{35} + x_{36} + x_{37} + x_{38} + x_{40} + x_{47} + x_{51} + x_{52} + x_{53} + x_{54} + x_{55} + x_{57} + x_{59} + x_{64}$

$y_{14} = x_{1}x_{2} + x_{1}x_{4} + x_{1}x_{6} + x_{1}x_{7} + x_{1}x_{9} + x_{1}x_{14} + x_{1}x_{15} + x_{1}x_{17} + x_{1}x_{18} + x_{1}x_{20} + x_{1}x_{22} + x_{1}x_{23} + x_{1}x_{24} + x_{1}x_{26} + x_{1}x_{30} + x_{1}x_{32} + x_{1}x_{34} + x_{1}x_{35} + x_{1}x_{37} + x_{1}x_{38} + x_{1}x_{42} + x_{1}x_{44} + x_{1}x_{51} + x_{1}x_{53} + x_{1}x_{55} + x_{1}x_{57} + x_{2}x_{3} + x_{2}x_{4} + x_{2}x_{6} + x_{2}x_{8} + x_{2}x_{9} + x_{2}x_{11} + x_{2}x_{13} + x_{2}x_{17} + x_{2}x_{19} + x_{2}x_{21} + x_{2}x_{24} + x_{2}x_{26} + x_{2}x_{27} + x_{2}x_{31} + x_{2}x_{33} + x_{2}x_{34} + x_{2}x_{35} + x_{2}x_{43} + x_{2}x_{44} + x_{2}x_{45} + x_{2}x_{47} + x_{2}x_{51} + x_{2}x_{52} + x_{2}x_{54} + x_{2}x_{56} + x_{2}x_{58} + x_{2}x_{60} + x_{2}x_{61} + x_{2}x_{62} + x_{2}x_{63} + x_{3}x_{4} + x_{3}x_{5} + x_{3}x_{7} + x_{3}x_{10} + x_{3}x_{12} + x_{3}x_{15} + x_{3}x_{17} + x_{3}x_{18} + x_{3}x_{20} + x_{3}x_{21} + x_{3}x_{22} + x_{3}x_{23} + x_{3}x_{25} + x_{3}x_{26} + x_{3}x_{28} + x_{3}x_{29} + x_{3}x_{30} + x_{3}x_{33} + x_{3}x_{35} + x_{3}x_{38} + x_{3}x_{39} + x_{3}x_{43} + x_{3}x_{45} + x_{3}x_{47} + x_{3}x_{48} + x_{3}x_{52} + x_{3}x_{53} + x_{3}x_{55} + x_{3}x_{56} + x_{3}x_{57} + x_{3}x_{59} + x_{3}x_{60} + x_{3}x_{62} + x_{3}x_{63} + x_{4}x_{5} + x_{4}x_{6} + x_{4}x_{13} + x_{4}x_{14} + x_{4}x_{16} + x_{4}x_{17} + x_{4}x_{20} + x_{4}x_{21} + x_{4}x_{24} + x_{4}x_{27} + x_{4}x_{28} + x_{4}x_{31} + x_{4}x_{32} + x_{4}x_{33} + x_{4}x_{36} + x_{4}x_{37} + x_{4}x_{40} + x_{4}x_{42} + x_{4}x_{46} + x_{4}x_{50} + x_{4}x_{52} + x_{4}x_{53} + x_{4}x_{54} + x_{4}x_{55} + x_{4}x_{58} + x_{4}x_{59} + x_{4}x_{61} + x_{4}x_{62} + x_{5}x_{6} + x_{5}x_{7} + x_{5}x_{8} + x_{5}x_{10} + x_{5}x_{12} + x_{5}x_{15} + x_{5}x_{17} + x_{5}x_{18} + x_{5}x_{20} + x_{5}x_{21} + x_{5}x_{24} + x_{5}x_{26} + x_{5}x_{34} + x_{5}x_{38} + x_{5}x_{39} + x_{5}x_{40} + x_{5}x_{45} + x_{5}x_{46} + x_{5}x_{49} + x_{5}x_{52} + x_{5}x_{53} + x_{5}x_{54} + x_{5}x_{55} + x_{5}x_{56} + x_{5}x_{58} + x_{5}x_{59} + x_{5}x_{60} + x_{5}x_{62} + x_{5}x_{63} + x_{5}x_{64} + x_{6}x_{7} + x_{6}x_{8} + x_{6}x_{9} + x_{6}x_{10} + x_{6}x_{11} + x_{6}x_{13} + x_{6}x_{14} + x_{6}x_{15} + x_{6}x_{16} + x_{6}x_{17} + x_{6}x_{20} + x_{6}x_{21} + x_{6}x_{22} + x_{6}x_{24} + x_{6}x_{30} + x_{6}x_{32} + x_{6}x_{34} + x_{6}x_{35} + x_{6}x_{36} + x_{6}x_{40} + x_{6}x_{41} + x_{6}x_{43} + x_{6}x_{52} + x_{6}x_{53} + x_{6}x_{57} + x_{6}x_{58} + x_{6}x_{60} + x_{6}x_{61} + x_{6}x_{62} + x_{6}x_{63} + x_{7}x_{8} + x_{7}x_{9} + x_{7}x_{10} + x_{7}x_{12} + x_{7}x_{13} + x_{7}x_{14} + x_{7}x_{15} + x_{7}x_{17} + x_{7}x_{18} + x_{7}x_{22} + x_{7}x_{24} + x_{7}x_{26} + x_{7}x_{27} + x_{7}x_{28} + x_{7}x_{33} + x_{7}x_{34} + x_{7}x_{37} + x_{7}x_{38} + x_{7}x_{39} + x_{7}x_{40} + x_{7}x_{43} + x_{7}x_{44} + x_{7}x_{45} + x_{7}x_{48} + x_{7}x_{49} + x_{7}x_{50} + x_{7}x_{52} + x_{7}x_{53} + x_{7}x_{54} + x_{7}x_{59} + x_{7}x_{60} + x_{7}x_{62} + x_{7}x_{63} + x_{7}x_{64} + x_{8}x_{10} + x_{8}x_{12} + x_{8}x_{19} + x_{8}x_{21} + x_{8}x_{22} + x_{8}x_{23} + x_{8}x_{25} + x_{8}x_{26} + x_{8}x_{28} + x_{8}x_{30} + x_{8}x_{32} + x_{8}x_{37} + x_{8}x_{45} + x_{8}x_{46} + x_{8}x_{47} + x_{8}x_{48} + x_{8}x_{54} + x_{8}x_{55} + x_{8}x_{57} + x_{8}x_{59} + x_{8}x_{61} + x_{8}x_{62} + x_{9}x_{11} + x_{9}x_{15} + x_{9}x_{16} + x_{9}x_{19} + x_{9}x_{21} + x_{9}x_{22} + x_{9}x_{26} + x_{9}x_{27} + x_{9}x_{31} + x_{9}x_{33} + x_{9}x_{41} + x_{9}x_{42} + x_{9}x_{47} + x_{9}x_{48} + x_{9}x_{50} + x_{9}x_{51} + x_{9}x_{52} + x_{9}x_{54} + x_{9}x_{55} + x_{9}x_{57} + x_{9}x_{58} + x_{9}x_{59} + x_{9}x_{61} + x_{9}x_{62} + x_{9}x_{64} + x_{10}x_{11} + x_{10}x_{14} + x_{10}x_{15} + x_{10}x_{16} + x_{10}x_{17} + x_{10}x_{18} + x_{10}x_{21} + x_{10}x_{22} + x_{10}x_{25} + x_{10}x_{26} + x_{10}x_{27} + x_{10}x_{29} + x_{10}x_{32} + x_{10}x_{34} + x_{10}x_{37} + x_{10}x_{39} + x_{10}x_{40} + x_{10}x_{42} + x_{10}x_{43} + x_{10}x_{47} + x_{10}x_{49} + x_{10}x_{50} + x_{10}x_{53} + x_{10}x_{54} + x_{10}x_{55} + x_{10}x_{57} + x_{10}x_{58} + x_{10}x_{59} + x_{10}x_{62} + x_{10}x_{63} + x_{10}x_{64} + x_{11}x_{13} + x_{11}x_{14} + x_{11}x_{15} + x_{11}x_{16} + x_{11}x_{17} + x_{11}x_{20} + x_{11}x_{21} + x_{11}x_{22} + x_{11}x_{23} + x_{11}x_{24} + x_{11}x_{26} + x_{11}x_{28} + x_{11}x_{29} + x_{11}x_{30} + x_{11}x_{31} + x_{11}x_{32} + x_{11}x_{33} + x_{11}x_{35} + x_{11}x_{37} + x_{11}x_{39} + x_{11}x_{41} + x_{11}x_{42} + x_{11}x_{43} + x_{11}x_{45} + x_{11}x_{47} + x_{11}x_{50} + x_{11}x_{51} + x_{11}x_{54} + x_{11}x_{55} + x_{11}x_{57} + x_{11}x_{59} + x_{11}x_{61} + x_{11}x_{62} + x_{11}x_{63} + x_{12}x_{13} + x_{12}x_{14} + x_{12}x_{15} + x_{12}x_{17} + x_{12}x_{19} + x_{12}x_{20} + x_{12}x_{25} + x_{12}x_{26} + x_{12}x_{28} + x_{12}x_{29} + x_{12}x_{31} + x_{12}x_{33} + x_{12}x_{36} + x_{12}x_{37} + x_{12}x_{38} + x_{12}x_{39} + x_{12}x_{41} + x_{12}x_{42} + x_{12}x_{44} + x_{12}x_{48} + x_{12}x_{50} + x_{12}x_{51} + x_{12}x_{52} + x_{12}x_{54} + x_{12}x_{56} + x_{12}x_{59} + x_{12}x_{61} + x_{12}x_{63} + x_{13}x_{14} + x_{13}x_{16} + x_{13}x_{17} + x_{13}x_{18} + x_{13}x_{19} + x_{13}x_{21} + x_{13}x_{25} + x_{13}x_{26} + x_{13}x_{27} + x_{13}x_{29} + x_{13}x_{31} + x_{13}x_{33} + x_{13}x_{35} + x_{13}x_{36} + x_{13}x_{37} + x_{13}x_{40} + x_{13}x_{41} + x_{13}x_{43} + x_{13}x_{44} + x_{13}x_{45} + x_{13}x_{46} + x_{13}x_{47} + x_{13}x_{48} + x_{13}x_{50} + x_{13}x_{51} + x_{13}x_{52} + x_{13}x_{54} + x_{13}x_{55} + x_{13}x_{57} + x_{13}x_{58} + x_{13}x_{59} + x_{13}x_{60} + x_{14}x_{25} + x_{14}x_{27} + x_{14}x_{32} + x_{14}x_{34} + x_{14}x_{35} + x_{14}x_{36} + x_{14}x_{39} + x_{14}x_{45} + x_{14}x_{46} + x_{14}x_{47} + x_{14}x_{48} + x_{14}x_{52} + x_{14}x_{55} + x_{14}x_{57} + x_{14}x_{59} + x_{14}x_{61} + x_{14}x_{63} + x_{15}x_{17} + x_{15}x_{18} + x_{15}x_{19} + x_{15}x_{21} + x_{15}x_{22} + x_{15}x_{23} + x_{15}x_{24} + x_{15}x_{25} + x_{15}x_{28} + x_{15}x_{29} + x_{15}x_{31} + x_{15}x_{33} + x_{15}x_{36} + x_{15}x_{37} + x_{15}x_{39} + x_{15}x_{43} + x_{15}x_{45} + x_{15}x_{46} + x_{15}x_{48} + x_{15}x_{51} + x_{15}x_{53} + x_{15}x_{55} + x_{15}x_{57} + x_{15}x_{59} + x_{15}x_{61} + x_{15}x_{63} + x_{16}x_{19} + x_{16}x_{21} + x_{16}x_{22} + x_{16}x_{24} + x_{16}x_{26} + x_{16}x_{27} + x_{16}x_{30} + x_{16}x_{31} + x_{16}x_{36} + x_{16}x_{40} + x_{16}x_{43} + x_{16}x_{45} + x_{16}x_{46} + x_{16}x_{47} + x_{16}x_{48} + x_{16}x_{49} + x_{16}x_{51} + x_{16}x_{53} + x_{16}x_{54} + x_{16}x_{56} + x_{16}x_{58} + x_{16}x_{60} + x_{16}x_{62} + x_{16}x_{64} + x_{17}x_{20} + x_{17}x_{22} + x_{17}x_{26} + x_{17}x_{27} + x_{17}x_{28} + x_{17}x_{29} + x_{17}x_{31} + x_{17}x_{33} + x_{17}x_{37} + x_{17}x_{42} + x_{17}x_{43} + x_{17}x_{45} + x_{17}x_{46} + x_{17}x_{48} + x_{17}x_{50} + x_{17}x_{52} + x_{17}x_{54} + x_{17}x_{58} + x_{17}x_{61} + x_{17}x_{62} + x_{17}x_{63} + x_{17}x_{64} + x_{18}x_{20} + x_{18}x_{23} + x_{18}x_{24} + x_{18}x_{25} + x_{18}x_{26} + x_{18}x_{29} + x_{18}x_{32} + x_{18}x_{33} + x_{18}x_{35} + x_{18}x_{36} + x_{18}x_{39} + x_{18}x_{40} + x_{18}x_{42} + x_{18}x_{46} + x_{18}x_{47} + x_{18}x_{49} + x_{18}x_{50} + x_{18}x_{53} + x_{18}x_{55} + x_{18}x_{58} + x_{18}x_{59} + x_{18}x_{61} + x_{19}x_{20} + x_{19}x_{23} + x_{19}x_{24} + x_{19}x_{25} + x_{19}x_{26} + x_{19}x_{28} + x_{19}x_{30} + x_{19}x_{32} + x_{19}x_{34} + x_{19}x_{35} + x_{19}x_{37} + x_{19}x_{39} + x_{19}x_{40} + x_{19}x_{42} + x_{19}x_{45} + x_{19}x_{49} + x_{19}x_{50} + x_{19}x_{51} + x_{19}x_{52} + x_{19}x_{54} + x_{19}x_{55} + x_{19}x_{58} + x_{19}x_{62} + x_{19}x_{63} + x_{20}x_{22} + x_{20}x_{25} + x_{20}x_{26} + x_{20}x_{29} + x_{20}x_{30} + x_{20}x_{31} + x_{20}x_{35} + x_{20}x_{36} + x_{20}x_{37} + x_{20}x_{39} + x_{20}x_{40} + x_{20}x_{44} + x_{20}x_{45} + x_{20}x_{46} + x_{20}x_{48} + x_{20}x_{52} + x_{20}x_{53} + x_{20}x_{55} + x_{20}x_{56} + x_{20}x_{57} + x_{20}x_{60} + x_{20}x_{61} + x_{20}x_{62} + x_{20}x_{63} + x_{20}x_{64} + x_{21}x_{22} + x_{21}x_{24} + x_{21}x_{25} + x_{21}x_{27} + x_{21}x_{30} + x_{21}x_{31} + x_{21}x_{32} + x_{21}x_{34} + x_{21}x_{36} + x_{21}x_{40} + x_{21}x_{45} + x_{21}x_{46} + x_{21}x_{48} + x_{21}x_{53} + x_{21}x_{55} + x_{21}x_{57} + x_{21}x_{58} + x_{21}x_{61} + x_{21}x_{62} + x_{22}x_{23} + x_{22}x_{24} + x_{22}x_{28} + x_{22}x_{29} + x_{22}x_{30} + x_{22}x_{31} + x_{22}x_{32} + x_{22}x_{34} + x_{22}x_{37} + x_{22}x_{40} + x_{22}x_{47} + x_{22}x_{48} + x_{22}x_{49} + x_{22}x_{50} + x_{22}x_{52} + x_{22}x_{53} + x_{22}x_{58} + x_{22}x_{59} + x_{22}x_{64} + x_{23}x_{27} + x_{23}x_{28} + x_{23}x_{30} + x_{23}x_{32} + x_{23}x_{37} + x_{23}x_{38} + x_{23}x_{41} + x_{23}x_{43} + x_{23}x_{48} + x_{23}x_{50} + x_{23}x_{52} + x_{23}x_{53} + x_{23}x_{54} + x_{23}x_{56} + x_{23}x_{57} + x_{23}x_{60} + x_{23}x_{62} + x_{23}x_{63} + x_{23}x_{64} + x_{24}x_{25} + x_{24}x_{26} + x_{24}x_{27} + x_{24}x_{28} + x_{24}x_{31} + x_{24}x_{32} + x_{24}x_{33} + x_{24}x_{36} + x_{24}x_{37} + x_{24}x_{38} + x_{24}x_{39} + x_{24}x_{47} + x_{24}x_{56} + x_{24}x_{58} + x_{24}x_{60} + x_{24}x_{61} + x_{25}x_{30} + x_{25}x_{31} + x_{25}x_{34} + x_{25}x_{35} + x_{25}x_{38} + x_{25}x_{39} + x_{25}x_{40} + x_{25}x_{42} + x_{25}x_{43} + x_{25}x_{49} + x_{25}x_{50} + x_{25}x_{51} + x_{25}x_{53} + x_{25}x_{54} + x_{25}x_{55} + x_{25}x_{56} + x_{25}x_{57} + x_{25}x_{61} + x_{26}x_{31} + x_{26}x_{32} + x_{26}x_{34} + x_{26}x_{35} + x_{26}x_{36} + x_{26}x_{40} + x_{26}x_{42} + x_{26}x_{48} + x_{26}x_{49} + x_{26}x_{51} + x_{26}x_{53} + x_{26}x_{54} + x_{26}x_{58} + x_{26}x_{61} + x_{26}x_{62} + x_{27}x_{29} + x_{27}x_{31} + x_{27}x_{32} + x_{27}x_{34} + x_{27}x_{37} + x_{27}x_{39} + x_{27}x_{41} + x_{27}x_{43} + x_{27}x_{44} + x_{27}x_{47} + x_{27}x_{48} + x_{27}x_{51} + x_{27}x_{53} + x_{27}x_{54} + x_{27}x_{56} + x_{27}x_{59} + x_{27}x_{61} + x_{28}x_{31} + x_{28}x_{36} + x_{28}x_{38} + x_{28}x_{39} + x_{28}x_{42} + x_{28}x_{43} + x_{28}x_{44} + x_{28}x_{46} + x_{28}x_{47} + x_{28}x_{50} + x_{28}x_{51} + x_{28}x_{52} + x_{28}x_{55} + x_{28}x_{57} + x_{28}x_{59} + x_{28}x_{60} + x_{28}x_{62} + x_{28}x_{64} + x_{29}x_{30} + x_{29}x_{32} + x_{29}x_{34} + x_{29}x_{35} + x_{29}x_{36} + x_{29}x_{43} + x_{29}x_{44} + x_{29}x_{46} + x_{29}x_{52} + x_{29}x_{53} + x_{29}x_{56} + x_{29}x_{59} + x_{29}x_{60} + x_{29}x_{61} + x_{29}x_{63} + x_{29}x_{64} + x_{30}x_{32} + x_{30}x_{33} + x_{30}x_{34} + x_{30}x_{35} + x_{30}x_{36} + x_{30}x_{41} + x_{30}x_{46} + x_{30}x_{47} + x_{30}x_{48} + x_{30}x_{50} + x_{30}x_{52} + x_{30}x_{54} + x_{30}x_{55} + x_{30}x_{56} + x_{30}x_{59} + x_{30}x_{62} + x_{31}x_{32} + x_{31}x_{34} + x_{31}x_{35} + x_{31}x_{39} + x_{31}x_{40} + x_{31}x_{41} + x_{31}x_{42} + x_{31}x_{43} + x_{31}x_{44} + x_{31}x_{45} + x_{31}x_{46} + x_{31}x_{49} + x_{31}x_{51} + x_{31}x_{55} + x_{31}x_{56} + x_{31}x_{58} + x_{31}x_{59} + x_{31}x_{60} + x_{31}x_{61} + x_{31}x_{62} + x_{31}x_{63} + x_{32}x_{36} + x_{32}x_{37} + x_{32}x_{40} + x_{32}x_{41} + x_{32}x_{43} + x_{32}x_{44} + x_{32}x_{50} + x_{32}x_{53} + x_{32}x_{54} + x_{32}x_{55} + x_{32}x_{56} + x_{32}x_{59} + x_{32}x_{60} + x_{32}x_{61} + x_{32}x_{63} + x_{32}x_{64} + x_{33}x_{34} + x_{33}x_{35} + x_{33}x_{36} + x_{33}x_{37} + x_{33}x_{38} + x_{33}x_{40} + x_{33}x_{41} + x_{33}x_{44} + x_{33}x_{45} + x_{33}x_{48} + x_{33}x_{50} + x_{33}x_{52} + x_{33}x_{53} + x_{33}x_{58} + x_{33}x_{60} + x_{33}x_{61} + x_{33}x_{63} + x_{34}x_{37} + x_{34}x_{38} + x_{34}x_{39} + x_{34}x_{40} + x_{34}x_{43} + x_{34}x_{44} + x_{34}x_{46} + x_{34}x_{48} + x_{34}x_{49} + x_{34}x_{50} + x_{34}x_{51} + x_{34}x_{55} + x_{34}x_{57} + x_{34}x_{60} + x_{34}x_{61} + x_{34}x_{62} + x_{34}x_{63} + x_{35}x_{37} + x_{35}x_{38} + x_{35}x_{39} + x_{35}x_{41} + x_{35}x_{42} + x_{35}x_{43} + x_{35}x_{44} + x_{35}x_{46} + x_{35}x_{47} + x_{35}x_{48} + x_{35}x_{50} + x_{35}x_{51} + x_{35}x_{54} + x_{35}x_{55} + x_{35}x_{56} + x_{35}x_{59} + x_{35}x_{62} + x_{36}x_{37} + x_{36}x_{38} + x_{36}x_{39} + x_{36}x_{40} + x_{36}x_{42} + x_{36}x_{52} + x_{36}x_{53} + x_{36}x_{54} + x_{36}x_{55} + x_{36}x_{56} + x_{36}x_{57} + x_{36}x_{62} + x_{36}x_{64} + x_{37}x_{38} + x_{37}x_{40} + x_{37}x_{41} + x_{37}x_{43} + x_{37}x_{56} + x_{37}x_{62} + x_{38}x_{40} + x_{38}x_{42} + x_{38}x_{46} + x_{38}x_{47} + x_{38}x_{49} + x_{38}x_{51} + x_{38}x_{52} + x_{38}x_{53} + x_{38}x_{55} + x_{38}x_{57} + x_{38}x_{61} + x_{38}x_{63} + x_{38}x_{64} + x_{39}x_{40} + x_{39}x_{41} + x_{39}x_{42} + x_{39}x_{48} + x_{39}x_{53} + x_{39}x_{54} + x_{39}x_{58} + x_{39}x_{60} + x_{39}x_{63} + x_{39}x_{64} + x_{40}x_{41} + x_{40}x_{46} + x_{40}x_{49} + x_{40}x_{52} + x_{40}x_{57} + x_{40}x_{62} + x_{40}x_{64} + x_{41}x_{42} + x_{41}x_{43} + x_{41}x_{44} + x_{41}x_{47} + x_{41}x_{49} + x_{41}x_{53} + x_{41}x_{55} + x_{41}x_{58} + x_{41}x_{59} + x_{41}x_{62} + x_{41}x_{63} + x_{41}x_{64} + x_{42}x_{44} + x_{42}x_{45} + x_{42}x_{46} + x_{42}x_{47} + x_{42}x_{48} + x_{42}x_{49} + x_{42}x_{50} + x_{42}x_{53} + x_{42}x_{55} + x_{42}x_{58} + x_{42}x_{59} + x_{43}x_{44} + x_{43}x_{45} + x_{43}x_{47} + x_{43}x_{48} + x_{43}x_{52} + x_{43}x_{54} + x_{43}x_{58} + x_{43}x_{61} + x_{43}x_{64} + x_{44}x_{46} + x_{44}x_{48} + x_{44}x_{51} + x_{44}x_{52} + x_{44}x_{53} + x_{44}x_{56} + x_{44}x_{57} + x_{44}x_{58} + x_{44}x_{60} + x_{44}x_{62} + x_{44}x_{64} + x_{45}x_{47} + x_{45}x_{51} + x_{45}x_{53} + x_{45}x_{55} + x_{45}x_{56} + x_{45}x_{62} + x_{45}x_{64} + x_{46}x_{48} + x_{46}x_{49} + x_{46}x_{51} + x_{46}x_{53} + x_{46}x_{56} + x_{46}x_{59} + x_{46}x_{61} + x_{46}x_{63} + x_{47}x_{48} + x_{47}x_{49} + x_{47}x_{50} + x_{47}x_{51} + x_{47}x_{52} + x_{47}x_{53} + x_{47}x_{57} + x_{47}x_{59} + x_{47}x_{60} + x_{47}x_{61} + x_{47}x_{63} + x_{48}x_{49} + x_{48}x_{52} + x_{48}x_{54} + x_{48}x_{56} + x_{48}x_{58} + x_{48}x_{60} + x_{48}x_{61} + x_{48}x_{62} + x_{48}x_{64} + x_{49}x_{52} + x_{49}x_{54} + x_{49}x_{55} + x_{49}x_{58} + x_{49}x_{59} + x_{49}x_{60} + x_{49}x_{62} + x_{49}x_{63} + x_{50}x_{52} + x_{50}x_{53} + x_{50}x_{58} + x_{50}x_{60} + x_{51}x_{53} + x_{51}x_{54} + x_{51}x_{56} + x_{51}x_{57} + x_{51}x_{59} + x_{51}x_{64} + x_{52}x_{53} + x_{52}x_{56} + x_{52}x_{57} + x_{52}x_{60} + x_{52}x_{62} + x_{52}x_{63} + x_{53}x_{54} + x_{53}x_{55} + x_{53}x_{59} + x_{53}x_{62} + x_{53}x_{64} + x_{54}x_{55} + x_{54}x_{58} + x_{54}x_{59} + x_{54}x_{60} + x_{54}x_{61} + x_{54}x_{64} + x_{55}x_{57} + x_{55}x_{58} + x_{55}x_{59} + x_{55}x_{61} + x_{55}x_{64} + x_{56}x_{57} + x_{56}x_{61} + x_{56}x_{63} + x_{56}x_{64} + x_{57}x_{58} + x_{57}x_{59} + x_{57}x_{62} + x_{57}x_{64} + x_{58}x_{59} + x_{58}x_{61} + x_{58}x_{63} + x_{58}x_{64} + x_{59}x_{60} + x_{59}x_{61} + x_{60}x_{61} + x_{61}x_{63} + x_{61}x_{64} + x_{62}x_{64} + x_{63}x_{64} + x_{3} + x_{6} + x_{7} + x_{8} + x_{9} + x_{10} + x_{11} + x_{12} + x_{15} + x_{16} + x_{17} + x_{18} + x_{27} + x_{29} + x_{32} + x_{33} + x_{34} + x_{37} + x_{38} + x_{43} + x_{45} + x_{46} + x_{49} + x_{50} + x_{51} + x_{52} + x_{54} + x_{57} + x_{58} + x_{60} + x_{64} + 1$

$y_{15} = x_{1}x_{2} + x_{1}x_{3} + x_{1}x_{5} + x_{1}x_{6} + x_{1}x_{7} + x_{1}x_{8} + x_{1}x_{9} + x_{1}x_{11} + x_{1}x_{12} + x_{1}x_{15} + x_{1}x_{19} + x_{1}x_{22} + x_{1}x_{25} + x_{1}x_{26} + x_{1}x_{27} + x_{1}x_{28} + x_{1}x_{29} + x_{1}x_{30} + x_{1}x_{34} + x_{1}x_{36} + x_{1}x_{38} + x_{1}x_{40} + x_{1}x_{43} + x_{1}x_{44} + x_{1}x_{45} + x_{1}x_{48} + x_{1}x_{51} + x_{1}x_{54} + x_{1}x_{55} + x_{1}x_{56} + x_{1}x_{57} + x_{1}x_{59} + x_{1}x_{60} + x_{1}x_{63} + x_{2}x_{3} + x_{2}x_{4} + x_{2}x_{6} + x_{2}x_{8} + x_{2}x_{9} + x_{2}x_{11} + x_{2}x_{12} + x_{2}x_{13} + x_{2}x_{14} + x_{2}x_{15} + x_{2}x_{17} + x_{2}x_{18} + x_{2}x_{20} + x_{2}x_{22} + x_{2}x_{25} + x_{2}x_{27} + x_{2}x_{31} + x_{2}x_{32} + x_{2}x_{34} + x_{2}x_{35} + x_{2}x_{37} + x_{2}x_{39} + x_{2}x_{40} + x_{2}x_{41} + x_{2}x_{42} + x_{2}x_{45} + x_{2}x_{46} + x_{2}x_{51} + x_{2}x_{52} + x_{2}x_{53} + x_{2}x_{56} + x_{2}x_{58} + x_{2}x_{59} + x_{2}x_{61} + x_{2}x_{62} + x_{2}x_{63} + x_{2}x_{64} + x_{3}x_{4} + x_{3}x_{6} + x_{3}x_{10} + x_{3}x_{11} + x_{3}x_{19} + x_{3}x_{20} + x_{3}x_{21} + x_{3}x_{26} + x_{3}x_{28} + x_{3}x_{29} + x_{3}x_{32} + x_{3}x_{36} + x_{3}x_{37} + x_{3}x_{38} + x_{3}x_{39} + x_{3}x_{45} + x_{3}x_{46} + x_{3}x_{49} + x_{3}x_{50} + x_{3}x_{51} + x_{3}x_{52} + x_{3}x_{56} + x_{3}x_{57} + x_{3}x_{60} + x_{3}x_{62} + x_{4}x_{7} + x_{4}x_{8} + x_{4}x_{10} + x_{4}x_{11} + x_{4}x_{12} + x_{4}x_{14} + x_{4}x_{15} + x_{4}x_{16} + x_{4}x_{18} + x_{4}x_{19} + x_{4}x_{22} + x_{4}x_{30} + x_{4}x_{33} + x_{4}x_{35} + x_{4}x_{36} + x_{4}x_{38} + x_{4}x_{39} + x_{4}x_{41} + x_{4}x_{44} + x_{4}x_{46} + x_{4}x_{48} + x_{4}x_{55} + x_{4}x_{56} + x_{4}x_{60} + x_{4}x_{64} + x_{5}x_{9} + x_{5}x_{11} + x_{5}x_{13} + x_{5}x_{16} + x_{5}x_{25} + x_{5}x_{27} + x_{5}x_{30} + x_{5}x_{31} + x_{5}x_{37} + x_{5}x_{38} + x_{5}x_{40} + x_{5}x_{44} + x_{5}x_{46} + x_{5}x_{47} + x_{5}x_{48} + x_{5}x_{52} + x_{5}x_{53} + x_{5}x_{55} + x_{5}x_{56} + x_{5}x_{58} + x_{5}x_{61} + x_{5}x_{63} + x_{5}x_{64} + x_{6}x_{7} + x_{6}x_{8} + x_{6}x_{10} + x_{6}x_{11} + x_{6}x_{12} + x_{6}x_{13} + x_{6}x_{15} + x_{6}x_{17} + x_{6}x_{19} + x_{6}x_{21} + x_{6}x_{23} + x_{6}x_{25} + x_{6}x_{26} + x_{6}x_{27} + x_{6}x_{28} + x_{6}x_{31} + x_{6}x_{32} + x_{6}x_{33} + x_{6}x_{37} + x_{6}x_{39} + x_{6}x_{40} + x_{6}x_{41} + x_{6}x_{42} + x_{6}x_{43} + x_{6}x_{47} + x_{6}x_{48} + x_{6}x_{50} + x_{6}x_{51} + x_{6}x_{52} + x_{6}x_{55} + x_{6}x_{57} + x_{6}x_{59} + x_{6}x_{60} + x_{6}x_{61} + x_{6}x_{64} + x_{7}x_{14} + x_{7}x_{16} + x_{7}x_{20} + x_{7}x_{21} + x_{7}x_{24} + x_{7}x_{25} + x_{7}x_{26} + x_{7}x_{27} + x_{7}x_{30} + x_{7}x_{33} + x_{7}x_{34} + x_{7}x_{35} + x_{7}x_{37} + x_{7}x_{38} + x_{7}x_{42} + x_{7}x_{44} + x_{7}x_{45} + x_{7}x_{46} + x_{7}x_{47} + x_{7}x_{48} + x_{7}x_{49} + x_{7}x_{50} + x_{7}x_{51} + x_{7}x_{52} + x_{7}x_{53} + x_{7}x_{56} + x_{7}x_{58} + x_{7}x_{59} + x_{7}x_{60} + x_{7}x_{62} + x_{7}x_{63} + x_{7}x_{64} + x_{8}x_{9} + x_{8}x_{10} + x_{8}x_{11} + x_{8}x_{12} + x_{8}x_{15} + x_{8}x_{16} + x_{8}x_{17} + x_{8}x_{19} + x_{8}x_{20} + x_{8}x_{22} + x_{8}x_{25} + x_{8}x_{26} + x_{8}x_{28} + x_{8}x_{29} + x_{8}x_{31} + x_{8}x_{32} + x_{8}x_{36} + x_{8}x_{37} + x_{8}x_{38} + x_{8}x_{39} + x_{8}x_{41} + x_{8}x_{42} + x_{8}x_{46} + x_{8}x_{48} + x_{8}x_{49} + x_{8}x_{50} + x_{8}x_{55} + x_{8}x_{56} + x_{8}x_{57} + x_{8}x_{58} + x_{8}x_{59} + x_{8}x_{62} + x_{8}x_{63} + x_{9}x_{12} + x_{9}x_{15} + x_{9}x_{16} + x_{9}x_{17} + x_{9}x_{18} + x_{9}x_{22} + x_{9}x_{24} + x_{9}x_{26} + x_{9}x_{27} + x_{9}x_{29} + x_{9}x_{32} + x_{9}x_{33} + x_{9}x_{34} + x_{9}x_{35} + x_{9}x_{36} + x_{9}x_{39} + x_{9}x_{40} + x_{9}x_{41} + x_{9}x_{43} + x_{9}x_{45} + x_{9}x_{46} + x_{9}x_{47} + x_{9}x_{49} + x_{9}x_{51} + x_{9}x_{52} + x_{9}x_{55} + x_{9}x_{57} + x_{9}x_{58} + x_{9}x_{60} + x_{9}x_{61} + x_{9}x_{62} + x_{9}x_{63} + x_{10}x_{11} + x_{10}x_{13} + x_{10}x_{14} + x_{10}x_{20} + x_{10}x_{26} + x_{10}x_{27} + x_{10}x_{28} + x_{10}x_{29} + x_{10}x_{31} + x_{10}x_{32} + x_{10}x_{35} + x_{10}x_{37} + x_{10}x_{40} + x_{10}x_{41} + x_{10}x_{42} + x_{10}x_{43} + x_{10}x_{45} + x_{10}x_{47} + x_{10}x_{49} + x_{10}x_{50} + x_{10}x_{51} + x_{10}x_{52} + x_{10}x_{58} + x_{10}x_{61} + x_{10}x_{62} + x_{10}x_{63} + x_{11}x_{12} + x_{11}x_{15} + x_{11}x_{18} + x_{11}x_{21} + x_{11}x_{22} + x_{11}x_{24} + x_{11}x_{30} + x_{11}x_{31} + x_{11}x_{33} + x_{11}x_{34} + x_{11}x_{35} + x_{11}x_{37} + x_{11}x_{39} + x_{11}x_{41} + x_{11}x_{45} + x_{11}x_{46} + x_{11}x_{47} + x_{11}x_{48} + x_{11}x_{49} + x_{11}x_{50} + x_{11}x_{52} + x_{11}x_{53} + x_{11}x_{54} + x_{11}x_{57} + x_{12}x_{19} + x_{12}x_{22} + x_{12}x_{24} + x_{12}x_{25} + x_{12}x_{26} + x_{12}x_{28} + x_{12}x_{30} + x_{12}x_{32} + x_{12}x_{33} + x_{12}x_{34} + x_{12}x_{37} + x_{12}x_{38} + x_{12}x_{40} + x_{12}x_{41} + x_{12}x_{45} + x_{12}x_{46} + x_{12}x_{48} + x_{12}x_{49} + x_{12}x_{52} + x_{12}x_{59} + x_{12}x_{63} + x_{12}x_{64} + x_{13}x_{14} + x_{13}x_{15} + x_{13}x_{20} + x_{13}x_{23} + x_{13}x_{24} + x_{13}x_{25} + x_{13}x_{26} + x_{13}x_{30} + x_{13}x_{33} + x_{13}x_{34} + x_{13}x_{36} + x_{13}x_{37} + x_{13}x_{38} + x_{13}x_{41} + x_{13}x_{42} + x_{13}x_{46} + x_{13}x_{47} + x_{13}x_{49} + x_{13}x_{50} + x_{13}x_{51} + x_{13}x_{52} + x_{13}x_{53} + x_{13}x_{54} + x_{13}x_{59} + x_{13}x_{60} + x_{13}x_{61} + x_{13}x_{63} + x_{13}x_{64} + x_{14}x_{16} + x_{14}x_{18} + x_{14}x_{19} + x_{14}x_{21} + x_{14}x_{22} + x_{14}x_{24} + x_{14}x_{25} + x_{14}x_{28} + x_{14}x_{29} + x_{14}x_{30} + x_{14}x_{32} + x_{14}x_{33} + x_{14}x_{34} + x_{14}x_{35} + x_{14}x_{37} + x_{14}x_{38} + x_{14}x_{41} + x_{14}x_{42} + x_{14}x_{43} + x_{14}x_{45} + x_{14}x_{47} + x_{14}x_{49} + x_{14}x_{50} + x_{14}x_{52} + x_{14}x_{54} + x_{14}x_{55} + x_{14}x_{57} + x_{14}x_{59} + x_{14}x_{61} + x_{14}x_{62} + x_{14}x_{64} + x_{15}x_{16} + x_{15}x_{19} + x_{15}x_{20} + x_{15}x_{21} + x_{15}x_{25} + x_{15}x_{28} + x_{15}x_{31} + x_{15}x_{32} + x_{15}x_{33} + x_{15}x_{35} + x_{15}x_{37} + x_{15}x_{38} + x_{15}x_{39} + x_{15}x_{40} + x_{15}x_{41} + x_{15}x_{44} + x_{15}x_{46} + x_{15}x_{47} + x_{15}x_{50} + x_{15}x_{51} + x_{15}x_{53} + x_{15}x_{54} + x_{15}x_{59} + x_{15}x_{60} + x_{15}x_{61} + x_{16}x_{17} + x_{16}x_{19} + x_{16}x_{21} + x_{16}x_{22} + x_{16}x_{26} + x_{16}x_{27} + x_{16}x_{30} + x_{16}x_{31} + x_{16}x_{32} + x_{16}x_{33} + x_{16}x_{34} + x_{16}x_{35} + x_{16}x_{36} + x_{16}x_{37} + x_{16}x_{43} + x_{16}x_{48} + x_{16}x_{50} + x_{16}x_{53} + x_{16}x_{57} + x_{16}x_{60} + x_{16}x_{64} + x_{17}x_{18} + x_{17}x_{19} + x_{17}x_{20} + x_{17}x_{21} + x_{17}x_{24} + x_{17}x_{25} + x_{17}x_{26} + x_{17}x_{27} + x_{17}x_{28} + x_{17}x_{29} + x_{17}x_{31} + x_{17}x_{32} + x_{17}x_{33} + x_{17}x_{34} + x_{17}x_{35} + x_{17}x_{37} + x_{17}x_{40} + x_{17}x_{41} + x_{17}x_{42} + x_{17}x_{44} + x_{17}x_{47} + x_{17}x_{50} + x_{17}x_{51} + x_{17}x_{58} + x_{17}x_{60} + x_{17}x_{62} + x_{17}x_{63} + x_{17}x_{64} + x_{18}x_{19} + x_{18}x_{23} + x_{18}x_{24} + x_{18}x_{25} + x_{18}x_{27} + x_{18}x_{30} + x_{18}x_{31} + x_{18}x_{32} + x_{18}x_{35} + x_{18}x_{36} + x_{18}x_{38} + x_{18}x_{40} + x_{18}x_{41} + x_{18}x_{42} + x_{18}x_{43} + x_{18}x_{45} + x_{18}x_{46} + x_{18}x_{48} + x_{18}x_{49} + x_{18}x_{50} + x_{18}x_{51} + x_{18}x_{52} + x_{18}x_{55} + x_{18}x_{56} + x_{18}x_{57} + x_{18}x_{61} + x_{18}x_{62} + x_{18}x_{64} + x_{19}x_{22} + x_{19}x_{23} + x_{19}x_{27} + x_{19}x_{31} + x_{19}x_{33} + x_{19}x_{36} + x_{19}x_{37} + x_{19}x_{39} + x_{19}x_{40} + x_{19}x_{41} + x_{19}x_{42} + x_{19}x_{43} + x_{19}x_{44} + x_{19}x_{46} + x_{19}x_{50} + x_{19}x_{52} + x_{19}x_{53} + x_{19}x_{55} + x_{19}x_{58} + x_{19}x_{60} + x_{19}x_{61} + x_{19}x_{64} + x_{20}x_{21} + x_{20}x_{24} + x_{20}x_{26} + x_{20}x_{27} + x_{20}x_{28} + x_{20}x_{29} + x_{20}x_{30} + x_{20}x_{31} + x_{20}x_{32} + x_{20}x_{33} + x_{20}x_{35} + x_{20}x_{40} + x_{20}x_{43} + x_{20}x_{44} + x_{20}x_{46} + x_{20}x_{48} + x_{20}x_{51} + x_{20}x_{52} + x_{20}x_{54} + x_{20}x_{55} + x_{20}x_{56} + x_{20}x_{64} + x_{21}x_{22} + x_{21}x_{23} + x_{21}x_{29} + x_{21}x_{30} + x_{21}x_{31} + x_{21}x_{34} + x_{21}x_{35} + x_{21}x_{36} + x_{21}x_{37} + x_{21}x_{42} + x_{21}x_{43} + x_{21}x_{44} + x_{21}x_{47} + x_{21}x_{49} + x_{21}x_{50} + x_{21}x_{51} + x_{21}x_{52} + x_{21}x_{53} + x_{21}x_{54} + x_{21}x_{57} + x_{21}x_{58} + x_{22}x_{23} + x_{22}x_{24} + x_{22}x_{28} + x_{22}x_{29} + x_{22}x_{30} + x_{22}x_{31} + x_{22}x_{33} + x_{22}x_{34} + x_{22}x_{40} + x_{22}x_{43} + x_{22}x_{45} + x_{22}x_{47} + x_{22}x_{52} + x_{22}x_{56} + x_{22}x_{57} + x_{22}x_{59} + x_{22}x_{60} + x_{22}x_{61} + x_{23}x_{26} + x_{23}x_{27} + x_{23}x_{28} + x_{23}x_{32} + x_{23}x_{34} + x_{23}x_{36} + x_{23}x_{37} + x_{23}x_{40} + x_{23}x_{41} + x_{23}x_{45} + x_{23}x_{46} + x_{23}x_{48} + x_{23}x_{49} + x_{23}x_{50} + x_{23}x_{52} + x_{23}x_{55} + x_{23}x_{57} + x_{23}x_{58} + x_{23}x_{59} + x_{23}x_{60} + x_{23}x_{61} + x_{23}x_{62} + x_{23}x_{63} + x_{23}x_{64} + x_{24}x_{25} + x_{24}x_{26} + x_{24}x_{27} + x_{24}x_{32} + x_{24}x_{33} + x_{24}x_{35} + x_{24}x_{36} + x_{24}x_{37} + x_{24}x_{39} + x_{24}x_{42} + x_{24}x_{44} + x_{24}x_{46} + x_{24}x_{48} + x_{24}x_{49} + x_{24}x_{50} + x_{24}x_{54} + x_{24}x_{57} + x_{24}x_{60} + x_{24}x_{64} + x_{25}x_{26} + x_{25}x_{27} + x_{25}x_{29} + x_{25}x_{33} + x_{25}x_{36} + x_{25}x_{40} + x_{25}x_{45} + x_{25}x_{50} + x_{25}x_{51} + x_{25}x_{53} + x_{25}x_{55} + x_{25}x_{61} + x_{25}x_{62} + x_{25}x_{63} + x_{26}x_{28} + x_{26}x_{30} + x_{26}x_{34} + x_{26}x_{40} + x_{26}x_{41} + x_{26}x_{42} + x_{26}x_{45} + x_{26}x_{46} + x_{26}x_{49} + x_{26}x_{50} + x_{26}x_{51} + x_{26}x_{53} + x_{26}x_{54} + x_{26}x_{57} + x_{26}x_{62} + x_{27}x_{28} + x_{27}x_{31} + x_{27}x_{34} + x_{27}x_{35} + x_{27}x_{36} + x_{27}x_{37} + x_{27}x_{39} + x_{27}x_{40} + x_{27}x_{42} + x_{27}x_{46} + x_{27}x_{47} + x_{27}x_{50} + x_{27}x_{51} + x_{27}x_{52} + x_{27}x_{53} + x_{27}x_{54} + x_{27}x_{56} + x_{27}x_{64} + x_{28}x_{31} + x_{28}x_{32} + x_{28}x_{34} + x_{28}x_{42} + x_{28}x_{44} + x_{28}x_{46} + x_{28}x_{48} + x_{28}x_{49} + x_{28}x_{53} + x_{28}x_{55} + x_{28}x_{58} + x_{28}x_{60} + x_{28}x_{61} + x_{28}x_{62} + x_{28}x_{63} + x_{28}x_{64} + x_{29}x_{32} + x_{29}x_{34} + x_{29}x_{36} + x_{29}x_{37} + x_{29}x_{41} + x_{29}x_{44} + x_{29}x_{46} + x_{29}x_{48} + x_{29}x_{51} + x_{29}x_{52} + x_{29}x_{53} + x_{29}x_{56} + x_{29}x_{58} + x_{29}x_{61} + x_{29}x_{62} + x_{30}x_{31} + x_{30}x_{32} + x_{30}x_{38} + x_{30}x_{45} + x_{30}x_{48} + x_{30}x_{52} + x_{30}x_{53} + x_{30}x_{54} + x_{30}x_{56} + x_{30}x_{60} + x_{30}x_{61} + x_{31}x_{32} + x_{31}x_{34} + x_{31}x_{37} + x_{31}x_{38} + x_{31}x_{39} + x_{31}x_{40} + x_{31}x_{41} + x_{31}x_{42} + x_{31}x_{44} + x_{31}x_{46} + x_{31}x_{48} + x_{31}x_{51} + x_{31}x_{56} + x_{31}x_{57} + x_{31}x_{60} + x_{31}x_{61} + x_{31}x_{62} + x_{31}x_{63} + x_{32}x_{33} + x_{32}x_{35} + x_{32}x_{39} + x_{32}x_{41} + x_{32}x_{43} + x_{32}x_{44} + x_{32}x_{47} + x_{32}x_{48} + x_{32}x_{52} + x_{32}x_{54} + x_{32}x_{55} + x_{32}x_{56} + x_{32}x_{57} + x_{32}x_{60} + x_{32}x_{61} + x_{32}x_{62} + x_{33}x_{34} + x_{33}x_{36} + x_{33}x_{37} + x_{33}x_{39} + x_{33}x_{40} + x_{33}x_{41} + x_{33}x_{42} + x_{33}x_{45} + x_{33}x_{47} + x_{33}x_{48} + x_{33}x_{49} + x_{33}x_{52} + x_{33}x_{53} + x_{33}x_{55} + x_{33}x_{59} + x_{33}x_{60} + x_{33}x_{61} + x_{33}x_{62} + x_{33}x_{64} + x_{34}x_{35} + x_{34}x_{39} + x_{34}x_{40} + x_{34}x_{41} + x_{34}x_{42} + x_{34}x_{43} + x_{34}x_{46} + x_{34}x_{47} + x_{34}x_{48} + x_{34}x_{52} + x_{34}x_{53} + x_{34}x_{57} + x_{34}x_{59} + x_{35}x_{38} + x_{35}x_{41} + x_{35}x_{42} + x_{35}x_{47} + x_{35}x_{48} + x_{35}x_{54} + x_{35}x_{56} + x_{35}x_{57} + x_{35}x_{60} + x_{35}x_{62} + x_{35}x_{63} + x_{35}x_{64} + x_{36}x_{37} + x_{36}x_{38} + x_{36}x_{39} + x_{36}x_{43} + x_{36}x_{45} + x_{36}x_{46} + x_{36}x_{47} + x_{36}x_{48} + x_{36}x_{52} + x_{36}x_{54} + x_{36}x_{56} + x_{36}x_{58} + x_{36}x_{59} + x_{36}x_{60} + x_{36}x_{61} + x_{36}x_{63} + x_{37}x_{40} + x_{37}x_{41} + x_{37}x_{42} + x_{37}x_{45} + x_{37}x_{48} + x_{37}x_{51} + x_{37}x_{53} + x_{37}x_{54} + x_{37}x_{55} + x_{37}x_{58} + x_{37}x_{61} + x_{37}x_{64} + x_{38}x_{40} + x_{38}x_{41} + x_{38}x_{43} + x_{38}x_{44} + x_{38}x_{46} + x_{38}x_{48} + x_{38}x_{50} + x_{38}x_{53} + x_{38}x_{55} + x_{38}x_{58} + x_{38}x_{59} + x_{38}x_{61} + x_{38}x_{63} + x_{39}x_{40} + x_{39}x_{44} + x_{39}x_{46} + x_{39}x_{50} + x_{39}x_{51} + x_{39}x_{52} + x_{39}x_{56} + x_{39}x_{57} + x_{39}x_{59} + x_{39}x_{61} + x_{40}x_{41} + x_{40}x_{42} + x_{40}x_{43} + x_{40}x_{45} + x_{40}x_{48} + x_{40}x_{49} + x_{40}x_{53} + x_{40}x_{54} + x_{40}x_{56} + x_{40}x_{58} + x_{40}x_{60} + x_{40}x_{61} + x_{40}x_{63} + x_{41}x_{46} + x_{41}x_{48} + x_{41}x_{49} + x_{41}x_{50} + x_{41}x_{52} + x_{41}x_{53} + x_{41}x_{54} + x_{41}x_{55} + x_{41}x_{58} + x_{41}x_{59} + x_{41}x_{62} + x_{41}x_{63} + x_{41}x_{64} + x_{42}x_{43} + x_{42}x_{47} + x_{42}x_{48} + x_{42}x_{49} + x_{42}x_{55} + x_{42}x_{56} + x_{42}x_{59} + x_{42}x_{60} + x_{42}x_{61} + x_{42}x_{62} + x_{42}x_{64} + x_{43}x_{46} + x_{43}x_{51} + x_{43}x_{52} + x_{43}x_{53} + x_{43}x_{54} + x_{43}x_{55} + x_{43}x_{60} + x_{43}x_{62} + x_{43}x_{63} + x_{44}x_{47} + x_{44}x_{50} + x_{44}x_{58} + x_{44}x_{59} + x_{44}x_{60} + x_{44}x_{62} + x_{44}x_{64} + x_{45}x_{47} + x_{45}x_{48} + x_{45}x_{49} + x_{45}x_{50} + x_{45}x_{51} + x_{45}x_{54} + x_{45}x_{55} + x_{45}x_{56} + x_{45}x_{57} + x_{45}x_{63} + x_{46}x_{48} + x_{46}x_{52} + x_{46}x_{54} + x_{46}x_{58} + x_{46}x_{60} + x_{46}x_{63} + x_{46}x_{64} + x_{47}x_{48} + x_{47}x_{57} + x_{47}x_{61} + x_{47}x_{62} + x_{47}x_{64} + x_{48}x_{49} + x_{48}x_{52} + x_{48}x_{55} + x_{48}x_{56} + x_{48}x_{57} + x_{48}x_{59} + x_{48}x_{60} + x_{48}x_{64} + x_{49}x_{51} + x_{49}x_{52} + x_{49}x_{57} + x_{49}x_{62} + x_{49}x_{63} + x_{49}x_{64} + x_{50}x_{51} + x_{50}x_{58} + x_{50}x_{59} + x_{50}x_{61} + x_{50}x_{62} + x_{50}x_{63} + x_{50}x_{64} + x_{51}x_{53} + x_{51}x_{54} + x_{51}x_{55} + x_{51}x_{56} + x_{51}x_{57} + x_{51}x_{58} + x_{51}x_{59} + x_{51}x_{60} + x_{51}x_{61} + x_{51}x_{62} + x_{51}x_{63} + x_{51}x_{64} + x_{52}x_{53} + x_{52}x_{58} + x_{52}x_{60} + x_{52}x_{61} + x_{52}x_{62} + x_{52}x_{64} + x_{53}x_{54} + x_{53}x_{55} + x_{53}x_{56} + x_{53}x_{57} + x_{53}x_{58} + x_{53}x_{60} + x_{53}x_{61} + x_{54}x_{58} + x_{54}x_{59} + x_{54}x_{60} + x_{54}x_{62} + x_{54}x_{64} + x_{55}x_{56} + x_{55}x_{59} + x_{55}x_{60} + x_{55}x_{62} + x_{55}x_{64} + x_{56}x_{58} + x_{56}x_{63} + x_{56}x_{64} + x_{57}x_{59} + x_{57}x_{60} + x_{57}x_{61} + x_{57}x_{62} + x_{58}x_{59} + x_{58}x_{62} + x_{58}x_{63} + x_{58}x_{64} + x_{59}x_{61} + x_{59}x_{62} + x_{60}x_{62} + x_{60}x_{64} + x_{61}x_{63} + x_{61}x_{64} + x_{62}x_{63} + x_{1} + x_{5} + x_{6} + x_{7} + x_{8} + x_{9} + x_{12} + x_{16} + x_{19} + x_{20} + x_{21} + x_{23} + x_{24} + x_{27} + x_{29} + x_{31} + x_{34} + x_{36} + x_{42} + x_{43} + x_{44} + x_{46} + x_{47} + x_{49} + x_{51} + x_{52} + x_{55} + x_{56} + x_{58} + x_{61} + x_{63} + x_{64} + 1$

$y_{16} = x_{1}x_{2} + x_{1}x_{3} + x_{1}x_{6} + x_{1}x_{7} + x_{1}x_{8} + x_{1}x_{10} + x_{1}x_{14} + x_{1}x_{15} + x_{1}x_{18} + x_{1}x_{19} + x_{1}x_{23} + x_{1}x_{25} + x_{1}x_{26} + x_{1}x_{28} + x_{1}x_{35} + x_{1}x_{41} + x_{1}x_{42} + x_{1}x_{43} + x_{1}x_{49} + x_{1}x_{50} + x_{1}x_{52} + x_{1}x_{53} + x_{1}x_{54} + x_{1}x_{55} + x_{1}x_{59} + x_{1}x_{62} + x_{1}x_{63} + x_{2}x_{3} + x_{2}x_{6} + x_{2}x_{7} + x_{2}x_{10} + x_{2}x_{11} + x_{2}x_{15} + x_{2}x_{18} + x_{2}x_{21} + x_{2}x_{25} + x_{2}x_{26} + x_{2}x_{27} + x_{2}x_{28} + x_{2}x_{29} + x_{2}x_{30} + x_{2}x_{31} + x_{2}x_{32} + x_{2}x_{33} + x_{2}x_{34} + x_{2}x_{41} + x_{2}x_{42} + x_{2}x_{45} + x_{2}x_{46} + x_{2}x_{48} + x_{2}x_{49} + x_{2}x_{50} + x_{2}x_{52} + x_{2}x_{55} + x_{2}x_{57} + x_{2}x_{58} + x_{2}x_{64} + x_{3}x_{6} + x_{3}x_{7} + x_{3}x_{8} + x_{3}x_{10} + x_{3}x_{11} + x_{3}x_{12} + x_{3}x_{14} + x_{3}x_{15} + x_{3}x_{18} + x_{3}x_{20} + x_{3}x_{21} + x_{3}x_{23} + x_{3}x_{24} + x_{3}x_{26} + x_{3}x_{27} + x_{3}x_{31} + x_{3}x_{32} + x_{3}x_{34} + x_{3}x_{35} + x_{3}x_{36} + x_{3}x_{37} + x_{3}x_{38} + x_{3}x_{39} + x_{3}x_{40} + x_{3}x_{41} + x_{3}x_{43} + x_{3}x_{44} + x_{3}x_{46} + x_{3}x_{50} + x_{3}x_{51} + x_{3}x_{54} + x_{3}x_{57} + x_{3}x_{58} + x_{3}x_{59} + x_{3}x_{60} + x_{3}x_{63} + x_{3}x_{64} + x_{4}x_{6} + x_{4}x_{7} + x_{4}x_{10} + x_{4}x_{12} + x_{4}x_{13} + x_{4}x_{14} + x_{4}x_{15} + x_{4}x_{18} + x_{4}x_{20} + x_{4}x_{21} + x_{4}x_{24} + x_{4}x_{26} + x_{4}x_{27} + x_{4}x_{28} + x_{4}x_{32} + x_{4}x_{33} + x_{4}x_{34} + x_{4}x_{35} + x_{4}x_{36} + x_{4}x_{39} + x_{4}x_{40} + x_{4}x_{41} + x_{4}x_{43} + x_{4}x_{47} + x_{4}x_{48} + x_{4}x_{53} + x_{4}x_{54} + x_{4}x_{61} + x_{4}x_{63} + x_{4}x_{64} + x_{5}x_{6} + x_{5}x_{7} + x_{5}x_{9} + x_{5}x_{13} + x_{5}x_{14} + x_{5}x_{15} + x_{5}x_{19} + x_{5}x_{21} + x_{5}x_{24} + x_{5}x_{30} + x_{5}x_{31} + x_{5}x_{36} + x_{5}x_{37} + x_{5}x_{42} + x_{5}x_{44} + x_{5}x_{45} + x_{5}x_{47} + x_{5}x_{48} + x_{5}x_{49} + x_{5}x_{50} + x_{5}x_{51} + x_{5}x_{52} + x_{5}x_{53} + x_{5}x_{57} + x_{5}x_{58} + x_{5}x_{60} + x_{5}x_{61} + x_{6}x_{7} + x_{6}x_{13} + x_{6}x_{16} + x_{6}x_{18} + x_{6}x_{19} + x_{6}x_{21} + x_{6}x_{24} + x_{6}x_{27} + x_{6}x_{28} + x_{6}x_{31} + x_{6}x_{32} + x_{6}x_{33} + x_{6}x_{34} + x_{6}x_{36} + x_{6}x_{38} + x_{6}x_{39} + x_{6}x_{41} + x_{6}x_{42} + x_{6}x_{46} + x_{6}x_{48} + x_{6}x_{49} + x_{6}x_{50} + x_{6}x_{55} + x_{6}x_{56} + x_{6}x_{57} + x_{6}x_{59} + x_{7}x_{8} + x_{7}x_{11} + x_{7}x_{13} + x_{7}x_{15} + x_{7}x_{16} + x_{7}x_{17} + x_{7}x_{18} + x_{7}x_{19} + x_{7}x_{21} + x_{7}x_{22} + x_{7}x_{27} + x_{7}x_{28} + x_{7}x_{33} + x_{7}x_{34} + x_{7}x_{42} + x_{7}x_{45} + x_{7}x_{46} + x_{7}x_{48} + x_{7}x_{50} + x_{7}x_{51} + x_{7}x_{52} + x_{7}x_{53} + x_{7}x_{58} + x_{8}x_{9} + x_{8}x_{11} + x_{8}x_{12} + x_{8}x_{15} + x_{8}x_{17} + x_{8}x_{18} + x_{8}x_{19} + x_{8}x_{20} + x_{8}x_{21} + x_{8}x_{24} + x_{8}x_{25} + x_{8}x_{31} + x_{8}x_{33} + x_{8}x_{35} + x_{8}x_{37} + x_{8}x_{38} + x_{8}x_{39} + x_{8}x_{40} + x_{8}x_{41} + x_{8}x_{42} + x_{8}x_{45} + x_{8}x_{46} + x_{8}x_{51} + x_{8}x_{52} + x_{8}x_{54} + x_{8}x_{55} + x_{8}x_{57} + x_{8}x_{60} + x_{8}x_{63} + x_{8}x_{64} + x_{9}x_{10} + x_{9}x_{11} + x_{9}x_{12} + x_{9}x_{14} + x_{9}x_{15} + x_{9}x_{16} + x_{9}x_{17} + x_{9}x_{18} + x_{9}x_{20} + x_{9}x_{21} + x_{9}x_{22} + x_{9}x_{25} + x_{9}x_{28} + x_{9}x_{29} + x_{9}x_{30} + x_{9}x_{31} + x_{9}x_{32} + x_{9}x_{34} + x_{9}x_{36} + x_{9}x_{38} + x_{9}x_{43} + x_{9}x_{45} + x_{9}x_{46} + x_{9}x_{49} + x_{9}x_{50} + x_{9}x_{53} + x_{9}x_{56} + x_{9}x_{57} + x_{9}x_{58} + x_{9}x_{59} + x_{9}x_{60} + x_{9}x_{61} + x_{9}x_{62} + x_{10}x_{12} + x_{10}x_{14} + x_{10}x_{15} + x_{10}x_{16} + x_{10}x_{19} + x_{10}x_{23} + x_{10}x_{24} + x_{10}x_{26} + x_{10}x_{29} + x_{10}x_{32} + x_{10}x_{36} + x_{10}x_{37} + x_{10}x_{41} + x_{10}x_{42} + x_{10}x_{43} + x_{10}x_{45} + x_{10}x_{46} + x_{10}x_{47} + x_{10}x_{50} + x_{10}x_{51} + x_{10}x_{52} + x_{10}x_{54} + x_{10}x_{56} + x_{10}x_{61} + x_{10}x_{62} + x_{10}x_{64} + x_{11}x_{12} + x_{11}x_{17} + x_{11}x_{21} + x_{11}x_{22} + x_{11}x_{23} + x_{11}x_{24} + x_{11}x_{28} + x_{11}x_{31} + x_{11}x_{32} + x_{11}x_{36} + x_{11}x_{37} + x_{11}x_{39} + x_{11}x_{42} + x_{11}x_{43} + x_{11}x_{45} + x_{11}x_{46} + x_{11}x_{47} + x_{11}x_{49} + x_{11}x_{60} + x_{11}x_{61} + x_{11}x_{62} + x_{11}x_{63} + x_{11}x_{64} + x_{12}x_{13} + x_{12}x_{14} + x_{12}x_{16} + x_{12}x_{17} + x_{12}x_{20} + x_{12}x_{22} + x_{12}x_{24} + x_{12}x_{25} + x_{12}x_{28} + x_{12}x_{36} + x_{12}x_{38} + x_{12}x_{39} + x_{12}x_{40} + x_{12}x_{45} + x_{12}x_{46} + x_{12}x_{48} + x_{12}x_{50} + x_{12}x_{52} + x_{12}x_{53} + x_{12}x_{58} + x_{12}x_{59} + x_{12}x_{60} + x_{12}x_{63} + x_{13}x_{14} + x_{13}x_{15} + x_{13}x_{16} + x_{13}x_{18} + x_{13}x_{21} + x_{13}x_{28} + x_{13}x_{30} + x_{13}x_{31} + x_{13}x_{35} + x_{13}x_{36} + x_{13}x_{38} + x_{13}x_{40} + x_{13}x_{41} + x_{13}x_{44} + x_{13}x_{45} + x_{13}x_{49} + x_{13}x_{54} + x_{13}x_{58} + x_{13}x_{60} + x_{13}x_{62} + x_{13}x_{63} + x_{14}x_{15} + x_{14}x_{16} + x_{14}x_{18} + x_{14}x_{20} + x_{14}x_{22} + x_{14}x_{23} + x_{14}x_{24} + x_{14}x_{25} + x_{14}x_{26} + x_{14}x_{29} + x_{14}x_{30} + x_{14}x_{33} + x_{14}x_{35} + x_{14}x_{36} + x_{14}x_{37} + x_{14}x_{40} + x_{14}x_{41} + x_{14}x_{43} + x_{14}x_{45} + x_{14}x_{46} + x_{14}x_{47} + x_{14}x_{48} + x_{14}x_{49} + x_{14}x_{54} + x_{14}x_{55} + x_{14}x_{58} + x_{14}x_{59} + x_{14}x_{61} + x_{14}x_{63} + x_{15}x_{16} + x_{15}x_{17} + x_{15}x_{18} + x_{15}x_{19} + x_{15}x_{21} + x_{15}x_{22} + x_{15}x_{24} + x_{15}x_{25} + x_{15}x_{26} + x_{15}x_{27} + x_{15}x_{28} + x_{15}x_{29} + x_{15}x_{30} + x_{15}x_{31} + x_{15}x_{32} + x_{15}x_{33} + x_{15}x_{35} + x_{15}x_{37} + x_{15}x_{38} + x_{15}x_{39} + x_{15}x_{41} + x_{15}x_{43} + x_{15}x_{50} + x_{15}x_{53} + x_{15}x_{55} + x_{15}x_{57} + x_{15}x_{58} + x_{15}x_{59} + x_{15}x_{61} + x_{15}x_{63} + x_{16}x_{17} + x_{16}x_{19} + x_{16}x_{21} + x_{16}x_{28} + x_{16}x_{29} + x_{16}x_{30} + x_{16}x_{31} + x_{16}x_{37} + x_{16}x_{38} + x_{16}x_{39} + x_{16}x_{44} + x_{16}x_{45} + x_{16}x_{46} + x_{16}x_{47} + x_{16}x_{49} + x_{16}x_{52} + x_{16}x_{55} + x_{16}x_{56} + x_{16}x_{57} + x_{16}x_{58} + x_{16}x_{60} + x_{16}x_{61} + x_{17}x_{18} + x_{17}x_{22} + x_{17}x_{23} + x_{17}x_{24} + x_{17}x_{25} + x_{17}x_{26} + x_{17}x_{27} + x_{17}x_{28} + x_{17}x_{29} + x_{17}x_{31} + x_{17}x_{35} + x_{17}x_{37} + x_{17}x_{38} + x_{17}x_{39} + x_{17}x_{40} + x_{17}x_{42} + x_{17}x_{43} + x_{17}x_{45} + x_{17}x_{47} + x_{17}x_{55} + x_{17}x_{56} + x_{17}x_{58} + x_{18}x_{19} + x_{18}x_{21} + x_{18}x_{22} + x_{18}x_{23} + x_{18}x_{24} + x_{18}x_{28} + x_{18}x_{30} + x_{18}x_{31} + x_{18}x_{32} + x_{18}x_{34} + x_{18}x_{35} + x_{18}x_{36} + x_{18}x_{37} + x_{18}x_{39} + x_{18}x_{43} + x_{18}x_{44} + x_{18}x_{46} + x_{18}x_{47} + x_{18}x_{48} + x_{18}x_{50} + x_{18}x_{51} + x_{18}x_{54} + x_{18}x_{55} + x_{18}x_{60} + x_{18}x_{62} + x_{18}x_{64} + x_{19}x_{23} + x_{19}x_{27} + x_{19}x_{28} + x_{19}x_{29} + x_{19}x_{31} + x_{19}x_{33} + x_{19}x_{34} + x_{19}x_{35} + x_{19}x_{36} + x_{19}x_{37} + x_{19}x_{38} + x_{19}x_{40} + x_{19}x_{41} + x_{19}x_{45} + x_{19}x_{46} + x_{19}x_{50} + x_{19}x_{51} + x_{19}x_{52} + x_{19}x_{55} + x_{19}x_{59} + x_{19}x_{62} + x_{19}x_{64} + x_{20}x_{21} + x_{20}x_{25} + x_{20}x_{27} + x_{20}x_{28} + x_{20}x_{29} + x_{20}x_{31} + x_{20}x_{32} + x_{20}x_{36} + x_{20}x_{39} + x_{20}x_{41} + x_{20}x_{42} + x_{20}x_{43} + x_{20}x_{45} + x_{20}x_{46} + x_{20}x_{47} + x_{20}x_{48} + x_{20}x_{49} + x_{20}x_{50} + x_{20}x_{51} + x_{20}x_{52} + x_{20}x_{54} + x_{20}x_{57} + x_{20}x_{58} + x_{20}x_{59} + x_{20}x_{60} + x_{20}x_{61} + x_{20}x_{62} + x_{20}x_{63} + x_{20}x_{64} + x_{21}x_{24} + x_{21}x_{27} + x_{21}x_{28} + x_{21}x_{29} + x_{21}x_{30} + x_{21}x_{31} + x_{21}x_{33} + x_{21}x_{34} + x_{21}x_{35} + x_{21}x_{36} + x_{21}x_{38} + x_{21}x_{40} + x_{21}x_{41} + x_{21}x_{42} + x_{21}x_{47} + x_{21}x_{48} + x_{21}x_{49} + x_{21}x_{50} + x_{21}x_{53} + x_{21}x_{54} + x_{21}x_{59} + x_{21}x_{64} + x_{22}x_{28} + x_{22}x_{30} + x_{22}x_{31} + x_{22}x_{32} + x_{22}x_{33} + x_{22}x_{34} + x_{22}x_{44} + x_{22}x_{45} + x_{22}x_{46} + x_{22}x_{48} + x_{22}x_{49} + x_{22}x_{50} + x_{22}x_{55} + x_{22}x_{60} + x_{22}x_{62} + x_{22}x_{64} + x_{23}x_{24} + x_{23}x_{27} + x_{23}x_{29} + x_{23}x_{31} + x_{23}x_{33} + x_{23}x_{34} + x_{23}x_{35} + x_{23}x_{37} + x_{23}x_{38} + x_{23}x_{40} + x_{23}x_{42} + x_{23}x_{43} + x_{23}x_{44} + x_{23}x_{45} + x_{23}x_{47} + x_{23}x_{48} + x_{23}x_{49} + x_{23}x_{51} + x_{23}x_{52} + x_{23}x_{53} + x_{23}x_{54} + x_{23}x_{56} + x_{23}x_{57} + x_{23}x_{59} + x_{23}x_{60} + x_{23}x_{63} + x_{24}x_{25} + x_{24}x_{28} + x_{24}x_{29} + x_{24}x_{30} + x_{24}x_{31} + x_{24}x_{33} + x_{24}x_{35} + x_{24}x_{39} + x_{24}x_{40} + x_{24}x_{41} + x_{24}x_{42} + x_{24}x_{44} + x_{24}x_{46} + x_{24}x_{47} + x_{24}x_{52} + x_{24}x_{53} + x_{24}x_{56} + x_{24}x_{57} + x_{24}x_{58} + x_{24}x_{59} + x_{24}x_{60} + x_{24}x_{62} + x_{24}x_{64} + x_{25}x_{26} + x_{25}x_{28} + x_{25}x_{30} + x_{25}x_{31} + x_{25}x_{35} + x_{25}x_{38} + x_{25}x_{44} + x_{25}x_{48} + x_{25}x_{50} + x_{25}x_{53} + x_{25}x_{54} + x_{25}x_{55} + x_{25}x_{56} + x_{25}x_{57} + x_{25}x_{58} + x_{25}x_{59} + x_{25}x_{63} + x_{26}x_{27} + x_{26}x_{28} + x_{26}x_{31} + x_{26}x_{35} + x_{26}x_{37} + x_{26}x_{38} + x_{26}x_{45} + x_{26}x_{48} + x_{26}x_{49} + x_{26}x_{53} + x_{26}x_{54} + x_{26}x_{56} + x_{26}x_{57} + x_{26}x_{58} + x_{26}x_{60} + x_{27}x_{28} + x_{27}x_{31} + x_{27}x_{36} + x_{27}x_{43} + x_{27}x_{44} + x_{27}x_{46} + x_{27}x_{47} + x_{27}x_{50} + x_{27}x_{52} + x_{27}x_{53} + x_{27}x_{58} + x_{27}x_{59} + x_{27}x_{60} + x_{27}x_{61} + x_{27}x_{62} + x_{27}x_{64} + x_{28}x_{29} + x_{28}x_{31} + x_{28}x_{32} + x_{28}x_{34} + x_{28}x_{35} + x_{28}x_{37} + x_{28}x_{39} + x_{28}x_{45} + x_{28}x_{46} + x_{28}x_{47} + x_{28}x_{48} + x_{28}x_{49} + x_{28}x_{50} + x_{28}x_{52} + x_{28}x_{53} + x_{28}x_{56} + x_{28}x_{64} + x_{29}x_{31} + x_{29}x_{32} + x_{29}x_{33} + x_{29}x_{34} + x_{29}x_{39} + x_{29}x_{41} + x_{29}x_{42} + x_{29}x_{46} + x_{29}x_{47} + x_{29}x_{51} + x_{29}x_{53} + x_{29}x_{55} + x_{29}x_{56} + x_{29}x_{57} + x_{29}x_{58} + x_{29}x_{60} + x_{29}x_{63} + x_{30}x_{33} + x_{30}x_{34} + x_{30}x_{35} + x_{30}x_{37} + x_{30}x_{39} + x_{30}x_{40} + x_{30}x_{41} + x_{30}x_{44} + x_{30}x_{45} + x_{30}x_{47} + x_{30}x_{49} + x_{30}x_{54} + x_{30}x_{55} + x_{30}x_{56} + x_{30}x_{57} + x_{30}x_{60} + x_{30}x_{62} + x_{30}x_{63} + x_{31}x_{35} + x_{31}x_{36} + x_{31}x_{38} + x_{31}x_{39} + x_{31}x_{41} + x_{31}x_{42} + x_{31}x_{44} + x_{31}x_{45} + x_{31}x_{46} + x_{31}x_{47} + x_{31}x_{48} + x_{31}x_{49} + x_{31}x_{52} + x_{31}x_{55} + x_{31}x_{56} + x_{31}x_{57} + x_{31}x_{59} + x_{31}x_{61} + x_{31}x_{62} + x_{31}x_{63} + x_{31}x_{64} + x_{32}x_{33} + x_{32}x_{34} + x_{32}x_{36} + x_{32}x_{38} + x_{32}x_{45} + x_{32}x_{46} + x_{32}x_{52} + x_{32}x_{54} + x_{32}x_{57} + x_{32}x_{58} + x_{32}x_{60} + x_{32}x_{61} + x_{32}x_{64} + x_{33}x_{34} + x_{33}x_{37} + x_{33}x_{38} + x_{33}x_{39} + x_{33}x_{40} + x_{33}x_{41} + x_{33}x_{43} + x_{33}x_{47} + x_{33}x_{49} + x_{33}x_{50} + x_{33}x_{51} + x_{33}x_{54} + x_{33}x_{55} + x_{33}x_{57} + x_{33}x_{59} + x_{33}x_{60} + x_{33}x_{64} + x_{34}x_{35} + x_{34}x_{36} + x_{34}x_{37} + x_{34}x_{42} + x_{34}x_{44} + x_{34}x_{45} + x_{34}x_{47} + x_{34}x_{48} + x_{34}x_{57} + x_{34}x_{62} + x_{34}x_{64} + x_{35}x_{37} + x_{35}x_{38} + x_{35}x_{39} + x_{35}x_{42} + x_{35}x_{47} + x_{35}x_{49} + x_{35}x_{50} + x_{35}x_{51} + x_{35}x_{53} + x_{35}x_{54} + x_{35}x_{56} + x_{35}x_{57} + x_{35}x_{63} + x_{36}x_{37} + x_{36}x_{38} + x_{36}x_{39} + x_{36}x_{46} + x_{36}x_{48} + x_{36}x_{51} + x_{36}x_{52} + x_{36}x_{53} + x_{36}x_{56} + x_{36}x_{58} + x_{36}x_{59} + x_{36}x_{60} + x_{36}x_{62} + x_{36}x_{64} + x_{37}x_{38} + x_{37}x_{40} + x_{37}x_{41} + x_{37}x_{43} + x_{37}x_{46} + x_{37}x_{47} + x_{37}x_{50} + x_{37}x_{51} + x_{37}x_{52} + x_{37}x_{54} + x_{37}x_{56} + x_{37}x_{57} + x_{37}x_{62} + x_{37}x_{64} + x_{38}x_{39} + x_{38}x_{42} + x_{38}x_{45} + x_{38}x_{48} + x_{38}x_{49} + x_{38}x_{50} + x_{38}x_{53} + x_{38}x_{54} + x_{38}x_{55} + x_{38}x_{57} + x_{38}x_{58} + x_{38}x_{59} + x_{38}x_{60} + x_{39}x_{40} + x_{39}x_{41} + x_{39}x_{42} + x_{39}x_{43} + x_{39}x_{46} + x_{39}x_{47} + x_{39}x_{51} + x_{39}x_{53} + x_{39}x_{54} + x_{39}x_{56} + x_{39}x_{58} + x_{39}x_{60} + x_{40}x_{41} + x_{40}x_{42} + x_{40}x_{43} + x_{40}x_{44} + x_{40}x_{45} + x_{40}x_{46} + x_{40}x_{47} + x_{40}x_{49} + x_{40}x_{50} + x_{40}x_{52} + x_{40}x_{53} + x_{40}x_{54} + x_{40}x_{55} + x_{40}x_{58} + x_{40}x_{62} + x_{40}x_{63} + x_{41}x_{44} + x_{41}x_{47} + x_{41}x_{49} + x_{41}x_{50} + x_{41}x_{51} + x_{41}x_{53} + x_{41}x_{55} + x_{41}x_{59} + x_{41}x_{60} + x_{41}x_{61} + x_{41}x_{62} + x_{41}x_{63} + x_{42}x_{45} + x_{42}x_{47} + x_{42}x_{48} + x_{42}x_{50} + x_{42}x_{54} + x_{42}x_{56} + x_{42}x_{57} + x_{42}x_{59} + x_{42}x_{63} + x_{43}x_{46} + x_{43}x_{47} + x_{43}x_{50} + x_{43}x_{54} + x_{43}x_{56} + x_{43}x_{59} + x_{43}x_{64} + x_{44}x_{47} + x_{44}x_{48} + x_{44}x_{51} + x_{44}x_{52} + x_{44}x_{54} + x_{44}x_{55} + x_{44}x_{59} + x_{44}x_{61} + x_{44}x_{62} + x_{44}x_{63} + x_{45}x_{48} + x_{45}x_{49} + x_{45}x_{50} + x_{45}x_{51} + x_{45}x_{52} + x_{45}x_{57} + x_{45}x_{59} + x_{45}x_{62} + x_{45}x_{64} + x_{46}x_{48} + x_{46}x_{50} + x_{46}x_{54} + x_{46}x_{55} + x_{46}x_{56} + x_{46}x_{58} + x_{46}x_{59} + x_{46}x_{61} + x_{46}x_{62} + x_{46}x_{64} + x_{47}x_{48} + x_{47}x_{49} + x_{47}x_{50} + x_{47}x_{51} + x_{47}x_{52} + x_{47}x_{56} + x_{47}x_{58} + x_{47}x_{60} + x_{47}x_{61} + x_{47}x_{62} + x_{47}x_{64} + x_{48}x_{49} + x_{48}x_{50} + x_{48}x_{52} + x_{48}x_{53} + x_{48}x_{57} + x_{48}x_{60} + x_{48}x_{61} + x_{48}x_{63} + x_{48}x_{64} + x_{49}x_{50} + x_{49}x_{54} + x_{49}x_{55} + x_{49}x_{57} + x_{49}x_{58} + x_{49}x_{59} + x_{49}x_{63} + x_{50}x_{52} + x_{50}x_{53} + x_{50}x_{54} + x_{50}x_{55} + x_{50}x_{56} + x_{50}x_{57} + x_{50}x_{58} + x_{50}x_{60} + x_{50}x_{61} + x_{50}x_{63} + x_{50}x_{64} + x_{51}x_{52} + x_{51}x_{55} + x_{51}x_{56} + x_{51}x_{60} + x_{51}x_{64} + x_{52}x_{53} + x_{52}x_{54} + x_{52}x_{55} + x_{52}x_{59} + x_{52}x_{61} + x_{52}x_{63} + x_{53}x_{55} + x_{53}x_{56} + x_{53}x_{57} + x_{53}x_{58} + x_{53}x_{61} + x_{53}x_{62} + x_{53}x_{63} + x_{53}x_{64} + x_{54}x_{57} + x_{54}x_{58} + x_{54}x_{59} + x_{54}x_{62} + x_{54}x_{64} + x_{55}x_{58} + x_{55}x_{60} + x_{55}x_{61} + x_{55}x_{62} + x_{56}x_{58} + x_{56}x_{59} + x_{56}x_{61} + x_{56}x_{63} + x_{57}x_{60} + x_{57}x_{61} + x_{57}x_{62} + x_{57}x_{64} + x_{58}x_{59} + x_{58}x_{61} + x_{58}x_{63} + x_{60}x_{63} + x_{60}x_{64} + x_{61}x_{63} + x_{61}x_{64} + x_{62}x_{63} + x_{62}x_{64} + x_{1} + x_{2} + x_{3} + x_{6} + x_{7} + x_{11} + x_{12} + x_{13} + x_{18} + x_{21} + x_{22} + x_{23} + x_{24} + x_{28} + x_{29} + x_{30} + x_{33} + x_{36} + x_{37} + x_{38} + x_{39} + x_{40} + x_{41} + x_{43} + x_{45} + x_{47} + x_{50} + x_{51} + x_{54} + x_{56} + x_{58} + x_{59} + x_{61} + x_{62} + x_{63} + x_{64} + 1$

$y_{17} = x_{1}x_{4} + x_{1}x_{6} + x_{1}x_{8} + x_{1}x_{9} + x_{1}x_{10} + x_{1}x_{12} + x_{1}x_{13} + x_{1}x_{14} + x_{1}x_{15} + x_{1}x_{18} + x_{1}x_{19} + x_{1}x_{22} + x_{1}x_{23} + x_{1}x_{24} + x_{1}x_{26} + x_{1}x_{28} + x_{1}x_{33} + x_{1}x_{34} + x_{1}x_{35} + x_{1}x_{36} + x_{1}x_{37} + x_{1}x_{39} + x_{1}x_{44} + x_{1}x_{47} + x_{1}x_{48} + x_{1}x_{49} + x_{1}x_{52} + x_{1}x_{53} + x_{1}x_{56} + x_{1}x_{60} + x_{1}x_{61} + x_{1}x_{62} + x_{1}x_{63} + x_{1}x_{64} + x_{2}x_{5} + x_{2}x_{7} + x_{2}x_{10} + x_{2}x_{11} + x_{2}x_{14} + x_{2}x_{16} + x_{2}x_{17} + x_{2}x_{19} + x_{2}x_{20} + x_{2}x_{22} + x_{2}x_{23} + x_{2}x_{24} + x_{2}x_{25} + x_{2}x_{26} + x_{2}x_{28} + x_{2}x_{36} + x_{2}x_{37} + x_{2}x_{38} + x_{2}x_{39} + x_{2}x_{45} + x_{2}x_{47} + x_{2}x_{49} + x_{2}x_{50} + x_{2}x_{51} + x_{2}x_{53} + x_{2}x_{57} + x_{2}x_{58} + x_{2}x_{61} + x_{2}x_{62} + x_{2}x_{63} + x_{2}x_{64} + x_{3}x_{6} + x_{3}x_{9} + x_{3}x_{10} + x_{3}x_{11} + x_{3}x_{12} + x_{3}x_{14} + x_{3}x_{15} + x_{3}x_{17} + x_{3}x_{18} + x_{3}x_{19} + x_{3}x_{20} + x_{3}x_{21} + x_{3}x_{22} + x_{3}x_{23} + x_{3}x_{26} + x_{3}x_{28} + x_{3}x_{31} + x_{3}x_{32} + x_{3}x_{34} + x_{3}x_{36} + x_{3}x_{37} + x_{3}x_{39} + x_{3}x_{40} + x_{3}x_{42} + x_{3}x_{43} + x_{3}x_{44} + x_{3}x_{46} + x_{3}x_{48} + x_{3}x_{49} + x_{3}x_{50} + x_{3}x_{52} + x_{3}x_{53} + x_{3}x_{58} + x_{3}x_{59} + x_{3}x_{60} + x_{3}x_{61} + x_{3}x_{63} + x_{3}x_{64} + x_{4}x_{5} + x_{4}x_{7} + x_{4}x_{8} + x_{4}x_{9} + x_{4}x_{10} + x_{4}x_{18} + x_{4}x_{19} + x_{4}x_{23} + x_{4}x_{24} + x_{4}x_{25} + x_{4}x_{26} + x_{4}x_{29} + x_{4}x_{31} + x_{4}x_{33} + x_{4}x_{34} + x_{4}x_{35} + x_{4}x_{36} + x_{4}x_{37} + x_{4}x_{39} + x_{4}x_{40} + x_{4}x_{41} + x_{4}x_{43} + x_{4}x_{49} + x_{4}x_{50} + x_{4}x_{53} + x_{4}x_{54} + x_{4}x_{57} + x_{4}x_{58} + x_{4}x_{59} + x_{4}x_{61} + x_{4}x_{63} + x_{5}x_{6} + x_{5}x_{7} + x_{5}x_{11} + x_{5}x_{13} + x_{5}x_{17} + x_{5}x_{18} + x_{5}x_{19} + x_{5}x_{21} + x_{5}x_{22} + x_{5}x_{24} + x_{5}x_{25} + x_{5}x_{26} + x_{5}x_{27} + x_{5}x_{31} + x_{5}x_{32} + x_{5}x_{36} + x_{5}x_{37} + x_{5}x_{41} + x_{5}x_{45} + x_{5}x_{46} + x_{5}x_{47} + x_{5}x_{48} + x_{5}x_{49} + x_{5}x_{50} + x_{5}x_{51} + x_{5}x_{54} + x_{5}x_{57} + x_{5}x_{59} + x_{5}x_{60} + x_{5}x_{61} + x_{5}x_{62} + x_{5}x_{63} + x_{6}x_{7} + x_{6}x_{8} + x_{6}x_{9} + x_{6}x_{11} + x_{6}x_{12} + x_{6}x_{13} + x_{6}x_{15} + x_{6}x_{16} + x_{6}x_{17} + x_{6}x_{18} + x_{6}x_{19} + x_{6}x_{20} + x_{6}x_{21} + x_{6}x_{22} + x_{6}x_{23} + x_{6}x_{25} + x_{6}x_{26} + x_{6}x_{27} + x_{6}x_{28} + x_{6}x_{31} + x_{6}x_{37} + x_{6}x_{38} + x_{6}x_{41} + x_{6}x_{42} + x_{6}x_{45} + x_{6}x_{46} + x_{6}x_{47} + x_{6}x_{48} + x_{6}x_{49} + x_{6}x_{52} + x_{6}x_{55} + x_{6}x_{59} + x_{6}x_{60} + x_{6}x_{61} + x_{6}x_{62} + x_{6}x_{63} + x_{7}x_{9} + x_{7}x_{12} + x_{7}x_{13} + x_{7}x_{14} + x_{7}x_{16} + x_{7}x_{17} + x_{7}x_{20} + x_{7}x_{24} + x_{7}x_{25} + x_{7}x_{26} + x_{7}x_{28} + x_{7}x_{29} + x_{7}x_{30} + x_{7}x_{32} + x_{7}x_{34} + x_{7}x_{37} + x_{7}x_{38} + x_{7}x_{43} + x_{7}x_{44} + x_{7}x_{46} + x_{7}x_{48} + x_{7}x_{50} + x_{7}x_{51} + x_{7}x_{56} + x_{7}x_{58} + x_{7}x_{63} + x_{8}x_{9} + x_{8}x_{16} + x_{8}x_{22} + x_{8}x_{23} + x_{8}x_{24} + x_{8}x_{25} + x_{8}x_{26} + x_{8}x_{27} + x_{8}x_{32} + x_{8}x_{33} + x_{8}x_{34} + x_{8}x_{36} + x_{8}x_{37} + x_{8}x_{39} + x_{8}x_{40} + x_{8}x_{48} + x_{8}x_{49} + x_{8}x_{51} + x_{8}x_{53} + x_{8}x_{54} + x_{8}x_{56} + x_{8}x_{57} + x_{8}x_{58} + x_{8}x_{59} + x_{8}x_{60} + x_{8}x_{61} + x_{8}x_{62} + x_{9}x_{10} + x_{9}x_{11} + x_{9}x_{12} + x_{9}x_{14} + x_{9}x_{16} + x_{9}x_{22} + x_{9}x_{24} + x_{9}x_{25} + x_{9}x_{26} + x_{9}x_{27} + x_{9}x_{30} + x_{9}x_{32} + x_{9}x_{34} + x_{9}x_{36} + x_{9}x_{40} + x_{9}x_{42} + x_{9}x_{44} + x_{9}x_{45} + x_{9}x_{49} + x_{9}x_{50} + x_{9}x_{51} + x_{9}x_{52} + x_{9}x_{56} + x_{9}x_{58} + x_{9}x_{61} + x_{9}x_{62} + x_{9}x_{63} + x_{10}x_{13} + x_{10}x_{14} + x_{10}x_{17} + x_{10}x_{18} + x_{10}x_{25} + x_{10}x_{26} + x_{10}x_{28} + x_{10}x_{29} + x_{10}x_{30} + x_{10}x_{32} + x_{10}x_{33} + x_{10}x_{34} + x_{10}x_{37} + x_{10}x_{38} + x_{10}x_{39} + x_{10}x_{40} + x_{10}x_{43} + x_{10}x_{48} + x_{10}x_{49} + x_{10}x_{51} + x_{10}x_{53} + x_{10}x_{54} + x_{10}x_{60} + x_{10}x_{61} + x_{10}x_{62} + x_{10}x_{63} + x_{10}x_{64} + x_{11}x_{13} + x_{11}x_{14} + x_{11}x_{15} + x_{11}x_{19} + x_{11}x_{20} + x_{11}x_{21} + x_{11}x_{22} + x_{11}x_{23} + x_{11}x_{25} + x_{11}x_{26} + x_{11}x_{30} + x_{11}x_{34} + x_{11}x_{35} + x_{11}x_{36} + x_{11}x_{37} + x_{11}x_{39} + x_{11}x_{41} + x_{11}x_{42} + x_{11}x_{44} + x_{11}x_{45} + x_{11}x_{47} + x_{11}x_{49} + x_{11}x_{54} + x_{11}x_{55} + x_{11}x_{58} + x_{11}x_{59} + x_{11}x_{61} + x_{11}x_{62} + x_{11}x_{64} + x_{12}x_{13} + x_{12}x_{15} + x_{12}x_{17} + x_{12}x_{18} + x_{12}x_{20} + x_{12}x_{21} + x_{12}x_{22} + x_{12}x_{25} + x_{12}x_{26} + x_{12}x_{27} + x_{12}x_{30} + x_{12}x_{32} + x_{12}x_{36} + x_{12}x_{39} + x_{12}x_{41} + x_{12}x_{42} + x_{12}x_{45} + x_{12}x_{47} + x_{12}x_{48} + x_{12}x_{50} + x_{12}x_{51} + x_{12}x_{52} + x_{12}x_{53} + x_{12}x_{56} + x_{12}x_{57} + x_{12}x_{58} + x_{13}x_{15} + x_{13}x_{16} + x_{13}x_{21} + x_{13}x_{22} + x_{13}x_{24} + x_{13}x_{33} + x_{13}x_{34} + x_{13}x_{35} + x_{13}x_{36} + x_{13}x_{40} + x_{13}x_{42} + x_{13}x_{43} + x_{13}x_{44} + x_{13}x_{45} + x_{13}x_{46} + x_{13}x_{47} + x_{13}x_{48} + x_{13}x_{50} + x_{13}x_{53} + x_{13}x_{54} + x_{13}x_{56} + x_{13}x_{57} + x_{13}x_{58} + x_{13}x_{59} + x_{13}x_{61} + x_{13}x_{62} + x_{13}x_{64} + x_{14}x_{15} + x_{14}x_{16} + x_{14}x_{17} + x_{14}x_{18} + x_{14}x_{26} + x_{14}x_{27} + x_{14}x_{28} + x_{14}x_{31} + x_{14}x_{35} + x_{14}x_{36} + x_{14}x_{38} + x_{14}x_{42} + x_{14}x_{44} + x_{14}x_{45} + x_{14}x_{50} + x_{14}x_{52} + x_{14}x_{56} + x_{14}x_{58} + x_{14}x_{59} + x_{14}x_{60} + x_{14}x_{63} + x_{15}x_{18} + x_{15}x_{19} + x_{15}x_{21} + x_{15}x_{24} + x_{15}x_{25} + x_{15}x_{26} + x_{15}x_{31} + x_{15}x_{36} + x_{15}x_{40} + x_{15}x_{43} + x_{15}x_{48} + x_{15}x_{50} + x_{15}x_{51} + x_{15}x_{52} + x_{15}x_{54} + x_{15}x_{58} + x_{15}x_{60} + x_{15}x_{62} + x_{15}x_{63} + x_{16}x_{20} + x_{16}x_{23} + x_{16}x_{25} + x_{16}x_{26} + x_{16}x_{27} + x_{16}x_{30} + x_{16}x_{31} + x_{16}x_{32} + x_{16}x_{33} + x_{16}x_{34} + x_{16}x_{37} + x_{16}x_{39} + x_{16}x_{40} + x_{16}x_{42} + x_{16}x_{44} + x_{16}x_{46} + x_{16}x_{47} + x_{16}x_{50} + x_{16}x_{51} + x_{16}x_{52} + x_{16}x_{54} + x_{16}x_{56} + x_{16}x_{60} + x_{16}x_{64} + x_{17}x_{18} + x_{17}x_{19} + x_{17}x_{23} + x_{17}x_{25} + x_{17}x_{26} + x_{17}x_{27} + x_{17}x_{28} + x_{17}x_{29} + x_{17}x_{30} + x_{17}x_{31} + x_{17}x_{32} + x_{17}x_{33} + x_{17}x_{34} + x_{17}x_{35} + x_{17}x_{37} + x_{17}x_{40} + x_{17}x_{42} + x_{17}x_{43} + x_{17}x_{44} + x_{17}x_{45} + x_{17}x_{47} + x_{17}x_{50} + x_{17}x_{51} + x_{17}x_{54} + x_{17}x_{57} + x_{17}x_{60} + x_{17}x_{62} + x_{17}x_{63} + x_{18}x_{19} + x_{18}x_{21} + x_{18}x_{23} + x_{18}x_{25} + x_{18}x_{26} + x_{18}x_{32} + x_{18}x_{33} + x_{18}x_{35} + x_{18}x_{37} + x_{18}x_{38} + x_{18}x_{41} + x_{18}x_{46} + x_{18}x_{47} + x_{18}x_{48} + x_{18}x_{49} + x_{18}x_{50} + x_{18}x_{51} + x_{18}x_{52} + x_{18}x_{54} + x_{18}x_{56} + x_{18}x_{57} + x_{18}x_{58} + x_{18}x_{61} + x_{18}x_{62} + x_{18}x_{63} + x_{19}x_{20} + x_{19}x_{21} + x_{19}x_{22} + x_{19}x_{23} + x_{19}x_{25} + x_{19}x_{28} + x_{19}x_{29} + x_{19}x_{30} + x_{19}x_{32} + x_{19}x_{33} + x_{19}x_{35} + x_{19}x_{36} + x_{19}x_{37} + x_{19}x_{38} + x_{19}x_{39} + x_{19}x_{42} + x_{19}x_{44} + x_{19}x_{45} + x_{19}x_{48} + x_{19}x_{49} + x_{19}x_{53} + x_{19}x_{57} + x_{19}x_{60} + x_{19}x_{62} + x_{19}x_{63} + x_{20}x_{21} + x_{20}x_{22} + x_{20}x_{23} + x_{20}x_{24} + x_{20}x_{26} + x_{20}x_{27} + x_{20}x_{33} + x_{20}x_{35} + x_{20}x_{36} + x_{20}x_{37} + x_{20}x_{38} + x_{20}x_{39} + x_{20}x_{40} + x_{20}x_{41} + x_{20}x_{47} + x_{20}x_{48} + x_{20}x_{49} + x_{20}x_{50} + x_{20}x_{51} + x_{20}x_{54} + x_{20}x_{55} + x_{20}x_{63} + x_{20}x_{64} + x_{21}x_{24} + x_{21}x_{25} + x_{21}x_{29} + x_{21}x_{30} + x_{21}x_{31} + x_{21}x_{32} + x_{21}x_{33} + x_{21}x_{35} + x_{21}x_{36} + x_{21}x_{38} + x_{21}x_{40} + x_{21}x_{41} + x_{21}x_{43} + x_{21}x_{44} + x_{21}x_{45} + x_{21}x_{47} + x_{21}x_{48} + x_{21}x_{51} + x_{21}x_{52} + x_{21}x_{53} + x_{21}x_{56} + x_{21}x_{57} + x_{21}x_{61} + x_{22}x_{25} + x_{22}x_{28} + x_{22}x_{29} + x_{22}x_{30} + x_{22}x_{32} + x_{22}x_{36} + x_{22}x_{37} + x_{22}x_{38} + x_{22}x_{39} + x_{22}x_{40} + x_{22}x_{43} + x_{22}x_{44} + x_{22}x_{45} + x_{22}x_{46} + x_{22}x_{47} + x_{22}x_{49} + x_{22}x_{53} + x_{22}x_{54} + x_{22}x_{56} + x_{22}x_{59} + x_{22}x_{60} + x_{22}x_{62} + x_{22}x_{63} + x_{23}x_{25} + x_{23}x_{26} + x_{23}x_{27} + x_{23}x_{28} + x_{23}x_{29} + x_{23}x_{32} + x_{23}x_{33} + x_{23}x_{35} + x_{23}x_{38} + x_{23}x_{39} + x_{23}x_{41} + x_{23}x_{44} + x_{23}x_{45} + x_{23}x_{46} + x_{23}x_{47} + x_{23}x_{48} + x_{23}x_{50} + x_{23}x_{51} + x_{23}x_{52} + x_{23}x_{53} + x_{23}x_{54} + x_{23}x_{56} + x_{23}x_{57} + x_{23}x_{58} + x_{23}x_{59} + x_{23}x_{63} + x_{23}x_{64} + x_{24}x_{28} + x_{24}x_{30} + x_{24}x_{31} + x_{24}x_{33} + x_{24}x_{35} + x_{24}x_{38} + x_{24}x_{39} + x_{24}x_{44} + x_{24}x_{50} + x_{24}x_{51} + x_{24}x_{53} + x_{24}x_{54} + x_{24}x_{58} + x_{24}x_{60} + x_{24}x_{61} + x_{24}x_{63} + x_{25}x_{27} + x_{25}x_{30} + x_{25}x_{31} + x_{25}x_{35} + x_{25}x_{36} + x_{25}x_{42} + x_{25}x_{43} + x_{25}x_{44} + x_{25}x_{45} + x_{25}x_{47} + x_{25}x_{49} + x_{25}x_{50} + x_{25}x_{51} + x_{25}x_{53} + x_{25}x_{54} + x_{25}x_{55} + x_{25}x_{59} + x_{25}x_{60} + x_{26}x_{27} + x_{26}x_{30} + x_{26}x_{34} + x_{26}x_{35} + x_{26}x_{36} + x_{26}x_{38} + x_{26}x_{39} + x_{26}x_{40} + x_{26}x_{41} + x_{26}x_{46} + x_{26}x_{47} + x_{26}x_{48} + x_{26}x_{52} + x_{26}x_{54} + x_{26}x_{55} + x_{26}x_{57} + x_{26}x_{58} + x_{26}x_{59} + x_{26}x_{61} + x_{26}x_{63} + x_{27}x_{28} + x_{27}x_{29} + x_{27}x_{30} + x_{27}x_{32} + x_{27}x_{34} + x_{27}x_{36} + x_{27}x_{37} + x_{27}x_{38} + x_{27}x_{40} + x_{27}x_{41} + x_{27}x_{42} + x_{27}x_{45} + x_{27}x_{46} + x_{27}x_{47} + x_{27}x_{50} + x_{27}x_{53} + x_{27}x_{54} + x_{27}x_{55} + x_{27}x_{56} + x_{27}x_{57} + x_{27}x_{62} + x_{27}x_{63} + x_{28}x_{29} + x_{28}x_{33} + x_{28}x_{34} + x_{28}x_{37} + x_{28}x_{38} + x_{28}x_{42} + x_{28}x_{45} + x_{28}x_{46} + x_{28}x_{47} + x_{28}x_{48} + x_{28}x_{51} + x_{28}x_{52} + x_{28}x_{53} + x_{28}x_{54} + x_{28}x_{55} + x_{28}x_{56} + x_{28}x_{58} + x_{28}x_{61} + x_{28}x_{63} + x_{28}x_{64} + x_{29}x_{30} + x_{29}x_{31} + x_{29}x_{32} + x_{29}x_{33} + x_{29}x_{34} + x_{29}x_{36} + x_{29}x_{39} + x_{29}x_{40} + x_{29}x_{41} + x_{29}x_{42} + x_{29}x_{49} + x_{29}x_{51} + x_{29}x_{53} + x_{29}x_{55} + x_{29}x_{56} + x_{29}x_{59} + x_{29}x_{62} + x_{29}x_{64} + x_{30}x_{33} + x_{30}x_{34} + x_{30}x_{35} + x_{30}x_{36} + x_{30}x_{41} + x_{30}x_{44} + x_{30}x_{46} + x_{30}x_{48} + x_{30}x_{51} + x_{30}x_{58} + x_{30}x_{59} + x_{30}x_{60} + x_{30}x_{61} + x_{30}x_{63} + x_{31}x_{33} + x_{31}x_{36} + x_{31}x_{41} + x_{31}x_{44} + x_{31}x_{48} + x_{31}x_{49} + x_{31}x_{51} + x_{31}x_{52} + x_{31}x_{53} + x_{31}x_{54} + x_{31}x_{55} + x_{31}x_{58} + x_{31}x_{61} + x_{31}x_{62} + x_{32}x_{33} + x_{32}x_{37} + x_{32}x_{38} + x_{32}x_{39} + x_{32}x_{40} + x_{32}x_{42} + x_{32}x_{43} + x_{32}x_{44} + x_{32}x_{49} + x_{32}x_{50} + x_{32}x_{52} + x_{32}x_{55} + x_{32}x_{56} + x_{32}x_{58} + x_{32}x_{59} + x_{32}x_{60} + x_{32}x_{61} + x_{32}x_{64} + x_{33}x_{34} + x_{33}x_{37} + x_{33}x_{38} + x_{33}x_{40} + x_{33}x_{44} + x_{33}x_{45} + x_{33}x_{47} + x_{33}x_{48} + x_{33}x_{49} + x_{33}x_{50} + x_{33}x_{52} + x_{33}x_{55} + x_{33}x_{56} + x_{33}x_{57} + x_{33}x_{58} + x_{33}x_{60} + x_{33}x_{62} + x_{34}x_{40} + x_{34}x_{45} + x_{34}x_{48} + x_{34}x_{49} + x_{34}x_{50} + x_{34}x_{51} + x_{34}x_{52} + x_{34}x_{53} + x_{34}x_{55} + x_{34}x_{56} + x_{34}x_{58} + x_{34}x_{60} + x_{34}x_{61} + x_{34}x_{63} + x_{34}x_{64} + x_{35}x_{36} + x_{35}x_{38} + x_{35}x_{40} + x_{35}x_{43} + x_{35}x_{44} + x_{35}x_{45} + x_{35}x_{49} + x_{35}x_{54} + x_{35}x_{58} + x_{35}x_{60} + x_{35}x_{62} + x_{36}x_{42} + x_{36}x_{44} + x_{36}x_{46} + x_{36}x_{47} + x_{36}x_{50} + x_{36}x_{51} + x_{36}x_{53} + x_{36}x_{54} + x_{36}x_{57} + x_{36}x_{58} + x_{36}x_{62} + x_{36}x_{64} + x_{37}x_{43} + x_{37}x_{44} + x_{37}x_{46} + x_{37}x_{47} + x_{37}x_{48} + x_{37}x_{53} + x_{37}x_{54} + x_{37}x_{55} + x_{37}x_{59} + x_{37}x_{63} + x_{37}x_{64} + x_{38}x_{40} + x_{38}x_{42} + x_{38}x_{43} + x_{38}x_{44} + x_{38}x_{45} + x_{38}x_{47} + x_{38}x_{48} + x_{38}x_{54} + x_{38}x_{60} + x_{38}x_{61} + x_{38}x_{62} + x_{39}x_{40} + x_{39}x_{42} + x_{39}x_{43} + x_{39}x_{46} + x_{39}x_{47} + x_{39}x_{48} + x_{39}x_{49} + x_{39}x_{51} + x_{39}x_{52} + x_{39}x_{54} + x_{39}x_{55} + x_{39}x_{56} + x_{39}x_{57} + x_{39}x_{59} + x_{39}x_{60} + x_{39}x_{62} + x_{40}x_{42} + x_{40}x_{43} + x_{40}x_{44} + x_{40}x_{45} + x_{40}x_{49} + x_{40}x_{52} + x_{40}x_{53} + x_{40}x_{54} + x_{40}x_{56} + x_{40}x_{57} + x_{40}x_{59} + x_{40}x_{60} + x_{40}x_{61} + x_{40}x_{62} + x_{40}x_{63} + x_{40}x_{64} + x_{41}x_{44} + x_{41}x_{45} + x_{41}x_{46} + x_{41}x_{49} + x_{41}x_{50} + x_{41}x_{51} + x_{41}x_{56} + x_{41}x_{58} + x_{41}x_{59} + x_{41}x_{62} + x_{41}x_{63} + x_{42}x_{44} + x_{42}x_{45} + x_{42}x_{47} + x_{42}x_{48} + x_{42}x_{50} + x_{42}x_{51} + x_{42}x_{52} + x_{42}x_{54} + x_{42}x_{56} + x_{42}x_{58} + x_{42}x_{60} + x_{42}x_{61} + x_{42}x_{62} + x_{43}x_{44} + x_{43}x_{45} + x_{43}x_{46} + x_{43}x_{48} + x_{43}x_{50} + x_{43}x_{55} + x_{43}x_{56} + x_{43}x_{58} + x_{43}x_{59} + x_{43}x_{61} + x_{43}x_{62} + x_{43}x_{63} + x_{44}x_{45} + x_{44}x_{46} + x_{44}x_{47} + x_{44}x_{48} + x_{44}x_{50} + x_{44}x_{51} + x_{44}x_{53} + x_{44}x_{54} + x_{44}x_{55} + x_{44}x_{56} + x_{44}x_{57} + x_{44}x_{59} + x_{44}x_{61} + x_{44}x_{64} + x_{45}x_{46} + x_{45}x_{49} + x_{45}x_{51} + x_{45}x_{54} + x_{45}x_{57} + x_{45}x_{60} + x_{45}x_{61} + x_{45}x_{63} + x_{45}x_{64} + x_{46}x_{49} + x_{46}x_{51} + x_{46}x_{54} + x_{46}x_{58} + x_{46}x_{61} + x_{46}x_{62} + x_{46}x_{64} + x_{47}x_{48} + x_{47}x_{50} + x_{47}x_{52} + x_{47}x_{54} + x_{47}x_{56} + x_{47}x_{58} + x_{47}x_{59} + x_{47}x_{60} + x_{47}x_{61} + x_{47}x_{64} + x_{48}x_{49} + x_{48}x_{50} + x_{48}x_{52} + x_{48}x_{54} + x_{48}x_{55} + x_{48}x_{56} + x_{48}x_{57} + x_{48}x_{61} + x_{48}x_{62} + x_{48}x_{63} + x_{48}x_{64} + x_{49}x_{52} + x_{49}x_{54} + x_{49}x_{55} + x_{49}x_{57} + x_{49}x_{59} + x_{50}x_{52} + x_{50}x_{55} + x_{50}x_{57} + x_{50}x_{58} + x_{50}x_{59} + x_{50}x_{62} + x_{50}x_{63} + x_{50}x_{64} + x_{51}x_{52} + x_{51}x_{54} + x_{51}x_{56} + x_{51}x_{57} + x_{51}x_{61} + x_{51}x_{62} + x_{51}x_{63} + x_{52}x_{54} + x_{52}x_{60} + x_{52}x_{61} + x_{52}x_{63} + x_{53}x_{55} + x_{53}x_{56} + x_{53}x_{59} + x_{53}x_{60} + x_{53}x_{61} + x_{53}x_{62} + x_{54}x_{59} + x_{55}x_{56} + x_{55}x_{58} + x_{55}x_{61} + x_{55}x_{62} + x_{55}x_{63} + x_{56}x_{59} + x_{56}x_{64} + x_{57}x_{58} + x_{57}x_{60} + x_{57}x_{62} + x_{57}x_{64} + x_{58}x_{61} + x_{58}x_{62} + x_{59}x_{61} + x_{59}x_{62} + x_{59}x_{63} + x_{61}x_{62} + x_{61}x_{63} + x_{61}x_{64} + x_{62}x_{63} + x_{62}x_{64} + x_{2} + x_{3} + x_{5} + x_{7} + x_{11} + x_{15} + x_{23} + x_{24} + x_{25} + x_{27} + x_{28} + x_{30} + x_{31} + x_{33} + x_{36} + x_{40} + x_{41} + x_{44} + x_{47} + x_{49} + x_{51} + x_{54} + x_{58} + x_{59} + x_{60} + x_{61} + x_{63} + 1$

$y_{18} = x_{1}x_{2} + x_{1}x_{5} + x_{1}x_{6} + x_{1}x_{7} + x_{1}x_{8} + x_{1}x_{13} + x_{1}x_{15} + x_{1}x_{16} + x_{1}x_{21} + x_{1}x_{23} + x_{1}x_{25} + x_{1}x_{26} + x_{1}x_{27} + x_{1}x_{35} + x_{1}x_{36} + x_{1}x_{40} + x_{1}x_{44} + x_{1}x_{46} + x_{1}x_{47} + x_{1}x_{49} + x_{1}x_{50} + x_{1}x_{52} + x_{1}x_{53} + x_{1}x_{54} + x_{1}x_{55} + x_{1}x_{56} + x_{1}x_{59} + x_{1}x_{62} + x_{1}x_{63} + x_{1}x_{64} + x_{2}x_{5} + x_{2}x_{7} + x_{2}x_{11} + x_{2}x_{12} + x_{2}x_{13} + x_{2}x_{16} + x_{2}x_{19} + x_{2}x_{21} + x_{2}x_{23} + x_{2}x_{24} + x_{2}x_{27} + x_{2}x_{29} + x_{2}x_{30} + x_{2}x_{31} + x_{2}x_{32} + x_{2}x_{35} + x_{2}x_{36} + x_{2}x_{39} + x_{2}x_{43} + x_{2}x_{45} + x_{2}x_{46} + x_{2}x_{47} + x_{2}x_{48} + x_{2}x_{50} + x_{2}x_{54} + x_{2}x_{55} + x_{2}x_{56} + x_{2}x_{57} + x_{2}x_{63} + x_{2}x_{64} + x_{3}x_{7} + x_{3}x_{10} + x_{3}x_{12} + x_{3}x_{13} + x_{3}x_{16} + x_{3}x_{18} + x_{3}x_{19} + x_{3}x_{22} + x_{3}x_{24} + x_{3}x_{25} + x_{3}x_{28} + x_{3}x_{29} + x_{3}x_{30} + x_{3}x_{31} + x_{3}x_{33} + x_{3}x_{34} + x_{3}x_{36} + x_{3}x_{37} + x_{3}x_{38} + x_{3}x_{41} + x_{3}x_{44} + x_{3}x_{46} + x_{3}x_{50} + x_{3}x_{53} + x_{3}x_{54} + x_{3}x_{56} + x_{3}x_{58} + x_{3}x_{60} + x_{3}x_{61} + x_{3}x_{62} + x_{3}x_{63} + x_{4}x_{5} + x_{4}x_{7} + x_{4}x_{8} + x_{4}x_{10} + x_{4}x_{15} + x_{4}x_{19} + x_{4}x_{20} + x_{4}x_{22} + x_{4}x_{25} + x_{4}x_{29} + x_{4}x_{30} + x_{4}x_{31} + x_{4}x_{34} + x_{4}x_{35} + x_{4}x_{37} + x_{4}x_{40} + x_{4}x_{42} + x_{4}x_{43} + x_{4}x_{48} + x_{4}x_{50} + x_{4}x_{51} + x_{4}x_{53} + x_{4}x_{55} + x_{4}x_{56} + x_{4}x_{59} + x_{4}x_{62} + x_{4}x_{63} + x_{4}x_{64} + x_{5}x_{6} + x_{5}x_{12} + x_{5}x_{17} + x_{5}x_{19} + x_{5}x_{21} + x_{5}x_{22} + x_{5}x_{25} + x_{5}x_{26} + x_{5}x_{28} + x_{5}x_{32} + x_{5}x_{33} + x_{5}x_{34} + x_{5}x_{35} + x_{5}x_{36} + x_{5}x_{37} + x_{5}x_{39} + x_{5}x_{40} + x_{5}x_{42} + x_{5}x_{43} + x_{5}x_{44} + x_{5}x_{45} + x_{5}x_{46} + x_{5}x_{47} + x_{5}x_{50} + x_{5}x_{51} + x_{5}x_{55} + x_{5}x_{58} + x_{5}x_{60} + x_{5}x_{61} + x_{6}x_{8} + x_{6}x_{9} + x_{6}x_{10} + x_{6}x_{11} + x_{6}x_{16} + x_{6}x_{19} + x_{6}x_{22} + x_{6}x_{24} + x_{6}x_{28} + x_{6}x_{29} + x_{6}x_{32} + x_{6}x_{36} + x_{6}x_{38} + x_{6}x_{41} + x_{6}x_{43} + x_{6}x_{44} + x_{6}x_{47} + x_{6}x_{48} + x_{6}x_{52} + x_{6}x_{53} + x_{6}x_{54} + x_{6}x_{58} + x_{6}x_{61} + x_{6}x_{62} + x_{6}x_{63} + x_{7}x_{11} + x_{7}x_{12} + x_{7}x_{16} + x_{7}x_{18} + x_{7}x_{20} + x_{7}x_{21} + x_{7}x_{22} + x_{7}x_{25} + x_{7}x_{26} + x_{7}x_{27} + x_{7}x_{33} + x_{7}x_{34} + x_{7}x_{38} + x_{7}x_{41} + x_{7}x_{42} + x_{7}x_{44} + x_{7}x_{47} + x_{7}x_{49} + x_{7}x_{51} + x_{7}x_{52} + x_{7}x_{53} + x_{7}x_{55} + x_{7}x_{58} + x_{7}x_{59} + x_{7}x_{62} + x_{8}x_{9} + x_{8}x_{10} + x_{8}x_{11} + x_{8}x_{13} + x_{8}x_{21} + x_{8}x_{22} + x_{8}x_{25} + x_{8}x_{26} + x_{8}x_{28} + x_{8}x_{29} + x_{8}x_{31} + x_{8}x_{32} + x_{8}x_{33} + x_{8}x_{35} + x_{8}x_{39} + x_{8}x_{40} + x_{8}x_{41} + x_{8}x_{42} + x_{8}x_{43} + x_{8}x_{49} + x_{8}x_{50} + x_{8}x_{51} + x_{8}x_{53} + x_{8}x_{54} + x_{8}x_{57} + x_{8}x_{59} + x_{8}x_{60} + x_{8}x_{63} + x_{8}x_{64} + x_{9}x_{12} + x_{9}x_{18} + x_{9}x_{21} + x_{9}x_{23} + x_{9}x_{25} + x_{9}x_{28} + x_{9}x_{29} + x_{9}x_{30} + x_{9}x_{31} + x_{9}x_{36} + x_{9}x_{38} + x_{9}x_{40} + x_{9}x_{42} + x_{9}x_{44} + x_{9}x_{45} + x_{9}x_{46} + x_{9}x_{48} + x_{9}x_{49} + x_{9}x_{52} + x_{9}x_{54} + x_{9}x_{55} + x_{9}x_{57} + x_{9}x_{59} + x_{9}x_{61} + x_{10}x_{11} + x_{10}x_{17} + x_{10}x_{23} + x_{10}x_{30} + x_{10}x_{32} + x_{10}x_{38} + x_{10}x_{41} + x_{10}x_{42} + x_{10}x_{43} + x_{10}x_{45} + x_{10}x_{46} + x_{10}x_{47} + x_{10}x_{48} + x_{10}x_{50} + x_{10}x_{51} + x_{10}x_{57} + x_{10}x_{58} + x_{10}x_{60} + x_{10}x_{61} + x_{10}x_{63} + x_{11}x_{14} + x_{11}x_{15} + x_{11}x_{16} + x_{11}x_{17} + x_{11}x_{18} + x_{11}x_{19} + x_{11}x_{20} + x_{11}x_{21} + x_{11}x_{22} + x_{11}x_{23} + x_{11}x_{24} + x_{11}x_{28} + x_{11}x_{33} + x_{11}x_{36} + x_{11}x_{38} + x_{11}x_{42} + x_{11}x_{43} + x_{11}x_{44} + x_{11}x_{45} + x_{11}x_{46} + x_{11}x_{47} + x_{11}x_{50} + x_{11}x_{53} + x_{11}x_{54} + x_{11}x_{55} + x_{11}x_{56} + x_{11}x_{59} + x_{11}x_{61} + x_{11}x_{63} + x_{12}x_{13} + x_{12}x_{14} + x_{12}x_{20} + x_{12}x_{21} + x_{12}x_{22} + x_{12}x_{23} + x_{12}x_{25} + x_{12}x_{27} + x_{12}x_{30} + x_{12}x_{31} + x_{12}x_{32} + x_{12}x_{33} + x_{12}x_{36} + x_{12}x_{39} + x_{12}x_{41} + x_{12}x_{44} + x_{12}x_{50} + x_{12}x_{52} + x_{12}x_{53} + x_{12}x_{55} + x_{12}x_{56} + x_{12}x_{59} + x_{12}x_{60} + x_{12}x_{62} + x_{12}x_{63} + x_{13}x_{14} + x_{13}x_{15} + x_{13}x_{19} + x_{13}x_{21} + x_{13}x_{23} + x_{13}x_{24} + x_{13}x_{25} + x_{13}x_{26} + x_{13}x_{27} + x_{13}x_{32} + x_{13}x_{34} + x_{13}x_{36} + x_{13}x_{41} + x_{13}x_{42} + x_{13}x_{44} + x_{13}x_{48} + x_{13}x_{49} + x_{13}x_{50} + x_{13}x_{53} + x_{13}x_{54} + x_{13}x_{56} + x_{13}x_{57} + x_{13}x_{59} + x_{13}x_{60} + x_{14}x_{17} + x_{14}x_{20} + x_{14}x_{22} + x_{14}x_{23} + x_{14}x_{24} + x_{14}x_{27} + x_{14}x_{28} + x_{14}x_{29} + x_{14}x_{33} + x_{14}x_{34} + x_{14}x_{37} + x_{14}x_{38} + x_{14}x_{39} + x_{14}x_{41} + x_{14}x_{42} + x_{14}x_{43} + x_{14}x_{44} + x_{14}x_{45} + x_{14}x_{50} + x_{14}x_{51} + x_{14}x_{54} + x_{14}x_{55} + x_{14}x_{56} + x_{14}x_{61} + x_{14}x_{63} + x_{15}x_{16} + x_{15}x_{19} + x_{15}x_{20} + x_{15}x_{26} + x_{15}x_{27} + x_{15}x_{28} + x_{15}x_{30} + x_{15}x_{32} + x_{15}x_{34} + x_{15}x_{35} + x_{15}x_{36} + x_{15}x_{37} + x_{15}x_{42} + x_{15}x_{45} + x_{15}x_{46} + x_{15}x_{47} + x_{15}x_{49} + x_{15}x_{52} + x_{15}x_{53} + x_{15}x_{56} + x_{15}x_{58} + x_{15}x_{59} + x_{15}x_{60} + x_{15}x_{61} + x_{15}x_{63} + x_{15}x_{64} + x_{16}x_{18} + x_{16}x_{19} + x_{16}x_{22} + x_{16}x_{23} + x_{16}x_{25} + x_{16}x_{29} + x_{16}x_{32} + x_{16}x_{36} + x_{16}x_{37} + x_{16}x_{40} + x_{16}x_{41} + x_{16}x_{42} + x_{16}x_{45} + x_{16}x_{46} + x_{16}x_{48} + x_{16}x_{49} + x_{16}x_{50} + x_{16}x_{52} + x_{16}x_{53} + x_{16}x_{54} + x_{16}x_{55} + x_{16}x_{57} + x_{16}x_{59} + x_{16}x_{61} + x_{16}x_{63} + x_{17}x_{18} + x_{17}x_{19} + x_{17}x_{24} + x_{17}x_{26} + x_{17}x_{28} + x_{17}x_{31} + x_{17}x_{34} + x_{17}x_{35} + x_{17}x_{36} + x_{17}x_{37} + x_{17}x_{39} + x_{17}x_{41} + x_{17}x_{42} + x_{17}x_{44} + x_{17}x_{45} + x_{17}x_{47} + x_{17}x_{52} + x_{17}x_{59} + x_{17}x_{60} + x_{17}x_{61} + x_{17}x_{62} + x_{17}x_{63} + x_{17}x_{64} + x_{18}x_{19} + x_{18}x_{20} + x_{18}x_{24} + x_{18}x_{25} + x_{18}x_{26} + x_{18}x_{30} + x_{18}x_{32} + x_{18}x_{33} + x_{18}x_{36} + x_{18}x_{37} + x_{18}x_{39} + x_{18}x_{41} + x_{18}x_{44} + x_{18}x_{45} + x_{18}x_{46} + x_{18}x_{48} + x_{18}x_{49} + x_{18}x_{51} + x_{18}x_{52} + x_{18}x_{53} + x_{18}x_{54} + x_{18}x_{55} + x_{18}x_{56} + x_{18}x_{57} + x_{18}x_{58} + x_{18}x_{59} + x_{18}x_{61} + x_{18}x_{62} + x_{18}x_{63} + x_{19}x_{20} + x_{19}x_{21} + x_{19}x_{24} + x_{19}x_{26} + x_{19}x_{27} + x_{19}x_{28} + x_{19}x_{29} + x_{19}x_{31} + x_{19}x_{35} + x_{19}x_{38} + x_{19}x_{41} + x_{19}x_{42} + x_{19}x_{45} + x_{19}x_{46} + x_{19}x_{47} + x_{19}x_{52} + x_{19}x_{53} + x_{19}x_{54} + x_{19}x_{56} + x_{19}x_{57} + x_{19}x_{59} + x_{19}x_{61} + x_{19}x_{62} + x_{19}x_{63} + x_{19}x_{64} + x_{20}x_{24} + x_{20}x_{25} + x_{20}x_{26} + x_{20}x_{27} + x_{20}x_{31} + x_{20}x_{33} + x_{20}x_{34} + x_{20}x_{35} + x_{20}x_{37} + x_{20}x_{38} + x_{20}x_{40} + x_{20}x_{41} + x_{20}x_{42} + x_{20}x_{43} + x_{20}x_{44} + x_{20}x_{45} + x_{20}x_{47} + x_{20}x_{49} + x_{20}x_{51} + x_{20}x_{52} + x_{20}x_{54} + x_{20}x_{56} + x_{20}x_{57} + x_{20}x_{58} + x_{20}x_{59} + x_{20}x_{64} + x_{21}x_{22} + x_{21}x_{26} + x_{21}x_{27} + x_{21}x_{28} + x_{21}x_{32} + x_{21}x_{34} + x_{21}x_{35} + x_{21}x_{36} + x_{21}x_{37} + x_{21}x_{41} + x_{21}x_{42} + x_{21}x_{43} + x_{21}x_{44} + x_{21}x_{45} + x_{21}x_{46} + x_{21}x_{47} + x_{21}x_{48} + x_{21}x_{50} + x_{21}x_{51} + x_{21}x_{54} + x_{21}x_{55} + x_{21}x_{56} + x_{21}x_{57} + x_{21}x_{58} + x_{21}x_{61} + x_{21}x_{62} + x_{22}x_{23} + x_{22}x_{25} + x_{22}x_{26} + x_{22}x_{28} + x_{22}x_{29} + x_{22}x_{30} + x_{22}x_{33} + x_{22}x_{36} + x_{22}x_{41} + x_{22}x_{43} + x_{22}x_{46} + x_{22}x_{48} + x_{22}x_{49} + x_{22}x_{50} + x_{22}x_{52} + x_{22}x_{53} + x_{22}x_{54} + x_{22}x_{55} + x_{22}x_{56} + x_{22}x_{58} + x_{22}x_{63} + x_{22}x_{64} + x_{23}x_{26} + x_{23}x_{27} + x_{23}x_{28} + x_{23}x_{31} + x_{23}x_{34} + x_{23}x_{36} + x_{23}x_{38} + x_{23}x_{39} + x_{23}x_{42} + x_{23}x_{43} + x_{23}x_{45} + x_{23}x_{47} + x_{23}x_{48} + x_{23}x_{49} + x_{23}x_{50} + x_{23}x_{51} + x_{23}x_{52} + x_{23}x_{56} + x_{23}x_{61} + x_{23}x_{62} + x_{23}x_{64} + x_{24}x_{25} + x_{24}x_{26} + x_{24}x_{29} + x_{24}x_{30} + x_{24}x_{31} + x_{24}x_{33} + x_{24}x_{34} + x_{24}x_{35} + x_{24}x_{37} + x_{24}x_{38} + x_{24}x_{45} + x_{24}x_{47} + x_{24}x_{48} + x_{24}x_{49} + x_{24}x_{50} + x_{24}x_{51} + x_{24}x_{54} + x_{24}x_{55} + x_{24}x_{56} + x_{24}x_{60} + x_{24}x_{62} + x_{25}x_{27} + x_{25}x_{29} + x_{25}x_{33} + x_{25}x_{35} + x_{25}x_{36} + x_{25}x_{38} + x_{25}x_{39} + x_{25}x_{40} + x_{25}x_{42} + x_{25}x_{43} + x_{25}x_{44} + x_{25}x_{46} + x_{25}x_{47} + x_{25}x_{48} + x_{25}x_{49} + x_{25}x_{50} + x_{25}x_{52} + x_{25}x_{53} + x_{25}x_{54} + x_{25}x_{56} + x_{25}x_{61} + x_{26}x_{28} + x_{26}x_{30} + x_{26}x_{31} + x_{26}x_{32} + x_{26}x_{34} + x_{26}x_{35} + x_{26}x_{36} + x_{26}x_{39} + x_{26}x_{42} + x_{26}x_{44} + x_{26}x_{47} + x_{26}x_{49} + x_{26}x_{50} + x_{26}x_{52} + x_{26}x_{53} + x_{26}x_{59} + x_{26}x_{60} + x_{26}x_{62} + x_{27}x_{29} + x_{27}x_{31} + x_{27}x_{32} + x_{27}x_{34} + x_{27}x_{35} + x_{27}x_{41} + x_{27}x_{42} + x_{27}x_{44} + x_{27}x_{48} + x_{27}x_{49} + x_{27}x_{51} + x_{27}x_{54} + x_{27}x_{55} + x_{27}x_{58} + x_{27}x_{59} + x_{27}x_{61} + x_{28}x_{31} + x_{28}x_{33} + x_{28}x_{36} + x_{28}x_{39} + x_{28}x_{40} + x_{28}x_{42} + x_{28}x_{43} + x_{28}x_{45} + x_{28}x_{47} + x_{28}x_{48} + x_{28}x_{49} + x_{28}x_{51} + x_{28}x_{52} + x_{28}x_{55} + x_{28}x_{57} + x_{28}x_{58} + x_{28}x_{59} + x_{28}x_{60} + x_{28}x_{61} + x_{28}x_{63} + x_{28}x_{64} + x_{29}x_{30} + x_{29}x_{31} + x_{29}x_{32} + x_{29}x_{34} + x_{29}x_{35} + x_{29}x_{36} + x_{29}x_{38} + x_{29}x_{40} + x_{29}x_{44} + x_{29}x_{45} + x_{29}x_{50} + x_{29}x_{54} + x_{29}x_{55} + x_{29}x_{56} + x_{29}x_{59} + x_{29}x_{61} + x_{29}x_{64} + x_{30}x_{32} + x_{30}x_{34} + x_{30}x_{35} + x_{30}x_{37} + x_{30}x_{39} + x_{30}x_{40} + x_{30}x_{41} + x_{30}x_{42} + x_{30}x_{45} + x_{30}x_{47} + x_{30}x_{49} + x_{30}x_{51} + x_{30}x_{52} + x_{30}x_{53} + x_{30}x_{54} + x_{30}x_{55} + x_{30}x_{57} + x_{30}x_{58} + x_{30}x_{60} + x_{30}x_{61} + x_{30}x_{64} + x_{31}x_{33} + x_{31}x_{34} + x_{31}x_{37} + x_{31}x_{41} + x_{31}x_{43} + x_{31}x_{44} + x_{31}x_{45} + x_{31}x_{46} + x_{31}x_{47} + x_{31}x_{50} + x_{31}x_{51} + x_{31}x_{52} + x_{31}x_{53} + x_{31}x_{54} + x_{31}x_{60} + x_{31}x_{62} + x_{31}x_{63} + x_{32}x_{34} + x_{32}x_{35} + x_{32}x_{36} + x_{32}x_{41} + x_{32}x_{43} + x_{32}x_{46} + x_{32}x_{47} + x_{32}x_{50} + x_{32}x_{52} + x_{32}x_{54} + x_{32}x_{56} + x_{32}x_{57} + x_{32}x_{59} + x_{32}x_{62} + x_{32}x_{63} + x_{32}x_{64} + x_{33}x_{35} + x_{33}x_{42} + x_{33}x_{47} + x_{33}x_{48} + x_{33}x_{49} + x_{33}x_{51} + x_{33}x_{52} + x_{33}x_{55} + x_{33}x_{60} + x_{33}x_{61} + x_{33}x_{62} + x_{33}x_{63} + x_{33}x_{64} + x_{34}x_{37} + x_{34}x_{39} + x_{34}x_{41} + x_{34}x_{42} + x_{34}x_{44} + x_{34}x_{45} + x_{34}x_{46} + x_{34}x_{47} + x_{34}x_{48} + x_{34}x_{49} + x_{34}x_{50} + x_{34}x_{53} + x_{34}x_{54} + x_{34}x_{58} + x_{34}x_{60} + x_{34}x_{62} + x_{34}x_{63} + x_{35}x_{37} + x_{35}x_{38} + x_{35}x_{40} + x_{35}x_{41} + x_{35}x_{42} + x_{35}x_{43} + x_{35}x_{45} + x_{35}x_{47} + x_{35}x_{49} + x_{35}x_{50} + x_{35}x_{51} + x_{35}x_{52} + x_{35}x_{55} + x_{35}x_{56} + x_{35}x_{57} + x_{35}x_{61} + x_{35}x_{64} + x_{36}x_{37} + x_{36}x_{38} + x_{36}x_{40} + x_{36}x_{44} + x_{36}x_{45} + x_{36}x_{47} + x_{36}x_{48} + x_{36}x_{50} + x_{36}x_{52} + x_{36}x_{54} + x_{36}x_{56} + x_{36}x_{57} + x_{36}x_{58} + x_{36}x_{59} + x_{36}x_{63} + x_{37}x_{38} + x_{37}x_{39} + x_{37}x_{42} + x_{37}x_{44} + x_{37}x_{45} + x_{37}x_{46} + x_{37}x_{47} + x_{37}x_{48} + x_{37}x_{49} + x_{37}x_{50} + x_{37}x_{51} + x_{37}x_{54} + x_{37}x_{58} + x_{37}x_{62} + x_{38}x_{39} + x_{38}x_{40} + x_{38}x_{41} + x_{38}x_{42} + x_{38}x_{44} + x_{38}x_{46} + x_{38}x_{47} + x_{38}x_{49} + x_{38}x_{51} + x_{38}x_{57} + x_{38}x_{59} + x_{38}x_{63} + x_{38}x_{64} + x_{39}x_{40} + x_{39}x_{41} + x_{39}x_{42} + x_{39}x_{45} + x_{39}x_{47} + x_{39}x_{49} + x_{39}x_{50} + x_{39}x_{53} + x_{39}x_{55} + x_{39}x_{57} + x_{39}x_{61} + x_{39}x_{62} + x_{39}x_{63} + x_{40}x_{42} + x_{40}x_{44} + x_{40}x_{45} + x_{40}x_{48} + x_{40}x_{50} + x_{40}x_{51} + x_{40}x_{52} + x_{40}x_{57} + x_{40}x_{60} + x_{40}x_{61} + x_{40}x_{63} + x_{41}x_{42} + x_{41}x_{43} + x_{41}x_{45} + x_{41}x_{46} + x_{41}x_{48} + x_{41}x_{49} + x_{41}x_{51} + x_{41}x_{53} + x_{41}x_{54} + x_{41}x_{55} + x_{41}x_{56} + x_{41}x_{57} + x_{41}x_{58} + x_{41}x_{60} + x_{41}x_{61} + x_{41}x_{62} + x_{41}x_{63} + x_{42}x_{47} + x_{42}x_{48} + x_{42}x_{50} + x_{42}x_{52} + x_{42}x_{54} + x_{42}x_{56} + x_{42}x_{57} + x_{42}x_{58} + x_{42}x_{59} + x_{42}x_{60} + x_{42}x_{61} + x_{42}x_{63} + x_{42}x_{64} + x_{43}x_{47} + x_{43}x_{49} + x_{43}x_{51} + x_{43}x_{53} + x_{43}x_{54} + x_{43}x_{56} + x_{43}x_{58} + x_{43}x_{59} + x_{43}x_{61} + x_{43}x_{62} + x_{43}x_{63} + x_{43}x_{64} + x_{44}x_{46} + x_{44}x_{47} + x_{44}x_{49} + x_{44}x_{50} + x_{44}x_{51} + x_{44}x_{55} + x_{44}x_{56} + x_{44}x_{60} + x_{45}x_{46} + x_{45}x_{47} + x_{45}x_{48} + x_{45}x_{51} + x_{45}x_{52} + x_{45}x_{54} + x_{45}x_{59} + x_{45}x_{60} + x_{45}x_{61} + x_{45}x_{62} + x_{45}x_{63} + x_{45}x_{64} + x_{46}x_{47} + x_{46}x_{48} + x_{46}x_{50} + x_{46}x_{52} + x_{46}x_{55} + x_{46}x_{57} + x_{46}x_{58} + x_{46}x_{59} + x_{46}x_{63} + x_{47}x_{49} + x_{47}x_{50} + x_{47}x_{51} + x_{47}x_{52} + x_{47}x_{55} + x_{47}x_{56} + x_{47}x_{59} + x_{47}x_{61} + x_{47}x_{64} + x_{48}x_{50} + x_{48}x_{52} + x_{48}x_{54} + x_{48}x_{55} + x_{48}x_{56} + x_{48}x_{57} + x_{48}x_{59} + x_{48}x_{60} + x_{48}x_{61} + x_{48}x_{62} + x_{48}x_{63} + x_{49}x_{50} + x_{49}x_{51} + x_{49}x_{53} + x_{49}x_{57} + x_{49}x_{59} + x_{49}x_{60} + x_{49}x_{61} + x_{49}x_{62} + x_{50}x_{51} + x_{50}x_{54} + x_{50}x_{55} + x_{50}x_{58} + x_{50}x_{60} + x_{50}x_{61} + x_{50}x_{62} + x_{50}x_{64} + x_{51}x_{53} + x_{51}x_{55} + x_{51}x_{56} + x_{51}x_{57} + x_{51}x_{58} + x_{51}x_{61} + x_{51}x_{62} + x_{51}x_{64} + x_{52}x_{53} + x_{52}x_{54} + x_{52}x_{55} + x_{52}x_{56} + x_{52}x_{62} + x_{52}x_{63} + x_{53}x_{55} + x_{53}x_{60} + x_{53}x_{61} + x_{53}x_{62} + x_{53}x_{63} + x_{54}x_{55} + x_{54}x_{56} + x_{54}x_{58} + x_{54}x_{59} + x_{54}x_{63} + x_{55}x_{57} + x_{55}x_{58} + x_{55}x_{59} + x_{55}x_{60} + x_{55}x_{62} + x_{55}x_{63} + x_{56}x_{57} + x_{56}x_{59} + x_{56}x_{60} + x_{56}x_{62} + x_{56}x_{63} + x_{57}x_{59} + x_{57}x_{62} + x_{57}x_{64} + x_{58}x_{60} + x_{58}x_{61} + x_{58}x_{62} + x_{58}x_{63} + x_{59}x_{60} + x_{59}x_{62} + x_{59}x_{64} + x_{60}x_{63} + x_{61}x_{63} + x_{62}x_{63} + x_{1} + x_{3} + x_{5} + x_{7} + x_{9} + x_{10} + x_{11} + x_{12} + x_{14} + x_{16} + x_{17} + x_{18} + x_{22} + x_{24} + x_{25} + x_{26} + x_{29} + x_{30} + x_{31} + x_{32} + x_{34} + x_{35} + x_{38} + x_{39} + x_{44} + x_{45} + x_{46} + x_{48} + x_{50} + x_{53} + x_{54} + x_{55} + x_{57} + x_{58} + x_{62}$

$y_{19} = x_{1}x_{2} + x_{1}x_{3} + x_{1}x_{4} + x_{1}x_{6} + x_{1}x_{8} + x_{1}x_{9} + x_{1}x_{11} + x_{1}x_{12} + x_{1}x_{14} + x_{1}x_{16} + x_{1}x_{21} + x_{1}x_{23} + x_{1}x_{25} + x_{1}x_{29} + x_{1}x_{31} + x_{1}x_{32} + x_{1}x_{34} + x_{1}x_{36} + x_{1}x_{41} + x_{1}x_{43} + x_{1}x_{46} + x_{1}x_{47} + x_{1}x_{51} + x_{1}x_{52} + x_{1}x_{54} + x_{1}x_{57} + x_{1}x_{58} + x_{1}x_{59} + x_{1}x_{60} + x_{1}x_{63} + x_{2}x_{3} + x_{2}x_{5} + x_{2}x_{13} + x_{2}x_{14} + x_{2}x_{16} + x_{2}x_{17} + x_{2}x_{18} + x_{2}x_{20} + x_{2}x_{22} + x_{2}x_{23} + x_{2}x_{27} + x_{2}x_{29} + x_{2}x_{31} + x_{2}x_{32} + x_{2}x_{34} + x_{2}x_{35} + x_{2}x_{36} + x_{2}x_{39} + x_{2}x_{40} + x_{2}x_{41} + x_{2}x_{42} + x_{2}x_{43} + x_{2}x_{44} + x_{2}x_{45} + x_{2}x_{47} + x_{2}x_{49} + x_{2}x_{50} + x_{2}x_{52} + x_{2}x_{53} + x_{2}x_{54} + x_{2}x_{57} + x_{2}x_{58} + x_{2}x_{60} + x_{2}x_{61} + x_{3}x_{4} + x_{3}x_{6} + x_{3}x_{7} + x_{3}x_{8} + x_{3}x_{12} + x_{3}x_{13} + x_{3}x_{14} + x_{3}x_{17} + x_{3}x_{18} + x_{3}x_{19} + x_{3}x_{21} + x_{3}x_{23} + x_{3}x_{24} + x_{3}x_{25} + x_{3}x_{26} + x_{3}x_{27} + x_{3}x_{28} + x_{3}x_{36} + x_{3}x_{38} + x_{3}x_{39} + x_{3}x_{40} + x_{3}x_{41} + x_{3}x_{44} + x_{3}x_{45} + x_{3}x_{47} + x_{3}x_{48} + x_{3}x_{50} + x_{3}x_{51} + x_{3}x_{52} + x_{3}x_{53} + x_{3}x_{55} + x_{3}x_{56} + x_{3}x_{59} + x_{3}x_{61} + x_{4}x_{6} + x_{4}x_{7} + x_{4}x_{12} + x_{4}x_{14} + x_{4}x_{15} + x_{4}x_{18} + x_{4}x_{19} + x_{4}x_{21} + x_{4}x_{24} + x_{4}x_{27} + x_{4}x_{32} + x_{4}x_{34} + x_{4}x_{36} + x_{4}x_{40} + x_{4}x_{43} + x_{4}x_{44} + x_{4}x_{45} + x_{4}x_{46} + x_{4}x_{49} + x_{4}x_{50} + x_{4}x_{51} + x_{4}x_{52} + x_{4}x_{56} + x_{4}x_{57} + x_{4}x_{58} + x_{4}x_{59} + x_{4}x_{61} + x_{4}x_{62} + x_{5}x_{6} + x_{5}x_{8} + x_{5}x_{9} + x_{5}x_{13} + x_{5}x_{14} + x_{5}x_{15} + x_{5}x_{18} + x_{5}x_{19} + x_{5}x_{20} + x_{5}x_{21} + x_{5}x_{24} + x_{5}x_{25} + x_{5}x_{26} + x_{5}x_{27} + x_{5}x_{28} + x_{5}x_{30} + x_{5}x_{32} + x_{5}x_{35} + x_{5}x_{36} + x_{5}x_{37} + x_{5}x_{39} + x_{5}x_{40} + x_{5}x_{41} + x_{5}x_{42} + x_{5}x_{43} + x_{5}x_{44} + x_{5}x_{45} + x_{5}x_{47} + x_{5}x_{49} + x_{5}x_{53} + x_{5}x_{55} + x_{5}x_{56} + x_{5}x_{58} + x_{5}x_{61} + x_{6}x_{7} + x_{6}x_{8} + x_{6}x_{10} + x_{6}x_{11} + x_{6}x_{16} + x_{6}x_{17} + x_{6}x_{23} + x_{6}x_{28} + x_{6}x_{29} + x_{6}x_{30} + x_{6}x_{34} + x_{6}x_{37} + x_{6}x_{38} + x_{6}x_{39} + x_{6}x_{40} + x_{6}x_{42} + x_{6}x_{44} + x_{6}x_{45} + x_{6}x_{50} + x_{6}x_{52} + x_{6}x_{57} + x_{6}x_{60} + x_{6}x_{61} + x_{6}x_{63} + x_{7}x_{9} + x_{7}x_{10} + x_{7}x_{13} + x_{7}x_{14} + x_{7}x_{15} + x_{7}x_{18} + x_{7}x_{20} + x_{7}x_{23} + x_{7}x_{24} + x_{7}x_{30} + x_{7}x_{31} + x_{7}x_{33} + x_{7}x_{36} + x_{7}x_{37} + x_{7}x_{39} + x_{7}x_{43} + x_{7}x_{44} + x_{7}x_{45} + x_{7}x_{46} + x_{7}x_{47} + x_{7}x_{51} + x_{7}x_{52} + x_{7}x_{53} + x_{7}x_{55} + x_{7}x_{58} + x_{7}x_{59} + x_{7}x_{61} + x_{7}x_{63} + x_{7}x_{64} + x_{8}x_{10} + x_{8}x_{11} + x_{8}x_{14} + x_{8}x_{15} + x_{8}x_{17} + x_{8}x_{20} + x_{8}x_{22} + x_{8}x_{24} + x_{8}x_{25} + x_{8}x_{27} + x_{8}x_{28} + x_{8}x_{29} + x_{8}x_{36} + x_{8}x_{37} + x_{8}x_{39} + x_{8}x_{41} + x_{8}x_{45} + x_{8}x_{50} + x_{8}x_{52} + x_{8}x_{54} + x_{8}x_{55} + x_{8}x_{58} + x_{8}x_{62} + x_{8}x_{63} + x_{9}x_{10} + x_{9}x_{11} + x_{9}x_{12} + x_{9}x_{15} + x_{9}x_{17} + x_{9}x_{18} + x_{9}x_{19} + x_{9}x_{20} + x_{9}x_{21} + x_{9}x_{25} + x_{9}x_{26} + x_{9}x_{29} + x_{9}x_{30} + x_{9}x_{32} + x_{9}x_{34} + x_{9}x_{37} + x_{9}x_{44} + x_{9}x_{47} + x_{9}x_{51} + x_{9}x_{53} + x_{9}x_{54} + x_{9}x_{55} + x_{9}x_{56} + x_{9}x_{57} + x_{9}x_{58} + x_{9}x_{59} + x_{9}x_{60} + x_{9}x_{64} + x_{10}x_{11} + x_{10}x_{12} + x_{10}x_{14} + x_{10}x_{18} + x_{10}x_{23} + x_{10}x_{24} + x_{10}x_{25} + x_{10}x_{27} + x_{10}x_{29} + x_{10}x_{31} + x_{10}x_{32} + x_{10}x_{33} + x_{10}x_{36} + x_{10}x_{38} + x_{10}x_{42} + x_{10}x_{43} + x_{10}x_{45} + x_{10}x_{46} + x_{10}x_{47} + x_{10}x_{49} + x_{10}x_{53} + x_{10}x_{56} + x_{10}x_{63} + x_{10}x_{64} + x_{11}x_{13} + x_{11}x_{17} + x_{11}x_{22} + x_{11}x_{23} + x_{11}x_{24} + x_{11}x_{28} + x_{11}x_{31} + x_{11}x_{32} + x_{11}x_{34} + x_{11}x_{35} + x_{11}x_{36} + x_{11}x_{38} + x_{11}x_{39} + x_{11}x_{40} + x_{11}x_{41} + x_{11}x_{44} + x_{11}x_{45} + x_{11}x_{46} + x_{11}x_{47} + x_{11}x_{50} + x_{11}x_{52} + x_{11}x_{53} + x_{11}x_{54} + x_{11}x_{58} + x_{11}x_{60} + x_{11}x_{61} + x_{12}x_{13} + x_{12}x_{17} + x_{12}x_{18} + x_{12}x_{19} + x_{12}x_{20} + x_{12}x_{22} + x_{12}x_{23} + x_{12}x_{24} + x_{12}x_{27} + x_{12}x_{29} + x_{12}x_{31} + x_{12}x_{40} + x_{12}x_{41} + x_{12}x_{42} + x_{12}x_{45} + x_{12}x_{46} + x_{12}x_{47} + x_{12}x_{49} + x_{12}x_{55} + x_{12}x_{56} + x_{12}x_{58} + x_{12}x_{61} + x_{12}x_{63} + x_{13}x_{21} + x_{13}x_{23} + x_{13}x_{26} + x_{13}x_{27} + x_{13}x_{28} + x_{13}x_{30} + x_{13}x_{32} + x_{13}x_{34} + x_{13}x_{35} + x_{13}x_{36} + x_{13}x_{37} + x_{13}x_{39} + x_{13}x_{43} + x_{13}x_{46} + x_{13}x_{47} + x_{13}x_{49} + x_{13}x_{50} + x_{13}x_{52} + x_{13}x_{53} + x_{13}x_{57} + x_{13}x_{58} + x_{13}x_{59} + x_{13}x_{62} + x_{13}x_{64} + x_{14}x_{17} + x_{14}x_{19} + x_{14}x_{21} + x_{14}x_{24} + x_{14}x_{25} + x_{14}x_{27} + x_{14}x_{29} + x_{14}x_{31} + x_{14}x_{33} + x_{14}x_{34} + x_{14}x_{35} + x_{14}x_{37} + x_{14}x_{43} + x_{14}x_{48} + x_{14}x_{51} + x_{14}x_{52} + x_{14}x_{53} + x_{14}x_{54} + x_{14}x_{56} + x_{14}x_{58} + x_{14}x_{59} + x_{14}x_{60} + x_{14}x_{61} + x_{14}x_{62} + x_{14}x_{63} + x_{15}x_{16} + x_{15}x_{18} + x_{15}x_{20} + x_{15}x_{22} + x_{15}x_{24} + x_{15}x_{25} + x_{15}x_{26} + x_{15}x_{27} + x_{15}x_{30} + x_{15}x_{32} + x_{15}x_{33} + x_{15}x_{35} + x_{15}x_{36} + x_{15}x_{40} + x_{15}x_{41} + x_{15}x_{42} + x_{15}x_{46} + x_{15}x_{47} + x_{15}x_{48} + x_{15}x_{49} + x_{15}x_{50} + x_{15}x_{54} + x_{15}x_{59} + x_{15}x_{60} + x_{15}x_{63} + x_{16}x_{18} + x_{16}x_{20} + x_{16}x_{22} + x_{16}x_{23} + x_{16}x_{26} + x_{16}x_{27} + x_{16}x_{28} + x_{16}x_{29} + x_{16}x_{32} + x_{16}x_{34} + x_{16}x_{35} + x_{16}x_{36} + x_{16}x_{37} + x_{16}x_{38} + x_{16}x_{40} + x_{16}x_{41} + x_{16}x_{42} + x_{16}x_{44} + x_{16}x_{45} + x_{16}x_{49} + x_{16}x_{50} + x_{16}x_{52} + x_{16}x_{54} + x_{16}x_{56} + x_{16}x_{58} + x_{16}x_{60} + x_{16}x_{61} + x_{16}x_{62} + x_{17}x_{19} + x_{17}x_{21} + x_{17}x_{22} + x_{17}x_{23} + x_{17}x_{24} + x_{17}x_{27} + x_{17}x_{28} + x_{17}x_{30} + x_{17}x_{31} + x_{17}x_{34} + x_{17}x_{35} + x_{17}x_{38} + x_{17}x_{39} + x_{17}x_{40} + x_{17}x_{41} + x_{17}x_{43} + x_{17}x_{45} + x_{17}x_{47} + x_{17}x_{51} + x_{17}x_{52} + x_{17}x_{55} + x_{17}x_{56} + x_{17}x_{57} + x_{17}x_{58} + x_{17}x_{61} + x_{17}x_{62} + x_{17}x_{63} + x_{18}x_{19} + x_{18}x_{21} + x_{18}x_{22} + x_{18}x_{23} + x_{18}x_{25} + x_{18}x_{28} + x_{18}x_{29} + x_{18}x_{31} + x_{18}x_{33} + x_{18}x_{36} + x_{18}x_{43} + x_{18}x_{44} + x_{18}x_{46} + x_{18}x_{49} + x_{18}x_{52} + x_{18}x_{56} + x_{18}x_{57} + x_{18}x_{59} + x_{18}x_{60} + x_{18}x_{62} + x_{18}x_{63} + x_{18}x_{64} + x_{19}x_{20} + x_{19}x_{22} + x_{19}x_{23} + x_{19}x_{25} + x_{19}x_{27} + x_{19}x_{30} + x_{19}x_{32} + x_{19}x_{33} + x_{19}x_{34} + x_{19}x_{35} + x_{19}x_{37} + x_{19}x_{38} + x_{19}x_{39} + x_{19}x_{40} + x_{19}x_{41} + x_{19}x_{43} + x_{19}x_{47} + x_{19}x_{48} + x_{19}x_{49} + x_{19}x_{51} + x_{19}x_{54} + x_{19}x_{55} + x_{19}x_{56} + x_{19}x_{59} + x_{19}x_{61} + x_{19}x_{62} + x_{19}x_{64} + x_{20}x_{21} + x_{20}x_{22} + x_{20}x_{23} + x_{20}x_{24} + x_{20}x_{25} + x_{20}x_{27} + x_{20}x_{30} + x_{20}x_{31} + x_{20}x_{32} + x_{20}x_{35} + x_{20}x_{36} + x_{20}x_{37} + x_{20}x_{38} + x_{20}x_{39} + x_{20}x_{40} + x_{20}x_{44} + x_{20}x_{45} + x_{20}x_{46} + x_{20}x_{50} + x_{20}x_{52} + x_{20}x_{53} + x_{20}x_{55} + x_{20}x_{57} + x_{20}x_{59} + x_{20}x_{60} + x_{20}x_{61} + x_{20}x_{63} + x_{20}x_{64} + x_{21}x_{23} + x_{21}x_{24} + x_{21}x_{25} + x_{21}x_{27} + x_{21}x_{30} + x_{21}x_{31} + x_{21}x_{32} + x_{21}x_{33} + x_{21}x_{35} + x_{21}x_{36} + x_{21}x_{37} + x_{21}x_{42} + x_{21}x_{43} + x_{21}x_{44} + x_{21}x_{47} + x_{21}x_{48} + x_{21}x_{53} + x_{21}x_{57} + x_{21}x_{59} + x_{21}x_{61} + x_{21}x_{62} + x_{22}x_{25} + x_{22}x_{26} + x_{22}x_{28} + x_{22}x_{30} + x_{22}x_{32} + x_{22}x_{35} + x_{22}x_{39} + x_{22}x_{43} + x_{22}x_{44} + x_{22}x_{45} + x_{22}x_{47} + x_{22}x_{50} + x_{22}x_{52} + x_{22}x_{54} + x_{22}x_{57} + x_{22}x_{59} + x_{22}x_{60} + x_{22}x_{63} + x_{23}x_{25} + x_{23}x_{26} + x_{23}x_{27} + x_{23}x_{30} + x_{23}x_{33} + x_{23}x_{35} + x_{23}x_{36} + x_{23}x_{39} + x_{23}x_{40} + x_{23}x_{41} + x_{23}x_{44} + x_{23}x_{47} + x_{23}x_{48} + x_{23}x_{49} + x_{23}x_{52} + x_{23}x_{55} + x_{23}x_{56} + x_{23}x_{57} + x_{23}x_{59} + x_{23}x_{61} + x_{23}x_{63} + x_{23}x_{64} + x_{24}x_{28} + x_{24}x_{29} + x_{24}x_{30} + x_{24}x_{31} + x_{24}x_{33} + x_{24}x_{34} + x_{24}x_{36} + x_{24}x_{39} + x_{24}x_{40} + x_{24}x_{41} + x_{24}x_{42} + x_{24}x_{43} + x_{24}x_{44} + x_{24}x_{46} + x_{24}x_{49} + x_{24}x_{55} + x_{24}x_{56} + x_{24}x_{57} + x_{24}x_{59} + x_{24}x_{62} + x_{24}x_{63} + x_{24}x_{64} + x_{25}x_{26} + x_{25}x_{29} + x_{25}x_{32} + x_{25}x_{34} + x_{25}x_{35} + x_{25}x_{38} + x_{25}x_{40} + x_{25}x_{42} + x_{25}x_{46} + x_{25}x_{47} + x_{25}x_{49} + x_{25}x_{50} + x_{25}x_{52} + x_{25}x_{53} + x_{25}x_{54} + x_{25}x_{56} + x_{25}x_{60} + x_{25}x_{62} + x_{25}x_{63} + x_{25}x_{64} + x_{26}x_{30} + x_{26}x_{31} + x_{26}x_{32} + x_{26}x_{33} + x_{26}x_{36} + x_{26}x_{37} + x_{26}x_{38} + x_{26}x_{39} + x_{26}x_{42} + x_{26}x_{43} + x_{26}x_{44} + x_{26}x_{45} + x_{26}x_{46} + x_{26}x_{47} + x_{26}x_{48} + x_{26}x_{49} + x_{26}x_{52} + x_{26}x_{53} + x_{26}x_{54} + x_{26}x_{55} + x_{26}x_{56} + x_{26}x_{57} + x_{26}x_{59} + x_{26}x_{63} + x_{26}x_{64} + x_{27}x_{28} + x_{27}x_{29} + x_{27}x_{31} + x_{27}x_{34} + x_{27}x_{37} + x_{27}x_{38} + x_{27}x_{42} + x_{27}x_{44} + x_{27}x_{46} + x_{27}x_{47} + x_{27}x_{49} + x_{27}x_{51} + x_{27}x_{52} + x_{27}x_{57} + x_{27}x_{58} + x_{27}x_{61} + x_{27}x_{62} + x_{28}x_{33} + x_{28}x_{34} + x_{28}x_{36} + x_{28}x_{39} + x_{28}x_{40} + x_{28}x_{41} + x_{28}x_{42} + x_{28}x_{48} + x_{28}x_{50} + x_{28}x_{51} + x_{28}x_{53} + x_{28}x_{54} + x_{28}x_{57} + x_{28}x_{59} + x_{28}x_{61} + x_{28}x_{63} + x_{29}x_{31} + x_{29}x_{33} + x_{29}x_{34} + x_{29}x_{40} + x_{29}x_{46} + x_{29}x_{47} + x_{29}x_{50} + x_{29}x_{52} + x_{29}x_{54} + x_{29}x_{58} + x_{29}x_{62} + x_{29}x_{63} + x_{30}x_{35} + x_{30}x_{36} + x_{30}x_{37} + x_{30}x_{38} + x_{30}x_{39} + x_{30}x_{41} + x_{30}x_{43} + x_{30}x_{44} + x_{30}x_{47} + x_{30}x_{49} + x_{30}x_{50} + x_{30}x_{53} + x_{30}x_{56} + x_{30}x_{60} + x_{30}x_{63} + x_{30}x_{64} + x_{31}x_{33} + x_{31}x_{34} + x_{31}x_{35} + x_{31}x_{37} + x_{31}x_{40} + x_{31}x_{41} + x_{31}x_{44} + x_{31}x_{49} + x_{31}x_{51} + x_{31}x_{53} + x_{31}x_{58} + x_{31}x_{59} + x_{31}x_{60} + x_{31}x_{61} + x_{32}x_{33} + x_{32}x_{34} + x_{32}x_{35} + x_{32}x_{36} + x_{32}x_{37} + x_{32}x_{40} + x_{32}x_{42} + x_{32}x_{44} + x_{32}x_{48} + x_{32}x_{52} + x_{32}x_{53} + x_{32}x_{57} + x_{32}x_{58} + x_{32}x_{60} + x_{33}x_{36} + x_{33}x_{37} + x_{33}x_{39} + x_{33}x_{41} + x_{33}x_{45} + x_{33}x_{52} + x_{33}x_{54} + x_{33}x_{57} + x_{33}x_{58} + x_{33}x_{61} + x_{34}x_{35} + x_{34}x_{36} + x_{34}x_{37} + x_{34}x_{38} + x_{34}x_{41} + x_{34}x_{43} + x_{34}x_{44} + x_{34}x_{46} + x_{34}x_{47} + x_{34}x_{48} + x_{34}x_{50} + x_{34}x_{52} + x_{34}x_{59} + x_{34}x_{60} + x_{34}x_{61} + x_{34}x_{63} + x_{34}x_{64} + x_{35}x_{36} + x_{35}x_{37} + x_{35}x_{39} + x_{35}x_{40} + x_{35}x_{41} + x_{35}x_{42} + x_{35}x_{43} + x_{35}x_{44} + x_{35}x_{46} + x_{35}x_{47} + x_{35}x_{48} + x_{35}x_{49} + x_{35}x_{54} + x_{35}x_{56} + x_{35}x_{62} + x_{35}x_{63} + x_{35}x_{64} + x_{36}x_{38} + x_{36}x_{39} + x_{36}x_{41} + x_{36}x_{42} + x_{36}x_{43} + x_{36}x_{44} + x_{36}x_{45} + x_{36}x_{50} + x_{36}x_{51} + x_{36}x_{53} + x_{36}x_{54} + x_{36}x_{55} + x_{36}x_{58} + x_{36}x_{60} + x_{36}x_{61} + x_{36}x_{63} + x_{37}x_{39} + x_{37}x_{40} + x_{37}x_{43} + x_{37}x_{44} + x_{37}x_{48} + x_{37}x_{53} + x_{37}x_{55} + x_{37}x_{56} + x_{37}x_{59} + x_{37}x_{60} + x_{37}x_{61} + x_{37}x_{62} + x_{37}x_{63} + x_{37}x_{64} + x_{38}x_{39} + x_{38}x_{41} + x_{38}x_{42} + x_{38}x_{43} + x_{38}x_{47} + x_{38}x_{48} + x_{38}x_{49} + x_{38}x_{51} + x_{38}x_{54} + x_{38}x_{56} + x_{38}x_{58} + x_{38}x_{60} + x_{38}x_{61} + x_{38}x_{62} + x_{39}x_{40} + x_{39}x_{43} + x_{39}x_{44} + x_{39}x_{45} + x_{39}x_{49} + x_{39}x_{50} + x_{39}x_{51} + x_{39}x_{53} + x_{39}x_{54} + x_{39}x_{60} + x_{39}x_{61} + x_{39}x_{63} + x_{39}x_{64} + x_{40}x_{42} + x_{40}x_{45} + x_{40}x_{46} + x_{40}x_{47} + x_{40}x_{48} + x_{40}x_{49} + x_{40}x_{51} + x_{40}x_{53} + x_{40}x_{57} + x_{40}x_{58} + x_{40}x_{59} + x_{40}x_{60} + x_{40}x_{61} + x_{40}x_{64} + x_{41}x_{42} + x_{41}x_{44} + x_{41}x_{45} + x_{41}x_{46} + x_{41}x_{47} + x_{41}x_{49} + x_{41}x_{53} + x_{41}x_{54} + x_{41}x_{56} + x_{41}x_{57} + x_{41}x_{58} + x_{41}x_{59} + x_{41}x_{61} + x_{41}x_{63} + x_{41}x_{64} + x_{42}x_{45} + x_{42}x_{49} + x_{42}x_{50} + x_{42}x_{51} + x_{42}x_{54} + x_{42}x_{56} + x_{42}x_{57} + x_{42}x_{58} + x_{42}x_{61} + x_{42}x_{62} + x_{42}x_{64} + x_{43}x_{44} + x_{43}x_{45} + x_{43}x_{46} + x_{43}x_{47} + x_{43}x_{48} + x_{43}x_{49} + x_{43}x_{52} + x_{43}x_{55} + x_{43}x_{56} + x_{43}x_{57} + x_{43}x_{58} + x_{43}x_{62} + x_{44}x_{45} + x_{44}x_{48} + x_{44}x_{49} + x_{44}x_{50} + x_{44}x_{51} + x_{44}x_{55} + x_{44}x_{57} + x_{44}x_{59} + x_{44}x_{61} + x_{44}x_{62} + x_{45}x_{47} + x_{45}x_{51} + x_{45}x_{54} + x_{45}x_{56} + x_{45}x_{57} + x_{45}x_{59} + x_{45}x_{61} + x_{45}x_{62} + x_{46}x_{48} + x_{46}x_{51} + x_{46}x_{54} + x_{46}x_{55} + x_{46}x_{56} + x_{46}x_{63} + x_{47}x_{51} + x_{47}x_{53} + x_{47}x_{54} + x_{47}x_{56} + x_{47}x_{57} + x_{47}x_{60} + x_{47}x_{62} + x_{47}x_{63} + x_{48}x_{49} + x_{48}x_{54} + x_{48}x_{56} + x_{48}x_{57} + x_{48}x_{62} + x_{48}x_{64} + x_{49}x_{53} + x_{49}x_{55} + x_{49}x_{56} + x_{49}x_{57} + x_{49}x_{58} + x_{49}x_{59} + x_{49}x_{60} + x_{49}x_{61} + x_{50}x_{52} + x_{50}x_{54} + x_{50}x_{56} + x_{50}x_{57} + x_{50}x_{58} + x_{50}x_{59} + x_{50}x_{60} + x_{50}x_{61} + x_{50}x_{63} + x_{51}x_{52} + x_{51}x_{53} + x_{51}x_{56} + x_{51}x_{57} + x_{51}x_{58} + x_{51}x_{59} + x_{51}x_{61} + x_{51}x_{62} + x_{51}x_{63} + x_{51}x_{64} + x_{52}x_{53} + x_{52}x_{55} + x_{52}x_{57} + x_{52}x_{58} + x_{52}x_{60} + x_{52}x_{62} + x_{52}x_{63} + x_{53}x_{57} + x_{53}x_{58} + x_{53}x_{61} + x_{53}x_{62} + x_{53}x_{63} + x_{53}x_{64} + x_{54}x_{55} + x_{54}x_{56} + x_{54}x_{57} + x_{54}x_{58} + x_{54}x_{60} + x_{54}x_{61} + x_{54}x_{62} + x_{54}x_{64} + x_{55}x_{60} + x_{55}x_{61} + x_{55}x_{63} + x_{56}x_{57} + x_{56}x_{59} + x_{56}x_{61} + x_{56}x_{62} + x_{57}x_{59} + x_{57}x_{60} + x_{57}x_{62} + x_{57}x_{63} + x_{57}x_{64} + x_{58}x_{59} + x_{58}x_{60} + x_{58}x_{61} + x_{58}x_{62} + x_{58}x_{63} + x_{58}x_{64} + x_{59}x_{62} + x_{59}x_{63} + x_{60}x_{63} + x_{61}x_{62} + x_{61}x_{64} + x_{2} + x_{3} + x_{4} + x_{7} + x_{8} + x_{9} + x_{10} + x_{12} + x_{13} + x_{14} + x_{16} + x_{17} + x_{18} + x_{19} + x_{20} + x_{21} + x_{22} + x_{23} + x_{24} + x_{29} + x_{30} + x_{31} + x_{32} + x_{34} + x_{36} + x_{37} + x_{38} + x_{40} + x_{41} + x_{44} + x_{45} + x_{47} + x_{51} + x_{53} + x_{54} + x_{55} + x_{60} + x_{61} + x_{63}$

$y_{20} = x_{1}x_{3} + x_{1}x_{4} + x_{1}x_{7} + x_{1}x_{8} + x_{1}x_{9} + x_{1}x_{11} + x_{1}x_{13} + x_{1}x_{18} + x_{1}x_{21} + x_{1}x_{22} + x_{1}x_{23} + x_{1}x_{25} + x_{1}x_{27} + x_{1}x_{28} + x_{1}x_{29} + x_{1}x_{31} + x_{1}x_{32} + x_{1}x_{34} + x_{1}x_{35} + x_{1}x_{36} + x_{1}x_{37} + x_{1}x_{39} + x_{1}x_{46} + x_{1}x_{47} + x_{1}x_{52} + x_{1}x_{54} + x_{1}x_{55} + x_{1}x_{56} + x_{1}x_{57} + x_{1}x_{58} + x_{1}x_{60} + x_{1}x_{61} + x_{2}x_{5} + x_{2}x_{6} + x_{2}x_{9} + x_{2}x_{12} + x_{2}x_{14} + x_{2}x_{15} + x_{2}x_{18} + x_{2}x_{19} + x_{2}x_{20} + x_{2}x_{22} + x_{2}x_{24} + x_{2}x_{27} + x_{2}x_{29} + x_{2}x_{31} + x_{2}x_{32} + x_{2}x_{33} + x_{2}x_{36} + x_{2}x_{38} + x_{2}x_{41} + x_{2}x_{42} + x_{2}x_{45} + x_{2}x_{49} + x_{2}x_{50} + x_{2}x_{52} + x_{2}x_{54} + x_{2}x_{55} + x_{2}x_{57} + x_{2}x_{58} + x_{2}x_{60} + x_{2}x_{62} + x_{2}x_{63} + x_{2}x_{64} + x_{3}x_{4} + x_{3}x_{5} + x_{3}x_{6} + x_{3}x_{7} + x_{3}x_{8} + x_{3}x_{9} + x_{3}x_{11} + x_{3}x_{12} + x_{3}x_{17} + x_{3}x_{20} + x_{3}x_{24} + x_{3}x_{26} + x_{3}x_{27} + x_{3}x_{33} + x_{3}x_{35} + x_{3}x_{39} + x_{3}x_{42} + x_{3}x_{43} + x_{3}x_{45} + x_{3}x_{52} + x_{3}x_{54} + x_{3}x_{55} + x_{3}x_{56} + x_{3}x_{62} + x_{3}x_{63} + x_{4}x_{5} + x_{4}x_{6} + x_{4}x_{7} + x_{4}x_{8} + x_{4}x_{10} + x_{4}x_{13} + x_{4}x_{14} + x_{4}x_{15} + x_{4}x_{18} + x_{4}x_{19} + x_{4}x_{20} + x_{4}x_{21} + x_{4}x_{23} + x_{4}x_{24} + x_{4}x_{25} + x_{4}x_{27} + x_{4}x_{29} + x_{4}x_{30} + x_{4}x_{32} + x_{4}x_{33} + x_{4}x_{34} + x_{4}x_{36} + x_{4}x_{39} + x_{4}x_{40} + x_{4}x_{42} + x_{4}x_{46} + x_{4}x_{47} + x_{4}x_{49} + x_{4}x_{50} + x_{4}x_{51} + x_{4}x_{52} + x_{4}x_{54} + x_{4}x_{56} + x_{4}x_{57} + x_{4}x_{60} + x_{4}x_{63} + x_{4}x_{64} + x_{5}x_{6} + x_{5}x_{11} + x_{5}x_{12} + x_{5}x_{14} + x_{5}x_{15} + x_{5}x_{16} + x_{5}x_{17} + x_{5}x_{19} + x_{5}x_{20} + x_{5}x_{25} + x_{5}x_{29} + x_{5}x_{33} + x_{5}x_{36} + x_{5}x_{40} + x_{5}x_{42} + x_{5}x_{43} + x_{5}x_{45} + x_{5}x_{46} + x_{5}x_{48} + x_{5}x_{49} + x_{5}x_{51} + x_{5}x_{53} + x_{5}x_{54} + x_{5}x_{55} + x_{5}x_{59} + x_{5}x_{61} + x_{5}x_{62} + x_{5}x_{64} + x_{6}x_{9} + x_{6}x_{11} + x_{6}x_{14} + x_{6}x_{15} + x_{6}x_{17} + x_{6}x_{18} + x_{6}x_{19} + x_{6}x_{20} + x_{6}x_{21} + x_{6}x_{23} + x_{6}x_{24} + x_{6}x_{26} + x_{6}x_{30} + x_{6}x_{31} + x_{6}x_{32} + x_{6}x_{34} + x_{6}x_{42} + x_{6}x_{43} + x_{6}x_{45} + x_{6}x_{46} + x_{6}x_{48} + x_{6}x_{49} + x_{6}x_{51} + x_{6}x_{52} + x_{6}x_{55} + x_{6}x_{56} + x_{6}x_{57} + x_{6}x_{58} + x_{6}x_{59} + x_{6}x_{62} + x_{6}x_{63} + x_{7}x_{8} + x_{7}x_{11} + x_{7}x_{12} + x_{7}x_{16} + x_{7}x_{17} + x_{7}x_{18} + x_{7}x_{19} + x_{7}x_{20} + x_{7}x_{22} + x_{7}x_{24} + x_{7}x_{26} + x_{7}x_{28} + x_{7}x_{29} + x_{7}x_{30} + x_{7}x_{39} + x_{7}x_{40} + x_{7}x_{41} + x_{7}x_{42} + x_{7}x_{43} + x_{7}x_{45} + x_{7}x_{48} + x_{7}x_{49} + x_{7}x_{51} + x_{7}x_{55} + x_{7}x_{57} + x_{7}x_{60} + x_{7}x_{61} + x_{7}x_{64} + x_{8}x_{10} + x_{8}x_{11} + x_{8}x_{14} + x_{8}x_{15} + x_{8}x_{16} + x_{8}x_{18} + x_{8}x_{20} + x_{8}x_{32} + x_{8}x_{39} + x_{8}x_{43} + x_{8}x_{44} + x_{8}x_{46} + x_{8}x_{47} + x_{8}x_{48} + x_{8}x_{50} + x_{8}x_{56} + x_{8}x_{57} + x_{8}x_{58} + x_{8}x_{61} + x_{9}x_{10} + x_{9}x_{12} + x_{9}x_{14} + x_{9}x_{15} + x_{9}x_{16} + x_{9}x_{17} + x_{9}x_{19} + x_{9}x_{21} + x_{9}x_{27} + x_{9}x_{28} + x_{9}x_{32} + x_{9}x_{33} + x_{9}x_{36} + x_{9}x_{39} + x_{9}x_{41} + x_{9}x_{45} + x_{9}x_{48} + x_{9}x_{55} + x_{9}x_{56} + x_{9}x_{59} + x_{9}x_{62} + x_{9}x_{63} + x_{10}x_{12} + x_{10}x_{15} + x_{10}x_{17} + x_{10}x_{21} + x_{10}x_{23} + x_{10}x_{29} + x_{10}x_{30} + x_{10}x_{31} + x_{10}x_{32} + x_{10}x_{34} + x_{10}x_{37} + x_{10}x_{39} + x_{10}x_{40} + x_{10}x_{42} + x_{10}x_{45} + x_{10}x_{46} + x_{10}x_{48} + x_{10}x_{49} + x_{10}x_{50} + x_{10}x_{55} + x_{10}x_{56} + x_{10}x_{58} + x_{10}x_{60} + x_{10}x_{61} + x_{11}x_{12} + x_{11}x_{16} + x_{11}x_{17} + x_{11}x_{18} + x_{11}x_{19} + x_{11}x_{22} + x_{11}x_{23} + x_{11}x_{24} + x_{11}x_{26} + x_{11}x_{28} + x_{11}x_{29} + x_{11}x_{30} + x_{11}x_{32} + x_{11}x_{34} + x_{11}x_{35} + x_{11}x_{39} + x_{11}x_{40} + x_{11}x_{41} + x_{11}x_{42} + x_{11}x_{43} + x_{11}x_{44} + x_{11}x_{45} + x_{11}x_{47} + x_{11}x_{48} + x_{11}x_{52} + x_{11}x_{53} + x_{11}x_{56} + x_{11}x_{61} + x_{11}x_{62} + x_{11}x_{63} + x_{11}x_{64} + x_{12}x_{13} + x_{12}x_{14} + x_{12}x_{15} + x_{12}x_{16} + x_{12}x_{17} + x_{12}x_{18} + x_{12}x_{19} + x_{12}x_{20} + x_{12}x_{27} + x_{12}x_{28} + x_{12}x_{30} + x_{12}x_{31} + x_{12}x_{34} + x_{12}x_{35} + x_{12}x_{42} + x_{12}x_{43} + x_{12}x_{44} + x_{12}x_{46} + x_{12}x_{47} + x_{12}x_{50} + x_{12}x_{51} + x_{12}x_{53} + x_{12}x_{54} + x_{12}x_{56} + x_{12}x_{57} + x_{12}x_{60} + x_{12}x_{61} + x_{12}x_{63} + x_{13}x_{15} + x_{13}x_{17} + x_{13}x_{18} + x_{13}x_{19} + x_{13}x_{20} + x_{13}x_{21} + x_{13}x_{22} + x_{13}x_{23} + x_{13}x_{26} + x_{13}x_{27} + x_{13}x_{29} + x_{13}x_{32} + x_{13}x_{33} + x_{13}x_{34} + x_{13}x_{35} + x_{13}x_{36} + x_{13}x_{39} + x_{13}x_{43} + x_{13}x_{44} + x_{13}x_{45} + x_{13}x_{48} + x_{13}x_{51} + x_{13}x_{52} + x_{13}x_{54} + x_{13}x_{55} + x_{13}x_{57} + x_{13}x_{58} + x_{13}x_{59} + x_{13}x_{63} + x_{14}x_{15} + x_{14}x_{18} + x_{14}x_{19} + x_{14}x_{20} + x_{14}x_{24} + x_{14}x_{25} + x_{14}x_{26} + x_{14}x_{30} + x_{14}x_{31} + x_{14}x_{37} + x_{14}x_{39} + x_{14}x_{41} + x_{14}x_{43} + x_{14}x_{44} + x_{14}x_{45} + x_{14}x_{46} + x_{14}x_{47} + x_{14}x_{50} + x_{14}x_{52} + x_{14}x_{53} + x_{14}x_{54} + x_{14}x_{55} + x_{14}x_{56} + x_{14}x_{61} + x_{14}x_{62} + x_{14}x_{63} + x_{15}x_{16} + x_{15}x_{17} + x_{15}x_{19} + x_{15}x_{22} + x_{15}x_{23} + x_{15}x_{26} + x_{15}x_{32} + x_{15}x_{34} + x_{15}x_{35} + x_{15}x_{38} + x_{15}x_{39} + x_{15}x_{40} + x_{15}x_{42} + x_{15}x_{44} + x_{15}x_{45} + x_{15}x_{46} + x_{15}x_{48} + x_{15}x_{49} + x_{15}x_{50} + x_{15}x_{56} + x_{15}x_{57} + x_{15}x_{58} + x_{15}x_{63} + x_{16}x_{17} + x_{16}x_{18} + x_{16}x_{20} + x_{16}x_{22} + x_{16}x_{24} + x_{16}x_{25} + x_{16}x_{27} + x_{16}x_{28} + x_{16}x_{29} + x_{16}x_{30} + x_{16}x_{31} + x_{16}x_{33} + x_{16}x_{34} + x_{16}x_{35} + x_{16}x_{37} + x_{16}x_{39} + x_{16}x_{40} + x_{16}x_{41} + x_{16}x_{43} + x_{16}x_{44} + x_{16}x_{45} + x_{16}x_{47} + x_{16}x_{49} + x_{16}x_{54} + x_{16}x_{58} + x_{16}x_{60} + x_{16}x_{61} + x_{16}x_{63} + x_{17}x_{18} + x_{17}x_{21} + x_{17}x_{22} + x_{17}x_{24} + x_{17}x_{29} + x_{17}x_{31} + x_{17}x_{32} + x_{17}x_{34} + x_{17}x_{37} + x_{17}x_{38} + x_{17}x_{41} + x_{17}x_{42} + x_{17}x_{46} + x_{17}x_{48} + x_{17}x_{49} + x_{17}x_{51} + x_{17}x_{52} + x_{17}x_{54} + x_{17}x_{56} + x_{17}x_{57} + x_{17}x_{62} + x_{17}x_{64} + x_{18}x_{19} + x_{18}x_{25} + x_{18}x_{27} + x_{18}x_{28} + x_{18}x_{29} + x_{18}x_{31} + x_{18}x_{32} + x_{18}x_{33} + x_{18}x_{35} + x_{18}x_{37} + x_{18}x_{39} + x_{18}x_{42} + x_{18}x_{47} + x_{18}x_{49} + x_{18}x_{53} + x_{18}x_{54} + x_{18}x_{56} + x_{18}x_{62} + x_{18}x_{63} + x_{18}x_{64} + x_{19}x_{25} + x_{19}x_{26} + x_{19}x_{29} + x_{19}x_{30} + x_{19}x_{32} + x_{19}x_{39} + x_{19}x_{40} + x_{19}x_{41} + x_{19}x_{47} + x_{19}x_{48} + x_{19}x_{50} + x_{19}x_{52} + x_{19}x_{54} + x_{19}x_{59} + x_{19}x_{60} + x_{19}x_{61} + x_{19}x_{64} + x_{20}x_{21} + x_{20}x_{25} + x_{20}x_{27} + x_{20}x_{29} + x_{20}x_{30} + x_{20}x_{39} + x_{20}x_{43} + x_{20}x_{46} + x_{20}x_{47} + x_{20}x_{48} + x_{20}x_{51} + x_{20}x_{53} + x_{20}x_{54} + x_{20}x_{55} + x_{20}x_{56} + x_{20}x_{59} + x_{20}x_{60} + x_{20}x_{61} + x_{20}x_{62} + x_{20}x_{63} + x_{20}x_{64} + x_{21}x_{22} + x_{21}x_{23} + x_{21}x_{24} + x_{21}x_{25} + x_{21}x_{26} + x_{21}x_{27} + x_{21}x_{28} + x_{21}x_{30} + x_{21}x_{34} + x_{21}x_{39} + x_{21}x_{40} + x_{21}x_{42} + x_{21}x_{44} + x_{21}x_{46} + x_{21}x_{52} + x_{21}x_{53} + x_{21}x_{59} + x_{21}x_{60} + x_{21}x_{61} + x_{21}x_{62} + x_{21}x_{63} + x_{21}x_{64} + x_{22}x_{23} + x_{22}x_{25} + x_{22}x_{26} + x_{22}x_{28} + x_{22}x_{31} + x_{22}x_{32} + x_{22}x_{39} + x_{22}x_{40} + x_{22}x_{42} + x_{22}x_{43} + x_{22}x_{44} + x_{22}x_{47} + x_{22}x_{48} + x_{22}x_{52} + x_{22}x_{54} + x_{22}x_{56} + x_{22}x_{61} + x_{22}x_{62} + x_{22}x_{63} + x_{22}x_{64} + x_{23}x_{29} + x_{23}x_{30} + x_{23}x_{31} + x_{23}x_{32} + x_{23}x_{38} + x_{23}x_{39} + x_{23}x_{40} + x_{23}x_{45} + x_{23}x_{49} + x_{23}x_{50} + x_{23}x_{51} + x_{23}x_{55} + x_{23}x_{56} + x_{23}x_{57} + x_{23}x_{58} + x_{23}x_{59} + x_{23}x_{60} + x_{23}x_{61} + x_{23}x_{62} + x_{23}x_{63} + x_{24}x_{25} + x_{24}x_{26} + x_{24}x_{28} + x_{24}x_{29} + x_{24}x_{30} + x_{24}x_{31} + x_{24}x_{33} + x_{24}x_{39} + x_{24}x_{41} + x_{24}x_{42} + x_{24}x_{44} + x_{24}x_{45} + x_{24}x_{48} + x_{24}x_{50} + x_{24}x_{51} + x_{24}x_{52} + x_{24}x_{53} + x_{24}x_{55} + x_{24}x_{56} + x_{24}x_{58} + x_{24}x_{60} + x_{24}x_{61} + x_{24}x_{62} + x_{25}x_{26} + x_{25}x_{32} + x_{25}x_{33} + x_{25}x_{36} + x_{25}x_{38} + x_{25}x_{40} + x_{25}x_{42} + x_{25}x_{43} + x_{25}x_{45} + x_{25}x_{48} + x_{25}x_{49} + x_{25}x_{51} + x_{25}x_{55} + x_{25}x_{56} + x_{25}x_{57} + x_{25}x_{58} + x_{25}x_{60} + x_{25}x_{62} + x_{25}x_{63} + x_{25}x_{64} + x_{26}x_{33} + x_{26}x_{42} + x_{26}x_{43} + x_{26}x_{44} + x_{26}x_{45} + x_{26}x_{46} + x_{26}x_{47} + x_{26}x_{50} + x_{26}x_{55} + x_{26}x_{60} + x_{26}x_{62} + x_{26}x_{64} + x_{27}x_{29} + x_{27}x_{31} + x_{27}x_{32} + x_{27}x_{36} + x_{27}x_{37} + x_{27}x_{40} + x_{27}x_{44} + x_{27}x_{46} + x_{27}x_{48} + x_{27}x_{50} + x_{27}x_{51} + x_{27}x_{52} + x_{27}x_{56} + x_{27}x_{58} + x_{27}x_{59} + x_{28}x_{29} + x_{28}x_{30} + x_{28}x_{31} + x_{28}x_{33} + x_{28}x_{35} + x_{28}x_{39} + x_{28}x_{40} + x_{28}x_{42} + x_{28}x_{43} + x_{28}x_{44} + x_{28}x_{45} + x_{28}x_{46} + x_{28}x_{47} + x_{28}x_{48} + x_{28}x_{49} + x_{28}x_{50} + x_{28}x_{51} + x_{28}x_{53} + x_{28}x_{54} + x_{28}x_{56} + x_{28}x_{58} + x_{28}x_{60} + x_{28}x_{62} + x_{28}x_{63} + x_{29}x_{33} + x_{29}x_{34} + x_{29}x_{35} + x_{29}x_{36} + x_{29}x_{38} + x_{29}x_{40} + x_{29}x_{41} + x_{29}x_{42} + x_{29}x_{44} + x_{29}x_{45} + x_{29}x_{48} + x_{29}x_{49} + x_{29}x_{50} + x_{29}x_{51} + x_{29}x_{52} + x_{29}x_{53} + x_{29}x_{54} + x_{29}x_{55} + x_{29}x_{59} + x_{29}x_{61} + x_{29}x_{62} + x_{29}x_{63} + x_{29}x_{64} + x_{30}x_{31} + x_{30}x_{34} + x_{30}x_{36} + x_{30}x_{37} + x_{30}x_{38} + x_{30}x_{39} + x_{30}x_{41} + x_{30}x_{43} + x_{30}x_{45} + x_{30}x_{48} + x_{30}x_{53} + x_{30}x_{55} + x_{30}x_{56} + x_{30}x_{58} + x_{30}x_{59} + x_{30}x_{60} + x_{30}x_{62} + x_{30}x_{63} + x_{31}x_{32} + x_{31}x_{36} + x_{31}x_{37} + x_{31}x_{40} + x_{31}x_{42} + x_{31}x_{43} + x_{31}x_{44} + x_{31}x_{46} + x_{31}x_{53} + x_{31}x_{54} + x_{31}x_{56} + x_{31}x_{58} + x_{31}x_{59} + x_{32}x_{34} + x_{32}x_{35} + x_{32}x_{36} + x_{32}x_{37} + x_{32}x_{39} + x_{32}x_{41} + x_{32}x_{43} + x_{32}x_{44} + x_{32}x_{46} + x_{32}x_{51} + x_{32}x_{52} + x_{32}x_{53} + x_{32}x_{54} + x_{32}x_{57} + x_{32}x_{59} + x_{32}x_{60} + x_{32}x_{61} + x_{32}x_{62} + x_{32}x_{64} + x_{33}x_{34} + x_{33}x_{35} + x_{33}x_{37} + x_{33}x_{39} + x_{33}x_{40} + x_{33}x_{41} + x_{33}x_{42} + x_{33}x_{43} + x_{33}x_{44} + x_{33}x_{47} + x_{33}x_{50} + x_{33}x_{52} + x_{33}x_{53} + x_{33}x_{56} + x_{33}x_{59} + x_{33}x_{63} + x_{34}x_{37} + x_{34}x_{39} + x_{34}x_{44} + x_{34}x_{47} + x_{34}x_{49} + x_{34}x_{50} + x_{34}x_{52} + x_{34}x_{56} + x_{34}x_{57} + x_{34}x_{59} + x_{34}x_{62} + x_{34}x_{64} + x_{35}x_{36} + x_{35}x_{37} + x_{35}x_{38} + x_{35}x_{40} + x_{35}x_{41} + x_{35}x_{45} + x_{35}x_{46} + x_{35}x_{47} + x_{35}x_{48} + x_{35}x_{49} + x_{35}x_{51} + x_{35}x_{52} + x_{35}x_{54} + x_{35}x_{56} + x_{35}x_{59} + x_{35}x_{61} + x_{35}x_{62} + x_{36}x_{41} + x_{36}x_{43} + x_{36}x_{44} + x_{36}x_{45} + x_{36}x_{48} + x_{36}x_{49} + x_{36}x_{51} + x_{36}x_{52} + x_{36}x_{53} + x_{36}x_{56} + x_{36}x_{60} + x_{36}x_{61} + x_{36}x_{64} + x_{37}x_{39} + x_{37}x_{40} + x_{37}x_{44} + x_{37}x_{46} + x_{37}x_{49} + x_{37}x_{54} + x_{37}x_{59} + x_{37}x_{60} + x_{37}x_{62} + x_{37}x_{63} + x_{37}x_{64} + x_{38}x_{39} + x_{38}x_{42} + x_{38}x_{43} + x_{38}x_{45} + x_{38}x_{49} + x_{38}x_{50} + x_{38}x_{53} + x_{38}x_{54} + x_{38}x_{55} + x_{38}x_{56} + x_{38}x_{57} + x_{38}x_{58} + x_{38}x_{59} + x_{38}x_{61} + x_{38}x_{63} + x_{38}x_{64} + x_{39}x_{40} + x_{39}x_{41} + x_{39}x_{43} + x_{39}x_{44} + x_{39}x_{49} + x_{39}x_{50} + x_{39}x_{53} + x_{39}x_{56} + x_{39}x_{58} + x_{39}x_{59} + x_{39}x_{60} + x_{39}x_{63} + x_{40}x_{43} + x_{40}x_{45} + x_{40}x_{46} + x_{40}x_{48} + x_{40}x_{50} + x_{40}x_{52} + x_{40}x_{55} + x_{40}x_{56} + x_{40}x_{57} + x_{40}x_{58} + x_{40}x_{62} + x_{40}x_{63} + x_{40}x_{64} + x_{41}x_{45} + x_{41}x_{47} + x_{41}x_{48} + x_{41}x_{50} + x_{41}x_{57} + x_{41}x_{58} + x_{41}x_{59} + x_{41}x_{61} + x_{41}x_{62} + x_{41}x_{63} + x_{42}x_{43} + x_{42}x_{44} + x_{42}x_{45} + x_{42}x_{47} + x_{42}x_{48} + x_{42}x_{50} + x_{42}x_{53} + x_{42}x_{55} + x_{42}x_{60} + x_{43}x_{45} + x_{43}x_{48} + x_{43}x_{49} + x_{43}x_{52} + x_{43}x_{56} + x_{43}x_{59} + x_{43}x_{60} + x_{43}x_{61} + x_{43}x_{63} + x_{44}x_{46} + x_{44}x_{47} + x_{44}x_{48} + x_{44}x_{49} + x_{44}x_{50} + x_{44}x_{52} + x_{44}x_{53} + x_{44}x_{57} + x_{44}x_{58} + x_{44}x_{59} + x_{44}x_{60} + x_{44}x_{61} + x_{44}x_{64} + x_{45}x_{47} + x_{45}x_{49} + x_{45}x_{52} + x_{45}x_{55} + x_{45}x_{57} + x_{45}x_{58} + x_{45}x_{61} + x_{45}x_{62} + x_{45}x_{63} + x_{46}x_{49} + x_{46}x_{50} + x_{46}x_{51} + x_{46}x_{53} + x_{46}x_{54} + x_{46}x_{57} + x_{46}x_{60} + x_{46}x_{62} + x_{46}x_{64} + x_{47}x_{48} + x_{47}x_{49} + x_{47}x_{51} + x_{47}x_{55} + x_{47}x_{56} + x_{47}x_{57} + x_{47}x_{59} + x_{47}x_{61} + x_{47}x_{63} + x_{47}x_{64} + x_{48}x_{53} + x_{48}x_{54} + x_{48}x_{55} + x_{48}x_{57} + x_{48}x_{58} + x_{48}x_{59} + x_{48}x_{63} + x_{49}x_{51} + x_{49}x_{52} + x_{49}x_{54} + x_{49}x_{55} + x_{49}x_{57} + x_{49}x_{58} + x_{49}x_{59} + x_{49}x_{60} + x_{49}x_{62} + x_{49}x_{64} + x_{50}x_{51} + x_{50}x_{54} + x_{50}x_{56} + x_{50}x_{57} + x_{50}x_{58} + x_{50}x_{60} + x_{50}x_{63} + x_{50}x_{64} + x_{51}x_{55} + x_{51}x_{57} + x_{51}x_{59} + x_{51}x_{60} + x_{51}x_{62} + x_{51}x_{63} + x_{52}x_{56} + x_{52}x_{57} + x_{52}x_{61} + x_{52}x_{62} + x_{52}x_{63} + x_{53}x_{56} + x_{53}x_{58} + x_{53}x_{59} + x_{53}x_{62} + x_{54}x_{55} + x_{54}x_{57} + x_{54}x_{58} + x_{54}x_{61} + x_{54}x_{63} + x_{55}x_{56} + x_{55}x_{59} + x_{55}x_{60} + x_{55}x_{61} + x_{56}x_{57} + x_{56}x_{58} + x_{56}x_{59} + x_{56}x_{60} + x_{56}x_{62} + x_{56}x_{63} + x_{56}x_{64} + x_{57}x_{58} + x_{57}x_{59} + x_{57}x_{62} + x_{57}x_{63} + x_{58}x_{60} + x_{58}x_{62} + x_{58}x_{63} + x_{58}x_{64} + x_{59}x_{63} + x_{59}x_{64} + x_{60}x_{61} + x_{60}x_{64} + x_{61}x_{62} + x_{61}x_{64} + x_{62}x_{63} + x_{62}x_{64} + x_{2} + x_{3} + x_{5} + x_{6} + x_{7} + x_{8} + x_{10} + x_{14} + x_{16} + x_{18} + x_{20} + x_{21} + x_{22} + x_{23} + x_{24} + x_{28} + x_{29} + x_{33} + x_{36} + x_{40} + x_{41} + x_{42} + x_{44} + x_{47} + x_{49} + x_{50} + x_{51} + x_{53} + x_{54} + x_{55} + x_{56} + x_{57} + x_{61} + x_{63}$

$y_{21} = x_{1}x_{3} + x_{1}x_{4} + x_{1}x_{5} + x_{1}x_{6} + x_{1}x_{11} + x_{1}x_{12} + x_{1}x_{13} + x_{1}x_{15} + x_{1}x_{17} + x_{1}x_{18} + x_{1}x_{25} + x_{1}x_{30} + x_{1}x_{32} + x_{1}x_{33} + x_{1}x_{34} + x_{1}x_{35} + x_{1}x_{37} + x_{1}x_{39} + x_{1}x_{41} + x_{1}x_{42} + x_{1}x_{44} + x_{1}x_{46} + x_{1}x_{50} + x_{1}x_{52} + x_{1}x_{57} + x_{1}x_{58} + x_{1}x_{60} + x_{1}x_{64} + x_{2}x_{4} + x_{2}x_{6} + x_{2}x_{8} + x_{2}x_{9} + x_{2}x_{12} + x_{2}x_{19} + x_{2}x_{25} + x_{2}x_{26} + x_{2}x_{27} + x_{2}x_{28} + x_{2}x_{33} + x_{2}x_{34} + x_{2}x_{35} + x_{2}x_{37} + x_{2}x_{38} + x_{2}x_{39} + x_{2}x_{40} + x_{2}x_{41} + x_{2}x_{42} + x_{2}x_{44} + x_{2}x_{45} + x_{2}x_{46} + x_{2}x_{48} + x_{2}x_{50} + x_{2}x_{53} + x_{2}x_{54} + x_{2}x_{55} + x_{2}x_{58} + x_{2}x_{59} + x_{2}x_{61} + x_{2}x_{62} + x_{2}x_{64} + x_{3}x_{4} + x_{3}x_{5} + x_{3}x_{7} + x_{3}x_{10} + x_{3}x_{15} + x_{3}x_{16} + x_{3}x_{18} + x_{3}x_{19} + x_{3}x_{20} + x_{3}x_{23} + x_{3}x_{24} + x_{3}x_{26} + x_{3}x_{27} + x_{3}x_{28} + x_{3}x_{31} + x_{3}x_{34} + x_{3}x_{35} + x_{3}x_{36} + x_{3}x_{40} + x_{3}x_{44} + x_{3}x_{46} + x_{3}x_{50} + x_{3}x_{51} + x_{3}x_{52} + x_{3}x_{54} + x_{3}x_{55} + x_{3}x_{56} + x_{3}x_{59} + x_{3}x_{61} + x_{3}x_{62} + x_{3}x_{64} + x_{4}x_{6} + x_{4}x_{8} + x_{4}x_{10} + x_{4}x_{13} + x_{4}x_{14} + x_{4}x_{15} + x_{4}x_{16} + x_{4}x_{20} + x_{4}x_{25} + x_{4}x_{32} + x_{4}x_{35} + x_{4}x_{39} + x_{4}x_{40} + x_{4}x_{41} + x_{4}x_{42} + x_{4}x_{45} + x_{4}x_{48} + x_{4}x_{49} + x_{4}x_{56} + x_{4}x_{57} + x_{4}x_{59} + x_{4}x_{60} + x_{4}x_{62} + x_{4}x_{64} + x_{5}x_{6} + x_{5}x_{7} + x_{5}x_{8} + x_{5}x_{10} + x_{5}x_{15} + x_{5}x_{18} + x_{5}x_{20} + x_{5}x_{21} + x_{5}x_{23} + x_{5}x_{24} + x_{5}x_{27} + x_{5}x_{29} + x_{5}x_{32} + x_{5}x_{33} + x_{5}x_{35} + x_{5}x_{38} + x_{5}x_{42} + x_{5}x_{46} + x_{5}x_{47} + x_{5}x_{50} + x_{5}x_{51} + x_{5}x_{53} + x_{5}x_{55} + x_{5}x_{57} + x_{5}x_{59} + x_{5}x_{60} + x_{6}x_{7} + x_{6}x_{8} + x_{6}x_{10} + x_{6}x_{11} + x_{6}x_{13} + x_{6}x_{15} + x_{6}x_{16} + x_{6}x_{17} + x_{6}x_{18} + x_{6}x_{19} + x_{6}x_{20} + x_{6}x_{21} + x_{6}x_{23} + x_{6}x_{25} + x_{6}x_{27} + x_{6}x_{30} + x_{6}x_{32} + x_{6}x_{34} + x_{6}x_{35} + x_{6}x_{36} + x_{6}x_{39} + x_{6}x_{40} + x_{6}x_{44} + x_{6}x_{45} + x_{6}x_{46} + x_{6}x_{48} + x_{6}x_{63} + x_{6}x_{64} + x_{7}x_{8} + x_{7}x_{12} + x_{7}x_{14} + x_{7}x_{17} + x_{7}x_{18} + x_{7}x_{19} + x_{7}x_{21} + x_{7}x_{22} + x_{7}x_{24} + x_{7}x_{25} + x_{7}x_{26} + x_{7}x_{27} + x_{7}x_{28} + x_{7}x_{29} + x_{7}x_{30} + x_{7}x_{31} + x_{7}x_{32} + x_{7}x_{36} + x_{7}x_{38} + x_{7}x_{40} + x_{7}x_{41} + x_{7}x_{44} + x_{7}x_{45} + x_{7}x_{47} + x_{7}x_{48} + x_{7}x_{49} + x_{7}x_{50} + x_{7}x_{51} + x_{7}x_{52} + x_{7}x_{53} + x_{7}x_{57} + x_{7}x_{60} + x_{7}x_{61} + x_{7}x_{64} + x_{8}x_{9} + x_{8}x_{11} + x_{8}x_{12} + x_{8}x_{13} + x_{8}x_{17} + x_{8}x_{18} + x_{8}x_{20} + x_{8}x_{26} + x_{8}x_{28} + x_{8}x_{29} + x_{8}x_{32} + x_{8}x_{33} + x_{8}x_{35} + x_{8}x_{39} + x_{8}x_{40} + x_{8}x_{41} + x_{8}x_{42} + x_{8}x_{44} + x_{8}x_{47} + x_{8}x_{50} + x_{8}x_{51} + x_{8}x_{53} + x_{8}x_{55} + x_{8}x_{56} + x_{8}x_{60} + x_{8}x_{62} + x_{8}x_{63} + x_{8}x_{64} + x_{9}x_{11} + x_{9}x_{15} + x_{9}x_{17} + x_{9}x_{18} + x_{9}x_{19} + x_{9}x_{20} + x_{9}x_{22} + x_{9}x_{23} + x_{9}x_{25} + x_{9}x_{28} + x_{9}x_{29} + x_{9}x_{30} + x_{9}x_{31} + x_{9}x_{32} + x_{9}x_{33} + x_{9}x_{35} + x_{9}x_{37} + x_{9}x_{40} + x_{9}x_{41} + x_{9}x_{45} + x_{9}x_{47} + x_{9}x_{49} + x_{9}x_{51} + x_{9}x_{52} + x_{9}x_{53} + x_{9}x_{54} + x_{9}x_{55} + x_{9}x_{56} + x_{9}x_{57} + x_{9}x_{58} + x_{9}x_{59} + x_{9}x_{60} + x_{9}x_{62} + x_{10}x_{11} + x_{10}x_{14} + x_{10}x_{15} + x_{10}x_{16} + x_{10}x_{17} + x_{10}x_{18} + x_{10}x_{19} + x_{10}x_{24} + x_{10}x_{26} + x_{10}x_{28} + x_{10}x_{30} + x_{10}x_{31} + x_{10}x_{33} + x_{10}x_{35} + x_{10}x_{36} + x_{10}x_{43} + x_{10}x_{44} + x_{10}x_{45} + x_{10}x_{47} + x_{10}x_{48} + x_{10}x_{49} + x_{10}x_{52} + x_{10}x_{53} + x_{10}x_{54} + x_{10}x_{57} + x_{10}x_{58} + x_{10}x_{60} + x_{10}x_{63} + x_{11}x_{12} + x_{11}x_{13} + x_{11}x_{15} + x_{11}x_{16} + x_{11}x_{18} + x_{11}x_{19} + x_{11}x_{21} + x_{11}x_{22} + x_{11}x_{24} + x_{11}x_{26} + x_{11}x_{29} + x_{11}x_{32} + x_{11}x_{35} + x_{11}x_{36} + x_{11}x_{37} + x_{11}x_{38} + x_{11}x_{40} + x_{11}x_{42} + x_{11}x_{44} + x_{11}x_{45} + x_{11}x_{46} + x_{11}x_{47} + x_{11}x_{48} + x_{11}x_{50} + x_{11}x_{53} + x_{11}x_{57} + x_{11}x_{61} + x_{11}x_{62} + x_{11}x_{64} + x_{12}x_{14} + x_{12}x_{16} + x_{12}x_{19} + x_{12}x_{21} + x_{12}x_{25} + x_{12}x_{26} + x_{12}x_{27} + x_{12}x_{28} + x_{12}x_{29} + x_{12}x_{31} + x_{12}x_{32} + x_{12}x_{35} + x_{12}x_{37} + x_{12}x_{39} + x_{12}x_{40} + x_{12}x_{41} + x_{12}x_{42} + x_{12}x_{43} + x_{12}x_{44} + x_{12}x_{45} + x_{12}x_{47} + x_{12}x_{50} + x_{12}x_{59} + x_{12}x_{60} + x_{12}x_{64} + x_{13}x_{14} + x_{13}x_{15} + x_{13}x_{17} + x_{13}x_{19} + x_{13}x_{20} + x_{13}x_{21} + x_{13}x_{22} + x_{13}x_{24} + x_{13}x_{25} + x_{13}x_{26} + x_{13}x_{27} + x_{13}x_{29} + x_{13}x_{31} + x_{13}x_{35} + x_{13}x_{37} + x_{13}x_{39} + x_{13}x_{42} + x_{13}x_{43} + x_{13}x_{47} + x_{13}x_{48} + x_{13}x_{50} + x_{13}x_{53} + x_{13}x_{56} + x_{13}x_{59} + x_{13}x_{60} + x_{13}x_{61} + x_{14}x_{16} + x_{14}x_{17} + x_{14}x_{18} + x_{14}x_{19} + x_{14}x_{22} + x_{14}x_{24} + x_{14}x_{26} + x_{14}x_{27} + x_{14}x_{31} + x_{14}x_{32} + x_{14}x_{35} + x_{14}x_{36} + x_{14}x_{41} + x_{14}x_{45} + x_{14}x_{46} + x_{14}x_{47} + x_{14}x_{49} + x_{14}x_{52} + x_{14}x_{55} + x_{14}x_{56} + x_{14}x_{59} + x_{14}x_{63} + x_{15}x_{17} + x_{15}x_{18} + x_{15}x_{19} + x_{15}x_{21} + x_{15}x_{23} + x_{15}x_{24} + x_{15}x_{25} + x_{15}x_{27} + x_{15}x_{31} + x_{15}x_{35} + x_{15}x_{38} + x_{15}x_{39} + x_{15}x_{40} + x_{15}x_{43} + x_{15}x_{44} + x_{15}x_{45} + x_{15}x_{46} + x_{15}x_{47} + x_{15}x_{48} + x_{15}x_{50} + x_{15}x_{51} + x_{15}x_{52} + x_{15}x_{54} + x_{15}x_{55} + x_{15}x_{56} + x_{15}x_{58} + x_{15}x_{59} + x_{15}x_{62} + x_{15}x_{63} + x_{16}x_{20} + x_{16}x_{23} + x_{16}x_{25} + x_{16}x_{27} + x_{16}x_{28} + x_{16}x_{31} + x_{16}x_{33} + x_{16}x_{38} + x_{16}x_{40} + x_{16}x_{41} + x_{16}x_{45} + x_{16}x_{47} + x_{16}x_{48} + x_{16}x_{51} + x_{16}x_{52} + x_{16}x_{54} + x_{16}x_{55} + x_{16}x_{57} + x_{16}x_{59} + x_{16}x_{60} + x_{16}x_{63} + x_{16}x_{64} + x_{17}x_{19} + x_{17}x_{20} + x_{17}x_{22} + x_{17}x_{26} + x_{17}x_{29} + x_{17}x_{31} + x_{17}x_{33} + x_{17}x_{34} + x_{17}x_{35} + x_{17}x_{36} + x_{17}x_{37} + x_{17}x_{41} + x_{17}x_{43} + x_{17}x_{45} + x_{17}x_{50} + x_{17}x_{53} + x_{17}x_{54} + x_{17}x_{55} + x_{17}x_{56} + x_{17}x_{57} + x_{17}x_{60} + x_{17}x_{62} + x_{17}x_{64} + x_{18}x_{20} + x_{18}x_{25} + x_{18}x_{26} + x_{18}x_{27} + x_{18}x_{30} + x_{18}x_{31} + x_{18}x_{34} + x_{18}x_{36} + x_{18}x_{37} + x_{18}x_{38} + x_{18}x_{39} + x_{18}x_{40} + x_{18}x_{42} + x_{18}x_{43} + x_{18}x_{44} + x_{18}x_{45} + x_{18}x_{46} + x_{18}x_{49} + x_{18}x_{51} + x_{18}x_{56} + x_{18}x_{57} + x_{18}x_{58} + x_{18}x_{60} + x_{18}x_{62} + x_{18}x_{64} + x_{19}x_{20} + x_{19}x_{25} + x_{19}x_{29} + x_{19}x_{34} + x_{19}x_{38} + x_{19}x_{39} + x_{19}x_{41} + x_{19}x_{42} + x_{19}x_{43} + x_{19}x_{44} + x_{19}x_{45} + x_{19}x_{46} + x_{19}x_{47} + x_{19}x_{48} + x_{19}x_{50} + x_{19}x_{53} + x_{19}x_{54} + x_{19}x_{57} + x_{19}x_{58} + x_{19}x_{60} + x_{19}x_{61} + x_{19}x_{62} + x_{19}x_{64} + x_{20}x_{21} + x_{20}x_{22} + x_{20}x_{23} + x_{20}x_{26} + x_{20}x_{29} + x_{20}x_{37} + x_{20}x_{38} + x_{20}x_{40} + x_{20}x_{42} + x_{20}x_{43} + x_{20}x_{44} + x_{20}x_{45} + x_{20}x_{48} + x_{20}x_{50} + x_{20}x_{52} + x_{20}x_{56} + x_{20}x_{58} + x_{20}x_{59} + x_{20}x_{63} + x_{21}x_{26} + x_{21}x_{27} + x_{21}x_{28} + x_{21}x_{31} + x_{21}x_{32} + x_{21}x_{34} + x_{21}x_{37} + x_{21}x_{38} + x_{21}x_{41} + x_{21}x_{42} + x_{21}x_{44} + x_{21}x_{45} + x_{21}x_{46} + x_{21}x_{47} + x_{21}x_{50} + x_{21}x_{51} + x_{21}x_{52} + x_{21}x_{55} + x_{21}x_{56} + x_{21}x_{57} + x_{21}x_{58} + x_{21}x_{61} + x_{21}x_{63} + x_{21}x_{64} + x_{22}x_{24} + x_{22}x_{29} + x_{22}x_{37} + x_{22}x_{41} + x_{22}x_{44} + x_{22}x_{48} + x_{22}x_{49} + x_{22}x_{52} + x_{22}x_{53} + x_{22}x_{54} + x_{22}x_{55} + x_{22}x_{56} + x_{22}x_{59} + x_{22}x_{61} + x_{22}x_{62} + x_{23}x_{26} + x_{23}x_{28} + x_{23}x_{29} + x_{23}x_{30} + x_{23}x_{37} + x_{23}x_{39} + x_{23}x_{40} + x_{23}x_{43} + x_{23}x_{45} + x_{23}x_{46} + x_{23}x_{47} + x_{23}x_{48} + x_{23}x_{50} + x_{23}x_{51} + x_{23}x_{54} + x_{23}x_{57} + x_{23}x_{59} + x_{23}x_{60} + x_{23}x_{62} + x_{23}x_{63} + x_{23}x_{64} + x_{24}x_{26} + x_{24}x_{27} + x_{24}x_{28} + x_{24}x_{29} + x_{24}x_{30} + x_{24}x_{31} + x_{24}x_{34} + x_{24}x_{35} + x_{24}x_{37} + x_{24}x_{41} + x_{24}x_{43} + x_{24}x_{45} + x_{24}x_{48} + x_{24}x_{52} + x_{24}x_{54} + x_{24}x_{56} + x_{24}x_{57} + x_{24}x_{58} + x_{24}x_{62} + x_{24}x_{63} + x_{25}x_{28} + x_{25}x_{29} + x_{25}x_{30} + x_{25}x_{31} + x_{25}x_{32} + x_{25}x_{33} + x_{25}x_{34} + x_{25}x_{36} + x_{25}x_{37} + x_{25}x_{40} + x_{25}x_{43} + x_{25}x_{44} + x_{25}x_{46} + x_{25}x_{47} + x_{25}x_{48} + x_{25}x_{51} + x_{25}x_{52} + x_{25}x_{54} + x_{25}x_{56} + x_{25}x_{58} + x_{25}x_{59} + x_{25}x_{61} + x_{25}x_{63} + x_{26}x_{28} + x_{26}x_{30} + x_{26}x_{35} + x_{26}x_{39} + x_{26}x_{40} + x_{26}x_{41} + x_{26}x_{43} + x_{26}x_{47} + x_{26}x_{49} + x_{26}x_{50} + x_{26}x_{51} + x_{26}x_{57} + x_{26}x_{58} + x_{26}x_{60} + x_{26}x_{62} + x_{26}x_{64} + x_{27}x_{28} + x_{27}x_{29} + x_{27}x_{34} + x_{27}x_{35} + x_{27}x_{36} + x_{27}x_{38} + x_{27}x_{40} + x_{27}x_{41} + x_{27}x_{42} + x_{27}x_{44} + x_{27}x_{46} + x_{27}x_{47} + x_{27}x_{48} + x_{27}x_{49} + x_{27}x_{51} + x_{27}x_{52} + x_{27}x_{56} + x_{27}x_{63} + x_{27}x_{64} + x_{28}x_{31} + x_{28}x_{34} + x_{28}x_{38} + x_{28}x_{41} + x_{28}x_{42} + x_{28}x_{45} + x_{28}x_{50} + x_{28}x_{51} + x_{28}x_{54} + x_{28}x_{55} + x_{28}x_{56} + x_{28}x_{59} + x_{28}x_{61} + x_{29}x_{35} + x_{29}x_{36} + x_{29}x_{39} + x_{29}x_{40} + x_{29}x_{42} + x_{29}x_{44} + x_{29}x_{45} + x_{29}x_{47} + x_{29}x_{49} + x_{29}x_{51} + x_{29}x_{53} + x_{29}x_{55} + x_{29}x_{56} + x_{29}x_{60} + x_{29}x_{61} + x_{29}x_{62} + x_{29}x_{63} + x_{30}x_{33} + x_{30}x_{35} + x_{30}x_{36} + x_{30}x_{39} + x_{30}x_{44} + x_{30}x_{45} + x_{30}x_{46} + x_{30}x_{49} + x_{30}x_{50} + x_{30}x_{54} + x_{30}x_{55} + x_{30}x_{56} + x_{30}x_{61} + x_{30}x_{63} + x_{31}x_{34} + x_{31}x_{35} + x_{31}x_{39} + x_{31}x_{41} + x_{31}x_{43} + x_{31}x_{44} + x_{31}x_{45} + x_{31}x_{46} + x_{31}x_{48} + x_{31}x_{50} + x_{31}x_{52} + x_{31}x_{57} + x_{31}x_{61} + x_{31}x_{62} + x_{32}x_{33} + x_{32}x_{35} + x_{32}x_{39} + x_{32}x_{41} + x_{32}x_{43} + x_{32}x_{46} + x_{32}x_{50} + x_{32}x_{53} + x_{32}x_{54} + x_{32}x_{55} + x_{32}x_{59} + x_{32}x_{61} + x_{33}x_{34} + x_{33}x_{35} + x_{33}x_{37} + x_{33}x_{39} + x_{33}x_{42} + x_{33}x_{43} + x_{33}x_{51} + x_{33}x_{56} + x_{33}x_{57} + x_{33}x_{59} + x_{33}x_{64} + x_{34}x_{35} + x_{34}x_{37} + x_{34}x_{38} + x_{34}x_{42} + x_{34}x_{43} + x_{34}x_{49} + x_{34}x_{50} + x_{34}x_{51} + x_{34}x_{53} + x_{34}x_{56} + x_{34}x_{58} + x_{35}x_{38} + x_{35}x_{39} + x_{35}x_{40} + x_{35}x_{42} + x_{35}x_{43} + x_{35}x_{45} + x_{35}x_{46} + x_{35}x_{47} + x_{35}x_{51} + x_{35}x_{53} + x_{35}x_{55} + x_{35}x_{56} + x_{35}x_{57} + x_{35}x_{58} + x_{35}x_{59} + x_{35}x_{60} + x_{35}x_{63} + x_{35}x_{64} + x_{36}x_{38} + x_{36}x_{39} + x_{36}x_{46} + x_{36}x_{47} + x_{36}x_{48} + x_{36}x_{49} + x_{36}x_{51} + x_{36}x_{52} + x_{36}x_{53} + x_{36}x_{54} + x_{36}x_{57} + x_{36}x_{58} + x_{36}x_{59} + x_{36}x_{62} + x_{36}x_{63} + x_{36}x_{64} + x_{37}x_{43} + x_{37}x_{44} + x_{37}x_{45} + x_{37}x_{47} + x_{37}x_{56} + x_{37}x_{58} + x_{37}x_{60} + x_{37}x_{63} + x_{37}x_{64} + x_{38}x_{40} + x_{38}x_{42} + x_{38}x_{46} + x_{38}x_{47} + x_{38}x_{48} + x_{38}x_{51} + x_{38}x_{52} + x_{38}x_{54} + x_{38}x_{58} + x_{38}x_{60} + x_{38}x_{61} + x_{38}x_{63} + x_{39}x_{40} + x_{39}x_{41} + x_{39}x_{42} + x_{39}x_{45} + x_{39}x_{49} + x_{39}x_{51} + x_{39}x_{55} + x_{39}x_{57} + x_{39}x_{58} + x_{39}x_{60} + x_{39}x_{61} + x_{39}x_{62} + x_{39}x_{63} + x_{39}x_{64} + x_{40}x_{46} + x_{40}x_{47} + x_{40}x_{52} + x_{40}x_{53} + x_{40}x_{55} + x_{40}x_{56} + x_{40}x_{58} + x_{40}x_{60} + x_{40}x_{63} + x_{40}x_{64} + x_{41}x_{43} + x_{41}x_{45} + x_{41}x_{46} + x_{41}x_{47} + x_{41}x_{49} + x_{41}x_{50} + x_{41}x_{51} + x_{41}x_{52} + x_{41}x_{57} + x_{41}x_{61} + x_{41}x_{62} + x_{41}x_{64} + x_{42}x_{44} + x_{42}x_{50} + x_{42}x_{51} + x_{42}x_{52} + x_{42}x_{53} + x_{42}x_{57} + x_{42}x_{61} + x_{42}x_{64} + x_{43}x_{44} + x_{43}x_{46} + x_{43}x_{47} + x_{43}x_{49} + x_{43}x_{50} + x_{43}x_{51} + x_{43}x_{52} + x_{43}x_{54} + x_{43}x_{55} + x_{43}x_{56} + x_{43}x_{57} + x_{43}x_{58} + x_{43}x_{59} + x_{43}x_{60} + x_{43}x_{61} + x_{43}x_{62} + x_{43}x_{63} + x_{44}x_{46} + x_{44}x_{49} + x_{44}x_{51} + x_{44}x_{52} + x_{44}x_{54} + x_{44}x_{55} + x_{44}x_{56} + x_{44}x_{57} + x_{44}x_{60} + x_{44}x_{61} + x_{44}x_{62} + x_{44}x_{63} + x_{44}x_{64} + x_{45}x_{50} + x_{45}x_{51} + x_{45}x_{52} + x_{45}x_{53} + x_{45}x_{54} + x_{45}x_{55} + x_{45}x_{56} + x_{45}x_{57} + x_{45}x_{59} + x_{45}x_{60} + x_{45}x_{61} + x_{45}x_{62} + x_{45}x_{64} + x_{46}x_{48} + x_{46}x_{51} + x_{46}x_{52} + x_{46}x_{53} + x_{46}x_{55} + x_{46}x_{57} + x_{46}x_{58} + x_{46}x_{59} + x_{46}x_{60} + x_{46}x_{64} + x_{47}x_{49} + x_{47}x_{50} + x_{47}x_{51} + x_{47}x_{52} + x_{47}x_{54} + x_{47}x_{55} + x_{47}x_{56} + x_{47}x_{57} + x_{47}x_{58} + x_{47}x_{60} + x_{47}x_{62} + x_{47}x_{63} + x_{47}x_{64} + x_{48}x_{49} + x_{48}x_{50} + x_{48}x_{51} + x_{48}x_{52} + x_{48}x_{53} + x_{48}x_{54} + x_{48}x_{56} + x_{48}x_{57} + x_{48}x_{62} + x_{48}x_{64} + x_{49}x_{50} + x_{49}x_{54} + x_{49}x_{56} + x_{49}x_{57} + x_{49}x_{60} + x_{49}x_{61} + x_{49}x_{62} + x_{49}x_{63} + x_{50}x_{51} + x_{50}x_{52} + x_{50}x_{53} + x_{50}x_{54} + x_{50}x_{59} + x_{50}x_{60} + x_{50}x_{62} + x_{50}x_{63} + x_{50}x_{64} + x_{51}x_{52} + x_{51}x_{56} + x_{51}x_{62} + x_{52}x_{53} + x_{52}x_{54} + x_{52}x_{55} + x_{52}x_{58} + x_{52}x_{59} + x_{52}x_{63} + x_{52}x_{64} + x_{53}x_{54} + x_{53}x_{55} + x_{53}x_{56} + x_{53}x_{58} + x_{53}x_{59} + x_{53}x_{60} + x_{53}x_{61} + x_{53}x_{62} + x_{54}x_{55} + x_{54}x_{57} + x_{54}x_{58} + x_{54}x_{59} + x_{54}x_{61} + x_{54}x_{62} + x_{55}x_{57} + x_{55}x_{60} + x_{55}x_{61} + x_{55}x_{62} + x_{55}x_{64} + x_{56}x_{57} + x_{56}x_{58} + x_{56}x_{60} + x_{56}x_{61} + x_{56}x_{62} + x_{56}x_{63} + x_{56}x_{64} + x_{57}x_{60} + x_{57}x_{61} + x_{57}x_{64} + x_{58}x_{61} + x_{58}x_{63} + x_{58}x_{64} + x_{59}x_{60} + x_{59}x_{61} + x_{59}x_{62} + x_{59}x_{64} + x_{60}x_{63} + x_{60}x_{64} + x_{61}x_{63} + x_{61}x_{64} + x_{62}x_{63} + x_{62}x_{64} + x_{1} + x_{3} + x_{4} + x_{6} + x_{8} + x_{9} + x_{12} + x_{19} + x_{22} + x_{23} + x_{24} + x_{25} + x_{27} + x_{30} + x_{33} + x_{34} + x_{36} + x_{37} + x_{40} + x_{44} + x_{45} + x_{48} + x_{50} + x_{51} + x_{52} + x_{58} + x_{59} + x_{60} + x_{63} + x_{64}$

$y_{22} = x_{1}x_{5} + x_{1}x_{6} + x_{1}x_{7} + x_{1}x_{11} + x_{1}x_{14} + x_{1}x_{15} + x_{1}x_{17} + x_{1}x_{19} + x_{1}x_{20} + x_{1}x_{21} + x_{1}x_{22} + x_{1}x_{24} + x_{1}x_{26} + x_{1}x_{29} + x_{1}x_{32} + x_{1}x_{34} + x_{1}x_{36} + x_{1}x_{37} + x_{1}x_{39} + x_{1}x_{43} + x_{1}x_{47} + x_{1}x_{48} + x_{1}x_{51} + x_{1}x_{52} + x_{1}x_{53} + x_{1}x_{54} + x_{1}x_{55} + x_{1}x_{57} + x_{1}x_{58} + x_{1}x_{62} + x_{1}x_{64} + x_{2}x_{4} + x_{2}x_{5} + x_{2}x_{7} + x_{2}x_{9} + x_{2}x_{13} + x_{2}x_{14} + x_{2}x_{15} + x_{2}x_{17} + x_{2}x_{19} + x_{2}x_{20} + x_{2}x_{23} + x_{2}x_{24} + x_{2}x_{25} + x_{2}x_{27} + x_{2}x_{30} + x_{2}x_{31} + x_{2}x_{34} + x_{2}x_{36} + x_{2}x_{37} + x_{2}x_{40} + x_{2}x_{48} + x_{2}x_{52} + x_{2}x_{56} + x_{2}x_{57} + x_{2}x_{58} + x_{2}x_{60} + x_{2}x_{61} + x_{2}x_{62} + x_{2}x_{64} + x_{3}x_{4} + x_{3}x_{6} + x_{3}x_{8} + x_{3}x_{9} + x_{3}x_{11} + x_{3}x_{12} + x_{3}x_{13} + x_{3}x_{14} + x_{3}x_{15} + x_{3}x_{18} + x_{3}x_{19} + x_{3}x_{20} + x_{3}x_{22} + x_{3}x_{23} + x_{3}x_{24} + x_{3}x_{25} + x_{3}x_{27} + x_{3}x_{29} + x_{3}x_{30} + x_{3}x_{32} + x_{3}x_{33} + x_{3}x_{34} + x_{3}x_{35} + x_{3}x_{36} + x_{3}x_{40} + x_{3}x_{41} + x_{3}x_{42} + x_{3}x_{43} + x_{3}x_{47} + x_{3}x_{49} + x_{3}x_{50} + x_{3}x_{51} + x_{3}x_{52} + x_{3}x_{54} + x_{3}x_{56} + x_{3}x_{57} + x_{3}x_{61} + x_{3}x_{63} + x_{3}x_{64} + x_{4}x_{6} + x_{4}x_{7} + x_{4}x_{10} + x_{4}x_{11} + x_{4}x_{12} + x_{4}x_{14} + x_{4}x_{15} + x_{4}x_{17} + x_{4}x_{19} + x_{4}x_{22} + x_{4}x_{23} + x_{4}x_{26} + x_{4}x_{27} + x_{4}x_{29} + x_{4}x_{30} + x_{4}x_{31} + x_{4}x_{34} + x_{4}x_{35} + x_{4}x_{37} + x_{4}x_{38} + x_{4}x_{39} + x_{4}x_{41} + x_{4}x_{42} + x_{4}x_{44} + x_{4}x_{45} + x_{4}x_{46} + x_{4}x_{48} + x_{4}x_{51} + x_{4}x_{54} + x_{4}x_{56} + x_{4}x_{57} + x_{4}x_{62} + x_{4}x_{63} + x_{4}x_{64} + x_{5}x_{7} + x_{5}x_{8} + x_{5}x_{9} + x_{5}x_{10} + x_{5}x_{13} + x_{5}x_{16} + x_{5}x_{18} + x_{5}x_{19} + x_{5}x_{21} + x_{5}x_{25} + x_{5}x_{29} + x_{5}x_{31} + x_{5}x_{38} + x_{5}x_{39} + x_{5}x_{40} + x_{5}x_{41} + x_{5}x_{43} + x_{5}x_{46} + x_{5}x_{47} + x_{5}x_{48} + x_{5}x_{52} + x_{5}x_{53} + x_{5}x_{55} + x_{5}x_{57} + x_{5}x_{58} + x_{5}x_{60} + x_{5}x_{61} + x_{5}x_{62} + x_{6}x_{9} + x_{6}x_{13} + x_{6}x_{14} + x_{6}x_{15} + x_{6}x_{16} + x_{6}x_{18} + x_{6}x_{21} + x_{6}x_{23} + x_{6}x_{24} + x_{6}x_{26} + x_{6}x_{28} + x_{6}x_{30} + x_{6}x_{35} + x_{6}x_{37} + x_{6}x_{38} + x_{6}x_{39} + x_{6}x_{42} + x_{6}x_{43} + x_{6}x_{44} + x_{6}x_{49} + x_{6}x_{50} + x_{6}x_{55} + x_{6}x_{62} + x_{6}x_{63} + x_{6}x_{64} + x_{7}x_{8} + x_{7}x_{9} + x_{7}x_{10} + x_{7}x_{12} + x_{7}x_{13} + x_{7}x_{14} + x_{7}x_{16} + x_{7}x_{19} + x_{7}x_{22} + x_{7}x_{23} + x_{7}x_{25} + x_{7}x_{26} + x_{7}x_{27} + x_{7}x_{29} + x_{7}x_{32} + x_{7}x_{34} + x_{7}x_{35} + x_{7}x_{36} + x_{7}x_{38} + x_{7}x_{39} + x_{7}x_{40} + x_{7}x_{42} + x_{7}x_{44} + x_{7}x_{45} + x_{7}x_{49} + x_{7}x_{50} + x_{7}x_{53} + x_{7}x_{59} + x_{7}x_{60} + x_{7}x_{62} + x_{7}x_{63} + x_{8}x_{10} + x_{8}x_{11} + x_{8}x_{13} + x_{8}x_{15} + x_{8}x_{17} + x_{8}x_{18} + x_{8}x_{21} + x_{8}x_{22} + x_{8}x_{27} + x_{8}x_{29} + x_{8}x_{33} + x_{8}x_{34} + x_{8}x_{36} + x_{8}x_{37} + x_{8}x_{38} + x_{8}x_{39} + x_{8}x_{41} + x_{8}x_{43} + x_{8}x_{44} + x_{8}x_{49} + x_{8}x_{50} + x_{8}x_{52} + x_{8}x_{56} + x_{8}x_{57} + x_{8}x_{58} + x_{8}x_{59} + x_{8}x_{62} + x_{9}x_{10} + x_{9}x_{13} + x_{9}x_{15} + x_{9}x_{19} + x_{9}x_{20} + x_{9}x_{21} + x_{9}x_{23} + x_{9}x_{25} + x_{9}x_{30} + x_{9}x_{36} + x_{9}x_{41} + x_{9}x_{45} + x_{9}x_{48} + x_{9}x_{49} + x_{9}x_{52} + x_{9}x_{53} + x_{9}x_{54} + x_{9}x_{60} + x_{9}x_{62} + x_{9}x_{63} + x_{10}x_{11} + x_{10}x_{14} + x_{10}x_{15} + x_{10}x_{16} + x_{10}x_{18} + x_{10}x_{19} + x_{10}x_{20} + x_{10}x_{21} + x_{10}x_{25} + x_{10}x_{26} + x_{10}x_{30} + x_{10}x_{31} + x_{10}x_{32} + x_{10}x_{34} + x_{10}x_{35} + x_{10}x_{36} + x_{10}x_{37} + x_{10}x_{38} + x_{10}x_{39} + x_{10}x_{40} + x_{10}x_{41} + x_{10}x_{44} + x_{10}x_{45} + x_{10}x_{48} + x_{10}x_{56} + x_{10}x_{57} + x_{10}x_{58} + x_{10}x_{59} + x_{10}x_{60} + x_{10}x_{61} + x_{10}x_{62} + x_{10}x_{63} + x_{10}x_{64} + x_{11}x_{13} + x_{11}x_{14} + x_{11}x_{17} + x_{11}x_{19} + x_{11}x_{20} + x_{11}x_{21} + x_{11}x_{22} + x_{11}x_{23} + x_{11}x_{24} + x_{11}x_{25} + x_{11}x_{27} + x_{11}x_{33} + x_{11}x_{34} + x_{11}x_{35} + x_{11}x_{36} + x_{11}x_{37} + x_{11}x_{39} + x_{11}x_{43} + x_{11}x_{44} + x_{11}x_{45} + x_{11}x_{46} + x_{11}x_{47} + x_{11}x_{48} + x_{11}x_{49} + x_{11}x_{50} + x_{11}x_{51} + x_{11}x_{53} + x_{11}x_{55} + x_{11}x_{58} + x_{11}x_{59} + x_{11}x_{62} + x_{11}x_{63} + x_{12}x_{14} + x_{12}x_{15} + x_{12}x_{17} + x_{12}x_{23} + x_{12}x_{24} + x_{12}x_{25} + x_{12}x_{26} + x_{12}x_{30} + x_{12}x_{31} + x_{12}x_{33} + x_{12}x_{34} + x_{12}x_{35} + x_{12}x_{37} + x_{12}x_{42} + x_{12}x_{48} + x_{12}x_{50} + x_{12}x_{52} + x_{12}x_{53} + x_{12}x_{56} + x_{12}x_{58} + x_{12}x_{59} + x_{13}x_{18} + x_{13}x_{19} + x_{13}x_{22} + x_{13}x_{23} + x_{13}x_{25} + x_{13}x_{26} + x_{13}x_{30} + x_{13}x_{34} + x_{13}x_{35} + x_{13}x_{37} + x_{13}x_{39} + x_{13}x_{42} + x_{13}x_{43} + x_{13}x_{44} + x_{13}x_{45} + x_{13}x_{46} + x_{13}x_{47} + x_{13}x_{49} + x_{13}x_{50} + x_{13}x_{53} + x_{13}x_{54} + x_{13}x_{55} + x_{13}x_{56} + x_{13}x_{58} + x_{13}x_{60} + x_{13}x_{61} + x_{13}x_{62} + x_{13}x_{64} + x_{14}x_{15} + x_{14}x_{16} + x_{14}x_{19} + x_{14}x_{20} + x_{14}x_{22} + x_{14}x_{25} + x_{14}x_{27} + x_{14}x_{28} + x_{14}x_{33} + x_{14}x_{35} + x_{14}x_{37} + x_{14}x_{39} + x_{14}x_{43} + x_{14}x_{45} + x_{14}x_{46} + x_{14}x_{51} + x_{14}x_{53} + x_{14}x_{54} + x_{14}x_{57} + x_{14}x_{59} + x_{14}x_{61} + x_{14}x_{62} + x_{15}x_{16} + x_{15}x_{17} + x_{15}x_{18} + x_{15}x_{21} + x_{15}x_{22} + x_{15}x_{23} + x_{15}x_{24} + x_{15}x_{25} + x_{15}x_{28} + x_{15}x_{29} + x_{15}x_{31} + x_{15}x_{33} + x_{15}x_{37} + x_{15}x_{41} + x_{15}x_{43} + x_{15}x_{45} + x_{15}x_{46} + x_{15}x_{48} + x_{15}x_{49} + x_{15}x_{51} + x_{15}x_{52} + x_{15}x_{54} + x_{15}x_{55} + x_{15}x_{57} + x_{15}x_{58} + x_{15}x_{60} + x_{15}x_{61} + x_{15}x_{62} + x_{15}x_{63} + x_{16}x_{17} + x_{16}x_{18} + x_{16}x_{20} + x_{16}x_{21} + x_{16}x_{23} + x_{16}x_{27} + x_{16}x_{29} + x_{16}x_{30} + x_{16}x_{33} + x_{16}x_{37} + x_{16}x_{38} + x_{16}x_{39} + x_{16}x_{40} + x_{16}x_{44} + x_{16}x_{47} + x_{16}x_{48} + x_{16}x_{49} + x_{16}x_{50} + x_{16}x_{55} + x_{16}x_{57} + x_{16}x_{58} + x_{16}x_{59} + x_{16}x_{60} + x_{16}x_{64} + x_{17}x_{18} + x_{17}x_{21} + x_{17}x_{23} + x_{17}x_{24} + x_{17}x_{25} + x_{17}x_{26} + x_{17}x_{27} + x_{17}x_{30} + x_{17}x_{31} + x_{17}x_{35} + x_{17}x_{37} + x_{17}x_{40} + x_{17}x_{41} + x_{17}x_{48} + x_{17}x_{50} + x_{17}x_{52} + x_{17}x_{54} + x_{17}x_{57} + x_{17}x_{58} + x_{17}x_{64} + x_{18}x_{19} + x_{18}x_{21} + x_{18}x_{22} + x_{18}x_{23} + x_{18}x_{24} + x_{18}x_{26} + x_{18}x_{27} + x_{18}x_{30} + x_{18}x_{32} + x_{18}x_{35} + x_{18}x_{37} + x_{18}x_{40} + x_{18}x_{41} + x_{18}x_{42} + x_{18}x_{43} + x_{18}x_{44} + x_{18}x_{48} + x_{18}x_{52} + x_{18}x_{53} + x_{18}x_{54} + x_{18}x_{56} + x_{18}x_{60} + x_{18}x_{61} + x_{18}x_{64} + x_{19}x_{20} + x_{19}x_{21} + x_{19}x_{22} + x_{19}x_{26} + x_{19}x_{28} + x_{19}x_{29} + x_{19}x_{33} + x_{19}x_{35} + x_{19}x_{36} + x_{19}x_{41} + x_{19}x_{43} + x_{19}x_{46} + x_{19}x_{49} + x_{19}x_{50} + x_{19}x_{52} + x_{19}x_{54} + x_{19}x_{55} + x_{19}x_{57} + x_{19}x_{59} + x_{19}x_{60} + x_{19}x_{61} + x_{19}x_{62} + x_{19}x_{64} + x_{20}x_{22} + x_{20}x_{23} + x_{20}x_{25} + x_{20}x_{26} + x_{20}x_{29} + x_{20}x_{30} + x_{20}x_{31} + x_{20}x_{32} + x_{20}x_{35} + x_{20}x_{36} + x_{20}x_{37} + x_{20}x_{38} + x_{20}x_{40} + x_{20}x_{44} + x_{20}x_{45} + x_{20}x_{46} + x_{20}x_{50} + x_{20}x_{51} + x_{20}x_{54} + x_{20}x_{55} + x_{20}x_{56} + x_{20}x_{58} + x_{20}x_{60} + x_{20}x_{62} + x_{20}x_{63} + x_{21}x_{25} + x_{21}x_{26} + x_{21}x_{27} + x_{21}x_{28} + x_{21}x_{29} + x_{21}x_{30} + x_{21}x_{32} + x_{21}x_{36} + x_{21}x_{38} + x_{21}x_{40} + x_{21}x_{42} + x_{21}x_{43} + x_{21}x_{44} + x_{21}x_{48} + x_{21}x_{49} + x_{21}x_{52} + x_{21}x_{54} + x_{21}x_{55} + x_{21}x_{56} + x_{21}x_{57} + x_{21}x_{58} + x_{21}x_{59} + x_{21}x_{61} + x_{21}x_{62} + x_{21}x_{64} + x_{22}x_{23} + x_{22}x_{25} + x_{22}x_{26} + x_{22}x_{30} + x_{22}x_{32} + x_{22}x_{34} + x_{22}x_{35} + x_{22}x_{36} + x_{22}x_{37} + x_{22}x_{38} + x_{22}x_{39} + x_{22}x_{41} + x_{22}x_{42} + x_{22}x_{45} + x_{22}x_{46} + x_{22}x_{48} + x_{22}x_{49} + x_{22}x_{54} + x_{22}x_{57} + x_{22}x_{58} + x_{22}x_{59} + x_{22}x_{61} + x_{22}x_{64} + x_{23}x_{25} + x_{23}x_{29} + x_{23}x_{33} + x_{23}x_{38} + x_{23}x_{41} + x_{23}x_{42} + x_{23}x_{44} + x_{23}x_{45} + x_{23}x_{49} + x_{23}x_{50} + x_{23}x_{51} + x_{23}x_{53} + x_{23}x_{54} + x_{23}x_{56} + x_{23}x_{58} + x_{23}x_{62} + x_{23}x_{63} + x_{24}x_{26} + x_{24}x_{28} + x_{24}x_{29} + x_{24}x_{33} + x_{24}x_{35} + x_{24}x_{36} + x_{24}x_{37} + x_{24}x_{40} + x_{24}x_{41} + x_{24}x_{42} + x_{24}x_{43} + x_{24}x_{49} + x_{24}x_{51} + x_{24}x_{52} + x_{24}x_{54} + x_{24}x_{55} + x_{24}x_{56} + x_{24}x_{57} + x_{24}x_{58} + x_{24}x_{59} + x_{24}x_{60} + x_{24}x_{63} + x_{24}x_{64} + x_{25}x_{26} + x_{25}x_{28} + x_{25}x_{30} + x_{25}x_{32} + x_{25}x_{33} + x_{25}x_{36} + x_{25}x_{37} + x_{25}x_{39} + x_{25}x_{44} + x_{25}x_{48} + x_{25}x_{50} + x_{25}x_{52} + x_{25}x_{53} + x_{25}x_{54} + x_{25}x_{55} + x_{25}x_{56} + x_{25}x_{57} + x_{25}x_{58} + x_{25}x_{60} + x_{25}x_{61} + x_{26}x_{28} + x_{26}x_{29} + x_{26}x_{31} + x_{26}x_{36} + x_{26}x_{40} + x_{26}x_{41} + x_{26}x_{43} + x_{26}x_{45} + x_{26}x_{46} + x_{26}x_{51} + x_{26}x_{52} + x_{26}x_{54} + x_{26}x_{55} + x_{26}x_{56} + x_{26}x_{57} + x_{26}x_{58} + x_{26}x_{61} + x_{26}x_{62} + x_{27}x_{28} + x_{27}x_{33} + x_{27}x_{34} + x_{27}x_{37} + x_{27}x_{38} + x_{27}x_{39} + x_{27}x_{41} + x_{27}x_{42} + x_{27}x_{43} + x_{27}x_{44} + x_{27}x_{45} + x_{27}x_{46} + x_{27}x_{47} + x_{27}x_{49} + x_{27}x_{50} + x_{27}x_{57} + x_{27}x_{58} + x_{27}x_{61} + x_{27}x_{62} + x_{27}x_{64} + x_{28}x_{30} + x_{28}x_{34} + x_{28}x_{35} + x_{28}x_{36} + x_{28}x_{37} + x_{28}x_{39} + x_{28}x_{40} + x_{28}x_{43} + x_{28}x_{44} + x_{28}x_{45} + x_{28}x_{47} + x_{28}x_{48} + x_{28}x_{50} + x_{28}x_{51} + x_{28}x_{52} + x_{28}x_{53} + x_{28}x_{54} + x_{28}x_{56} + x_{28}x_{58} + x_{28}x_{62} + x_{28}x_{63} + x_{29}x_{31} + x_{29}x_{34} + x_{29}x_{37} + x_{29}x_{38} + x_{29}x_{42} + x_{29}x_{45} + x_{29}x_{46} + x_{29}x_{53} + x_{29}x_{56} + x_{29}x_{57} + x_{29}x_{62} + x_{29}x_{63} + x_{29}x_{64} + x_{30}x_{32} + x_{30}x_{34} + x_{30}x_{38} + x_{30}x_{43} + x_{30}x_{44} + x_{30}x_{46} + x_{30}x_{48} + x_{30}x_{49} + x_{30}x_{50} + x_{30}x_{51} + x_{30}x_{55} + x_{30}x_{56} + x_{30}x_{57} + x_{30}x_{62} + x_{30}x_{63} + x_{30}x_{64} + x_{31}x_{32} + x_{31}x_{34} + x_{31}x_{36} + x_{31}x_{37} + x_{31}x_{39} + x_{31}x_{41} + x_{31}x_{43} + x_{31}x_{44} + x_{31}x_{45} + x_{31}x_{49} + x_{31}x_{50} + x_{31}x_{52} + x_{31}x_{54} + x_{31}x_{55} + x_{31}x_{57} + x_{31}x_{58} + x_{31}x_{59} + x_{31}x_{61} + x_{31}x_{62} + x_{31}x_{63} + x_{32}x_{33} + x_{32}x_{38} + x_{32}x_{39} + x_{32}x_{40} + x_{32}x_{41} + x_{32}x_{43} + x_{32}x_{46} + x_{32}x_{47} + x_{32}x_{50} + x_{32}x_{52} + x_{32}x_{53} + x_{32}x_{54} + x_{32}x_{56} + x_{32}x_{57} + x_{32}x_{60} + x_{33}x_{34} + x_{33}x_{35} + x_{33}x_{36} + x_{33}x_{38} + x_{33}x_{45} + x_{33}x_{47} + x_{33}x_{48} + x_{33}x_{49} + x_{33}x_{50} + x_{33}x_{52} + x_{33}x_{54} + x_{33}x_{58} + x_{33}x_{59} + x_{33}x_{62} + x_{33}x_{63} + x_{34}x_{35} + x_{34}x_{36} + x_{34}x_{38} + x_{34}x_{39} + x_{34}x_{41} + x_{34}x_{42} + x_{34}x_{43} + x_{34}x_{45} + x_{34}x_{48} + x_{34}x_{49} + x_{34}x_{51} + x_{34}x_{52} + x_{34}x_{53} + x_{34}x_{58} + x_{34}x_{59} + x_{34}x_{60} + x_{34}x_{62} + x_{34}x_{64} + x_{35}x_{38} + x_{35}x_{40} + x_{35}x_{41} + x_{35}x_{42} + x_{35}x_{43} + x_{35}x_{44} + x_{35}x_{45} + x_{35}x_{48} + x_{35}x_{51} + x_{35}x_{52} + x_{35}x_{53} + x_{35}x_{55} + x_{35}x_{56} + x_{35}x_{57} + x_{35}x_{58} + x_{35}x_{59} + x_{35}x_{60} + x_{35}x_{61} + x_{35}x_{64} + x_{36}x_{39} + x_{36}x_{41} + x_{36}x_{42} + x_{36}x_{46} + x_{36}x_{49} + x_{36}x_{50} + x_{36}x_{53} + x_{36}x_{54} + x_{36}x_{55} + x_{36}x_{59} + x_{36}x_{60} + x_{36}x_{61} + x_{37}x_{40} + x_{37}x_{43} + x_{37}x_{48} + x_{37}x_{49} + x_{37}x_{51} + x_{37}x_{52} + x_{37}x_{53} + x_{37}x_{54} + x_{37}x_{55} + x_{37}x_{56} + x_{37}x_{57} + x_{37}x_{58} + x_{37}x_{59} + x_{37}x_{60} + x_{37}x_{61} + x_{38}x_{40} + x_{38}x_{45} + x_{38}x_{48} + x_{38}x_{49} + x_{38}x_{50} + x_{38}x_{52} + x_{38}x_{53} + x_{38}x_{54} + x_{38}x_{56} + x_{38}x_{57} + x_{38}x_{58} + x_{38}x_{60} + x_{38}x_{61} + x_{39}x_{41} + x_{39}x_{43} + x_{39}x_{45} + x_{39}x_{47} + x_{39}x_{48} + x_{39}x_{49} + x_{39}x_{50} + x_{39}x_{52} + x_{39}x_{53} + x_{39}x_{54} + x_{39}x_{55} + x_{39}x_{58} + x_{39}x_{59} + x_{39}x_{62} + x_{39}x_{63} + x_{39}x_{64} + x_{40}x_{42} + x_{40}x_{44} + x_{40}x_{45} + x_{40}x_{49} + x_{40}x_{50} + x_{40}x_{51} + x_{40}x_{56} + x_{40}x_{57} + x_{40}x_{58} + x_{40}x_{60} + x_{40}x_{61} + x_{40}x_{62} + x_{40}x_{64} + x_{41}x_{43} + x_{41}x_{44} + x_{41}x_{45} + x_{41}x_{47} + x_{41}x_{48} + x_{41}x_{52} + x_{41}x_{56} + x_{41}x_{57} + x_{41}x_{58} + x_{41}x_{60} + x_{41}x_{61} + x_{41}x_{62} + x_{41}x_{63} + x_{42}x_{43} + x_{42}x_{44} + x_{42}x_{45} + x_{42}x_{51} + x_{42}x_{52} + x_{42}x_{54} + x_{42}x_{55} + x_{42}x_{56} + x_{42}x_{57} + x_{42}x_{59} + x_{42}x_{62} + x_{43}x_{45} + x_{43}x_{52} + x_{43}x_{53} + x_{43}x_{54} + x_{43}x_{55} + x_{43}x_{56} + x_{43}x_{57} + x_{43}x_{61} + x_{43}x_{62} + x_{43}x_{63} + x_{44}x_{46} + x_{44}x_{48} + x_{44}x_{50} + x_{44}x_{52} + x_{44}x_{53} + x_{44}x_{55} + x_{44}x_{56} + x_{44}x_{61} + x_{44}x_{63} + x_{44}x_{64} + x_{45}x_{47} + x_{45}x_{48} + x_{45}x_{49} + x_{45}x_{51} + x_{45}x_{52} + x_{45}x_{53} + x_{45}x_{54} + x_{45}x_{57} + x_{45}x_{59} + x_{45}x_{61} + x_{45}x_{62} + x_{45}x_{63} + x_{46}x_{47} + x_{46}x_{49} + x_{46}x_{50} + x_{46}x_{51} + x_{46}x_{54} + x_{46}x_{58} + x_{46}x_{59} + x_{46}x_{61} + x_{46}x_{63} + x_{46}x_{64} + x_{47}x_{48} + x_{47}x_{49} + x_{47}x_{50} + x_{47}x_{52} + x_{47}x_{53} + x_{47}x_{54} + x_{47}x_{62} + x_{47}x_{63} + x_{47}x_{64} + x_{48}x_{49} + x_{48}x_{50} + x_{48}x_{52} + x_{48}x_{53} + x_{48}x_{57} + x_{48}x_{62} + x_{48}x_{63} + x_{48}x_{64} + x_{49}x_{50} + x_{49}x_{52} + x_{49}x_{53} + x_{49}x_{54} + x_{49}x_{56} + x_{49}x_{57} + x_{49}x_{63} + x_{50}x_{52} + x_{50}x_{53} + x_{50}x_{54} + x_{50}x_{55} + x_{50}x_{57} + x_{50}x_{58} + x_{50}x_{59} + x_{50}x_{61} + x_{50}x_{62} + x_{51}x_{53} + x_{51}x_{55} + x_{51}x_{57} + x_{51}x_{59} + x_{51}x_{61} + x_{51}x_{63} + x_{52}x_{53} + x_{52}x_{56} + x_{52}x_{58} + x_{52}x_{60} + x_{52}x_{61} + x_{52}x_{63} + x_{53}x_{54} + x_{53}x_{58} + x_{53}x_{59} + x_{53}x_{62} + x_{53}x_{63} + x_{54}x_{57} + x_{54}x_{58} + x_{54}x_{62} + x_{55}x_{56} + x_{55}x_{58} + x_{55}x_{59} + x_{55}x_{60} + x_{56}x_{57} + x_{56}x_{58} + x_{56}x_{60} + x_{56}x_{61} + x_{56}x_{63} + x_{57}x_{59} + x_{57}x_{62} + x_{57}x_{64} + x_{58}x_{61} + x_{58}x_{62} + x_{58}x_{64} + x_{59}x_{62} + x_{59}x_{63} + x_{59}x_{64} + x_{60}x_{61} + x_{60}x_{62} + x_{60}x_{64} + x_{61}x_{64} + x_{62}x_{63} + x_{2} + x_{8} + x_{9} + x_{10} + x_{11} + x_{12} + x_{14} + x_{17} + x_{19} + x_{20} + x_{21} + x_{24} + x_{25} + x_{27} + x_{28} + x_{29} + x_{33} + x_{38} + x_{41} + x_{43} + x_{44} + x_{45} + x_{48} + x_{49} + x_{51} + x_{52} + x_{55} + x_{56} + x_{57} + x_{58} + x_{60} + x_{61} + x_{64} + 1$

$y_{23} = x_{1}x_{2} + x_{1}x_{3} + x_{1}x_{4} + x_{1}x_{5} + x_{1}x_{7} + x_{1}x_{8} + x_{1}x_{13} + x_{1}x_{15} + x_{1}x_{16} + x_{1}x_{17} + x_{1}x_{21} + x_{1}x_{23} + x_{1}x_{26} + x_{1}x_{28} + x_{1}x_{32} + x_{1}x_{33} + x_{1}x_{36} + x_{1}x_{37} + x_{1}x_{44} + x_{1}x_{45} + x_{1}x_{48} + x_{1}x_{49} + x_{1}x_{52} + x_{1}x_{55} + x_{1}x_{56} + x_{1}x_{57} + x_{1}x_{58} + x_{1}x_{60} + x_{1}x_{61} + x_{1}x_{62} + x_{1}x_{63} + x_{2}x_{3} + x_{2}x_{5} + x_{2}x_{6} + x_{2}x_{8} + x_{2}x_{14} + x_{2}x_{16} + x_{2}x_{17} + x_{2}x_{18} + x_{2}x_{20} + x_{2}x_{21} + x_{2}x_{25} + x_{2}x_{26} + x_{2}x_{30} + x_{2}x_{31} + x_{2}x_{32} + x_{2}x_{36} + x_{2}x_{37} + x_{2}x_{38} + x_{2}x_{41} + x_{2}x_{42} + x_{2}x_{45} + x_{2}x_{46} + x_{2}x_{47} + x_{2}x_{50} + x_{2}x_{52} + x_{2}x_{55} + x_{2}x_{61} + x_{3}x_{4} + x_{3}x_{5} + x_{3}x_{6} + x_{3}x_{11} + x_{3}x_{24} + x_{3}x_{28} + x_{3}x_{29} + x_{3}x_{31} + x_{3}x_{32} + x_{3}x_{33} + x_{3}x_{38} + x_{3}x_{40} + x_{3}x_{41} + x_{3}x_{44} + x_{3}x_{45} + x_{3}x_{49} + x_{3}x_{51} + x_{3}x_{54} + x_{3}x_{59} + x_{3}x_{60} + x_{3}x_{62} + x_{3}x_{63} + x_{3}x_{64} + x_{4}x_{7} + x_{4}x_{8} + x_{4}x_{10} + x_{4}x_{11} + x_{4}x_{12} + x_{4}x_{13} + x_{4}x_{14} + x_{4}x_{17} + x_{4}x_{18} + x_{4}x_{19} + x_{4}x_{21} + x_{4}x_{23} + x_{4}x_{25} + x_{4}x_{26} + x_{4}x_{28} + x_{4}x_{29} + x_{4}x_{31} + x_{4}x_{33} + x_{4}x_{35} + x_{4}x_{38} + x_{4}x_{39} + x_{4}x_{41} + x_{4}x_{43} + x_{4}x_{45} + x_{4}x_{47} + x_{4}x_{48} + x_{4}x_{49} + x_{4}x_{53} + x_{4}x_{54} + x_{4}x_{62} + x_{4}x_{63} + x_{5}x_{7} + x_{5}x_{8} + x_{5}x_{11} + x_{5}x_{12} + x_{5}x_{13} + x_{5}x_{21} + x_{5}x_{24} + x_{5}x_{26} + x_{5}x_{29} + x_{5}x_{31} + x_{5}x_{33} + x_{5}x_{34} + x_{5}x_{38} + x_{5}x_{40} + x_{5}x_{41} + x_{5}x_{42} + x_{5}x_{46} + x_{5}x_{50} + x_{5}x_{52} + x_{5}x_{56} + x_{5}x_{61} + x_{6}x_{7} + x_{6}x_{9} + x_{6}x_{11} + x_{6}x_{12} + x_{6}x_{13} + x_{6}x_{16} + x_{6}x_{17} + x_{6}x_{18} + x_{6}x_{19} + x_{6}x_{20} + x_{6}x_{23} + x_{6}x_{24} + x_{6}x_{25} + x_{6}x_{28} + x_{6}x_{29} + x_{6}x_{31} + x_{6}x_{34} + x_{6}x_{37} + x_{6}x_{41} + x_{6}x_{43} + x_{6}x_{44} + x_{6}x_{45} + x_{6}x_{47} + x_{6}x_{49} + x_{6}x_{50} + x_{6}x_{51} + x_{6}x_{52} + x_{6}x_{53} + x_{6}x_{54} + x_{6}x_{59} + x_{6}x_{64} + x_{7}x_{9} + x_{7}x_{11} + x_{7}x_{12} + x_{7}x_{18} + x_{7}x_{22} + x_{7}x_{25} + x_{7}x_{29} + x_{7}x_{30} + x_{7}x_{31} + x_{7}x_{34} + x_{7}x_{35} + x_{7}x_{36} + x_{7}x_{38} + x_{7}x_{39} + x_{7}x_{42} + x_{7}x_{43} + x_{7}x_{45} + x_{7}x_{46} + x_{7}x_{47} + x_{7}x_{48} + x_{7}x_{50} + x_{7}x_{55} + x_{7}x_{57} + x_{7}x_{58} + x_{7}x_{59} + x_{7}x_{62} + x_{8}x_{9} + x_{8}x_{10} + x_{8}x_{11} + x_{8}x_{13} + x_{8}x_{14} + x_{8}x_{16} + x_{8}x_{17} + x_{8}x_{18} + x_{8}x_{20} + x_{8}x_{21} + x_{8}x_{24} + x_{8}x_{25} + x_{8}x_{26} + x_{8}x_{29} + x_{8}x_{31} + x_{8}x_{32} + x_{8}x_{34} + x_{8}x_{35} + x_{8}x_{36} + x_{8}x_{38} + x_{8}x_{39} + x_{8}x_{41} + x_{8}x_{46} + x_{8}x_{48} + x_{8}x_{53} + x_{8}x_{55} + x_{8}x_{57} + x_{8}x_{58} + x_{8}x_{61} + x_{8}x_{62} + x_{8}x_{64} + x_{9}x_{10} + x_{9}x_{14} + x_{9}x_{19} + x_{9}x_{20} + x_{9}x_{22} + x_{9}x_{24} + x_{9}x_{25} + x_{9}x_{27} + x_{9}x_{28} + x_{9}x_{29} + x_{9}x_{31} + x_{9}x_{34} + x_{9}x_{35} + x_{9}x_{36} + x_{9}x_{37} + x_{9}x_{38} + x_{9}x_{40} + x_{9}x_{41} + x_{9}x_{44} + x_{9}x_{45} + x_{9}x_{46} + x_{9}x_{47} + x_{9}x_{52} + x_{9}x_{54} + x_{9}x_{57} + x_{9}x_{58} + x_{9}x_{61} + x_{9}x_{62} + x_{9}x_{64} + x_{10}x_{11} + x_{10}x_{12} + x_{10}x_{14} + x_{10}x_{16} + x_{10}x_{18} + x_{10}x_{19} + x_{10}x_{21} + x_{10}x_{22} + x_{10}x_{26} + x_{10}x_{28} + x_{10}x_{29} + x_{10}x_{32} + x_{10}x_{33} + x_{10}x_{35} + x_{10}x_{36} + x_{10}x_{37} + x_{10}x_{39} + x_{10}x_{40} + x_{10}x_{41} + x_{10}x_{42} + x_{10}x_{43} + x_{10}x_{44} + x_{10}x_{48} + x_{10}x_{52} + x_{10}x_{55} + x_{10}x_{56} + x_{10}x_{57} + x_{10}x_{61} + x_{10}x_{63} + x_{11}x_{12} + x_{11}x_{13} + x_{11}x_{14} + x_{11}x_{16} + x_{11}x_{17} + x_{11}x_{18} + x_{11}x_{20} + x_{11}x_{23} + x_{11}x_{26} + x_{11}x_{29} + x_{11}x_{32} + x_{11}x_{37} + x_{11}x_{42} + x_{11}x_{43} + x_{11}x_{44} + x_{11}x_{47} + x_{11}x_{48} + x_{11}x_{49} + x_{11}x_{51} + x_{11}x_{52} + x_{11}x_{53} + x_{11}x_{60} + x_{11}x_{61} + x_{11}x_{62} + x_{11}x_{64} + x_{12}x_{15} + x_{12}x_{16} + x_{12}x_{18} + x_{12}x_{19} + x_{12}x_{20} + x_{12}x_{21} + x_{12}x_{22} + x_{12}x_{24} + x_{12}x_{26} + x_{12}x_{27} + x_{12}x_{29} + x_{12}x_{32} + x_{12}x_{34} + x_{12}x_{36} + x_{12}x_{38} + x_{12}x_{39} + x_{12}x_{40} + x_{12}x_{42} + x_{12}x_{43} + x_{12}x_{44} + x_{12}x_{46} + x_{12}x_{48} + x_{12}x_{50} + x_{12}x_{51} + x_{12}x_{53} + x_{12}x_{55} + x_{12}x_{56} + x_{12}x_{57} + x_{12}x_{60} + x_{12}x_{61} + x_{12}x_{62} + x_{12}x_{64} + x_{13}x_{15} + x_{13}x_{16} + x_{13}x_{20} + x_{13}x_{21} + x_{13}x_{23} + x_{13}x_{25} + x_{13}x_{26} + x_{13}x_{27} + x_{13}x_{28} + x_{13}x_{29} + x_{13}x_{31} + x_{13}x_{33} + x_{13}x_{40} + x_{13}x_{45} + x_{13}x_{47} + x_{13}x_{48} + x_{13}x_{49} + x_{13}x_{50} + x_{13}x_{51} + x_{13}x_{52} + x_{13}x_{53} + x_{13}x_{55} + x_{13}x_{56} + x_{13}x_{57} + x_{13}x_{58} + x_{13}x_{60} + x_{13}x_{62} + x_{14}x_{15} + x_{14}x_{16} + x_{14}x_{20} + x_{14}x_{25} + x_{14}x_{28} + x_{14}x_{29} + x_{14}x_{30} + x_{14}x_{31} + x_{14}x_{33} + x_{14}x_{40} + x_{14}x_{44} + x_{14}x_{47} + x_{14}x_{48} + x_{14}x_{52} + x_{14}x_{53} + x_{14}x_{57} + x_{14}x_{58} + x_{14}x_{61} + x_{14}x_{62} + x_{14}x_{63} + x_{14}x_{64} + x_{15}x_{16} + x_{15}x_{20} + x_{15}x_{21} + x_{15}x_{24} + x_{15}x_{25} + x_{15}x_{28} + x_{15}x_{29} + x_{15}x_{34} + x_{15}x_{42} + x_{15}x_{48} + x_{15}x_{50} + x_{15}x_{51} + x_{15}x_{52} + x_{15}x_{54} + x_{15}x_{55} + x_{15}x_{61} + x_{15}x_{64} + x_{16}x_{17} + x_{16}x_{18} + x_{16}x_{19} + x_{16}x_{21} + x_{16}x_{22} + x_{16}x_{23} + x_{16}x_{25} + x_{16}x_{26} + x_{16}x_{27} + x_{16}x_{29} + x_{16}x_{30} + x_{16}x_{31} + x_{16}x_{32} + x_{16}x_{33} + x_{16}x_{34} + x_{16}x_{36} + x_{16}x_{37} + x_{16}x_{38} + x_{16}x_{40} + x_{16}x_{43} + x_{16}x_{44} + x_{16}x_{51} + x_{16}x_{52} + x_{16}x_{53} + x_{16}x_{55} + x_{16}x_{56} + x_{16}x_{63} + x_{17}x_{18} + x_{17}x_{19} + x_{17}x_{26} + x_{17}x_{27} + x_{17}x_{29} + x_{17}x_{31} + x_{17}x_{35} + x_{17}x_{37} + x_{17}x_{42} + x_{17}x_{43} + x_{17}x_{44} + x_{17}x_{46} + x_{17}x_{47} + x_{17}x_{50} + x_{17}x_{54} + x_{17}x_{56} + x_{17}x_{57} + x_{17}x_{58} + x_{17}x_{62} + x_{17}x_{63} + x_{18}x_{20} + x_{18}x_{21} + x_{18}x_{24} + x_{18}x_{25} + x_{18}x_{26} + x_{18}x_{28} + x_{18}x_{29} + x_{18}x_{33} + x_{18}x_{35} + x_{18}x_{39} + x_{18}x_{40} + x_{18}x_{41} + x_{18}x_{42} + x_{18}x_{43} + x_{18}x_{45} + x_{18}x_{48} + x_{18}x_{49} + x_{18}x_{53} + x_{18}x_{54} + x_{18}x_{56} + x_{18}x_{57} + x_{18}x_{58} + x_{18}x_{59} + x_{18}x_{63} + x_{19}x_{20} + x_{19}x_{22} + x_{19}x_{23} + x_{19}x_{27} + x_{19}x_{29} + x_{19}x_{30} + x_{19}x_{31} + x_{19}x_{37} + x_{19}x_{38} + x_{19}x_{40} + x_{19}x_{43} + x_{19}x_{45} + x_{19}x_{47} + x_{19}x_{49} + x_{19}x_{51} + x_{19}x_{52} + x_{19}x_{53} + x_{19}x_{55} + x_{19}x_{56} + x_{19}x_{58} + x_{19}x_{61} + x_{19}x_{63} + x_{19}x_{64} + x_{20}x_{21} + x_{20}x_{23} + x_{20}x_{24} + x_{20}x_{25} + x_{20}x_{27} + x_{20}x_{28} + x_{20}x_{29} + x_{20}x_{30} + x_{20}x_{31} + x_{20}x_{33} + x_{20}x_{35} + x_{20}x_{36} + x_{20}x_{37} + x_{20}x_{40} + x_{20}x_{46} + x_{20}x_{47} + x_{20}x_{48} + x_{20}x_{49} + x_{20}x_{50} + x_{20}x_{53} + x_{20}x_{55} + x_{20}x_{57} + x_{20}x_{58} + x_{20}x_{59} + x_{20}x_{60} + x_{21}x_{23} + x_{21}x_{24} + x_{21}x_{26} + x_{21}x_{28} + x_{21}x_{29} + x_{21}x_{32} + x_{21}x_{33} + x_{21}x_{35} + x_{21}x_{41} + x_{21}x_{43} + x_{21}x_{44} + x_{21}x_{45} + x_{21}x_{46} + x_{21}x_{47} + x_{21}x_{48} + x_{21}x_{49} + x_{21}x_{50} + x_{21}x_{52} + x_{21}x_{53} + x_{21}x_{56} + x_{21}x_{61} + x_{21}x_{64} + x_{22}x_{25} + x_{22}x_{27} + x_{22}x_{28} + x_{22}x_{30} + x_{22}x_{32} + x_{22}x_{36} + x_{22}x_{37} + x_{22}x_{38} + x_{22}x_{39} + x_{22}x_{40} + x_{22}x_{41} + x_{22}x_{42} + x_{22}x_{43} + x_{22}x_{46} + x_{22}x_{47} + x_{22}x_{49} + x_{22}x_{50} + x_{22}x_{52} + x_{22}x_{54} + x_{22}x_{56} + x_{22}x_{58} + x_{22}x_{59} + x_{22}x_{61} + x_{22}x_{62} + x_{22}x_{63} + x_{23}x_{30} + x_{23}x_{33} + x_{23}x_{34} + x_{23}x_{38} + x_{23}x_{39} + x_{23}x_{41} + x_{23}x_{42} + x_{23}x_{43} + x_{23}x_{44} + x_{23}x_{45} + x_{23}x_{46} + x_{23}x_{47} + x_{23}x_{49} + x_{23}x_{51} + x_{23}x_{53} + x_{23}x_{54} + x_{23}x_{57} + x_{23}x_{58} + x_{23}x_{59} + x_{23}x_{60} + x_{23}x_{62} + x_{23}x_{63} + x_{23}x_{64} + x_{24}x_{25} + x_{24}x_{26} + x_{24}x_{28} + x_{24}x_{29} + x_{24}x_{32} + x_{24}x_{34} + x_{24}x_{35} + x_{24}x_{42} + x_{24}x_{43} + x_{24}x_{45} + x_{24}x_{47} + x_{24}x_{48} + x_{24}x_{49} + x_{24}x_{51} + x_{24}x_{52} + x_{24}x_{53} + x_{24}x_{55} + x_{24}x_{58} + x_{24}x_{62} + x_{24}x_{63} + x_{24}x_{64} + x_{25}x_{26} + x_{25}x_{30} + x_{25}x_{33} + x_{25}x_{34} + x_{25}x_{37} + x_{25}x_{39} + x_{25}x_{41} + x_{25}x_{43} + x_{25}x_{44} + x_{25}x_{47} + x_{25}x_{48} + x_{25}x_{50} + x_{25}x_{54} + x_{25}x_{59} + x_{25}x_{61} + x_{25}x_{62} + x_{26}x_{29} + x_{26}x_{30} + x_{26}x_{31} + x_{26}x_{32} + x_{26}x_{35} + x_{26}x_{39} + x_{26}x_{40} + x_{26}x_{41} + x_{26}x_{43} + x_{26}x_{46} + x_{26}x_{48} + x_{26}x_{53} + x_{26}x_{57} + x_{26}x_{62} + x_{26}x_{63} + x_{26}x_{64} + x_{27}x_{29} + x_{27}x_{30} + x_{27}x_{31} + x_{27}x_{35} + x_{27}x_{38} + x_{27}x_{41} + x_{27}x_{43} + x_{27}x_{44} + x_{27}x_{47} + x_{27}x_{51} + x_{27}x_{56} + x_{27}x_{59} + x_{27}x_{61} + x_{27}x_{62} + x_{27}x_{63} + x_{28}x_{29} + x_{28}x_{35} + x_{28}x_{36} + x_{28}x_{38} + x_{28}x_{40} + x_{28}x_{41} + x_{28}x_{42} + x_{28}x_{43} + x_{28}x_{45} + x_{28}x_{46} + x_{28}x_{47} + x_{28}x_{49} + x_{28}x_{51} + x_{28}x_{52} + x_{28}x_{55} + x_{28}x_{57} + x_{28}x_{58} + x_{28}x_{62} + x_{29}x_{30} + x_{29}x_{31} + x_{29}x_{32} + x_{29}x_{33} + x_{29}x_{34} + x_{29}x_{35} + x_{29}x_{36} + x_{29}x_{38} + x_{29}x_{40} + x_{29}x_{41} + x_{29}x_{42} + x_{29}x_{43} + x_{29}x_{45} + x_{29}x_{52} + x_{29}x_{54} + x_{29}x_{55} + x_{29}x_{56} + x_{29}x_{58} + x_{29}x_{62} + x_{30}x_{33} + x_{30}x_{40} + x_{30}x_{41} + x_{30}x_{48} + x_{30}x_{57} + x_{30}x_{58} + x_{30}x_{59} + x_{30}x_{60} + x_{30}x_{61} + x_{30}x_{62} + x_{30}x_{63} + x_{31}x_{32} + x_{31}x_{35} + x_{31}x_{36} + x_{31}x_{37} + x_{31}x_{38} + x_{31}x_{39} + x_{31}x_{41} + x_{31}x_{43} + x_{31}x_{44} + x_{31}x_{46} + x_{31}x_{48} + x_{31}x_{49} + x_{31}x_{50} + x_{31}x_{52} + x_{31}x_{54} + x_{31}x_{55} + x_{31}x_{61} + x_{31}x_{62} + x_{31}x_{63} + x_{32}x_{33} + x_{32}x_{36} + x_{32}x_{37} + x_{32}x_{40} + x_{32}x_{43} + x_{32}x_{45} + x_{32}x_{46} + x_{32}x_{50} + x_{32}x_{51} + x_{32}x_{52} + x_{32}x_{55} + x_{32}x_{56} + x_{32}x_{58} + x_{32}x_{62} + x_{32}x_{63} + x_{32}x_{64} + x_{33}x_{34} + x_{33}x_{35} + x_{33}x_{36} + x_{33}x_{38} + x_{33}x_{39} + x_{33}x_{41} + x_{33}x_{43} + x_{33}x_{46} + x_{33}x_{48} + x_{33}x_{51} + x_{33}x_{54} + x_{33}x_{57} + x_{33}x_{58} + x_{33}x_{59} + x_{33}x_{60} + x_{33}x_{63} + x_{33}x_{64} + x_{34}x_{37} + x_{34}x_{39} + x_{34}x_{42} + x_{34}x_{48} + x_{34}x_{50} + x_{34}x_{51} + x_{34}x_{52} + x_{34}x_{55} + x_{34}x_{56} + x_{34}x_{57} + x_{34}x_{58} + x_{34}x_{59} + x_{34}x_{62} + x_{34}x_{63} + x_{35}x_{38} + x_{35}x_{43} + x_{35}x_{48} + x_{35}x_{50} + x_{35}x_{51} + x_{35}x_{52} + x_{35}x_{55} + x_{35}x_{56} + x_{35}x_{58} + x_{35}x_{60} + x_{35}x_{63} + x_{36}x_{40} + x_{36}x_{43} + x_{36}x_{44} + x_{36}x_{49} + x_{36}x_{50} + x_{36}x_{51} + x_{36}x_{56} + x_{36}x_{57} + x_{36}x_{61} + x_{36}x_{64} + x_{37}x_{39} + x_{37}x_{41} + x_{37}x_{42} + x_{37}x_{43} + x_{37}x_{45} + x_{37}x_{46} + x_{37}x_{47} + x_{37}x_{49} + x_{37}x_{51} + x_{37}x_{53} + x_{37}x_{54} + x_{37}x_{55} + x_{37}x_{58} + x_{37}x_{59} + x_{37}x_{61} + x_{37}x_{63} + x_{38}x_{41} + x_{38}x_{42} + x_{38}x_{45} + x_{38}x_{46} + x_{38}x_{47} + x_{38}x_{48} + x_{38}x_{52} + x_{38}x_{55} + x_{38}x_{56} + x_{38}x_{58} + x_{38}x_{60} + x_{38}x_{63} + x_{39}x_{40} + x_{39}x_{41} + x_{39}x_{42} + x_{39}x_{46} + x_{39}x_{47} + x_{39}x_{48} + x_{39}x_{50} + x_{39}x_{51} + x_{39}x_{53} + x_{39}x_{54} + x_{39}x_{55} + x_{39}x_{56} + x_{39}x_{57} + x_{39}x_{60} + x_{39}x_{62} + x_{40}x_{41} + x_{40}x_{42} + x_{40}x_{46} + x_{40}x_{47} + x_{40}x_{48} + x_{40}x_{53} + x_{40}x_{55} + x_{40}x_{58} + x_{40}x_{59} + x_{40}x_{60} + x_{40}x_{61} + x_{40}x_{63} + x_{40}x_{64} + x_{41}x_{45} + x_{41}x_{46} + x_{41}x_{48} + x_{41}x_{50} + x_{41}x_{53} + x_{41}x_{55} + x_{41}x_{60} + x_{41}x_{61} + x_{42}x_{43} + x_{42}x_{44} + x_{42}x_{45} + x_{42}x_{50} + x_{42}x_{51} + x_{42}x_{52} + x_{42}x_{54} + x_{42}x_{55} + x_{42}x_{57} + x_{42}x_{59} + x_{42}x_{60} + x_{42}x_{64} + x_{43}x_{45} + x_{43}x_{46} + x_{43}x_{47} + x_{43}x_{48} + x_{43}x_{49} + x_{43}x_{50} + x_{43}x_{52} + x_{43}x_{53} + x_{43}x_{54} + x_{43}x_{55} + x_{43}x_{56} + x_{43}x_{57} + x_{43}x_{64} + x_{44}x_{45} + x_{44}x_{47} + x_{44}x_{54} + x_{44}x_{55} + x_{44}x_{57} + x_{44}x_{59} + x_{44}x_{62} + x_{44}x_{63} + x_{45}x_{46} + x_{45}x_{47} + x_{45}x_{50} + x_{45}x_{55} + x_{45}x_{59} + x_{45}x_{61} + x_{46}x_{47} + x_{46}x_{48} + x_{46}x_{53} + x_{46}x_{54} + x_{46}x_{55} + x_{46}x_{57} + x_{46}x_{59} + x_{46}x_{61} + x_{46}x_{62} + x_{46}x_{64} + x_{47}x_{48} + x_{47}x_{49} + x_{47}x_{51} + x_{47}x_{54} + x_{47}x_{55} + x_{47}x_{56} + x_{47}x_{58} + x_{47}x_{59} + x_{47}x_{61} + x_{47}x_{62} + x_{48}x_{49} + x_{48}x_{50} + x_{48}x_{52} + x_{48}x_{53} + x_{48}x_{54} + x_{48}x_{58} + x_{48}x_{64} + x_{49}x_{50} + x_{49}x_{51} + x_{49}x_{52} + x_{49}x_{54} + x_{49}x_{55} + x_{49}x_{60} + x_{49}x_{61} + x_{49}x_{64} + x_{50}x_{53} + x_{50}x_{56} + x_{50}x_{58} + x_{50}x_{62} + x_{50}x_{63} + x_{51}x_{52} + x_{51}x_{54} + x_{51}x_{55} + x_{51}x_{56} + x_{51}x_{59} + x_{51}x_{61} + x_{51}x_{62} + x_{51}x_{63} + x_{51}x_{64} + x_{52}x_{54} + x_{52}x_{55} + x_{52}x_{56} + x_{52}x_{57} + x_{52}x_{58} + x_{52}x_{61} + x_{52}x_{62} + x_{53}x_{54} + x_{53}x_{55} + x_{53}x_{59} + x_{53}x_{60} + x_{53}x_{61} + x_{53}x_{63} + x_{54}x_{56} + x_{54}x_{57} + x_{54}x_{58} + x_{54}x_{63} + x_{55}x_{56} + x_{55}x_{59} + x_{55}x_{61} + x_{55}x_{64} + x_{56}x_{60} + x_{56}x_{61} + x_{56}x_{63} + x_{56}x_{64} + x_{57}x_{59} + x_{57}x_{60} + x_{57}x_{61} + x_{58}x_{59} + x_{58}x_{60} + x_{58}x_{63} + x_{58}x_{64} + x_{59}x_{60} + x_{59}x_{63} + x_{59}x_{64} + x_{60}x_{61} + x_{60}x_{62} + x_{61}x_{62} + x_{62}x_{64} + x_{1} + x_{2} + x_{5} + x_{7} + x_{8} + x_{10} + x_{11} + x_{13} + x_{14} + x_{15} + x_{16} + x_{18} + x_{20} + x_{23} + x_{24} + x_{32} + x_{33} + x_{34} + x_{35} + x_{36} + x_{39} + x_{40} + x_{42} + x_{46} + x_{49} + x_{50} + x_{51} + x_{56} + x_{59} + x_{60} + x_{62} + x_{64} + 1$

$y_{24} = x_{1}x_{7} + x_{1}x_{8} + x_{1}x_{9} + x_{1}x_{12} + x_{1}x_{13} + x_{1}x_{14} + x_{1}x_{18} + x_{1}x_{19} + x_{1}x_{21} + x_{1}x_{22} + x_{1}x_{23} + x_{1}x_{25} + x_{1}x_{26} + x_{1}x_{27} + x_{1}x_{28} + x_{1}x_{31} + x_{1}x_{32} + x_{1}x_{33} + x_{1}x_{36} + x_{1}x_{41} + x_{1}x_{42} + x_{1}x_{43} + x_{1}x_{44} + x_{1}x_{46} + x_{1}x_{49} + x_{1}x_{50} + x_{1}x_{51} + x_{1}x_{55} + x_{1}x_{57} + x_{1}x_{62} + x_{1}x_{63} + x_{1}x_{64} + x_{2}x_{3} + x_{2}x_{4} + x_{2}x_{5} + x_{2}x_{6} + x_{2}x_{7} + x_{2}x_{8} + x_{2}x_{9} + x_{2}x_{11} + x_{2}x_{15} + x_{2}x_{16} + x_{2}x_{17} + x_{2}x_{19} + x_{2}x_{20} + x_{2}x_{23} + x_{2}x_{26} + x_{2}x_{27} + x_{2}x_{30} + x_{2}x_{36} + x_{2}x_{41} + x_{2}x_{43} + x_{2}x_{44} + x_{2}x_{47} + x_{2}x_{49} + x_{2}x_{53} + x_{2}x_{56} + x_{2}x_{57} + x_{2}x_{60} + x_{2}x_{61} + x_{2}x_{62} + x_{2}x_{64} + x_{3}x_{5} + x_{3}x_{6} + x_{3}x_{7} + x_{3}x_{10} + x_{3}x_{11} + x_{3}x_{12} + x_{3}x_{13} + x_{3}x_{14} + x_{3}x_{18} + x_{3}x_{19} + x_{3}x_{26} + x_{3}x_{28} + x_{3}x_{29} + x_{3}x_{30} + x_{3}x_{33} + x_{3}x_{35} + x_{3}x_{38} + x_{3}x_{40} + x_{3}x_{41} + x_{3}x_{42} + x_{3}x_{44} + x_{3}x_{45} + x_{3}x_{47} + x_{3}x_{49} + x_{3}x_{51} + x_{3}x_{53} + x_{3}x_{54} + x_{3}x_{57} + x_{3}x_{58} + x_{3}x_{59} + x_{3}x_{62} + x_{3}x_{64} + x_{4}x_{6} + x_{4}x_{9} + x_{4}x_{13} + x_{4}x_{15} + x_{4}x_{18} + x_{4}x_{20} + x_{4}x_{24} + x_{4}x_{26} + x_{4}x_{29} + x_{4}x_{32} + x_{4}x_{33} + x_{4}x_{37} + x_{4}x_{39} + x_{4}x_{41} + x_{4}x_{44} + x_{4}x_{45} + x_{4}x_{46} + x_{4}x_{47} + x_{4}x_{49} + x_{4}x_{50} + x_{4}x_{51} + x_{4}x_{52} + x_{4}x_{55} + x_{4}x_{58} + x_{4}x_{60} + x_{4}x_{61} + x_{4}x_{62} + x_{4}x_{64} + x_{5}x_{6} + x_{5}x_{9} + x_{5}x_{11} + x_{5}x_{12} + x_{5}x_{17} + x_{5}x_{18} + x_{5}x_{19} + x_{5}x_{20} + x_{5}x_{21} + x_{5}x_{24} + x_{5}x_{28} + x_{5}x_{32} + x_{5}x_{33} + x_{5}x_{34} + x_{5}x_{36} + x_{5}x_{38} + x_{5}x_{40} + x_{5}x_{41} + x_{5}x_{43} + x_{5}x_{44} + x_{5}x_{46} + x_{5}x_{47} + x_{5}x_{48} + x_{5}x_{49} + x_{5}x_{50} + x_{5}x_{52} + x_{5}x_{54} + x_{5}x_{55} + x_{5}x_{56} + x_{5}x_{60} + x_{5}x_{62} + x_{5}x_{63} + x_{6}x_{7} + x_{6}x_{13} + x_{6}x_{14} + x_{6}x_{16} + x_{6}x_{17} + x_{6}x_{18} + x_{6}x_{20} + x_{6}x_{21} + x_{6}x_{24} + x_{6}x_{27} + x_{6}x_{29} + x_{6}x_{30} + x_{6}x_{32} + x_{6}x_{35} + x_{6}x_{37} + x_{6}x_{39} + x_{6}x_{42} + x_{6}x_{43} + x_{6}x_{45} + x_{6}x_{46} + x_{6}x_{47} + x_{6}x_{48} + x_{6}x_{50} + x_{6}x_{53} + x_{6}x_{57} + x_{6}x_{58} + x_{6}x_{60} + x_{6}x_{61} + x_{6}x_{62} + x_{6}x_{64} + x_{7}x_{18} + x_{7}x_{19} + x_{7}x_{20} + x_{7}x_{22} + x_{7}x_{26} + x_{7}x_{27} + x_{7}x_{31} + x_{7}x_{32} + x_{7}x_{34} + x_{7}x_{36} + x_{7}x_{37} + x_{7}x_{38} + x_{7}x_{39} + x_{7}x_{40} + x_{7}x_{41} + x_{7}x_{43} + x_{7}x_{45} + x_{7}x_{46} + x_{7}x_{48} + x_{7}x_{50} + x_{7}x_{52} + x_{7}x_{53} + x_{7}x_{54} + x_{7}x_{55} + x_{7}x_{56} + x_{7}x_{61} + x_{7}x_{62} + x_{7}x_{64} + x_{8}x_{10} + x_{8}x_{11} + x_{8}x_{12} + x_{8}x_{15} + x_{8}x_{17} + x_{8}x_{20} + x_{8}x_{21} + x_{8}x_{22} + x_{8}x_{23} + x_{8}x_{24} + x_{8}x_{25} + x_{8}x_{30} + x_{8}x_{33} + x_{8}x_{35} + x_{8}x_{36} + x_{8}x_{37} + x_{8}x_{42} + x_{8}x_{45} + x_{8}x_{48} + x_{8}x_{51} + x_{8}x_{53} + x_{8}x_{55} + x_{8}x_{56} + x_{8}x_{59} + x_{8}x_{64} + x_{9}x_{11} + x_{9}x_{14} + x_{9}x_{19} + x_{9}x_{20} + x_{9}x_{21} + x_{9}x_{22} + x_{9}x_{23} + x_{9}x_{27} + x_{9}x_{28} + x_{9}x_{32} + x_{9}x_{36} + x_{9}x_{37} + x_{9}x_{38} + x_{9}x_{39} + x_{9}x_{40} + x_{9}x_{41} + x_{9}x_{47} + x_{9}x_{48} + x_{9}x_{51} + x_{9}x_{52} + x_{9}x_{54} + x_{9}x_{55} + x_{9}x_{57} + x_{9}x_{58} + x_{9}x_{59} + x_{9}x_{60} + x_{9}x_{62} + x_{10}x_{12} + x_{10}x_{15} + x_{10}x_{18} + x_{10}x_{23} + x_{10}x_{24} + x_{10}x_{27} + x_{10}x_{28} + x_{10}x_{30} + x_{10}x_{31} + x_{10}x_{35} + x_{10}x_{37} + x_{10}x_{39} + x_{10}x_{43} + x_{10}x_{46} + x_{10}x_{49} + x_{10}x_{50} + x_{10}x_{51} + x_{10}x_{53} + x_{10}x_{55} + x_{10}x_{56} + x_{10}x_{57} + x_{10}x_{60} + x_{10}x_{62} + x_{10}x_{64} + x_{11}x_{12} + x_{11}x_{14} + x_{11}x_{15} + x_{11}x_{17} + x_{11}x_{18} + x_{11}x_{19} + x_{11}x_{21} + x_{11}x_{23} + x_{11}x_{24} + x_{11}x_{25} + x_{11}x_{26} + x_{11}x_{27} + x_{11}x_{29} + x_{11}x_{30} + x_{11}x_{33} + x_{11}x_{36} + x_{11}x_{38} + x_{11}x_{39} + x_{11}x_{40} + x_{11}x_{42} + x_{11}x_{43} + x_{11}x_{44} + x_{11}x_{47} + x_{11}x_{48} + x_{11}x_{50} + x_{11}x_{51} + x_{11}x_{53} + x_{11}x_{57} + x_{11}x_{58} + x_{11}x_{59} + x_{12}x_{16} + x_{12}x_{19} + x_{12}x_{20} + x_{12}x_{22} + x_{12}x_{27} + x_{12}x_{29} + x_{12}x_{31} + x_{12}x_{33} + x_{12}x_{35} + x_{12}x_{36} + x_{12}x_{37} + x_{12}x_{39} + x_{12}x_{40} + x_{12}x_{41} + x_{12}x_{42} + x_{12}x_{43} + x_{12}x_{45} + x_{12}x_{47} + x_{12}x_{49} + x_{12}x_{51} + x_{12}x_{53} + x_{12}x_{55} + x_{12}x_{56} + x_{12}x_{57} + x_{12}x_{58} + x_{12}x_{60} + x_{12}x_{63} + x_{12}x_{64} + x_{13}x_{14} + x_{13}x_{15} + x_{13}x_{17} + x_{13}x_{18} + x_{13}x_{21} + x_{13}x_{23} + x_{13}x_{25} + x_{13}x_{26} + x_{13}x_{29} + x_{13}x_{33} + x_{13}x_{35} + x_{13}x_{39} + x_{13}x_{42} + x_{13}x_{43} + x_{13}x_{45} + x_{13}x_{47} + x_{13}x_{51} + x_{13}x_{52} + x_{13}x_{55} + x_{13}x_{57} + x_{13}x_{59} + x_{13}x_{60} + x_{13}x_{61} + x_{13}x_{62} + x_{13}x_{63} + x_{13}x_{64} + x_{14}x_{18} + x_{14}x_{19} + x_{14}x_{20} + x_{14}x_{21} + x_{14}x_{23} + x_{14}x_{27} + x_{14}x_{30} + x_{14}x_{32} + x_{14}x_{33} + x_{14}x_{36} + x_{14}x_{37} + x_{14}x_{46} + x_{14}x_{47} + x_{14}x_{51} + x_{14}x_{52} + x_{14}x_{54} + x_{14}x_{55} + x_{14}x_{56} + x_{14}x_{57} + x_{14}x_{58} + x_{14}x_{61} + x_{14}x_{62} + x_{15}x_{17} + x_{15}x_{20} + x_{15}x_{21} + x_{15}x_{22} + x_{15}x_{25} + x_{15}x_{27} + x_{15}x_{28} + x_{15}x_{30} + x_{15}x_{33} + x_{15}x_{35} + x_{15}x_{36} + x_{15}x_{39} + x_{15}x_{40} + x_{15}x_{41} + x_{15}x_{42} + x_{15}x_{44} + x_{15}x_{45} + x_{15}x_{46} + x_{15}x_{47} + x_{15}x_{48} + x_{15}x_{49} + x_{15}x_{50} + x_{15}x_{51} + x_{15}x_{54} + x_{15}x_{55} + x_{15}x_{56} + x_{15}x_{60} + x_{15}x_{63} + x_{16}x_{17} + x_{16}x_{18} + x_{16}x_{21} + x_{16}x_{22} + x_{16}x_{23} + x_{16}x_{24} + x_{16}x_{26} + x_{16}x_{28} + x_{16}x_{29} + x_{16}x_{31} + x_{16}x_{32} + x_{16}x_{33} + x_{16}x_{38} + x_{16}x_{40} + x_{16}x_{41} + x_{16}x_{44} + x_{16}x_{46} + x_{16}x_{47} + x_{16}x_{48} + x_{16}x_{49} + x_{16}x_{53} + x_{16}x_{54} + x_{16}x_{56} + x_{16}x_{57} + x_{16}x_{59} + x_{16}x_{60} + x_{16}x_{61} + x_{16}x_{62} + x_{16}x_{63} + x_{17}x_{18} + x_{17}x_{20} + x_{17}x_{23} + x_{17}x_{24} + x_{17}x_{25} + x_{17}x_{27} + x_{17}x_{29} + x_{17}x_{32} + x_{17}x_{33} + x_{17}x_{34} + x_{17}x_{35} + x_{17}x_{36} + x_{17}x_{37} + x_{17}x_{38} + x_{17}x_{39} + x_{17}x_{42} + x_{17}x_{43} + x_{17}x_{44} + x_{17}x_{45} + x_{17}x_{46} + x_{17}x_{50} + x_{17}x_{51} + x_{17}x_{53} + x_{17}x_{54} + x_{17}x_{56} + x_{17}x_{57} + x_{17}x_{61} + x_{17}x_{62} + x_{17}x_{63} + x_{18}x_{19} + x_{18}x_{21} + x_{18}x_{24} + x_{18}x_{25} + x_{18}x_{26} + x_{18}x_{28} + x_{18}x_{29} + x_{18}x_{30} + x_{18}x_{31} + x_{18}x_{32} + x_{18}x_{33} + x_{18}x_{34} + x_{18}x_{37} + x_{18}x_{41} + x_{18}x_{43} + x_{18}x_{44} + x_{18}x_{45} + x_{18}x_{51} + x_{18}x_{52} + x_{18}x_{54} + x_{18}x_{55} + x_{18}x_{56} + x_{18}x_{57} + x_{18}x_{63} + x_{18}x_{64} + x_{19}x_{20} + x_{19}x_{21} + x_{19}x_{22} + x_{19}x_{42} + x_{19}x_{51} + x_{19}x_{52} + x_{19}x_{56} + x_{19}x_{58} + x_{19}x_{59} + x_{19}x_{60} + x_{19}x_{61} + x_{19}x_{62} + x_{19}x_{63} + x_{20}x_{22} + x_{20}x_{23} + x_{20}x_{24} + x_{20}x_{26} + x_{20}x_{27} + x_{20}x_{29} + x_{20}x_{36} + x_{20}x_{38} + x_{20}x_{39} + x_{20}x_{43} + x_{20}x_{52} + x_{20}x_{55} + x_{20}x_{56} + x_{20}x_{59} + x_{20}x_{61} + x_{20}x_{63} + x_{21}x_{22} + x_{21}x_{24} + x_{21}x_{26} + x_{21}x_{30} + x_{21}x_{31} + x_{21}x_{35} + x_{21}x_{36} + x_{21}x_{37} + x_{21}x_{38} + x_{21}x_{41} + x_{21}x_{45} + x_{21}x_{46} + x_{21}x_{48} + x_{21}x_{49} + x_{21}x_{51} + x_{21}x_{52} + x_{21}x_{53} + x_{21}x_{56} + x_{21}x_{58} + x_{21}x_{59} + x_{21}x_{60} + x_{21}x_{61} + x_{21}x_{62} + x_{21}x_{64} + x_{22}x_{24} + x_{22}x_{26} + x_{22}x_{27} + x_{22}x_{28} + x_{22}x_{29} + x_{22}x_{33} + x_{22}x_{35} + x_{22}x_{37} + x_{22}x_{38} + x_{22}x_{39} + x_{22}x_{43} + x_{22}x_{44} + x_{22}x_{45} + x_{22}x_{46} + x_{22}x_{48} + x_{22}x_{50} + x_{22}x_{51} + x_{22}x_{54} + x_{22}x_{55} + x_{22}x_{56} + x_{22}x_{60} + x_{22}x_{61} + x_{22}x_{64} + x_{23}x_{24} + x_{23}x_{27} + x_{23}x_{28} + x_{23}x_{31} + x_{23}x_{34} + x_{23}x_{36} + x_{23}x_{40} + x_{23}x_{41} + x_{23}x_{46} + x_{23}x_{48} + x_{23}x_{51} + x_{23}x_{53} + x_{23}x_{54} + x_{23}x_{56} + x_{23}x_{58} + x_{23}x_{59} + x_{23}x_{60} + x_{23}x_{64} + x_{24}x_{25} + x_{24}x_{29} + x_{24}x_{30} + x_{24}x_{32} + x_{24}x_{33} + x_{24}x_{34} + x_{24}x_{35} + x_{24}x_{36} + x_{24}x_{38} + x_{24}x_{43} + x_{24}x_{46} + x_{24}x_{50} + x_{24}x_{52} + x_{24}x_{53} + x_{24}x_{54} + x_{24}x_{55} + x_{24}x_{56} + x_{24}x_{57} + x_{24}x_{58} + x_{24}x_{60} + x_{24}x_{62} + x_{25}x_{26} + x_{25}x_{28} + x_{25}x_{31} + x_{25}x_{36} + x_{25}x_{37} + x_{25}x_{39} + x_{25}x_{40} + x_{25}x_{42} + x_{25}x_{43} + x_{25}x_{48} + x_{25}x_{51} + x_{25}x_{53} + x_{25}x_{57} + x_{25}x_{61} + x_{25}x_{64} + x_{26}x_{28} + x_{26}x_{29} + x_{26}x_{30} + x_{26}x_{32} + x_{26}x_{33} + x_{26}x_{35} + x_{26}x_{36} + x_{26}x_{39} + x_{26}x_{44} + x_{26}x_{45} + x_{26}x_{47} + x_{26}x_{48} + x_{26}x_{51} + x_{26}x_{52} + x_{26}x_{54} + x_{26}x_{56} + x_{26}x_{57} + x_{26}x_{58} + x_{26}x_{64} + x_{27}x_{28} + x_{27}x_{29} + x_{27}x_{33} + x_{27}x_{34} + x_{27}x_{35} + x_{27}x_{36} + x_{27}x_{37} + x_{27}x_{38} + x_{27}x_{41} + x_{27}x_{43} + x_{27}x_{45} + x_{27}x_{46} + x_{27}x_{49} + x_{27}x_{50} + x_{27}x_{51} + x_{27}x_{58} + x_{27}x_{61} + x_{27}x_{63} + x_{28}x_{29} + x_{28}x_{31} + x_{28}x_{32} + x_{28}x_{34} + x_{28}x_{36} + x_{28}x_{38} + x_{28}x_{40} + x_{28}x_{42} + x_{28}x_{43} + x_{28}x_{45} + x_{28}x_{46} + x_{28}x_{47} + x_{28}x_{49} + x_{28}x_{50} + x_{28}x_{54} + x_{28}x_{57} + x_{28}x_{58} + x_{28}x_{59} + x_{28}x_{61} + x_{28}x_{63} + x_{28}x_{64} + x_{29}x_{30} + x_{29}x_{32} + x_{29}x_{33} + x_{29}x_{34} + x_{29}x_{35} + x_{29}x_{36} + x_{29}x_{37} + x_{29}x_{38} + x_{29}x_{40} + x_{29}x_{44} + x_{29}x_{45} + x_{29}x_{46} + x_{29}x_{49} + x_{29}x_{50} + x_{29}x_{54} + x_{29}x_{57} + x_{29}x_{59} + x_{29}x_{60} + x_{29}x_{61} + x_{29}x_{62} + x_{29}x_{63} + x_{29}x_{64} + x_{30}x_{35} + x_{30}x_{38} + x_{30}x_{39} + x_{30}x_{40} + x_{30}x_{42} + x_{30}x_{44} + x_{30}x_{45} + x_{30}x_{48} + x_{30}x_{51} + x_{30}x_{54} + x_{30}x_{55} + x_{30}x_{57} + x_{30}x_{58} + x_{30}x_{59} + x_{30}x_{60} + x_{30}x_{63} + x_{30}x_{64} + x_{31}x_{33} + x_{31}x_{34} + x_{31}x_{35} + x_{31}x_{36} + x_{31}x_{37} + x_{31}x_{39} + x_{31}x_{43} + x_{31}x_{44} + x_{31}x_{45} + x_{31}x_{46} + x_{31}x_{53} + x_{31}x_{56} + x_{31}x_{58} + x_{31}x_{63} + x_{31}x_{64} + x_{32}x_{34} + x_{32}x_{37} + x_{32}x_{40} + x_{32}x_{42} + x_{32}x_{44} + x_{32}x_{45} + x_{32}x_{47} + x_{32}x_{50} + x_{32}x_{51} + x_{32}x_{55} + x_{32}x_{56} + x_{32}x_{61} + x_{32}x_{64} + x_{33}x_{35} + x_{33}x_{36} + x_{33}x_{45} + x_{33}x_{47} + x_{33}x_{50} + x_{33}x_{52} + x_{33}x_{53} + x_{33}x_{58} + x_{33}x_{60} + x_{33}x_{61} + x_{33}x_{62} + x_{33}x_{63} + x_{34}x_{36} + x_{34}x_{38} + x_{34}x_{40} + x_{34}x_{41} + x_{34}x_{45} + x_{34}x_{47} + x_{34}x_{48} + x_{34}x_{49} + x_{34}x_{53} + x_{34}x_{59} + x_{34}x_{60} + x_{34}x_{63} + x_{34}x_{64} + x_{35}x_{37} + x_{35}x_{38} + x_{35}x_{40} + x_{35}x_{41} + x_{35}x_{42} + x_{35}x_{44} + x_{35}x_{45} + x_{35}x_{46} + x_{35}x_{47} + x_{35}x_{49} + x_{35}x_{50} + x_{35}x_{51} + x_{35}x_{53} + x_{35}x_{54} + x_{35}x_{55} + x_{35}x_{56} + x_{35}x_{59} + x_{35}x_{61} + x_{36}x_{38} + x_{36}x_{39} + x_{36}x_{40} + x_{36}x_{42} + x_{36}x_{46} + x_{36}x_{48} + x_{36}x_{54} + x_{36}x_{57} + x_{36}x_{59} + x_{36}x_{61} + x_{36}x_{63} + x_{36}x_{64} + x_{37}x_{39} + x_{37}x_{40} + x_{37}x_{44} + x_{37}x_{45} + x_{37}x_{49} + x_{37}x_{50} + x_{37}x_{52} + x_{37}x_{54} + x_{37}x_{55} + x_{37}x_{58} + x_{37}x_{61} + x_{37}x_{63} + x_{38}x_{39} + x_{38}x_{40} + x_{38}x_{41} + x_{38}x_{42} + x_{38}x_{43} + x_{38}x_{44} + x_{38}x_{48} + x_{38}x_{51} + x_{38}x_{53} + x_{38}x_{55} + x_{38}x_{58} + x_{38}x_{59} + x_{38}x_{60} + x_{38}x_{62} + x_{38}x_{63} + x_{39}x_{41} + x_{39}x_{42} + x_{39}x_{45} + x_{39}x_{46} + x_{39}x_{47} + x_{39}x_{50} + x_{39}x_{54} + x_{39}x_{56} + x_{39}x_{58} + x_{39}x_{60} + x_{39}x_{64} + x_{40}x_{41} + x_{40}x_{43} + x_{40}x_{44} + x_{40}x_{45} + x_{40}x_{48} + x_{40}x_{49} + x_{40}x_{51} + x_{40}x_{52} + x_{40}x_{55} + x_{40}x_{61} + x_{41}x_{42} + x_{41}x_{43} + x_{41}x_{45} + x_{41}x_{48} + x_{41}x_{51} + x_{41}x_{53} + x_{41}x_{54} + x_{41}x_{55} + x_{41}x_{60} + x_{41}x_{61} + x_{41}x_{62} + x_{42}x_{43} + x_{42}x_{45} + x_{42}x_{46} + x_{42}x_{47} + x_{42}x_{49} + x_{42}x_{50} + x_{42}x_{51} + x_{42}x_{53} + x_{42}x_{54} + x_{42}x_{56} + x_{42}x_{58} + x_{42}x_{60} + x_{42}x_{63} + x_{43}x_{44} + x_{43}x_{45} + x_{43}x_{47} + x_{43}x_{48} + x_{43}x_{50} + x_{43}x_{52} + x_{43}x_{53} + x_{43}x_{57} + x_{43}x_{58} + x_{43}x_{59} + x_{43}x_{60} + x_{43}x_{61} + x_{44}x_{46} + x_{44}x_{48} + x_{44}x_{49} + x_{44}x_{50} + x_{44}x_{52} + x_{44}x_{54} + x_{44}x_{64} + x_{45}x_{46} + x_{45}x_{47} + x_{45}x_{48} + x_{45}x_{49} + x_{45}x_{50} + x_{45}x_{53} + x_{45}x_{54} + x_{45}x_{55} + x_{45}x_{58} + x_{45}x_{59} + x_{45}x_{62} + x_{45}x_{64} + x_{46}x_{47} + x_{46}x_{49} + x_{46}x_{50} + x_{46}x_{51} + x_{46}x_{53} + x_{46}x_{55} + x_{46}x_{58} + x_{46}x_{59} + x_{46}x_{62} + x_{46}x_{63} + x_{46}x_{64} + x_{47}x_{48} + x_{47}x_{50} + x_{47}x_{51} + x_{47}x_{57} + x_{47}x_{58} + x_{47}x_{59} + x_{47}x_{60} + x_{47}x_{61} + x_{47}x_{63} + x_{48}x_{50} + x_{48}x_{51} + x_{48}x_{53} + x_{48}x_{55} + x_{48}x_{57} + x_{48}x_{58} + x_{48}x_{59} + x_{48}x_{61} + x_{48}x_{62} + x_{48}x_{63} + x_{49}x_{51} + x_{49}x_{52} + x_{49}x_{54} + x_{49}x_{56} + x_{49}x_{58} + x_{49}x_{59} + x_{49}x_{60} + x_{49}x_{61} + x_{49}x_{63} + x_{49}x_{64} + x_{50}x_{51} + x_{50}x_{52} + x_{50}x_{53} + x_{50}x_{54} + x_{50}x_{56} + x_{50}x_{57} + x_{50}x_{60} + x_{50}x_{62} + x_{50}x_{64} + x_{51}x_{52} + x_{51}x_{53} + x_{51}x_{54} + x_{51}x_{55} + x_{51}x_{56} + x_{51}x_{57} + x_{51}x_{58} + x_{51}x_{59} + x_{51}x_{61} + x_{52}x_{53} + x_{52}x_{55} + x_{52}x_{56} + x_{52}x_{57} + x_{52}x_{60} + x_{53}x_{55} + x_{53}x_{57} + x_{53}x_{61} + x_{53}x_{62} + x_{53}x_{64} + x_{54}x_{55} + x_{54}x_{56} + x_{54}x_{57} + x_{54}x_{62} + x_{54}x_{64} + x_{55}x_{56} + x_{55}x_{57} + x_{55}x_{62} + x_{55}x_{64} + x_{56}x_{58} + x_{56}x_{61} + x_{56}x_{64} + x_{57}x_{59} + x_{57}x_{61} + x_{57}x_{63} + x_{58}x_{61} + x_{58}x_{62} + x_{58}x_{63} + x_{58}x_{64} + x_{59}x_{60} + x_{59}x_{61} + x_{59}x_{63} + x_{60}x_{61} + x_{60}x_{62} + x_{60}x_{63} + x_{61}x_{62} + x_{62}x_{63} + x_{2} + x_{4} + x_{5} + x_{8} + x_{9} + x_{10} + x_{13} + x_{14} + x_{15} + x_{17} + x_{18} + x_{19} + x_{24} + x_{25} + x_{26} + x_{29} + x_{32} + x_{35} + x_{37} + x_{38} + x_{40} + x_{41} + x_{42} + x_{43} + x_{44} + x_{49} + x_{50} + x_{51} + x_{56} + x_{57} + x_{58} + x_{60} + x_{63}$

$y_{25} = x_{1}x_{2} + x_{1}x_{3} + x_{1}x_{4} + x_{1}x_{5} + x_{1}x_{6} + x_{1}x_{7} + x_{1}x_{8} + x_{1}x_{9} + x_{1}x_{11} + x_{1}x_{13} + x_{1}x_{17} + x_{1}x_{18} + x_{1}x_{20} + x_{1}x_{21} + x_{1}x_{22} + x_{1}x_{25} + x_{1}x_{26} + x_{1}x_{28} + x_{1}x_{29} + x_{1}x_{30} + x_{1}x_{32} + x_{1}x_{33} + x_{1}x_{37} + x_{1}x_{38} + x_{1}x_{39} + x_{1}x_{42} + x_{1}x_{45} + x_{1}x_{46} + x_{1}x_{51} + x_{1}x_{52} + x_{1}x_{55} + x_{1}x_{56} + x_{1}x_{57} + x_{1}x_{58} + x_{2}x_{3} + x_{2}x_{4} + x_{2}x_{9} + x_{2}x_{12} + x_{2}x_{13} + x_{2}x_{14} + x_{2}x_{16} + x_{2}x_{18} + x_{2}x_{19} + x_{2}x_{20} + x_{2}x_{21} + x_{2}x_{27} + x_{2}x_{29} + x_{2}x_{33} + x_{2}x_{34} + x_{2}x_{35} + x_{2}x_{36} + x_{2}x_{37} + x_{2}x_{38} + x_{2}x_{39} + x_{2}x_{40} + x_{2}x_{42} + x_{2}x_{45} + x_{2}x_{46} + x_{2}x_{47} + x_{2}x_{48} + x_{2}x_{49} + x_{2}x_{52} + x_{2}x_{53} + x_{2}x_{54} + x_{2}x_{59} + x_{2}x_{61} + x_{2}x_{63} + x_{3}x_{10} + x_{3}x_{12} + x_{3}x_{13} + x_{3}x_{14} + x_{3}x_{15} + x_{3}x_{17} + x_{3}x_{18} + x_{3}x_{19} + x_{3}x_{20} + x_{3}x_{23} + x_{3}x_{24} + x_{3}x_{25} + x_{3}x_{26} + x_{3}x_{27} + x_{3}x_{28} + x_{3}x_{29} + x_{3}x_{32} + x_{3}x_{36} + x_{3}x_{41} + x_{3}x_{42} + x_{3}x_{44} + x_{3}x_{45} + x_{3}x_{47} + x_{3}x_{48} + x_{3}x_{49} + x_{3}x_{51} + x_{3}x_{54} + x_{3}x_{55} + x_{3}x_{57} + x_{3}x_{59} + x_{3}x_{60} + x_{3}x_{62} + x_{3}x_{63} + x_{4}x_{9} + x_{4}x_{13} + x_{4}x_{16} + x_{4}x_{17} + x_{4}x_{19} + x_{4}x_{21} + x_{4}x_{23} + x_{4}x_{25} + x_{4}x_{29} + x_{4}x_{30} + x_{4}x_{31} + x_{4}x_{32} + x_{4}x_{33} + x_{4}x_{34} + x_{4}x_{35} + x_{4}x_{39} + x_{4}x_{43} + x_{4}x_{46} + x_{4}x_{53} + x_{4}x_{56} + x_{4}x_{60} + x_{4}x_{62} + x_{4}x_{63} + x_{4}x_{64} + x_{5}x_{14} + x_{5}x_{15} + x_{5}x_{16} + x_{5}x_{19} + x_{5}x_{22} + x_{5}x_{25} + x_{5}x_{26} + x_{5}x_{28} + x_{5}x_{29} + x_{5}x_{31} + x_{5}x_{34} + x_{5}x_{36} + x_{5}x_{37} + x_{5}x_{38} + x_{5}x_{41} + x_{5}x_{43} + x_{5}x_{44} + x_{5}x_{46} + x_{5}x_{47} + x_{5}x_{48} + x_{5}x_{50} + x_{5}x_{52} + x_{5}x_{53} + x_{5}x_{54} + x_{5}x_{60} + x_{5}x_{61} + x_{5}x_{63} + x_{5}x_{64} + x_{6}x_{8} + x_{6}x_{9} + x_{6}x_{11} + x_{6}x_{13} + x_{6}x_{14} + x_{6}x_{17} + x_{6}x_{18} + x_{6}x_{19} + x_{6}x_{22} + x_{6}x_{24} + x_{6}x_{27} + x_{6}x_{29} + x_{6}x_{31} + x_{6}x_{33} + x_{6}x_{34} + x_{6}x_{36} + x_{6}x_{39} + x_{6}x_{40} + x_{6}x_{41} + x_{6}x_{44} + x_{6}x_{47} + x_{6}x_{50} + x_{6}x_{51} + x_{6}x_{52} + x_{6}x_{57} + x_{6}x_{58} + x_{6}x_{59} + x_{6}x_{61} + x_{6}x_{62} + x_{6}x_{64} + x_{7}x_{9} + x_{7}x_{10} + x_{7}x_{11} + x_{7}x_{12} + x_{7}x_{14} + x_{7}x_{16} + x_{7}x_{17} + x_{7}x_{19} + x_{7}x_{20} + x_{7}x_{23} + x_{7}x_{25} + x_{7}x_{26} + x_{7}x_{27} + x_{7}x_{32} + x_{7}x_{34} + x_{7}x_{37} + x_{7}x_{39} + x_{7}x_{40} + x_{7}x_{44} + x_{7}x_{45} + x_{7}x_{47} + x_{7}x_{48} + x_{7}x_{53} + x_{7}x_{56} + x_{7}x_{58} + x_{7}x_{59} + x_{7}x_{60} + x_{7}x_{62} + x_{7}x_{63} + x_{7}x_{64} + x_{8}x_{9} + x_{8}x_{10} + x_{8}x_{12} + x_{8}x_{15} + x_{8}x_{17} + x_{8}x_{19} + x_{8}x_{21} + x_{8}x_{23} + x_{8}x_{24} + x_{8}x_{25} + x_{8}x_{27} + x_{8}x_{29} + x_{8}x_{31} + x_{8}x_{32} + x_{8}x_{34} + x_{8}x_{35} + x_{8}x_{37} + x_{8}x_{38} + x_{8}x_{39} + x_{8}x_{41} + x_{8}x_{44} + x_{8}x_{46} + x_{8}x_{50} + x_{8}x_{51} + x_{8}x_{52} + x_{8}x_{53} + x_{8}x_{54} + x_{8}x_{56} + x_{8}x_{58} + x_{8}x_{59} + x_{8}x_{60} + x_{9}x_{11} + x_{9}x_{12} + x_{9}x_{13} + x_{9}x_{15} + x_{9}x_{18} + x_{9}x_{20} + x_{9}x_{21} + x_{9}x_{23} + x_{9}x_{24} + x_{9}x_{25} + x_{9}x_{27} + x_{9}x_{28} + x_{9}x_{30} + x_{9}x_{31} + x_{9}x_{32} + x_{9}x_{33} + x_{9}x_{34} + x_{9}x_{35} + x_{9}x_{36} + x_{9}x_{39} + x_{9}x_{46} + x_{9}x_{47} + x_{9}x_{50} + x_{9}x_{51} + x_{9}x_{55} + x_{9}x_{56} + x_{9}x_{57} + x_{9}x_{61} + x_{9}x_{62} + x_{9}x_{63} + x_{9}x_{64} + x_{10}x_{11} + x_{10}x_{13} + x_{10}x_{14} + x_{10}x_{15} + x_{10}x_{20} + x_{10}x_{22} + x_{10}x_{23} + x_{10}x_{24} + x_{10}x_{26} + x_{10}x_{27} + x_{10}x_{30} + x_{10}x_{31} + x_{10}x_{33} + x_{10}x_{34} + x_{10}x_{35} + x_{10}x_{37} + x_{10}x_{39} + x_{10}x_{40} + x_{10}x_{41} + x_{10}x_{42} + x_{10}x_{43} + x_{10}x_{46} + x_{10}x_{47} + x_{10}x_{48} + x_{10}x_{50} + x_{10}x_{57} + x_{10}x_{58} + x_{10}x_{59} + x_{10}x_{60} + x_{10}x_{61} + x_{10}x_{63} + x_{10}x_{64} + x_{11}x_{12} + x_{11}x_{17} + x_{11}x_{18} + x_{11}x_{19} + x_{11}x_{20} + x_{11}x_{26} + x_{11}x_{27} + x_{11}x_{28} + x_{11}x_{29} + x_{11}x_{31} + x_{11}x_{34} + x_{11}x_{35} + x_{11}x_{37} + x_{11}x_{39} + x_{11}x_{41} + x_{11}x_{42} + x_{11}x_{45} + x_{11}x_{46} + x_{11}x_{48} + x_{11}x_{49} + x_{11}x_{54} + x_{11}x_{56} + x_{11}x_{57} + x_{11}x_{59} + x_{11}x_{62} + x_{11}x_{63} + x_{12}x_{14} + x_{12}x_{15} + x_{12}x_{16} + x_{12}x_{17} + x_{12}x_{18} + x_{12}x_{20} + x_{12}x_{22} + x_{12}x_{23} + x_{12}x_{26} + x_{12}x_{27} + x_{12}x_{29} + x_{12}x_{30} + x_{12}x_{34} + x_{12}x_{35} + x_{12}x_{36} + x_{12}x_{38} + x_{12}x_{40} + x_{12}x_{45} + x_{12}x_{46} + x_{12}x_{47} + x_{12}x_{48} + x_{12}x_{49} + x_{12}x_{50} + x_{12}x_{51} + x_{12}x_{54} + x_{12}x_{59} + x_{13}x_{14} + x_{13}x_{17} + x_{13}x_{20} + x_{13}x_{23} + x_{13}x_{25} + x_{13}x_{26} + x_{13}x_{27} + x_{13}x_{28} + x_{13}x_{29} + x_{13}x_{30} + x_{13}x_{31} + x_{13}x_{33} + x_{13}x_{35} + x_{13}x_{36} + x_{13}x_{37} + x_{13}x_{38} + x_{13}x_{39} + x_{13}x_{40} + x_{13}x_{43} + x_{13}x_{44} + x_{13}x_{46} + x_{13}x_{52} + x_{13}x_{56} + x_{13}x_{63} + x_{13}x_{64} + x_{14}x_{15} + x_{14}x_{18} + x_{14}x_{20} + x_{14}x_{21} + x_{14}x_{22} + x_{14}x_{23} + x_{14}x_{24} + x_{14}x_{25} + x_{14}x_{27} + x_{14}x_{31} + x_{14}x_{32} + x_{14}x_{34} + x_{14}x_{35} + x_{14}x_{37} + x_{14}x_{38} + x_{14}x_{39} + x_{14}x_{49} + x_{14}x_{50} + x_{14}x_{52} + x_{14}x_{53} + x_{14}x_{58} + x_{14}x_{59} + x_{14}x_{61} + x_{14}x_{62} + x_{14}x_{64} + x_{15}x_{16} + x_{15}x_{18} + x_{15}x_{19} + x_{15}x_{21} + x_{15}x_{23} + x_{15}x_{24} + x_{15}x_{25} + x_{15}x_{26} + x_{15}x_{29} + x_{15}x_{32} + x_{15}x_{33} + x_{15}x_{36} + x_{15}x_{37} + x_{15}x_{41} + x_{15}x_{42} + x_{15}x_{44} + x_{15}x_{45} + x_{15}x_{46} + x_{15}x_{52} + x_{15}x_{55} + x_{15}x_{56} + x_{15}x_{58} + x_{15}x_{59} + x_{15}x_{62} + x_{15}x_{63} + x_{15}x_{64} + x_{16}x_{18} + x_{16}x_{19} + x_{16}x_{25} + x_{16}x_{28} + x_{16}x_{30} + x_{16}x_{31} + x_{16}x_{35} + x_{16}x_{39} + x_{16}x_{40} + x_{16}x_{42} + x_{16}x_{43} + x_{16}x_{47} + x_{16}x_{48} + x_{16}x_{50} + x_{16}x_{51} + x_{16}x_{53} + x_{16}x_{54} + x_{16}x_{57} + x_{16}x_{58} + x_{16}x_{59} + x_{16}x_{60} + x_{16}x_{61} + x_{16}x_{62} + x_{17}x_{23} + x_{17}x_{26} + x_{17}x_{29} + x_{17}x_{33} + x_{17}x_{34} + x_{17}x_{35} + x_{17}x_{36} + x_{17}x_{37} + x_{17}x_{38} + x_{17}x_{39} + x_{17}x_{40} + x_{17}x_{41} + x_{17}x_{42} + x_{17}x_{45} + x_{17}x_{48} + x_{17}x_{49} + x_{17}x_{50} + x_{17}x_{51} + x_{17}x_{52} + x_{17}x_{55} + x_{17}x_{58} + x_{17}x_{60} + x_{17}x_{61} + x_{17}x_{62} + x_{17}x_{63} + x_{18}x_{20} + x_{18}x_{21} + x_{18}x_{22} + x_{18}x_{27} + x_{18}x_{29} + x_{18}x_{30} + x_{18}x_{31} + x_{18}x_{34} + x_{18}x_{36} + x_{18}x_{39} + x_{18}x_{43} + x_{18}x_{44} + x_{18}x_{46} + x_{18}x_{47} + x_{18}x_{48} + x_{18}x_{49} + x_{18}x_{50} + x_{18}x_{51} + x_{18}x_{54} + x_{18}x_{56} + x_{18}x_{57} + x_{18}x_{59} + x_{18}x_{62} + x_{18}x_{63} + x_{18}x_{64} + x_{19}x_{21} + x_{19}x_{22} + x_{19}x_{23} + x_{19}x_{27} + x_{19}x_{28} + x_{19}x_{30} + x_{19}x_{36} + x_{19}x_{37} + x_{19}x_{39} + x_{19}x_{41} + x_{19}x_{43} + x_{19}x_{45} + x_{19}x_{47} + x_{19}x_{49} + x_{19}x_{51} + x_{19}x_{54} + x_{19}x_{56} + x_{19}x_{58} + x_{19}x_{60} + x_{19}x_{61} + x_{19}x_{63} + x_{19}x_{64} + x_{20}x_{22} + x_{20}x_{23} + x_{20}x_{24} + x_{20}x_{32} + x_{20}x_{33} + x_{20}x_{37} + x_{20}x_{39} + x_{20}x_{41} + x_{20}x_{43} + x_{20}x_{45} + x_{20}x_{48} + x_{20}x_{51} + x_{20}x_{52} + x_{20}x_{54} + x_{20}x_{57} + x_{20}x_{61} + x_{21}x_{22} + x_{21}x_{25} + x_{21}x_{29} + x_{21}x_{30} + x_{21}x_{31} + x_{21}x_{32} + x_{21}x_{35} + x_{21}x_{36} + x_{21}x_{40} + x_{21}x_{41} + x_{21}x_{42} + x_{21}x_{43} + x_{21}x_{47} + x_{21}x_{48} + x_{21}x_{50} + x_{21}x_{53} + x_{21}x_{57} + x_{21}x_{59} + x_{21}x_{61} + x_{21}x_{64} + x_{22}x_{23} + x_{22}x_{24} + x_{22}x_{26} + x_{22}x_{27} + x_{22}x_{32} + x_{22}x_{33} + x_{22}x_{37} + x_{22}x_{38} + x_{22}x_{40} + x_{22}x_{41} + x_{22}x_{42} + x_{22}x_{46} + x_{22}x_{47} + x_{22}x_{48} + x_{22}x_{49} + x_{22}x_{51} + x_{22}x_{53} + x_{22}x_{55} + x_{22}x_{57} + x_{22}x_{60} + x_{22}x_{63} + x_{22}x_{64} + x_{23}x_{25} + x_{23}x_{29} + x_{23}x_{30} + x_{23}x_{31} + x_{23}x_{33} + x_{23}x_{36} + x_{23}x_{37} + x_{23}x_{38} + x_{23}x_{41} + x_{23}x_{42} + x_{23}x_{43} + x_{23}x_{44} + x_{23}x_{47} + x_{23}x_{49} + x_{23}x_{52} + x_{23}x_{53} + x_{23}x_{55} + x_{23}x_{56} + x_{23}x_{57} + x_{23}x_{60} + x_{23}x_{61} + x_{23}x_{62} + x_{23}x_{63} + x_{24}x_{25} + x_{24}x_{28} + x_{24}x_{31} + x_{24}x_{33} + x_{24}x_{34} + x_{24}x_{35} + x_{24}x_{39} + x_{24}x_{43} + x_{24}x_{45} + x_{24}x_{46} + x_{24}x_{50} + x_{24}x_{51} + x_{24}x_{56} + x_{24}x_{64} + x_{25}x_{27} + x_{25}x_{29} + x_{25}x_{31} + x_{25}x_{33} + x_{25}x_{34} + x_{25}x_{35} + x_{25}x_{36} + x_{25}x_{37} + x_{25}x_{38} + x_{25}x_{39} + x_{25}x_{40} + x_{25}x_{42} + x_{25}x_{48} + x_{25}x_{50} + x_{25}x_{52} + x_{25}x_{53} + x_{25}x_{54} + x_{25}x_{55} + x_{25}x_{58} + x_{25}x_{59} + x_{25}x_{61} + x_{25}x_{63} + x_{26}x_{28} + x_{26}x_{29} + x_{26}x_{31} + x_{26}x_{32} + x_{26}x_{34} + x_{26}x_{35} + x_{26}x_{39} + x_{26}x_{40} + x_{26}x_{44} + x_{26}x_{47} + x_{26}x_{48} + x_{26}x_{52} + x_{26}x_{53} + x_{26}x_{54} + x_{26}x_{55} + x_{26}x_{56} + x_{26}x_{57} + x_{26}x_{59} + x_{27}x_{28} + x_{27}x_{29} + x_{27}x_{30} + x_{27}x_{34} + x_{27}x_{40} + x_{27}x_{43} + x_{27}x_{45} + x_{27}x_{46} + x_{27}x_{49} + x_{27}x_{51} + x_{27}x_{52} + x_{27}x_{55} + x_{27}x_{56} + x_{27}x_{59} + x_{27}x_{62} + x_{27}x_{63} + x_{28}x_{36} + x_{28}x_{37} + x_{28}x_{39} + x_{28}x_{40} + x_{28}x_{41} + x_{28}x_{42} + x_{28}x_{43} + x_{28}x_{44} + x_{28}x_{48} + x_{28}x_{49} + x_{28}x_{50} + x_{28}x_{51} + x_{28}x_{52} + x_{28}x_{54} + x_{28}x_{58} + x_{28}x_{59} + x_{28}x_{60} + x_{28}x_{62} + x_{28}x_{63} + x_{28}x_{64} + x_{29}x_{30} + x_{29}x_{31} + x_{29}x_{32} + x_{29}x_{33} + x_{29}x_{36} + x_{29}x_{37} + x_{29}x_{38} + x_{29}x_{39} + x_{29}x_{41} + x_{29}x_{42} + x_{29}x_{43} + x_{29}x_{44} + x_{29}x_{45} + x_{29}x_{48} + x_{29}x_{51} + x_{29}x_{52} + x_{29}x_{53} + x_{29}x_{54} + x_{29}x_{55} + x_{29}x_{56} + x_{29}x_{60} + x_{29}x_{62} + x_{30}x_{34} + x_{30}x_{37} + x_{30}x_{38} + x_{30}x_{39} + x_{30}x_{41} + x_{30}x_{42} + x_{30}x_{43} + x_{30}x_{45} + x_{30}x_{49} + x_{30}x_{50} + x_{30}x_{51} + x_{30}x_{56} + x_{30}x_{57} + x_{30}x_{58} + x_{30}x_{61} + x_{31}x_{32} + x_{31}x_{33} + x_{31}x_{35} + x_{31}x_{36} + x_{31}x_{37} + x_{31}x_{39} + x_{31}x_{47} + x_{31}x_{50} + x_{31}x_{51} + x_{31}x_{52} + x_{31}x_{55} + x_{31}x_{58} + x_{31}x_{62} + x_{31}x_{63} + x_{32}x_{34} + x_{32}x_{36} + x_{32}x_{37} + x_{32}x_{38} + x_{32}x_{39} + x_{32}x_{41} + x_{32}x_{43} + x_{32}x_{47} + x_{32}x_{48} + x_{32}x_{50} + x_{32}x_{52} + x_{32}x_{57} + x_{32}x_{58} + x_{32}x_{59} + x_{32}x_{60} + x_{32}x_{63} + x_{33}x_{35} + x_{33}x_{36} + x_{33}x_{38} + x_{33}x_{40} + x_{33}x_{41} + x_{33}x_{43} + x_{33}x_{44} + x_{33}x_{49} + x_{33}x_{50} + x_{33}x_{54} + x_{33}x_{55} + x_{33}x_{60} + x_{33}x_{64} + x_{34}x_{36} + x_{34}x_{37} + x_{34}x_{43} + x_{34}x_{46} + x_{34}x_{50} + x_{34}x_{53} + x_{34}x_{54} + x_{34}x_{55} + x_{34}x_{56} + x_{34}x_{58} + x_{35}x_{37} + x_{35}x_{38} + x_{35}x_{42} + x_{35}x_{45} + x_{35}x_{47} + x_{35}x_{48} + x_{35}x_{49} + x_{35}x_{50} + x_{35}x_{51} + x_{35}x_{52} + x_{35}x_{54} + x_{35}x_{55} + x_{35}x_{56} + x_{35}x_{57} + x_{35}x_{58} + x_{35}x_{59} + x_{35}x_{60} + x_{35}x_{63} + x_{35}x_{64} + x_{36}x_{37} + x_{36}x_{38} + x_{36}x_{40} + x_{36}x_{42} + x_{36}x_{45} + x_{36}x_{46} + x_{36}x_{48} + x_{36}x_{52} + x_{36}x_{53} + x_{36}x_{56} + x_{36}x_{57} + x_{36}x_{59} + x_{36}x_{62} + x_{36}x_{64} + x_{37}x_{38} + x_{37}x_{40} + x_{37}x_{41} + x_{37}x_{42} + x_{37}x_{43} + x_{37}x_{44} + x_{37}x_{50} + x_{37}x_{54} + x_{37}x_{56} + x_{37}x_{57} + x_{37}x_{59} + x_{37}x_{60} + x_{37}x_{61} + x_{37}x_{63} + x_{37}x_{64} + x_{38}x_{41} + x_{38}x_{42} + x_{38}x_{46} + x_{38}x_{47} + x_{38}x_{48} + x_{38}x_{49} + x_{38}x_{53} + x_{38}x_{54} + x_{38}x_{55} + x_{38}x_{58} + x_{38}x_{61} + x_{38}x_{63} + x_{38}x_{64} + x_{39}x_{40} + x_{39}x_{42} + x_{39}x_{43} + x_{39}x_{45} + x_{39}x_{46} + x_{39}x_{49} + x_{39}x_{51} + x_{39}x_{52} + x_{39}x_{55} + x_{39}x_{57} + x_{39}x_{58} + x_{39}x_{60} + x_{39}x_{61} + x_{39}x_{62} + x_{39}x_{63} + x_{39}x_{64} + x_{40}x_{41} + x_{40}x_{44} + x_{40}x_{47} + x_{40}x_{50} + x_{40}x_{51} + x_{40}x_{55} + x_{40}x_{57} + x_{40}x_{58} + x_{40}x_{60} + x_{40}x_{62} + x_{40}x_{63} + x_{41}x_{42} + x_{41}x_{44} + x_{41}x_{45} + x_{41}x_{47} + x_{41}x_{48} + x_{41}x_{50} + x_{41}x_{53} + x_{41}x_{55} + x_{41}x_{59} + x_{41}x_{61} + x_{41}x_{62} + x_{41}x_{64} + x_{42}x_{43} + x_{42}x_{44} + x_{42}x_{45} + x_{42}x_{47} + x_{42}x_{50} + x_{42}x_{53} + x_{42}x_{55} + x_{42}x_{58} + x_{42}x_{62} + x_{43}x_{44} + x_{43}x_{45} + x_{43}x_{47} + x_{43}x_{48} + x_{43}x_{50} + x_{43}x_{53} + x_{43}x_{58} + x_{43}x_{59} + x_{43}x_{62} + x_{43}x_{63} + x_{44}x_{46} + x_{44}x_{47} + x_{44}x_{48} + x_{44}x_{49} + x_{44}x_{51} + x_{44}x_{53} + x_{44}x_{54} + x_{44}x_{55} + x_{44}x_{57} + x_{44}x_{59} + x_{44}x_{60} + x_{44}x_{63} + x_{45}x_{51} + x_{45}x_{52} + x_{45}x_{53} + x_{45}x_{56} + x_{45}x_{57} + x_{45}x_{58} + x_{45}x_{63} + x_{46}x_{49} + x_{46}x_{51} + x_{46}x_{52} + x_{46}x_{53} + x_{46}x_{59} + x_{46}x_{60} + x_{46}x_{62} + x_{46}x_{63} + x_{47}x_{48} + x_{47}x_{49} + x_{47}x_{50} + x_{47}x_{53} + x_{47}x_{54} + x_{47}x_{57} + x_{47}x_{59} + x_{47}x_{60} + x_{47}x_{64} + x_{48}x_{49} + x_{48}x_{50} + x_{48}x_{52} + x_{48}x_{54} + x_{48}x_{56} + x_{48}x_{57} + x_{48}x_{58} + x_{48}x_{59} + x_{48}x_{62} + x_{48}x_{64} + x_{49}x_{53} + x_{49}x_{56} + x_{49}x_{57} + x_{49}x_{59} + x_{49}x_{60} + x_{49}x_{64} + x_{50}x_{51} + x_{50}x_{52} + x_{50}x_{54} + x_{50}x_{57} + x_{50}x_{58} + x_{50}x_{59} + x_{50}x_{63} + x_{50}x_{64} + x_{51}x_{52} + x_{51}x_{57} + x_{51}x_{59} + x_{51}x_{60} + x_{51}x_{62} + x_{52}x_{53} + x_{52}x_{56} + x_{52}x_{57} + x_{52}x_{59} + x_{52}x_{60} + x_{52}x_{62} + x_{52}x_{64} + x_{53}x_{56} + x_{53}x_{59} + x_{53}x_{60} + x_{53}x_{61} + x_{53}x_{64} + x_{54}x_{55} + x_{54}x_{56} + x_{54}x_{57} + x_{55}x_{56} + x_{55}x_{58} + x_{55}x_{63} + x_{55}x_{64} + x_{56}x_{57} + x_{56}x_{58} + x_{56}x_{62} + x_{57}x_{59} + x_{57}x_{63} + x_{58}x_{59} + x_{58}x_{60} + x_{58}x_{62} + x_{58}x_{63} + x_{59}x_{60} + x_{59}x_{63} + x_{59}x_{64} + x_{60}x_{61} + x_{60}x_{62} + x_{60}x_{63} + x_{60}x_{64} + x_{61}x_{63} + x_{62}x_{63} + x_{2} + x_{3} + x_{5} + x_{7} + x_{10} + x_{11} + x_{12} + x_{14} + x_{16} + x_{19} + x_{20} + x_{21} + x_{23} + x_{24} + x_{25} + x_{26} + x_{28} + x_{30} + x_{31} + x_{34} + x_{37} + x_{38} + x_{39} + x_{40} + x_{45} + x_{46} + x_{48} + x_{50} + x_{51} + x_{54} + x_{59} + x_{61} + x_{63}$

$y_{26} = x_{1}x_{2} + x_{1}x_{4} + x_{1}x_{5} + x_{1}x_{6} + x_{1}x_{10} + x_{1}x_{12} + x_{1}x_{13} + x_{1}x_{15} + x_{1}x_{16} + x_{1}x_{17} + x_{1}x_{19} + x_{1}x_{21} + x_{1}x_{23} + x_{1}x_{24} + x_{1}x_{26} + x_{1}x_{29} + x_{1}x_{31} + x_{1}x_{35} + x_{1}x_{36} + x_{1}x_{38} + x_{1}x_{39} + x_{1}x_{41} + x_{1}x_{42} + x_{1}x_{47} + x_{1}x_{48} + x_{1}x_{49} + x_{1}x_{50} + x_{1}x_{51} + x_{1}x_{53} + x_{1}x_{62} + x_{1}x_{63} + x_{2}x_{5} + x_{2}x_{6} + x_{2}x_{7} + x_{2}x_{11} + x_{2}x_{14} + x_{2}x_{18} + x_{2}x_{19} + x_{2}x_{20} + x_{2}x_{21} + x_{2}x_{26} + x_{2}x_{27} + x_{2}x_{28} + x_{2}x_{29} + x_{2}x_{32} + x_{2}x_{34} + x_{2}x_{37} + x_{2}x_{39} + x_{2}x_{40} + x_{2}x_{41} + x_{2}x_{42} + x_{2}x_{43} + x_{2}x_{44} + x_{2}x_{48} + x_{2}x_{50} + x_{2}x_{52} + x_{2}x_{53} + x_{2}x_{54} + x_{2}x_{56} + x_{2}x_{57} + x_{2}x_{58} + x_{2}x_{59} + x_{2}x_{60} + x_{2}x_{61} + x_{3}x_{4} + x_{3}x_{7} + x_{3}x_{14} + x_{3}x_{15} + x_{3}x_{17} + x_{3}x_{18} + x_{3}x_{19} + x_{3}x_{20} + x_{3}x_{22} + x_{3}x_{25} + x_{3}x_{26} + x_{3}x_{27} + x_{3}x_{28} + x_{3}x_{30} + x_{3}x_{33} + x_{3}x_{35} + x_{3}x_{37} + x_{3}x_{38} + x_{3}x_{40} + x_{3}x_{42} + x_{3}x_{43} + x_{3}x_{47} + x_{3}x_{49} + x_{3}x_{53} + x_{3}x_{55} + x_{3}x_{56} + x_{3}x_{57} + x_{3}x_{61} + x_{3}x_{62} + x_{4}x_{6} + x_{4}x_{8} + x_{4}x_{9} + x_{4}x_{11} + x_{4}x_{12} + x_{4}x_{15} + x_{4}x_{16} + x_{4}x_{17} + x_{4}x_{18} + x_{4}x_{19} + x_{4}x_{22} + x_{4}x_{26} + x_{4}x_{27} + x_{4}x_{29} + x_{4}x_{30} + x_{4}x_{32} + x_{4}x_{33} + x_{4}x_{35} + x_{4}x_{38} + x_{4}x_{39} + x_{4}x_{40} + x_{4}x_{43} + x_{4}x_{46} + x_{4}x_{47} + x_{4}x_{48} + x_{4}x_{52} + x_{4}x_{53} + x_{4}x_{55} + x_{4}x_{56} + x_{4}x_{57} + x_{4}x_{58} + x_{4}x_{59} + x_{4}x_{61} + x_{4}x_{64} + x_{5}x_{7} + x_{5}x_{10} + x_{5}x_{11} + x_{5}x_{12} + x_{5}x_{14} + x_{5}x_{17} + x_{5}x_{22} + x_{5}x_{23} + x_{5}x_{24} + x_{5}x_{25} + x_{5}x_{27} + x_{5}x_{37} + x_{5}x_{38} + x_{5}x_{39} + x_{5}x_{41} + x_{5}x_{42} + x_{5}x_{43} + x_{5}x_{50} + x_{5}x_{51} + x_{5}x_{54} + x_{5}x_{55} + x_{5}x_{57} + x_{5}x_{58} + x_{5}x_{59} + x_{5}x_{60} + x_{5}x_{61} + x_{5}x_{62} + x_{6}x_{8} + x_{6}x_{10} + x_{6}x_{13} + x_{6}x_{14} + x_{6}x_{18} + x_{6}x_{23} + x_{6}x_{24} + x_{6}x_{25} + x_{6}x_{26} + x_{6}x_{28} + x_{6}x_{30} + x_{6}x_{31} + x_{6}x_{32} + x_{6}x_{35} + x_{6}x_{38} + x_{6}x_{39} + x_{6}x_{44} + x_{6}x_{45} + x_{6}x_{46} + x_{6}x_{47} + x_{6}x_{48} + x_{6}x_{52} + x_{6}x_{53} + x_{6}x_{54} + x_{6}x_{55} + x_{6}x_{58} + x_{6}x_{59} + x_{6}x_{60} + x_{6}x_{61} + x_{6}x_{64} + x_{7}x_{9} + x_{7}x_{10} + x_{7}x_{13} + x_{7}x_{14} + x_{7}x_{15} + x_{7}x_{17} + x_{7}x_{18} + x_{7}x_{22} + x_{7}x_{24} + x_{7}x_{25} + x_{7}x_{30} + x_{7}x_{31} + x_{7}x_{35} + x_{7}x_{37} + x_{7}x_{42} + x_{7}x_{44} + x_{7}x_{49} + x_{7}x_{54} + x_{7}x_{56} + x_{7}x_{57} + x_{7}x_{58} + x_{7}x_{62} + x_{8}x_{9} + x_{8}x_{10} + x_{8}x_{12} + x_{8}x_{15} + x_{8}x_{16} + x_{8}x_{20} + x_{8}x_{21} + x_{8}x_{24} + x_{8}x_{29} + x_{8}x_{31} + x_{8}x_{32} + x_{8}x_{34} + x_{8}x_{35} + x_{8}x_{36} + x_{8}x_{38} + x_{8}x_{39} + x_{8}x_{41} + x_{8}x_{44} + x_{8}x_{45} + x_{8}x_{47} + x_{8}x_{48} + x_{8}x_{50} + x_{8}x_{51} + x_{8}x_{52} + x_{8}x_{53} + x_{8}x_{54} + x_{8}x_{55} + x_{8}x_{60} + x_{8}x_{64} + x_{9}x_{10} + x_{9}x_{16} + x_{9}x_{18} + x_{9}x_{21} + x_{9}x_{22} + x_{9}x_{24} + x_{9}x_{27} + x_{9}x_{29} + x_{9}x_{31} + x_{9}x_{32} + x_{9}x_{34} + x_{9}x_{35} + x_{9}x_{38} + x_{9}x_{40} + x_{9}x_{44} + x_{9}x_{45} + x_{9}x_{47} + x_{9}x_{48} + x_{9}x_{49} + x_{9}x_{52} + x_{9}x_{54} + x_{9}x_{58} + x_{9}x_{60} + x_{9}x_{62} + x_{9}x_{64} + x_{10}x_{11} + x_{10}x_{13} + x_{10}x_{16} + x_{10}x_{17} + x_{10}x_{18} + x_{10}x_{19} + x_{10}x_{21} + x_{10}x_{23} + x_{10}x_{24} + x_{10}x_{25} + x_{10}x_{27} + x_{10}x_{30} + x_{10}x_{35} + x_{10}x_{37} + x_{10}x_{38} + x_{10}x_{39} + x_{10}x_{44} + x_{10}x_{45} + x_{10}x_{46} + x_{10}x_{48} + x_{10}x_{49} + x_{10}x_{50} + x_{10}x_{56} + x_{10}x_{58} + x_{10}x_{59} + x_{10}x_{62} + x_{10}x_{64} + x_{11}x_{12} + x_{11}x_{14} + x_{11}x_{17} + x_{11}x_{20} + x_{11}x_{21} + x_{11}x_{26} + x_{11}x_{28} + x_{11}x_{30} + x_{11}x_{31} + x_{11}x_{33} + x_{11}x_{34} + x_{11}x_{36} + x_{11}x_{37} + x_{11}x_{38} + x_{11}x_{40} + x_{11}x_{42} + x_{11}x_{43} + x_{11}x_{44} + x_{11}x_{46} + x_{11}x_{49} + x_{11}x_{51} + x_{11}x_{53} + x_{11}x_{55} + x_{11}x_{56} + x_{11}x_{57} + x_{11}x_{59} + x_{11}x_{60} + x_{11}x_{63} + x_{11}x_{64} + x_{12}x_{13} + x_{12}x_{14} + x_{12}x_{15} + x_{12}x_{16} + x_{12}x_{18} + x_{12}x_{19} + x_{12}x_{21} + x_{12}x_{27} + x_{12}x_{31} + x_{12}x_{34} + x_{12}x_{35} + x_{12}x_{36} + x_{12}x_{39} + x_{12}x_{40} + x_{12}x_{43} + x_{12}x_{44} + x_{12}x_{45} + x_{12}x_{46} + x_{12}x_{47} + x_{12}x_{50} + x_{12}x_{52} + x_{12}x_{55} + x_{12}x_{56} + x_{12}x_{57} + x_{12}x_{58} + x_{12}x_{60} + x_{13}x_{18} + x_{13}x_{19} + x_{13}x_{21} + x_{13}x_{22} + x_{13}x_{23} + x_{13}x_{29} + x_{13}x_{30} + x_{13}x_{34} + x_{13}x_{36} + x_{13}x_{37} + x_{13}x_{39} + x_{13}x_{44} + x_{13}x_{45} + x_{13}x_{46} + x_{13}x_{49} + x_{13}x_{51} + x_{13}x_{52} + x_{13}x_{53} + x_{13}x_{56} + x_{13}x_{57} + x_{13}x_{58} + x_{13}x_{60} + x_{13}x_{61} + x_{13}x_{62} + x_{13}x_{63} + x_{14}x_{15} + x_{14}x_{20} + x_{14}x_{26} + x_{14}x_{27} + x_{14}x_{30} + x_{14}x_{31} + x_{14}x_{34} + x_{14}x_{36} + x_{14}x_{38} + x_{14}x_{39} + x_{14}x_{41} + x_{14}x_{43} + x_{14}x_{44} + x_{14}x_{45} + x_{14}x_{48} + x_{14}x_{49} + x_{14}x_{50} + x_{14}x_{51} + x_{14}x_{52} + x_{14}x_{53} + x_{14}x_{55} + x_{14}x_{57} + x_{14}x_{60} + x_{14}x_{61} + x_{14}x_{62} + x_{14}x_{63} + x_{15}x_{16} + x_{15}x_{17} + x_{15}x_{18} + x_{15}x_{19} + x_{15}x_{22} + x_{15}x_{23} + x_{15}x_{26} + x_{15}x_{36} + x_{15}x_{38} + x_{15}x_{41} + x_{15}x_{44} + x_{15}x_{46} + x_{15}x_{47} + x_{15}x_{50} + x_{15}x_{51} + x_{15}x_{52} + x_{15}x_{53} + x_{15}x_{54} + x_{15}x_{56} + x_{15}x_{58} + x_{15}x_{60} + x_{15}x_{63} + x_{15}x_{64} + x_{16}x_{17} + x_{16}x_{23} + x_{16}x_{26} + x_{16}x_{27} + x_{16}x_{29} + x_{16}x_{30} + x_{16}x_{31} + x_{16}x_{33} + x_{16}x_{34} + x_{16}x_{35} + x_{16}x_{36} + x_{16}x_{38} + x_{16}x_{41} + x_{16}x_{43} + x_{16}x_{46} + x_{16}x_{47} + x_{16}x_{49} + x_{16}x_{50} + x_{16}x_{51} + x_{16}x_{52} + x_{16}x_{54} + x_{16}x_{56} + x_{16}x_{57} + x_{16}x_{58} + x_{16}x_{60} + x_{16}x_{62} + x_{16}x_{63} + x_{17}x_{18} + x_{17}x_{19} + x_{17}x_{21} + x_{17}x_{24} + x_{17}x_{25} + x_{17}x_{27} + x_{17}x_{28} + x_{17}x_{29} + x_{17}x_{30} + x_{17}x_{31} + x_{17}x_{33} + x_{17}x_{34} + x_{17}x_{40} + x_{17}x_{41} + x_{17}x_{42} + x_{17}x_{43} + x_{17}x_{44} + x_{17}x_{45} + x_{17}x_{46} + x_{17}x_{47} + x_{17}x_{49} + x_{17}x_{51} + x_{17}x_{57} + x_{17}x_{58} + x_{17}x_{59} + x_{17}x_{60} + x_{18}x_{19} + x_{18}x_{20} + x_{18}x_{21} + x_{18}x_{23} + x_{18}x_{24} + x_{18}x_{25} + x_{18}x_{26} + x_{18}x_{29} + x_{18}x_{31} + x_{18}x_{32} + x_{18}x_{35} + x_{18}x_{36} + x_{18}x_{37} + x_{18}x_{40} + x_{18}x_{41} + x_{18}x_{42} + x_{18}x_{44} + x_{18}x_{45} + x_{18}x_{46} + x_{18}x_{47} + x_{18}x_{49} + x_{18}x_{50} + x_{18}x_{52} + x_{18}x_{55} + x_{18}x_{61} + x_{18}x_{62} + x_{18}x_{63} + x_{19}x_{20} + x_{19}x_{23} + x_{19}x_{25} + x_{19}x_{26} + x_{19}x_{28} + x_{19}x_{29} + x_{19}x_{31} + x_{19}x_{33} + x_{19}x_{34} + x_{19}x_{36} + x_{19}x_{37} + x_{19}x_{38} + x_{19}x_{40} + x_{19}x_{42} + x_{19}x_{46} + x_{19}x_{47} + x_{19}x_{48} + x_{19}x_{52} + x_{19}x_{53} + x_{19}x_{54} + x_{19}x_{55} + x_{19}x_{60} + x_{19}x_{61} + x_{19}x_{64} + x_{20}x_{22} + x_{20}x_{26} + x_{20}x_{28} + x_{20}x_{31} + x_{20}x_{42} + x_{20}x_{43} + x_{20}x_{46} + x_{20}x_{47} + x_{20}x_{50} + x_{20}x_{51} + x_{20}x_{53} + x_{20}x_{54} + x_{20}x_{55} + x_{20}x_{56} + x_{20}x_{57} + x_{20}x_{58} + x_{20}x_{61} + x_{20}x_{62} + x_{20}x_{64} + x_{21}x_{24} + x_{21}x_{29} + x_{21}x_{30} + x_{21}x_{31} + x_{21}x_{32} + x_{21}x_{33} + x_{21}x_{35} + x_{21}x_{36} + x_{21}x_{38} + x_{21}x_{39} + x_{21}x_{41} + x_{21}x_{42} + x_{21}x_{43} + x_{21}x_{44} + x_{21}x_{46} + x_{21}x_{48} + x_{21}x_{49} + x_{21}x_{50} + x_{21}x_{54} + x_{21}x_{55} + x_{21}x_{62} + x_{21}x_{63} + x_{22}x_{23} + x_{22}x_{25} + x_{22}x_{27} + x_{22}x_{28} + x_{22}x_{32} + x_{22}x_{34} + x_{22}x_{36} + x_{22}x_{38} + x_{22}x_{40} + x_{22}x_{42} + x_{22}x_{43} + x_{22}x_{46} + x_{22}x_{48} + x_{22}x_{50} + x_{22}x_{53} + x_{22}x_{56} + x_{22}x_{57} + x_{22}x_{61} + x_{22}x_{62} + x_{22}x_{64} + x_{23}x_{28} + x_{23}x_{29} + x_{23}x_{30} + x_{23}x_{31} + x_{23}x_{32} + x_{23}x_{35} + x_{23}x_{37} + x_{23}x_{39} + x_{23}x_{40} + x_{23}x_{43} + x_{23}x_{45} + x_{23}x_{47} + x_{23}x_{48} + x_{23}x_{50} + x_{23}x_{52} + x_{23}x_{53} + x_{23}x_{54} + x_{23}x_{55} + x_{23}x_{58} + x_{23}x_{60} + x_{23}x_{61} + x_{23}x_{62} + x_{23}x_{64} + x_{24}x_{25} + x_{24}x_{26} + x_{24}x_{30} + x_{24}x_{31} + x_{24}x_{33} + x_{24}x_{34} + x_{24}x_{36} + x_{24}x_{37} + x_{24}x_{39} + x_{24}x_{42} + x_{24}x_{44} + x_{24}x_{45} + x_{24}x_{47} + x_{24}x_{48} + x_{24}x_{49} + x_{24}x_{50} + x_{24}x_{54} + x_{24}x_{56} + x_{24}x_{58} + x_{24}x_{64} + x_{25}x_{27} + x_{25}x_{29} + x_{25}x_{30} + x_{25}x_{32} + x_{25}x_{34} + x_{25}x_{35} + x_{25}x_{38} + x_{25}x_{42} + x_{25}x_{43} + x_{25}x_{45} + x_{25}x_{48} + x_{25}x_{49} + x_{25}x_{50} + x_{25}x_{55} + x_{25}x_{57} + x_{25}x_{59} + x_{25}x_{60} + x_{26}x_{27} + x_{26}x_{28} + x_{26}x_{29} + x_{26}x_{34} + x_{26}x_{36} + x_{26}x_{37} + x_{26}x_{42} + x_{26}x_{44} + x_{26}x_{47} + x_{26}x_{48} + x_{26}x_{51} + x_{26}x_{52} + x_{26}x_{53} + x_{26}x_{54} + x_{26}x_{55} + x_{26}x_{56} + x_{26}x_{57} + x_{26}x_{59} + x_{26}x_{63} + x_{26}x_{64} + x_{27}x_{30} + x_{27}x_{31} + x_{27}x_{32} + x_{27}x_{34} + x_{27}x_{36} + x_{27}x_{37} + x_{27}x_{40} + x_{27}x_{41} + x_{27}x_{42} + x_{27}x_{43} + x_{27}x_{44} + x_{27}x_{45} + x_{27}x_{46} + x_{27}x_{47} + x_{27}x_{49} + x_{27}x_{52} + x_{27}x_{53} + x_{27}x_{56} + x_{27}x_{57} + x_{27}x_{58} + x_{27}x_{59} + x_{27}x_{60} + x_{27}x_{61} + x_{28}x_{30} + x_{28}x_{34} + x_{28}x_{35} + x_{28}x_{38} + x_{28}x_{42} + x_{28}x_{44} + x_{28}x_{46} + x_{28}x_{48} + x_{28}x_{50} + x_{28}x_{51} + x_{28}x_{59} + x_{28}x_{61} + x_{28}x_{63} + x_{29}x_{30} + x_{29}x_{31} + x_{29}x_{32} + x_{29}x_{37} + x_{29}x_{39} + x_{29}x_{40} + x_{29}x_{44} + x_{29}x_{45} + x_{29}x_{46} + x_{29}x_{49} + x_{29}x_{51} + x_{29}x_{52} + x_{29}x_{54} + x_{29}x_{55} + x_{29}x_{56} + x_{29}x_{57} + x_{29}x_{60} + x_{29}x_{63} + x_{30}x_{32} + x_{30}x_{33} + x_{30}x_{34} + x_{30}x_{35} + x_{30}x_{36} + x_{30}x_{37} + x_{30}x_{38} + x_{30}x_{40} + x_{30}x_{44} + x_{30}x_{46} + x_{30}x_{48} + x_{30}x_{50} + x_{30}x_{53} + x_{30}x_{54} + x_{30}x_{55} + x_{30}x_{57} + x_{30}x_{59} + x_{30}x_{60} + x_{30}x_{62} + x_{30}x_{63} + x_{30}x_{64} + x_{31}x_{34} + x_{31}x_{36} + x_{31}x_{38} + x_{31}x_{40} + x_{31}x_{42} + x_{31}x_{45} + x_{31}x_{49} + x_{31}x_{51} + x_{31}x_{52} + x_{31}x_{53} + x_{31}x_{55} + x_{31}x_{56} + x_{31}x_{57} + x_{31}x_{58} + x_{31}x_{60} + x_{31}x_{61} + x_{31}x_{62} + x_{31}x_{64} + x_{32}x_{33} + x_{32}x_{38} + x_{32}x_{39} + x_{32}x_{40} + x_{32}x_{42} + x_{32}x_{43} + x_{32}x_{47} + x_{32}x_{48} + x_{32}x_{50} + x_{32}x_{51} + x_{32}x_{52} + x_{32}x_{54} + x_{32}x_{55} + x_{32}x_{56} + x_{32}x_{58} + x_{32}x_{59} + x_{32}x_{60} + x_{32}x_{61} + x_{32}x_{63} + x_{32}x_{64} + x_{33}x_{34} + x_{33}x_{37} + x_{33}x_{38} + x_{33}x_{39} + x_{33}x_{41} + x_{33}x_{42} + x_{33}x_{43} + x_{33}x_{44} + x_{33}x_{45} + x_{33}x_{46} + x_{33}x_{47} + x_{33}x_{49} + x_{33}x_{50} + x_{33}x_{54} + x_{33}x_{55} + x_{33}x_{56} + x_{33}x_{57} + x_{33}x_{58} + x_{33}x_{60} + x_{33}x_{63} + x_{33}x_{64} + x_{34}x_{35} + x_{34}x_{36} + x_{34}x_{37} + x_{34}x_{41} + x_{34}x_{45} + x_{34}x_{47} + x_{34}x_{48} + x_{34}x_{50} + x_{34}x_{52} + x_{34}x_{54} + x_{34}x_{55} + x_{34}x_{56} + x_{34}x_{59} + x_{34}x_{61} + x_{34}x_{62} + x_{34}x_{63} + x_{35}x_{37} + x_{35}x_{38} + x_{35}x_{39} + x_{35}x_{42} + x_{35}x_{47} + x_{35}x_{48} + x_{35}x_{49} + x_{35}x_{50} + x_{35}x_{55} + x_{35}x_{59} + x_{35}x_{61} + x_{35}x_{62} + x_{36}x_{37} + x_{36}x_{41} + x_{36}x_{42} + x_{36}x_{43} + x_{36}x_{45} + x_{36}x_{46} + x_{36}x_{47} + x_{36}x_{49} + x_{36}x_{52} + x_{36}x_{54} + x_{36}x_{55} + x_{36}x_{56} + x_{36}x_{58} + x_{36}x_{59} + x_{36}x_{62} + x_{36}x_{63} + x_{37}x_{38} + x_{37}x_{40} + x_{37}x_{41} + x_{37}x_{42} + x_{37}x_{43} + x_{37}x_{44} + x_{37}x_{47} + x_{37}x_{48} + x_{37}x_{49} + x_{37}x_{50} + x_{37}x_{52} + x_{37}x_{54} + x_{37}x_{56} + x_{37}x_{58} + x_{37}x_{59} + x_{37}x_{60} + x_{37}x_{64} + x_{38}x_{39} + x_{38}x_{45} + x_{38}x_{47} + x_{38}x_{48} + x_{38}x_{49} + x_{38}x_{50} + x_{38}x_{54} + x_{38}x_{55} + x_{38}x_{56} + x_{38}x_{57} + x_{38}x_{58} + x_{38}x_{61} + x_{38}x_{64} + x_{39}x_{40} + x_{39}x_{41} + x_{39}x_{45} + x_{39}x_{46} + x_{39}x_{47} + x_{39}x_{48} + x_{39}x_{49} + x_{39}x_{52} + x_{39}x_{54} + x_{39}x_{61} + x_{39}x_{63} + x_{40}x_{41} + x_{40}x_{43} + x_{40}x_{52} + x_{40}x_{53} + x_{40}x_{55} + x_{40}x_{57} + x_{40}x_{61} + x_{40}x_{62} + x_{40}x_{63} + x_{40}x_{64} + x_{41}x_{42} + x_{41}x_{43} + x_{41}x_{44} + x_{41}x_{45} + x_{41}x_{46} + x_{41}x_{49} + x_{41}x_{51} + x_{41}x_{55} + x_{41}x_{58} + x_{41}x_{60} + x_{41}x_{64} + x_{42}x_{43} + x_{42}x_{44} + x_{42}x_{46} + x_{42}x_{49} + x_{42}x_{50} + x_{42}x_{53} + x_{42}x_{54} + x_{42}x_{55} + x_{42}x_{56} + x_{42}x_{58} + x_{42}x_{60} + x_{42}x_{61} + x_{42}x_{64} + x_{43}x_{46} + x_{43}x_{48} + x_{43}x_{56} + x_{43}x_{62} + x_{44}x_{46} + x_{44}x_{47} + x_{44}x_{51} + x_{44}x_{52} + x_{44}x_{53} + x_{44}x_{54} + x_{44}x_{55} + x_{44}x_{56} + x_{44}x_{57} + x_{44}x_{59} + x_{44}x_{61} + x_{44}x_{64} + x_{45}x_{46} + x_{45}x_{47} + x_{45}x_{48} + x_{45}x_{49} + x_{45}x_{50} + x_{45}x_{56} + x_{45}x_{57} + x_{45}x_{60} + x_{45}x_{61} + x_{45}x_{64} + x_{46}x_{51} + x_{46}x_{55} + x_{46}x_{56} + x_{46}x_{57} + x_{46}x_{63} + x_{46}x_{64} + x_{47}x_{49} + x_{47}x_{51} + x_{47}x_{53} + x_{47}x_{55} + x_{47}x_{58} + x_{47}x_{60} + x_{47}x_{61} + x_{48}x_{51} + x_{48}x_{54} + x_{48}x_{55} + x_{48}x_{57} + x_{48}x_{58} + x_{48}x_{61} + x_{48}x_{62} + x_{48}x_{63} + x_{49}x_{51} + x_{49}x_{54} + x_{49}x_{55} + x_{49}x_{56} + x_{49}x_{61} + x_{49}x_{63} + x_{50}x_{51} + x_{50}x_{53} + x_{50}x_{54} + x_{50}x_{55} + x_{50}x_{56} + x_{50}x_{59} + x_{50}x_{60} + x_{50}x_{61} + x_{51}x_{54} + x_{51}x_{55} + x_{51}x_{56} + x_{51}x_{57} + x_{51}x_{58} + x_{51}x_{60} + x_{51}x_{61} + x_{51}x_{62} + x_{52}x_{53} + x_{52}x_{60} + x_{52}x_{62} + x_{53}x_{55} + x_{53}x_{56} + x_{53}x_{57} + x_{53}x_{58} + x_{53}x_{60} + x_{53}x_{62} + x_{53}x_{63} + x_{54}x_{55} + x_{54}x_{57} + x_{54}x_{59} + x_{54}x_{60} + x_{54}x_{63} + x_{55}x_{56} + x_{55}x_{60} + x_{56}x_{57} + x_{56}x_{60} + x_{56}x_{62} + x_{56}x_{63} + x_{56}x_{64} + x_{57}x_{58} + x_{57}x_{60} + x_{57}x_{62} + x_{58}x_{59} + x_{58}x_{61} + x_{58}x_{63} + x_{58}x_{64} + x_{59}x_{61} + x_{60}x_{64} + x_{61}x_{62} + x_{61}x_{64} + x_{62}x_{63} + x_{62}x_{64} + x_{63}x_{64} + x_{1} + x_{2} + x_{6} + x_{7} + x_{8} + x_{11} + x_{12} + x_{13} + x_{14} + x_{19} + x_{21} + x_{23} + x_{25} + x_{28} + x_{32} + x_{35} + x_{36} + x_{39} + x_{40} + x_{41} + x_{43} + x_{44} + x_{49} + x_{50} + x_{51} + x_{53} + x_{54} + x_{58} + x_{63} + 1$

$y_{27} = x_{1}x_{2} + x_{1}x_{3} + x_{1}x_{5} + x_{1}x_{7} + x_{1}x_{8} + x_{1}x_{9} + x_{1}x_{11} + x_{1}x_{13} + x_{1}x_{17} + x_{1}x_{18} + x_{1}x_{23} + x_{1}x_{24} + x_{1}x_{30} + x_{1}x_{31} + x_{1}x_{32} + x_{1}x_{35} + x_{1}x_{38} + x_{1}x_{39} + x_{1}x_{40} + x_{1}x_{41} + x_{1}x_{42} + x_{1}x_{44} + x_{1}x_{45} + x_{1}x_{47} + x_{1}x_{48} + x_{1}x_{54} + x_{1}x_{55} + x_{1}x_{61} + x_{1}x_{64} + x_{2}x_{7} + x_{2}x_{11} + x_{2}x_{12} + x_{2}x_{13} + x_{2}x_{15} + x_{2}x_{16} + x_{2}x_{17} + x_{2}x_{19} + x_{2}x_{21} + x_{2}x_{23} + x_{2}x_{26} + x_{2}x_{27} + x_{2}x_{28} + x_{2}x_{29} + x_{2}x_{30} + x_{2}x_{33} + x_{2}x_{36} + x_{2}x_{38} + x_{2}x_{39} + x_{2}x_{40} + x_{2}x_{43} + x_{2}x_{44} + x_{2}x_{46} + x_{2}x_{48} + x_{2}x_{52} + x_{2}x_{54} + x_{2}x_{56} + x_{2}x_{57} + x_{2}x_{59} + x_{2}x_{60} + x_{3}x_{4} + x_{3}x_{6} + x_{3}x_{9} + x_{3}x_{11} + x_{3}x_{12} + x_{3}x_{19} + x_{3}x_{20} + x_{3}x_{22} + x_{3}x_{27} + x_{3}x_{30} + x_{3}x_{31} + x_{3}x_{35} + x_{3}x_{36} + x_{3}x_{37} + x_{3}x_{38} + x_{3}x_{39} + x_{3}x_{40} + x_{3}x_{41} + x_{3}x_{43} + x_{3}x_{44} + x_{3}x_{46} + x_{3}x_{52} + x_{3}x_{53} + x_{3}x_{55} + x_{3}x_{58} + x_{3}x_{59} + x_{3}x_{61} + x_{3}x_{62} + x_{3}x_{64} + x_{4}x_{5} + x_{4}x_{6} + x_{4}x_{7} + x_{4}x_{9} + x_{4}x_{10} + x_{4}x_{11} + x_{4}x_{12} + x_{4}x_{13} + x_{4}x_{14} + x_{4}x_{15} + x_{4}x_{16} + x_{4}x_{18} + x_{4}x_{19} + x_{4}x_{22} + x_{4}x_{23} + x_{4}x_{24} + x_{4}x_{25} + x_{4}x_{26} + x_{4}x_{28} + x_{4}x_{29} + x_{4}x_{31} + x_{4}x_{33} + x_{4}x_{34} + x_{4}x_{37} + x_{4}x_{43} + x_{4}x_{45} + x_{4}x_{46} + x_{4}x_{55} + x_{4}x_{58} + x_{4}x_{63} + x_{4}x_{64} + x_{5}x_{8} + x_{5}x_{12} + x_{5}x_{14} + x_{5}x_{16} + x_{5}x_{19} + x_{5}x_{20} + x_{5}x_{22} + x_{5}x_{23} + x_{5}x_{24} + x_{5}x_{25} + x_{5}x_{28} + x_{5}x_{29} + x_{5}x_{30} + x_{5}x_{31} + x_{5}x_{32} + x_{5}x_{33} + x_{5}x_{36} + x_{5}x_{37} + x_{5}x_{38} + x_{5}x_{39} + x_{5}x_{41} + x_{5}x_{42} + x_{5}x_{45} + x_{5}x_{46} + x_{5}x_{48} + x_{5}x_{52} + x_{5}x_{53} + x_{5}x_{56} + x_{5}x_{57} + x_{5}x_{58} + x_{5}x_{60} + x_{5}x_{61} + x_{5}x_{62} + x_{5}x_{63} + x_{5}x_{64} + x_{6}x_{7} + x_{6}x_{8} + x_{6}x_{9} + x_{6}x_{10} + x_{6}x_{11} + x_{6}x_{12} + x_{6}x_{13} + x_{6}x_{14} + x_{6}x_{16} + x_{6}x_{19} + x_{6}x_{20} + x_{6}x_{22} + x_{6}x_{23} + x_{6}x_{26} + x_{6}x_{28} + x_{6}x_{29} + x_{6}x_{30} + x_{6}x_{32} + x_{6}x_{35} + x_{6}x_{36} + x_{6}x_{37} + x_{6}x_{39} + x_{6}x_{40} + x_{6}x_{42} + x_{6}x_{44} + x_{6}x_{45} + x_{6}x_{48} + x_{6}x_{49} + x_{6}x_{51} + x_{6}x_{58} + x_{6}x_{60} + x_{6}x_{62} + x_{6}x_{63} + x_{6}x_{64} + x_{7}x_{9} + x_{7}x_{11} + x_{7}x_{12} + x_{7}x_{13} + x_{7}x_{14} + x_{7}x_{15} + x_{7}x_{17} + x_{7}x_{18} + x_{7}x_{20} + x_{7}x_{21} + x_{7}x_{22} + x_{7}x_{29} + x_{7}x_{30} + x_{7}x_{32} + x_{7}x_{33} + x_{7}x_{35} + x_{7}x_{36} + x_{7}x_{38} + x_{7}x_{39} + x_{7}x_{41} + x_{7}x_{43} + x_{7}x_{44} + x_{7}x_{45} + x_{7}x_{47} + x_{7}x_{48} + x_{7}x_{49} + x_{7}x_{51} + x_{7}x_{53} + x_{7}x_{57} + x_{7}x_{58} + x_{7}x_{59} + x_{7}x_{60} + x_{7}x_{64} + x_{8}x_{10} + x_{8}x_{11} + x_{8}x_{12} + x_{8}x_{13} + x_{8}x_{14} + x_{8}x_{18} + x_{8}x_{20} + x_{8}x_{21} + x_{8}x_{24} + x_{8}x_{28} + x_{8}x_{30} + x_{8}x_{32} + x_{8}x_{37} + x_{8}x_{38} + x_{8}x_{41} + x_{8}x_{42} + x_{8}x_{44} + x_{8}x_{45} + x_{8}x_{46} + x_{8}x_{48} + x_{8}x_{51} + x_{8}x_{52} + x_{8}x_{53} + x_{8}x_{56} + x_{8}x_{57} + x_{8}x_{60} + x_{8}x_{63} + x_{8}x_{64} + x_{9}x_{10} + x_{9}x_{11} + x_{9}x_{14} + x_{9}x_{15} + x_{9}x_{17} + x_{9}x_{18} + x_{9}x_{19} + x_{9}x_{20} + x_{9}x_{22} + x_{9}x_{24} + x_{9}x_{25} + x_{9}x_{26} + x_{9}x_{27} + x_{9}x_{29} + x_{9}x_{31} + x_{9}x_{33} + x_{9}x_{34} + x_{9}x_{40} + x_{9}x_{42} + x_{9}x_{43} + x_{9}x_{45} + x_{9}x_{47} + x_{9}x_{48} + x_{9}x_{49} + x_{9}x_{52} + x_{9}x_{53} + x_{9}x_{55} + x_{9}x_{56} + x_{9}x_{62} + x_{10}x_{15} + x_{10}x_{17} + x_{10}x_{18} + x_{10}x_{20} + x_{10}x_{24} + x_{10}x_{25} + x_{10}x_{27} + x_{10}x_{28} + x_{10}x_{30} + x_{10}x_{31} + x_{10}x_{32} + x_{10}x_{34} + x_{10}x_{37} + x_{10}x_{39} + x_{10}x_{40} + x_{10}x_{42} + x_{10}x_{46} + x_{10}x_{48} + x_{10}x_{52} + x_{10}x_{53} + x_{10}x_{54} + x_{10}x_{59} + x_{10}x_{64} + x_{11}x_{17} + x_{11}x_{20} + x_{11}x_{22} + x_{11}x_{28} + x_{11}x_{32} + x_{11}x_{35} + x_{11}x_{40} + x_{11}x_{41} + x_{11}x_{42} + x_{11}x_{43} + x_{11}x_{44} + x_{11}x_{46} + x_{11}x_{49} + x_{11}x_{50} + x_{11}x_{52} + x_{11}x_{53} + x_{11}x_{54} + x_{11}x_{57} + x_{11}x_{59} + x_{11}x_{60} + x_{11}x_{64} + x_{12}x_{13} + x_{12}x_{14} + x_{12}x_{17} + x_{12}x_{21} + x_{12}x_{23} + x_{12}x_{27} + x_{12}x_{29} + x_{12}x_{32} + x_{12}x_{34} + x_{12}x_{35} + x_{12}x_{39} + x_{12}x_{40} + x_{12}x_{41} + x_{12}x_{42} + x_{12}x_{46} + x_{12}x_{48} + x_{12}x_{53} + x_{12}x_{54} + x_{12}x_{58} + x_{12}x_{64} + x_{13}x_{14} + x_{13}x_{16} + x_{13}x_{18} + x_{13}x_{19} + x_{13}x_{23} + x_{13}x_{25} + x_{13}x_{30} + x_{13}x_{32} + x_{13}x_{33} + x_{13}x_{39} + x_{13}x_{42} + x_{13}x_{44} + x_{13}x_{45} + x_{13}x_{46} + x_{13}x_{48} + x_{13}x_{49} + x_{13}x_{52} + x_{13}x_{55} + x_{13}x_{59} + x_{13}x_{61} + x_{14}x_{16} + x_{14}x_{19} + x_{14}x_{20} + x_{14}x_{22} + x_{14}x_{23} + x_{14}x_{24} + x_{14}x_{26} + x_{14}x_{28} + x_{14}x_{29} + x_{14}x_{30} + x_{14}x_{34} + x_{14}x_{36} + x_{14}x_{39} + x_{14}x_{41} + x_{14}x_{43} + x_{14}x_{44} + x_{14}x_{45} + x_{14}x_{46} + x_{14}x_{47} + x_{14}x_{49} + x_{14}x_{52} + x_{14}x_{53} + x_{14}x_{55} + x_{14}x_{56} + x_{14}x_{59} + x_{14}x_{60} + x_{14}x_{64} + x_{15}x_{16} + x_{15}x_{19} + x_{15}x_{20} + x_{15}x_{21} + x_{15}x_{25} + x_{15}x_{26} + x_{15}x_{27} + x_{15}x_{28} + x_{15}x_{30} + x_{15}x_{31} + x_{15}x_{32} + x_{15}x_{33} + x_{15}x_{36} + x_{15}x_{38} + x_{15}x_{41} + x_{15}x_{44} + x_{15}x_{45} + x_{15}x_{46} + x_{15}x_{50} + x_{15}x_{51} + x_{15}x_{52} + x_{15}x_{53} + x_{15}x_{54} + x_{15}x_{55} + x_{15}x_{60} + x_{15}x_{62} + x_{15}x_{63} + x_{16}x_{17} + x_{16}x_{20} + x_{16}x_{22} + x_{16}x_{25} + x_{16}x_{30} + x_{16}x_{31} + x_{16}x_{33} + x_{16}x_{35} + x_{16}x_{36} + x_{16}x_{37} + x_{16}x_{38} + x_{16}x_{42} + x_{16}x_{43} + x_{16}x_{46} + x_{16}x_{47} + x_{16}x_{48} + x_{16}x_{49} + x_{16}x_{50} + x_{16}x_{56} + x_{16}x_{57} + x_{16}x_{58} + x_{16}x_{59} + x_{16}x_{64} + x_{17}x_{19} + x_{17}x_{20} + x_{17}x_{25} + x_{17}x_{27} + x_{17}x_{30} + x_{17}x_{32} + x_{17}x_{35} + x_{17}x_{36} + x_{17}x_{37} + x_{17}x_{38} + x_{17}x_{39} + x_{17}x_{40} + x_{17}x_{41} + x_{17}x_{42} + x_{17}x_{44} + x_{17}x_{45} + x_{17}x_{47} + x_{17}x_{49} + x_{17}x_{50} + x_{17}x_{54} + x_{17}x_{55} + x_{17}x_{56} + x_{17}x_{58} + x_{17}x_{60} + x_{17}x_{62} + x_{17}x_{63} + x_{17}x_{64} + x_{18}x_{19} + x_{18}x_{21} + x_{18}x_{22} + x_{18}x_{23} + x_{18}x_{27} + x_{18}x_{28} + x_{18}x_{29} + x_{18}x_{35} + x_{18}x_{36} + x_{18}x_{40} + x_{18}x_{41} + x_{18}x_{46} + x_{18}x_{47} + x_{18}x_{51} + x_{18}x_{52} + x_{18}x_{55} + x_{18}x_{58} + x_{18}x_{60} + x_{18}x_{61} + x_{18}x_{62} + x_{18}x_{63} + x_{18}x_{64} + x_{19}x_{20} + x_{19}x_{21} + x_{19}x_{22} + x_{19}x_{23} + x_{19}x_{25} + x_{19}x_{27} + x_{19}x_{28} + x_{19}x_{29} + x_{19}x_{31} + x_{19}x_{33} + x_{19}x_{35} + x_{19}x_{37} + x_{19}x_{39} + x_{19}x_{40} + x_{19}x_{41} + x_{19}x_{42} + x_{19}x_{43} + x_{19}x_{44} + x_{19}x_{45} + x_{19}x_{46} + x_{19}x_{48} + x_{19}x_{49} + x_{19}x_{50} + x_{19}x_{52} + x_{19}x_{53} + x_{19}x_{54} + x_{19}x_{55} + x_{19}x_{56} + x_{19}x_{58} + x_{19}x_{59} + x_{19}x_{60} + x_{19}x_{61} + x_{19}x_{62} + x_{20}x_{22} + x_{20}x_{23} + x_{20}x_{26} + x_{20}x_{33} + x_{20}x_{36} + x_{20}x_{40} + x_{20}x_{41} + x_{20}x_{44} + x_{20}x_{45} + x_{20}x_{48} + x_{20}x_{49} + x_{20}x_{50} + x_{20}x_{52} + x_{20}x_{58} + x_{20}x_{60} + x_{20}x_{64} + x_{21}x_{23} + x_{21}x_{24} + x_{21}x_{25} + x_{21}x_{28} + x_{21}x_{29} + x_{21}x_{32} + x_{21}x_{34} + x_{21}x_{36} + x_{21}x_{37} + x_{21}x_{38} + x_{21}x_{39} + x_{21}x_{40} + x_{21}x_{44} + x_{21}x_{46} + x_{21}x_{48} + x_{21}x_{52} + x_{21}x_{55} + x_{21}x_{62} + x_{21}x_{64} + x_{22}x_{28} + x_{22}x_{29} + x_{22}x_{30} + x_{22}x_{34} + x_{22}x_{35} + x_{22}x_{36} + x_{22}x_{37} + x_{22}x_{42} + x_{22}x_{45} + x_{22}x_{46} + x_{22}x_{47} + x_{22}x_{48} + x_{22}x_{49} + x_{22}x_{50} + x_{22}x_{54} + x_{22}x_{56} + x_{22}x_{57} + x_{22}x_{61} + x_{22}x_{62} + x_{22}x_{63} + x_{23}x_{25} + x_{23}x_{29} + x_{23}x_{30} + x_{23}x_{32} + x_{23}x_{33} + x_{23}x_{34} + x_{23}x_{36} + x_{23}x_{37} + x_{23}x_{38} + x_{23}x_{39} + x_{23}x_{41} + x_{23}x_{44} + x_{23}x_{46} + x_{23}x_{49} + x_{23}x_{52} + x_{23}x_{53} + x_{23}x_{58} + x_{23}x_{61} + x_{23}x_{62} + x_{23}x_{63} + x_{23}x_{64} + x_{24}x_{26} + x_{24}x_{27} + x_{24}x_{29} + x_{24}x_{31} + x_{24}x_{33} + x_{24}x_{34} + x_{24}x_{39} + x_{24}x_{40} + x_{24}x_{42} + x_{24}x_{43} + x_{24}x_{44} + x_{24}x_{45} + x_{24}x_{47} + x_{24}x_{54} + x_{24}x_{57} + x_{24}x_{61} + x_{24}x_{62} + x_{24}x_{63} + x_{25}x_{27} + x_{25}x_{33} + x_{25}x_{35} + x_{25}x_{36} + x_{25}x_{38} + x_{25}x_{39} + x_{25}x_{40} + x_{25}x_{41} + x_{25}x_{42} + x_{25}x_{44} + x_{25}x_{45} + x_{25}x_{47} + x_{25}x_{48} + x_{25}x_{49} + x_{25}x_{50} + x_{25}x_{51} + x_{25}x_{52} + x_{25}x_{53} + x_{25}x_{54} + x_{25}x_{56} + x_{25}x_{61} + x_{25}x_{62} + x_{25}x_{63} + x_{26}x_{29} + x_{26}x_{31} + x_{26}x_{32} + x_{26}x_{34} + x_{26}x_{35} + x_{26}x_{36} + x_{26}x_{37} + x_{26}x_{38} + x_{26}x_{40} + x_{26}x_{42} + x_{26}x_{45} + x_{26}x_{46} + x_{26}x_{47} + x_{26}x_{48} + x_{26}x_{51} + x_{26}x_{52} + x_{26}x_{53} + x_{26}x_{55} + x_{26}x_{56} + x_{26}x_{57} + x_{26}x_{58} + x_{26}x_{60} + x_{27}x_{28} + x_{27}x_{29} + x_{27}x_{31} + x_{27}x_{32} + x_{27}x_{33} + x_{27}x_{34} + x_{27}x_{36} + x_{27}x_{43} + x_{27}x_{44} + x_{27}x_{45} + x_{27}x_{46} + x_{27}x_{47} + x_{27}x_{49} + x_{27}x_{50} + x_{27}x_{53} + x_{27}x_{56} + x_{27}x_{59} + x_{27}x_{61} + x_{27}x_{62} + x_{27}x_{63} + x_{27}x_{64} + x_{28}x_{29} + x_{28}x_{30} + x_{28}x_{31} + x_{28}x_{34} + x_{28}x_{36} + x_{28}x_{38} + x_{28}x_{39} + x_{28}x_{43} + x_{28}x_{45} + x_{28}x_{47} + x_{28}x_{48} + x_{28}x_{50} + x_{28}x_{52} + x_{28}x_{54} + x_{28}x_{55} + x_{28}x_{59} + x_{28}x_{60} + x_{28}x_{61} + x_{28}x_{62} + x_{29}x_{31} + x_{29}x_{32} + x_{29}x_{33} + x_{29}x_{34} + x_{29}x_{35} + x_{29}x_{36} + x_{29}x_{41} + x_{29}x_{42} + x_{29}x_{44} + x_{29}x_{45} + x_{29}x_{47} + x_{29}x_{48} + x_{29}x_{51} + x_{29}x_{52} + x_{29}x_{53} + x_{29}x_{54} + x_{29}x_{64} + x_{30}x_{31} + x_{30}x_{36} + x_{30}x_{37} + x_{30}x_{38} + x_{30}x_{40} + x_{30}x_{43} + x_{30}x_{45} + x_{30}x_{48} + x_{30}x_{49} + x_{30}x_{50} + x_{30}x_{52} + x_{30}x_{53} + x_{30}x_{54} + x_{30}x_{55} + x_{30}x_{56} + x_{30}x_{57} + x_{30}x_{63} + x_{30}x_{64} + x_{31}x_{33} + x_{31}x_{34} + x_{31}x_{37} + x_{31}x_{39} + x_{31}x_{42} + x_{31}x_{45} + x_{31}x_{46} + x_{31}x_{48} + x_{31}x_{49} + x_{31}x_{50} + x_{31}x_{52} + x_{31}x_{54} + x_{31}x_{55} + x_{31}x_{59} + x_{31}x_{60} + x_{31}x_{61} + x_{31}x_{63} + x_{31}x_{64} + x_{32}x_{34} + x_{32}x_{35} + x_{32}x_{36} + x_{32}x_{37} + x_{32}x_{38} + x_{32}x_{39} + x_{32}x_{40} + x_{32}x_{44} + x_{32}x_{45} + x_{32}x_{46} + x_{32}x_{48} + x_{32}x_{50} + x_{32}x_{51} + x_{32}x_{52} + x_{32}x_{53} + x_{32}x_{54} + x_{32}x_{56} + x_{32}x_{57} + x_{32}x_{61} + x_{32}x_{62} + x_{32}x_{63} + x_{32}x_{64} + x_{33}x_{34} + x_{33}x_{36} + x_{33}x_{37} + x_{33}x_{40} + x_{33}x_{41} + x_{33}x_{45} + x_{33}x_{46} + x_{33}x_{47} + x_{33}x_{49} + x_{33}x_{54} + x_{33}x_{55} + x_{33}x_{57} + x_{33}x_{59} + x_{33}x_{62} + x_{34}x_{35} + x_{34}x_{36} + x_{34}x_{39} + x_{34}x_{40} + x_{34}x_{43} + x_{34}x_{44} + x_{34}x_{45} + x_{34}x_{46} + x_{34}x_{47} + x_{34}x_{50} + x_{34}x_{51} + x_{34}x_{52} + x_{34}x_{53} + x_{34}x_{57} + x_{34}x_{61} + x_{34}x_{63} + x_{35}x_{37} + x_{35}x_{40} + x_{35}x_{42} + x_{35}x_{44} + x_{35}x_{47} + x_{35}x_{48} + x_{35}x_{53} + x_{35}x_{55} + x_{35}x_{58} + x_{35}x_{61} + x_{35}x_{63} + x_{36}x_{38} + x_{36}x_{39} + x_{36}x_{42} + x_{36}x_{44} + x_{36}x_{45} + x_{36}x_{48} + x_{36}x_{51} + x_{36}x_{53} + x_{36}x_{54} + x_{36}x_{55} + x_{36}x_{60} + x_{36}x_{62} + x_{36}x_{63} + x_{37}x_{40} + x_{37}x_{43} + x_{37}x_{47} + x_{37}x_{49} + x_{37}x_{52} + x_{37}x_{53} + x_{37}x_{54} + x_{37}x_{55} + x_{37}x_{57} + x_{37}x_{58} + x_{37}x_{59} + x_{37}x_{61} + x_{37}x_{62} + x_{37}x_{63} + x_{37}x_{64} + x_{38}x_{39} + x_{38}x_{41} + x_{38}x_{42} + x_{38}x_{44} + x_{38}x_{45} + x_{38}x_{47} + x_{38}x_{48} + x_{38}x_{50} + x_{38}x_{52} + x_{38}x_{53} + x_{38}x_{54} + x_{38}x_{56} + x_{38}x_{57} + x_{38}x_{60} + x_{38}x_{62} + x_{38}x_{63} + x_{38}x_{64} + x_{39}x_{40} + x_{39}x_{43} + x_{39}x_{44} + x_{39}x_{47} + x_{39}x_{48} + x_{39}x_{49} + x_{39}x_{50} + x_{39}x_{52} + x_{39}x_{54} + x_{39}x_{57} + x_{39}x_{63} + x_{39}x_{64} + x_{40}x_{41} + x_{40}x_{44} + x_{40}x_{46} + x_{40}x_{49} + x_{40}x_{52} + x_{40}x_{55} + x_{40}x_{57} + x_{40}x_{60} + x_{40}x_{61} + x_{40}x_{63} + x_{40}x_{64} + x_{41}x_{49} + x_{41}x_{50} + x_{41}x_{51} + x_{41}x_{56} + x_{41}x_{57} + x_{41}x_{60} + x_{41}x_{61} + x_{41}x_{62} + x_{41}x_{64} + x_{42}x_{43} + x_{42}x_{45} + x_{42}x_{46} + x_{42}x_{48} + x_{42}x_{51} + x_{42}x_{52} + x_{42}x_{54} + x_{42}x_{55} + x_{42}x_{56} + x_{42}x_{57} + x_{42}x_{59} + x_{42}x_{64} + x_{43}x_{44} + x_{43}x_{45} + x_{43}x_{46} + x_{43}x_{49} + x_{43}x_{50} + x_{43}x_{51} + x_{43}x_{55} + x_{43}x_{57} + x_{43}x_{60} + x_{43}x_{62} + x_{43}x_{63} + x_{43}x_{64} + x_{44}x_{45} + x_{44}x_{48} + x_{44}x_{49} + x_{44}x_{50} + x_{44}x_{52} + x_{44}x_{53} + x_{44}x_{55} + x_{44}x_{56} + x_{44}x_{57} + x_{44}x_{58} + x_{44}x_{60} + x_{44}x_{61} + x_{44}x_{63} + x_{44}x_{64} + x_{45}x_{46} + x_{45}x_{48} + x_{45}x_{50} + x_{45}x_{52} + x_{45}x_{60} + x_{45}x_{62} + x_{45}x_{64} + x_{46}x_{47} + x_{46}x_{48} + x_{46}x_{49} + x_{46}x_{51} + x_{46}x_{53} + x_{46}x_{56} + x_{46}x_{64} + x_{47}x_{51} + x_{47}x_{53} + x_{47}x_{57} + x_{47}x_{59} + x_{47}x_{60} + x_{47}x_{62} + x_{47}x_{63} + x_{47}x_{64} + x_{48}x_{50} + x_{48}x_{51} + x_{48}x_{56} + x_{48}x_{58} + x_{48}x_{62} + x_{48}x_{64} + x_{49}x_{54} + x_{49}x_{57} + x_{49}x_{58} + x_{49}x_{59} + x_{49}x_{62} + x_{49}x_{63} + x_{50}x_{51} + x_{50}x_{52} + x_{50}x_{53} + x_{50}x_{54} + x_{50}x_{55} + x_{50}x_{56} + x_{50}x_{57} + x_{50}x_{58} + x_{50}x_{61} + x_{50}x_{63} + x_{51}x_{56} + x_{51}x_{57} + x_{51}x_{60} + x_{52}x_{54} + x_{52}x_{57} + x_{52}x_{59} + x_{52}x_{60} + x_{52}x_{61} + x_{52}x_{62} + x_{52}x_{63} + x_{52}x_{64} + x_{53}x_{54} + x_{53}x_{56} + x_{53}x_{58} + x_{53}x_{62} + x_{53}x_{63} + x_{54}x_{55} + x_{54}x_{56} + x_{54}x_{60} + x_{55}x_{57} + x_{55}x_{58} + x_{55}x_{61} + x_{55}x_{64} + x_{56}x_{57} + x_{56}x_{58} + x_{56}x_{59} + x_{56}x_{60} + x_{56}x_{64} + x_{57}x_{58} + x_{57}x_{61} + x_{57}x_{62} + x_{58}x_{63} + x_{58}x_{64} + x_{59}x_{62} + x_{59}x_{63} + x_{60}x_{62} + x_{60}x_{64} + x_{61}x_{63} + x_{62}x_{63} + x_{62}x_{64} + x_{2} + x_{5} + x_{6} + x_{8} + x_{10} + x_{15} + x_{16} + x_{17} + x_{18} + x_{19} + x_{20} + x_{24} + x_{25} + x_{29} + x_{32} + x_{34} + x_{35} + x_{38} + x_{41} + x_{42} + x_{43} + x_{45} + x_{46} + x_{47} + x_{48} + x_{49} + x_{50} + x_{51} + x_{53} + x_{54} + x_{57} + x_{58} + x_{63}$

$y_{28} = x_{1}x_{3} + x_{1}x_{10} + x_{1}x_{12} + x_{1}x_{15} + x_{1}x_{18} + x_{1}x_{22} + x_{1}x_{24} + x_{1}x_{25} + x_{1}x_{27} + x_{1}x_{30} + x_{1}x_{34} + x_{1}x_{35} + x_{1}x_{38} + x_{1}x_{40} + x_{1}x_{44} + x_{1}x_{46} + x_{1}x_{48} + x_{1}x_{50} + x_{1}x_{51} + x_{1}x_{52} + x_{1}x_{53} + x_{1}x_{54} + x_{1}x_{55} + x_{1}x_{57} + x_{1}x_{60} + x_{1}x_{62} + x_{1}x_{63} + x_{2}x_{4} + x_{2}x_{7} + x_{2}x_{8} + x_{2}x_{10} + x_{2}x_{11} + x_{2}x_{13} + x_{2}x_{14} + x_{2}x_{17} + x_{2}x_{19} + x_{2}x_{20} + x_{2}x_{22} + x_{2}x_{25} + x_{2}x_{26} + x_{2}x_{30} + x_{2}x_{31} + x_{2}x_{33} + x_{2}x_{35} + x_{2}x_{37} + x_{2}x_{39} + x_{2}x_{40} + x_{2}x_{41} + x_{2}x_{44} + x_{2}x_{45} + x_{2}x_{46} + x_{2}x_{48} + x_{2}x_{49} + x_{2}x_{50} + x_{2}x_{52} + x_{2}x_{53} + x_{2}x_{54} + x_{2}x_{57} + x_{2}x_{58} + x_{2}x_{60} + x_{2}x_{63} + x_{2}x_{64} + x_{3}x_{4} + x_{3}x_{5} + x_{3}x_{9} + x_{3}x_{10} + x_{3}x_{11} + x_{3}x_{12} + x_{3}x_{13} + x_{3}x_{15} + x_{3}x_{20} + x_{3}x_{22} + x_{3}x_{23} + x_{3}x_{25} + x_{3}x_{26} + x_{3}x_{30} + x_{3}x_{31} + x_{3}x_{34} + x_{3}x_{35} + x_{3}x_{36} + x_{3}x_{40} + x_{3}x_{41} + x_{3}x_{42} + x_{3}x_{43} + x_{3}x_{46} + x_{3}x_{49} + x_{3}x_{50} + x_{3}x_{51} + x_{3}x_{53} + x_{3}x_{55} + x_{3}x_{57} + x_{3}x_{58} + x_{3}x_{59} + x_{3}x_{61} + x_{3}x_{62} + x_{3}x_{64} + x_{4}x_{5} + x_{4}x_{7} + x_{4}x_{9} + x_{4}x_{12} + x_{4}x_{13} + x_{4}x_{14} + x_{4}x_{15} + x_{4}x_{16} + x_{4}x_{20} + x_{4}x_{22} + x_{4}x_{23} + x_{4}x_{25} + x_{4}x_{26} + x_{4}x_{27} + x_{4}x_{28} + x_{4}x_{29} + x_{4}x_{30} + x_{4}x_{32} + x_{4}x_{35} + x_{4}x_{39} + x_{4}x_{42} + x_{4}x_{43} + x_{4}x_{46} + x_{4}x_{47} + x_{4}x_{49} + x_{4}x_{50} + x_{4}x_{52} + x_{4}x_{57} + x_{4}x_{59} + x_{4}x_{60} + x_{4}x_{62} + x_{5}x_{6} + x_{5}x_{7} + x_{5}x_{14} + x_{5}x_{16} + x_{5}x_{17} + x_{5}x_{18} + x_{5}x_{21} + x_{5}x_{22} + x_{5}x_{24} + x_{5}x_{25} + x_{5}x_{30} + x_{5}x_{31} + x_{5}x_{33} + x_{5}x_{38} + x_{5}x_{42} + x_{5}x_{43} + x_{5}x_{44} + x_{5}x_{46} + x_{5}x_{47} + x_{5}x_{51} + x_{5}x_{53} + x_{5}x_{57} + x_{5}x_{59} + x_{5}x_{61} + x_{5}x_{62} + x_{5}x_{64} + x_{6}x_{8} + x_{6}x_{9} + x_{6}x_{11} + x_{6}x_{12} + x_{6}x_{15} + x_{6}x_{16} + x_{6}x_{19} + x_{6}x_{20} + x_{6}x_{24} + x_{6}x_{25} + x_{6}x_{26} + x_{6}x_{29} + x_{6}x_{30} + x_{6}x_{31} + x_{6}x_{32} + x_{6}x_{34} + x_{6}x_{35} + x_{6}x_{39} + x_{6}x_{40} + x_{6}x_{41} + x_{6}x_{43} + x_{6}x_{44} + x_{6}x_{45} + x_{6}x_{47} + x_{6}x_{48} + x_{6}x_{49} + x_{6}x_{50} + x_{6}x_{51} + x_{6}x_{53} + x_{6}x_{54} + x_{6}x_{56} + x_{6}x_{59} + x_{6}x_{61} + x_{6}x_{62} + x_{7}x_{9} + x_{7}x_{10} + x_{7}x_{11} + x_{7}x_{14} + x_{7}x_{15} + x_{7}x_{16} + x_{7}x_{17} + x_{7}x_{18} + x_{7}x_{25} + x_{7}x_{29} + x_{7}x_{33} + x_{7}x_{35} + x_{7}x_{36} + x_{7}x_{39} + x_{7}x_{40} + x_{7}x_{41} + x_{7}x_{48} + x_{7}x_{49} + x_{7}x_{51} + x_{7}x_{54} + x_{7}x_{55} + x_{7}x_{56} + x_{7}x_{57} + x_{7}x_{59} + x_{7}x_{62} + x_{7}x_{64} + x_{8}x_{9} + x_{8}x_{12} + x_{8}x_{13} + x_{8}x_{14} + x_{8}x_{15} + x_{8}x_{18} + x_{8}x_{19} + x_{8}x_{21} + x_{8}x_{23} + x_{8}x_{29} + x_{8}x_{30} + x_{8}x_{31} + x_{8}x_{32} + x_{8}x_{34} + x_{8}x_{36} + x_{8}x_{41} + x_{8}x_{46} + x_{8}x_{48} + x_{8}x_{49} + x_{8}x_{52} + x_{8}x_{54} + x_{8}x_{56} + x_{8}x_{57} + x_{8}x_{58} + x_{8}x_{61} + x_{8}x_{62} + x_{8}x_{64} + x_{9}x_{11} + x_{9}x_{15} + x_{9}x_{16} + x_{9}x_{19} + x_{9}x_{20} + x_{9}x_{21} + x_{9}x_{23} + x_{9}x_{26} + x_{9}x_{28} + x_{9}x_{30} + x_{9}x_{31} + x_{9}x_{33} + x_{9}x_{36} + x_{9}x_{38} + x_{9}x_{40} + x_{9}x_{41} + x_{9}x_{46} + x_{9}x_{47} + x_{9}x_{49} + x_{9}x_{50} + x_{9}x_{54} + x_{9}x_{55} + x_{9}x_{57} + x_{9}x_{58} + x_{9}x_{59} + x_{9}x_{61} + x_{9}x_{63} + x_{10}x_{13} + x_{10}x_{16} + x_{10}x_{19} + x_{10}x_{20} + x_{10}x_{21} + x_{10}x_{22} + x_{10}x_{24} + x_{10}x_{25} + x_{10}x_{26} + x_{10}x_{27} + x_{10}x_{28} + x_{10}x_{30} + x_{10}x_{33} + x_{10}x_{34} + x_{10}x_{35} + x_{10}x_{37} + x_{10}x_{39} + x_{10}x_{40} + x_{10}x_{41} + x_{10}x_{42} + x_{10}x_{43} + x_{10}x_{45} + x_{10}x_{46} + x_{10}x_{47} + x_{10}x_{48} + x_{10}x_{51} + x_{10}x_{58} + x_{10}x_{61} + x_{10}x_{64} + x_{11}x_{15} + x_{11}x_{16} + x_{11}x_{17} + x_{11}x_{18} + x_{11}x_{19} + x_{11}x_{30} + x_{11}x_{31} + x_{11}x_{37} + x_{11}x_{41} + x_{11}x_{42} + x_{11}x_{43} + x_{11}x_{45} + x_{11}x_{47} + x_{11}x_{51} + x_{11}x_{52} + x_{11}x_{55} + x_{11}x_{56} + x_{11}x_{57} + x_{11}x_{59} + x_{11}x_{62} + x_{11}x_{63} + x_{12}x_{16} + x_{12}x_{20} + x_{12}x_{21} + x_{12}x_{25} + x_{12}x_{31} + x_{12}x_{32} + x_{12}x_{33} + x_{12}x_{35} + x_{12}x_{37} + x_{12}x_{39} + x_{12}x_{41} + x_{12}x_{42} + x_{12}x_{43} + x_{12}x_{44} + x_{12}x_{47} + x_{12}x_{53} + x_{12}x_{57} + x_{12}x_{58} + x_{12}x_{60} + x_{13}x_{15} + x_{13}x_{18} + x_{13}x_{20} + x_{13}x_{21} + x_{13}x_{22} + x_{13}x_{23} + x_{13}x_{25} + x_{13}x_{28} + x_{13}x_{29} + x_{13}x_{33} + x_{13}x_{34} + x_{13}x_{36} + x_{13}x_{38} + x_{13}x_{39} + x_{13}x_{42} + x_{13}x_{46} + x_{13}x_{47} + x_{13}x_{50} + x_{13}x_{51} + x_{13}x_{52} + x_{13}x_{54} + x_{13}x_{57} + x_{13}x_{58} + x_{13}x_{61} + x_{13}x_{63} + x_{13}x_{64} + x_{14}x_{15} + x_{14}x_{16} + x_{14}x_{18} + x_{14}x_{19} + x_{14}x_{20} + x_{14}x_{22} + x_{14}x_{23} + x_{14}x_{25} + x_{14}x_{27} + x_{14}x_{29} + x_{14}x_{30} + x_{14}x_{32} + x_{14}x_{34} + x_{14}x_{38} + x_{14}x_{39} + x_{14}x_{40} + x_{14}x_{42} + x_{14}x_{46} + x_{14}x_{47} + x_{14}x_{48} + x_{14}x_{49} + x_{14}x_{53} + x_{14}x_{55} + x_{14}x_{56} + x_{14}x_{60} + x_{14}x_{62} + x_{14}x_{63} + x_{14}x_{64} + x_{15}x_{18} + x_{15}x_{21} + x_{15}x_{23} + x_{15}x_{26} + x_{15}x_{28} + x_{15}x_{33} + x_{15}x_{38} + x_{15}x_{41} + x_{15}x_{42} + x_{15}x_{44} + x_{15}x_{45} + x_{15}x_{46} + x_{15}x_{47} + x_{15}x_{49} + x_{15}x_{50} + x_{15}x_{54} + x_{15}x_{55} + x_{15}x_{56} + x_{15}x_{57} + x_{15}x_{59} + x_{15}x_{60} + x_{15}x_{61} + x_{15}x_{64} + x_{16}x_{17} + x_{16}x_{20} + x_{16}x_{21} + x_{16}x_{24} + x_{16}x_{28} + x_{16}x_{29} + x_{16}x_{30} + x_{16}x_{31} + x_{16}x_{33} + x_{16}x_{34} + x_{16}x_{37} + x_{16}x_{38} + x_{16}x_{39} + x_{16}x_{40} + x_{16}x_{44} + x_{16}x_{46} + x_{16}x_{47} + x_{16}x_{48} + x_{16}x_{49} + x_{16}x_{51} + x_{16}x_{53} + x_{16}x_{55} + x_{16}x_{56} + x_{16}x_{59} + x_{16}x_{60} + x_{16}x_{61} + x_{16}x_{63} + x_{17}x_{18} + x_{17}x_{20} + x_{17}x_{23} + x_{17}x_{25} + x_{17}x_{27} + x_{17}x_{28} + x_{17}x_{29} + x_{17}x_{30} + x_{17}x_{31} + x_{17}x_{33} + x_{17}x_{34} + x_{17}x_{36} + x_{17}x_{38} + x_{17}x_{42} + x_{17}x_{44} + x_{17}x_{46} + x_{17}x_{48} + x_{17}x_{51} + x_{17}x_{52} + x_{17}x_{53} + x_{17}x_{54} + x_{17}x_{56} + x_{17}x_{57} + x_{17}x_{62} + x_{18}x_{22} + x_{18}x_{24} + x_{18}x_{25} + x_{18}x_{28} + x_{18}x_{29} + x_{18}x_{30} + x_{18}x_{32} + x_{18}x_{33} + x_{18}x_{34} + x_{18}x_{40} + x_{18}x_{41} + x_{18}x_{46} + x_{18}x_{47} + x_{18}x_{48} + x_{18}x_{50} + x_{18}x_{51} + x_{18}x_{52} + x_{18}x_{53} + x_{18}x_{54} + x_{18}x_{55} + x_{18}x_{61} + x_{18}x_{62} + x_{18}x_{64} + x_{19}x_{20} + x_{19}x_{23} + x_{19}x_{25} + x_{19}x_{26} + x_{19}x_{27} + x_{19}x_{28} + x_{19}x_{30} + x_{19}x_{31} + x_{19}x_{35} + x_{19}x_{37} + x_{19}x_{38} + x_{19}x_{41} + x_{19}x_{42} + x_{19}x_{43} + x_{19}x_{44} + x_{19}x_{45} + x_{19}x_{46} + x_{19}x_{51} + x_{19}x_{52} + x_{19}x_{55} + x_{19}x_{60} + x_{19}x_{61} + x_{19}x_{62} + x_{20}x_{22} + x_{20}x_{23} + x_{20}x_{24} + x_{20}x_{27} + x_{20}x_{28} + x_{20}x_{29} + x_{20}x_{32} + x_{20}x_{33} + x_{20}x_{34} + x_{20}x_{35} + x_{20}x_{37} + x_{20}x_{38} + x_{20}x_{39} + x_{20}x_{41} + x_{20}x_{42} + x_{20}x_{50} + x_{20}x_{51} + x_{20}x_{55} + x_{20}x_{57} + x_{20}x_{58} + x_{20}x_{59} + x_{21}x_{26} + x_{21}x_{29} + x_{21}x_{32} + x_{21}x_{34} + x_{21}x_{35} + x_{21}x_{36} + x_{21}x_{37} + x_{21}x_{38} + x_{21}x_{42} + x_{21}x_{45} + x_{21}x_{46} + x_{21}x_{47} + x_{21}x_{48} + x_{21}x_{50} + x_{21}x_{51} + x_{21}x_{52} + x_{21}x_{53} + x_{21}x_{54} + x_{21}x_{56} + x_{21}x_{57} + x_{21}x_{58} + x_{21}x_{61} + x_{21}x_{62} + x_{21}x_{64} + x_{22}x_{25} + x_{22}x_{30} + x_{22}x_{32} + x_{22}x_{34} + x_{22}x_{38} + x_{22}x_{42} + x_{22}x_{43} + x_{22}x_{44} + x_{22}x_{45} + x_{22}x_{46} + x_{22}x_{50} + x_{22}x_{51} + x_{22}x_{57} + x_{22}x_{63} + x_{22}x_{64} + x_{23}x_{24} + x_{23}x_{25} + x_{23}x_{27} + x_{23}x_{28} + x_{23}x_{29} + x_{23}x_{30} + x_{23}x_{31} + x_{23}x_{32} + x_{23}x_{33} + x_{23}x_{34} + x_{23}x_{37} + x_{23}x_{39} + x_{23}x_{42} + x_{23}x_{43} + x_{23}x_{44} + x_{23}x_{45} + x_{23}x_{46} + x_{23}x_{47} + x_{23}x_{48} + x_{23}x_{49} + x_{23}x_{52} + x_{23}x_{55} + x_{23}x_{56} + x_{23}x_{57} + x_{23}x_{58} + x_{23}x_{59} + x_{23}x_{60} + x_{23}x_{62} + x_{23}x_{63} + x_{23}x_{64} + x_{24}x_{28} + x_{24}x_{31} + x_{24}x_{32} + x_{24}x_{35} + x_{24}x_{37} + x_{24}x_{43} + x_{24}x_{44} + x_{24}x_{45} + x_{24}x_{46} + x_{24}x_{47} + x_{24}x_{48} + x_{24}x_{49} + x_{24}x_{51} + x_{24}x_{60} + x_{24}x_{61} + x_{24}x_{62} + x_{25}x_{26} + x_{25}x_{31} + x_{25}x_{34} + x_{25}x_{35} + x_{25}x_{37} + x_{25}x_{47} + x_{25}x_{49} + x_{25}x_{52} + x_{25}x_{55} + x_{25}x_{57} + x_{25}x_{60} + x_{25}x_{61} + x_{25}x_{63} + x_{26}x_{29} + x_{26}x_{30} + x_{26}x_{35} + x_{26}x_{36} + x_{26}x_{38} + x_{26}x_{39} + x_{26}x_{40} + x_{26}x_{41} + x_{26}x_{44} + x_{26}x_{46} + x_{26}x_{48} + x_{26}x_{50} + x_{26}x_{53} + x_{26}x_{54} + x_{26}x_{55} + x_{26}x_{59} + x_{26}x_{61} + x_{26}x_{62} + x_{26}x_{64} + x_{27}x_{28} + x_{27}x_{29} + x_{27}x_{30} + x_{27}x_{31} + x_{27}x_{32} + x_{27}x_{33} + x_{27}x_{38} + x_{27}x_{43} + x_{27}x_{45} + x_{27}x_{46} + x_{27}x_{47} + x_{27}x_{48} + x_{27}x_{50} + x_{27}x_{52} + x_{27}x_{53} + x_{27}x_{54} + x_{27}x_{56} + x_{27}x_{57} + x_{27}x_{60} + x_{27}x_{61} + x_{27}x_{64} + x_{28}x_{30} + x_{28}x_{33} + x_{28}x_{35} + x_{28}x_{38} + x_{28}x_{39} + x_{28}x_{41} + x_{28}x_{42} + x_{28}x_{45} + x_{28}x_{46} + x_{28}x_{47} + x_{28}x_{48} + x_{28}x_{50} + x_{28}x_{51} + x_{28}x_{52} + x_{28}x_{53} + x_{28}x_{57} + x_{28}x_{60} + x_{28}x_{62} + x_{29}x_{31} + x_{29}x_{35} + x_{29}x_{37} + x_{29}x_{40} + x_{29}x_{41} + x_{29}x_{42} + x_{29}x_{44} + x_{29}x_{45} + x_{29}x_{47} + x_{29}x_{49} + x_{29}x_{51} + x_{29}x_{53} + x_{29}x_{54} + x_{29}x_{57} + x_{29}x_{59} + x_{29}x_{64} + x_{30}x_{31} + x_{30}x_{32} + x_{30}x_{35} + x_{30}x_{39} + x_{30}x_{40} + x_{30}x_{42} + x_{30}x_{47} + x_{30}x_{49} + x_{30}x_{51} + x_{30}x_{52} + x_{30}x_{53} + x_{30}x_{54} + x_{30}x_{56} + x_{30}x_{57} + x_{30}x_{58} + x_{30}x_{63} + x_{30}x_{64} + x_{31}x_{32} + x_{31}x_{33} + x_{31}x_{35} + x_{31}x_{36} + x_{31}x_{38} + x_{31}x_{40} + x_{31}x_{42} + x_{31}x_{47} + x_{31}x_{48} + x_{31}x_{50} + x_{31}x_{51} + x_{31}x_{52} + x_{31}x_{53} + x_{31}x_{55} + x_{31}x_{56} + x_{31}x_{57} + x_{31}x_{61} + x_{31}x_{64} + x_{32}x_{33} + x_{32}x_{38} + x_{32}x_{39} + x_{32}x_{40} + x_{32}x_{41} + x_{32}x_{42} + x_{32}x_{44} + x_{32}x_{48} + x_{32}x_{50} + x_{32}x_{51} + x_{32}x_{55} + x_{32}x_{57} + x_{32}x_{58} + x_{32}x_{59} + x_{32}x_{60} + x_{32}x_{61} + x_{32}x_{62} + x_{33}x_{35} + x_{33}x_{37} + x_{33}x_{38} + x_{33}x_{39} + x_{33}x_{41} + x_{33}x_{44} + x_{33}x_{46} + x_{33}x_{49} + x_{33}x_{50} + x_{33}x_{52} + x_{33}x_{54} + x_{33}x_{55} + x_{33}x_{57} + x_{33}x_{59} + x_{33}x_{60} + x_{33}x_{62} + x_{34}x_{35} + x_{34}x_{37} + x_{34}x_{38} + x_{34}x_{40} + x_{34}x_{41} + x_{34}x_{42} + x_{34}x_{43} + x_{34}x_{44} + x_{34}x_{48} + x_{34}x_{52} + x_{34}x_{53} + x_{34}x_{54} + x_{34}x_{56} + x_{34}x_{59} + x_{34}x_{62} + x_{35}x_{36} + x_{35}x_{39} + x_{35}x_{41} + x_{35}x_{42} + x_{35}x_{44} + x_{35}x_{46} + x_{35}x_{48} + x_{35}x_{49} + x_{35}x_{58} + x_{35}x_{59} + x_{35}x_{61} + x_{35}x_{62} + x_{35}x_{63} + x_{35}x_{64} + x_{36}x_{37} + x_{36}x_{39} + x_{36}x_{43} + x_{36}x_{44} + x_{36}x_{48} + x_{36}x_{49} + x_{36}x_{50} + x_{36}x_{53} + x_{36}x_{54} + x_{36}x_{55} + x_{36}x_{57} + x_{36}x_{58} + x_{36}x_{60} + x_{36}x_{61} + x_{36}x_{62} + x_{36}x_{64} + x_{37}x_{39} + x_{37}x_{43} + x_{37}x_{45} + x_{37}x_{47} + x_{37}x_{48} + x_{37}x_{49} + x_{37}x_{53} + x_{37}x_{54} + x_{37}x_{55} + x_{37}x_{57} + x_{37}x_{58} + x_{37}x_{60} + x_{37}x_{61} + x_{37}x_{64} + x_{38}x_{41} + x_{38}x_{43} + x_{38}x_{51} + x_{38}x_{57} + x_{38}x_{58} + x_{38}x_{61} + x_{38}x_{62} + x_{38}x_{63} + x_{38}x_{64} + x_{39}x_{40} + x_{39}x_{41} + x_{39}x_{43} + x_{39}x_{45} + x_{39}x_{48} + x_{39}x_{49} + x_{39}x_{50} + x_{39}x_{52} + x_{39}x_{55} + x_{39}x_{58} + x_{40}x_{42} + x_{40}x_{44} + x_{40}x_{45} + x_{40}x_{47} + x_{40}x_{50} + x_{40}x_{52} + x_{40}x_{61} + x_{40}x_{63} + x_{41}x_{42} + x_{41}x_{43} + x_{41}x_{44} + x_{41}x_{46} + x_{41}x_{49} + x_{41}x_{50} + x_{41}x_{51} + x_{41}x_{52} + x_{41}x_{53} + x_{41}x_{54} + x_{41}x_{56} + x_{41}x_{57} + x_{41}x_{59} + x_{41}x_{61} + x_{41}x_{62} + x_{42}x_{46} + x_{42}x_{49} + x_{42}x_{50} + x_{42}x_{51} + x_{42}x_{52} + x_{42}x_{53} + x_{42}x_{55} + x_{42}x_{56} + x_{42}x_{57} + x_{42}x_{59} + x_{42}x_{61} + x_{42}x_{64} + x_{43}x_{44} + x_{43}x_{46} + x_{43}x_{48} + x_{43}x_{49} + x_{43}x_{50} + x_{43}x_{51} + x_{43}x_{52} + x_{43}x_{53} + x_{43}x_{56} + x_{43}x_{57} + x_{43}x_{58} + x_{43}x_{59} + x_{43}x_{61} + x_{44}x_{45} + x_{44}x_{46} + x_{44}x_{53} + x_{44}x_{54} + x_{44}x_{55} + x_{44}x_{57} + x_{44}x_{59} + x_{44}x_{60} + x_{44}x_{62} + x_{44}x_{64} + x_{45}x_{46} + x_{45}x_{47} + x_{45}x_{52} + x_{45}x_{55} + x_{45}x_{56} + x_{45}x_{57} + x_{45}x_{58} + x_{45}x_{59} + x_{46}x_{48} + x_{46}x_{49} + x_{46}x_{50} + x_{46}x_{51} + x_{46}x_{53} + x_{46}x_{60} + x_{46}x_{61} + x_{46}x_{62} + x_{46}x_{63} + x_{46}x_{64} + x_{47}x_{48} + x_{47}x_{49} + x_{47}x_{50} + x_{47}x_{51} + x_{47}x_{53} + x_{47}x_{54} + x_{47}x_{55} + x_{47}x_{56} + x_{47}x_{57} + x_{47}x_{58} + x_{47}x_{59} + x_{47}x_{61} + x_{47}x_{64} + x_{48}x_{49} + x_{48}x_{51} + x_{48}x_{52} + x_{48}x_{55} + x_{48}x_{56} + x_{48}x_{57} + x_{48}x_{62} + x_{48}x_{64} + x_{49}x_{50} + x_{49}x_{51} + x_{49}x_{53} + x_{49}x_{54} + x_{49}x_{57} + x_{49}x_{58} + x_{49}x_{59} + x_{49}x_{62} + x_{50}x_{53} + x_{50}x_{55} + x_{50}x_{57} + x_{51}x_{54} + x_{51}x_{56} + x_{51}x_{57} + x_{51}x_{59} + x_{51}x_{60} + x_{51}x_{61} + x_{51}x_{64} + x_{52}x_{55} + x_{52}x_{56} + x_{52}x_{58} + x_{52}x_{60} + x_{52}x_{61} + x_{52}x_{62} + x_{52}x_{63} + x_{53}x_{54} + x_{53}x_{55} + x_{53}x_{56} + x_{53}x_{57} + x_{53}x_{58} + x_{53}x_{59} + x_{53}x_{60} + x_{53}x_{64} + x_{54}x_{55} + x_{54}x_{56} + x_{54}x_{57} + x_{54}x_{58} + x_{54}x_{60} + x_{54}x_{62} + x_{54}x_{63} + x_{54}x_{64} + x_{55}x_{56} + x_{55}x_{57} + x_{55}x_{58} + x_{55}x_{61} + x_{55}x_{62} + x_{55}x_{64} + x_{56}x_{57} + x_{56}x_{63} + x_{56}x_{64} + x_{57}x_{60} + x_{57}x_{61} + x_{57}x_{62} + x_{57}x_{64} + x_{58}x_{59} + x_{58}x_{63} + x_{58}x_{64} + x_{59}x_{60} + x_{59}x_{63} + x_{60}x_{61} + x_{60}x_{62} + x_{2} + x_{3} + x_{4} + x_{6} + x_{7} + x_{9} + x_{12} + x_{14} + x_{18} + x_{19} + x_{21} + x_{23} + x_{24} + x_{25} + x_{28} + x_{30} + x_{31} + x_{35} + x_{36} + x_{40} + x_{41} + x_{43} + x_{46} + x_{47} + x_{49} + x_{51} + x_{52} + x_{54} + x_{55} + x_{62} + x_{63} + x_{64}$

$y_{29} = x_{1}x_{2} + x_{1}x_{4} + x_{1}x_{5} + x_{1}x_{7} + x_{1}x_{8} + x_{1}x_{11} + x_{1}x_{12} + x_{1}x_{13} + x_{1}x_{14} + x_{1}x_{17} + x_{1}x_{20} + x_{1}x_{21} + x_{1}x_{23} + x_{1}x_{25} + x_{1}x_{27} + x_{1}x_{28} + x_{1}x_{29} + x_{1}x_{31} + x_{1}x_{32} + x_{1}x_{33} + x_{1}x_{43} + x_{1}x_{44} + x_{1}x_{46} + x_{1}x_{47} + x_{1}x_{48} + x_{1}x_{50} + x_{1}x_{51} + x_{1}x_{56} + x_{1}x_{60} + x_{2}x_{3} + x_{2}x_{7} + x_{2}x_{10} + x_{2}x_{12} + x_{2}x_{13} + x_{2}x_{15} + x_{2}x_{18} + x_{2}x_{21} + x_{2}x_{22} + x_{2}x_{23} + x_{2}x_{24} + x_{2}x_{25} + x_{2}x_{27} + x_{2}x_{28} + x_{2}x_{29} + x_{2}x_{31} + x_{2}x_{32} + x_{2}x_{33} + x_{2}x_{35} + x_{2}x_{36} + x_{2}x_{37} + x_{2}x_{41} + x_{2}x_{43} + x_{2}x_{44} + x_{2}x_{45} + x_{2}x_{49} + x_{2}x_{52} + x_{2}x_{53} + x_{2}x_{56} + x_{2}x_{58} + x_{2}x_{59} + x_{2}x_{61} + x_{2}x_{63} + x_{2}x_{64} + x_{3}x_{5} + x_{3}x_{10} + x_{3}x_{11} + x_{3}x_{12} + x_{3}x_{13} + x_{3}x_{19} + x_{3}x_{20} + x_{3}x_{21} + x_{3}x_{22} + x_{3}x_{23} + x_{3}x_{24} + x_{3}x_{25} + x_{3}x_{28} + x_{3}x_{29} + x_{3}x_{31} + x_{3}x_{32} + x_{3}x_{33} + x_{3}x_{35} + x_{3}x_{37} + x_{3}x_{38} + x_{3}x_{40} + x_{3}x_{41} + x_{3}x_{43} + x_{3}x_{44} + x_{3}x_{45} + x_{3}x_{46} + x_{3}x_{47} + x_{3}x_{48} + x_{3}x_{51} + x_{3}x_{54} + x_{3}x_{56} + x_{3}x_{59} + x_{3}x_{60} + x_{4}x_{7} + x_{4}x_{8} + x_{4}x_{9} + x_{4}x_{14} + x_{4}x_{21} + x_{4}x_{24} + x_{4}x_{27} + x_{4}x_{34} + x_{4}x_{36} + x_{4}x_{37} + x_{4}x_{38} + x_{4}x_{40} + x_{4}x_{42} + x_{4}x_{43} + x_{4}x_{44} + x_{4}x_{45} + x_{4}x_{48} + x_{4}x_{49} + x_{4}x_{50} + x_{4}x_{53} + x_{4}x_{55} + x_{4}x_{56} + x_{4}x_{57} + x_{4}x_{61} + x_{4}x_{63} + x_{5}x_{7} + x_{5}x_{9} + x_{5}x_{11} + x_{5}x_{12} + x_{5}x_{14} + x_{5}x_{16} + x_{5}x_{17} + x_{5}x_{18} + x_{5}x_{21} + x_{5}x_{23} + x_{5}x_{25} + x_{5}x_{27} + x_{5}x_{30} + x_{5}x_{32} + x_{5}x_{33} + x_{5}x_{34} + x_{5}x_{35} + x_{5}x_{36} + x_{5}x_{37} + x_{5}x_{38} + x_{5}x_{39} + x_{5}x_{41} + x_{5}x_{48} + x_{5}x_{50} + x_{5}x_{52} + x_{5}x_{55} + x_{5}x_{57} + x_{5}x_{58} + x_{5}x_{60} + x_{5}x_{62} + x_{5}x_{64} + x_{6}x_{7} + x_{6}x_{8} + x_{6}x_{10} + x_{6}x_{14} + x_{6}x_{15} + x_{6}x_{16} + x_{6}x_{18} + x_{6}x_{21} + x_{6}x_{22} + x_{6}x_{23} + x_{6}x_{25} + x_{6}x_{26} + x_{6}x_{30} + x_{6}x_{32} + x_{6}x_{33} + x_{6}x_{35} + x_{6}x_{36} + x_{6}x_{39} + x_{6}x_{41} + x_{6}x_{42} + x_{6}x_{43} + x_{6}x_{44} + x_{6}x_{46} + x_{6}x_{48} + x_{6}x_{52} + x_{6}x_{53} + x_{6}x_{54} + x_{6}x_{57} + x_{6}x_{58} + x_{6}x_{60} + x_{6}x_{62} + x_{6}x_{63} + x_{6}x_{64} + x_{7}x_{8} + x_{7}x_{12} + x_{7}x_{13} + x_{7}x_{14} + x_{7}x_{16} + x_{7}x_{17} + x_{7}x_{18} + x_{7}x_{20} + x_{7}x_{21} + x_{7}x_{24} + x_{7}x_{26} + x_{7}x_{29} + x_{7}x_{31} + x_{7}x_{32} + x_{7}x_{33} + x_{7}x_{34} + x_{7}x_{35} + x_{7}x_{40} + x_{7}x_{43} + x_{7}x_{46} + x_{7}x_{48} + x_{7}x_{49} + x_{7}x_{50} + x_{7}x_{57} + x_{7}x_{58} + x_{7}x_{64} + x_{8}x_{10} + x_{8}x_{11} + x_{8}x_{14} + x_{8}x_{15} + x_{8}x_{16} + x_{8}x_{17} + x_{8}x_{18} + x_{8}x_{19} + x_{8}x_{20} + x_{8}x_{21} + x_{8}x_{22} + x_{8}x_{23} + x_{8}x_{24} + x_{8}x_{25} + x_{8}x_{26} + x_{8}x_{27} + x_{8}x_{29} + x_{8}x_{32} + x_{8}x_{33} + x_{8}x_{35} + x_{8}x_{37} + x_{8}x_{38} + x_{8}x_{39} + x_{8}x_{41} + x_{8}x_{42} + x_{8}x_{43} + x_{8}x_{44} + x_{8}x_{45} + x_{8}x_{48} + x_{8}x_{50} + x_{8}x_{52} + x_{8}x_{53} + x_{8}x_{54} + x_{8}x_{56} + x_{8}x_{59} + x_{8}x_{62} + x_{8}x_{64} + x_{9}x_{12} + x_{9}x_{13} + x_{9}x_{15} + x_{9}x_{17} + x_{9}x_{18} + x_{9}x_{19} + x_{9}x_{23} + x_{9}x_{24} + x_{9}x_{25} + x_{9}x_{28} + x_{9}x_{29} + x_{9}x_{31} + x_{9}x_{34} + x_{9}x_{35} + x_{9}x_{36} + x_{9}x_{44} + x_{9}x_{45} + x_{9}x_{46} + x_{9}x_{48} + x_{9}x_{50} + x_{9}x_{51} + x_{9}x_{55} + x_{9}x_{57} + x_{9}x_{58} + x_{9}x_{61} + x_{10}x_{15} + x_{10}x_{16} + x_{10}x_{17} + x_{10}x_{18} + x_{10}x_{21} + x_{10}x_{23} + x_{10}x_{24} + x_{10}x_{27} + x_{10}x_{31} + x_{10}x_{34} + x_{10}x_{36} + x_{10}x_{37} + x_{10}x_{42} + x_{10}x_{46} + x_{10}x_{47} + x_{10}x_{51} + x_{10}x_{54} + x_{10}x_{55} + x_{10}x_{56} + x_{10}x_{57} + x_{10}x_{58} + x_{10}x_{59} + x_{10}x_{60} + x_{10}x_{63} + x_{11}x_{15} + x_{11}x_{17} + x_{11}x_{19} + x_{11}x_{23} + x_{11}x_{24} + x_{11}x_{26} + x_{11}x_{27} + x_{11}x_{29} + x_{11}x_{31} + x_{11}x_{33} + x_{11}x_{38} + x_{11}x_{40} + x_{11}x_{41} + x_{11}x_{42} + x_{11}x_{44} + x_{11}x_{48} + x_{11}x_{50} + x_{11}x_{51} + x_{11}x_{52} + x_{11}x_{54} + x_{11}x_{60} + x_{11}x_{61} + x_{11}x_{62} + x_{11}x_{63} + x_{12}x_{14} + x_{12}x_{15} + x_{12}x_{16} + x_{12}x_{18} + x_{12}x_{19} + x_{12}x_{20} + x_{12}x_{25} + x_{12}x_{29} + x_{12}x_{30} + x_{12}x_{34} + x_{12}x_{35} + x_{12}x_{37} + x_{12}x_{42} + x_{12}x_{43} + x_{12}x_{44} + x_{12}x_{45} + x_{12}x_{47} + x_{12}x_{48} + x_{12}x_{50} + x_{12}x_{52} + x_{12}x_{53} + x_{12}x_{54} + x_{12}x_{55} + x_{12}x_{56} + x_{12}x_{57} + x_{12}x_{61} + x_{12}x_{62} + x_{12}x_{64} + x_{13}x_{14} + x_{13}x_{16} + x_{13}x_{18} + x_{13}x_{20} + x_{13}x_{22} + x_{13}x_{26} + x_{13}x_{28} + x_{13}x_{29} + x_{13}x_{30} + x_{13}x_{31} + x_{13}x_{33} + x_{13}x_{34} + x_{13}x_{36} + x_{13}x_{42} + x_{13}x_{43} + x_{13}x_{44} + x_{13}x_{50} + x_{13}x_{53} + x_{13}x_{54} + x_{13}x_{55} + x_{13}x_{56} + x_{13}x_{60} + x_{13}x_{61} + x_{13}x_{62} + x_{13}x_{63} + x_{13}x_{64} + x_{14}x_{17} + x_{14}x_{21} + x_{14}x_{22} + x_{14}x_{25} + x_{14}x_{27} + x_{14}x_{28} + x_{14}x_{29} + x_{14}x_{30} + x_{14}x_{31} + x_{14}x_{34} + x_{14}x_{35} + x_{14}x_{36} + x_{14}x_{42} + x_{14}x_{44} + x_{14}x_{46} + x_{14}x_{49} + x_{14}x_{52} + x_{14}x_{55} + x_{14}x_{56} + x_{14}x_{58} + x_{14}x_{62} + x_{14}x_{63} + x_{15}x_{16} + x_{15}x_{17} + x_{15}x_{18} + x_{15}x_{19} + x_{15}x_{21} + x_{15}x_{22} + x_{15}x_{23} + x_{15}x_{26} + x_{15}x_{27} + x_{15}x_{28} + x_{15}x_{29} + x_{15}x_{30} + x_{15}x_{33} + x_{15}x_{36} + x_{15}x_{37} + x_{15}x_{38} + x_{15}x_{39} + x_{15}x_{41} + x_{15}x_{43} + x_{15}x_{44} + x_{15}x_{45} + x_{15}x_{47} + x_{15}x_{48} + x_{15}x_{49} + x_{15}x_{50} + x_{15}x_{53} + x_{15}x_{60} + x_{15}x_{61} + x_{15}x_{62} + x_{15}x_{63} + x_{15}x_{64} + x_{16}x_{22} + x_{16}x_{24} + x_{16}x_{25} + x_{16}x_{26} + x_{16}x_{29} + x_{16}x_{32} + x_{16}x_{35} + x_{16}x_{36} + x_{16}x_{37} + x_{16}x_{38} + x_{16}x_{40} + x_{16}x_{41} + x_{16}x_{43} + x_{16}x_{45} + x_{16}x_{46} + x_{16}x_{48} + x_{16}x_{49} + x_{16}x_{50} + x_{16}x_{53} + x_{16}x_{54} + x_{16}x_{56} + x_{16}x_{58} + x_{16}x_{62} + x_{16}x_{64} + x_{17}x_{20} + x_{17}x_{22} + x_{17}x_{23} + x_{17}x_{25} + x_{17}x_{27} + x_{17}x_{28} + x_{17}x_{29} + x_{17}x_{30} + x_{17}x_{34} + x_{17}x_{36} + x_{17}x_{37} + x_{17}x_{39} + x_{17}x_{40} + x_{17}x_{41} + x_{17}x_{42} + x_{17}x_{43} + x_{17}x_{46} + x_{17}x_{48} + x_{17}x_{49} + x_{17}x_{51} + x_{17}x_{54} + x_{17}x_{55} + x_{17}x_{56} + x_{18}x_{19} + x_{18}x_{20} + x_{18}x_{21} + x_{18}x_{22} + x_{18}x_{25} + x_{18}x_{28} + x_{18}x_{29} + x_{18}x_{30} + x_{18}x_{31} + x_{18}x_{32} + x_{18}x_{35} + x_{18}x_{42} + x_{18}x_{43} + x_{18}x_{45} + x_{18}x_{47} + x_{18}x_{51} + x_{18}x_{54} + x_{18}x_{56} + x_{18}x_{64} + x_{19}x_{20} + x_{19}x_{24} + x_{19}x_{26} + x_{19}x_{30} + x_{19}x_{31} + x_{19}x_{33} + x_{19}x_{34} + x_{19}x_{35} + x_{19}x_{39} + x_{19}x_{42} + x_{19}x_{43} + x_{19}x_{44} + x_{19}x_{46} + x_{19}x_{50} + x_{19}x_{51} + x_{19}x_{53} + x_{19}x_{54} + x_{19}x_{58} + x_{19}x_{59} + x_{19}x_{60} + x_{19}x_{64} + x_{20}x_{22} + x_{20}x_{23} + x_{20}x_{24} + x_{20}x_{25} + x_{20}x_{27} + x_{20}x_{29} + x_{20}x_{32} + x_{20}x_{34} + x_{20}x_{37} + x_{20}x_{38} + x_{20}x_{39} + x_{20}x_{40} + x_{20}x_{42} + x_{20}x_{43} + x_{20}x_{45} + x_{20}x_{47} + x_{20}x_{49} + x_{20}x_{53} + x_{20}x_{55} + x_{20}x_{58} + x_{20}x_{62} + x_{20}x_{63} + x_{21}x_{23} + x_{21}x_{24} + x_{21}x_{25} + x_{21}x_{27} + x_{21}x_{28} + x_{21}x_{30} + x_{21}x_{31} + x_{21}x_{36} + x_{21}x_{37} + x_{21}x_{39} + x_{21}x_{40} + x_{21}x_{41} + x_{21}x_{44} + x_{21}x_{48} + x_{21}x_{49} + x_{21}x_{50} + x_{21}x_{51} + x_{21}x_{52} + x_{21}x_{54} + x_{21}x_{59} + x_{21}x_{61} + x_{21}x_{62} + x_{21}x_{64} + x_{22}x_{23} + x_{22}x_{24} + x_{22}x_{25} + x_{22}x_{26} + x_{22}x_{30} + x_{22}x_{33} + x_{22}x_{34} + x_{22}x_{35} + x_{22}x_{37} + x_{22}x_{38} + x_{22}x_{39} + x_{22}x_{41} + x_{22}x_{42} + x_{22}x_{44} + x_{22}x_{46} + x_{22}x_{47} + x_{22}x_{48} + x_{22}x_{54} + x_{22}x_{55} + x_{22}x_{57} + x_{22}x_{59} + x_{22}x_{61} + x_{22}x_{64} + x_{23}x_{25} + x_{23}x_{27} + x_{23}x_{32} + x_{23}x_{33} + x_{23}x_{35} + x_{23}x_{36} + x_{23}x_{37} + x_{23}x_{38} + x_{23}x_{41} + x_{23}x_{42} + x_{23}x_{43} + x_{23}x_{44} + x_{23}x_{45} + x_{23}x_{50} + x_{23}x_{52} + x_{23}x_{54} + x_{23}x_{56} + x_{23}x_{57} + x_{23}x_{59} + x_{23}x_{61} + x_{23}x_{62} + x_{23}x_{64} + x_{24}x_{26} + x_{24}x_{28} + x_{24}x_{30} + x_{24}x_{31} + x_{24}x_{32} + x_{24}x_{33} + x_{24}x_{34} + x_{24}x_{38} + x_{24}x_{39} + x_{24}x_{40} + x_{24}x_{42} + x_{24}x_{43} + x_{24}x_{44} + x_{24}x_{45} + x_{24}x_{47} + x_{24}x_{48} + x_{24}x_{49} + x_{24}x_{50} + x_{24}x_{51} + x_{24}x_{52} + x_{24}x_{54} + x_{24}x_{60} + x_{24}x_{63} + x_{24}x_{64} + x_{25}x_{26} + x_{25}x_{27} + x_{25}x_{32} + x_{25}x_{34} + x_{25}x_{38} + x_{25}x_{40} + x_{25}x_{43} + x_{25}x_{44} + x_{25}x_{48} + x_{25}x_{52} + x_{25}x_{54} + x_{25}x_{56} + x_{25}x_{60} + x_{25}x_{63} + x_{25}x_{64} + x_{26}x_{27} + x_{26}x_{29} + x_{26}x_{30} + x_{26}x_{31} + x_{26}x_{32} + x_{26}x_{34} + x_{26}x_{37} + x_{26}x_{38} + x_{26}x_{39} + x_{26}x_{40} + x_{26}x_{42} + x_{26}x_{44} + x_{26}x_{45} + x_{26}x_{47} + x_{26}x_{48} + x_{26}x_{49} + x_{26}x_{50} + x_{26}x_{51} + x_{26}x_{52} + x_{26}x_{53} + x_{26}x_{54} + x_{26}x_{55} + x_{26}x_{56} + x_{26}x_{58} + x_{26}x_{59} + x_{26}x_{62} + x_{27}x_{28} + x_{27}x_{30} + x_{27}x_{31} + x_{27}x_{32} + x_{27}x_{33} + x_{27}x_{36} + x_{27}x_{38} + x_{27}x_{39} + x_{27}x_{41} + x_{27}x_{44} + x_{27}x_{45} + x_{27}x_{46} + x_{27}x_{48} + x_{27}x_{52} + x_{27}x_{54} + x_{27}x_{56} + x_{27}x_{57} + x_{27}x_{58} + x_{27}x_{59} + x_{27}x_{60} + x_{27}x_{61} + x_{27}x_{62} + x_{28}x_{29} + x_{28}x_{30} + x_{28}x_{31} + x_{28}x_{34} + x_{28}x_{35} + x_{28}x_{36} + x_{28}x_{37} + x_{28}x_{39} + x_{28}x_{41} + x_{28}x_{44} + x_{28}x_{45} + x_{28}x_{46} + x_{28}x_{47} + x_{28}x_{48} + x_{28}x_{49} + x_{28}x_{54} + x_{28}x_{60} + x_{28}x_{62} + x_{28}x_{63} + x_{29}x_{31} + x_{29}x_{33} + x_{29}x_{34} + x_{29}x_{37} + x_{29}x_{40} + x_{29}x_{42} + x_{29}x_{44} + x_{29}x_{45} + x_{29}x_{48} + x_{29}x_{49} + x_{29}x_{53} + x_{29}x_{54} + x_{29}x_{59} + x_{30}x_{31} + x_{30}x_{33} + x_{30}x_{34} + x_{30}x_{35} + x_{30}x_{39} + x_{30}x_{41} + x_{30}x_{42} + x_{30}x_{45} + x_{30}x_{46} + x_{30}x_{52} + x_{30}x_{54} + x_{30}x_{57} + x_{30}x_{61} + x_{30}x_{63} + x_{31}x_{32} + x_{31}x_{35} + x_{31}x_{36} + x_{31}x_{37} + x_{31}x_{39} + x_{31}x_{41} + x_{31}x_{42} + x_{31}x_{43} + x_{31}x_{44} + x_{31}x_{45} + x_{31}x_{47} + x_{31}x_{50} + x_{31}x_{53} + x_{31}x_{55} + x_{31}x_{58} + x_{31}x_{59} + x_{31}x_{60} + x_{31}x_{61} + x_{32}x_{33} + x_{32}x_{35} + x_{32}x_{36} + x_{32}x_{37} + x_{32}x_{38} + x_{32}x_{40} + x_{32}x_{42} + x_{32}x_{43} + x_{32}x_{44} + x_{32}x_{45} + x_{32}x_{46} + x_{32}x_{50} + x_{32}x_{56} + x_{32}x_{58} + x_{32}x_{59} + x_{32}x_{61} + x_{32}x_{64} + x_{33}x_{34} + x_{33}x_{35} + x_{33}x_{38} + x_{33}x_{39} + x_{33}x_{40} + x_{33}x_{41} + x_{33}x_{42} + x_{33}x_{43} + x_{33}x_{50} + x_{33}x_{53} + x_{33}x_{56} + x_{33}x_{59} + x_{33}x_{61} + x_{33}x_{62} + x_{33}x_{63} + x_{33}x_{64} + x_{34}x_{36} + x_{34}x_{42} + x_{34}x_{43} + x_{34}x_{44} + x_{34}x_{45} + x_{34}x_{46} + x_{34}x_{47} + x_{34}x_{54} + x_{34}x_{55} + x_{34}x_{56} + x_{34}x_{57} + x_{34}x_{62} + x_{35}x_{36} + x_{35}x_{41} + x_{35}x_{42} + x_{35}x_{44} + x_{35}x_{47} + x_{35}x_{48} + x_{35}x_{52} + x_{35}x_{53} + x_{35}x_{57} + x_{35}x_{58} + x_{35}x_{59} + x_{35}x_{61} + x_{35}x_{62} + x_{35}x_{63} + x_{36}x_{38} + x_{36}x_{39} + x_{36}x_{40} + x_{36}x_{42} + x_{36}x_{43} + x_{36}x_{45} + x_{36}x_{46} + x_{36}x_{49} + x_{36}x_{50} + x_{36}x_{54} + x_{36}x_{55} + x_{36}x_{56} + x_{36}x_{57} + x_{36}x_{59} + x_{36}x_{61} + x_{36}x_{62} + x_{36}x_{63} + x_{36}x_{64} + x_{37}x_{41} + x_{37}x_{43} + x_{37}x_{44} + x_{37}x_{48} + x_{37}x_{49} + x_{37}x_{55} + x_{37}x_{58} + x_{37}x_{60} + x_{37}x_{62} + x_{37}x_{64} + x_{38}x_{40} + x_{38}x_{43} + x_{38}x_{44} + x_{38}x_{47} + x_{38}x_{48} + x_{38}x_{52} + x_{38}x_{55} + x_{38}x_{56} + x_{38}x_{57} + x_{38}x_{59} + x_{38}x_{60} + x_{38}x_{61} + x_{38}x_{62} + x_{38}x_{63} + x_{39}x_{43} + x_{39}x_{44} + x_{39}x_{45} + x_{39}x_{47} + x_{39}x_{49} + x_{39}x_{50} + x_{39}x_{52} + x_{39}x_{53} + x_{39}x_{54} + x_{39}x_{56} + x_{39}x_{58} + x_{39}x_{59} + x_{40}x_{41} + x_{40}x_{45} + x_{40}x_{47} + x_{40}x_{48} + x_{40}x_{50} + x_{40}x_{51} + x_{40}x_{57} + x_{40}x_{61} + x_{40}x_{62} + x_{40}x_{64} + x_{41}x_{43} + x_{41}x_{44} + x_{41}x_{45} + x_{41}x_{46} + x_{41}x_{48} + x_{41}x_{51} + x_{41}x_{52} + x_{41}x_{56} + x_{41}x_{58} + x_{41}x_{59} + x_{41}x_{63} + x_{42}x_{43} + x_{42}x_{50} + x_{42}x_{51} + x_{42}x_{54} + x_{42}x_{55} + x_{42}x_{57} + x_{42}x_{58} + x_{42}x_{59} + x_{42}x_{60} + x_{42}x_{61} + x_{42}x_{63} + x_{42}x_{64} + x_{43}x_{44} + x_{43}x_{45} + x_{43}x_{46} + x_{43}x_{48} + x_{43}x_{50} + x_{43}x_{54} + x_{43}x_{61} + x_{43}x_{63} + x_{44}x_{46} + x_{44}x_{49} + x_{44}x_{50} + x_{44}x_{51} + x_{44}x_{55} + x_{44}x_{58} + x_{44}x_{60} + x_{44}x_{62} + x_{45}x_{53} + x_{45}x_{55} + x_{45}x_{58} + x_{45}x_{62} + x_{45}x_{63} + x_{46}x_{47} + x_{46}x_{48} + x_{46}x_{53} + x_{46}x_{54} + x_{46}x_{55} + x_{46}x_{56} + x_{46}x_{57} + x_{46}x_{58} + x_{46}x_{60} + x_{46}x_{62} + x_{46}x_{63} + x_{46}x_{64} + x_{47}x_{48} + x_{47}x_{49} + x_{47}x_{50} + x_{47}x_{53} + x_{47}x_{55} + x_{47}x_{58} + x_{47}x_{61} + x_{47}x_{62} + x_{48}x_{49} + x_{48}x_{51} + x_{48}x_{54} + x_{48}x_{55} + x_{48}x_{56} + x_{48}x_{59} + x_{48}x_{62} + x_{48}x_{63} + x_{48}x_{64} + x_{49}x_{50} + x_{49}x_{52} + x_{49}x_{53} + x_{49}x_{56} + x_{49}x_{59} + x_{49}x_{62} + x_{49}x_{64} + x_{50}x_{51} + x_{50}x_{53} + x_{50}x_{56} + x_{50}x_{57} + x_{50}x_{59} + x_{50}x_{60} + x_{50}x_{62} + x_{50}x_{64} + x_{51}x_{52} + x_{51}x_{59} + x_{51}x_{60} + x_{51}x_{61} + x_{51}x_{62} + x_{52}x_{54} + x_{52}x_{56} + x_{52}x_{57} + x_{52}x_{58} + x_{52}x_{61} + x_{52}x_{62} + x_{52}x_{63} + x_{53}x_{55} + x_{53}x_{56} + x_{53}x_{58} + x_{53}x_{60} + x_{53}x_{61} + x_{53}x_{62} + x_{53}x_{63} + x_{53}x_{64} + x_{54}x_{55} + x_{54}x_{57} + x_{54}x_{58} + x_{54}x_{60} + x_{55}x_{56} + x_{55}x_{59} + x_{55}x_{61} + x_{55}x_{63} + x_{56}x_{57} + x_{56}x_{60} + x_{56}x_{61} + x_{56}x_{63} + x_{56}x_{64} + x_{57}x_{58} + x_{57}x_{60} + x_{57}x_{61} + x_{57}x_{62} + x_{57}x_{63} + x_{57}x_{64} + x_{58}x_{59} + x_{58}x_{60} + x_{58}x_{62} + x_{59}x_{62} + x_{59}x_{63} + x_{59}x_{64} + x_{60}x_{64} + x_{61}x_{62} + x_{63}x_{64} + x_{1} + x_{4} + x_{6} + x_{7} + x_{8} + x_{10} + x_{11} + x_{14} + x_{15} + x_{18} + x_{19} + x_{20} + x_{23} + x_{25} + x_{26} + x_{29} + x_{30} + x_{33} + x_{34} + x_{37} + x_{38} + x_{40} + x_{41} + x_{44} + x_{48} + x_{53} + x_{54} + x_{58} + x_{59} + x_{60} + x_{61}$

$y_{30} = x_{1}x_{2} + x_{1}x_{3} + x_{1}x_{7} + x_{1}x_{8} + x_{1}x_{11} + x_{1}x_{13} + x_{1}x_{17} + x_{1}x_{23} + x_{1}x_{24} + x_{1}x_{28} + x_{1}x_{30} + x_{1}x_{31} + x_{1}x_{32} + x_{1}x_{38} + x_{1}x_{39} + x_{1}x_{40} + x_{1}x_{42} + x_{1}x_{45} + x_{1}x_{46} + x_{1}x_{49} + x_{1}x_{50} + x_{1}x_{52} + x_{1}x_{54} + x_{1}x_{55} + x_{1}x_{56} + x_{1}x_{58} + x_{1}x_{59} + x_{1}x_{61} + x_{1}x_{63} + x_{2}x_{6} + x_{2}x_{7} + x_{2}x_{9} + x_{2}x_{17} + x_{2}x_{19} + x_{2}x_{20} + x_{2}x_{23} + x_{2}x_{24} + x_{2}x_{26} + x_{2}x_{27} + x_{2}x_{28} + x_{2}x_{33} + x_{2}x_{34} + x_{2}x_{35} + x_{2}x_{36} + x_{2}x_{37} + x_{2}x_{38} + x_{2}x_{39} + x_{2}x_{40} + x_{2}x_{46} + x_{2}x_{48} + x_{2}x_{52} + x_{2}x_{54} + x_{2}x_{59} + x_{2}x_{62} + x_{3}x_{4} + x_{3}x_{5} + x_{3}x_{6} + x_{3}x_{8} + x_{3}x_{10} + x_{3}x_{11} + x_{3}x_{12} + x_{3}x_{14} + x_{3}x_{15} + x_{3}x_{16} + x_{3}x_{17} + x_{3}x_{23} + x_{3}x_{25} + x_{3}x_{26} + x_{3}x_{28} + x_{3}x_{29} + x_{3}x_{31} + x_{3}x_{32} + x_{3}x_{34} + x_{3}x_{36} + x_{3}x_{40} + x_{3}x_{41} + x_{3}x_{46} + x_{3}x_{49} + x_{3}x_{50} + x_{3}x_{52} + x_{3}x_{54} + x_{3}x_{62} + x_{3}x_{63} + x_{4}x_{10} + x_{4}x_{14} + x_{4}x_{17} + x_{4}x_{18} + x_{4}x_{20} + x_{4}x_{21} + x_{4}x_{22} + x_{4}x_{23} + x_{4}x_{24} + x_{4}x_{27} + x_{4}x_{29} + x_{4}x_{31} + x_{4}x_{36} + x_{4}x_{38} + x_{4}x_{39} + x_{4}x_{40} + x_{4}x_{42} + x_{4}x_{43} + x_{4}x_{46} + x_{4}x_{50} + x_{4}x_{51} + x_{4}x_{53} + x_{4}x_{58} + x_{4}x_{62} + x_{5}x_{6} + x_{5}x_{7} + x_{5}x_{8} + x_{5}x_{11} + x_{5}x_{12} + x_{5}x_{13} + x_{5}x_{14} + x_{5}x_{17} + x_{5}x_{18} + x_{5}x_{20} + x_{5}x_{22} + x_{5}x_{24} + x_{5}x_{25} + x_{5}x_{26} + x_{5}x_{27} + x_{5}x_{28} + x_{5}x_{29} + x_{5}x_{30} + x_{5}x_{31} + x_{5}x_{32} + x_{5}x_{39} + x_{5}x_{43} + x_{5}x_{45} + x_{5}x_{50} + x_{5}x_{51} + x_{5}x_{52} + x_{5}x_{53} + x_{5}x_{54} + x_{5}x_{57} + x_{5}x_{59} + x_{5}x_{60} + x_{5}x_{61} + x_{6}x_{9} + x_{6}x_{12} + x_{6}x_{13} + x_{6}x_{15} + x_{6}x_{17} + x_{6}x_{18} + x_{6}x_{22} + x_{6}x_{23} + x_{6}x_{24} + x_{6}x_{25} + x_{6}x_{27} + x_{6}x_{28} + x_{6}x_{31} + x_{6}x_{35} + x_{6}x_{36} + x_{6}x_{37} + x_{6}x_{38} + x_{6}x_{39} + x_{6}x_{40} + x_{6}x_{42} + x_{6}x_{44} + x_{6}x_{45} + x_{6}x_{46} + x_{6}x_{47} + x_{6}x_{49} + x_{6}x_{52} + x_{6}x_{53} + x_{6}x_{55} + x_{6}x_{57} + x_{6}x_{59} + x_{6}x_{62} + x_{6}x_{63} + x_{7}x_{11} + x_{7}x_{14} + x_{7}x_{19} + x_{7}x_{21} + x_{7}x_{22} + x_{7}x_{23} + x_{7}x_{24} + x_{7}x_{25} + x_{7}x_{28} + x_{7}x_{29} + x_{7}x_{30} + x_{7}x_{35} + x_{7}x_{37} + x_{7}x_{38} + x_{7}x_{40} + x_{7}x_{41} + x_{7}x_{42} + x_{7}x_{43} + x_{7}x_{44} + x_{7}x_{45} + x_{7}x_{48} + x_{7}x_{49} + x_{7}x_{50} + x_{7}x_{51} + x_{7}x_{56} + x_{7}x_{62} + x_{8}x_{10} + x_{8}x_{12} + x_{8}x_{13} + x_{8}x_{15} + x_{8}x_{16} + x_{8}x_{21} + x_{8}x_{24} + x_{8}x_{26} + x_{8}x_{27} + x_{8}x_{31} + x_{8}x_{33} + x_{8}x_{35} + x_{8}x_{36} + x_{8}x_{40} + x_{8}x_{43} + x_{8}x_{45} + x_{8}x_{46} + x_{8}x_{47} + x_{8}x_{48} + x_{8}x_{49} + x_{8}x_{50} + x_{8}x_{52} + x_{8}x_{54} + x_{8}x_{55} + x_{8}x_{57} + x_{8}x_{59} + x_{8}x_{62} + x_{8}x_{63} + x_{9}x_{13} + x_{9}x_{14} + x_{9}x_{15} + x_{9}x_{16} + x_{9}x_{19} + x_{9}x_{23} + x_{9}x_{24} + x_{9}x_{25} + x_{9}x_{27} + x_{9}x_{31} + x_{9}x_{35} + x_{9}x_{36} + x_{9}x_{39} + x_{9}x_{40} + x_{9}x_{41} + x_{9}x_{47} + x_{9}x_{51} + x_{9}x_{55} + x_{9}x_{56} + x_{9}x_{59} + x_{9}x_{60} + x_{9}x_{64} + x_{10}x_{11} + x_{10}x_{16} + x_{10}x_{17} + x_{10}x_{19} + x_{10}x_{21} + x_{10}x_{24} + x_{10}x_{25} + x_{10}x_{26} + x_{10}x_{30} + x_{10}x_{32} + x_{10}x_{33} + x_{10}x_{34} + x_{10}x_{35} + x_{10}x_{36} + x_{10}x_{38} + x_{10}x_{39} + x_{10}x_{40} + x_{10}x_{41} + x_{10}x_{44} + x_{10}x_{46} + x_{10}x_{47} + x_{10}x_{48} + x_{10}x_{49} + x_{10}x_{52} + x_{10}x_{53} + x_{10}x_{54} + x_{10}x_{55} + x_{10}x_{56} + x_{10}x_{60} + x_{10}x_{61} + x_{10}x_{62} + x_{11}x_{13} + x_{11}x_{15} + x_{11}x_{21} + x_{11}x_{23} + x_{11}x_{24} + x_{11}x_{30} + x_{11}x_{32} + x_{11}x_{33} + x_{11}x_{34} + x_{11}x_{36} + x_{11}x_{38} + x_{11}x_{43} + x_{11}x_{44} + x_{11}x_{46} + x_{11}x_{50} + x_{11}x_{52} + x_{11}x_{54} + x_{11}x_{56} + x_{11}x_{58} + x_{11}x_{59} + x_{11}x_{60} + x_{11}x_{61} + x_{11}x_{63} + x_{11}x_{64} + x_{12}x_{13} + x_{12}x_{14} + x_{12}x_{18} + x_{12}x_{19} + x_{12}x_{20} + x_{12}x_{21} + x_{12}x_{24} + x_{12}x_{27} + x_{12}x_{28} + x_{12}x_{29} + x_{12}x_{31} + x_{12}x_{33} + x_{12}x_{36} + x_{12}x_{37} + x_{12}x_{38} + x_{12}x_{41} + x_{12}x_{42} + x_{12}x_{48} + x_{12}x_{50} + x_{12}x_{53} + x_{12}x_{57} + x_{12}x_{58} + x_{12}x_{60} + x_{12}x_{62} + x_{13}x_{14} + x_{13}x_{15} + x_{13}x_{17} + x_{13}x_{19} + x_{13}x_{20} + x_{13}x_{21} + x_{13}x_{24} + x_{13}x_{26} + x_{13}x_{28} + x_{13}x_{31} + x_{13}x_{32} + x_{13}x_{34} + x_{13}x_{35} + x_{13}x_{36} + x_{13}x_{37} + x_{13}x_{38} + x_{13}x_{39} + x_{13}x_{42} + x_{13}x_{44} + x_{13}x_{46} + x_{13}x_{47} + x_{13}x_{50} + x_{13}x_{51} + x_{13}x_{52} + x_{13}x_{55} + x_{13}x_{62} + x_{13}x_{63} + x_{14}x_{16} + x_{14}x_{18} + x_{14}x_{19} + x_{14}x_{21} + x_{14}x_{22} + x_{14}x_{23} + x_{14}x_{25} + x_{14}x_{27} + x_{14}x_{29} + x_{14}x_{30} + x_{14}x_{31} + x_{14}x_{33} + x_{14}x_{35} + x_{14}x_{37} + x_{14}x_{39} + x_{14}x_{40} + x_{14}x_{42} + x_{14}x_{44} + x_{14}x_{45} + x_{14}x_{46} + x_{14}x_{47} + x_{14}x_{49} + x_{14}x_{53} + x_{14}x_{54} + x_{14}x_{55} + x_{14}x_{56} + x_{14}x_{57} + x_{14}x_{62} + x_{14}x_{63} + x_{15}x_{16} + x_{15}x_{18} + x_{15}x_{19} + x_{15}x_{20} + x_{15}x_{21} + x_{15}x_{22} + x_{15}x_{23} + x_{15}x_{24} + x_{15}x_{25} + x_{15}x_{26} + x_{15}x_{27} + x_{15}x_{31} + x_{15}x_{33} + x_{15}x_{36} + x_{15}x_{38} + x_{15}x_{42} + x_{15}x_{43} + x_{15}x_{45} + x_{15}x_{46} + x_{15}x_{47} + x_{15}x_{51} + x_{15}x_{52} + x_{15}x_{54} + x_{15}x_{57} + x_{15}x_{58} + x_{15}x_{60} + x_{15}x_{62} + x_{15}x_{63} + x_{16}x_{20} + x_{16}x_{21} + x_{16}x_{22} + x_{16}x_{23} + x_{16}x_{24} + x_{16}x_{25} + x_{16}x_{27} + x_{16}x_{29} + x_{16}x_{31} + x_{16}x_{32} + x_{16}x_{33} + x_{16}x_{35} + x_{16}x_{36} + x_{16}x_{37} + x_{16}x_{39} + x_{16}x_{49} + x_{16}x_{52} + x_{16}x_{53} + x_{16}x_{54} + x_{16}x_{56} + x_{16}x_{57} + x_{16}x_{61} + x_{17}x_{18} + x_{17}x_{19} + x_{17}x_{20} + x_{17}x_{24} + x_{17}x_{29} + x_{17}x_{31} + x_{17}x_{32} + x_{17}x_{35} + x_{17}x_{37} + x_{17}x_{40} + x_{17}x_{41} + x_{17}x_{44} + x_{17}x_{49} + x_{17}x_{51} + x_{17}x_{52} + x_{17}x_{53} + x_{17}x_{58} + x_{17}x_{60} + x_{17}x_{61} + x_{17}x_{62} + x_{18}x_{20} + x_{18}x_{21} + x_{18}x_{25} + x_{18}x_{27} + x_{18}x_{34} + x_{18}x_{35} + x_{18}x_{36} + x_{18}x_{37} + x_{18}x_{38} + x_{18}x_{39} + x_{18}x_{41} + x_{18}x_{42} + x_{18}x_{44} + x_{18}x_{45} + x_{18}x_{51} + x_{18}x_{52} + x_{18}x_{56} + x_{18}x_{57} + x_{18}x_{59} + x_{18}x_{60} + x_{18}x_{63} + x_{19}x_{22} + x_{19}x_{23} + x_{19}x_{24} + x_{19}x_{26} + x_{19}x_{27} + x_{19}x_{28} + x_{19}x_{29} + x_{19}x_{33} + x_{19}x_{34} + x_{19}x_{36} + x_{19}x_{39} + x_{19}x_{40} + x_{19}x_{41} + x_{19}x_{42} + x_{19}x_{46} + x_{19}x_{50} + x_{19}x_{51} + x_{19}x_{53} + x_{19}x_{59} + x_{19}x_{60} + x_{19}x_{61} + x_{19}x_{62} + x_{20}x_{21} + x_{20}x_{27} + x_{20}x_{28} + x_{20}x_{29} + x_{20}x_{35} + x_{20}x_{36} + x_{20}x_{37} + x_{20}x_{38} + x_{20}x_{43} + x_{20}x_{50} + x_{20}x_{51} + x_{20}x_{52} + x_{20}x_{54} + x_{20}x_{55} + x_{20}x_{57} + x_{20}x_{59} + x_{20}x_{60} + x_{20}x_{61} + x_{20}x_{62} + x_{20}x_{63} + x_{21}x_{22} + x_{21}x_{25} + x_{21}x_{28} + x_{21}x_{29} + x_{21}x_{32} + x_{21}x_{33} + x_{21}x_{35} + x_{21}x_{36} + x_{21}x_{37} + x_{21}x_{45} + x_{21}x_{46} + x_{21}x_{47} + x_{21}x_{48} + x_{21}x_{49} + x_{21}x_{53} + x_{21}x_{54} + x_{21}x_{55} + x_{21}x_{57} + x_{21}x_{58} + x_{21}x_{59} + x_{21}x_{60} + x_{21}x_{62} + x_{21}x_{64} + x_{22}x_{23} + x_{22}x_{24} + x_{22}x_{25} + x_{22}x_{26} + x_{22}x_{28} + x_{22}x_{29} + x_{22}x_{31} + x_{22}x_{32} + x_{22}x_{35} + x_{22}x_{38} + x_{22}x_{43} + x_{22}x_{44} + x_{22}x_{45} + x_{22}x_{46} + x_{22}x_{51} + x_{22}x_{52} + x_{22}x_{56} + x_{22}x_{59} + x_{22}x_{60} + x_{22}x_{64} + x_{23}x_{26} + x_{23}x_{28} + x_{23}x_{32} + x_{23}x_{33} + x_{23}x_{34} + x_{23}x_{35} + x_{23}x_{36} + x_{23}x_{37} + x_{23}x_{38} + x_{23}x_{39} + x_{23}x_{40} + x_{23}x_{43} + x_{23}x_{45} + x_{23}x_{46} + x_{23}x_{49} + x_{23}x_{51} + x_{23}x_{53} + x_{23}x_{55} + x_{23}x_{56} + x_{23}x_{59} + x_{23}x_{64} + x_{24}x_{25} + x_{24}x_{32} + x_{24}x_{33} + x_{24}x_{35} + x_{24}x_{38} + x_{24}x_{41} + x_{24}x_{48} + x_{24}x_{51} + x_{24}x_{52} + x_{24}x_{54} + x_{24}x_{55} + x_{24}x_{56} + x_{24}x_{57} + x_{24}x_{60} + x_{24}x_{61} + x_{24}x_{63} + x_{25}x_{26} + x_{25}x_{29} + x_{25}x_{30} + x_{25}x_{35} + x_{25}x_{36} + x_{25}x_{37} + x_{25}x_{38} + x_{25}x_{39} + x_{25}x_{40} + x_{25}x_{44} + x_{25}x_{47} + x_{25}x_{50} + x_{25}x_{54} + x_{25}x_{56} + x_{25}x_{60} + x_{26}x_{28} + x_{26}x_{30} + x_{26}x_{31} + x_{26}x_{32} + x_{26}x_{33} + x_{26}x_{34} + x_{26}x_{35} + x_{26}x_{37} + x_{26}x_{40} + x_{26}x_{41} + x_{26}x_{42} + x_{26}x_{45} + x_{26}x_{46} + x_{26}x_{48} + x_{26}x_{49} + x_{26}x_{53} + x_{26}x_{57} + x_{26}x_{62} + x_{26}x_{64} + x_{27}x_{29} + x_{27}x_{36} + x_{27}x_{41} + x_{27}x_{44} + x_{27}x_{45} + x_{27}x_{47} + x_{27}x_{48} + x_{27}x_{49} + x_{27}x_{52} + x_{27}x_{53} + x_{27}x_{54} + x_{27}x_{55} + x_{27}x_{58} + x_{27}x_{59} + x_{27}x_{61} + x_{27}x_{63} + x_{27}x_{64} + x_{28}x_{29} + x_{28}x_{30} + x_{28}x_{31} + x_{28}x_{32} + x_{28}x_{33} + x_{28}x_{36} + x_{28}x_{37} + x_{28}x_{40} + x_{28}x_{41} + x_{28}x_{43} + x_{28}x_{45} + x_{28}x_{47} + x_{28}x_{52} + x_{28}x_{54} + x_{28}x_{55} + x_{28}x_{58} + x_{28}x_{59} + x_{28}x_{60} + x_{28}x_{61} + x_{28}x_{62} + x_{28}x_{63} + x_{29}x_{30} + x_{29}x_{32} + x_{29}x_{34} + x_{29}x_{37} + x_{29}x_{39} + x_{29}x_{41} + x_{29}x_{43} + x_{29}x_{45} + x_{29}x_{46} + x_{29}x_{48} + x_{29}x_{49} + x_{29}x_{50} + x_{29}x_{52} + x_{29}x_{53} + x_{29}x_{59} + x_{29}x_{61} + x_{30}x_{32} + x_{30}x_{33} + x_{30}x_{36} + x_{30}x_{40} + x_{30}x_{42} + x_{30}x_{44} + x_{30}x_{45} + x_{30}x_{46} + x_{30}x_{47} + x_{30}x_{49} + x_{30}x_{50} + x_{30}x_{51} + x_{30}x_{54} + x_{30}x_{56} + x_{30}x_{60} + x_{30}x_{62} + x_{30}x_{64} + x_{31}x_{32} + x_{31}x_{38} + x_{31}x_{40} + x_{31}x_{41} + x_{31}x_{43} + x_{31}x_{44} + x_{31}x_{47} + x_{31}x_{48} + x_{31}x_{49} + x_{31}x_{50} + x_{31}x_{53} + x_{31}x_{61} + x_{31}x_{62} + x_{31}x_{63} + x_{32}x_{34} + x_{32}x_{36} + x_{32}x_{37} + x_{32}x_{39} + x_{32}x_{40} + x_{32}x_{42} + x_{32}x_{43} + x_{32}x_{44} + x_{32}x_{46} + x_{32}x_{49} + x_{32}x_{52} + x_{32}x_{53} + x_{32}x_{55} + x_{32}x_{56} + x_{32}x_{58} + x_{32}x_{59} + x_{32}x_{60} + x_{32}x_{61} + x_{33}x_{34} + x_{33}x_{36} + x_{33}x_{38} + x_{33}x_{40} + x_{33}x_{43} + x_{33}x_{45} + x_{33}x_{46} + x_{33}x_{52} + x_{33}x_{53} + x_{33}x_{54} + x_{33}x_{56} + x_{33}x_{57} + x_{33}x_{58} + x_{33}x_{60} + x_{33}x_{61} + x_{33}x_{62} + x_{33}x_{63} + x_{33}x_{64} + x_{34}x_{35} + x_{34}x_{37} + x_{34}x_{38} + x_{34}x_{40} + x_{34}x_{41} + x_{34}x_{43} + x_{34}x_{44} + x_{34}x_{46} + x_{34}x_{47} + x_{34}x_{48} + x_{34}x_{54} + x_{34}x_{57} + x_{34}x_{59} + x_{34}x_{62} + x_{35}x_{38} + x_{35}x_{39} + x_{35}x_{40} + x_{35}x_{41} + x_{35}x_{48} + x_{35}x_{49} + x_{35}x_{50} + x_{35}x_{54} + x_{35}x_{60} + x_{35}x_{64} + x_{36}x_{37} + x_{36}x_{38} + x_{36}x_{43} + x_{36}x_{46} + x_{36}x_{49} + x_{36}x_{51} + x_{36}x_{52} + x_{36}x_{53} + x_{36}x_{55} + x_{36}x_{57} + x_{36}x_{58} + x_{36}x_{61} + x_{36}x_{62} + x_{36}x_{63} + x_{37}x_{38} + x_{37}x_{39} + x_{37}x_{40} + x_{37}x_{41} + x_{37}x_{43} + x_{37}x_{44} + x_{37}x_{45} + x_{37}x_{46} + x_{37}x_{47} + x_{37}x_{48} + x_{37}x_{52} + x_{37}x_{56} + x_{37}x_{59} + x_{37}x_{60} + x_{37}x_{62} + x_{37}x_{64} + x_{38}x_{41} + x_{38}x_{44} + x_{38}x_{45} + x_{38}x_{47} + x_{38}x_{50} + x_{38}x_{52} + x_{38}x_{53} + x_{38}x_{54} + x_{38}x_{56} + x_{38}x_{57} + x_{38}x_{61} + x_{38}x_{62} + x_{39}x_{41} + x_{39}x_{42} + x_{39}x_{45} + x_{39}x_{47} + x_{39}x_{48} + x_{39}x_{49} + x_{39}x_{51} + x_{39}x_{52} + x_{39}x_{53} + x_{39}x_{54} + x_{39}x_{55} + x_{39}x_{57} + x_{39}x_{60} + x_{39}x_{64} + x_{40}x_{42} + x_{40}x_{44} + x_{40}x_{45} + x_{40}x_{47} + x_{40}x_{51} + x_{40}x_{53} + x_{40}x_{54} + x_{40}x_{55} + x_{40}x_{56} + x_{40}x_{57} + x_{40}x_{59} + x_{40}x_{62} + x_{40}x_{64} + x_{41}x_{44} + x_{41}x_{45} + x_{41}x_{46} + x_{41}x_{47} + x_{41}x_{48} + x_{41}x_{49} + x_{41}x_{51} + x_{41}x_{52} + x_{41}x_{53} + x_{41}x_{55} + x_{41}x_{59} + x_{41}x_{61} + x_{42}x_{43} + x_{42}x_{48} + x_{42}x_{49} + x_{42}x_{52} + x_{42}x_{53} + x_{42}x_{56} + x_{42}x_{57} + x_{42}x_{58} + x_{42}x_{59} + x_{42}x_{60} + x_{42}x_{63} + x_{42}x_{64} + x_{43}x_{45} + x_{43}x_{47} + x_{43}x_{49} + x_{43}x_{50} + x_{43}x_{51} + x_{43}x_{54} + x_{43}x_{57} + x_{43}x_{59} + x_{43}x_{61} + x_{43}x_{62} + x_{43}x_{63} + x_{44}x_{50} + x_{44}x_{52} + x_{44}x_{54} + x_{44}x_{58} + x_{44}x_{60} + x_{44}x_{61} + x_{45}x_{47} + x_{45}x_{48} + x_{45}x_{49} + x_{45}x_{52} + x_{45}x_{55} + x_{45}x_{56} + x_{45}x_{58} + x_{45}x_{59} + x_{46}x_{49} + x_{46}x_{50} + x_{46}x_{51} + x_{46}x_{52} + x_{46}x_{55} + x_{46}x_{56} + x_{46}x_{57} + x_{46}x_{59} + x_{46}x_{60} + x_{46}x_{61} + x_{46}x_{62} + x_{46}x_{64} + x_{47}x_{48} + x_{47}x_{49} + x_{47}x_{50} + x_{47}x_{52} + x_{47}x_{56} + x_{47}x_{58} + x_{47}x_{59} + x_{47}x_{60} + x_{47}x_{63} + x_{48}x_{49} + x_{48}x_{50} + x_{48}x_{52} + x_{48}x_{53} + x_{48}x_{54} + x_{48}x_{55} + x_{48}x_{57} + x_{48}x_{58} + x_{48}x_{60} + x_{48}x_{61} + x_{48}x_{62} + x_{49}x_{50} + x_{49}x_{52} + x_{49}x_{60} + x_{49}x_{61} + x_{49}x_{63} + x_{49}x_{64} + x_{50}x_{51} + x_{50}x_{52} + x_{50}x_{55} + x_{50}x_{56} + x_{50}x_{58} + x_{50}x_{59} + x_{50}x_{62} + x_{50}x_{64} + x_{51}x_{52} + x_{51}x_{53} + x_{51}x_{55} + x_{51}x_{56} + x_{51}x_{59} + x_{51}x_{60} + x_{51}x_{63} + x_{51}x_{64} + x_{52}x_{53} + x_{52}x_{55} + x_{52}x_{58} + x_{52}x_{59} + x_{52}x_{61} + x_{52}x_{63} + x_{52}x_{64} + x_{53}x_{54} + x_{53}x_{55} + x_{53}x_{56} + x_{53}x_{57} + x_{53}x_{58} + x_{53}x_{59} + x_{53}x_{60} + x_{53}x_{62} + x_{53}x_{63} + x_{53}x_{64} + x_{54}x_{55} + x_{54}x_{56} + x_{54}x_{59} + x_{54}x_{61} + x_{55}x_{57} + x_{55}x_{58} + x_{55}x_{60} + x_{55}x_{61} + x_{55}x_{62} + x_{56}x_{57} + x_{56}x_{58} + x_{56}x_{59} + x_{56}x_{63} + x_{56}x_{64} + x_{57}x_{58} + x_{57}x_{59} + x_{57}x_{61} + x_{57}x_{64} + x_{58}x_{62} + x_{58}x_{63} + x_{59}x_{64} + x_{60}x_{61} + x_{60}x_{62} + x_{60}x_{64} + x_{61}x_{62} + x_{61}x_{64} + x_{62}x_{63} + x_{3} + x_{4} + x_{5} + x_{7} + x_{8} + x_{9} + x_{10} + x_{12} + x_{13} + x_{14} + x_{16} + x_{17} + x_{18} + x_{22} + x_{24} + x_{25} + x_{26} + x_{28} + x_{29} + x_{31} + x_{32} + x_{34} + x_{35} + x_{37} + x_{39} + x_{40} + x_{42} + x_{44} + x_{45} + x_{46} + x_{48} + x_{49} + x_{50} + x_{51} + x_{54} + x_{55} + x_{56} + x_{60} + x_{62} + 1$

$y_{31} = x_{1}x_{2} + x_{1}x_{3} + x_{1}x_{4} + x_{1}x_{5} + x_{1}x_{7} + x_{1}x_{9} + x_{1}x_{13} + x_{1}x_{14} + x_{1}x_{15} + x_{1}x_{16} + x_{1}x_{17} + x_{1}x_{19} + x_{1}x_{21} + x_{1}x_{22} + x_{1}x_{24} + x_{1}x_{27} + x_{1}x_{28} + x_{1}x_{29} + x_{1}x_{34} + x_{1}x_{35} + x_{1}x_{39} + x_{1}x_{42} + x_{1}x_{48} + x_{1}x_{50} + x_{1}x_{52} + x_{1}x_{54} + x_{1}x_{55} + x_{1}x_{56} + x_{1}x_{61} + x_{1}x_{62} + x_{2}x_{4} + x_{2}x_{5} + x_{2}x_{7} + x_{2}x_{9} + x_{2}x_{10} + x_{2}x_{11} + x_{2}x_{14} + x_{2}x_{16} + x_{2}x_{17} + x_{2}x_{19} + x_{2}x_{25} + x_{2}x_{26} + x_{2}x_{27} + x_{2}x_{28} + x_{2}x_{29} + x_{2}x_{30} + x_{2}x_{32} + x_{2}x_{35} + x_{2}x_{38} + x_{2}x_{43} + x_{2}x_{46} + x_{2}x_{47} + x_{2}x_{48} + x_{2}x_{49} + x_{2}x_{50} + x_{2}x_{52} + x_{2}x_{53} + x_{2}x_{56} + x_{2}x_{59} + x_{2}x_{60} + x_{2}x_{63} + x_{2}x_{64} + x_{3}x_{4} + x_{3}x_{7} + x_{3}x_{8} + x_{3}x_{9} + x_{3}x_{10} + x_{3}x_{11} + x_{3}x_{12} + x_{3}x_{13} + x_{3}x_{14} + x_{3}x_{15} + x_{3}x_{16} + x_{3}x_{17} + x_{3}x_{21} + x_{3}x_{23} + x_{3}x_{24} + x_{3}x_{27} + x_{3}x_{29} + x_{3}x_{30} + x_{3}x_{35} + x_{3}x_{36} + x_{3}x_{37} + x_{3}x_{39} + x_{3}x_{41} + x_{3}x_{42} + x_{3}x_{45} + x_{3}x_{47} + x_{3}x_{50} + x_{3}x_{53} + x_{3}x_{54} + x_{3}x_{56} + x_{3}x_{63} + x_{3}x_{64} + x_{4}x_{5} + x_{4}x_{6} + x_{4}x_{7} + x_{4}x_{9} + x_{4}x_{10} + x_{4}x_{11} + x_{4}x_{14} + x_{4}x_{15} + x_{4}x_{16} + x_{4}x_{18} + x_{4}x_{19} + x_{4}x_{20} + x_{4}x_{22} + x_{4}x_{23} + x_{4}x_{24} + x_{4}x_{25} + x_{4}x_{26} + x_{4}x_{27} + x_{4}x_{28} + x_{4}x_{31} + x_{4}x_{33} + x_{4}x_{35} + x_{4}x_{42} + x_{4}x_{43} + x_{4}x_{44} + x_{4}x_{45} + x_{4}x_{49} + x_{4}x_{50} + x_{4}x_{52} + x_{4}x_{55} + x_{4}x_{56} + x_{4}x_{57} + x_{4}x_{58} + x_{4}x_{60} + x_{4}x_{61} + x_{5}x_{6} + x_{5}x_{9} + x_{5}x_{12} + x_{5}x_{15} + x_{5}x_{16} + x_{5}x_{17} + x_{5}x_{18} + x_{5}x_{19} + x_{5}x_{22} + x_{5}x_{24} + x_{5}x_{25} + x_{5}x_{28} + x_{5}x_{29} + x_{5}x_{30} + x_{5}x_{33} + x_{5}x_{34} + x_{5}x_{35} + x_{5}x_{37} + x_{5}x_{38} + x_{5}x_{44} + x_{5}x_{46} + x_{5}x_{47} + x_{5}x_{48} + x_{5}x_{49} + x_{5}x_{51} + x_{5}x_{52} + x_{5}x_{53} + x_{5}x_{54} + x_{5}x_{56} + x_{5}x_{57} + x_{5}x_{59} + x_{5}x_{60} + x_{5}x_{63} + x_{5}x_{64} + x_{6}x_{8} + x_{6}x_{10} + x_{6}x_{11} + x_{6}x_{12} + x_{6}x_{17} + x_{6}x_{18} + x_{6}x_{20} + x_{6}x_{23} + x_{6}x_{24} + x_{6}x_{25} + x_{6}x_{28} + x_{6}x_{30} + x_{6}x_{32} + x_{6}x_{36} + x_{6}x_{37} + x_{6}x_{41} + x_{6}x_{42} + x_{6}x_{45} + x_{6}x_{50} + x_{6}x_{52} + x_{6}x_{54} + x_{6}x_{57} + x_{6}x_{60} + x_{6}x_{63} + x_{6}x_{64} + x_{7}x_{19} + x_{7}x_{21} + x_{7}x_{23} + x_{7}x_{24} + x_{7}x_{26} + x_{7}x_{28} + x_{7}x_{29} + x_{7}x_{30} + x_{7}x_{31} + x_{7}x_{32} + x_{7}x_{33} + x_{7}x_{34} + x_{7}x_{35} + x_{7}x_{37} + x_{7}x_{38} + x_{7}x_{40} + x_{7}x_{42} + x_{7}x_{43} + x_{7}x_{44} + x_{7}x_{45} + x_{7}x_{47} + x_{7}x_{50} + x_{7}x_{51} + x_{7}x_{53} + x_{7}x_{56} + x_{7}x_{57} + x_{7}x_{58} + x_{7}x_{60} + x_{7}x_{62} + x_{7}x_{64} + x_{8}x_{10} + x_{8}x_{12} + x_{8}x_{13} + x_{8}x_{14} + x_{8}x_{16} + x_{8}x_{17} + x_{8}x_{18} + x_{8}x_{25} + x_{8}x_{27} + x_{8}x_{28} + x_{8}x_{29} + x_{8}x_{30} + x_{8}x_{31} + x_{8}x_{32} + x_{8}x_{33} + x_{8}x_{36} + x_{8}x_{37} + x_{8}x_{39} + x_{8}x_{43} + x_{8}x_{46} + x_{8}x_{49} + x_{8}x_{51} + x_{8}x_{55} + x_{8}x_{57} + x_{8}x_{60} + x_{8}x_{63} + x_{8}x_{64} + x_{9}x_{10} + x_{9}x_{12} + x_{9}x_{14} + x_{9}x_{18} + x_{9}x_{19} + x_{9}x_{22} + x_{9}x_{23} + x_{9}x_{24} + x_{9}x_{25} + x_{9}x_{26} + x_{9}x_{30} + x_{9}x_{31} + x_{9}x_{35} + x_{9}x_{37} + x_{9}x_{38} + x_{9}x_{39} + x_{9}x_{40} + x_{9}x_{42} + x_{9}x_{43} + x_{9}x_{49} + x_{9}x_{52} + x_{9}x_{54} + x_{9}x_{55} + x_{9}x_{56} + x_{9}x_{57} + x_{9}x_{58} + x_{9}x_{59} + x_{9}x_{61} + x_{9}x_{63} + x_{10}x_{11} + x_{10}x_{16} + x_{10}x_{18} + x_{10}x_{22} + x_{10}x_{25} + x_{10}x_{31} + x_{10}x_{32} + x_{10}x_{34} + x_{10}x_{35} + x_{10}x_{36} + x_{10}x_{38} + x_{10}x_{41} + x_{10}x_{42} + x_{10}x_{45} + x_{10}x_{46} + x_{10}x_{47} + x_{10}x_{48} + x_{10}x_{50} + x_{10}x_{51} + x_{10}x_{52} + x_{10}x_{53} + x_{10}x_{54} + x_{10}x_{55} + x_{10}x_{56} + x_{10}x_{57} + x_{10}x_{61} + x_{11}x_{13} + x_{11}x_{14} + x_{11}x_{16} + x_{11}x_{21} + x_{11}x_{22} + x_{11}x_{23} + x_{11}x_{25} + x_{11}x_{26} + x_{11}x_{27} + x_{11}x_{31} + x_{11}x_{32} + x_{11}x_{33} + x_{11}x_{34} + x_{11}x_{35} + x_{11}x_{36} + x_{11}x_{38} + x_{11}x_{39} + x_{11}x_{42} + x_{11}x_{44} + x_{11}x_{46} + x_{11}x_{47} + x_{11}x_{48} + x_{11}x_{49} + x_{11}x_{50} + x_{11}x_{54} + x_{11}x_{56} + x_{11}x_{57} + x_{11}x_{63} + x_{12}x_{13} + x_{12}x_{15} + x_{12}x_{17} + x_{12}x_{19} + x_{12}x_{20} + x_{12}x_{22} + x_{12}x_{25} + x_{12}x_{26} + x_{12}x_{27} + x_{12}x_{30} + x_{12}x_{31} + x_{12}x_{32} + x_{12}x_{33} + x_{12}x_{34} + x_{12}x_{36} + x_{12}x_{39} + x_{12}x_{40} + x_{12}x_{41} + x_{12}x_{42} + x_{12}x_{45} + x_{12}x_{46} + x_{12}x_{52} + x_{12}x_{59} + x_{12}x_{60} + x_{12}x_{62} + x_{12}x_{64} + x_{13}x_{14} + x_{13}x_{17} + x_{13}x_{18} + x_{13}x_{19} + x_{13}x_{24} + x_{13}x_{26} + x_{13}x_{27} + x_{13}x_{28} + x_{13}x_{29} + x_{13}x_{30} + x_{13}x_{32} + x_{13}x_{35} + x_{13}x_{37} + x_{13}x_{39} + x_{13}x_{40} + x_{13}x_{41} + x_{13}x_{42} + x_{13}x_{43} + x_{13}x_{45} + x_{13}x_{46} + x_{13}x_{47} + x_{13}x_{48} + x_{13}x_{49} + x_{13}x_{53} + x_{13}x_{54} + x_{13}x_{55} + x_{13}x_{56} + x_{13}x_{57} + x_{13}x_{58} + x_{13}x_{59} + x_{13}x_{64} + x_{14}x_{21} + x_{14}x_{23} + x_{14}x_{24} + x_{14}x_{30} + x_{14}x_{31} + x_{14}x_{33} + x_{14}x_{36} + x_{14}x_{39} + x_{14}x_{40} + x_{14}x_{41} + x_{14}x_{43} + x_{14}x_{44} + x_{14}x_{47} + x_{14}x_{48} + x_{14}x_{52} + x_{14}x_{53} + x_{14}x_{55} + x_{14}x_{57} + x_{14}x_{60} + x_{14}x_{61} + x_{14}x_{62} + x_{14}x_{63} + x_{14}x_{64} + x_{15}x_{16} + x_{15}x_{17} + x_{15}x_{18} + x_{15}x_{21} + x_{15}x_{22} + x_{15}x_{24} + x_{15}x_{26} + x_{15}x_{27} + x_{15}x_{28} + x_{15}x_{29} + x_{15}x_{30} + x_{15}x_{35} + x_{15}x_{36} + x_{15}x_{37} + x_{15}x_{38} + x_{15}x_{39} + x_{15}x_{41} + x_{15}x_{42} + x_{15}x_{43} + x_{15}x_{44} + x_{15}x_{46} + x_{15}x_{48} + x_{15}x_{49} + x_{15}x_{50} + x_{15}x_{52} + x_{15}x_{53} + x_{15}x_{54} + x_{15}x_{55} + x_{15}x_{56} + x_{15}x_{57} + x_{15}x_{59} + x_{15}x_{62} + x_{15}x_{63} + x_{16}x_{19} + x_{16}x_{22} + x_{16}x_{24} + x_{16}x_{25} + x_{16}x_{26} + x_{16}x_{27} + x_{16}x_{29} + x_{16}x_{31} + x_{16}x_{34} + x_{16}x_{35} + x_{16}x_{38} + x_{16}x_{41} + x_{16}x_{43} + x_{16}x_{44} + x_{16}x_{45} + x_{16}x_{46} + x_{16}x_{47} + x_{16}x_{50} + x_{16}x_{51} + x_{16}x_{52} + x_{16}x_{55} + x_{16}x_{56} + x_{16}x_{57} + x_{16}x_{58} + x_{16}x_{59} + x_{16}x_{62} + x_{16}x_{64} + x_{17}x_{21} + x_{17}x_{22} + x_{17}x_{23} + x_{17}x_{25} + x_{17}x_{26} + x_{17}x_{27} + x_{17}x_{32} + x_{17}x_{36} + x_{17}x_{37} + x_{17}x_{39} + x_{17}x_{42} + x_{17}x_{44} + x_{17}x_{48} + x_{17}x_{49} + x_{17}x_{51} + x_{17}x_{52} + x_{17}x_{53} + x_{17}x_{56} + x_{17}x_{57} + x_{17}x_{58} + x_{17}x_{60} + x_{17}x_{61} + x_{17}x_{62} + x_{18}x_{19} + x_{18}x_{22} + x_{18}x_{23} + x_{18}x_{24} + x_{18}x_{25} + x_{18}x_{29} + x_{18}x_{30} + x_{18}x_{31} + x_{18}x_{32} + x_{18}x_{34} + x_{18}x_{40} + x_{18}x_{42} + x_{18}x_{52} + x_{18}x_{55} + x_{18}x_{57} + x_{19}x_{20} + x_{19}x_{24} + x_{19}x_{25} + x_{19}x_{27} + x_{19}x_{29} + x_{19}x_{30} + x_{19}x_{31} + x_{19}x_{32} + x_{19}x_{35} + x_{19}x_{36} + x_{19}x_{38} + x_{19}x_{44} + x_{19}x_{45} + x_{19}x_{46} + x_{19}x_{48} + x_{19}x_{50} + x_{19}x_{51} + x_{19}x_{54} + x_{19}x_{55} + x_{19}x_{57} + x_{19}x_{60} + x_{19}x_{61} + x_{19}x_{64} + x_{20}x_{21} + x_{20}x_{23} + x_{20}x_{24} + x_{20}x_{26} + x_{20}x_{28} + x_{20}x_{29} + x_{20}x_{31} + x_{20}x_{33} + x_{20}x_{34} + x_{20}x_{36} + x_{20}x_{38} + x_{20}x_{40} + x_{20}x_{41} + x_{20}x_{45} + x_{20}x_{46} + x_{20}x_{47} + x_{20}x_{48} + x_{20}x_{49} + x_{20}x_{50} + x_{20}x_{51} + x_{20}x_{52} + x_{20}x_{53} + x_{20}x_{55} + x_{20}x_{60} + x_{20}x_{61} + x_{20}x_{62} + x_{20}x_{63} + x_{20}x_{64} + x_{21}x_{22} + x_{21}x_{24} + x_{21}x_{25} + x_{21}x_{27} + x_{21}x_{29} + x_{21}x_{31} + x_{21}x_{42} + x_{21}x_{43} + x_{21}x_{46} + x_{21}x_{47} + x_{21}x_{48} + x_{21}x_{49} + x_{21}x_{52} + x_{21}x_{59} + x_{21}x_{61} + x_{21}x_{62} + x_{21}x_{63} + x_{21}x_{64} + x_{22}x_{24} + x_{22}x_{25} + x_{22}x_{29} + x_{22}x_{30} + x_{22}x_{31} + x_{22}x_{32} + x_{22}x_{33} + x_{22}x_{36} + x_{22}x_{38} + x_{22}x_{43} + x_{22}x_{45} + x_{22}x_{47} + x_{22}x_{49} + x_{22}x_{52} + x_{22}x_{53} + x_{22}x_{54} + x_{22}x_{55} + x_{22}x_{56} + x_{22}x_{59} + x_{22}x_{61} + x_{22}x_{62} + x_{22}x_{64} + x_{23}x_{24} + x_{23}x_{25} + x_{23}x_{31} + x_{23}x_{32} + x_{23}x_{33} + x_{23}x_{34} + x_{23}x_{36} + x_{23}x_{37} + x_{23}x_{38} + x_{23}x_{41} + x_{23}x_{43} + x_{23}x_{44} + x_{23}x_{45} + x_{23}x_{46} + x_{23}x_{47} + x_{23}x_{48} + x_{23}x_{51} + x_{23}x_{52} + x_{23}x_{53} + x_{23}x_{54} + x_{23}x_{56} + x_{23}x_{58} + x_{23}x_{60} + x_{23}x_{64} + x_{24}x_{26} + x_{24}x_{28} + x_{24}x_{30} + x_{24}x_{34} + x_{24}x_{35} + x_{24}x_{36} + x_{24}x_{37} + x_{24}x_{39} + x_{24}x_{49} + x_{24}x_{50} + x_{24}x_{52} + x_{24}x_{54} + x_{24}x_{55} + x_{24}x_{56} + x_{24}x_{57} + x_{24}x_{59} + x_{24}x_{62} + x_{24}x_{63} + x_{24}x_{64} + x_{25}x_{26} + x_{25}x_{28} + x_{25}x_{33} + x_{25}x_{34} + x_{25}x_{38} + x_{25}x_{40} + x_{25}x_{41} + x_{25}x_{46} + x_{25}x_{47} + x_{25}x_{52} + x_{25}x_{53} + x_{25}x_{54} + x_{25}x_{58} + x_{25}x_{61} + x_{25}x_{63} + x_{26}x_{27} + x_{26}x_{29} + x_{26}x_{30} + x_{26}x_{32} + x_{26}x_{34} + x_{26}x_{36} + x_{26}x_{37} + x_{26}x_{41} + x_{26}x_{45} + x_{26}x_{46} + x_{26}x_{47} + x_{26}x_{50} + x_{26}x_{52} + x_{26}x_{53} + x_{26}x_{54} + x_{26}x_{57} + x_{26}x_{58} + x_{26}x_{59} + x_{26}x_{61} + x_{27}x_{30} + x_{27}x_{31} + x_{27}x_{34} + x_{27}x_{37} + x_{27}x_{40} + x_{27}x_{42} + x_{27}x_{43} + x_{27}x_{44} + x_{27}x_{45} + x_{27}x_{46} + x_{27}x_{47} + x_{27}x_{48} + x_{27}x_{49} + x_{27}x_{50} + x_{27}x_{51} + x_{27}x_{52} + x_{27}x_{53} + x_{27}x_{54} + x_{27}x_{57} + x_{27}x_{59} + x_{27}x_{60} + x_{27}x_{61} + x_{27}x_{62} + x_{27}x_{63} + x_{28}x_{29} + x_{28}x_{32} + x_{28}x_{33} + x_{28}x_{36} + x_{28}x_{37} + x_{28}x_{38} + x_{28}x_{40} + x_{28}x_{41} + x_{28}x_{42} + x_{28}x_{43} + x_{28}x_{44} + x_{28}x_{45} + x_{28}x_{48} + x_{28}x_{50} + x_{28}x_{54} + x_{28}x_{55} + x_{28}x_{56} + x_{28}x_{57} + x_{28}x_{61} + x_{28}x_{62} + x_{28}x_{64} + x_{29}x_{30} + x_{29}x_{34} + x_{29}x_{35} + x_{29}x_{38} + x_{29}x_{41} + x_{29}x_{42} + x_{29}x_{43} + x_{29}x_{44} + x_{29}x_{46} + x_{29}x_{47} + x_{29}x_{49} + x_{29}x_{51} + x_{29}x_{53} + x_{29}x_{55} + x_{29}x_{57} + x_{29}x_{59} + x_{29}x_{60} + x_{29}x_{63} + x_{29}x_{64} + x_{30}x_{32} + x_{30}x_{33} + x_{30}x_{35} + x_{30}x_{37} + x_{30}x_{38} + x_{30}x_{40} + x_{30}x_{41} + x_{30}x_{43} + x_{30}x_{46} + x_{30}x_{53} + x_{30}x_{54} + x_{30}x_{56} + x_{30}x_{57} + x_{30}x_{58} + x_{30}x_{59} + x_{30}x_{62} + x_{31}x_{38} + x_{31}x_{40} + x_{31}x_{41} + x_{31}x_{44} + x_{31}x_{46} + x_{31}x_{47} + x_{31}x_{51} + x_{31}x_{54} + x_{31}x_{60} + x_{31}x_{61} + x_{31}x_{62} + x_{31}x_{64} + x_{32}x_{33} + x_{32}x_{34} + x_{32}x_{38} + x_{32}x_{40} + x_{32}x_{41} + x_{32}x_{43} + x_{32}x_{46} + x_{32}x_{47} + x_{32}x_{49} + x_{32}x_{50} + x_{32}x_{52} + x_{32}x_{53} + x_{32}x_{56} + x_{32}x_{58} + x_{32}x_{59} + x_{32}x_{61} + x_{32}x_{64} + x_{33}x_{35} + x_{33}x_{39} + x_{33}x_{40} + x_{33}x_{42} + x_{33}x_{44} + x_{33}x_{49} + x_{33}x_{50} + x_{33}x_{52} + x_{33}x_{54} + x_{33}x_{55} + x_{33}x_{57} + x_{33}x_{58} + x_{34}x_{35} + x_{34}x_{37} + x_{34}x_{41} + x_{34}x_{43} + x_{34}x_{44} + x_{34}x_{46} + x_{34}x_{48} + x_{34}x_{49} + x_{34}x_{52} + x_{34}x_{53} + x_{34}x_{56} + x_{34}x_{57} + x_{34}x_{58} + x_{34}x_{60} + x_{34}x_{61} + x_{34}x_{62} + x_{34}x_{64} + x_{35}x_{37} + x_{35}x_{39} + x_{35}x_{43} + x_{35}x_{44} + x_{35}x_{45} + x_{35}x_{46} + x_{35}x_{47} + x_{35}x_{49} + x_{35}x_{51} + x_{35}x_{53} + x_{35}x_{54} + x_{35}x_{55} + x_{35}x_{57} + x_{35}x_{61} + x_{35}x_{62} + x_{35}x_{63} + x_{36}x_{37} + x_{36}x_{38} + x_{36}x_{39} + x_{36}x_{41} + x_{36}x_{42} + x_{36}x_{44} + x_{36}x_{49} + x_{36}x_{50} + x_{36}x_{51} + x_{36}x_{53} + x_{36}x_{54} + x_{36}x_{55} + x_{36}x_{57} + x_{36}x_{58} + x_{36}x_{61} + x_{36}x_{62} + x_{36}x_{63} + x_{37}x_{38} + x_{37}x_{40} + x_{37}x_{41} + x_{37}x_{42} + x_{37}x_{45} + x_{37}x_{46} + x_{37}x_{48} + x_{37}x_{49} + x_{37}x_{50} + x_{37}x_{52} + x_{37}x_{54} + x_{37}x_{55} + x_{37}x_{56} + x_{37}x_{57} + x_{37}x_{60} + x_{37}x_{61} + x_{37}x_{63} + x_{37}x_{64} + x_{38}x_{40} + x_{38}x_{41} + x_{38}x_{42} + x_{38}x_{44} + x_{38}x_{47} + x_{38}x_{50} + x_{38}x_{54} + x_{38}x_{56} + x_{38}x_{58} + x_{38}x_{59} + x_{38}x_{60} + x_{39}x_{40} + x_{39}x_{41} + x_{39}x_{44} + x_{39}x_{45} + x_{39}x_{46} + x_{39}x_{48} + x_{39}x_{49} + x_{39}x_{55} + x_{39}x_{57} + x_{39}x_{63} + x_{40}x_{42} + x_{40}x_{43} + x_{40}x_{44} + x_{40}x_{45} + x_{40}x_{46} + x_{40}x_{47} + x_{40}x_{48} + x_{40}x_{50} + x_{40}x_{52} + x_{40}x_{55} + x_{40}x_{61} + x_{40}x_{64} + x_{41}x_{44} + x_{41}x_{46} + x_{41}x_{48} + x_{41}x_{50} + x_{41}x_{53} + x_{41}x_{58} + x_{41}x_{59} + x_{41}x_{63} + x_{41}x_{64} + x_{42}x_{44} + x_{42}x_{48} + x_{42}x_{50} + x_{42}x_{53} + x_{42}x_{54} + x_{42}x_{58} + x_{42}x_{59} + x_{42}x_{60} + x_{42}x_{61} + x_{42}x_{63} + x_{42}x_{64} + x_{43}x_{44} + x_{43}x_{48} + x_{43}x_{49} + x_{43}x_{50} + x_{43}x_{52} + x_{43}x_{53} + x_{43}x_{55} + x_{44}x_{47} + x_{44}x_{48} + x_{44}x_{49} + x_{44}x_{51} + x_{44}x_{52} + x_{44}x_{53} + x_{44}x_{55} + x_{44}x_{56} + x_{44}x_{57} + x_{44}x_{60} + x_{45}x_{47} + x_{45}x_{48} + x_{45}x_{50} + x_{45}x_{51} + x_{45}x_{54} + x_{45}x_{60} + x_{45}x_{62} + x_{46}x_{47} + x_{46}x_{52} + x_{46}x_{53} + x_{46}x_{56} + x_{46}x_{57} + x_{46}x_{58} + x_{46}x_{59} + x_{46}x_{61} + x_{46}x_{63} + x_{46}x_{64} + x_{47}x_{48} + x_{47}x_{49} + x_{47}x_{51} + x_{47}x_{52} + x_{47}x_{54} + x_{47}x_{55} + x_{47}x_{60} + x_{47}x_{62} + x_{48}x_{49} + x_{48}x_{51} + x_{48}x_{52} + x_{48}x_{56} + x_{48}x_{62} + x_{48}x_{63} + x_{48}x_{64} + x_{49}x_{52} + x_{49}x_{53} + x_{49}x_{54} + x_{49}x_{55} + x_{49}x_{57} + x_{49}x_{61} + x_{49}x_{62} + x_{49}x_{63} + x_{50}x_{51} + x_{50}x_{54} + x_{50}x_{55} + x_{50}x_{56} + x_{50}x_{57} + x_{50}x_{59} + x_{50}x_{60} + x_{50}x_{64} + x_{51}x_{52} + x_{51}x_{53} + x_{51}x_{55} + x_{51}x_{57} + x_{51}x_{58} + x_{51}x_{59} + x_{51}x_{61} + x_{51}x_{64} + x_{52}x_{55} + x_{52}x_{57} + x_{52}x_{59} + x_{52}x_{62} + x_{52}x_{63} + x_{52}x_{64} + x_{53}x_{55} + x_{53}x_{58} + x_{53}x_{59} + x_{53}x_{60} + x_{53}x_{61} + x_{54}x_{55} + x_{54}x_{56} + x_{54}x_{58} + x_{54}x_{59} + x_{54}x_{62} + x_{54}x_{63} + x_{54}x_{64} + x_{55}x_{56} + x_{55}x_{57} + x_{55}x_{59} + x_{55}x_{61} + x_{55}x_{64} + x_{56}x_{57} + x_{56}x_{59} + x_{56}x_{61} + x_{56}x_{62} + x_{56}x_{64} + x_{57}x_{59} + x_{57}x_{60} + x_{57}x_{61} + x_{57}x_{62} + x_{57}x_{63} + x_{57}x_{64} + x_{58}x_{59} + x_{58}x_{60} + x_{58}x_{61} + x_{58}x_{62} + x_{59}x_{60} + x_{59}x_{61} + x_{59}x_{62} + x_{59}x_{64} + x_{60}x_{62} + x_{60}x_{63} + x_{61}x_{62} + x_{61}x_{63} + x_{63}x_{64} + x_{1} + x_{2} + x_{3} + x_{4} + x_{5} + x_{6} + x_{7} + x_{8} + x_{9} + x_{11} + x_{19} + x_{23} + x_{25} + x_{26} + x_{27} + x_{29} + x_{30} + x_{31} + x_{32} + x_{35} + x_{36} + x_{38} + x_{40} + x_{41} + x_{42} + x_{44} + x_{45} + x_{47} + x_{49} + x_{52} + x_{53} + x_{54} + x_{55} + x_{58} + x_{61} + 1$

$y_{32} = x_{1}x_{3} + x_{1}x_{6} + x_{1}x_{8} + x_{1}x_{11} + x_{1}x_{13} + x_{1}x_{15} + x_{1}x_{16} + x_{1}x_{18} + x_{1}x_{19} + x_{1}x_{20} + x_{1}x_{21} + x_{1}x_{23} + x_{1}x_{24} + x_{1}x_{28} + x_{1}x_{30} + x_{1}x_{32} + x_{1}x_{34} + x_{1}x_{36} + x_{1}x_{44} + x_{1}x_{47} + x_{1}x_{52} + x_{1}x_{56} + x_{1}x_{57} + x_{1}x_{58} + x_{1}x_{60} + x_{1}x_{63} + x_{1}x_{64} + x_{2}x_{3} + x_{2}x_{4} + x_{2}x_{5} + x_{2}x_{6} + x_{2}x_{7} + x_{2}x_{10} + x_{2}x_{11} + x_{2}x_{17} + x_{2}x_{19} + x_{2}x_{24} + x_{2}x_{25} + x_{2}x_{26} + x_{2}x_{27} + x_{2}x_{29} + x_{2}x_{30} + x_{2}x_{35} + x_{2}x_{36} + x_{2}x_{38} + x_{2}x_{39} + x_{2}x_{40} + x_{2}x_{43} + x_{2}x_{44} + x_{2}x_{45} + x_{2}x_{48} + x_{2}x_{49} + x_{2}x_{53} + x_{2}x_{54} + x_{2}x_{55} + x_{2}x_{56} + x_{2}x_{57} + x_{2}x_{60} + x_{2}x_{61} + x_{2}x_{64} + x_{3}x_{8} + x_{3}x_{9} + x_{3}x_{12} + x_{3}x_{14} + x_{3}x_{15} + x_{3}x_{16} + x_{3}x_{19} + x_{3}x_{21} + x_{3}x_{26} + x_{3}x_{28} + x_{3}x_{29} + x_{3}x_{32} + x_{3}x_{35} + x_{3}x_{36} + x_{3}x_{38} + x_{3}x_{40} + x_{3}x_{43} + x_{3}x_{44} + x_{3}x_{45} + x_{3}x_{47} + x_{3}x_{48} + x_{3}x_{49} + x_{3}x_{50} + x_{3}x_{53} + x_{3}x_{54} + x_{3}x_{55} + x_{3}x_{56} + x_{3}x_{57} + x_{3}x_{58} + x_{3}x_{60} + x_{3}x_{64} + x_{4}x_{7} + x_{4}x_{9} + x_{4}x_{10} + x_{4}x_{14} + x_{4}x_{16} + x_{4}x_{17} + x_{4}x_{18} + x_{4}x_{19} + x_{4}x_{21} + x_{4}x_{22} + x_{4}x_{23} + x_{4}x_{25} + x_{4}x_{26} + x_{4}x_{27} + x_{4}x_{29} + x_{4}x_{30} + x_{4}x_{31} + x_{4}x_{32} + x_{4}x_{33} + x_{4}x_{34} + x_{4}x_{35} + x_{4}x_{42} + x_{4}x_{43} + x_{4}x_{44} + x_{4}x_{46} + x_{4}x_{47} + x_{4}x_{50} + x_{4}x_{51} + x_{4}x_{57} + x_{4}x_{61} + x_{4}x_{62} + x_{4}x_{63} + x_{5}x_{6} + x_{5}x_{7} + x_{5}x_{8} + x_{5}x_{11} + x_{5}x_{14} + x_{5}x_{17} + x_{5}x_{18} + x_{5}x_{19} + x_{5}x_{27} + x_{5}x_{29} + x_{5}x_{37} + x_{5}x_{38} + x_{5}x_{39} + x_{5}x_{41} + x_{5}x_{42} + x_{5}x_{45} + x_{5}x_{49} + x_{5}x_{52} + x_{5}x_{55} + x_{5}x_{59} + x_{5}x_{61} + x_{5}x_{62} + x_{6}x_{8} + x_{6}x_{10} + x_{6}x_{12} + x_{6}x_{15} + x_{6}x_{18} + x_{6}x_{23} + x_{6}x_{24} + x_{6}x_{25} + x_{6}x_{27} + x_{6}x_{28} + x_{6}x_{29} + x_{6}x_{40} + x_{6}x_{43} + x_{6}x_{44} + x_{6}x_{45} + x_{6}x_{46} + x_{6}x_{49} + x_{6}x_{50} + x_{6}x_{53} + x_{6}x_{55} + x_{6}x_{56} + x_{6}x_{57} + x_{6}x_{58} + x_{6}x_{59} + x_{6}x_{63} + x_{6}x_{64} + x_{7}x_{11} + x_{7}x_{13} + x_{7}x_{15} + x_{7}x_{18} + x_{7}x_{19} + x_{7}x_{20} + x_{7}x_{21} + x_{7}x_{22} + x_{7}x_{26} + x_{7}x_{28} + x_{7}x_{29} + x_{7}x_{31} + x_{7}x_{33} + x_{7}x_{35} + x_{7}x_{38} + x_{7}x_{39} + x_{7}x_{43} + x_{7}x_{46} + x_{7}x_{47} + x_{7}x_{49} + x_{7}x_{51} + x_{7}x_{53} + x_{7}x_{55} + x_{7}x_{56} + x_{7}x_{58} + x_{7}x_{59} + x_{7}x_{60} + x_{7}x_{61} + x_{7}x_{62} + x_{7}x_{63} + x_{8}x_{10} + x_{8}x_{12} + x_{8}x_{13} + x_{8}x_{14} + x_{8}x_{19} + x_{8}x_{20} + x_{8}x_{21} + x_{8}x_{23} + x_{8}x_{25} + x_{8}x_{27} + x_{8}x_{31} + x_{8}x_{34} + x_{8}x_{37} + x_{8}x_{38} + x_{8}x_{40} + x_{8}x_{41} + x_{8}x_{42} + x_{8}x_{43} + x_{8}x_{44} + x_{8}x_{47} + x_{8}x_{50} + x_{8}x_{51} + x_{8}x_{53} + x_{8}x_{57} + x_{8}x_{62} + x_{9}x_{10} + x_{9}x_{11} + x_{9}x_{12} + x_{9}x_{13} + x_{9}x_{14} + x_{9}x_{15} + x_{9}x_{17} + x_{9}x_{18} + x_{9}x_{20} + x_{9}x_{24} + x_{9}x_{26} + x_{9}x_{28} + x_{9}x_{30} + x_{9}x_{31} + x_{9}x_{32} + x_{9}x_{36} + x_{9}x_{41} + x_{9}x_{43} + x_{9}x_{44} + x_{9}x_{45} + x_{9}x_{46} + x_{9}x_{48} + x_{9}x_{51} + x_{9}x_{54} + x_{9}x_{55} + x_{9}x_{56} + x_{9}x_{58} + x_{9}x_{59} + x_{9}x_{63} + x_{10}x_{12} + x_{10}x_{13} + x_{10}x_{15} + x_{10}x_{18} + x_{10}x_{19} + x_{10}x_{27} + x_{10}x_{31} + x_{10}x_{32} + x_{10}x_{36} + x_{10}x_{37} + x_{10}x_{38} + x_{10}x_{39} + x_{10}x_{40} + x_{10}x_{43} + x_{10}x_{44} + x_{10}x_{46} + x_{10}x_{48} + x_{10}x_{52} + x_{10}x_{54} + x_{10}x_{55} + x_{10}x_{59} + x_{10}x_{64} + x_{11}x_{13} + x_{11}x_{17} + x_{11}x_{18} + x_{11}x_{19} + x_{11}x_{20} + x_{11}x_{21} + x_{11}x_{22} + x_{11}x_{23} + x_{11}x_{24} + x_{11}x_{27} + x_{11}x_{30} + x_{11}x_{31} + x_{11}x_{32} + x_{11}x_{33} + x_{11}x_{36} + x_{11}x_{37} + x_{11}x_{39} + x_{11}x_{41} + x_{11}x_{45} + x_{11}x_{46} + x_{11}x_{47} + x_{11}x_{48} + x_{11}x_{49} + x_{11}x_{52} + x_{11}x_{58} + x_{11}x_{59} + x_{11}x_{61} + x_{12}x_{13} + x_{12}x_{15} + x_{12}x_{16} + x_{12}x_{17} + x_{12}x_{18} + x_{12}x_{20} + x_{12}x_{21} + x_{12}x_{22} + x_{12}x_{23} + x_{12}x_{25} + x_{12}x_{28} + x_{12}x_{29} + x_{12}x_{32} + x_{12}x_{33} + x_{12}x_{34} + x_{12}x_{35} + x_{12}x_{38} + x_{12}x_{42} + x_{12}x_{48} + x_{12}x_{50} + x_{12}x_{51} + x_{12}x_{55} + x_{12}x_{58} + x_{12}x_{60} + x_{12}x_{61} + x_{12}x_{62} + x_{12}x_{64} + x_{13}x_{16} + x_{13}x_{17} + x_{13}x_{19} + x_{13}x_{20} + x_{13}x_{22} + x_{13}x_{24} + x_{13}x_{26} + x_{13}x_{28} + x_{13}x_{30} + x_{13}x_{32} + x_{13}x_{33} + x_{13}x_{34} + x_{13}x_{35} + x_{13}x_{38} + x_{13}x_{39} + x_{13}x_{42} + x_{13}x_{43} + x_{13}x_{44} + x_{13}x_{47} + x_{13}x_{48} + x_{13}x_{50} + x_{13}x_{51} + x_{13}x_{52} + x_{13}x_{56} + x_{13}x_{57} + x_{13}x_{58} + x_{13}x_{63} + x_{14}x_{15} + x_{14}x_{17} + x_{14}x_{18} + x_{14}x_{20} + x_{14}x_{22} + x_{14}x_{23} + x_{14}x_{24} + x_{14}x_{25} + x_{14}x_{26} + x_{14}x_{27} + x_{14}x_{28} + x_{14}x_{29} + x_{14}x_{31} + x_{14}x_{33} + x_{14}x_{36} + x_{14}x_{37} + x_{14}x_{43} + x_{14}x_{45} + x_{14}x_{46} + x_{14}x_{47} + x_{14}x_{48} + x_{14}x_{49} + x_{14}x_{50} + x_{14}x_{51} + x_{14}x_{52} + x_{14}x_{54} + x_{14}x_{56} + x_{14}x_{57} + x_{14}x_{58} + x_{14}x_{60} + x_{14}x_{64} + x_{15}x_{18} + x_{15}x_{19} + x_{15}x_{21} + x_{15}x_{22} + x_{15}x_{26} + x_{15}x_{28} + x_{15}x_{35} + x_{15}x_{39} + x_{15}x_{41} + x_{15}x_{43} + x_{15}x_{44} + x_{15}x_{46} + x_{15}x_{47} + x_{15}x_{48} + x_{15}x_{49} + x_{15}x_{51} + x_{15}x_{54} + x_{15}x_{55} + x_{15}x_{58} + x_{15}x_{59} + x_{15}x_{60} + x_{15}x_{63} + x_{16}x_{17} + x_{16}x_{18} + x_{16}x_{19} + x_{16}x_{20} + x_{16}x_{22} + x_{16}x_{23} + x_{16}x_{24} + x_{16}x_{26} + x_{16}x_{28} + x_{16}x_{29} + x_{16}x_{31} + x_{16}x_{33} + x_{16}x_{34} + x_{16}x_{37} + x_{16}x_{38} + x_{16}x_{40} + x_{16}x_{44} + x_{16}x_{45} + x_{16}x_{46} + x_{16}x_{47} + x_{16}x_{48} + x_{16}x_{49} + x_{16}x_{50} + x_{16}x_{51} + x_{16}x_{52} + x_{16}x_{55} + x_{16}x_{57} + x_{16}x_{59} + x_{16}x_{60} + x_{16}x_{61} + x_{16}x_{62} + x_{16}x_{63} + x_{16}x_{64} + x_{17}x_{19} + x_{17}x_{20} + x_{17}x_{27} + x_{17}x_{28} + x_{17}x_{30} + x_{17}x_{32} + x_{17}x_{36} + x_{17}x_{40} + x_{17}x_{41} + x_{17}x_{43} + x_{17}x_{45} + x_{17}x_{46} + x_{17}x_{47} + x_{17}x_{48} + x_{17}x_{51} + x_{17}x_{58} + x_{17}x_{62} + x_{17}x_{63} + x_{17}x_{64} + x_{18}x_{19} + x_{18}x_{22} + x_{18}x_{23} + x_{18}x_{30} + x_{18}x_{32} + x_{18}x_{33} + x_{18}x_{34} + x_{18}x_{35} + x_{18}x_{40} + x_{18}x_{43} + x_{18}x_{44} + x_{18}x_{47} + x_{18}x_{48} + x_{18}x_{49} + x_{18}x_{54} + x_{18}x_{55} + x_{18}x_{56} + x_{18}x_{58} + x_{18}x_{59} + x_{18}x_{60} + x_{18}x_{61} + x_{18}x_{64} + x_{19}x_{22} + x_{19}x_{23} + x_{19}x_{26} + x_{19}x_{27} + x_{19}x_{29} + x_{19}x_{30} + x_{19}x_{32} + x_{19}x_{33} + x_{19}x_{34} + x_{19}x_{35} + x_{19}x_{37} + x_{19}x_{41} + x_{19}x_{42} + x_{19}x_{44} + x_{19}x_{46} + x_{19}x_{48} + x_{19}x_{51} + x_{19}x_{52} + x_{19}x_{53} + x_{19}x_{57} + x_{19}x_{59} + x_{19}x_{60} + x_{19}x_{63} + x_{20}x_{21} + x_{20}x_{23} + x_{20}x_{24} + x_{20}x_{30} + x_{20}x_{32} + x_{20}x_{34} + x_{20}x_{35} + x_{20}x_{36} + x_{20}x_{37} + x_{20}x_{38} + x_{20}x_{39} + x_{20}x_{41} + x_{20}x_{43} + x_{20}x_{44} + x_{20}x_{45} + x_{20}x_{47} + x_{20}x_{48} + x_{20}x_{49} + x_{20}x_{50} + x_{20}x_{52} + x_{20}x_{54} + x_{20}x_{55} + x_{20}x_{56} + x_{20}x_{57} + x_{20}x_{60} + x_{20}x_{61} + x_{21}x_{23} + x_{21}x_{24} + x_{21}x_{25} + x_{21}x_{26} + x_{21}x_{27} + x_{21}x_{30} + x_{21}x_{31} + x_{21}x_{33} + x_{21}x_{34} + x_{21}x_{35} + x_{21}x_{37} + x_{21}x_{39} + x_{21}x_{40} + x_{21}x_{41} + x_{21}x_{42} + x_{21}x_{44} + x_{21}x_{45} + x_{21}x_{46} + x_{21}x_{47} + x_{21}x_{51} + x_{21}x_{56} + x_{21}x_{57} + x_{21}x_{59} + x_{21}x_{60} + x_{21}x_{62} + x_{21}x_{63} + x_{22}x_{23} + x_{22}x_{24} + x_{22}x_{25} + x_{22}x_{26} + x_{22}x_{31} + x_{22}x_{32} + x_{22}x_{33} + x_{22}x_{37} + x_{22}x_{38} + x_{22}x_{39} + x_{22}x_{41} + x_{22}x_{44} + x_{22}x_{45} + x_{22}x_{48} + x_{22}x_{54} + x_{22}x_{57} + x_{22}x_{58} + x_{22}x_{59} + x_{22}x_{60} + x_{22}x_{62} + x_{22}x_{63} + x_{22}x_{64} + x_{23}x_{25} + x_{23}x_{28} + x_{23}x_{30} + x_{23}x_{32} + x_{23}x_{33} + x_{23}x_{34} + x_{23}x_{36} + x_{23}x_{38} + x_{23}x_{42} + x_{23}x_{43} + x_{23}x_{47} + x_{23}x_{48} + x_{23}x_{51} + x_{23}x_{52} + x_{23}x_{53} + x_{23}x_{54} + x_{23}x_{56} + x_{23}x_{58} + x_{23}x_{60} + x_{23}x_{61} + x_{24}x_{26} + x_{24}x_{27} + x_{24}x_{28} + x_{24}x_{31} + x_{24}x_{32} + x_{24}x_{33} + x_{24}x_{34} + x_{24}x_{37} + x_{24}x_{38} + x_{24}x_{39} + x_{24}x_{44} + x_{24}x_{45} + x_{24}x_{46} + x_{24}x_{49} + x_{24}x_{50} + x_{24}x_{51} + x_{24}x_{52} + x_{24}x_{53} + x_{24}x_{55} + x_{24}x_{57} + x_{24}x_{58} + x_{24}x_{59} + x_{24}x_{62} + x_{24}x_{64} + x_{25}x_{26} + x_{25}x_{27} + x_{25}x_{28} + x_{25}x_{29} + x_{25}x_{31} + x_{25}x_{33} + x_{25}x_{34} + x_{25}x_{36} + x_{25}x_{38} + x_{25}x_{42} + x_{25}x_{43} + x_{25}x_{45} + x_{25}x_{46} + x_{25}x_{47} + x_{25}x_{48} + x_{25}x_{52} + x_{25}x_{54} + x_{25}x_{59} + x_{25}x_{60} + x_{26}x_{28} + x_{26}x_{30} + x_{26}x_{32} + x_{26}x_{33} + x_{26}x_{35} + x_{26}x_{36} + x_{26}x_{38} + x_{26}x_{40} + x_{26}x_{41} + x_{26}x_{42} + x_{26}x_{44} + x_{26}x_{45} + x_{26}x_{46} + x_{26}x_{49} + x_{26}x_{52} + x_{26}x_{54} + x_{26}x_{56} + x_{26}x_{57} + x_{26}x_{58} + x_{26}x_{59} + x_{26}x_{60} + x_{26}x_{61} + x_{26}x_{62} + x_{26}x_{64} + x_{27}x_{31} + x_{27}x_{33} + x_{27}x_{34} + x_{27}x_{35} + x_{27}x_{37} + x_{27}x_{41} + x_{27}x_{42} + x_{27}x_{43} + x_{27}x_{44} + x_{27}x_{45} + x_{27}x_{46} + x_{27}x_{47} + x_{27}x_{48} + x_{27}x_{50} + x_{27}x_{52} + x_{27}x_{53} + x_{27}x_{55} + x_{27}x_{59} + x_{27}x_{60} + x_{27}x_{62} + x_{27}x_{63} + x_{28}x_{29} + x_{28}x_{33} + x_{28}x_{35} + x_{28}x_{36} + x_{28}x_{37} + x_{28}x_{39} + x_{28}x_{40} + x_{28}x_{42} + x_{28}x_{45} + x_{28}x_{49} + x_{28}x_{50} + x_{28}x_{52} + x_{28}x_{55} + x_{28}x_{56} + x_{28}x_{59} + x_{28}x_{64} + x_{29}x_{30} + x_{29}x_{31} + x_{29}x_{32} + x_{29}x_{33} + x_{29}x_{35} + x_{29}x_{37} + x_{29}x_{38} + x_{29}x_{40} + x_{29}x_{41} + x_{29}x_{42} + x_{29}x_{43} + x_{29}x_{47} + x_{29}x_{49} + x_{29}x_{51} + x_{29}x_{52} + x_{29}x_{53} + x_{29}x_{55} + x_{29}x_{60} + x_{29}x_{61} + x_{29}x_{64} + x_{30}x_{31} + x_{30}x_{32} + x_{30}x_{34} + x_{30}x_{35} + x_{30}x_{40} + x_{30}x_{42} + x_{30}x_{47} + x_{30}x_{48} + x_{30}x_{49} + x_{30}x_{50} + x_{30}x_{51} + x_{30}x_{52} + x_{30}x_{55} + x_{30}x_{57} + x_{30}x_{59} + x_{30}x_{63} + x_{30}x_{64} + x_{31}x_{33} + x_{31}x_{35} + x_{31}x_{37} + x_{31}x_{42} + x_{31}x_{44} + x_{31}x_{45} + x_{31}x_{47} + x_{31}x_{48} + x_{31}x_{50} + x_{31}x_{51} + x_{31}x_{52} + x_{31}x_{54} + x_{31}x_{55} + x_{31}x_{57} + x_{31}x_{59} + x_{31}x_{60} + x_{31}x_{61} + x_{31}x_{62} + x_{31}x_{63} + x_{31}x_{64} + x_{32}x_{33} + x_{32}x_{35} + x_{32}x_{37} + x_{32}x_{39} + x_{32}x_{42} + x_{32}x_{43} + x_{32}x_{45} + x_{32}x_{47} + x_{32}x_{48} + x_{32}x_{51} + x_{32}x_{52} + x_{32}x_{53} + x_{32}x_{54} + x_{32}x_{55} + x_{32}x_{56} + x_{32}x_{57} + x_{32}x_{58} + x_{32}x_{59} + x_{32}x_{60} + x_{32}x_{61} + x_{32}x_{62} + x_{33}x_{36} + x_{33}x_{39} + x_{33}x_{40} + x_{33}x_{42} + x_{33}x_{43} + x_{33}x_{46} + x_{33}x_{48} + x_{33}x_{49} + x_{33}x_{50} + x_{33}x_{52} + x_{33}x_{54} + x_{33}x_{58} + x_{33}x_{59} + x_{33}x_{61} + x_{33}x_{63} + x_{33}x_{64} + x_{34}x_{35} + x_{34}x_{36} + x_{34}x_{37} + x_{34}x_{40} + x_{34}x_{44} + x_{34}x_{45} + x_{34}x_{47} + x_{34}x_{48} + x_{34}x_{50} + x_{34}x_{51} + x_{34}x_{52} + x_{34}x_{54} + x_{34}x_{55} + x_{34}x_{56} + x_{34}x_{57} + x_{34}x_{58} + x_{34}x_{60} + x_{34}x_{61} + x_{34}x_{63} + x_{34}x_{64} + x_{35}x_{37} + x_{35}x_{44} + x_{35}x_{48} + x_{35}x_{51} + x_{35}x_{55} + x_{35}x_{58} + x_{35}x_{63} + x_{35}x_{64} + x_{36}x_{37} + x_{36}x_{41} + x_{36}x_{42} + x_{36}x_{44} + x_{36}x_{45} + x_{36}x_{47} + x_{36}x_{50} + x_{36}x_{52} + x_{36}x_{53} + x_{36}x_{55} + x_{36}x_{56} + x_{36}x_{58} + x_{36}x_{59} + x_{36}x_{60} + x_{36}x_{61} + x_{37}x_{38} + x_{37}x_{39} + x_{37}x_{41} + x_{37}x_{42} + x_{37}x_{44} + x_{37}x_{45} + x_{37}x_{47} + x_{37}x_{48} + x_{37}x_{51} + x_{37}x_{52} + x_{37}x_{53} + x_{37}x_{54} + x_{37}x_{55} + x_{37}x_{57} + x_{37}x_{60} + x_{38}x_{42} + x_{38}x_{43} + x_{38}x_{44} + x_{38}x_{47} + x_{38}x_{49} + x_{38}x_{50} + x_{38}x_{51} + x_{38}x_{54} + x_{38}x_{56} + x_{38}x_{57} + x_{38}x_{60} + x_{38}x_{61} + x_{38}x_{63} + x_{38}x_{64} + x_{39}x_{40} + x_{39}x_{42} + x_{39}x_{43} + x_{39}x_{44} + x_{39}x_{46} + x_{39}x_{47} + x_{39}x_{49} + x_{39}x_{50} + x_{39}x_{51} + x_{39}x_{54} + x_{39}x_{59} + x_{39}x_{60} + x_{39}x_{63} + x_{40}x_{41} + x_{40}x_{42} + x_{40}x_{43} + x_{40}x_{45} + x_{40}x_{47} + x_{40}x_{48} + x_{40}x_{52} + x_{40}x_{53} + x_{40}x_{55} + x_{40}x_{56} + x_{40}x_{59} + x_{40}x_{60} + x_{41}x_{42} + x_{41}x_{46} + x_{41}x_{47} + x_{41}x_{49} + x_{41}x_{53} + x_{41}x_{55} + x_{41}x_{58} + x_{41}x_{60} + x_{41}x_{63} + x_{42}x_{43} + x_{42}x_{45} + x_{42}x_{47} + x_{42}x_{52} + x_{42}x_{55} + x_{42}x_{56} + x_{42}x_{58} + x_{42}x_{60} + x_{42}x_{63} + x_{42}x_{64} + x_{43}x_{46} + x_{43}x_{47} + x_{43}x_{48} + x_{43}x_{49} + x_{43}x_{50} + x_{43}x_{52} + x_{43}x_{54} + x_{43}x_{55} + x_{43}x_{56} + x_{43}x_{57} + x_{43}x_{58} + x_{43}x_{59} + x_{43}x_{60} + x_{43}x_{61} + x_{43}x_{62} + x_{44}x_{45} + x_{44}x_{46} + x_{44}x_{47} + x_{44}x_{52} + x_{44}x_{61} + x_{44}x_{62} + x_{44}x_{63} + x_{44}x_{64} + x_{45}x_{46} + x_{45}x_{47} + x_{45}x_{48} + x_{45}x_{49} + x_{45}x_{52} + x_{45}x_{53} + x_{45}x_{55} + x_{45}x_{57} + x_{45}x_{59} + x_{45}x_{61} + x_{45}x_{63} + x_{45}x_{64} + x_{46}x_{53} + x_{46}x_{54} + x_{46}x_{55} + x_{46}x_{56} + x_{46}x_{60} + x_{46}x_{62} + x_{47}x_{48} + x_{47}x_{49} + x_{47}x_{50} + x_{47}x_{53} + x_{47}x_{54} + x_{47}x_{55} + x_{47}x_{59} + x_{47}x_{61} + x_{47}x_{62} + x_{48}x_{50} + x_{48}x_{51} + x_{48}x_{54} + x_{48}x_{55} + x_{48}x_{57} + x_{48}x_{60} + x_{48}x_{62} + x_{48}x_{63} + x_{49}x_{51} + x_{49}x_{52} + x_{49}x_{53} + x_{49}x_{54} + x_{49}x_{56} + x_{49}x_{58} + x_{49}x_{59} + x_{49}x_{60} + x_{50}x_{54} + x_{50}x_{57} + x_{50}x_{59} + x_{50}x_{60} + x_{50}x_{63} + x_{50}x_{64} + x_{51}x_{53} + x_{51}x_{56} + x_{51}x_{57} + x_{51}x_{63} + x_{51}x_{64} + x_{52}x_{54} + x_{52}x_{58} + x_{52}x_{62} + x_{52}x_{64} + x_{53}x_{58} + x_{53}x_{59} + x_{53}x_{60} + x_{53}x_{61} + x_{53}x_{63} + x_{54}x_{56} + x_{54}x_{58} + x_{54}x_{60} + x_{54}x_{62} + x_{54}x_{63} + x_{54}x_{64} + x_{55}x_{56} + x_{55}x_{63} + x_{55}x_{64} + x_{56}x_{58} + x_{56}x_{59} + x_{56}x_{61} + x_{56}x_{62} + x_{56}x_{63} + x_{57}x_{60} + x_{57}x_{61} + x_{57}x_{64} + x_{58}x_{60} + x_{58}x_{64} + x_{59}x_{62} + x_{60}x_{64} + x_{61}x_{62} + x_{61}x_{63} + x_{62}x_{63} + x_{2} + x_{3} + x_{4} + x_{9} + x_{10} + x_{11} + x_{12} + x_{16} + x_{17} + x_{18} + x_{19} + x_{20} + x_{23} + x_{26} + x_{27} + x_{31} + x_{33} + x_{35} + x_{36} + x_{41} + x_{44} + x_{46} + x_{51} + x_{53} + x_{54} + x_{55} + x_{56} + x_{58} + x_{59} + x_{60} + x_{62} + x_{63}$

\end{document}